\pdfoutput=1

\documentclass{article}

\usepackage{microtype}
\usepackage{graphicx}
\usepackage{subcaption}
\usepackage{booktabs}
\usepackage{array}  %
\usepackage{hyperref}

\usepackage[preprint]{icml2026}
\makeatletter
\renewcommand{\ICML@preprint}{\textit{Preprint.}}
\makeatother
\hypersetup{pdfsubject={}}

\usepackage{amsmath}
\usepackage{amssymb}
\usepackage{mathtools}
\usepackage{amsthm}
\usepackage[capitalize,noabbrev]{cleveref}

\graphicspath{{figures/}}

\newif\ifgatesixmarks
\gatesixmarkstrue

\theoremstyle{plain}
\newtheorem{theorem}{Theorem}[section]
\newtheorem{proposition}[theorem]{Proposition}
\newtheorem{lemma}[theorem]{Lemma}
\newtheorem{corollary}[theorem]{Corollary}
\theoremstyle{definition}
\newtheorem{definition}[theorem]{Definition}
\newtheorem{example}[theorem]{Example}
\theoremstyle{remark}
\newtheorem{remark}[theorem]{Remark}

\newcommand{\reals}{\mathbb{R}}
\newcommand{\complexes}{\mathbb{C}}
\newcommand{\eps}{\varepsilon}
\newcommand{\ahi}{a^{\rm hi}}
\newcommand{\alo}{a^{\rm lo}}
\newcommand{\acen}{a^{\rm c}}
\newcommand{\arad}{a^{\rm r}}
\newcommand{\ucen}{u^{\rm c}}
\newcommand{\urad}{u^{\rm r}}
\newcommand{\xcen}{x^{\rm c}}
\newcommand{\xrad}{x^{\rm r}}
\newcommand{\denh}{\mathrm{den}_{\rm h}}
\newcommand{\pih}{\pi_{\rm h}}
\newcommand{\pic}{\pi_{\rm c}}
\newcommand{\cfree}{c_{\rm free}}
\newcommand{\Gh}{\Gamma^{\rm h}}
\newcommand{\Gc}{\Gamma^{\rm c}}
\newcommand{\Reach}{\mathrm{Reach}}
\newcommand{\hullD}{\mathrm{hull}_{\mathcal D}}
\newcommand{\hullbox}{\mathrm{hull}_{\rm box}}
\newcommand{\hull}{\mathrm{hull}}
\newcommand{\disk}{\mathbb{D}}
\newcommand{\Binf}{\mathcal{B}_\infty}
\newcommand{\Bcup}{\mathcal{B}_\cup}
\newcommand{\gbox}{g_{\rm box}}
\newcommand{\tsat}{t_{\rm sat}}
\newcommand{\one}{\mathbf{1}}
\newcommand{\Rot}{\mathrm{Rot}}
\newcommand{\sgn}{\mathrm{sgn}}
\newcommand{\diag}{\mathrm{diag}}
\newcommand{\softplus}{\mathrm{softplus}}
\newcommand{\abs}[1]{\lvert#1\rvert}
\newcommand{\norm}[1]{\lVert#1\rVert}
\newcommand{\famname}[1]{\texttt{#1}}

\icmltitlerunning{Interval Bound Propagation over Unbounded Contexts}

\begin{document}

\twocolumn[
  \icmltitle{Interval Bound Propagation over Unbounded Contexts:\\
    When the Abstract Domain Matches the Recurrence}

  \begin{icmlauthorlist}
    \icmlauthor{Peng Xie}{tum}
    \icmlauthor{Amr Alanwar}{tum}
  \end{icmlauthorlist}
  \icmlaffiliation{tum}{TUM School of Computation, Information and Technology, Department of Computer Engineering,
    Technical University of Munich, 74076 Heilbronn, Germany}
  \icmlcorrespondingauthor{Peng Xie}{p.xie@tum.de}
  \icmlcorrespondingauthor{Amr Alanwar}{alanwar@tum.de}

  \icmlkeywords{interval bound propagation, verification, recurrent neural networks, state space models,
    abstract interpretation}

  \vskip 0.3in
]

\printAffiliationsAndNotice{}  %

\begin{abstract}
State-space models (SSMs) such as Mamba read contexts of any length with a fixed-size state, but sound verifiers for sequence
models unroll the input, at a cost that grows with its length. We ask when interval bound propagation (IBP), which streams at a
constant cost per token, stays informative over unbounded contexts. The answer depends on how the abstract domain interacts with
the state transition, not on the architecture family. For nonnegative diagonal transitions (Mamba, real-diagonal S4D, minGRU),
boxes are exact in time given the per-step coefficient boxes: the recurrence adds nothing to what IBP loses in computing those
boxes. For stacks of such layers we derive closed-form certificates whose cost does not depend on the context length.
One covers every number of repetitions of a perturbed context period, or every context over a token set (informative in our
tests mainly for a single repeated token); a tighter one covers only the infinite-context limit. On complex-diagonal LTI layers
boxes wrap and can diverge although every mode contracts, while per-mode disks lose at most a factor $\pi/2$ in circumradius. On
gated cells the interval map can expand while the cell contracts, and in float64 zonotopes reach farther the larger their budget.
Robust objectives computed on finite inputs are blind to the slow channels of Mamba and minGRU; on one task, small minGRU-type
models trained with the periodic certificate as the robust term from the start keep it within a few points of their bare-document
certificates at over 80\% clean accuracy on every seed, while Mamba falls below 78\% on two of three seeds. We test each statement
against pre-registered criteria and report every failure.
\end{abstract}

\section{Introduction}\label{sec:intro}

State-space models (SSMs) and linear recurrent networks stream contexts of any length with a constant-size state, at a constant cost
per token \citep{gu2022a,gu2023,orvieto2023,feng2024}. The context is also where an adversary can act: universal
adversarial triggers, many-shot jailbreaking and hidden-state poisoning of Mamba all work through it
\citep{wallace2019,anil2024,lemercier2026}. A certificate for such a model should hold however long the perturbed context is, at a
cost that does not grow with its length. Verifiers for recurrent and other sequence models address inputs of a given length
\citep{ko2019,du2021,ryou2021,shi2020,bonaert2021}, so their cost grows with it, or fix the horizon of their invariants
in advance \citep{jacoby2020}. We ask when interval bound propagation (IBP) stays informative over unbounded contexts.

IBP propagates a box through each operation by its natural interval extension \citep{moore1966,gowal2018,mirman2018}; it is
sound, streams at a constant cost per token, and underlies most certified training. Along time, interval arithmetic suffers from
the wrapping effect \citep{moore1966,neumaier1993}: a
rotated box needs a larger axis-aligned box around it, and the enlargement compounds from step to step, so boxes can grow while the
true dynamics contracts.

\begin{figure*}[t]
\centering
\includegraphics[width=\textwidth]{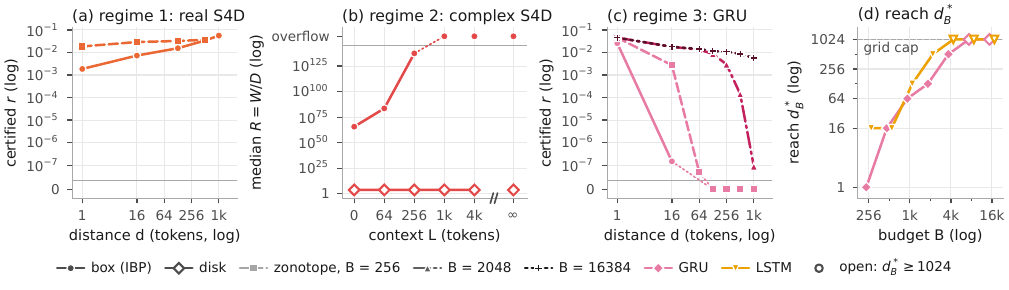}
\caption{The three regimes; medians over the documents. (a)~Regime~1, real-diagonal S4D-type model (seed~0, 200 documents):
certified radius $r(d)$ of one token at distance $d$ before the readout, for the box (plain IBP) and the zonotope with budget
$B=256$ (run up to $d=512$). (b)~Regime~2, the complex-diagonal control (two seeds, mean of their medians): the same weights,
documents and forward pass in real boxes and in complex disks; looseness $R$ at $\eps=10^{-8}$ against the context length $L$;
overflow strip: non-finite bounds. (c)~Regime~3, GRU
(seed~0): $r(d)$ as in (a), $r=0$ in the floor strip, for the box (50 documents) and the
zonotopes of the budget sweep (49 documents). (d)~Reach $d^*_B$ (\cref{sec:exp-gate5}) against $B$; open marker: end of the grid,
$d^*_B\ge1024$.}
\label{fig:regimes}
\end{figure*}

A recurrent layer updates its state as $h_t=T_th_{t-1}+u_t$. Over a long context IBP can lose precision in two ways: within a
step, when the box of one update is larger than the update's true image, and across steps, when the best box at step $t-1$,
mapped forward, is no longer the best box at step $t$, so that the loss compounds. We call a domain \emph{exact in time} when the
second loss never occurs (\cref{def:exact}). Whether it occurs depends on the domain and the transition together, which gives three
regimes (\cref{fig:regimes}). \emph{Nonnegative diagonal transitions} (Mamba, real-diagonal S4D, minGRU;
\citealp{gu2023,gu2022b,feng2024}): given the coefficient boxes of each step, boxes are exact in time (\cref{lem:boxes}), so the
looseness of IBP, large in our natural models, comes from the coefficient boxes, not the recurrence; that loss is local in the
first layer but not necessarily in deeper ones (\cref{app:scope}). In our two-layer models the looseness is flat in $L$ at small
$\eps$ over the measured lengths, which no theorem predicts.
\emph{Complex-diagonal layers} (complex S4D, S5, LRU; \citealp{gu2022b,smith2023,orvieto2023}) rotate their state: boxes wrap and
can diverge although every mode contracts, while per-mode disks do not wrap and lose at most a factor $\pi/2$ in circumradius
(\cref{lem:disks}). \emph{Gated cells} (LSTM, GRU; \citealp{hochreiter1997,cho2014}): the interval map can expand while the cell
contracts (\cref{prop:exact}); in our trained models boxes saturate after a few dozen tokens, and zonotopes with larger budgets
reach farther.

\paragraph{What is new.}
(1)~\emph{Certificates for unbounded perturbed contexts} of deep nonnegative diagonal stacks whose coefficient boxes the network
computes from its input (\cref{sec:theory-periodic}): a K-uniform hull that covers every number of copies of a perturbed context
period, none and infinitely many included, or every context over a token set (\cref{thm:hull}, \cref{cor:anyctx}), and a tighter box
that covers the infinite-context limit and no finite context (\cref{thm:inf}). Both cost a fixed number of scans over the period.
(2)~\emph{A finite-horizon blind spot, and a remedy} (\cref{sec:theory-horizon,sec:exp-train}): robust objectives computed on finite
inputs cannot see channels slower than their horizon when the injection carries the forgetting factor, as in Mamba and minGRU but not
in our S4D-type layer, whose injection $xB$ carries no forgetting factor (under a zero-order-hold discretisation it would). With the
K-uniform hull as the robust term from the start of training, small minGRU-type models on one task keep certificates for
every number of copies of a context period within a few points of their bare-document certificates at over 80\% clean accuracy on
every seed, with or without a time-constant cap (a 512-token cap makes the certificate close to a one-period one; after 4096 tokens
these models carry no detectable evidence of the document); for Mamba a gentler schedule alone keeps most certificates at
$L=\infty$, but with the hull term two of three seeds fall below 78\% clean accuracy; on a memory task only a cap at twice the memory
span kept the certificates and stayed within 5 points of plain IBP's memory (\cref{app:memory}).
(3)~\emph{Where each domain stops working} for the recurrent layers in use (\cref{sec:theory-exact}), with a contracting linear
recurrence whose IBP box diverges (Moore's example, scaled to contract) and a new contracting LSTM whose limiting
IBP box keeps a width of at least 0.94 for every $\eps>0$, and pre-registered tests of the three regimes on trained models
(\cref{sec:exp}).

\section{Setting}\label{sec:setting}

\paragraph{Models.}
A $Q$-layer model maps input vectors $e_t\in\reals^{d_m}$ to logits. Layer $\ell$ applies RMSNorm \citep{zhang2019}, a linear map,
a causal depthwise convolution of kernel size $k_c=4$ and SiLU \citep{elfwing2018} to its input $v^\ell_t$ ($v^1_t=e_t$), which
gives vectors $x_t,z_t$, and runs for every state channel $i$ the scan
\begin{equation}\label{eq:scan}
h_{t,i}=a_{t,i}\,h_{t-1,i}+u_{t,i},\qquad (a_t,u_t)=\Phi(x_t),
\end{equation}
with $a_{t,i}\in[0,1)$, from $h=0$; a token-wise readout of $(h_t,x_t,z_t)$, a residual connection, a final RMSNorm and a linear head
on the last token follow. The coefficients depend on the layer input only through the window $v^\ell_{t-k_c+1},\dots,v^\ell_t$,
never on $h$, so the state Jacobian $\diag(a_t)$ is nonnegative, diagonal and independent of the state: the layer is
\emph{nonnegative diagonal}. Our three instances (\cref{tab:layers}) are Mamba \citep{gu2023} and two layers built from its block
with another scan, an S4D-type layer with real diagonal decays \citep{gu2022b} and a minGRU-type layer \citep{feng2024}; the controls
are a complex-diagonal S4D, LSTM and GRU \citep{hochreiter1997,cho2014}.

\begin{table}[t]
\caption{Layers, transitions and regimes (\cref{sec:intro}). $\Delta_t=\softplus(s_t)$ with a gate logit $s_t$ computed from $x_t$
($\Delta$ is a parameter for S4D-type); $\alpha>0$; $\sigma$: logistic function; ``fixed'': input-independent. ``$u_t$ carries
$1-a_t$'': every injection is at most a bounded factor times $1-a_t$ (\cref{lem:dc}).}
\label{tab:layers}
\centering\scriptsize
\setlength{\tabcolsep}{3pt}
\begin{tabular}{@{}lllcc@{}}
\toprule
layer & decay $a_t$ & injection $u_t$ & $u_t$ carries $1-a_t$ & regime\\
\midrule
Mamba & $e^{-\alpha\Delta_t}$ & $\Delta_t\,x_tB_t$ & yes & 1\\
S4D-type (real) & $e^{-\alpha\Delta}$, fixed & $x_tB$ & no & 1\\
minGRU-type & $1-\sigma(s_t)$ & $\sigma(s_t)\,W_hx_t$ & yes & 1\\
S4D (complex) & $\lambda\in\complexes$, fixed & $\bar B\,x_t$ & -- & 2\\
LSTM, GRU & gates read $h$ & & -- & 3\\
\bottomrule
\end{tabular}
\end{table}

\paragraph{Box arithmetic.}
IBP propagates every token-wise map and the convolution by its natural interval extension, the composition of the exact-range
interval versions of the elementary operations \citep{moore1966,gowal2018}. Per position and channel this gives a decay box
$A_t=[\alo_t,\ahi_t]\subset[0,1]$ with centre $\acen_t$ and radius $\arad_t$, and an injection box with centre $\ucen_t$ and radius
$\urad_t$, both computed from the window alone. The scan maps a state box $[c-r,c+r]$ in centre-radius form to
\begin{equation}\label{eq:cr}
c'=\acen c+\ucen,\qquad r'=\ahi r+\arad\abs{c}+\urad. \tag{CR}
\end{equation}

\paragraph{Threat models and probes.}
Every position $t$, in the context and in the document, has its own box $E_t$, and an admissible perturbation picks any
$\tilde e_t\in E_t$ independently at every position: (i)~$\ell_\infty$ boxes $[e_t-\eps\one,e_t+\eps\one]$; (ii)~synonym
substitution \citep{jia2019}, the box hull of the embeddings of the word at $t$ and of its up to 8 counter-fitted neighbours within
distance 0.5 \citep{mrksic2016}, scaled about $e_t$ by $\eps\in[0,1]$, which covers every combination of substitutions;
(iii)~probes (\cref{sec:exp-setup}), $\ell_\infty$ boxes that measure how sets propagate and are not threat models.

\paragraph{Contexts and limit semantics.}
The document $x$ of $n$ tokens occupies positions $0,\dots,n-1$ and the context the negative positions. For a period $p$ of $P$
tokens over the vocabulary $V$, $p^Kx$ is $K$ copies of $p$ followed by $x$, a context of $L=KP$ tokens, and $F_K(\tilde e)$ are
the logits of the model run on $(\tilde e_t)_{-KP\le t<n}$ from $h=0$ with zero convolution padding; $K=0$ is the bare document.
Only the boxes are periodic: each copy can be perturbed differently.

\begin{definition}[Limit semantics]\label{def:limit}
An admissible perturbation of the infinite context is a sequence $\tilde e=(\tilde e_t)_{t<n}$ with $\tilde e_t\in E_t$ for all
$t<n$. Its infinite-context logits are the limit of the finite-context logits, $F_\infty(\tilde e):=\lim_{K\to\infty}F_K(\tilde e)$,
where $F_K$ uses the restriction of $\tilde e$ to $[-KP,n)$.
\end{definition}

\paragraph{Certificates and arithmetic.}
A logit box $[\ell^{\rm lo},\ell^{\rm hi}]$ certifies a label $y$ on a set of inputs if it contains the logits of every input in
the set, all its bounds are finite, and $\ell^{\rm lo}_y>\max_{k\ne y}\ell^{\rm hi}_k$. Every statement is in real arithmetic:
``sound'' and ``certificate'' refer to the real-arithmetic network with the code's float64 constants, archived embedding table and
library softplus (\cref{rem:float}), not to a deployed float32 network. We compute in float64 and re-verify the reported
certificates with outward rounding, except those listed in \cref{app:numerics}, under one measured assumption on the math library.

\paragraph{Looseness of IBP.}\label{sec:theory-loss}
For a class $y$ and a set $S$ of positions perturbed in $\ell_\infty$ boxes of radius $\eps$, let $b(\eps)$ be the IBP lower bound of
the margin $m=\ell_y-\max_{k\ne y}\ell_k$, $W(\eps):=b(0)-b(\eps)$ the margin it loses, $D:=\sum_{t\in S}\norm{\nabla_{e_t}m}_1$ the
first-order true sensitivity, and $R(\eps):=W(\eps)/(\eps D)$ the looseness of IBP. As $\eps\to0$, $R(\eps)\to G/D\ge1$, where $G$ is
the margin loss per unit $\eps$ that IBP would report were every operation linear: the path sum of the absolute local
derivatives (\cref{prop:loss}). This is the classical first-order behaviour of interval extensions
\citep{moore1966,neumaier1990}, and $1/R$ is a margin-level form of the propagation tightness of \citet{mao2024} over the context.

\section{Theory}\label{sec:theory}

\subsection{One step, and the forgetting factor}\label{sec:theory-step}

For a decay box $A=[\alo,\ahi]$ with $\alo\ge0$, an injection box $U$ and a state box $H=[c-r,c+r]$, the box \eqref{eq:cr} contains
$ah+u$ for all $a\in A$, $u\in U$, $h\in H$, and it is inclusion monotone in $(H,A,U)$. It is exact for a point decay or $c=0$, and
otherwise exceeds the exact range in radius by $\arad\min(r,\abs{c})$, a second-order term when the boxes have radius $O(\eps)$:
the scan is exact to first order (\cref{lem:cr}; proofs in \cref{app:proofs}). With the natural interval extension of the token-wise
maps it gives a sound, inclusion-monotone IBP pass of the stack (\cref{cor:stack}).
In Mamba and minGRU everything a channel receives per step carries the factor $1-\ahi$ that also sets its forgetting
(\cref{lem:dc}), unlike our S4D-type injection $xB$. Its \emph{DC level} $(\arad\abs{c}+\urad)/(1-\ahi)$, the radius at which its
box settles under repeated coefficient boxes, stays bounded as the channel slows down. We call a channel
\emph{slow} when its time constant, about $1/(1-\ahi)$ tokens, exceeds the training documents' length.

\subsection{Exactness in time is domain-relative}\label{sec:theory-exact}

\begin{definition}[Abstract domain; exactness in time]\label{def:exact}
Fix one state channel with state space $X=\reals^m$ ($X=\complexes\cong\reals^2$ for a complex mode) and the recursion
$h_t=T_th_{t-1}+u_t$, $h_0=0$, with $(T_t,u_t)$ ranging over a set $\mathcal C_t$ of admissible coefficients, independently of
$h_{t-1}$ and across $t$. The reachable sets are $\Reach_0=\{0\}$ and $\Reach_t=\{Th+u:h\in\Reach_{t-1},(T,u)\in\mathcal C_t\}$. An
\emph{abstract domain} $\mathcal D$ is a family of closed convex sets that contains the points, with a map $\hullD$ that sends every
bounded set to an element containing it: the least box for boxes, the minimal disk for disks $\disk(c,\rho)\subset\complexes$. The \emph{best step} is $\mathcal F_t(Z):=\hullD\{Th+u:h\in Z,(T,u)\in\mathcal C_t\}$, and
$\mathcal D$ is \emph{exact in time} for $(\mathcal C_t)_t$ if $\mathcal F_t\circ\dots\circ\mathcal F_1(\{0\})=\hullD(\Reach_t)$
for every $t$.
\end{definition}

Given its coefficient boxes, a channel of \eqref{eq:scan} has this form with $\mathcal C_t=A_t\times U_t$. This is the semantics of
IBP: the decay and the injection of one position, which share an input, and the coefficients of different positions range
independently. Exactness in time is therefore relative to the per-step coefficient boxes; what IBP loses in computing them is a
separate loss, local in the first layer but not necessarily in deeper ones (\cref{prop:loss}; (A8) in \cref{app:scope}). In abstract
interpretation \citep{cousot1977}, \cref{def:exact} asks the best abstract step to be complete \citep{giacobazzi2000} along the
iterates from the zero state.

\begin{lemma}[Boxes]\label{lem:boxes}
(i)~If $m=1$ and $\mathcal C_t=A_t\times U_t$ with intervals (decays of either sign), $\Reach_t$ is an interval, which the endpoint
scan with exact interval products returns exactly. (ii)~Let $\mathcal C_t=\{T_t\}\times U_t$ with point matrices $T_t$ and boxes
$U_t$ of radius vectors $\urad_t$, and $\Phi(t,s):=T_t\cdots T_{s+1}$. The natural interval extension of a step is the best box
step; composed from $\{0\}$ it has radius vector $\sum_{s\le t}\abs{T_t}\cdots\abs{T_{s+1}}\urad_s$, while $\hullbox(\Reach_t)$ has
radius vector $\sum_{s\le t}\abs{\Phi(t,s)}\urad_s$. So boxes are exact in time iff $\big(\abs{T_t}\cdots\abs{T_{s+1}}-
\abs{\Phi(t,s)}\big)\urad_s=0$ for every $t$ and all $s\le t$, as for diagonal or nonnegative $T_t$. (iii)~If every admissible $T$
is entrywise nonnegative and ranges over a box of matrices, and $U_t$ is a box, boxes are exact in time.
\end{lemma}

Part~(iii) is the comparison principle for monotone systems \citep{muller1927,smith1995,angeli2003}, and part~(ii) characterises the
wrapping effect \citep{moore1966,neumaier1993} for point transitions. No layer in this paper needs more than diagonal transitions; a
signed class for which (ii) holds, and when it is necessary, is in \cref{rem:provisos}.

\begin{lemma}[Disks: a loss that does not compound]\label{lem:disks}
Let a complex mode evolve by $h_t=\lambda_th_{t-1}+b_tx_t$ with input-independent $\lambda_t,b_t\in\complexes$, real inputs
$x_t\in[\xcen_t-\xrad_t,\xcen_t+\xrad_t]$ independent across $t$, and $h_0=0$; let $c_t=\lambda_tc_{t-1}+b_t\xcen_t$ and
$\rho_t=\abs{\lambda_t}\rho_{t-1}+\abs{b_t}\xrad_t$ from $(c_0,\rho_0)=(0,0)$. (i)~$\disk(c_t,\rho_t)$ is the composition of the
best disk steps and contains $\Reach_t$. (ii)~$\rho_t\le\frac\pi2\Omega_t$ for every $t$, where $\Omega_t:=\max_{h\in\Reach_t}
\abs{h-c_t}$ is the circumradius of the reachable set: the loss is bounded by a constant and does not compound. (iii)~If
$\lambda_t=\lambda=\abs{\lambda}e^{i\theta}$ for all $t$, then in real coordinates $\lambda$ acts by $\abs{\lambda}\Rot(\theta)$,
whose entrywise absolute value has spectral radius $\gbox=\abs{\lambda}(\abs{\cos\theta}+\abs{\sin\theta})\in[\abs{\lambda},
\sqrt2\abs{\lambda}]$; if $\theta\notin\frac\pi2\mathbb Z$, real boxes wrap for every nonzero injection radius.
\end{lemma}

Disks are in general not exact in time, and $\pi/2$ bounds only the circumradius, not the looseness of a disk certificate in a
readout direction; with several real inputs per mode (S5, LRU) only (ii) carries over. For set-valued eigenvalues, as in a
selective complex-diagonal layer under IBP, no constant bound holds: over the family $\lambda_t\in\{1,e^{i\vartheta}\}$ the limit
ratio of the composed disk radius to the hull radius is unbounded as $\vartheta\to0$, though bounded in $t$ for each $\vartheta$
(\cref{rem:pihalf}). Disk arithmetic is classical \citep{gargantini1971}.

\begin{proposition}[How exactness in time fails]\label{prop:exact}
(a$'$)~\emph{Nonnegative diagonal stacks.} Let the embeddings lie in a compact set, every input box contain the clean embedding
and have radius at most $C_0\eps$, and the interval extensions be exact. There are $\eps_0>0$ and $C'<\infty$, depending only on the model, the embedding set and $C_0$,
such that for $\eps\le\eps_0$ every box of the plain IBP pass over every input sequence of every length, periodic or not, and of the
cascade of \cref{thm:inf} for every period, has radius at most $C'\eps$; so $W(\eps)\le4C'\eps$ uniformly in the context length.
(b$'$)~\emph{Wrapping.} Let $h_t=Uh_{t-1}+b\,x_t$ on $\reals^2$ with $U=0.6\begin{psmallmatrix}1&1\\-1&1\end{psmallmatrix}$,
$b=\frac12(1,1)^\top$, $x_t\in[-\eps,\eps]$, $h_0=0$. Then $\norm{U}_2=0.6\sqrt2<1$ and every reachable state has
$\norm{h_t}_2\le\eps/(\sqrt2(1-0.6\sqrt2))$, while the IBP box of $h_t$ has radius vector $2.5\,\eps\,(1.2^t-1)\one$, which diverges;
the disk recursion of \cref{lem:disks} has radius at most $\eps/(\sqrt2(1-0.6\sqrt2))$. In a complex-diagonal LTI layer in real
boxes, a mode with $\gbox\ge1$ and a nonzero injection radius over the period has no finite invariant box, so the $L=\infty$ box
certificate is vacuous for every $\eps>0$, while in disks the layer satisfies (a$'$).
(b)~\emph{A contracting LSTM.} There is a one-layer LSTM with two hidden units, constant gates and input $x_t\in[-\eps,\eps]$ whose
products of state Jacobians contract geometrically and whose reachable set has diameter $O(\eps)$, while for every $\eps>0$ the IBP
box of $h$, iterated from the zero box, converges to a box of width at least $2\nu^*=0.9433$; at $\eps=0$ the box is $\{0\}$.
(b$''$)~\emph{Time to saturation.} In (b) the box reaches half that width after $\tsat(\eps)<\infty$ steps, $\Theta(\log(1/\eps))$
as $\eps\downarrow0$, while at every finite $t$ it tends to $\{0\}$ as $\eps\to0$ (\cref{prop:lstmfull}).
\end{proposition}

In (b) the gates are constant and the width comes from the sign-indefinite recurrent matrix; in general, interval arithmetic follows a
matrix $M_t\ge\abs{J_t}$ built from a gated cell's primitives, which can expand while the true Jacobians $J_t$ contract
(\cref{prop:loss}), so the line is not between SSMs and RNNs. Part~(a$'$) concerns $\eps\to0$ in exact arithmetic, with a possibly
astronomical constant (\cref{rem:notsay}).

\subsection{Certificates for unbounded periodic contexts}\label{sec:theory-periodic}

Take one channel $h\leftarrow ah+u$ with $a\in[\alo,\ahi]\subset[0,1)$ and $u\in[\ucen\pm\urad]$. By \eqref{eq:cr}, sound but not
the best box step (\cref{lem:cr}(b)), its box evolves by $c\leftarrow\acen c+\ucen$ and $r\leftarrow\ahi r+\arad\abs{c}+\urad$. Both maps contract, so the box converges to
$c^*=\ucen/(1-\acen)$ and the steady-state radius $r^*=(\arad\abs{c^*}+\urad)/(1-\ahi)$, and $[c^*-r^*,c^*+r^*]$ is invariant: once
the state is inside it, every admissible input keeps it inside. For a period of $P$ tokens with coefficient boxes $(A_j,U_j)$,
$j=1,\dots,P$, the same holds for the composition of the $P$ steps, the map $\Psi(c,r)=(\pic c+\cfree,\ \pih r+\beta(c))$: $\pic$
and $\pih$ are the products of the $\acen_j$ and of the $\ahi_j$, $\cfree$ is the centre after one period from centre $0$, and
$\beta(c)$ is the radius accumulated over the period from $(c,0)$, with $q_j(c)$ the radius injected at phase $j$
(\cref{app:proofs-periodic}). It contracts if the \emph{period leak} $\denh:=1-\pih$ is positive.

\begin{lemma}[Period map]\label{lem:period}
If $\denh>0$, then $\Psi$ is a contraction of $\reals\times[0,\infty)$ in a weighted maximum norm, with the unique fixed point
$c^*=\cfree/(1-\pic)$, $r^*=\beta(c^*)/\denh$; if every $\ahi_j<1$, the \emph{DC form} writes $r^*$ as a convex combination of the
DC levels $q_j(c^*)/(1-\ahi_j)$ (\cref{rem:dcform}). The box $H^*=[c^*-r^*,c^*+r^*]$ is robustly positively invariant over one
period \citep{blanchini1999}.
\end{lemma}

The DC form stays bounded for slow channels, where $\denh$ and $\beta(c^*)$ both vanish (\cref{lem:dc}). In a stack, the
fixed-point phase boxes of layer $\ell$ are the periodic input boxes of layer $\ell+1$, with a periodic convolution: the
\emph{periodic cascade} (\cref{def:cascade}). Running \eqref{eq:cr} over the document from the last phase box gives the logit box
$\Binf(p,x)$.

\begin{theorem}[Robust invariant box of a periodic context]\label{thm:inf}
Let the model be a $Q$-layer stack of nonnegative diagonal layers, evaluated in real arithmetic with the natural interval extension
and \eqref{eq:cr}. Fix $p\in V^P$, $x\in V^n$ and input boxes that are $P$-periodic on the context, and assume $\denh>0$ in every
channel of every layer of the periodic cascade. Then for every admissible perturbation $\tilde e$: (i)~the limit
$F_\infty(\tilde e)=\lim_{K\to\infty}F_K(\tilde e)$ exists; (ii)~$F_\infty(\tilde e)\in\Binf(p,x)$, and the limiting states lie in
the corresponding cascade boxes.
\end{theorem}

The certificate concerns the given period and covers no finite context (\cref{ex:finite}). To cover every number of copies at once,
note two facts. After $k$ copies from the zero state the centre is $(1-\pic^k)c^*$, which lies between $0$ and $c^*$; and because the
radius map $\beta$ is convex in the centre, the radius stays below $\bar r:=\max\{r^*,\beta(0)/\denh\}$. So the hull of the
trajectories from the zero state and from the fixed point contains the state after every number of copies:

\begin{theorem}[One certificate for every number of copies]\label{thm:hull}
Let $\Bcup$ be the logit box of the hull construction (\cref{def:hullcons}): in every layer, the convolution inputs that the
period reads before its start are replaced by their hull with $0$; every channel gets at each phase the hull of its centre
trajectories from $0$ and from $c^*$, widened by a radius built from $\bar r$ that bounds every radius trajectory; the document starts
from the hull of the last phase box with $0$, its convolution reading the hull of the period's tail with $0$. Under the hypotheses of
\cref{thm:inf}, with $P\ge(k_c-1)Q$ and $\denh>0$ in every channel of every layer of this construction: (i)~$F_K(\tilde e)\in\Bcup$
for every $K\in\{0,1,2,\dots\}$ and every admissible perturbation of $p^Kx$; (ii)~$\Binf\subset\Bcup$.
\end{theorem}

As $\eps\to0$ the hull still contains the clean trajectories from $0$ and from $c^*$, so a K-uniform certificate needs a correct
clean prediction on $p^Kx$ for all $K$, and slow channels with large DC levels make it long (\cref{rem:price}). With the box hull
$E_S$ of the embeddings of a token set $S$ at every context position and one hull per residue modulo $P$, \cref{thm:hull} certifies
every context over $S$ of every length, and the infinite one (\cref{cor:anyctx}), at the price of a hull that covers far more than
these contexts. Both cost a fixed number of scans over the period and one over the document.
One certificate covers a horizon of trailing context (\cref{rem:startboxM}); none covers every length for both labels (\cref{prop:M1}).

\subsection{Finite-horizon objectives are blind to slow channels}\label{sec:theory-horizon}

Robust training evaluates its robust term on finite inputs: $K$ copies of a perturbed period, or the bare document of standard IBP
training. In Mamba and minGRU everything a channel receives is multiplied by $1-a$ (\cref{lem:dc}). Slowing a channel ($a\to1$)
therefore shrinks what it accumulates over any fixed number of tokens towards zero, while over an
infinite context it still reaches its DC level, which stays bounded. A finite-horizon robust loss cannot see such channels: training
can lower it by slowing channels without shrinking the infinite-context box. A channel of our S4D-type layer cannot hide in this
way, because its injection $xB$ carries no factor $1-a$; a real-diagonal S4D layer's zero-order hold would give it
$(1-a)/\alpha$ (\cref{rem:otherterms}). Formally, fix one channel with $P$-periodic coefficient
boxes, $\ahi_j<1$ and $\denh\in(0,1]$; let $Q_j(c):=q_j(c)/(1-\ahi_j)$ be the DC level of phase $j$,
$\bar Q:=\max_j\max\{Q_j(0),Q_j(c^*)\}$, and $r_K$ the radius after $K$ copies from the zero state.

\begin{proposition}[Finite-horizon robust objectives are exploitable]\label{prop:horizon}
(a)~\emph{Slowing a channel hides it.} In a Mamba channel, shift the gate logits of all phases by $-\omega$, with their widths and
the boxes of $x$ and $B$ fixed. As $\omega\to\infty$, $\denh\to0$ while the DC levels stay bounded and converge, so $r_K\to0$ for
every fixed $K$, whereas $r^*$ converges to a finite limit, positive whenever for some phase the interval product
$[1,e^{\eta_j}]\cdot x_j\cdot B_j$ has positive width. The minGRU-type layer behaves the same way; slowing an S4D-type channel at
fixed $B$ does not hide it.
(b)~\emph{The finite-horizon radius.} $r_K\le\bar Q(1-\pih^K)\le\bar Q\min\{1,K\denh\}$, while
$r_\infty=r^*\ge\min_jQ_j(c^*)$ by the DC form.
(c)~\emph{The document horizon.} On documents of at most $n$ tokens, as $\prod_{t\le n}\alo_t\to1$ at bounded DC levels, the
channel's box tends to $\{0\}$ at every position, and a standard IBP term on such documents tends to its value with the channel
removed, while the channel's infinite-context box keeps its DC-level width; $\prod_t\ahi_t\to1$ alone is not enough
(\cref{lem:dochorizon}).
\end{proposition}

\begin{corollary}[Train at the fixed point]\label{cor:fixed}
If a robust term is computed from the $L=\infty$ fixed point in the DC form of \cref{lem:period}, its radius $r^*$ converges along
the move of \cref{prop:horizon}(a) to the limit given there, positive under the condition stated there, instead of to $0$, so slowing
a channel down cannot reduce the term to its zero-radius value.
\end{corollary}

Hence an objective meant to control the infinite-context box should be evaluated at $K=\infty$ or uniformly in $K$
(\cref{thm:hull}), not at a finite horizon, unless every time constant is capped well below that horizon
(\cref{prop:horizon}(b), (c)).

\section{Experiments}\label{sec:exp}

\subsection{Setup and pre-registration}\label{sec:exp-setup}

\paragraph{Data and models.}
We use word-level IMDB \citep{maas2011}, truncated at $T=512$ tokens, with licence text as neutral context. All recurrent models have
two layers of width $d_m=64$: Mamba, the S4D-type and minGRU-type layers, LSTM and GRU, and, as the control of regime~2, a
complex-diagonal LTI S4D in the same block; a CNN, a bag-of-words model and a byte-level Mamba and LSTM ($T=1024$) complete the
natural models; at word level they reach 85.8--90.8\% clean accuracy on 1000 test documents, the recurrent ones 81.1--86.7\%
after 4096 neutral tokens (\cref{app:cleanctx}). For IBP training under synonym substitution, a
first study has three seeds of standard IBP and of \emph{fixed-point fine-tuning} (with the robust term at the $L=\infty$ fixed
point, \cref{cor:fixed}) per word-level architecture, and a second study three per arm (\cref{sec:exp-train}). Training, code and
compute: \cref{app:compute}.

\paragraph{Probes.}
The \emph{context-length probe} (200 documents) reads the document after $L\in\{0,64,256,1024,4096\}$ tokens of neutral text, after
$K\in\{1,2,4\}$ copies of a 1024-token period, and at $L=\infty$, with $\ell_\infty$ boxes of radius $\eps\in\{10^{-10},\dots,
10^{-4}\}$ on every embedding, for the class $y_0$ predicted on the bare document, and reports the looseness $R$
(\cref{sec:theory-loss}; \cref{app:gate3}). The
\emph{distance probe} perturbs only the token at distance $d$ before the readout of [1024 neutral tokens $\mid$ document], in the
$\ell_\infty$ ball of its embedding. A method's \emph{certified radius} $r(d)$ is the largest $\eps\in[10^{-12},10^2]$ at which it
certifies the clean prediction ($r=0$ if none), a radius of the input ball, not of a state box. The methods are the box (plain IBP) and
zonotopes, exact to first order, whose in-flight values become fresh symbols of which a Girard-type reduction keeps a budget $B$
(\cref{app:gate5}). Certificates were checked against sampled inputs or at the box corner (\cref{app:audits}). Every experiment tests
one statement of \cref{sec:theory}, with criteria fixed in a specification before the production run and applied by a script
(\cref{tab:tests}); per-checkpoint 95\% intervals: \cref{app:cleanctx}.

\begin{table}[t]
\caption{Pre-registered tests: predictions and outcomes in brief. Record (\cref{app:prereg}; every criterion and outcome in
\cref{tab:prereg}): dates, criteria, amendments (three before the production runs), one corrected document count (no verdict
changed).}
\label{tab:tests}
\centering\scriptsize
\setlength{\tabcolsep}{2pt}
\renewcommand{\arraystretch}{1.05}
\begin{tabular}{@{}>{\raggedright\arraybackslash}p{0.25\columnwidth}>{\raggedright\arraybackslash}p{0.40\columnwidth}%
>{\raggedright\arraybackslash}p{0.316\columnwidth}@{}}
\toprule
test (section; seeds) & prediction & outcome\\
\midrule
context-length probe (\ref{sec:exp-gate3}; 3) & nonnegative diagonal: $R$ grows ${\le}10\times$ with $L$; gated: tenfold & held,
except one linearity statistic; gated: not met\\
complex-diagonal control (\ref{sec:exp-r2}; 2) & boxes wrap, disks do not & held\\
interval-map growth rates (\ref{sec:exp-r4}; 3) & median gated document: $\lambda_M>0>\lambda_J$ & held\\
certified radius against distance; budget sweep (\ref{sec:exp-gate5}; 3; 1) & gated: box collapses, zonotope keeps $r(1)/10$ to
$d=1024$; nonnegative: no collapse, zonotope near the box; reach grows with $B$ & held, except the zonotope (gated; S4D-type)\\
linear relaxations at finite context (\ref{sec:exp-gate5}; 3) & CROWN keeps $r(1)/10$ where boxes collapse or wrap; synonyms:
CROWN-IBP adds ${\le}3$ points; costs & held, except two cost predictions and, on Mamba, CROWN's completion (5 of 30)\\
IBP training under synonyms; slow-channel attribution (\ref{sec:exp-train}; 3) & first study: certified training transfers to
unbounded contexts; slow channels explain the failures & not met, stopped; attribution held but for two criteria on one seed each\\
token-set certificate (App.~\ref{app:anyctx}; 3) & over one token (S4D-type); fixed-point fine-tuning helps & the first held,
the second not\\
training remedy for the blind spot (\ref{sec:exp-train}; 3) & blind spot without remedy; cap + hull term K-uniform at
$\ge78\%$ clean; S4D-type keeps $L=\infty$ & held but on Mamba: most kept at $L=\infty$ unremedied, cap + hull ${<}78\%$
clean on two seeds\\
\bottomrule
\end{tabular}
\end{table}

\subsection{Regimes 1 and 2 over the context}\label{sec:exp-gate3}

\paragraph{Regime 1: the context-length probe.}
For all ten nonnegative diagonal models the median $R$ changed by factors of 0.69--1.06 from $L=256$ to $4096$ and 0.95--1.06 from
one to four periods: no geometric accumulation, on three seeds per architecture (\cref{tab:gate3}). The width
was linear in $\eps$ (median exponent in $\eps$: 0.997--1.000). On these natural models the looseness is large, from 508 to
$1.78\times10^5$ (1.65--2.01 after certified training, \cref{app:gate3-ctx}), but set by the network, not the context. At
$\eps=10^{-8}$ the $L=\infty$ certificate was finite on every document and certified 84--93.5\% of
them, but it covers no finite context; the K-uniform hull, which does, certified at most 0.5\% of the documents of any natural
model (the price of uniformity, \cref{rem:price}).

\paragraph{Regime 2: the complex-diagonal control.}\label{sec:exp-r2}
This tests \cref{lem:disks} and \cref{prop:exact}(b$'$) with the context-length probe on two trained complex-diagonal SSMs (clean
accuracy 88\% and 84.5\%), each propagated in real boxes and in complex disks with the same weights, documents and forward pass
(modes: \cref{app:gate3-r2}). The prediction held on both seeds (\cref{fig:regimes}b): in boxes $R$ was $4.8\times10^{65}$
already at $L=0$ (mean of the seeds' medians), the bound overflowed
float64 on almost every document from $L=1024$ on and the $L=\infty$ certificate was vacuous; in disks $R$ stayed between
$5.2\times10^3$ and $6.2\times10^3$ up to $L=\infty$, where 89\% and 93.5\% of the documents were certified (float64; disks,
zonotopes, gated box radii and auto\_LiRPA are not re-verified, \cref{app:numerics}). This shows a mismatch between transition
and domain, not that no box-like domain works (\cref{app:gate3}).

\subsection{Boxes against zonotopes and linear relaxations}\label{sec:exp-gate5}

\paragraph{Gated cells.}
The distance probe ran on the natural models (200 documents per seed; \cref{fig:regimes}a,~c;
\cref{app:gate5}); every outcome of seed~0 held on seeds 1 and 2. On the gated cells, within 16 steps the box's median certified radius
fell to at most $1.2\times10^{-4}$ of its adjacent-token value, and from $d=128$ on it certified none. The $B=2048$ zonotope kept
$r(1)/10$ up to $d=128$ (GRU) and $d=512$ (LSTM) on most documents, but to $d=1024$ on no GRU document and on 14--58.5\% of the
LSTM documents (80\% were predicted), although it certified a positive radius there on every document. In
the budget sweep (seed~0) the zonotope's \emph{reach} $d^*_B$, the largest grid distance up to which at least half of the documents
keep $r(1)/10$, grew with $B$ to the grid's end, 1024 (\cref{fig:regimes}d; \cref{app:sweep}).

\paragraph{Why gated boxes fail: interval-map growth rates.}\label{sec:exp-r4}
On three seeds the median gated document contracts along clean text (top Lyapunov exponent $\lambda_J<0$) while its interval
map $M_t\ge\abs{J_t}$ expands ($\lambda_M$ of 0.16--0.84 nats per token; \cref{tab:r4}): the mechanism of both collapses,
shown for one layer in Propositions~\ref{prop:exact}(b) and~\ref{prop:loss}(c1) (ours have two).

\paragraph{Nonnegative transitions, and what a relational domain changes.}
The box's certified radius did not collapse: from $d=1$ to $1024$ it grew by a median factor of 9.7--18 (Mamba) and 24--58
(S4D-type), as a distant token matters less. The zonotopes were exact to first order at every distance, at 34--110 box evaluations
per bound at $d=512$ (original runs); on seed~0 the $B=256$ zonotope's median certified radius was 110 times the box's at $d=1$
and 30 times at $d=512$ on Mamba, and 10 and 1.1 times on the S4D-type model, where it reached 0.99 times the box at $d=512$ on
only 28--52.5\% of the documents per seed. Its gain shrinks with $d$ as larger radii make its higher-order terms bind (a reading,
not a measured decomposition; on the S4D-type scan alone a reduced zonotope is never wider than boxes, \cref{lem:zono}). So, in
these runs, where boxes are exact in time the zonotope changed the per-step constant, not the behaviour in time, and did not
dominate the box; on gated cells the zonotope, and on the complex-diagonal control disks and CROWN, changed the behaviour in time
(limits: \cref{app:gate5}; an earlier study: \cref{app:charlm}).

\paragraph{Linear relaxations at finite context.}
On the distance probe and under synonym substitution we ran CROWN-IBP \citep{zhang2020} and CROWN \citep{zhang2018} in
auto\_LiRPA \citep{xu2020}, with two of its soundness gaps closed (\cref{app:lirpa}). On the gated cells (three seeds, 20
documents each) the collapse is one of interval intermediate bounds, which CROWN replaces by linear ones: the box and CROWN-IBP
certified none from $d=64$ on, while CROWN kept $r(1)/10$ at $d=64$ on every document, at 160--181\,s per bound (seed~0, four
threads). Under synonym substitution CROWN-IBP, whose
intermediate boxes are looser than ours (one linear relaxation, not the best), certified fewer documents than IBP on every seed of
the first study's standard-IBP Mamba, S4D- and minGRU-type models (9--48\% against 38.5--59\% at $L=0$) and, on the documents
classified correctly at both lengths, did not remove the minGRU-type decline from $L=0$ to $1024$.

\subsection{Training and token sets}\label{sec:exp-train}

\paragraph{IBP training under synonym substitution, and a remedy for the blind spot.}\label{sec:exp-anyctx}
In our first training study the standard-IBP Mamba and minGRU-type models lost their $L=\infty$ certificates, and a counterfactual
attributed most of their width failures to channels slower than the 512-token training documents, as \cref{prop:horizon} predicts
(\cref{app:gate2}). A second pre-registered study (\cref{tab:remedy}; three seeds; 95\% t-intervals) uses the
\emph{linear-ramp recipe} (after \citealp{jia2019}) and, from the first step, remedies: a robust term on half of each robust batch that sees the unbounded
context, the K-uniform hull of \cref{thm:hull} or the fixed point of \cref{cor:fixed}, and a cap $\tau_{\max}$ on the time
constants. With the hull term and $\tau_{\max}=512$ the minGRU-type model reached 83.7\%
[80.3, 87.1] clean accuracy and certificates for every number of copies of the context period on 72.2\% [69.6, 74.9] of the
documents (one copy 73.8\% [69.9, 77.7]; 1024 tokens 73.8\% [70.9, 76.6]), 2.8 [1.4, 4.2] points below $L=0$, at $-$0.3 [$-$3.9,
3.3] points of clean accuracy from the linear-ramp recipe; the hull term gave it without the cap as well. Mamba kept 78.5\% of its
certificates at $L=\infty$ with the linear-ramp recipe alone, so the collapse in our first study was partly its schedule; with the
hull term and $\tau_{\max}=512$ it kept 99.0\% for every number of copies, at $-$9.9 [$-$28.6, 8.8] points of clean accuracy (two of
three seeds below 78\%). On both models the fixed-point term kept ${\ge}80\%$ of the $L=0$ certificates at $L=\infty$, at $-$1.1
(minGRU-type) and $-$2.2 (Mamba) points of clean accuracy, and the alternative, $\tau_{\max}=128$
alone, kept at $L=\infty$ ${\ge}80\%$ of those whose clean prediction stays correct (width only) and ${\ge}70\%$ for every number
of copies, at $+$0.5 (minGRU-type) and $-$0.9 (Mamba), but on Mamba with no $L=\infty$ gain beyond
seed noise (\cref{app:remedy}). These are real-arithmetic certificates of two-layer width-64 models
on one task (re-verified with outward rounding, none changed; \cref{rem:float}); the certified capped models' K-uniform certificate
is close to a one-period one, and after $8\tau_{\max}$ neutral words they carry no detectable evidence of the document in 3 of 3
seeds (the cap 128 forgets it by 4096 words), unlike the linear-ramp minGRU-type model after 4096 words in 2 of 3 seeds
(\cref{app:remedy}). On a task that needs memory (\cref{app:memory}) only the cap at $\tau_{\max}=2048$ kept K-uniform
certificates and stayed within 5 points of plain IBP's memory, in 3 of 3 seeds of both models. Over one repeated neutral token, no
context of any length changed the cap + hull models' prediction, with the document in its substitution box, on 85.5\% [73.7, 97.3]
(Mamba) and 75.7\% [63.2, 88.2] (minGRU-type) of 200 documents; over 8 or 64 tokens, on none.

\section{Related work}\label{sec:related}

\paragraph{IBP and verifiers of sequence models.}
IBP \citep{gowal2018,mirman2018} and its refinements for certified training \citep{zhang2020,shi2021,muller2023,mao2023,depalma2024}
are cheap and differentiable; linear relaxations, zonotopes and polyhedra \citep{wong2018,zhang2018,gehr2018,singh2018,singh2019,
xu2020} and complete verifiers built on them \citep{xu2021,wang2021} are tighter and costlier (\cref{sec:exp-gate5}; on the first
study's models one CROWN-IBP bound at $L=256$ took 0.76--3.60\,s after a 7.9--42.4\,s graph build on four threads, and our
K-uniform hull 0.099--0.65\,s per document on one thread).
RNN verifiers unroll the sequence and relax every
step \citep{ko2019,du2021,ryou2021}, \citet{jacoby2020} replace unrolling for ReLU RNNs
by invariants up to a given horizon, and word-substitution certificates \citep{jia2019,huang2019} and transformer verifiers
\citep{shi2020,bonaert2021} also address inputs of a given length (other strands: \cref{app:related}).
Our certificates cover any number of copies of a perturbed period, or every context over a token set, at a cost independent of the
context length, and say nothing about other contexts.

\paragraph{Fixed points, invariance, and what is new.}
IBP certificates for deep equilibrium models compute an interval fixed point of an implicit layer \citep{wei2022,li2022}, and
contraction in non-Euclidean or mixed-monotone form makes implicit networks well posed and certifiable
\citep{jafarpour2021,jafarpour2022}; interval reachability certifies forward invariance with neural networks in the loop
\citep{everett2021,jafarpour2023,harapanahalli2023}. Our $L=\infty$ certificate is such an interval fixed point, but over time rather
than depth, for a cascade of layers; it exists for all weights, as the period leak is positive in exact arithmetic.
For monotone systems interval reasoning is tight \citep{muller1927,smith1995,angeli2003}, as in interval observers and
mixed-monotone reachability \citep{gouze2000,coogan2020}; interval arithmetic re-wraps boxes that the dynamics rotates
\citep{moore1966,neumaier1993}, which zonotope methods reduce \citep{kuhn1998,girard2005}; robust invariant sets
\citep{kolmanovsky1998,blanchini1999} and abstract domains for digital filters \citep{feret2004,monniaux2005,roux2012} are classical;
our $L=\infty$ box is a widening-free interval fixpoint \citep{su2005} whose existence condition, $\denh>0$, is per channel the
classical criterion $\rho(\abs{[A]})<1$ of interval iteration \citep{alefeld1983}, and the K-uniform hull an abstract acceleration of
the period \citep{gonnord2006,jeannet2014};
and \cref{def:exact} is the completeness of abstract interpretation \citep{cousot1977,giacobazzi2000} along the iterates. We use these
facts (\cref{lem:boxes}(iii), \cref{lem:disks,lem:period,lem:zono}, \cref{prop:loss}(a)) without claiming them. New are their
composition through deep networks that compute input-dependent coefficient boxes, the limit semantics with
the K-uniform hull and the token-set certificate, the domain-relative boundary for recurrent layers in use, and the
finite-horizon blind spot of robust objectives (\cref{app:related}).

\paragraph{State-space models.}
Control-theoretic analyses give the stability of Mamba's
recurrence and of selective SSMs \citep{halloran2025,zubic2026}, and robustness bounds for the selective scan whose terms grow
linearly with the sequence length \citep{cao2026}.
Closest to ours, \citet{thapliyal2026} certifies an LTI SSM classifier under $\ell_\infty$
perturbations of every embedding, with a penalty-enforced contraction condition and a steady-state bound on the state box, checked at
a finite length and without selection; we treat selective scans, with interval decays, and multi-layer stacks whose contraction is
architectural.

\section{Limitations and conclusion}\label{sec:conclusion}

\paragraph{Limitations.}
The probes are not threat models. Under synonym substitution the natural models certify nothing, so their true robust accuracy is
only known to lie between 0\% and about 50\% (attacks); in our first training study the standard-IBP models had low clean accuracy
(62--68\%), and only the S4D-type model kept its certificates at $L=\infty$; in the second (\cref{sec:exp-train}), the cap
with the hull term met its main pre-registered criterion only for the minGRU-type model, and these capped models carry no
detectable evidence of the document after 4096 tokens in 3 of 3 seeds. The looseness
$R$ of the natural models is $10^2$--$10^5$ at every length; only its shape in $L$ is claimed, and no theorem predicts the flat
profiles of our models. The periodic certificates cover one period at a time, the $L=\infty$ certificate covers no finite
context, and the token-set certificate certified a substantial share mainly over a single repeated token. Every model is small (two
layers, width 64) and not pretrained; the complex-diagonal control has two seeds, and the budget sweep one seed, 50 documents and
a grid ending at $d=1024$. Certificates are real-arithmetic statements,
not about float32 networks; all but those listed in \cref{app:numerics} were re-verified with outward
rounding under measured math-library accuracy. \emph{Open questions:} whether some fixed-size domain certifies gated transitions uniformly in
time, and which domain, if any, follows selective complex-diagonal transitions, where no constant disk bound holds.

\paragraph{Conclusion.}
Over unbounded contexts, the question for IBP is not SSM against RNN but whether the abstract domain can follow the transition. For
nonnegative diagonal stacks IBP certifies every number of copies of a perturbed period, or every context over a token set, at a cost
independent of their length, and since a finite-horizon robust term can be lowered by slowing channels down, a robust term for such
contexts should be evaluated at the fixed point or uniformly in the number of copies, unless every time constant is capped well
below the horizon.

\section*{Acknowledgements}
Experiments presented in this work were carried out using the CIT-TUM-HN cluster at TUM Campus Heilbronn.

\section*{Impact Statement}

This paper studies when interval bounds certify recurrent sequence models over long and unbounded contexts, with the aim of
making the verification of such models cheaper and better understood. Its certificates are real-arithmetic statements about small
models under box-shaped perturbations: they do not cover the deployed floating-point networks, inserted or deleted text, or
real-world contexts, and they should not be presented as guarantees in those settings; rounding alone can break real-arithmetic
certificates of neural networks \citep{jia2021fp,zombori2021}. We also show that a robust training
objective evaluated at a finite horizon can be reduced by slowing channels down without shrinking the infinite-context box, which
we see as a caution for practitioners of certified training. We do not see other societal consequences that need to be
highlighted here.

\bibliography{refs}

\newpage
\appendix
\onecolumn

\section*{Map of the appendix}
Each row names a claim of the main text and the parts of the appendix that state it in full, prove it, or report the experiment
behind it with its pre-registered criteria.

\begin{center}
\small
\renewcommand{\arraystretch}{1.12}
\begin{tabular}{@{}>{\raggedright\arraybackslash}p{0.16\textwidth}>{\raggedright\arraybackslash}p{0.41\textwidth}>{\raggedright\arraybackslash}p{0.37\textwidth}@{}}
\toprule
main text & claim & appendix\\
\midrule
\multicolumn{3}{@{}l}{\emph{Setting and theory}}\\
\cref{sec:setting}, \eqref{eq:cr} & the layers, the centre-radius scan, the limit semantics (\cref{def:limit}) & notation, layer
instances, two scans: \cref{app:numerics}; \cref{lem:cr,cor:stack}; the limit exists: \cref{lem:limit}\\
\cref{sec:setting} & real-arithmetic certificates, re-verified with outward rounding except where listed; the certified network &
\cref{rem:float} (with the trusted base and what was not re-verified); checks: \cref{app:audits}\\
\cref{sec:theory-loss} & the looseness $R$ and its first-order theory & \cref{app:proofs-loss}, \cref{prop:loss}\\
\cref{sec:theory-step} & \eqref{eq:cr} is sound and exact to first order; injections carry the forgetting factor; DC levels &
\cref{lem:cr,lem:dc}, \cref{rem:cancel}; the excess of \eqref{eq:cr}: \cref{app:numerics}, ``Two scans''\\
\cref{def:exact} & exactness in time, relative to the coefficient boxes and to the domain & \cref{app:proofs-exact},
\cref{rem:preservation}; assumption (A8) in \cref{app:scope}\\
\cref{lem:boxes} & boxes: interval channels, point transitions, nonnegative transitions & proof in \cref{app:proofs-exact}; the
sign class: \cref{rem:signclass}\\
\cref{lem:disks} & disks lose at most $\pi/2$ in circumradius & proof, \cref{lem:diskreadout}, \cref{rem:pihalf}\\
\cref{prop:exact} & how exactness fails: stacks, wrapping, a contracting LSTM, the time to saturation &
\cref{app:proofs-fail}: \cref{prop:exactac,prop:lstmfull}, constants in \cref{rem:bpp-constants}, \cref{rem:notsay}\\
\cref{lem:period} & the period map, its fixed point and the DC form & \cref{app:proofs-periodic}: \cref{rem:dcform}, proof,
\cref{rem:denzero,cor:monofp}\\
\cref{thm:inf} & the robust invariant box of a periodic context; it covers no finite context &
\cref{def:cascade,lem:implcascade,lem:asymp,lem:limit}, proof; \cref{ex:finite,rem:timescales}\\
\cref{thm:hull} & one certificate for every number of copies; every context over a token set &
\cref{def:hullcons,lem:hull1}, proof; \cref{rem:price}; \cref{thm:hullK}; \cref{cor:anyctx}; implementation: \cref{rem:impl}\\
\cref{prop:horizon,cor:fixed} & finite-horizon robust objectives are blind to slow channels; train at the fixed point &
\cref{app:proofs-horizon}: \cref{lem:dochorizon}, proofs, \cref{rem:otherterms,rem:injnorm}\\
\cref{sec:theory} & assumptions, scope, what is not proved & \cref{app:scope}: \cref{rem:notclaimed}, (A0)--(A8), the list of what
is not proved\\
\midrule
\multicolumn{3}{@{}l}{\emph{Experiments}}\\
\cref{sec:exp-setup}, \cref{tab:tests} & pre-registered predictions and outcomes & \cref{app:prereg}: \cref{tab:prereg}, criteria,
amendments, what the experiments do not show; code and compute: \cref{app:compute}; intervals: \cref{app:cleanctx}\\
\cref{sec:exp-gate3} & regime 1: the context-length probe & \cref{app:gate3-ctx}; clean accuracy over the context:
\cref{app:cleanctx}\\
\cref{sec:exp-r2} & regime 2: boxes against disks on the complex-diagonal control & \cref{app:gate3-r2}\\
\cref{sec:exp-gate5} & gated cells: certified radius against distance; zonotopes and their budget; reduced zonotopes on linear
recursions & \cref{app:gate5}, budget sweep: \cref{app:sweep}; \cref{lem:zono}\\
\cref{sec:exp-r4} & interval-map growth rates & \cref{app:gate3-r4}\\
\cref{sec:exp-gate5} & linear relaxations at finite context; cost & \cref{app:lirpa}; cost: \cref{app:wallclock}\\
\cref{sec:exp-train} & the first training study and the slow-channel attribution & \cref{app:gate2}; byte level: \cref{app:byte}\\
\cref{sec:exp-train} & the training remedy for the blind spot & \cref{app:remedy}\\
\cref{sec:exp-train} & certificates for every context over a token set & \cref{app:anyctx}; for the training remedy's models:
\cref{app:remedy}\\
\cref{sec:exp-train} & the memory task: a cap at twice the memory span keeps K-uniform certificates within 5 points of plain IBP's
memory; where the information goes; the zero-order-hold control & \cref{app:memory}; \cref{app:remedy}, ``Where the information goes'' and ``A
zero-order-hold S4D-type control''\\
\midrule
\cref{sec:related} & related work; further architectures & \cref{app:related}; an earlier study: \cref{app:charlm}\\
\cref{sec:conclusion} & limitations & \cref{app:scope}, \cref{app:prereg-notshown}, \cref{rem:float}\\
\bottomrule
\end{tabular}
\end{center}
\clearpage

\section{Proofs and further theory}\label{app:proofs}

This appendix proves every result of \cref{sec:setting,sec:theory} and states in full the results that the main text gives in
prose. \Cref{app:proofs-step} treats one step of the scan: the centre-radius step, the soundness of the stacked pass, and why every
injection of Mamba and minGRU carries the forgetting factor. \Cref{app:proofs-exact} proves the box and disk lemmas of
\cref{sec:theory-exact}: when composing best steps loses nothing, under which provisos the class $D_tN_tD_{t-1}$ is necessary, what
the factor $\pi/2$ bounds and what it does not, and the zonotope lemma. \Cref{app:proofs-fail} states \cref{prop:exact} in full and
proves it. \Cref{app:proofs-periodic} proves \cref{thm:inf,thm:hull}, the horizon-bounded variant and the token-set corollary.
\Cref{app:proofs-horizon} proves \cref{prop:horizon,cor:fixed}. \Cref{app:proofs-loss} states and proves \cref{prop:loss}, the
first-order theory of the looseness $R$ of \cref{sec:theory-loss}. \Cref{app:scope} collects the assumptions, what the results do
not claim, and what is not proved.

\subsection{One step, gate derivatives and the forgetting factor}\label{app:proofs-step}

\begin{lemma}[Centre-radius step]\label{lem:cr}
Let $A=[\alo,\ahi]$ with $0\le\alo\le\ahi$, $U=[\ucen-\urad,\ucen+\urad]$ and $H=[c-r,c+r]$ with $\urad,r\ge0$, and let $(c',r')$
be given by \eqref{eq:cr}. (a)~\emph{Soundness:} $ah+u\in[c'-r',c'+r']$ for all $a\in A$, $u\in U$, $h\in H$.
(b)~\emph{Excess:} the exact range $\{ah+u\}$ is an interval of radius $r'-\arad\min(r,\abs{c})$; so \eqref{eq:cr} is exact when
$\arad=0$ (a point decay, as in the S4D-type layer) or $c=0$. (c)~\emph{Inclusion monotonicity:} if $H'\subset H$, $A'\subset A$ and
$U'\subset U$, the box that \eqref{eq:cr} produces from $(H',A',U')$ lies in the one it produces from $(H,A,U)$.
(d)~\emph{Continuity:} $c'$ and $r'$ are continuous functions of $(c,r,\alo,\ahi,\ucen,\urad)$.
\end{lemma}

\begin{proof}
(a)~For $a\in A$, $u\in U$, $h\in H$, $ah+u-c'=(a-\acen)\,c+a\,(h-c)+(u-\ucen)$, and $\abs{a-\acen}\le\arad$, $0\le a\le\ahi$,
$\abs{h-c}\le r$, $\abs{u-\ucen}\le\urad$. No case split is needed: the sign of $c$, and whether $H$ contains $0$, play no role.

(b)~Because $a\ge0$, the exact range of $ah$ is $[\alo(c-r),\ahi(c+r)]$ if $c\ge r$, its mirror image if $c\le-r$, and
$[\ahi(c-r),\ahi(c+r)]$ if $\abs{c}<r$. Its radius is $\arad\abs{c}+\acen r$ in the first two cases and $\ahi r$ in the third.
Since $\ahi-\acen=\arad$, this equals $\ahi r+\arad\abs{c}-\arad\min(r,\abs{c})$ in every case. Adding $U$ adds $\urad$ to the
radius.

(c)~Write the upper end of the \eqref{eq:cr} box in terms of $lo=c-r$, $hi=c+r$ and $u^{\rm hi}=\ucen+\urad$:
\[
c'+r'=\begin{cases}
\ahi\,hi+u^{\rm hi}, & lo+hi\ge0,\\
\acen\,hi-\arad\,lo+u^{\rm hi}=\alo c+\ahi r+u^{\rm hi}, & lo+hi<0.
\end{cases}
\]
The two expressions agree on $lo+hi=0$, so the upper end is continuous in $(lo,hi)$. On each piece it is non-decreasing in $hi$
and non-increasing in $lo$ (coefficients $\ahi,\acen\ge0$ and $-\arad\le0$). Shrinking $H$ means raising $lo$ and lowering $hi$;
moving $hi$ first and $lo$ second keeps $lo\le hi$ along the way, so the upper end cannot rise. In the coefficients, on the first
piece it is non-decreasing in $\ahi$ (there $hi\ge0$) and in $u^{\rm hi}$ and does not depend on $\alo$. On the second piece it is
non-increasing in $\alo$ (there $c<0$) and non-decreasing in $\ahi$ and $u^{\rm hi}$. Shrinking $A$ and $U$ (raising $\alo$,
lowering $\ahi$ and $u^{\rm hi}$) therefore cannot raise the upper end either. The lower end $c'-r'$ is minus the upper end of the
mirrored problem $(h,u)\mapsto(-h,-u)$, which maps $H,U$ to $-H,-U$ and leaves $A$ unchanged, so it cannot fall. Applying the two
shrinkings one after the other gives the claim.

(d)~Evident from \eqref{eq:cr}.
\end{proof}

The endpoint scan, which bounds $ah$ by its exact range, is therefore tighter than \eqref{eq:cr} by $\arad\min(r,\abs{c})$ per step.
This excess is of second order when the boxes are $O(\eps)$, so the two scans agree to first order (\cref{prop:loss}(a)):
\eqref{eq:cr} is exact to first order.

\begin{corollary}[Centre-radius IBP of a stack]\label{cor:stack}
Propagate every token-wise map and the convolution by its natural interval extension and every scan by \eqref{eq:cr}, layer by
layer, from the zero state box and zero convolution padding. Then (i)~for every admissible perturbation, every intermediate value
of the model lies in its box (soundness); (ii)~shrinking any input box, or the start box, weakly shrinks every box computed from
it (inclusion monotonicity); (iii)~every box endpoint is a continuous function of the endpoints of the input boxes and of the
start box.
\end{corollary}

\begin{proof}
An exact-range interval operation returns the range of a continuous function over a box. The range contains the true value,
shrinks with the box, and has endpoints that depend continuously on the box endpoints, because the minimum and the maximum of a
continuous function over $\prod_i[lo_i,hi_i]$ are continuous in $(lo,hi)$ (uniform continuity on compact sets). \Cref{lem:cr}
gives the same three properties for the scan step. All three are preserved under composition. Induction over positions and layers.
\end{proof}

The same argument gives (i)--(iii) for any composition of exact-range primitives, for instance the endpoint scan, the LSTM and GRU
cells of our models and the disk scan of \cref{lem:disks}; \cref{app:proofs-loss} uses it in this form.

\begin{lemma}[Gate derivatives]\label{lem:gate}
For $\alpha>0$ let $\bar a(s)=\exp(-\alpha\softplus(s))=(1+e^s)^{-\alpha}$. For $k\in\{1,2,3\}$ and every $s\in\reals$,
\[
\Big|\frac{d^k\bar a}{ds^k}(s)\Big|<1-\bar a(s),
\]
and $\sup_s\abs{\bar a^{(k)}(s)}/(1-\bar a(s))=1$ for every $\alpha>0$; the supremum is approached as $s\to-\infty$ and is not
attained.
\end{lemma}

Only $k=1$ is used (\cref{lem:dc,prop:horizon}). A numerical scan over $s\in[-40,40]$ and $\alpha\in[10^{-3},10^{3}]$ found the
supremum $1.000$ for $k=1,2,3$.

\begin{proof}
\emph{Step 1 (a Gamma mixture).} Let $\xi\sim\mathrm{Gamma}(\alpha,1)$. Its Laplace transform is $\mathbb E[e^{-z\xi}]=(1+z)^{-\alpha}$
for $z>-1$, so with $Y:=e^s\xi>0$,
\[
\bar a(s)=\mathbb E\big[e^{-Y}\big],\qquad 1-\bar a(s)=\mathbb E\big[1-e^{-Y}\big].
\]

\emph{Step 2 (derivatives).} For fixed $g>0$ put $y=e^sg$. On functions of $y$, $\partial_s=y\,\partial_y$, and
$(y\partial_y)^k=\sum_{m=1}^kS(k,m)\,y^m\partial_y^m$ with the Stirling numbers of the second kind $S(k,m)$. Hence
\[
\partial_s^k\,e^{-e^sg}=T_k(-y)\,e^{-y},\qquad T_k(y):=\sum_{m=1}^kS(k,m)\,y^m,
\]
the Touchard polynomials: $T_1(-y)=-y$, $T_2(-y)=y^2-y$, $T_3(-y)=-y^3+3y^2-y$. Since $\abs{T_k(-y)}e^{-y}$ is bounded on $y\ge0$,
dominated convergence allows differentiation under the expectation: $\bar a^{(k)}(s)=\mathbb E\big[T_k(-Y)\,e^{-Y}\big]$.

\emph{Step 3 (a pointwise inequality).} We show $\abs{T_k(-y)}<e^y-1$ for all $y>0$ and $k=1,2,3$. For $k=1$: $y<e^y-1$. For $k=2$:
if $0<y\le1$, $\abs{y^2-y}=y-y^2<y<e^y-1$; if $y>1$, $y^2-y<y^2\le e^y-1$, because $f(y)=e^y-1-y^2$ has $f(0)=0$ and
$f'(y)=e^y-2y\ge2-2\ln2>0$. For $k=3$, upper side: $e^y-1\ge y+\frac{y^2}{2}+\frac{y^3}{6}$ for $y\ge0$, and
$y+\frac{y^2}{2}+\frac{y^3}{6}-(-y^3+3y^2-y)=y\big(\frac76y^2-\frac52y+2\big)$, whose quadratic factor has discriminant
$\frac{25}{4}-\frac{28}{3}=-\frac{37}{12}<0$ and is therefore positive. For $k=3$, lower side: $g(y):=e^y-1-(y^3-3y^2+y)$ has
$g(0)=g'(0)=0$ and $g''(y)=e^y-6y+6$; the minimum of $g''$ over $y\ge0$ is attained at $y=\ln6$ and equals $6(2-\ln6)>0$, since
$e^2>6$. So $g'$ increases from $0$, $g$ increases from $0$, and $g(y)>0$ for $y>0$.

\emph{Step 4 (the bound).} By Steps 2 and 3, $\abs{\bar a^{(k)}(s)}\le\mathbb E\big[\abs{T_k(-Y)}e^{-Y}\big]<
\mathbb E\big[(e^Y-1)e^{-Y}\big]=1-\bar a(s)$ for every $\alpha>0$; the inequality is strict because $Y>0$ almost surely.

\emph{Step 5 (sharpness).} Taking the expectation of Step 2 term by term with $\mathbb E[\xi^me^{-z\xi}]=(\alpha)_m(1+z)^{-\alpha-m}$
(rising factorial $(\alpha)_m$) and $\sigma:=e^s/(1+e^s)$ gives
\[
\bar a^{(k)}(s)=\bar a(s)\sum_{m=1}^k(-1)^mS(k,m)\,(\alpha)_m\,\sigma^m .
\]
As $s\to-\infty$: $\sigma\to0$, $\bar a\to1$, $\bar a^{(k)}=-\alpha\sigma+O(\sigma^2)$ and $1-\bar a=1-(1-\sigma)^\alpha=\alpha\sigma
+O(\sigma^2)$. The ratio tends to $1$.
\end{proof}

\begin{remark}[Higher orders]\label{rem:higher}
The same argument gives $\abs{\bar a^{(k)}}\le c_k\,(1-\bar a)$ for every $k$ and every $\alpha>0$, with
$c_k:=\sup_{y>0}\abs{T_k(-y)}/(e^y-1)$. The constant $c_k$ is also the supremum over $(\alpha,s)$: take $e^s=y_0/\alpha$, so that
$Y=y_0\xi/\alpha\to y_0$ in probability as $\alpha\to\infty$, and pass to the limit in both expectations by bounded convergence.
Computed numerically, $c_4=1.207$, $c_5=3.244$, $c_6=10.4$ and $c_7=46.3$, and $\abs{T_4(-y)}/(e^y-1)$ exceeds $1$ for some $y>0$;
the numerical scan limited to $\alpha\le10^3$ found at most $1.20$ for $k=4$, consistent with a supremum approached only as
$\alpha\to\infty$. Control by $1-\bar a$ thus holds at every order, with constants that grow fast. For the step itself,
$\abs{\Delta^{(k)}}\le\Delta$ was found numerically for $k\le6$.
\end{remark}

\begin{lemma}[Injections carry the forgetting factor]\label{lem:dc}
Let a gate logit range over $[s^{\rm lo},s^{\rm hi}]$, $\eta:=s^{\rm hi}-s^{\rm lo}$. (a)~The decay box
$[\alo,\ahi]=[\bar a(s^{\rm hi}),\bar a(s^{\rm lo})]$ satisfies $1-\alo\le e^\eta(1-\ahi)$, equivalently
$(\ahi-\alo)/(1-\ahi)\le e^\eta-1$, so $\arad\le\frac12(e^\eta-1)(1-\ahi)$.
(b)~\emph{Mamba.} Let $[\Delta^{\rm lo},\Delta^{\rm hi}]=[\softplus(s^{\rm lo}),\softplus(s^{\rm hi})]$, let $U$ be the
interval-product box of $\Delta xB$, and let $M$ be the product of the largest magnitudes in the boxes of $x$ and of $B$. Every element
of $U$, hence $\urad$ and $\abs{\ucen}$, is at most $e^\eta\varphi(\Delta^{\rm lo})M(1-\ahi)$ in magnitude, where
$\varphi(\Delta):=\Delta/(1-e^{-\alpha\Delta})$ satisfies $\Delta^{\rm lo}<\varphi(\Delta^{\rm lo})<\Delta^{\rm lo}+1/\alpha$.
Consequently the DC-normalised radius injection of one step satisfies
\[
\frac{\arad\abs{c}+\urad}{1-\ahi}\le\tfrac12(e^\eta-1)\abs{c}+e^\eta\varphi(\Delta^{\rm lo})M,
\]
a bound that stays finite as the channel slows down ($\Delta^{\rm lo}\to0$, $\varphi(\Delta^{\rm lo})\to1/\alpha$).
(c)~\emph{minGRU-type.} $\bar a=1-\sigma(s)$ is the case $\alpha=1$ of \cref{lem:gate}, so (a) holds verbatim. With $u=zg$,
$z\in[1-\ahi,1-\alo]$ and $M_g$ the largest magnitude in the box of $g=W_hx$, $\abs{u}\le(1-\alo)M_g\le e^\eta(1-\ahi)M_g$ and
$(\arad\abs{c}+\urad)/(1-\ahi)\le\frac12(e^\eta-1)\abs{c}+e^\eta M_g$. (d)~\emph{S4D-type.} The decay is a point ($\arad=0$,
$\eta=0$) and $u=xB$ with a free parameter $B$: the factor $1-\bar a$ is not part of the parametrisation.
\end{lemma}

\begin{proof}
(a)~$g(s):=\ln(1-\bar a(s))$ has $g'(s)=-\bar a'(s)/(1-\bar a(s))\in(0,1]$ by \cref{lem:gate} ($k=1$) and $\bar a'<0$; integrate
from $s^{\rm lo}$ to $s^{\rm hi}$. (b)~The endpoints of an interval product are products of endpoints, so every element of $U$ has
magnitude at most $\Delta^{\rm hi}M$. Softplus is log-1-Lipschitz: $(\ln\softplus)'(s)=\sigma(s)/\softplus(s)\in(0,1)$, because
$\ln(1+v)>v/(1+v)$ for $v=e^s>0$. Hence $\Delta^{\rm hi}\le e^\eta\Delta^{\rm lo}$, and $\Delta^{\rm lo}=\varphi(\Delta^{\rm lo})
(1-e^{-\alpha\Delta^{\rm lo}})=\varphi(\Delta^{\rm lo})(1-\ahi)$. For the bounds on $\varphi$, put $z=\alpha\Delta>0$:
$\varphi(\Delta)>\Delta$ because $1-e^{-z}<1$, and $\varphi(\Delta)<\Delta+1/\alpha$ is equivalent to $z<(z+1)(1-e^{-z})$, that is to
$e^z>1+z$; moreover $\varphi(\Delta)\to1/\alpha$ as $\Delta\to0$. Add (a) for $\arad$. (c)~The same argument with (a) at $\alpha=1$.
\end{proof}

\begin{remark}[The cancellation behind the slow channels]\label{rem:cancel}
\Cref{lem:dc} is the box form of a cancellation: everything a slow channel receives per step is scaled by the factor $1-\bar a$ that
also sets its forgetting. This keeps the fixed point of \cref{lem:period} finite and well scaled for slow channels (its DC form), and
it is what makes finite-horizon robust objectives exploitable (\cref{prop:horizon}).
\end{remark}

\subsection{Exactness in time: domains, boxes, disks and zonotopes}\label{app:proofs-exact}

\Cref{def:exact} fixes one channel and its sets $\mathcal C_t$ of admissible coefficients, which range independently of the state and
independently across $t$; for a scan of \cref{sec:setting} given its coefficient boxes, $\mathcal C_t=A_t\times U_t$, the semantics of
IBP. Exactness in time is therefore relative to the per-step coefficient boxes: the correlation between the decay and the injection
of one position, which share an input, and between the coefficients of different positions is lost before the definition applies
(\cref{prop:loss}(b2); the assumptions in \cref{app:scope}).
That state independence, which lets the natural interval extension of
a step coincide with the best step, is discussed in \cref{rem:archs}.
Where a domain is exact in time, the recurrence therefore adds nothing to what IBP loses in computing the coefficient boxes; that loss
is local in the first layer but not necessarily in deeper ones ((A8) in \cref{app:scope}; \cref{rem:gate3reading}, item 6).

\begin{remark}[Hull maps, preservation, and when composing best steps loses nothing]\label{rem:preservation}
\emph{Hull maps.} For boxes the hull map of \cref{def:exact} is the box hull, the least enclosing box; boxes are closed under
intersection, so $\hullbox$ is monotone and idempotent. For disks $\disk(c,\rho)\subset\complexes$ it is the minimal enclosing disk,
which is unique and least in radius but not least for inclusion, so the disk hull is not monotone: $\{-1,1\}$ has the minimal disk
$\disk(0,1)$, while $\{-1,1,10i\}$ has the minimal disk $\disk(99i/20,101/20)$, which does not contain $-i$.

\emph{Preservation.} The transition \emph{preserves} $\mathcal D$ if every admissible $T$ maps every element of $\mathcal D$ onto an
element of $\mathcal D$, and $\mathcal D$ is closed under Minkowski sums.

\emph{Point transitions.} Exactness in time is a property of the pair (recursion, domain), not of the architecture. Suppose the
transitions are points that preserve $\mathcal D$, and that $\hullD(Z+S)=Z+\hullD(S)$ for every $Z\in\mathcal D$ and every bounded
$S$. This holds for boxes, whose coordinate ranges add, and for disks, since $\disk(c,\rho)+S\subset\disk(z,R)$ iff
$S\subset\disk(z-c,R-\rho)$. Then $Z_t=T_tZ_{t-1}+\hullD(U_t)$ by induction, so the composed best steps equal the reachable set of the
recursion whose injection sets are replaced by their $\mathcal D$-hulls. The hypothesis is needed: \cref{def:exact} allows any
enclosing map, and with the interval hull doubled about its midpoint, $T_t=1$ and $U_1=U_2=[-1,1]$, the composed steps give $[-6,6]$
while the hulled-injection recursion gives $[-4,4]$. Boxes lose nothing when the injections are boxes (\cref{lem:boxes}(ii), with
$\abs{\Phi(t,s)}=\abs{T_t}\cdots\abs{T_{s+1}}$ on the injection radii). Disks lose the enclosure of each injection segment, and that
loss does not compound (\cref{lem:disks}(ii)); disks are therefore in general not exact in time (an instance follows the proof of
\cref{lem:disks}); they are exact when every $\Phi(t,s)b_s$ lies on one line.

\emph{Set-valued transitions.} Here preservation by every admissible $T$ is not sufficient. It is sufficient when $\hullD$ is a
closure operator (monotone and idempotent) that commutes with every admissible map $h\mapsto Th+u$, because then
$\hullD\big(\bigcup(T\,\hullD(S)+u)\big)=\hullD\big(\bigcup(TS+u)\big)$; the box hull commutes in this way with monomial $T$. Disks
fail the first condition: take $T_1=1$, $U_1=\disk(1,1)$, $T_2,T_3\in\{1,i\}$ and no further injection; then
$\hullD(\Reach_3)=\disk(0,2)$, while the composed best steps have radius $3/2+\sqrt2/2$. Boxes fail the second condition for singular
box-preserving maps. Take $T_t\in\{A,B\}$ for $t\ge2$ with $A=\begin{psmallmatrix}1&1\\0&0\end{psmallmatrix}$ and
$B=\begin{psmallmatrix}0&0\\1&1\end{psmallmatrix}$, $U_1=[0,1]^2$ and no further injection. Each of $A$, $B$ maps every box onto a box
($A$ maps $[a,b]\times[c,d]$ onto $[a+c,b+d]\times\{0\}$), and boxes are closed under intersection and Minkowski sums. Yet
$\Reach_t=[0,2]\times\{0\}\cup\{0\}\times[0,2]$ for $t\ge2$, whose box hull is $[0,2]^2$, while the best step maps $[0,a]^2$ onto
$[0,2a]^2$, so the composed best steps give $[0,2^{t-1}]^2$: a loss of a factor $2^{t-2}$, geometric in $t$. (\Cref{lem:boxes}(iii) is
not contradicted: $\{A,B\}$ is not a box of matrices.) Preservation is not necessary either: an entrywise nonnegative coupled
transition maps a box onto a parallelotope, not a box, yet it is exact in time for boxes by monotonicity (\cref{lem:boxes}(iii)), and
so is the transition of \cref{rem:provisos}, which maps boxes onto segments.
\end{remark}

\begin{proof}[Proof of \cref{lem:boxes}]
(i)~By induction $\Reach_{t-1}=H$ is an interval. $\Reach_t$ is the image of the connected compact set $A_t\times H\times U_t$
under the continuous map $(a,h,u)\mapsto ah+u$, hence an interval; its endpoints are the minimum and the maximum of a function
that is affine in each variable, attained at corners, which is what the four endpoint products and the sum return (for $a\ge0$,
the corner rule of a product with a nonnegative factor). If $A_t\subset[0,\infty)$, the \eqref{eq:cr} box contains $\Reach_t$, and
at each step it exceeds the exact image of its own previous box by $\arad_t\min(r_{t-1},\abs{c_{t-1}})$ (\cref{lem:cr}(a), (b)),
which is $O(\eps^2)$ when the boxes are $O(\eps)$; the excesses are carried forward and also shift later centres.

(ii)~For a point matrix $T$ and a box with centre $c$ and radius $r$, each coordinate $(Th)_i=\sum_jT_{ij}h_j$ ranges exactly over
$(Tc)_i\pm\sum_j\abs{T_{ij}}r_j$, so the natural extension returns the box hull of the image, and adding the box $U_t$ adds radii.
From $\{0\}$ the composed radius is the stated sum. The reachable set is $\Reach_t=\sum_{s\le t}\Phi(t,s)U_s$ (Minkowski sum, the
injections being independent), and coordinate $i$ of $\Phi(t,s)U_s$ ranges over its centre
$\pm\sum_j\abs{\Phi(t,s)_{ij}}\urad_{s,j}$; ranges of independent summands add. Since $\abs{XY}\le\abs{X}\,\abs{Y}$ entrywise,
$\abs{\Phi(t,s)}\le\abs{T_t}\cdots\abs{T_{s+1}}$, with equality of the radius vectors iff the stated condition holds; when it fails, the
composed box is strictly wider than the box hull of the reachable set, the wrapping effect. Products of diagonal matrices satisfy
$\abs{\Phi}=\prod\abs{T}$, and so do products of nonnegative ones. For $T_t=D_tN_tD_{t-1}$, since $D_k^2=I$,
$\Phi(t,s)=D_tN_t\cdots N_{s+1}D_s$, and $\abs{DMD'}=\abs{M}$ for sign matrices, so $\abs{\Phi(t,s)}=N_t\cdots N_{s+1}=
\abs{T_t}\cdots\abs{T_{s+1}}$. Real diagonal transitions of either sign and signed permutations are of this form
(\cref{rem:provisos}).

(iii)~We show by induction that the coordinatewise upper corner $h^{\rm hi}_{t-1}$ of $\hullbox(\Reach_{t-1})$ is itself
reachable, and likewise the lower corner; for $t-1=0$ both are $0$. For coordinate $i$ and a fixed $T\ge0$, $(Th)_i$ is
non-decreasing in every $h_j$, so its maximum over $\Reach_{t-1}$ is attained at $h^{\rm hi}_{t-1}$, which dominates every
reachable point. The maximum over $T\in\mathcal T_t$ of $\sum_jT_{ij}h^{\rm hi}_{t-1,j}$ takes, entry by entry, $T^{\rm hi}_{ij}$
if $h^{\rm hi}_{t-1,j}\ge0$ and $T^{\rm lo}_{ij}$ otherwise; these choices define one matrix $T^\star\in\mathcal T_t$ (the entries
are independent) that maximises every coordinate at once. So $T^\star h^{\rm hi}_{t-1}+u^{\rm hi}_t$ is reachable and is the upper
corner of $\Reach_t$. The natural interval extension computes coordinate $i$ as the sum over $j$ of the exact ranges of
$T_{ij}h_j$ over the box, whose maximum is the same expression; the lower corner is symmetric.
\end{proof}

Part~(iii) is the interval form of the comparison principle for monotone (cooperative) systems \citep{muller1927,smith1995}, which
underlies interval observers for positive systems \citep{farina2000,gouze2000}, and part~(ii) quantifies the wrapping effect
\citep{moore1966,neumaier1993}.

\begin{remark}[Orthant-monotone transitions: the box boundary is not nonnegativity]\label{rem:provisos}\label{rem:signclass}
\emph{The sign class of \cref{lem:boxes}(ii).} No layer of this paper needs more than diagonal transitions. The class
$T_t=D_tN_tD_{t-1}$, with sign matrices $D_t=\diag(\pm1)$ and $N_t\ge0$, satisfies the condition of \cref{lem:boxes}(ii) and contains
the real diagonal transitions of either sign; it is stated for completeness, and because under two provisos it is also necessary.

The matrix $T=\frac25\begin{psmallmatrix}1&-1\\-1&1\end{psmallmatrix}$ is neither diagonal nor nonnegative, and it does not preserve
boxes: it maps every box onto a segment of the line spanned by $(1,-1)$. But $T=DND$ with $D=\diag(1,-1)$ and
$N=\frac25\one\one^\top\ge0$, so $\abs{T^k}=N^k=\abs{T}^k$, and by \cref{lem:boxes}(ii) boxes are exact in time for
$h_t=Th_{t-1}+u_t$ with box injections; its spectral radius is $4/5$, so it contracts. More generally, a sequence
$T_t=D_tN_tD_{t-1}$ with signature matrices $D_t$ and $N_t\ge0$ satisfies the condition of \cref{lem:boxes}(ii): since $D_k^2=I$,
$\Phi(t,s)=D_tN_t\cdots N_{s+1}D_s$, and $\abs{DMD'}=\abs{M}$ for signature matrices, so $\abs{\Phi(t,s)}=N_t\cdots N_{s+1}=
\abs{T_t}\cdots\abs{T_{s+1}}$. With one common $D$ these are the systems that are monotone for an orthant order \citep{smith1995}.
Signed permutations ($D_t=PD_{t-1}\Pi^{-1}$ for $P=D'\Pi$) and real diagonal transitions of either sign ($D_0=I$,
$D_t=\diag(\sgn a_t)D_{t-1}$ with $\sgn0:=1$, $N_t=\diag\abs{a_t}$) are further examples, with $D_t$ changing from step to step;
signed permutations are not orthant-monotone with one common $D$ in general ($-I$ and
$\begin{psmallmatrix}0&1\\-1&0\end{psmallmatrix}$ are counterexamples).

\emph{Sufficient, and necessary only under provisos.} In general the class is not necessary. $T_1$ never matters, since $h_0=0$.
Zero entries: with $T_2=\begin{psmallmatrix}1&1\\1&-1\end{psmallmatrix}$ and $T_t=e_1e_1^\top$ for $t\ge3$, every product satisfies
$\abs{\Phi(t,s)}=\abs{T_t}\cdots\abs{T_{s+1}}$, so boxes are exact for every injection, yet the sign pattern of $T_2$ does not have
rank one, so $T_2$ is not of the form $DND'$. Zero radii, even without zero entries: with
$T_2=\begin{psmallmatrix}1&1\\1&-1\end{psmallmatrix}$, $T_t=\frac14\one\one^\top$ for $t\ge3$ and injection radii $(r_s,0)^\top$,
boxes are exact, since $\abs{\Phi(t,1)}e_1=(\frac12)^{t-2}\one=\abs{T_t}\cdots\abs{T_2}\,e_1$ and every other product is nonnegative,
yet $T_2$ is not of the form $DND'$; with positive radii they are not exact, since $\abs{\Phi(t,1)}e_2=0\ne(\frac12)^{t-2}\one$. So
each proviso below is needed.

Conversely, suppose every $T_t$ with $t\ge2$ has no zero entry and every injection radius is positive in every coordinate (over an
unbounded horizon). Then boxes are exact at every $t$ iff $T_t=D_tN_tD_{t-1}$ for all $t\ge2$. \emph{Proof:} at $(t,s)=(k+1,k-1)$ the
condition reads $\abs{T_{k+1}T_k}=\abs{T_{k+1}}\,\abs{T_k}$; equality in $\abs{\sum_px_p}\le\sum_p\abs{x_p}$ with every $x_p\ne0$
forces $\sgn(T_{k+1})_{ip}\,\sgn(T_k)_{pj}$ to be independent of $p$; hence $\sgn(T_k)_{pj}\,\sgn(T_k)_{qj}$ is independent of $j$, so
$\sgn T_k=\alpha_k\beta_k^\top$ has rank one, and the column signs of $T_{k+1}$ equal $\alpha_k$ up to one global sign. The rank-one
decomposition is unique up to sign, so $D_1=\diag\beta_2$ and $D_k=\diag\alpha_k$ for $k\ge2$ form one consistent sequence with
$T_k=D_kN_kD_{k-1}$, $N_k=\abs{T_k}$; the converse is the telescoping argument above.

The box boundary for point transitions is therefore the factorisation of \cref{lem:boxes}(ii) on the injection radii; neither
nonnegativity, nor diagonality, nor box preservation is necessary. The three nonnegative diagonal families of \cref{sec:setting}
are its simplest members.
\end{remark}

\begin{proof}[Proof of \cref{lem:disks}]
(i)--(ii)~Multiplication by $\lambda$ maps $\disk(c,\rho)$ onto $\disk(\lambda c,\abs{\lambda}\rho)$, since
$\abs{\lambda h-\lambda c}=\abs{\lambda}\abs{h-c}$, and $\disk(c_1,\rho_1)+\disk(c_2,\rho_2)=\disk(c_1+c_2,\rho_1+\rho_2)$; so the
transition preserves disks. The injection set $\{b_tx:x\in[\xcen_t\pm\xrad_t]\}$ is a segment with midpoint $b_t\xcen_t$ and
half-length $\abs{b_t}\xrad_t$, whose minimal enclosing disk is $\disk(b_t\xcen_t,\abs{b_t}\xrad_t)$. The exact image of
$\disk(c,\rho)$ under one step is the Minkowski sum of $\disk(\lambda_tc,\abs{\lambda_t}\rho)$ and that segment; its points farthest
from $\lambda_tc+b_t\xcen_t$ are at distance $\abs{\lambda_t}\rho+\abs{b_t}\xrad_t$, so the best disk step is the stated recursion.
Induction and linearity give $\disk(c_t,\rho_t)=\sum_{s\le t}\Phi(t,s)\,\disk(b_s\xcen_s,\abs{b_s}\xrad_s)$ with
$\Phi(t,s)=\lambda_t\cdots\lambda_{s+1}$: the disk recursion is exactly the reachable set of the recursion whose injection segments
are replaced by their minimal enclosing disks; this gives the containment. For the bound, $\Reach_t-c_t=\sum_s\Phi(t,s)b_s
[-\xrad_s,\xrad_s]$ is a centrally symmetric convex set, so $\Omega_t$ is the maximum over unit vectors $v$ of its support function
$\sigma(v)=\sum_s\abs{\Phi(t,s)b_s}\,\xrad_s\,\abs{\cos(\arg(\Phi(t,s)b_s)-\arg v)}$. The mean of $\abs{\cos}$ over a full turn is
$2/\pi$, so $\Omega_t\ge\frac2\pi\sum_s\abs{\Phi(t,s)}\,\abs{b_s}\,\xrad_s=\frac2\pi\rho_t$.

(iii)~The eigenvalues of the symmetric matrix $\abs{\lambda}\begin{psmallmatrix}\abs{\cos\theta}&\abs{\sin\theta}\\
\abs{\sin\theta}&\abs{\cos\theta}\end{psmallmatrix}$ are $\abs{\lambda}(\abs{\cos\theta}\pm\abs{\sin\theta})$, with eigenvectors
$(1,\pm1)$; so $\gbox$ is its spectral radius, with Perron vector $(1,1)$. If $\cos\theta\sin\theta\ne0$, then $\abs{\Rot(\theta)}^2$
has diagonal entries $\cos^2\theta+\sin^2\theta=1$, while $\abs{\Rot(2\theta)}$ has diagonal entries $\abs{\cos2\theta}<1$ unless
$\cos2\theta=\pm1$, that is $\theta\in\frac\pi2\mathbb Z$; the off-diagonal entries agree ($2\abs{\cos\theta\sin\theta}=
\abs{\sin2\theta}$), so $\big(\abs{\Rot(\theta)}^2-\abs{\Rot(2\theta)}\big)v=(1-\abs{\cos2\theta})\,v\ne0$ for every nonzero $v\ge0$,
and real boxes wrap by \cref{lem:boxes}(ii).

\emph{Several real inputs per mode.} For $b_tx_t=\sum_kb_{t,k}x_{t,k}$ the disk recursion with the injection disk
$\disk(\sum_kb_{t,k}\xcen_{t,k},\sum_k\abs{b_{t,k}}\xrad_{t,k})$, the disk sum of the input segments, is sound and satisfies
$\rho_t\le\frac\pi2\Omega_t$ by the same proof, the support-function argument summing over $(s,k)$. It is the composition of best
disk steps only if, at every $t$, the $b_{t,k}$ with $\xrad_{t,k}>0$ are real multiples of one another: otherwise the injection set is
a zonotope in $\complexes$ whose minimal disk has radius $\max_\sigma\abs{\sum_k\sigma_kb_{t,k}\xrad_{t,k}}<\sum_k\abs{b_{t,k}}
\xrad_{t,k}$; for $b=(1,i)$ and unit radii the injection set is a square, with minimal radius $\sqrt2<2$. The statements about best
steps and minimal disks in (i) and (ii) are for one real input per mode.
\end{proof}

\emph{Disks are not exact in time: an instance.} In \cref{prop:exact}(b$'$) at $t=2$ the disk radius is $(0.6+\sqrt2/2)\eps$, while
the circumradius of the reachable set is $\sqrt{1.46}\,\eps$, and $(0.6+\sqrt2/2)^2=0.86+0.6\sqrt2>1.46$: here
$\Reach_2-c_2=0.6\,\eps[-1,1]+\frac{1+i}{2}\,\eps[-1,1]$, whose farthest point is at distance $\abs{1.1+0.5i}\,\eps$.

\begin{lemma}[Readout of a disk]\label{lem:diskreadout}
Over $\disk(c,\rho)$, the readout $\mathrm{Re}(Ch)$ has the exact range $\mathrm{Re}(Cc)\pm\abs{C}\rho$.
\end{lemma}

\begin{proof}
$\mathrm{Re}(C(h-c))$ over $\abs{h-c}\le\rho$ is the real part of the disk $\disk(0,\abs{C}\rho)$.
\end{proof}

\begin{remark}[What the factor $\pi/2$ bounds, and what it does not]\label{rem:pihalf}
(1)~\emph{Only the circumradius.} In a fixed readout direction the disk can be arbitrarily looser than both the reachable set and
the box. Take one mode with $b=i$, $C=1$ and $x_1\in[-\eps,\eps]$: the exact readout is $\mathrm{Re}\,h_1=0$; the real-coordinate box
is $\{0\}\times[-\eps,\eps]$ and reads out exactly $0$; the disk $\disk(0,\eps)$ gives $\mathrm{Re}\,h_1\in[-\eps,\eps]$. With input
radii that vary over positions, a long horizon does not bound the directional ratio either: radius concentrated at lags where
$C\,\Phi(t,s)\,b$ is imaginary has no true effect on the readout. So $\pi/2$ is not a bound on the looseness of the disk certificate
($R$ or the margin). Nor do the measured $R$ isolate that looseness: $R$ measures the whole two-layer model, with every token-wise
loss; the real-diagonal S4D-type models, whose boxes are exact in time, had $R$ between 508 and 906 in the context-length probe, and
the disk arm of the complex-diagonal control had $R$ between $5.2\times10^{3}$ and $6.2\times10^{3}$ (\cref{sec:exp-gate3}).

(2)~\emph{Not monotone in time.} The ratio $\rho_t/\Omega_t$ grows at first, $\rho_1/\Omega_1=1<\rho_2/\Omega_2$, and it can also
fall. In \cref{prop:exact}(b$'$), in units of $\eps$, $\rho_t=\frac1{\sqrt2}\sum_{k<t}(0.6\sqrt2)^k$, and $\Omega_t$ is the largest
modulus of a vertex of the zonotope $\sum_{k<t}\lambda^kb\,[-1,1]$; then $\Omega_4=\abs{1.46+0.572i}$ and
$\Omega_5=\abs{1.7192+0.8312i}$, and $\rho_4/\Omega_4\approx1.434>\rho_5/\Omega_5\approx1.369$. What holds is $\rho_t/\Omega_t\le\pi/2$
for every $t$: the loss does not compound, but it is neither constant nor monotone.

(3)~\emph{Point coefficients only; the open boundary.} \Cref{lem:disks} assumes input-independent $\lambda_t$. Under IBP a selective
complex-diagonal layer has set-valued $\lambda_t$, and the example of \cref{rem:preservation} ($\lambda\in\{1,i\}$) shows that the
$\pi/2$ bound then fails although every admissible $\lambda$ preserves disks: with $T_1=1$, $U_1=\disk(1,1)$, $T_t\in\{1,i\}$ for all
$t\ge2$ and no further injection, the best step maps $\disk(c,\rho)$ to $\disk(c(1+i)/2,\ \rho+\abs{c}/\sqrt2)$, the composed radius
tends to $2+\sqrt2$, the hull radius is $2$ for $t\ge3$, and the ratio tends to $(2+\sqrt2)/2>\pi/2$, while it stays bounded. With
$T_t\in\{1,e^{i\vartheta}\}$ (same start: $T_1=1$, $U_1=\disk(1,1)$, no further injection) the best step gives
$\rho'=\rho+\abs{c}\sin(\vartheta/2)$ and $\abs{c'}=\abs{c}\cos(\vartheta/2)$, so the composed radius tends to $1+\cot(\vartheta/4)$,
while the hull radius is $2$ once $(t-1)\vartheta\ge\pi$; the ratio tends to $(1+\cot(\vartheta/4))/2$. With $\{a,a\,e^{i\vartheta}\}$,
$0<a<1$, every radius at step $t$ is multiplied by $a^{t-1}$, and the ratio is the same. So for input-dependent (selective)
complex-diagonal modes no constant bound holds in disks: over the family the limit ratio $(1+\cot(\vartheta/4))/2$ is unbounded as
$\vartheta\to0$, while for each fixed $\vartheta$ the ratio stays bounded in $t$. The arc $\{e^{i\theta}:0\le\theta\le\vartheta\le\pi\}$
has the same best step as $\{1,e^{i\vartheta}\}$, because the arc lies in the disk on its chord, so the example also holds for
interval-type eigenvalue sets. This is the open boundary of the disk result; whether another domain follows selective
complex-diagonal transitions is not treated here.

(4)~\emph{Dense input maps} (S5, LRU): soundness and the factor $\pi/2$ hold by the extension to several real inputs per mode in the
proof of \cref{lem:disks}; the best-step property does not.
\end{remark}

\begin{lemma}[Reduced zonotopes on linear point recursions]\label{lem:zono}
Let $h_t=T_th_{t-1}+u_t$ on $\reals^m$ with point matrices $T_t$ and $h_0=0$, where both domains receive the same injection set $U_t$:
a box with centre $\ucen_t$ and radius vector $\urad_t$, or a zonotope whose box hull is that box. Propagate
$Z_t=\{c_t+G_t\xi+\diag(\varrho_t)\zeta:\norm{\xi}_\infty\le1,\norm{\zeta}_\infty\le1\}$ by mapping it exactly through $T_t$ (its
remainder box $\diag(\varrho_{t-1})$ becomes the generators $T_t\diag(\varrho_{t-1})$), adding the injection, and applying any
Girard-type reduction: some generators are kept, and every dropped generator $g$ is boxed into the remainder,
$\varrho\mapsto\varrho+\abs{g}$. Let $r_t=\abs{T_t}r_{t-1}+\urad_t$, $r_0=0$, be the radius of the composed box steps
(\cref{lem:boxes}(ii)). Then for every $t$, every budget and every choice of the dropped generators: (i)~$\Reach_t\subset Z_t$, and
$Z_t$ and the composed box have the same centre $c_t=T_tc_{t-1}+\ucen_t$; (ii)~$\sum_j\abs{G_{t,ij}}+\varrho_{t,i}\le r_{t,i}$ for
every coordinate $i$, and $\sum_j\abs{w^\top G_{t,\cdot j}}+\abs{w}^\top\varrho_t\le\abs{w}^\top r_t$ for every linear readout $w$;
(iii)~$\hullbox(\Reach_t)\subset\hullbox(Z_t)\subset[c_t-r_t,c_t+r_t]$, so the interval hull of $Z_t$ is exact wherever boxes are
exact in time for this recursion (\cref{def:exact}; for box injections, the condition of \cref{lem:boxes}(ii)).
\end{lemma}

\begin{proof}
(i)~Each step encloses the exact image of the previous set, and boxing a generator only enlarges the set
($g\xi_g\in\diag(\abs{g})\,[-1,1]^m$); the affine maps and the reduction keep the centre. (ii)~Boxing a dropped generator $g$
moves $\abs{g_i}$ from the generator part to the remainder of coordinate $i$, so reduction keeps every coordinate radius
$\mathrm{rad}(Z)_i:=\sum_j\abs{G_{ij}}+\varrho_i$. A point transition maps coordinate radii to at most $\abs{T}$ times them:
$\sum_j\abs{(TG)_{ij}}+\sum_l\abs{T_{il}}\varrho_l\le\sum_l\abs{T_{il}}\big(\sum_j\abs{G_{lj}}+\varrho_l\big)=(\abs{T}\,
\mathrm{rad}(Z))_i$. The injection adds $\urad_t$. By induction and $\abs{T_t}\ge0$, $\mathrm{rad}(Z_t)\le\abs{T_t}r_{t-1}+
\urad_t=r_t$. For a readout, $\sum_j\abs{w^\top G_{\cdot j}}+\abs{w}^\top\varrho\le\abs{w}^\top\mathrm{rad}(Z)\le\abs{w}^\top r$.
(iii)~follows from (i), (ii) and the common centre.
\end{proof}

\Cref{lem:zono} is an elementary consequence of box-type order reduction, in which the dropped generators are enclosed by their
interval hull \citep{kuhn1998} after a ranking of the generators \citep{girard2005} (both attributions as in \citealp{kopetzki2017},
Sec.~II.A): the enclosure keeps every coordinate radius, and a point linear map widens interval radii by at most $\abs{T}$
\citep{moore1966,neumaier1990}. It does not say that a reduced zonotope can be better than boxes in time: with budget~0 the reduced
zonotope is the box, which diverges on \cref{prop:exact}(b$'$), where an unreduced zonotope is exact.

The bound also holds when the injections share symbols across steps (coordinate radii are subadditive), for any interleaving of
mapping, injecting and reducing, and for the settle and promote steps of the zonotope of the distance probe (\cref{sec:exp-gate5}),
whose remainders are independent between forms and receive the dropped symbols. With injection sets other than boxes, the condition
of \cref{lem:boxes}(ii) does not decide exactness: $T=\frac25\begin{psmallmatrix}1&-1\\-1&1\end{psmallmatrix}$ is box-exact by
\cref{lem:boxes}(ii), but with the segment $U_1=\{\xi(1,1)^\top:\abs{\xi}\le1\}$ and no further injection $\Reach_2=\{0\}$, while the
composed box has radius $(4/5,4/5)$.

\emph{Scope in the model.} The lemma concerns the linear part of a model. Against boxes a zonotope can lose only in its relaxations of
products and nonlinear maps, whose excess over the exact interval ranges is of higher order in $\eps$. In the affine arithmetic of the
distance probe, the product rule adds the product of the two radii to the radius of a product; the linear relaxations of sigmoid,
tanh, softplus, SiLU and the Mamba decay carry a fresh error term of second order in the input width (the chord remainder $w^2/8$ times
a bound on $f''$, or the slope rule of DeepZ, \citealp{singh2018}); and the new remainder of RMSNorm consists of products of radii and
a Taylor term of second order in the variation $v$ of the mean square, which becomes very loose as $v$ approaches $m_{\rm c}+
\epsilon_{\rm n}$, with $m_{\rm c}$ the mean square at the centre and $\epsilon_{\rm n}$ the RMSNorm constant. On the S4D-type layer the
scan is a linear point recursion, and \cref{lem:zono} applies to it for the same injection sets (in the model each domain computes its
own injections through the nonlinear maps); the SiLU gate, RMSNorm and the output map around it are not linear. On Mamba the decay is
input-dependent, so its scan contains products, and the lemma does not apply; nor does it apply to whole blocks.

\subsection{How exactness fails: Proposition 3.4 in full}\label{app:proofs-fail}

The main text states parts (a$'$), (b$'$), (b) and (b$''$) of \cref{prop:exact}. This subsection adds parts (a) and (c)
(\cref{prop:exactac}), states (b) and (b$''$) with all their constants (\cref{prop:lstmfull}), and proves every part. Part (a$'$)
rests on the following lemma.

\begin{lemma}[Uniform contraction is automatic]\label{lem:uniform}
In real arithmetic, let the model be a stack of $Q$ layers of one of the three nonnegative diagonal families of \cref{tab:layers},
and let every input embedding lie in a compact set $\mathcal E$ (for instance the finite embedding table). There are $\bar a<1$ and a
compact set $\mathcal K$ that contains every clean intermediate value of every input sequence of every length (zero state and zero
convolution padding at its start), such that the clean decays satisfy $a^0_{t,i}\le\bar a$ in every layer, channel and position.
\end{lemma}

\begin{proof}
Induction over the layers; the input $v^\ell_t$ of layer $\ell$ lies in a compact set $V_\ell$, with $V_1=\mathcal E$. RMSNorm (with
$\epsilon_{\rm n}>0$) and the input projection are continuous, so the convolution window ranges in a compact set (zero padding added),
and the coefficients $(a_t,u_t)$ and the quantities $x_t,z_t,B_t,C_t$ are continuous functions of the window and of $v^\ell_t$.
Pointwise $a<1$: $\Delta=\softplus(s)>0$ and $\alpha>0$ (Mamba), $1-\sigma(s)<1$ (minGRU-type), a constant in $(0,1)$ (S4D-type). By
compactness $\bar a_\ell:=\max a<1$ and $\abs{u}\le U_\ell$, so $\abs{h_t}\le\bar a_\ell\abs{h_{t-1}}+U_\ell$ and $h=0$ at the start
give $\abs{h_t}\le U_\ell/(1-\bar a_\ell)$ for every $t$. The readout, the output map and $v^{\ell+1}_t=v^\ell_t+o_t$ are continuous in
bounded arguments, so $v^{\ell+1}_t$ ranges in a compact set $V_{\ell+1}$. Take $\bar a=\max_\ell\bar a_\ell$ and for $\mathcal K$ the
closure of the union of all these ranges.
\end{proof}

The lemma is qualitative: $\bar a$ is a maximum over all contexts and can be extremely close to $1$ (\cref{rem:timescales}). The
clean periodic states of \cref{thm:inf} are limits of clean states of finite contexts, hence lie in $\mathcal K$.

\begin{proposition}[Further parts of \cref{prop:exact}]\label{prop:exactac}
(a)~\emph{Nonnegative diagonal stacks, the infinite periodic context.} Use exact interval extensions (the implementation's slack is
discussed in \cref{rem:silu}). Let the input boxes depend on $\eps\ge0$ with $E_t(0)=\{e_t\}$ and widths at most $C\eps$ (both threat
models qualify, with $\eps$ the radius or the box scale). Assume that the clean periodic cascade (\cref{def:cascade}) contracts, that
is $\pih<1$ at $\eps=0$ in every channel of every layer; in exact arithmetic this holds automatically for the three families
(\cref{lem:uniform}), so the hypothesis matters only in floating point. Then there are $\eps_0>0$ and $C'<\infty$ such that, for
$\eps\le\eps_0$, every box of the cascade (phase boxes in every layer, document boxes, and $\Binf$) has width at most $C'\eps$, and the
boxes converge to the clean infinite-context values as $\eps\to0$. The infinite-context box is $O(\eps)$ and continuous at $\eps=0$;
part (a$\infty$) of \cref{prop:loss} identifies the first-order constant.

(c)~\emph{Bounded invariant boxes exist for saturating gated RNNs.} For the LSTM and GRU cells of our implementation and any bounded
input boxes, there is a bounded box of states that the interval cell maps into itself, in every layer. What can fail is non-vacuity:
by \cref{prop:lstmfull}(b), the width of such a box need not vanish as $\eps\to0$.
\end{proposition}

\begin{proposition}[\cref{prop:exact}(b) and (b$''$) in full]\label{prop:lstmfull}
(b)~\emph{A contracting LSTM.} There is a one-layer LSTM with two hidden units and scalar input $x_t\in[-\eps,\eps]$ such that, with
$\gamma:=\sigma(8)$:
\begin{itemize}
\item the true dynamics contracts: every product of $t$ state Jacobians along any trajectory has spectral radius at most $\kappa^t$,
with $\kappa=1-\gamma+0.6\sqrt2\,\gamma^2<1$, and the reachable set of $h_t$ has diameter at most $\sqrt2\,\gamma\eps/(1-0.6\sqrt2\,
\gamma)$ for every $t$;
\item the entrywise absolute Jacobian at the zero state has spectral radius $\rho_A:=1-\gamma+1.2\gamma^2>1$;
\item for every $\eps>0$, the IBP box of $h$ obtained by iterating the interval cell from the zero box converges to a box of width at
least $2\nu^*>0$, where $\nu^*$ is the positive root of $\nu=\gamma\tanh(\tanh(1.2\nu))$, and this limit width tends to $2\nu^*$ as
$\eps\to0$. At $\eps=0$ the IBP box is $\{0\}$.
\end{itemize}
The infinite-horizon IBP width is therefore discontinuous at $\eps=0$, while the true spread is $O(\eps)$. The gates are constant,
and replacing $\tanh$ by the identity gives a linear, state-independent recurrence of the type of \cref{prop:exact}(b$'$), whose IBP
width diverges at the rate $\rho_A$ per step; $\tanh$ caps the width instead.

(b$''$)~\emph{Finite horizon: the time to saturation.} (i)~In \cref{prop:exact}(b$'$) the IBP radius reaches a level $w>0$ after
$\lceil\log(1+w/(2.5\eps))/\log1.2\rceil$ steps. (ii)~In (b), let $\nu_t$ be the radius of the $h$ box after $t$ interval steps from
the zero box and $\tsat(\eps):=\min\{t:\nu_t\ge\nu^*/2\}$. Then $\tsat(\eps)<\infty$ for every $\eps>0$,
\[
\frac{\log\big(1+(\rho_A-1)\nu^*/(\gamma^2\eps)\big)}{\log\rho_A}\le\tsat(\eps),\qquad\text{and}\qquad
\lim_{\eps\downarrow0}\frac{\tsat(\eps)\log\rho_A}{\log(1/\eps)}=1.
\]
(iii)~At every finite $t$ the box of (b) tends to $\{0\}$ as $\eps\to0$ (\cref{prop:loss}, part (c3)): the discontinuity of (b) is a
statement about $t=\infty$, and at a fixed $\eps$ the box saturates after $\Theta(\log(1/\eps))$ steps.
\end{proposition}

\begin{proof}[Proof of \cref{prop:exact}(a$'$)]
We also prove part (i) of the full statement: in every state channel, given its coefficient boxes, the endpoint scan returns exactly
the reachable set of the interval-coefficient recursion, and the \eqref{eq:cr} scan contains it; at each step the \eqref{eq:cr} box
exceeds the exact image of its own previous box by $\arad_t\min(r_{t-1},\abs{c_{t-1}})$, which is $O(\eps^2)$ when the boxes are
$O(\eps)$. This is \cref{lem:boxes}(i) for each channel, and \cref{lem:cr}(b).

For the radius bound, let $\mathcal K$, $\bar a$ and a bound $M_h$ of the clean states be as in \cref{lem:uniform} for the embedding
set $\mathcal E$. Every primitive of the pass is locally Lipschitz, and every decay map is continuous with values at most $\bar a$ at
clean arguments. By compactness there are $\delta>0$ and $\Lambda\ge1$ such that, on the closed $\delta$-neighbourhood $\mathcal
K_\delta$ of $\mathcal K$ (per argument, in the $\infty$-norm), every primitive is $\Lambda$-Lipschitz in the $\infty$-norm and every
decay map takes values at most $\bar a':=\frac12(1+\bar a)<1$. The exact-range extension of a $\Lambda$-Lipschitz map sends a box of
radius $\theta$ inside $\mathcal K_\delta$ to a box of radius at most $\Lambda\theta$ that contains the clean value (\cref{cor:stack}(i),
the clean input being admissible), hence lies within $2\Lambda\theta$ of it.

We show by induction over the layers: there are $C_\ell$ and $\eps_\ell>0$ such that for $\eps\le\eps_\ell$ every input box of layer
$\ell$, at every position of every sequence, has radius at most $C_\ell\eps$. For $\ell=1$, $C_1=C_0$. Assume the claim for $\ell$. The
token-wise computation from the layer input to the coefficient boxes is a fixed composition of finitely many, say $n_p$, primitives
(convolution window included); while every intermediate box has radius at most $\delta/2$ it lies in $\mathcal K_\delta$, so by
induction over these primitives the coefficient boxes have radius at most $c_\ell\eps$, $c_\ell:=\Lambda^{n_p}C_\ell$, provided
$c_\ell\eps\le\delta/2$; and $\ahi_t\le\bar a'$, because the upper end of a decay box is a value of the decay map in $\mathcal
K_\delta$. The \eqref{eq:cr} radius obeys
\[
r_t=\ahi_tr_{t-1}+\arad_t\abs{c_{t-1}}+\urad_t\le\bar a'r_{t-1}+c_\ell\eps\,(\abs{c_{t-1}}+1),
\]
and $\abs{c_{t-1}}\le M_h+r_{t-1}$, because the box contains the clean state. If moreover $c_\ell\eps\le(1-\bar a')/2$, then
$r_t\le\frac12(1+\bar a')\,r_{t-1}+c_\ell\eps(M_h+1)$, and from $r=0$ at the start of any context, $r_t\le2c_\ell\eps(M_h+1)/(1-\bar
a')$ for every $t$. The endpoint scan's boxes lie in the \eqref{eq:cr} boxes (\cref{lem:cr}(b), (c), induction over $t$). The readout,
the gate, the output map and the residual are again finitely many primitives, so the claim holds for $\ell+1$ after shrinking
$\eps_{\ell+1}$. After layer $Q$, the final RMSNorm and the head give logit boxes of radius at most $C'\eps$ that contain the clean
logits, so $\ell^{\rm lo}_y\ge\ell^0_y-2C'\eps$ and $\ell^{\rm hi}_k\le\ell^0_k+2C'\eps$, hence $b(\eps)\ge m^0-4C'\eps$, while
$b(0)=m^0$ for exact extensions: $W(\eps)\le4C'\eps$. No constant depends on the sequence or its length. The cascade boxes of
\cref{thm:inf} are limits of the boxes of $p^Kx$ as $K\to\infty$ (step (B) of the proof of \cref{thm:inf} in
\cref{app:proofs-periodic}), all bounded by $C'\eps$, so the limits are too.
\end{proof}

\Cref{prop:exact}(a$'$) is an exact-arithmetic statement about $\eps\to0$. Its constant contains $1/(1-\bar a')$ and $\Lambda^{n_p}$
per layer, where $\bar a$ is a worst case over all contexts, not a bound along a trajectory, and it can be astronomically large
(\cref{rem:notsay}). It says nothing about a float64 channel whose decay rounds to $1$, nor about the threat models of the
experiments: at box scale 1 under synonym substitution, the slow-channel attribution measured radius time constants
$\tau_{\rm h}=\infty$ for the natural Mamba and minGRU-type models (\cref{rem:timescales}), that is, boxes far outside the regime
$\eps\le\eps_0$.

\begin{proof}[Proof of \cref{prop:exactac}(a)]
Induction over the layers of the cascade. On a bounded set, the exact-range interval operation of a Lipschitz elementary function
maps a box to a box whose width is at most a constant times the width of the input box, and its endpoints are continuous
(\cref{cor:stack}(iii)). So the coefficient boxes $(A^\ell_j(\eps),U^\ell_j(\eps))$ have widths $O(\eps)$ and shrink to the clean
points as $\eps\to0$: $\arad_j=O(\eps)$, $\urad_j=O(\eps)$, and $\pih(\eps)\to\pih(0)<1$, so $\denh(\eps)\ge\frac12\denh(0)$ for
small $\eps$. Then $c^*(\eps)$ converges to the clean infinite-context state $c^*(0)$, and, in the notation of
\cref{app:proofs-periodic},
\[
r^*(\eps)=\frac{\beta_\eps(c^*_\eps)}{\denh(\eps)}\le\frac{2P\,\max_j\big(\arad_j\abs{c_{j-1}}+\urad_j\big)}{\denh(0)}=O(\eps).
\]
The phase radii $\Gh_jr^*+\beta_j(c^*)$ are $O(\eps)$, hence so are the output boxes, which converge to the clean outputs. The document
adds finitely many steps of \eqref{eq:cr}, each with $r'=\ahi r+\arad\abs{c}+\urad=O(\eps)$.
\end{proof}

\begin{proof}[Proof of \cref{prop:exact}(b$'$)]
(i)~$U=0.6\sqrt2\,O$ with $O$ the rotation by $-\pi/4$, so $\norm{U}_2=0.6\sqrt2$, and
$\norm{h_t}_2\le\sum_{k\ge0}\norm{U}_2^k\,\norm{b}_2\,\eps=\eps\norm{b}_2/(1-0.6\sqrt2)$ with $\norm{b}_2=1/\sqrt2$.
(ii)~By \cref{lem:boxes}(ii) the radius obeys $r_t=\abs{U}r_{t-1}+\abs{b}\eps$ with $\abs{U}=0.6\,\one\one^\top$ and
$\abs{b}\eps=\frac\eps2\one$. If $r_{t-1}=\alpha_{t-1}\one$ then $r_t=(1.2\,\alpha_{t-1}+\eps/2)\one$, and $\alpha_0=0$ gives
$\alpha_t=\frac\eps2\cdot\frac{1.2^t-1}{0.2}=2.5\,\eps\,(1.2^t-1)$.
(iii)~$0.6(1-i)(h_1+ih_2)=0.6(h_1+h_2)+0.6\,i\,(h_2-h_1)$, which is $Uh$ in complex form, and $0.6(1-i)=0.6\sqrt2\,e^{-i\pi/4}$; $b$
becomes $\frac12(1+i)$, of modulus $1/\sqrt2$. \Cref{lem:disks} gives $\rho_t=\frac{\eps}{\sqrt2}\sum_{k<t}(0.6\sqrt2)^k\le
\eps/(\sqrt2(1-0.6\sqrt2))$ and the factor $\pi/2$.
(iv)~The box radius map of one period is $\abs{M}^P$ with $\abs{M}$ the matrix of \cref{lem:disks}(iii); its least fixed point is
$r^*=\sum_{k\ge0}\abs{M}^{kP}r_{\rm free}$ when the series converges. The free radius is nonnegative and nonzero, so its component
along the Perron vector, $\langle(1,1),r_{\rm free}\rangle$, is positive, and $\langle(1,1),\abs{M}^{kP}r_{\rm free}\rangle=
\gbox^{kP}\langle(1,1),r_{\rm free}\rangle$ does not tend to $0$ when $\gbox\ge1$. So the series diverges and no finite radius $r$
satisfies $r\ge\abs{M}^Pr+r_{\rm free}$ (iterating it would give $r\ge\sum_{k<K}\abs{M}^{kP}r_{\rm free}$ for every $K$). In disks the
radius map is $\abs{\lambda}^P<1$, and the proofs of \cref{prop:exactac}(a) and of \cref{prop:exact}(a$'$) go through with
\cref{lem:disks} in place of \cref{lem:cr,lem:boxes}.
\end{proof}
In real arithmetic the vacuity of (iv) concerns a mode that is read out ($C_{d,n}\neq0$) or feeds a later layer; the implementation
returns $\pm\infty$ in any case, because infinities propagate.

\begin{proof}[Proof of \cref{prop:lstmfull}(b)]
\emph{Construction.} Take the LSTM cell of our implementation with input size 1 and $H=2$ (gate order $i,f,g,o$;
$c'=\sigma(f)\,c+\sigma(i)\tanh(g)$, $h'=\sigma(o)\tanh(c')$). Set all weights to zero except: the candidate rows,
$W_{hh}^{(g)}=U:=0.6\begin{psmallmatrix}1&1\\-1&1\end{psmallmatrix}$ and $W_{ih}^{(g)}=(0.5,0.5)^\top$; and the biases $8$, $-8$, $8$ of
the input, forget and output gates (candidate bias $0$). The gates are then constant, $\sigma(i)=\sigma(o)=\gamma$ and
$\sigma(f)=1-\gamma$, and
\[
c_t=(1-\gamma)\,c_{t-1}+\gamma\tanh\!\big(0.5\,x_t\one+Uh_{t-1}\big),\qquad h_t=\gamma\tanh(c_t),\qquad h_0=c_0=0 .
\]
Two facts are used throughout: $\gamma>5/6$, because $e^{-8}<1/5$; and $U=0.6\sqrt2\,O$ with
$O=\frac1{\sqrt2}\begin{psmallmatrix}1&1\\-1&1\end{psmallmatrix}$ orthogonal, so $\norm{U}_2=\rho(U)=0.6\sqrt2<1$, while
$\abs{U}=0.6\begin{psmallmatrix}1&1\\1&1\end{psmallmatrix}$ has $\rho(\abs{U})=1.2$.

\emph{True contraction.} Let $N(\delta h,\delta c):=\max(\norm{\delta h}_2/\gamma,\ \norm{\delta c}_2)$. At any state and input, the
Jacobian maps $(\delta h,\delta c)$ to $\delta c'=(1-\gamma)\delta c+\gamma D_gU\delta h$ and $\delta h'=\gamma D_{c'}\delta c'$, with
diagonal $D_g$, $D_{c'}$ whose entries ($\tanh'$ values) lie in $(0,1]$. Hence $\norm{\delta c'}_2\le(1-\gamma)\norm{\delta c}_2+
0.6\sqrt2\,\gamma\norm{\delta h}_2\le\kappa N$ and $\norm{\delta h'}_2/\gamma\le\norm{\delta c'}_2\le\kappa N$, that is
$N(Jv)\le\kappa N(v)$; and $\kappa<1$ iff $0.6\sqrt2\,\gamma<1$, which holds. A product of $t$ Jacobians has $N$-operator norm, and hence
spectral radius, at most $\kappa^t$. For two input sequences in $[-\eps,\eps]$, $\tanh$ being 1-Lipschitz and
$\norm{\delta h_{t-1}}_2\le\gamma\norm{\delta c_{t-1}}_2$ give $\norm{\delta c_t}_2\le\kappa\norm{\delta c_{t-1}}_2+\sqrt2\,\gamma\eps$,
so $\norm{\delta c_t}_2\le\sqrt2\gamma\eps/(1-\kappa)$ and $\norm{\delta h_t}_2\le\sqrt2\,\gamma^2\eps/(1-\kappa)=\sqrt2\,\gamma\eps/
(1-0.6\sqrt2\,\gamma)$, using $1-\kappa=\gamma(1-0.6\sqrt2\,\gamma)$.

\emph{Absolute Jacobian.} At $h=c=0$, $x=0$: $D_g=D_{c'}=I$ and, in the coordinates $(h,c)$,
\[
\abs{J_0}=\begin{pmatrix}\gamma^2\abs{U}&\gamma(1-\gamma)I\\ \gamma\abs{U}&(1-\gamma)I\end{pmatrix}
=\begin{pmatrix}\gamma I\\ I\end{pmatrix}\begin{pmatrix}\gamma\abs{U}&(1-\gamma)I\end{pmatrix}.
\]
Its nonzero eigenvalues are those of $\begin{pmatrix}\gamma\abs{U}&(1-\gamma)I\end{pmatrix}\begin{pmatrix}\gamma I\\ I\end{pmatrix}=
\gamma^2\abs{U}+(1-\gamma)I$, namely $1-\gamma+1.2\gamma^2$ and $1-\gamma$. The first exceeds $1$ iff $\gamma>5/6$.

\emph{The IBP recursion.} The interval cell propagates the pre-activations in centre-radius form, $\sigma$ and $\tanh$ at the
endpoints, and the gate products with the exact product rule for a factor in $[0,1]$. If the $h$ box is $[-\nu\one,\nu\one]$ and the
$c$ box is $[-C\one,C\one]$, the candidate pre-activation box has centre $0$ and radius $0.5\eps+\abs{U}\nu\one=(0.5\eps+1.2\nu)\one$
(the row sums of $\abs{U}$ are $1.2$); $\tanh$ is odd and increasing; and a point gate times a symmetric box is symmetric. So the next
boxes have the same form, with
\[
T_\eps(C,\nu)=\Big((1-\gamma)C+\gamma\tanh(0.5\eps+1.2\nu),\ \ \gamma\tanh\big((1-\gamma)C+\gamma\tanh(0.5\eps+1.2\nu)\big)\Big),
\]
starting from $(C_0,\nu_0)=(0,0)$. The width of the $h$ box is $2\nu_t$.

\emph{Limit of the IBP recursion.} $T_\eps$ is continuous, non-decreasing in both arguments, and maps $[0,1]^2$ into itself. From
$(0,0)\le T_\eps(0,0)$, the iterates are non-decreasing and bounded, so they converge to the least fixed point $(C_\eps,\nu_\eps)$ of
$T_\eps$ on $[0,\infty)^2$: by induction they stay below any other fixed point. Fixed points satisfy $C=\tanh(0.5\eps+1.2\nu)$ and
$\nu=\psi_\eps(\nu):=\gamma\tanh(\tanh(0.5\eps+1.2\nu))$. The function $\psi_0$ has $\psi_0(0)=0$, $\psi_0'(0)=1.2\gamma>1$,
$\psi_0<\gamma<1$, and is strictly concave on $[0,\infty)$, because for $u>0$
$(\tanh\circ\tanh)''(u)=\tanh''(\tanh u)\tanh'(u)^2+\tanh'(\tanh u)\tanh''(u)<0$. Hence $\psi_0(\nu)-\nu$ is strictly concave,
vanishes at $0$, is positive just right of $0$ and negative for $\nu\ge1$. It has exactly one positive zero $\nu^*\in(0,1)$, and
$\psi_0(\nu)>\nu$ on $(0,\nu^*)$. For $\eps>0$, $\psi_\eps\ge\psi_0$ and $\psi_\eps(0)>0$, so no $\nu\in[0,\nu^*)$ is a fixed point of
$\psi_\eps$; hence $\nu_\eps\ge\nu^*$. As $\eps\downarrow0$, the least fixed point is non-increasing (since $T_\eps$ increases with
$\eps$) and bounded below by $\nu^*$; its limit is a fixed point of $\psi_0$ by continuity, hence equals $\nu^*$. At $\eps=0$, $(0,0)$
is a fixed point of $T_0$ and the IBP box stays $\{0\}$. With $\tanh$ replaced by the identity, the recursion becomes
$C'=(1-\gamma)C+\gamma(0.5\eps+1.2\nu)$, $\nu'=\gamma C'$, and on $\nu=\gamma C$ it reads $C'=\rho_AC+0.5\gamma\eps$, which diverges;
the true Jacobians keep the bound $N(Jv)\le\kappa N(v)$ with $D_g=D_{c'}=I$.
\end{proof}

\begin{proof}[Proof of \cref{prop:lstmfull}(b$''$)]
(i)~From \cref{prop:exact}(b$'$), $2.5\eps(1.2^t-1)\ge w$ iff $1.2^t\ge1+w/(2.5\eps)$.

(ii)~\emph{Lower bound.} Part (c3) of \cref{prop:loss} gives $\nu_t\le\bar\nu_t:=\frac12\gamma^2\eps(\rho_A^t-1)/(\rho_A-1)$, and
$\bar\nu_t<\nu^*/2$ as long as $\rho_A^t<1+(\rho_A-1)\nu^*/(\gamma^2\eps)$.

\emph{Finiteness.} For every $\eps>0$, $\tsat(\eps)<\infty$ by (b): $\nu_t\to\nu_\eps\ge\nu^*>\nu^*/2$. The upper bound below needs
$\eps\le u_\delta$.

\emph{Upper bound.} Since $\gamma>5/6$, $\rho_A>1$. Fix $\delta\in(0,1)$ with $\rho_\delta:=1-\gamma+1.2\gamma^2(1-\delta)^2>1$, and
put $u_\delta:=\sqrt{3\delta}$. For $0\le u\le u_\delta$, $\tanh u\ge u-u^3/3\ge(1-\delta)u$ (the first inequality because
$f(u)=\tanh u-u+u^3/3$ has $f(0)=0$ and $f'(u)=u^2-\tanh^2u\ge0$). Let $\eps\le u_\delta$.

\emph{Phase 1.} Let $t_1:=\min\{t\ge0:\ C_t>u_\delta\ \text{or}\ 0.5\eps+1.2\nu_t>u_\delta\}$; $t_1\ge1$ because $C_0=\nu_0=0$ and
$\eps\le u_\delta$. For $1\le t\le t_1$ we have $C_{t-1}\le u_\delta$ and $0.5\eps+1.2\nu_{t-1}\le u_\delta$, so
$C_t\ge(1-\gamma)C_{t-1}+\gamma(1-\delta)(0.5\eps+1.2\nu_{t-1})$ and $\nu_{t-1}=\gamma\tanh C_{t-1}\ge\gamma(1-\delta)C_{t-1}$, hence
$C_t\ge\rho_\delta C_{t-1}+0.5\gamma(1-\delta)\eps$ and $C_t\ge0.5\gamma(1-\delta)\eps\,(\rho_\delta^t-1)/(\rho_\delta-1)$. As
$C_{t_1-1}\le u_\delta$,
\[
t_1\le1+\frac{\log\!\big(1+2(\rho_\delta-1)u_\delta/(\gamma(1-\delta)\eps)\big)}{\log\rho_\delta},
\]
which is finite. At $t_1$, either $C_{t_1}>u_\delta$ and then $\nu_{t_1}=\gamma\tanh C_{t_1}\ge\gamma\tanh u_\delta$, or
$1.2\nu_{t_1}>u_\delta-0.5\eps\ge u_\delta/2$. So $\nu_{t_1}\ge m_\delta:=\min\{\gamma\tanh u_\delta,\ u_\delta/2.4\}$, and
$C_{t_1}\ge\nu_{t_1}/\gamma$ because $\tanh C\le C$.

\emph{Phase 2.} For $s>0$ put $z_s:=(s,\gamma s)$. Choose $\delta'\in(0,1)$ with $(1-\delta')\big(1-\gamma+1.2\gamma^2(1-\delta')\big)
\ge1$ (possible: the left side tends to $\rho_A>1$ as $\delta'\to0$), and $s_0:=\min\{m_\delta/\gamma,\ u_{\delta'}/\rho_A,\
u_{\delta'}/(1.2\gamma)\}$. Then $T_0(z_{s_0})\ge z_{s_0}$: its first component is $C'=(1-\gamma)s_0+\gamma\tanh(1.2\gamma s_0)\ge
s_0\big(1-\gamma+1.2\gamma^2(1-\delta')\big)\ge s_0$, and its second is $\gamma\tanh C'\ge\gamma(1-\delta')C'\ge\gamma s_0$, using
$1.2\gamma s_0\le u_{\delta'}$ and $C'\le\rho_As_0\le u_{\delta'}$. Since $(C_{t_1},\nu_{t_1})\ge z_{s_0}$ and $T_\eps\ge T_0$ with both
maps monotone, $(C_{t_1+k},\nu_{t_1+k})\ge T_0^k(z_{s_0})$ for all $k\ge0$. The sequence $T_0^k(z_{s_0})$ is non-decreasing and bounded,
so it converges to a fixed point of $T_0$ that is at least $z_{s_0}>0$; positive fixed points have $\nu=\nu^*$ (proof of (b)). Hence
there is $k_0$, depending on $\gamma,\delta,\delta'$ but not on $\eps$, with $\nu_{t_1+k_0}\ge\nu^*/2$, and $\tsat(\eps)\le t_1+k_0$.

So $\log(1/\eps)/\log\rho_A+O(1)\le\tsat(\eps)\le\log(1/\eps)/\log\rho_\delta+O_\delta(1)$, and $\rho_\delta\to\rho_A$ as $\delta\to0$
gives the limit. (iii)~is part (c3) of \cref{prop:loss}.
\end{proof}

\begin{proof}[Proof of \cref{prop:exactac}(c)]
LSTM: $h'=\sigma(o)\tanh(c')$ is bounded by the product rule for a box in $[0,1]$ and a box in $[-1,1]$, so the $h$ box lies in
$[-1,1]^H$ after every step. With $h\in[-1,1]^H$ and bounded input boxes, the forget pre-activation box lies in some $[-\bar F,\bar F]$,
so $\sigma(f)\le\bar f:=\sigma(\bar F)<1$, and $\sigma(i)\tanh(g)\in[-1,1]$. With $\bar C:=1/(1-\bar f)$, a $c$ box inside
$[-\bar C,\bar C]^H$ is mapped into $[-(\bar f\bar C+1),\,\bar f\bar C+1]^H=[-\bar C,\bar C]^H$. So $[-1,1]^H\times[-\bar C,\bar C]^H$
is invariant under the interval cell, hence under the period map. GRU: the update $h'=(1-z)n+zh$ is bounded by the exact range of this
convex combination over the boxes of $z$, $n$ and $h$, which lies in the hull of the boxes of $n=\tanh(\cdot)\subset[-1,1]^H$ and of
$h$; so $[-1,1]^H$ is invariant. Deeper layers receive inputs in $[-1,1]^H$, which are bounded, and the same argument applies.
\end{proof}
There is no GRU analogue of \cref{prop:lstmfull}(b) in this paper; nothing here says that GRUs exhibit (b).

\begin{remark}[The construction of \cref{prop:lstmfull}, numerically]\label{rem:lstmnum}\label{rem:bpp-constants}
Here $\gamma=\sigma(8)=0.999664649870$ and $\rho_A=1.199530644769$; the positive root of $\nu=\gamma\tanh(\tanh(1.2\nu))$ is
$\nu^*=0.471667598642$ (bisection in float64, residual $5.6\times10^{-17}$), so the limit width of (b) tends to $2\nu^*=0.943335$ as
$\eps\to0$.
A 400-step interval run of the construction from the zero box gave the width $0.9433$ of the $h$ box at $\eps=10^{-8}$ and
$10^{-6}$, $0.9435$ at $10^{-4}$ and $0.9609$ at $10^{-2}$, while the per-coordinate range of $h$ at step 400 over 2000 random input
sequences, a lower bound on the width of the reachable set, was $3.736\times10^{-8}$, $3.552\times10^{-6}$, $3.550\times10^{-4}$ and
$3.356\times10^{-2}$; the computed spectral radii were $\rho(U)=0.849$ and $\rho(\abs{U})=1.200$. The widths at $10^{-8}$ and
$10^{-6}$ equal the limit $2\nu^*$ to the printed digits, and the larger widths at $10^{-4}$ and $10^{-2}$ are those of the least
fixed point $\nu_\eps>\nu^*$.
Iterating $T_\eps$ from $(0,0)$ gives $\tsat(\eps)=14,39,64,90,115,140$ at $\eps=10^{-2},10^{-4},\dots,10^{-12}$, against the lower
bound of (b$''$)(ii) of $12.9,37.6,63.0,88.3,113.6,138.9$ (one step above it after rounding up, in every row), and
$\tsat\log\rho_A/\log(1/\eps)=0.55,0.77,0.84,0.89,0.91,0.92$: the limit 1 of (b$''$)(ii) is approached slowly, because $\tsat$ stays
11--12 steps below $\log(1/\eps)/\log\rho_A$ (derived by hand from these values). In particular half the limit radius is reached after
90 steps at $\eps=10^{-8}$, well inside the 400 steps run. These are numbers for the construction of (b), a one-layer cell; they say
nothing about the measured saturation of the two-layer gated models (\cref{sec:exp-r4}).
\end{remark}

\begin{remark}[What (a) and (a$'$) do not say]\label{rem:notsay}
The constants of \cref{prop:exactac}(a) and \cref{prop:exact}(a$'$) contain $1/\denh(0)$, resp.\ $1/(1-\bar a')$ per layer, and
products of Lipschitz constants over the layers, and they can be huge. On a trained byte-level checkpoint, $r^*$ was linear in $\eps$
(a factor $10.0$--$10.1$ per decade from $10^{-6}$ to $10^{-4}$ in both layers), with gain $\max r^*/\eps$ about $1.5$ in the first
layer and about $711$ in the second.
On the natural nonnegative diagonal word-level models of the context-length probe (three seeds each) the margin looseness
$R=W/(\eps D)$ was between $508$ and $1.78\times10^{5}$ at every context length.
The separation is not between SSMs and RNNs. For a gated cell the relevant matrix is $M_t\ge\abs{J_t}$ (\cref{prop:loss}(c)), not
$\abs{J_t}$. An LSTM cell whose clean fixed point has $\rho(J)=\rho(\abs{J})<1$ can have $\rho(M)>1$, already with one unit, and then
no invariant box of width $O(\eps)$ contains that fixed point. An earlier two-unit GRU computation showed the same effect: a box gain
of $1.61$, the spectral radius of the path-sum matrix, against $\rho(\abs{J})=0.61$. An $O(\eps)$ invariant box can be expected when
the products of $M_t$ over the clean period contract; this is not proved here.
\end{remark}

\begin{remark}[The invariant-box search for LSTM and GRU]\label{rem:invbox}
For the gated models, the implementation iterates the interval period map from the zero box, keeps the running hull, widens a
stalled hull by $1\%$ (relative) at most three times, and stops after at most $64$ periods. A row without a certified invariant box is
reported as vacuous ($-\infty,+\infty$). By \cref{prop:exactac}(c), such a failure means that the search did not find an invariant box
within its budget, not that none exists.
In the IBP training under synonym substitution every IBP-trained LSTM and GRU collapsed to the majority class ($47.4\%$ of the test
subset; a pre-registered criterion of that study failed for this reason), so their $L=\infty$ to $L=0$ ratios (LSTM 0, GRU 1.0) carry
no information about \cref{prop:lstmfull}(b) or \cref{prop:exactac}(c).
On the natural two-layer models of the context-length probe the invariant box was vacuous on every document for the word-level LSTM
(three seeds) and the byte-level LSTM, and finite but saturated at $\eps=0$ for the GRU (median lower bound of the margin at $\eps=0$
of $-14.9$, $-15.9$ and $-14.5$ on seeds 0, 1 and 2; on seed 0 a median $19.6$ below the clean margins of the members it covers).
Over a single repeated token, the token-set evaluation, which uses this invariant box for LSTM and GRU, certified no gated model on any
document (\cref{app:anyctx}). An IBP-trained byte-level LSTM was certified at $0\%$ for every $\eps$ from $0.002$ to $0.02$ at its
training length; that is a finite-length result, on which \cref{prop:lstmfull}(b) has no bearing.
\end{remark}

Measurements on trained models that test these statements (the context-length probe, the complex-diagonal control, the interval-map
growth rates and the certified radius against distance) are reported in \cref{app:gate3,app:gate5}.

\subsection{Certificates for unbounded periodic contexts}\label{app:proofs-periodic}

This subsection proves \cref{lem:period}, \cref{thm:inf} and \cref{thm:hull}, states the periodic cascade and the hull
construction in full, and adds a horizon-bounded variant of \cref{thm:hull} (\cref{thm:hullK}) and the token-set corollary
(\cref{cor:anyctx}).

\paragraph{One period, unrolled.}
Fix one layer and one state channel with $P$-periodic coefficient boxes $(A_j,U_j)$, $j=1,\dots,P$. Running \eqref{eq:cr} over one
period from $(c_0,r_0)=(c,r)$ gives $c_j=\acen_jc_{j-1}+\ucen_j$ and $r_j=\ahi_jr_{j-1}+\arad_j\abs{c_{j-1}}+\urad_j$. The centre
recursion does not involve $r$, and the radius recursion is linear in $r$ for a given centre trajectory. Unrolling,
\[
c_j=\Gc_jc+c^0_j,\qquad r_j=\Gh_jr+\beta_j(c),\qquad c_P=\pic c+\cfree,\qquad r_P=\pih r+\beta(c),
\]
with the prefix products $\Gc_j:=\prod_{i\le j}\acen_i$ and $\Gh_j:=\prod_{i\le j}\ahi_i$ ($\Gc_0=\Gh_0=1$, so $\pic=\Gc_P$ and
$\pih=\Gh_P$), the centre trajectory from $0$, $c^0_j=\sum_{i\le j}\big(\prod_{i<m\le j}\acen_m\big)\ucen_i$,
$\cfree=c^0_P=\sum_jw^{\rm c}_j\ucen_j$ with $w^{\rm c}_j:=\prod_{i>j}\acen_i$, and
\[
\beta_j(c)=\sum_{i\le j}\Big(\prod_{i<m\le j}\ahi_m\Big)q_i(c),\qquad
\beta(c):=\beta_P(c)=\sum_{j=1}^Pw_j\,q_j(c),
\]
where $w_j:=\prod_{i>j}\ahi_i$ ($w_P=1$) and $q_j(c)=\arad_j\,\abs{\Gc_{j-1}c+c^0_{j-1}}+\urad_j$ is the radius injected at phase
$j$ when the period starts at centre $c$. \Cref{tab:notation} lists these symbols.

\begin{remark}[The DC form of \cref{lem:period}]\label{rem:dcform}
If every $\ahi_j<1$, put $\mu_j:=w_j(1-\ahi_j)/\denh$ and $Q_j:=q_j(c^*)/(1-\ahi_j)$, the \emph{DC level} of phase $j$: the radius at
which the recursion $r\leftarrow\ahi_jr+q_j(c^*)$ of that phase would settle if its injected radius $q_j(c^*)$ repeated at every
position.
Then $\mu_j\ge0$, $\sum_j\mu_j=1$ and
\[
r^*=\frac{\beta(c^*)}{\denh}=\sum_{j=1}^P\mu_jQ_j ,
\]
so $r^*$ is a convex combination of the DC levels (proof below, with the form for phases with $\ahi_j=1$); the same identity writes
$c^*$ as a convex combination of the $\ucen_j/(1-\acen_j)$. In a slowing Mamba or minGRU channel, $\denh$ and $\beta(c^*)$ tend to $0$
together while the DC levels stay bounded (\cref{lem:dc}): the quotient divides two vanishing quantities, and the DC form does not.
The implementation computes $r^*$ in this form (\cref{rem:impl}), and the outward-rounded re-verification encloses it without
amplification by $1/(1-\pih)$ (\cref{rem:float}).
\end{remark}

\begin{proof}[Proof of \cref{lem:period}]
\emph{Convexity.} $c_{j-1}(c)=\Gc_{j-1}c+c^0_{j-1}$ is affine in $c$ with slope $\Gc_{j-1}\ge0$, the absolute value is convex, and
$\beta$ and every $\beta_j$ are nonnegative combinations of the $q_i$; so they are convex, and $\beta$ is Lipschitz with constant
$\Lambda_\beta:=\sum_jw_j\arad_j\Gc_{j-1}$, since $\big|\abs{c_{j-1}(c)}-\abs{c_{j-1}(c')}\big|\le\Gc_{j-1}\abs{c-c'}$.

\emph{Contraction.} For $\kappa>\Lambda_\beta/\denh$ let $N_\kappa(c,r):=\max(\kappa\abs{c},\abs{r})$, and let $z=(c,r)$, $z'=(c',r')$.
The first component of $\Psi z-\Psi z'$ is $\pic(c-c')$, so $\kappa\abs{\pic(c-c')}\le\pic N_\kappa(z-z')$. The second has absolute
value at most $\pih\abs{r-r'}+\Lambda_\beta\abs{c-c'}\le(\pih+\Lambda_\beta/\kappa)N_\kappa(z-z')$. Since $\acen_j\le\ahi_j$,
$\pic\le\pih<1$, and $\pih+\Lambda_\beta/\kappa<1$ by the choice of $\kappa$; so $\Psi$ is a contraction in $N_\kappa$ with modulus
$\max(\pic,\pih+\Lambda_\beta/\kappa)<1$. $\Psi$ maps $\reals\times[0,\infty)$ into itself because $\beta\ge0$, and this set is closed
in $\reals^2$, hence complete.

\emph{Fixed point.} Banach's fixed-point theorem gives existence, uniqueness and geometric convergence from every start. Solving
$c=\pic c+\cfree$ (note $1-\pic\ge\denh>0$) and then $r=\pih r+\beta(c^*)$ gives the formulas. In particular the boxes
$(c_k,r_k):=\Psi^k(0,0)$ of $k$ periods from the zero state converge to $(c^*,r^*)$, at any rate larger than $\pih$ per period (let
$\kappa\to\infty$).

\emph{DC form} (\cref{rem:dcform}). With $w_{j-1}=\ahi_jw_j$, $w_0=\pih$ and $w_P=1$, the sum telescopes:
$\sum_{j=1}^Pw_j(1-\ahi_j)=\sum_{j=1}^P(w_j-w_{j-1})=1-\pih$. Hence $\mu_j\ge0$ and $\sum_j\mu_j=1$ when every $\ahi_j<1$, and
$r^*=\denh^{-1}\sum_jw_jq_j(c^*)=\sum_j\mu_jQ_j$. If some phases have $\ahi_j=1$ (in exact arithmetic this never happens for the three
families of \cref{tab:layers}), let $\mathcal J_0$ be these phases and $\mathcal J_+$ the others: their terms vanish in the
telescoping sum, $\sum_{j\in\mathcal J_+}\mu_j=1$, and
\[
r^*=\sum_{j\in\mathcal J_+}\mu_j\,\frac{q_j(c^*)}{1-\ahi_j}+\frac1{\denh}\sum_{j\in\mathcal J_0}w_j\,q_j(c^*).
\]
The same identity with $\acen$, $w^{\rm c}$ and $\ucen$ in place of $\ahi$, $w$ and $q$ writes $c^*$ as a convex combination of the
$\ucen_j/(1-\acen_j)$, plus the analogous term for phases with $\acen_j=1$.

\emph{Robust invariance.} \Cref{lem:cr}(a), phase by phase, and $\Psi(c^*,r^*)=(c^*,r^*)$: if $h_0\in H^*$ and $a_j\in A_j$,
$u_j\in U_j$ for $j=1,\dots,P$, then $h_P\in H^*$, and $h_j$ lies in the phase box $H^*_j:=[c^*_j-r^*_j,\,c^*_j+r^*_j]$ with
$(c^*_j,r^*_j)=(\Gc_jc^*+c^0_j,\ \Gh_jr^*+\beta_j(c^*))$. $H^*$ is the robustly positively invariant box at the fixed point of the
centre-radius box map \citep{blanchini1999}, not a minimal robustly positively invariant set of the true system.
\end{proof}

\begin{remark}[The structure of the period map, and a period that does not leak]\label{rem:denzero}
(i)~The Lipschitz matrix of $\Psi$ is lower triangular, $\begin{psmallmatrix}\pic&0\\ \Lambda_\beta&\pih\end{psmallmatrix}$, because
the centre never sees the radius; the weighted norm $N_\kappa$ of the proof is the textbook contraction argument for a
lower-triangular map with diagonal entries below one. (ii)~If $\denh=0$, the radius does not contract over a period:
$r_k=r_{k-1}+\beta(c_{k-1})$, so the boxes from the zero state grow without bound unless every $q_j$ vanishes along them. No finite
invariant box is produced, and the certificate is vacuous, though still sound.
\end{remark}

\begin{corollary}[Monotonicity of the fixed point]\label{cor:monofp}
If $A'_j\subset A_j$ and $U'_j\subset U_j$ for every phase and $\denh>0$ for the larger boxes, the fixed-point box and the phase boxes
of $(A',U')$ lie inside those of $(A,U)$.
\end{corollary}

\begin{proof}
The smaller boxes have $\pi'_{\rm h}\le\pih<1$. By \cref{lem:cr}(c) and induction, the box after $k$ periods from the zero state for
$(A',U')$ lies in the one for $(A,U)$. Both sequences converge (proof of \cref{lem:period}), and inclusion of closed intervals passes
to the limit.
\end{proof}

\paragraph{The periodic cascade and \cref{thm:inf}.}

\begin{definition}[Periodic cascade]\label{def:cascade}
Let $E^p_j$ be the input box of phase $j$. For $\ell=1,\dots,Q$ define: (i)~the layer-$\ell$ input boxes $V^\ell_j$ at the phases,
with $V^1_j=E^p_j$; (ii)~the coefficient boxes $(A^\ell_j,U^\ell_j)$ of every channel, obtained from $V^\ell$ by the natural interval
extension, with the convolution read periodically: phase $j$ reads phases $j-k_c+1,\dots,j$ modulo $P$; (iii)~for every channel, the
fixed point $(c^*,r^*)$ of \cref{lem:period} and the phase boxes $H^{*,\ell}_j$; (iv)~the output boxes, from $H^{*,\ell}_j$ and the
per-phase boxes of $x$, $z$ and $C$ through the interval extension of the readout, and $V^{\ell+1}_j$ as the residual sums. The
document is then processed layer by layer: the convolution at a document position $t<k_c-1$ reads the last phases of $V^\ell$, the
scan starts from $H^{*,\ell}_P$, and \eqref{eq:cr} runs through $t=0,\dots,n-1$. The final RMSNorm and the head at $t=n-1$ give the
logit box $\Binf(p,x)$, along with state boxes at every phase and at every document position. If $\denh=0$ in some channel, the box is
set to be infinite (\cref{rem:denzero}).
\end{definition}

\begin{lemma}[The implementation computes the cascade]\label{lem:implcascade}
In real arithmetic, the $L=\infty$ certificate of our implementation (for Mamba, the S4D-type and the minGRU-type models), applied to
the layout $[\,p_{P-P_0+1},\dots,p_P\mid p\mid x\,]$, returns $\Binf(p,x)$ and the cascade state boxes whenever $P\ge P_0\ge(k_c-1)Q$,
that is $P_0\ge3Q$ for $k_c=4$. The bound on $P_0$ is tight. (For the minGRU-type layer, up to the period decay above gate logit $20$,
\cref{app:numerics}.)
\end{lemma}

\begin{proof}
The layout is read with zero padding before its first position. Its first $P_0$ positions (the context positions) carry the tokens of
the last $P_0$ phases. Their state boxes are overwritten with the periodic phase boxes, but their readouts use the convolution outputs
computed at those positions. A layer-1 convolution output at layout position $t$ is the periodic one iff its window
$t-k_c+1,\dots,t$ avoids the padding, that is $t\ge k_c-1$; so layer-1 outputs are periodic from position $k_c-1$ on. The layer-2
convolution at $t$ reads layer-1 outputs at $t-k_c+1,\dots,t$, which are periodic iff $t\ge2(k_c-1)$. Inductively, the layer-$\ell$
convolution outputs are periodic from position $\ell(k_c-1)$ on, and in general not before. The period starts at position $P_0$, so
its coefficient boxes are the periodic ones in every layer if $P_0\ge Q(k_c-1)$. Then the period part is the cascade of
\cref{def:cascade}, and the document's convolution reads correct period tails. $P\ge P_0$ is needed because the context positions are a
suffix of the period. With $P_0=Q(k_c-1)-1$, the first period position of layer $Q$ reads one input corrupted by the padding, which in
general changes the boxes.
\end{proof}

The boundary is visible numerically: with two layers, $P_0=5$ changes the state boxes by $3.1\times10^{-4}$ and $P_0=6$ by
$2.8\times10^{-17}$; with three layers, $P_0=8$ changes the boxes of the third layer by $9.6\times10^{-4}$ and the logit box by
$1.9\times10^{-4}$, and $P_0=9$ by $3\times10^{-16}$. The implementation asserts $P\ge P_0\ge(k_c-1)Q$; every experiment uses $Q=2$ and
$P_0=8$, so $\Binf$ of \cref{thm:inf} is the output of the implementation.

The proof of \cref{thm:inf} has four steps: (A)~every finite context is covered by its own centre-radius IBP box
(\cref{cor:stack}(i)); (B)~these boxes converge to the cascade (\cref{lem:asymp}); (C)~the true logits converge (\cref{lem:limit});
(D)~the limit of points in converging closed boxes lies in the limit box.

\begin{lemma}[Asymptotically periodic contraction]\label{lem:asymp}
Let $\Psi_k$, $k\ge1$, be period maps of one channel built from coefficient boxes $(A_{k,j},U_{k,j})$ whose endpoints converge, as
$k\to\infty$, to those of periodic boxes $(A_j,U_j)$ with $\denh>0$. Let $\Psi$ be the period map of the limit boxes and $z^*$ its
fixed point. Then for every $z_0$ the sequence $z_k=\Psi_k(z_{k-1})$ converges to $z^*$, and the phase boxes of period $k$ converge to
the phase boxes $H^*_j$.
\end{lemma}

\begin{proof}
The constants $\pi_{{\rm c},k}$, $\pi_{{\rm h},k}$, $c_{{\rm free},k}$, $\Lambda_{\beta,k}$ and the coefficients of $\beta_k$ are
continuous in the box endpoints, so they converge to those of $\Psi$. Pick $\kappa$ with $\pih+\Lambda_\beta/\kappa<1$. By convergence
there are $k_0$ and $\bar\mu<1$ with $\pi_{{\rm c},k}\le\bar\mu$ and $\pi_{{\rm h},k}+\Lambda_{\beta,k}/\kappa\le\bar\mu$ for all
$k\ge k_0$, so every $\Psi_k$ with $k\ge k_0$ is a $\bar\mu$-contraction in $N_\kappa$ (proof of \cref{lem:period}). Then
\[
N_\kappa(z_k-z^*)\le N_\kappa(\Psi_kz_{k-1}-\Psi_kz^*)+N_\kappa(\Psi_kz^*-\Psi z^*)\le\bar\mu\,N_\kappa(z_{k-1}-z^*)+e_k ,
\]
with $e_k:=N_\kappa(\Psi_kz^*-\Psi z^*)\to0$. Hence
$N_\kappa(z_k-z^*)\le\bar\mu^{\,k-k_0}N_\kappa(z_{k_0}-z^*)+\sum_{i=k_0+1}^{k}\bar\mu^{\,k-i}e_i\to0$. The phase boxes of period $k$
are continuous functions of $z_{k-1}$ and of the period-$k$ coefficients (\cref{cor:stack}(iii)).
\end{proof}

\begin{lemma}[The infinite-context limit exists]\label{lem:limit}
Under the hypotheses of \cref{thm:inf}, $F_K(\tilde e)$ converges as $K\to\infty$ for every admissible $\tilde e$, and so do the states
of every layer at every fixed position $t<n$.
\end{lemma}

\begin{proof}
Let $\mathcal H_k$ denote the centre-radius IBP boxes, in all layers, of the $k$-th copy of the period counted from the start of a
context. By causality they do not depend on the total number of copies, as long as it is at least $k$.

\emph{Uniform bounds.} Layer 1 has exactly periodic coefficient boxes from the second copy on (only the first copy's convolution reads
the padding), so \cref{lem:asymp} applies to it. Layer $\ell+1$'s coefficient boxes in copy $k$ are continuous functions of layer
$\ell$'s boxes in copies $k-1$ and $k$, which converge; \cref{lem:asymp} applies again. Hence $\mathcal H_k$ converges to the cascade as
$k\to\infty$ and is bounded uniformly in $k$; the $n$ document positions add finitely many bounded boxes. By \cref{cor:stack}(i), every
true value of every context lies in one bounded set $\mathcal S$. On $\mathcal S$, the token-wise maps (RMSNorm with $\epsilon_{\rm
n}>0$, linear maps, SiLU, softplus, $\exp$, sigmoid, products) and the coefficient maps $\Phi$ are Lipschitz with a common constant
$\Lambda$, and $\abs{h}\le M_h$. Because the limit cascade has $\pih<1$ in every channel, there are $k_0$ and $\bar\pi<1$ such that, in
every layer and channel, the product of $\ahi$ over any copy $k\ge k_0$ is at most $\bar\pi$. True decays satisfy
$0\le a_t\le\ahi_t\le1$.

\emph{Induction over layers.} Let $K'>K\ge2^{Q+1}(k_0+1)$ and $W^\ell_K:=[-\lfloor K/2^{\ell-1}\rfloor P,\,n)$. A given copy has a
larger index in the $K'$-context than in the $K$-context, and for $\ell\ge2$ every copy inside $W^\ell_K$ has index at least $k_0$ in
both. We show by induction on $\ell$ that $\abs{v^{\ell,K'}_t-v^{\ell,K}_t}\le C_\ell\,\theta_\ell^{K}$ on $W^\ell_K$ for some
$C_\ell<\infty$, $\theta_\ell<1$. For $\ell=1$ the inputs coincide on $W^1_K=[-KP,n)$. Given the claim for $\ell$, the coefficients of
layer $\ell$ in the two contexts differ by at most $D_\ell\theta_\ell^K$, $D_\ell:=k_c\Lambda C_\ell$, at every position whose
convolution window lies in $W^\ell_K$. The state difference $\delta_t=h^{K'}_t-h^{K}_t$ obeys
\[
\delta_t=a^{K'}_t\,\delta_{t-1}+(a^{K'}_t-a^{K}_t)\,h^{K}_{t-1}+(u^{K'}_t-u^{K}_t),\qquad\text{so}\qquad
\abs{\delta_t}\le a^{K'}_t\,\abs{\delta_{t-1}}+(M_h+1)\,D_\ell\,\theta_\ell^{K}
\]
at the positions where the coefficients of the two contexts differ by at most $D_\ell\theta_\ell^K$.
For $\ell=1$ the coefficients coincide on the window, so the last term vanishes. Start $k_c$ positions after the left end of $W^\ell_K$
with $\abs{\delta}\le2M_h$. The homogeneous factor is at most $\bar\pi$ over each full copy of index at least $k_0$ and at most $1$
otherwise. If $m$ such copies separate the start from $t$,
\[
\abs{\delta_t}\le2M_h\,\bar\pi^{\,m}+\frac{(P+n)(M_h+1)\,D_\ell\,\theta_\ell^{K}}{1-\bar\pi}.
\]
For $t\in W^{\ell+1}_K$ we have $m\ge K/2^{\ell}-k_0-3$, so both terms decay geometrically in $K$. The outputs and the next inputs
$v^{\ell+1}_t=v^\ell_t+o_t$ are $\Lambda$-Lipschitz in $h_t$ and in the input window, which proves the claim for $\ell+1$. After the
last layer, the logits are a Lipschitz function of $v^{Q+1}_{n-1}$, and $n-1\in W^{Q+1}_K$. Hence
$\abs{F_{K'}(\tilde e)-F_K(\tilde e)}\le C\theta^K$ for all $K'>K$ large: the sequence is Cauchy. The same estimate gives the states.
\end{proof}

\begin{proof}[Proof of \cref{thm:inf}]
Let $\mathcal B^{(K)}$ be the centre-radius IBP logit box of $p^Kx$ from the zero state. (A)~$F_K(\tilde e)\in\mathcal B^{(K)}$ by
\cref{cor:stack}(i). (B)~The document part of $\mathcal B^{(K)}$ starts, in every layer, from the box at the end of the last copy and
reads the last copy's tail; as $K\to\infty$ these converge to $H^{*,\ell}_P$ and to the periodic tail (\cref{lem:asymp}, layer by
layer as in the proof of \cref{lem:limit}), so by \cref{cor:stack}(iii) $\mathcal B^{(K)}\to\Binf$, endpoint by endpoint.
(C)~$F_K(\tilde e)\to F_\infty(\tilde e)$ by \cref{lem:limit}. (D)~$\operatorname{dist}(F_\infty,\Binf)\le\abs{F_\infty-F_K}+
\operatorname{dist}(F_K,\mathcal B^{(K)})+d_H(\mathcal B^{(K)},\Binf)\to0$, with $d_H$ the Hausdorff distance, and $\Binf$ is closed.
The state boxes are handled identically.
\end{proof}

The certificate concerns the given period $p$ only, with independent perturbations at every position. It does not imply
$F_K(\tilde e)\in\Binf$ for any finite $K$:

\begin{example}[The infinite-context certificate does not cover finite contexts]\label{ex:finite}
One layer, one channel, $P=1$, point decay $a\equiv\frac12$, injection box $U=[0.9,1.1]$ at every position, a document of one token
with the same boxes, and the margin $h-1.45$ (certified iff positive). Infinite context: $c^*=1/(1-\frac12)=2$ and
$r^*=0.1/(1-\frac12)=0.2$; the box stays $[1.8,2.2]$ after the document token, and the margin is at least $0.35$, so \cref{thm:inf}
certifies it. One copy ($K=1$): after the period $h\in[0.9,1.1]$, after the document $h\in[1.35,1.65]$, and the margin lies in
$[-0.1,0.2]$, so $K=1$ is not certified. The constant choice $u\equiv0.9$ gives $h=1.35<1.45$, a real counterexample at $K=1$. The
infinite context has forgotten the zero initial state; a finite context has not.
\end{example}

\begin{remark}[When the infinite-context box does cover finite contexts]\label{rem:covers}
If, in every layer, the zero state lies in $H^{*,\ell}_P$ and the zero convolution history lies in the periodic convolution-input
boxes, then by \cref{cor:stack}(ii) and induction the centre-radius IBP boxes of $p^Kx$ lie in the cascade boxes for every $K$, and
$\Binf$ certifies every $K$. On a trained byte-level checkpoint at $\eps=0.002$ the condition fails: $0\in H^*$ in only $18$--$64\%$ of
the channels, and the $K=1$ state boxes leave the $L=\infty$ boxes in $1.2\%$ (first layer) and $15.5\%$ (second layer) of the channels.
Nor do finite-context certificates carry over to the limit: three standard IBP-trained byte-level models certified $43.7$, $47.6$ and
$38.6\%$ of the reviews after a 4096-byte licence-text prefix, but $0.0$, $0.9$ and $1.9\%$ at $L=\infty$ ($\eps=0.01$). That finite
prefix is contiguous licence text, not copies of $p$; these numbers were first computed with a float32 model around a float64 period
part, and the float64 re-evaluation of \cref{app:byte} reproduces them.
\end{remark}

\begin{remark}[Bi-infinite trajectories]\label{rem:biinf}
\Cref{def:limit} starts every finite context at $h=0$, as a deployed model does. If one asked instead for all bi-infinite solutions of
the recursions, uniqueness would fail: adding a solution $g$ of the homogeneous recursion $g_t=a_tg_{t-1}$ gives another solution, and
$g$ grows like $\pih^{-k}$ going $k$ copies back, so it leaves every box. Uniqueness holds among solutions with
$\abs{h_{-kP}}=o(\pih^{-k})$.
\end{remark}

\begin{remark}[Time scales]\label{rem:timescales}
By \cref{lem:asymp} the boxes approach $\Binf$ at rate $\pih$ per copy, so $\Binf$ describes contexts of order $1/\denh$ copies. In
trained models this can be astronomically long. A standard IBP-trained byte-level model had a second-layer channel with
$\denh=1.8\times10^{-5}$ per 1024-byte period (time constant about $6\times10^{7}$ tokens), and $17.6\%$ of its state channels had
$\denh<0.1$. Every $L=\infty$ number should therefore be read together with the distribution of the radius time constant
$\tau_{\rm h}=P/(-\log\pih)$ of the channels (\cref{app:gate2}).
The slow-channel attribution measured these distributions at each model's threat for 38 checkpoints (200 documents each, the periods
of the context-length probe). Under synonym substitution at box scale 1, the fraction of (document, channel) pairs with
$\tau_{\rm h}>4096$ tokens, maximised over the two layers, was $21.7\%$, $0.24\%$ and $31.9\%$ for the three standard-IBP Mamba seeds and
$69.5\%$, $46.0\%$ and $41.7\%$ for the three standard-IBP minGRU-type seeds; no channel of a non-degenerate standard-IBP S4D-type seed
exceeded 512 tokens (largest $\tau_{\rm h}$ 343 and 425 tokens). For the natural Mamba and minGRU-type models every second-layer pair
exceeded $10^8$ tokens at the threat (largest $\tau_{\rm h}=\infty$), because their gate-logit boxes at box scale 1 are wide, while no
clean time constant exceeded 4096 tokens. The byte-level fine-tunes with a finite-horizon curriculum reached $\tau_{\rm h}$ up to
$1.38\times10^{17}$, $1.24\times10^{18}$ and $3.45\times10^{14}$ tokens (\cref{app:byte}).
\end{remark}

\begin{remark}[Enclosures and the implementation]\label{rem:enclosure}
Suppose every implemented interval primitive returns a box that contains the exact range of its operation, as with the SiLU slack and
the softplus widening of \cref{rem:silu}. Then, layer by layer, the implemented coefficient boxes contain the ideal ones
(\cref{cor:stack}(ii)), the implemented fixed points contain the ideal ones (\cref{cor:monofp}), and the implemented logit box contains
$\Binf$, hence $F_\infty(\tilde e)$. Rounding to nearest is not an enclosure in this sense, as it can go either way, nor is the
minGRU-type period decay above gate logit $20$ (\cref{app:numerics}), a small error in the unsafe direction; outward rounding is
one, under the assumption of \cref{rem:float}.
\end{remark}

\paragraph{The hull construction and \cref{thm:hull}.}
\Cref{thm:inf} certifies only the limit. To certify $p^Kx$ for all $K$ at once, the start box of every copy is covered instead of the
fixed point alone.

\begin{definition}[Hull construction]\label{def:hullcons}
Let $P\ge(k_c-1)Q$, with $k_c$ the kernel size of the convolution and $Q$ the number of layers, and lay the period out as in
\cref{lem:implcascade}, with $P_0$ context positions and $P\ge P_0\ge(k_c-1)Q$.
In every layer, the convolution-input boxes of the last $k_c-1$ context positions are replaced by their hull with $0$, and the period's
convolution, hence its coefficient boxes $(\hat A_j,\hat U_j)$, are computed from these. With $(c^*,r^*)$ the fixed point of
$(\hat A,\hat U)$ in the DC form of \cref{lem:period}, every channel gets the phase boxes
\[
\hat H_j=\Big[\,c^0_j+\min(0,d_j)-\varrho_j,\ \ c^0_j+\max(0,d_j)+\varrho_j\,\Big],
\]
where $d_j=\Gc_j\,c^*$, $\varrho_j=\Gh_j\,\bar r+\max\{\beta_j(0),\beta_j(c^*)\}$ and $\bar r=\max\{r^*,\ \beta(0)/\denh\}$.
This is the hull of the centre trajectories from $0$ (namely $c^0_j$) and from $c^*$ (namely $c^0_j+d_j$), widened by a radius that
bounds every radius trajectory; $\beta(0)/\denh$ is evaluated in the DC form with the injections $q_j(0)$. A channel whose $c^*$ or
$\bar r$ is infinite gets the whole line. The context positions carry the boxes $\hat H_j$ of the last $P_0$ phases, and the outputs of
layer $\ell$ at the period positions are the period input boxes of layer $\ell+1$. In the document, the convolution reads the hull of
the period's tail with $0$, the scan starts from the hull of $\hat H_P$ with $0$, and \eqref{eq:cr} runs through $t=0,\dots,n-1$. The
final RMSNorm and the head give the logit box $\Bcup$.
\end{definition}

\begin{lemma}[One layer, uniform coefficients]\label{lem:hull1}
Consider one channel whose coefficient boxes in copy $k$ satisfy $(A_{k,j},U_{k,j})\subset(\hat A_j,\hat U_j)$ for all $k\ge1$ and all
phases $j$, with $\denh>0$ for $(\hat A,\hat U)$. Start \eqref{eq:cr} from $(0,0)$ at the beginning of copy 1. Then for every $k\ge1$
and every phase $j$, the \eqref{eq:cr} box at phase $j$ of copy $k$ lies in $\hat H_j$. The fixed-point phase boxes $H^*_j$ of
$(\hat A,\hat U)$ lie in $\hat H_j$ as well.
\end{lemma}

\begin{proof}
By \cref{lem:cr}(c) and induction over the phases it suffices to treat $(A_{k,j},U_{k,j})=(\hat A_j,\hat U_j)$ for all $k$. Write
$(c_k,r_k)$ for the box at the start of copy $k+1$, so that $(c_0,r_0)=(0,0)$.

\emph{Centres.} $c_k=\pic c_{k-1}+\cfree$ gives $c_k=(1-\pic^k)\,c^*$, by induction:
$\pic(1-\pic^{k-1})c^*+(1-\pic)c^*=(1-\pic^k)c^*$. Every start centre lies on the segment between $0$ and $c^*$.

\emph{Radii.} $r_0=0\le\bar r$, and if $r_{k-1}\le\bar r$ then
\[
r_k=\pih r_{k-1}+\beta(c_{k-1})\le\pih\bar r+\max\{\beta(0),\beta(c^*)\}\le\pih\bar r+\denh\,\bar r=\bar r,
\]
by convexity of $\beta$ on the segment (proof of \cref{lem:period}), $\beta(c^*)=\denh\,r^*$, and $\beta(0)\le\denh\,\bar r$. So
$r_k\le\bar r$ for every $k$.

\emph{Phases.} At phase $j$ of copy $k+1$ the centre is $c^0_j+\Gc_jc_k=c^0_j+(1-\pic^k)\,d_j$, which lies between
$c^0_j+\min(0,d_j)$ and $c^0_j+\max(0,d_j)$. The radius is $\Gh_jr_k+\beta_j(c_k)\le\Gh_j\bar r+\max\{\beta_j(0),\beta_j(c^*)\}=
\varrho_j$, by convexity of $\beta_j$. Hence the box lies in $\hat H_j$. The fixed point has centre $c^0_j+d_j$ and radius
$\Gh_jr^*+\beta_j(c^*)\le\varrho_j$.
\end{proof}

\begin{proof}[Proof of \cref{thm:hull}]
By hypothesis $P\ge(k_c-1)Q$, and the construction runs on the layout of \cref{lem:implcascade} with $P\ge P_0\ge(k_c-1)Q$, as the
implementation does (it asserts the inequality, so every reported hull satisfies it). Count layout positions from $0$, as in the
proof of \cref{lem:implcascade}: the context positions are $t=0,\dots,P_0-1$, and position $t$ carries phase $P-P_0+1+t$. The
hull's \emph{tail boxes} in layer $\ell$ are the convolution-input boxes that the period's convolution reads, hulled with $0$: those at
the last $k_c-1$ context positions, computed token-wise from the hull's layer-$\ell$ input boxes there. We show by induction over the
layers that, for every $K\ge0$, the centre-radius IBP boxes of $p^Kx$ from the zero state (\cref{cor:stack}) lie in the boxes of the hull construction in the following sense, and that the same holds for
the cascade of \cref{def:cascade}:
\begin{itemize}
\item[($\mathrm H_\ell$)] the layer-$\ell$ input box at phase $j$ of every copy lies in the hull's layer-$\ell$ input box at the period
position of phase $j$ and, if phase $j$ is carried by a context position $t\ge(\ell-1)(k_c-1)$, in its layer-$\ell$ input box at $t$;
and the layer-$\ell$ input box at every document position lies in the hull's there.
\end{itemize}
Since $P_0\ge(k_c-1)Q$, the last $k_c-1$ context positions have $t\ge P_0-k_c+1\ge(\ell-1)(k_c-1)$ for every $\ell\le Q$, so
($\mathrm H_\ell$) covers the positions of the tail boxes. ($\mathrm H_1$) holds: the layer-1 input boxes are the input boxes $E^p_j$
of the period and $E_t$ of the document, in every copy and in the hull.

\emph{Period part of layer $\ell$.} Assume ($\mathrm H_\ell$). The convolution at a phase $j<k_c$ reads positions before the copy:
zero padding in copy 1, and the tail of copy $k-1$ in copy $k\ge2$, whose convolution-input boxes lie in the hull's tail boxes by
($\mathrm H_\ell$) and the token-wise inclusion monotonicity. The hull uses at these positions the hull of the tail boxes with $0$,
which contains both cases. By
\cref{cor:stack}(ii), the coefficient boxes of every copy lie in $(\hat A_j,\hat U_j)$, and by \cref{lem:hull1} the state boxes at
every phase of every copy lie in $\hat H_j$. Readout, gate, output map and residual are token-wise and inclusion monotone, so the
outputs and the layer-$(\ell+1)$ input boxes at phase $j$ of every copy lie in the hull's at the period position of phase $j$. At a
context position $t\ge\ell(k_c-1)$, carrying phase $j$, the hull's convolution reads the context positions $t-k_c+1,\dots,t$, which lie
at or after $(\ell-1)(k_c-1)$ and carry the phases $j-k_c+1,\dots,j$ of one copy ($j\ge1+t\ge k_c$), so their boxes contain those of
every copy by ($\mathrm H_\ell$); the hull's state box there is $\hat H_j$, which contains those of every copy; and the same
monotonicity gives the layer-$(\ell+1)$ input boxes there.
This is the reach-back count of \cref{lem:implcascade}: the zero padding in front of the layout reaches the hull's layer-$(\ell+1)$
boxes only at context positions $t<\ell(k_c-1)$, which ($\mathrm H_{\ell+1}$) does not claim and the period does not read. For the
cascade, the coefficient boxes are those of the periodic convolution, which reads the cascade's tail, inside the hull's tail boxes by
($\mathrm H_\ell$); so they lie in $(\hat A_j,\hat U_j)$, its phase boxes lie in the $\hat H_j$ by \cref{cor:monofp} and the last
statement of \cref{lem:hull1}, and the same monotonicity carries its input boxes to layer $\ell+1$.

\emph{Document part of layer $\ell$.} For $K\ge1$, the convolution at a document position $t<k_c-1$ reads the tail of copy $K$, and the
scan starts from the state box at the end of copy $K$, which lies in $\hat H_P$. For $K=0$, the convolution reads zero padding and the
scan starts from $h=0$. The hull uses the hull of the period tail with $0$ and the hull of $\hat H_P$ with $0$, which contain every
case, including the cascade. \Cref{lem:cr}(c) and \cref{cor:stack}(ii) carry the containment through the document, which gives the
document part of ($\mathrm H_{\ell+1}$).

So the logit box of $p^Kx$ lies in $\Bcup$ for every $K\ge0$, and \cref{cor:stack}(i) gives (i). The cascade part gives
$\Binf\subset\Bcup$; the hull's $\denh>0$ implies the cascade's, since the cascade's decay boxes are smaller, and \cref{thm:inf} gives
$F_\infty(\tilde e)\in\Binf\subset\Bcup$ for every admissible perturbation of the infinite context.
\end{proof}

\begin{corollary}[Certificate for every context length]\label{cor:everyK}
If $\Bcup$ certifies $y$, the model predicts $y$ on every admissible perturbation of $p^Kx$, for every $K\in\{0,1,2,\dots\}$, and on
every admissible perturbation of the infinite context in the sense of \cref{def:limit}.
\end{corollary}

\begin{proof}
\Cref{thm:hull}, the last sentence of its proof, and the definition of certification.
\end{proof}

\begin{remark}[The price of uniformity]\label{rem:price}
As $\eps\to0$ the hull does not shrink to a point: it still contains the clean trajectories from $0$ and from $c^*$, that is, the clean
outputs for every $K$, with or without the context. A K-uniform certificate therefore requires the classifier to be correct on the
clean $p^Kx$ for all $K$ simultaneously, and its width does not vanish with $\eps$. Measured:
\begin{itemize}
\item On the fixed-point fine-tuned checkpoint of the byte-level study (32 hand-written reviews, one 1024-byte licence-text period per review,
$\eps=0.01$), the plain centre-radius boxes of $K=0,1,2,4$ and $\infty$ were all inside the hull boxes (worst float64 slack
$-6\times10^{-16}$, rounding level; the test tolerance is $-10^{-12}$ relative), the hull certified 27 of the 32 reviews, and its logit
width was $1.35$--$1.57$ times that of the plain $L=0$ box.
\item In the IBP training under synonym substitution (box scale 1, 1000 documents, float64; mean $\pm$ sd over three seeds) the hull
certified $2.8\pm4.7\%$ (standard-IBP Mamba), $0.9\pm0.9\%$ (minGRU-type) and $33.2\pm10.9\%$ (S4D-type), against $47.6\pm4.5\%$,
$52.3\pm6.6\%$ and $42.5\pm4.6\%$ at $L=0$ and $3.8\pm4.7\%$, $1.5\pm1.2\%$ and $39.8\pm8.3\%$ at $L=\infty$; after fixed-point
fine-tuning, $19.9\pm19.4\%$,
$10.6\pm7.5\%$ and $39.2\pm5.2\%$ (S4D-type: seeds 0 and 2; seed 1, evaluated later, $48.9\%$).
\item On the natural models of the context-length probe ($\ell_\infty$ probe), the hull was vacuous already at $\eps=0$ on
$87.5$--$99.5\%$ of the documents of the Mamba and minGRU-type models (three seeds each, and the byte-level Mamba), because slow
channels make the segment of start centres between $0$ and $c^*$ very long, and it was finite but far below the clean margin for the
S4D-type models (median lower bound of the margin at $\eps=0$ of $-3.39\times10^{4}$, $-2.4\times10^{4}$ and $-1.61\times10^{4}$ on
seeds 0, 1 and 2). At $\eps=10^{-8}$ it certified at most $0.5\%$ of the documents of any natural model, while the $L=\infty$
certificate of \cref{thm:inf} certified $84$--$93.5\%$ of them.
\item The certified-trained models run through the same probe were hull-certified at $\eps=10^{-8}$ on $85.5\%$ (S4D-type, standard
IBP, seed 0), $68.5\%$ (Mamba, fixed-point fine-tuning, seed 2), $28.5\%$ (Mamba, standard IBP, seed 2) and $18\%$ (minGRU-type,
fixed-point fine-tuning, seed 0) of
the documents, against $96\%$, $80\%$, $45\%$ and $86\%$ for the $L=\infty$ certificate (200 documents each).
\end{itemize}
No K-uniform certificate exceeded a member it covers, in any test: the integration tests of the training code ($K=0,\dots,64$) and the
production run of the context-length probe (its containment check at every $\eps$, $\eps=0$ included).
\end{remark}

\paragraph{A horizon-bounded variant.}

\begin{lemma}[Horizon-bounded uniform box]\label{lem:hullK}
Fix $K_{\max}\ge1$ and consider one channel whose coefficient boxes in copies $k=1,\dots,K_{\max}$ satisfy
$(A_{k,j},U_{k,j})\subset(\hat A_j,\hat U_j)$ for all phases $j$, with $\denh>0$ for $(\hat A,\hat U)$; nothing is assumed about later
copies. Start \eqref{eq:cr} from $(0,0)$ at the beginning of copy 1, and put $c_{K_{\max}}:=(1-\pic^{K_{\max}})\,c^*$,
\[
\bar r_{K_{\max}}:=\max\{\beta(0),\beta(c_{K_{\max}})\}\,\frac{1-\pih^{K_{\max}}}{\denh},\qquad
\varrho^{K_{\max}}_j:=\Gh_j\bar r_{K_{\max}}+\max\{\beta_j(0),\beta_j(c_{K_{\max}})\},
\]
\[
\hat H^{K_{\max}}_j:=\Big[c^0_j+\min(0,\Gc_jc_{K_{\max}})-\varrho^{K_{\max}}_j,\ \ c^0_j+\max(0,\Gc_jc_{K_{\max}})+\varrho^{K_{\max}}_j\Big].
\] Then for every
$k\in\{0,\dots,K_{\max}-1\}$ and every phase $j$ the \eqref{eq:cr} box at phase $j$ of copy $k+1$ lies in $\hat H^{K_{\max}}_j$; the box
at the end of copy $k$, for every $k\in\{0,\dots,K_{\max}\}$, lies in $[\min(0,c_{K_{\max}})-\bar r_{K_{\max}},\
\max(0,c_{K_{\max}})+\bar r_{K_{\max}}]$; and $\hat H^{K_{\max}}_j\subset\hat H_j$.
\end{lemma}

\begin{proof}
As in \cref{lem:hull1}, reduce to $(A_{k,j},U_{k,j})=(\hat A_j,\hat U_j)$. Since $0\le\pic<1$, $1-\pic^k$ increases in $k$, so
$c_k=(1-\pic^k)c^*$ lies on the segment between $0$ and $c_{K_{\max}}$ for $k\le K_{\max}$. Unrolling the radius recursion from
$r_0=0$, $r_k=\sum_{i<k}\pih^{k-1-i}\beta(c_i)\le\max_{c\in[0,c_{K_{\max}}]}\beta(c)\,(1-\pih^k)/\denh\le\bar r_{K_{\max}}$ for
$k\le K_{\max}$, by convexity of $\beta$ (proof of \cref{lem:period}), which puts the maximum at an end of the segment. The phases follow
as in \cref{lem:hull1}, with $\Gc_j\ge0$ and the convexity of $\beta_j$. Finally, $c_{K_{\max}}$ lies between $0$ and $c^*$, and
$\beta(c_{K_{\max}})\le\max\{\beta(0),\beta(c^*)\}=\denh\bar r$ by convexity, so $\bar r_{K_{\max}}\le\bar r$ and
$\beta_j(c_{K_{\max}})\le\max\{\beta_j(0),\beta_j(c^*)\}$, which gives $\hat H^{K_{\max}}_j\subset\hat H_j$.
\end{proof}

\Cref{thm:hullK} uses only the first statement of \cref{lem:hullK} (the phase boxes). The end-of-copy statement is correct but not
used: the box it gives puts the radius $\bar r_{K_{\max}}$ around the zero state as well (\cref{rem:startbox}). Written with the
geometric sums $c_{K_{\max}}=\cfree\sum_{i<K_{\max}}\pic^i$ and $\sum_{i<K_{\max}}\pih^i$ in place of $(1-\pih^{K_{\max}})/\denh$, the
first two statements hold without $\denh>0$ (the proof uses $\pic,\pih\ge0$ and the convexity shown in the proof of \cref{lem:period},
which does not use $\denh>0$); the inclusion in $\hat H_j$ needs it, because $\hat H_j$ does.

\begin{theorem}[Certificate for every number of copies up to a horizon]\label{thm:hullK}
Let the model, $p$, $x$ and the input boxes be as in \cref{thm:hull}. Run the hull construction with $\hat H^{K_{\max}}_j$ in place of
$\hat H_j$ in every layer (the coefficient boxes of layer $\ell+1$ computed from layer $\ell$'s boxes of this construction) and start
the document from $\hull(\hat H^{K_{\max}}_P,0)$, with the convolution reading the hull of the period's tail with $0$, as in
\cref{thm:hull}. Assume $\denh>0$ in every channel of every layer of this construction (from layer 2 on its coefficient boxes differ
from those of \cref{thm:hull}; in exact arithmetic the condition holds for finite boxes, as for \cref{lem:period}). Then the resulting
logit box $\mathcal B^{K_{\max}}_\cup$ contains $F_K(\tilde e)$ for every $K\in\{0,\dots,K_{\max}\}$ and every admissible perturbation
$\tilde e$ of $p^Kx$.
\end{theorem}

\begin{proof}
As for \cref{thm:hull}(i), by induction over the layers: if the \eqref{eq:cr} boxes of copies $1,\dots,K_{\max}$ of layer $\ell$ lie in
this construction's boxes, the coefficient boxes of those copies in layer $\ell+1$ lie in this construction's coefficient boxes
(\cref{cor:stack}(ii); the first copy's convolution reads zero padding, which the hull with $0$ covers), and \cref{lem:hullK}, whose
hypothesis concerns only copies $1,\dots,K_{\max}$, gives the containment in layer $\ell+1$. For $1\le K\le K_{\max}$ the scan of the
document starts from the box at the end of copy $K$, which is the box at phase $P$ of copy $K$ and lies in $\hat H^{K_{\max}}_P$
(\cref{lem:hullK}); for $K=0$ it starts from $h=0$. Both lie in $\hull(\hat H^{K_{\max}}_P,0)$. The document part is that of
\cref{thm:hull} with $\hat H^{K_{\max}}_P$ in place of $\hat H_P$.
\end{proof}

For one layer and the same coefficient boxes, $\hat H^{K_{\max}}_j\subset\hat H_j$, hence $\hull(\hat H^{K_{\max}}_P,0)\subset
\hull(\hat H_P,0)$ (\cref{lem:hullK}), and \cref{lem:cr}(c) with \cref{cor:stack}(ii) carries the inclusion through the document: in
that layer \cref{thm:hullK} is never wider than \cref{thm:hull}. Across layers we do not claim this inclusion: the coefficient boxes of
layer $\ell+1$ shrink, but the hull formula is not proved inclusion monotone in its coefficient boxes (the centre $c^*$ of a smaller
box can move away from $0$). The intersection $\mathcal B^{K_{\max}}_\cup\cap\Bcup$ of the two logit boxes is sound for every
$K\le K_{\max}$ and never wider than $\Bcup$. \Cref{thm:hullK} is not implemented and was not evaluated.

\begin{remark}[Why the document starts from $\hull(\hat H^{K_{\max}}_P,0)$]\label{rem:startbox}
Starting the document instead from the end-of-copy box of \cref{lem:hullK},
$[\min(0,c_{K_{\max}})-\bar r_{K_{\max}},\ \max(0,c_{K_{\max}})+\bar r_{K_{\max}}]$, would also be sound, but it puts the radius
$\bar r_{K_{\max}}$ around the zero state of $K=0$ as well, and then the construction can be wider than \cref{thm:hull} already in one
layer. \Cref{thm:hull} starts from $\hull(\hat H_P,0)$, whose lower end is $\min(0,\cfree-\bar r)$ when $c^*\ge0$, because the end of
every copy $k\ge1$ has centre at least $\cfree$. Take the channel of \cref{ex:finite} (one layer, $P=1$, point decay $a=1/2$,
$U=[9/10,11/10]$ at every position, a one-token document with the same boxes).
\begin{itemize}
\item \Cref{thm:hull}: $c^*=2$ and $r^*=\bar r=1/5$, so $\hat H_1=[4/5,11/5]$. The document starts from $[0,11/5]$ and ends in
$[9/10,11/5]$.
\item The end-of-copy start with $K_{\max}=1$: $c_1=1$ and $\bar r_1=1/10$. The document starts from $[-1/10,11/10]$ and ends in
$[17/20,33/20]$.
\item The true union over $K\in\{0,1\}$ is $[9/10,11/10]\cup[27/20,33/20]$ (for $K=0$ the state is $u$, for $K=1$ it is $u_1/2+u_2$);
its interval hull is $[9/10,33/20]$.
\end{itemize}
For the margin $h-7/8$, \cref{thm:hull} certifies every $K$, and the end-of-copy start does not certify $K\le1$. The construction of
\cref{thm:hullK} has $\hat H^1_1=[17/20,33/20]$; the document starts from $[0,33/20]$ and ends in $[9/10,77/40]\subset[9/10,11/5]$.
\end{remark}

\begin{remark}[An optional tightening by one copy]\label{rem:tighten}
\Cref{lem:hullK} is loose by one copy. The phase boxes of copies $1,\dots,K_{\max}$ start from $(c_k,r_k)$ with $k\le K_{\max}-1$, so
they may use $c_{K_{\max}-1}$ and $\bar r_{K_{\max}-1}$, written with the geometric sums $c_K=\cfree\sum_{i<K}\pic^i$ and
$\bar r_K=\max\{\beta(0),\beta(c_K)\}\sum_{i<K}\pih^i$, so that $c_0=0$ and $\bar r_0=0$. The proof is unchanged, and \cref{thm:hullK}
stays sound with $\hat H^{K_{\max}-1}_j$ in place of $\hat H^{K_{\max}}_j$. In the example of \cref{rem:startbox} ($K_{\max}=1$) the
document then starts from $[0,11/10]$ and ends in $[9/10,33/20]$, which is exactly the interval hull of the true union over
$K\in\{0,1\}$. For the same coefficient boxes $\hat H^{K-1}_j\subset\hat H^K_j$, so in one layer the tightened construction is never
wider. \Cref{thm:hullK} is stated with $K_{\max}$; the tightening is optional and not implemented.
\end{remark}

\Cref{lem:hullK} makes the time scale of \cref{rem:timescales} visible: for a channel with $\denh\ll1/K_{\max}$, the radius bound
$\bar r_{K_{\max}}\le\max\{\beta(0),\beta(c_{K_{\max}})\}\,K_{\max}$ (Bernoulli) can be much smaller than
$\bar r=\max\{\beta(0),\beta(c^*)\}/\denh$, and the width of $\mathcal B^{K_{\max}}_\cup$ can grow with $K_{\max}$ until $K_{\max}$
reaches the slowest time constants.

\paragraph{Memory after the information.}
\Cref{thm:hull,thm:hullK} cover the context before the document. The following results concern the context after it: the trailing
positions between the information that a prediction needs and the position where the prediction is read.

\begin{lemma}[Fading memory over all inputs]\label{lem:F}
Let the model be a $Q$-layer stack of one of the three nonnegative diagonal families of \cref{tab:layers}, in real arithmetic, and let
every input embedding lie in a compact set $\mathcal E$. Let $\bar a<1$ and the compact set $\mathcal K$ be those of \cref{lem:uniform}
for $\mathcal E$, let $M_h$ bound $\abs h$ on $\mathcal K$ in every layer and channel, and let $\Lambda\ge1$ be a common Lipschitz
constant on $\mathcal K$ of the token-wise maps, of the coefficient maps $\Phi$ of a convolution window, and of the readout and output
maps (they exist because these maps are locally Lipschitz: RMSNorm with $\epsilon_{\rm n}>0$, linear maps, SiLU, softplus, $\exp$,
sigmoid and products, as in the proof of \cref{lem:limit}). There is $C<\infty$, depending only on the model, $\mathcal E$, $\bar a$,
$\Lambda$, $M_h$, $Q$ and $k_c$, with the following property. Let $e$ and $e'$ be any two sequences over $\mathcal E$, of any lengths,
each read from $h=0$ with zero convolution padding, that agree on their last $d$ positions. Then the logits at their last positions
satisfy
\[
\norm{F(e)-F(e')}_\infty\le C\,(d+1)^{Q-1}\,\bar a^{\,d}.
\]
\end{lemma}

\begin{proof}
\emph{Setup.} Index the common suffix by $s=1,\dots,d$, with $s=d$ the last position. Every clean value of both inputs lies in
$\mathcal K$, and every clean decay of both is at most $\bar a$, in every layer, channel and position (\cref{lem:uniform}). Primes
denote $e'$.

\emph{Layer 1.} The inputs agree at every common position, and at $s\ge k_c$ the convolution window $s-k_c+1,\dots,s$ lies in the
suffix. So at $s\ge k_c$ the coefficients agree, $a_s=a'_s$ and $u_s=u'_s$, and so do the quantities computed from the window
($x_s$, $z_s$, and $B_s$, $C_s$ for Mamba). Per channel, $\delta_s:=h'_s-h_s$ (the sign of \cref{lem:fading}(ii) and of the proof of
\cref{lem:limit}) obeys $\delta_s=a_s\delta_{s-1}$ with $0\le a_s\le\bar a$, and $\abs{\delta_{k_c-1}}\le2M_h$; so
$\abs{\delta_s}\le2M_h\bar a^{\,s-k_c+1}$ for $s\ge k_c-1$. The output at $s$ is $\Lambda$-Lipschitz in $h_s$ and in the agreeing
quantities, and $v^2=v^1+o$, so the layer-2 inputs differ by at most $C_2\bar a^{\,s}$ for $s\ge s_2:=k_c$, with
$C_2:=2\Lambda M_h\bar a^{-(k_c-1)}$.

\emph{Induction.} Suppose the layer-$\ell$ inputs differ by at most $C_\ell\,s^{\ell-2}\bar a^{\,s}$ at every common $s\ge s_\ell$.
At $s\ge s_\ell+k_c-1$ the window lies in that range, so the coefficients differ by at most
$k_c\Lambda C_\ell\,s^{\ell-2}\bar a^{\,s-k_c+1}=:D_\ell\,s^{\ell-2}\bar a^{\,s}$. Then
$\delta_s=a'_s\delta_{s-1}+(a'_s-a_s)h_{s-1}+(u'_s-u_s)$, with $0\le a'_s\le\bar a$ and $\abs{h_{s-1}}\le M_h$, gives
$\abs{\delta_s}\le\bar a\abs{\delta_{s-1}}+(M_h+1)D_\ell\,s^{\ell-2}\bar a^{\,s}$. Starting from $\abs\delta\le2M_h$ at
$s_0:=s_\ell+k_c-2$,
\[
\abs{\delta_s}\le2M_h\bar a^{\,s-s_0}+\sum_{t=s_0+1}^{s}\bar a^{\,s-t}(M_h+1)D_\ell\,t^{\ell-2}\bar a^{\,t}
\le\big(2M_h\bar a^{-s_0}+(M_h+1)D_\ell\big)\,s^{\ell-1}\bar a^{\,s}.
\]
The outputs, and $v^{\ell+1}=v^\ell+o$, are $\Lambda$-Lipschitz in $h_s$ and in the window, so the layer-$(\ell+1)$ inputs differ by
at most $C_{\ell+1}\,s^{\ell-1}\bar a^{\,s}$ for $s\ge s_{\ell+1}:=s_\ell+k_c-1$.

\emph{Conclusion.} After $Q$ layers, the final RMSNorm and the head at $s=d$ are Lipschitz, which gives the claim for
$d\ge s_{Q+1}$. For the finitely many $d<s_{Q+1}$ the logits are bounded on $\mathcal K$, say by $M_F$, and $C\ge2M_F\bar a^{-s_{Q+1}}$
covers them.
\end{proof}

This is the argument of \cref{lem:fading}(ii) and of the induction over layers in the proof of \cref{lem:limit}, applied to two
inputs that agree on a suffix instead of two truncations of one context. \Cref{lem:fading} is stated under the hypotheses of
\cref{prop:loss}(b3), which include the smoothness assumption (A1) and $D_\infty>0$ (A3); its part (ii) uses only the uniform
contraction, the compact set and the Lipschitz constants, so \cref{lem:F} needs neither. Admissible perturbations are covered: with
$\mathcal E$ the union of all input boxes (compact, as a finite union of boxes), every admissible perturbation is a sequence over
$\mathcal E$, and so is every concrete substitution.

\begin{proposition}[No single certificate of memory over trailing contexts of every length]\label{prop:M1}
Let the model and $\mathcal E$ be as in \cref{lem:F}. Fix a readout segment $\mathsf r$ of $m$ tokens with input boxes; the logits are
read at its last position. For every $d$ in an unbounded set $\mathcal L\subset\mathbb N$ fix a nonempty family $\mathcal T_d$ of
admissible trailing inputs of length $d$ (all input boxes lie in $\mathcal E$), for example the admissible perturbations of $K$ copies of a period $p$, $d=KP$, each copy
perturbed independently. For a document $x$ with input boxes, let $\mathcal S(x)$ be the set of all inputs
$[\,\tilde x\mid w\mid\tilde{\mathsf r}\,]$ with $\tilde x$ and $\tilde{\mathsf r}$ admissible and
$w\in\bigcup_{d\in\mathcal L}\mathcal T_d$. If $x$ and $x'$ have labels $y\ne y'$, there are no two closed boxes
$\mathcal B\supseteq F(\mathcal S(x))$ and $\mathcal B'\supseteq F(\mathcal S(x'))$ such that $\mathcal B$ certifies $y$ and
$\mathcal B'$ certifies $y'$.
\end{proposition}

\begin{proof}
\emph{The margins.} Suppose there were such boxes. Certification is $\ell^{\rm lo}_y>\max_{k\ne y}\ell^{\rm hi}_k$ with finite
bounds (\cref{sec:setting}), so $\gamma:=\ell^{\rm lo}_y-\ell^{\rm hi}_{y'}>0$ on $\mathcal B$, and the margin $\ell_y-\ell_{y'}$ is
at least $\gamma$ at every point of $\mathcal B$. Likewise $\ell_{y'}-\ell_y\ge\gamma'>0$ on $\mathcal B'$.

\emph{The two inputs.} Fix admissible $\tilde x$, $\tilde x'$ and $\tilde{\mathsf r}$, and for $d\in\mathcal L$ any
$w\in\mathcal T_d$. The inputs $[\,\tilde x\mid w\mid\tilde{\mathsf r}\,]$ and $[\,\tilde x'\mid w\mid\tilde{\mathsf r}\,]$ agree on
their last $d+m$ positions.

\emph{Fading memory.} By \cref{lem:F} their logits differ by at most $C(d+m+1)^{Q-1}\bar a^{\,d+m}$ per coordinate, so their margins
$\ell_y-\ell_{y'}$ differ by at most twice that.

\emph{The contradiction.} The first margin is at least $\gamma$ (its logits lie in $\mathcal B$) and the second at most $-\gamma'$
(in $\mathcal B'$): a difference of at least $\gamma+\gamma'$. Since $\mathcal L$ is unbounded, some $d\in\mathcal L$ makes
$2C(d+m+1)^{Q-1}\bar a^{\,d+m}<\gamma+\gamma'$.
\end{proof}

\begin{remark}[What \cref{prop:M1} covers]\label{rem:M1what}
The proof uses only that one compact set contains the logits of every input of $\mathcal S(x)$ and that the margin is strictly
positive on it. So it holds for the paper's notion of a certificate (one logit box for the set, \cref{sec:setting}) and for every
method that certifies the whole set by one compact set of logits: boxes, zonotopes, linear relaxations over the set. It is a
statement in real arithmetic, for the nonnegative diagonal class. It uses no hypothesis on $\denh$ of any construction, only
\cref{lem:uniform}.
\end{remark}

\begin{remark}[What it does not cover]\label{rem:M1not}
Separate certificates, one box per trailing length $d$, can certify both labels at every finite length only if their margins tend to
$0$ as $d$ grows (\cref{lem:F} forces that). \Cref{prop:M1} does not exclude them (for example, point decays with zero injections on
every trailing box). Nor does it restrict the context before the information: \cref{thm:hull} covers every number of copies there,
$K=\infty$ included. A single-box certificate of memory can be unbounded before the information; after it, \cref{prop:M1} forces it
to bound the trailing length.
\end{remark}

\begin{corollary}[The limit of trailing copies]\label{cor:M1a}
Let $\mathcal T_d$ be the admissible perturbations of $K$ copies of a period $p$, $d=KP$, and index the copies backwards from the
readout segment (copy $1$ adjacent to $\mathsf r$), as \cref{def:limit} indexes the context backwards from the document. For every admissible sequence
$\tilde e$ of copy perturbations, the logits $F_K$ of $[\,\tilde x\mid\tilde p_K\cdots\tilde p_1\mid\tilde{\mathsf r}\,]$ converge
as $K\to\infty$. The limit depends neither on $\tilde x$ nor on $x$: it is the $F_\infty$ of \cref{def:limit} for the ``document''
$\tilde{\mathsf r}$ after the context $p$, the quantity that \cref{thm:inf} certifies for the document $\mathsf r$. Consequently a
closed box that contains every $F_K$ contains the limit, and it can certify only the label of the limit, $y_\infty(p,\mathsf r)$.
\end{corollary}

\begin{proof}
$F_K$ and $F_{K'}$ ($K'>K$) are logits of inputs that agree on their last $KP+m$ positions, so by \cref{lem:F} the sequence is Cauchy.
Replacing $\tilde x$ by the empty document changes $F_K$ by at most the same bound, so the limits coincide; with the empty document,
$F_K$ is the $F_K$ of \cref{def:limit} for the document $\tilde{\mathsf r}$. A closed box that contains every $F_K$ contains the
limit.
\end{proof}

\begin{corollary}[A K-uniform certificate started from the document's box]\label{cor:M1b}
Suppose the hull construction were started from the document's final box, to report a K-uniform certified accuracy for inputs
$[\,\text{document}\mid\text{perturbed context}\mid\text{readout}\,]$. By \cref{cor:M1a}, a single-box certificate that covers every
$K$ certifies document $i$ only if $y_i=y_\infty(p_i,\mathsf r)$: a rule that depends on document $i$ only through its period $p_i$.
If each document gets its own period from a generator that does not see the labels, as in the evaluations of \cref{app:remedy}, the
certified share is at most that rule's accuracy, which at $n=1000$ documents lies in $[42.2,57.8]\%$ with probability at least $0.99$
under the model of the draw as i.i.d.\ starts independent of the labels (the expectation lies in $[47.4,52.6]\%$ for the split's label
shares, and Hoeffding's inequality for the $n$ independent per-document terms allows a deviation above $5.15$ points with probability
at most $0.01$; bounds rounded outward).
\end{corollary}

\begin{remark}[A non-zero start box]\label{rem:startboxM}
After the information, a certificate can still cover a horizon of trailing context. The hull construction of \cref{def:hullcons} and
its horizon-bounded variant (\cref{lem:hullK,thm:hullK}) start every channel from the zero state; the lemma (M2) below starts it from a
box, the document's end box, the theorem (M3) then certifies $[\,x\mid p^K\mid\mathsf r\,]$ for every $K$ up to a horizon, and the
corollary (M4) adds a leading context of every length. All period quantities of \cref{tab:notation} are those of $(\hat A,\hat U)$,
and $S_K:=\sum_{i<K}\pih^i$ and $S^{\rm c}_K:=\sum_{i<K}\pic^i$ are the geometric sums of \cref{rem:tighten}.

\emph{(M2) One channel, a non-zero start box, a horizon.} Hypotheses: (H1)~one channel of a nonnegative diagonal layer; copies
$k=1,2,\dots$ of a $P$-phase period have coefficient boxes $(A_{k,j},U_{k,j})\subset(\hat A_j,\hat U_j)$, with $\hat A_j\subset[0,1)$;
(H2)~\eqref{eq:cr} starts at the beginning of copy $1$ from the box $(c_D,r_D)$, with $r_D\ge0$; (H3)~a horizon $K\in\{1,2,\dots\}$,
and (H1) for the copies $1,\dots,K$; only these copies are used; (H3$\infty$)~the variant $K=\infty$ assumes (H1) for every copy and
$\denh>0$ for $(\hat A,\hat U)$. Put
\[
c_K:=\pic^Kc_D+\cfree S^{\rm c}_K,\qquad M_K:=\max\{\beta(c_D),\beta(c_K)\},
\]
\[
\bar r_K(D):=\max\{r_D,\ \pih^Kr_D+M_KS_K\},\qquad
\varrho^K_j(D):=\Gh_j\bar r_K(D)+\max\{\beta_j(c_D),\beta_j(c_K)\},
\]
\[
\hat H^K_j(D):=\Big[c^0_j+\min(\Gc_jc_D,\Gc_jc_K)-\varrho^K_j(D),\ \ c^0_j+\max(\Gc_jc_D,\Gc_jc_K)+\varrho^K_j(D)\Big],
\]
and for $K=\infty$ put $c_\infty:=c^*$ and $\bar r_\infty(D):=\max\{r_D,\,M_\infty/\denh\}$ in these formulas.
Then: (a)~for every $k\in\{0,\dots,K-1\}$ with copies $1,\dots,k+1$ present and every phase $j$, the \eqref{eq:cr} box at phase $j$ of
copy $k+1$ lies in $\hat H^K_j(D)$; (b)~for $k\in\{0,\dots,K\}$, the box at the start of copy $k+1$ lies in
$[\min(c_D,c_K)-\bar r_K(D),\ \max(c_D,c_K)+\bar r_K(D)]$; (c)~(a) and (b) hold for \eqref{eq:cr} started from any box
$H_0\subset[c_D-r_D,c_D+r_D]$; (d)~under (H3$\infty$), (a) holds for every $k\ge0$, and the fixed-point phase boxes $H^*_j$ of
$(\hat A,\hat U)$ lie in $\hat H^\infty_j(D)$; (e)~\emph{reductions:} with $(c_D,r_D)=(0,0)$, $K=\infty$ gives the $\hat H_j$ of
\cref{def:hullcons}, and (a) and (d) are \cref{lem:hull1}; finite $K$ gives the $\hat H^{K_{\max}}_j$ of \cref{lem:hullK} with
$K=K_{\max}$, and (a) and (b) are its first two statements; (f)~\emph{width:} if every $\ahi_j<1$, set
$\bar Q_K(D):=\max_j\max\{Q_j(c_D),Q_j(c_K)\}$ with $Q_j(c)=q_j(c)/(1-\ahi_j)$; then
$\bar r_K(D)\le\max\{r_D,\ \pih^Kr_D+(1-\pih^K)\bar Q_K(D)\}$ and, when $\pic<1$, $\abs{c_K-c_D}\le(1-\pic^K)\abs{c^*-c_D}$.

\begin{proof}[Proof of (M2)]
\emph{Step 0 (reduction).} The \eqref{eq:cr} box map is inclusion monotone in the state box and in the coefficient boxes
(\cref{lem:cr}(c)). By induction over the positions of copies $1,\dots,K$, the actual box lies in the box of the same position
computed with $(\hat A_j,\hat U_j)$ from the same start, as in the first sentence of the proof of \cref{lem:hull1}. So it suffices to
treat $(A_{k,j},U_{k,j})=(\hat A_j,\hat U_j)$. Write $(c_k,r_k)$ for the box at the start of copy $k+1$, so that $(c_0,r_0)=(c_D,r_D)$.

\emph{Step 1 (centres).} One period maps the centre by $c\mapsto\pic c+\cfree$ (``One period, unrolled''). By induction,
$c_k=\pic^kc_D+\cfree S^{\rm c}_k$: indeed $\pic(\pic^{k-1}c_D+\cfree S^{\rm c}_{k-1})+\cfree=\pic^kc_D+\cfree S^{\rm c}_k$.
Subtracting two consecutive steps gives $c_{k+1}-c_k=\pic(c_k-c_{k-1})=\pic^k(c_1-c_0)$. Since $\pic\ge0$, every increment has the
sign of $c_1-c_0$ or is $0$, so $k\mapsto c_k$ is monotone, and $c_0,\dots,c_K$ lie on the segment between $c_D$ and $c_K$. For
$K=\infty$, $\pic\le\pih<1$ by (H3$\infty$), because $\acen_j\le\ahi_j$ (proof of \cref{lem:period}, ``Contraction''); then
$\cfree S^{\rm c}_k=(1-\pic^k)c^*$ and $c_k=c_D+(1-\pic^k)(c^*-c_D)$, and every $c_k$ lies on the segment $[c_D,c^*]$.

\emph{Step 2 (radii).} The radius maps by $r\mapsto\pih r+\beta(c)$ (``One period, unrolled''); unrolled,
$r_k=\pih^kr_D+\sum_{i<k}\pih^{k-1-i}\beta(c_i)$. $\beta$ is convex on $\reals$: a nonnegative combination of the $q_i$, each the
absolute value of an affine function plus a constant (proof of \cref{lem:period}, ``Convexity''). On the segment of Step 1,
$\beta(c_i)\le M_K$ for $i\le K$. So $r_k\le f(k):=\pih^kr_D+M_KS_k$ for $k\le K$. Now $f(k+1)-f(k)=\pih^k(M_K-\denh r_D)$ has the
same sign for every $k$, or vanishes, so $\max_{0\le k\le K}f(k)=\max\{f(0),f(K)\}=\bar r_K(D)$. No division occurs; $\denh$ may be $0$. For
$K=\infty$, by induction: $r_0=r_D\le\bar r:=\bar r_\infty(D)$, and $r_{k-1}\le\bar r$ gives
$r_k\le\pih\bar r+M_\infty\le\pih\bar r+\denh\bar r=\bar r$, because $M_\infty\le\denh\bar r$ by the definition of $\bar r$. This
proves (b).

\emph{Step 3 (phases).} At phase $j$ of copy $k+1$ the centre is $c^0_j+\Gc_jc_k$ and the radius $\Gh_jr_k+\beta_j(c_k)$ (``One
period, unrolled''). $\Gc_j\ge0$, so by Step 1 the centre lies between $c^0_j+\Gc_jc_D$ and $c^0_j+\Gc_jc_K$. $\beta_j$ is convex
(same argument), so $\beta_j(c_k)\le\max\{\beta_j(c_D),\beta_j(c_K)\}$; with Step 2, the radius is at most $\varrho^K_j(D)$. This
proves (a).

(c)~By \cref{lem:cr}(c), the \eqref{eq:cr} trajectory from $H_0\subset[c_D-r_D,c_D+r_D]$ stays, position by position, inside the
trajectory from $(c_D,r_D)$.

(d)~Steps 1--3 with $K=\infty$ give (a) for every $k$. The fixed-point phase box has centre $c^0_j+\Gc_jc^*$ and radius
$\Gh_jr^*+\beta_j(c^*)$ (proof of \cref{lem:period}, ``Robust invariance''). It lies inside, because
$r^*=\beta(c^*)/\denh\le M_\infty/\denh\le\bar r_\infty(D)$.

(e)~Finite $K$ at $(0,0)$: $c_K=\cfree S^{\rm c}_K=(1-\pic^K)c^*$ when $\pic<1$, which is the $c_{K_{\max}}$ of \cref{lem:hullK};
$\bar r_K(0)=\max\{0,M_KS_K\}=\max\{\beta(0),\beta(c_K)\}\,(1-\pih^K)/\denh$, its $\bar r_{K_{\max}}$; $\varrho$ and $\hat H$ then
coincide with its $\varrho^{K_{\max}}_j$ and $\hat H^{K_{\max}}_j$; and its first two statements hold without $\denh>0$ in the
geometric-sum form (the paragraph after the proof of \cref{lem:hullK}). $K=\infty$ at $(0,0)$:
$\bar r_\infty(0)=\max\{\beta(0),\beta(c^*)\}/\denh=\max\{\beta(0)/\denh,\,r^*\}$, the $\bar r$ of \cref{def:hullcons}, because
$\beta(c^*)=\denh r^*$ (\cref{lem:period}); and $\hat H^\infty_j(0)$ is the $\hat H_j$ of \cref{def:hullcons}.

(f)~The weights $w_j(1-\ahi_j)$ telescope to $\denh$, so $\beta(c)=\sum_jw_j(1-\ahi_j)\,Q_j(c)\le\denh\max_jQ_j(c)$ (proof of
\cref{prop:horizon}(b)); hence $M_KS_K\le\bar Q_K(D)\,(1-\pih^K)$. The centre bound is Step 1.
\end{proof}

Interpretation, used by no proof: per channel, the box after $k$ trailing copies has centre $c_D+(1-\pic^k)(c^*-c_D)$ (Step 1) and a
radius that moves from $r_D$ towards $r^*$ (Step 2); it interpolates between the document's box and the context's fixed-point box, at
the rates $\pic^k$ (centre) and $\pih^k$ (radius), and (M2)'s start box (claim (b)) is the hull of that path, which always contains the document's box.
By Bernoulli, $1-\pic^K\le K(1-\pic)$ and $1-\pih^K\le K\denh$: $K(1-\pic)\ll1$ keeps the centre,
$\abs{c_K-c_D}\le K(1-\pic)\abs{c^*-c_D}$, and $K\denh\ll1$ bounds the added radius by $K\denh\bar Q_K(D)$ (part (f)). Since
$\pic\le\pih$, $1-\pic\ge\denh$: wide decay boxes can give $K(1-\pic)=O(1)$ while $K\denh\ll1$, so the radius condition alone does not
keep the centre. Both conditions together are memory at a price linear in the horizon, \cref{prop:horizon}(b) read from the
document's box. A channel with $K\denh\gg1$ (hence $K(1-\pic)\gg1$) ends near the context's fixed-point box, which does not depend on
the document.

\emph{(M3) A memory certificate up to a horizon.} Let the model and the arithmetic be those of \cref{thm:inf}: a $Q$-layer stack of
nonnegative diagonal layers, real arithmetic, the natural interval extension and \eqref{eq:cr}. The input is
$[\,x\mid p^K\mid\mathsf r\,]$, read from $h=0$ with zero convolution padding: $x$ is a document of $n\ge1$ tokens with boxes $E_t$;
$p$ a period of $P\ge k_c-1$ tokens with boxes $E^p_j$, the same boxes in every copy, each copy perturbed independently; $\mathsf r$ a
readout segment of $m\ge1$ tokens with boxes $E^{\mathsf r}_\nu$, $\nu=1,\dots,m$; and $K_{\max}\ge1$ is a horizon. The construction
runs per layer $\ell=1,\dots,Q$ from the layer-$\ell$ input boxes $V^\ell_t$ at the document positions, $\hat V^\ell_j$ at the phases
$j=1,\dots,P$ and $\hat V^\ell_\nu$ at the readout positions ($V^1$ are the input boxes):
\begin{itemize}
\item[(D)] \emph{Document.} Run the plain \eqref{eq:cr} pass over $x$ from $h=0$ with zero padding. It gives, per channel, every state
box and the end box $(c_D,r_D)$ at $t=n-1$, and it gives the convolution-input boxes $X_t$ of every document position and the outputs
at the document positions.
\item[(P)] \emph{Period.} $\hat X_j$ are the convolution-input boxes at the phases, token-wise from $\hat V^\ell_j$. The boundary
boxes are $\hat X^\partial_i:=\hull(X_{n-i},\hat X_{P+1-i})$ for the offsets $i=1,\dots,k_c-1$, with $X_{n-i}:=\{0\}$ when $n-i<0$;
this is exact here, because the document starts from $h=0$, so the positions before it are the zero padding. The convolution at
phase $j$ reads $\hat X^\partial$ at the window offsets before the copy and $\hat X$ at the copy's own phases $\le j$; through the
natural interval extension this gives the coefficient boxes $(\hat A_j,\hat U_j)$ of every channel and the per-phase quantities of
the layer's readout. (M2) with start $(c_D,r_D)$ and horizon $K_{\max}$ gives the phase boxes $\hat H^{K_{\max}}_j(D)$. The outputs at
the phases are token-wise from $\hat H^{K_{\max}}_j(D)$ and the per-phase quantities, and $\hat V^{\ell+1}_j$ is $\hat V^\ell_j$ plus
the output.
\item[(R)] \emph{Readout segment.} The convolution at readout position $\nu$ reads $\hat X^\partial$ at the window offsets before the
segment and the segment's own convolution-input boxes at positions $\le\nu$. The scan starts from
$\hull\big(\hat H^{K_{\max}}_P(D),[c_D-r_D,c_D+r_D]\big)$ and runs \eqref{eq:cr} over $\nu=1,\dots,m$. This gives the outputs and
$\hat V^{\ell+1}_\nu$.
\item[(H)] \emph{Head.} After layer $Q$, the final RMSNorm and the head at $\nu=m$ give $\mathcal B^{K_{\max}}_{\rm mem}(x,p,\mathsf r)$.
\end{itemize}
\emph{Claim.} For every $K\in\{0,\dots,K_{\max}\}$ and every admissible perturbation of $[\,x\mid p^K\mid\mathsf r\,]$, the logits lie
in $\mathcal B^{K_{\max}}_{\rm mem}(x,p,\mathsf r)$. No hypothesis on $\denh$ is needed.

\begin{proof}[Proof of (M3)]
Fix $K\le K_{\max}$. Let $V^{\ell,K}$ denote the layer-$\ell$ input boxes of the plain \eqref{eq:cr} IBP pass of
$[\,x\mid p^K\mid\mathsf r\,]$ from the zero state (\cref{cor:stack}); by \cref{cor:stack}(i) every true value lies in its box. We show
by induction over the layers:
\begin{itemize}
\item[($\mathrm H'_\ell$)] (1)~$V^{\ell,K}_t=V^\ell_t$ at every document position; (2)~at phase $j$ of every copy $k=1,\dots,K$,
$V^{\ell,K}\subset\hat V^\ell_j$; (3)~at every readout position $\nu$, $V^{\ell,K}\subset\hat V^\ell_\nu$.
\end{itemize}
($\mathrm H'_1$) holds: in both, the layer-1 boxes are the input boxes, and the copies share the period's boxes.

\emph{Document part.} The plain pass is causal: at a document position it reads only document positions and the zero padding. Its
layer-$\ell$ computation there is therefore exactly (D). This gives ($\mathrm H'_{\ell+1}$)(1).

\emph{Period part.} Convolution-input boxes are token-wise and inclusion monotone in the input boxes (\cref{cor:stack}(ii)). At a
phase $j<k_c$, copy 1's window reads the document's last positions, $X_{n-i}$ (or the zero padding when $n-i<0$); copy $k\ge2$'s
window reads copy $k-1$'s last phases, $P+1-i\ge P+2-k_c\ge1$ because $P\ge k_c-1$, whose boxes lie in $\hat X_{P+1-i}$ by
($\mathrm H'_\ell$)(2). Both lie in $\hat X^\partial_i$. So by \cref{cor:stack}(ii) every copy's coefficient boxes lie in
$(\hat A_j,\hat U_j)$. The plain state box at the start of copy 1 is the document's end box $(c_D,r_D)$, by ($\mathrm H'_\ell$)(1).
(M2)(a), with the hypotheses (H1)--(H3) for the copies $1,\dots,K$ present and $K\le K_{\max}$, gives: the plain state box at phase
$j$ of copy $k+1$ lies in $\hat H^{K_{\max}}_j(D)$, for every $k+1\le K$. The layer's readout, gate, output map and residual are
token-wise and inclusion monotone, so the plain layer-$(\ell+1)$ input box at phase $j$ of every copy lies in $\hat V^{\ell+1}_j$.
This gives ($\mathrm H'_{\ell+1}$)(2).

\emph{Readout part.} The window offsets before the segment carry, for $K=0$, the document's tail $X_{n-i}$ (or the zero padding)
and, for $K\ge1$, copy $K$'s tail, which lies in $\hat X_{P+1-i}$ by ($\mathrm H'_\ell$)(2); both lie in $\hat X^\partial_i$. The plain
scan starts, for $K=0$, from $(c_D,r_D)$ and, for $K\ge1$, from the box at the end of copy $K$, which is the box at phase $P$ of copy
$K$ and lies in $\hat H^{K_{\max}}_P(D)$ by (M2)(a) with $k+1=K$. Both lie in the hull the construction starts from. \Cref{lem:cr}(c)
and \cref{cor:stack}(ii) carry the inclusion through the segment, and so through its outputs. This gives ($\mathrm H'_{\ell+1}$)(3).

\emph{Conclusion.} After layer $Q$ the final RMSNorm and the head are token-wise and inclusion monotone, so the plain logit box lies
in $\mathcal B^{K_{\max}}_{\rm mem}$, and \cref{cor:stack}(i) gives the claim. (M2) for finite $K$ needs no condition on $\denh$ (its
Step 2 uses $S_K$), so (M3) needs none either.
\end{proof}

Compared with \cref{thm:hullK}: that theorem runs the period first, from $0$, and the document after; (M3) runs the document first,
so the period's start box is the document's end box and its convolution boundary is the document's tail. The layout condition
$P_0\ge(k_c-1)Q$ of \cref{lem:implcascade} is not needed: the period's coefficient boxes need not be periodic here, and whatever copy
$1$ reads before its start (the document's tail, or the zero padding when $n<k_c-1$) enters the boundary boxes of (P) exactly;
$P\ge k_c-1$ makes a window reach at most one copy back. As for \cref{thm:hullK}, inclusion monotonicity in the coefficient
boxes across layers is not claimed (the paragraph after the proof of \cref{thm:hullK}). The variant $K_{\max}=\infty$ ((M2)(d)) is
sound, but its boxes contain the document-independent fixed-point boxes: by \cref{prop:M1} it certifies at most one label, a
certificate of forgetting, not of memory. The cost per layer is the closed-form period part (scans over the $P$ phases; $\pic^K$, $\pih^K$, $S^{\rm c}_K$ and
$S_K$ in closed form) and scans over the $n$ document and $m$ readout positions; it does not depend on $K_{\max}$.

\emph{(M4) The sandwich.} The input is $[\,p^K\mid x\mid p'^{K'}\mid\mathsf r\,]$. The leading period $p$ ($P$ tokens, $P_0$ context
positions) satisfies the hypotheses of \cref{thm:hull}: those of \cref{thm:inf}, $P\ge(k_c-1)Q$, and $\denh>0$ in every channel of
every layer of the leading construction. The trailing period $p'$ has $P'\ge k_c-1$ tokens, and $K'_{\max}\ge0$ is a horizon. Per
layer:
\begin{itemize}
\item[(L)] the period part of \cref{def:hullcons} on $[\,\text{context }(P_0)\mid p\,]$: the phase boxes $\hat H_j$;
\item[(D)] the document part of \cref{def:hullcons} on $x$: the scan starts from $\hull(\hat H_P,0)$, the convolution reads
$\hull(\text{period tail},0)$ at the offsets before $x$, and \eqref{eq:cr} runs over $x$. This gives the document boxes, the end box
$(c_D,r_D)$ and the document tail $X_{n-i}$. \emph{Short documents:} for $n-i<0$ the tail position lies before $x$, in the leading
part, and its box is (D)'s boundary box at that offset, $\hull(\text{period tail},0)$ at offset $i-n$. It is never $\{0\}$ alone: for
$K\ge1$ that position holds the period's tail; the rule $X_{n-i}:=\{0\}$ of (M3) applies only there, where the document starts from
$h=0$;
\item[(T)] the trailing part: for $K'_{\max}\ge1$, (M3)'s (P) and (R) on $p'$ and $\mathsf r$, with the start $(c_D,r_D)$, the
document tail of (D) and the horizon $K'_{\max}$; for $K'_{\max}=0$ there is no (P): the segment's scan starts from $(c_D,r_D)$, and
its convolution reads the document tail of (D) at the offsets before the segment.
\end{itemize}
\emph{Claim.} For every $K\in\{0,1,2,\dots\}$, every $K'\in\{0,\dots,K'_{\max}\}$ and every admissible perturbation of
$[\,p^K\mid x\mid p'^{K'}\mid\mathsf r\,]$, the logits lie in the resulting box; so, for each $K'$, does the limit $K\to\infty$ of
\cref{def:limit} with the ``document'' $[\,x\mid p'^{K'}\mid\mathsf r\,]$.

\begin{proof}[Proof of (M4)]
\emph{Leading part and document.} Apply the induction of the proof of \cref{thm:hull}, with its hypothesis ($\mathrm H_\ell$), to the
``document'' $[\,x\mid p'^{K'}\mid\mathsf r\,]$. Restricted to the positions of $x$, it gives, for every $K\ge0$ and for the cascade,
that the plain layer-$\ell$ input boxes and state boxes at the positions of $x$ lie in the construction's (D) boxes (its ``Document
part''). The same induction puts the plain boxes of the positions just before $x$ (zero padding for $K=0$, the period's tail for
$K\ge1$) inside $\hull(\text{period tail},0)$.

\emph{Trailing part.} For $K'_{\max}\ge1$, (M3)'s period and readout parts go through with ($\mathrm H'_\ell$)(1) weakened from
equality to inclusion. The plain start box at the beginning of the first trailing copy lies in $[c_D-r_D,c_D+r_D]$, and (M2)(c)
covers a start inside the construction's box. The plain document tail, including the positions before a short $x$, lies in the
construction's tail boxes, so the boundary hull still contains it. For $K'_{\max}=0$ the trailing part is the continuation of (D)
over $\mathsf r$, and the same inclusions carry through it (\cref{lem:cr}(c), \cref{cor:stack}(ii)); this is the document part of
\cref{thm:hull} for the document $[\,x\mid\mathsf r\,]$. This gives the containment for every $K$, the cascade included, and every
$K'\le K'_{\max}$.

\emph{The limit.} For fixed $K'$ and a fixed perturbation, the logits converge as $K\to\infty$ by \cref{lem:limit}, applied with the
document $[\,x\mid p'^{K'}\mid\mathsf r\,]$: its hypotheses are those of \cref{thm:inf}, and the leading construction's $\denh>0$
implies the cascade's (the last paragraph of the proof of \cref{thm:hull}). The box is closed, so the limit lies in it.
\end{proof}

With $K'_{\max}=0$ (no trailing period, as defined in (T)) the construction reduces exactly to \cref{thm:hull} for the document
$[\,x\mid\mathsf r\,]$. Its cost per layer is two closed-form period parts and scans over the $n$ and $m$ positions, independent of
$K$ and $K'_{\max}$. It certifies every number of copies before the information and every trailing length that is a multiple of $P'$
up to $K'_{\max}P'$ after it, at constant cost; by \cref{prop:M1}, no single box can do the same for every trailing length for
documents of both labels. This construction is evaluated in \cref{app:memory}.
\end{remark}

\paragraph{Every context over a token set.}

\begin{corollary}[Every context over a token set]\label{cor:anyctx}
Assume the hypotheses of \cref{thm:hull}, in particular $\denh>0$ in every channel of every layer of each hull construction below (for
the minGRU-type layer as implemented, see the caveat of \cref{app:numerics}). Let $S\subset V$ be a set of tokens and $E_S$ the box
hull of their embeddings (a point if $\abs{S}=1$); let the document $x$ have input boxes $E_t$, $t\ge0$. Take $P\ge P_0\ge(k_c-1)Q$
(\cref{lem:implcascade}) and, for each residue $r\in\{0,\dots,P-1\}$, let $\Bcup^{(r)}$ be the logit box of the hull construction with
every period box equal to $E_S$ and the document $x^{(r)}:=[\,E_S^{\,r}\mid x\,]$ ($r$ positions with box $E_S$, then $x$). If every
$\Bcup^{(r)}$ certifies $y$, the model predicts $y$ on $[\,w\mid\tilde x\,]$ for every context $w\in S^m$ of every length $m\ge0$, read
from the zero state, and every admissible perturbation $\tilde x$ of the document, and on every admissible perturbation of the infinite
context over $E_S$ in the sense of \cref{def:limit}.
\end{corollary}

\begin{proof}
Write $m=KP+r$ with $K\ge0$ and $0\le r<P$. Every token of $w$ has its embedding in $E_S$, so the embedded input $[\,w\mid\tilde x\,]$
is an admissible perturbation of $p^Kx^{(r)}$, where $p$ is the period of $P$ positions with box $E_S$: the first $KP$ tokens of $w$
fill the $K$ copies, and its last $r$ tokens the first $r$ positions of $x^{(r)}$. The constant boxes are $P$-periodic, and
\cref{thm:hull}(i) for $x^{(r)}$ gives $F_K\in\Bcup^{(r)}$, which certifies $y$. The infinite context is \cref{cor:everyK} with $r=0$.
\end{proof}

The implementation of the token-set certificate uses $P_0=8$ and $P\in\{8,64\}$, and the layout $[E_S^{P_0}\mid E_S^P\mid E_S^r\mid x]$;
for the three nonnegative diagonal families (the minGRU-type hull included, so the caveat above applies to its rows) it is an instance
of \cref{cor:anyctx}. For the bag-of-words model an exact closed form replaces it, and for the CNN the hull over its sets of
convolution windows. For LSTM and GRU the implementation replaces the hull by the invariant box of \cref{rem:invbox}: those rows are not
instances of \cref{cor:anyctx}; their uniformity in the number of copies follows from the invariance of that box, so a gated row that
certifies nothing is a statement about that construction (\cref{app:anyctx} reports what each construction certifies). The loop over
residues is forced by the layout, in which the context is a suffix of the period, so $P\ge P_0$; the argument itself does not need it
for constant boxes, but a one-run variant ($P=1$ with a context of $P_0$ copies) would need its own check, and we do not claim it. A box
hull of $\abs{S}\ge2$ embeddings consists almost entirely of points that embed no token, so the certificate covers far more than the
contexts in $S^m$; and the hull is not proved inclusion monotone in $E_S$.

\subsection{Finite-horizon objectives}\label{app:proofs-horizon}

This subsection proves \cref{prop:horizon} and \cref{cor:fixed}. As in \cref{sec:theory-horizon}, fix one channel of a nonnegative
diagonal layer with $P$-periodic coefficient boxes, $\ahi_j<1$ for all $j$ and $\denh\in(0,1]$; $Q_j(c):=q_j(c)/(1-\ahi_j)$ is the DC
level of phase $j$ when the period starts at centre $c$, $\bar Q:=\max_j\max\{Q_j(0),Q_j(c^*)\}$, and $r_K$ is the radius after $K$
copies from the zero state, the radius with which the document starts at horizon $K$. Standard IBP training on the bare document is
itself a finite-horizon objective, with the document as the horizon; the following lemma covers that case.

\begin{lemma}[Document horizon]\label{lem:dochorizon}
Run \eqref{eq:cr} for one channel of a nonnegative diagonal layer from $(c_0,r_0)=(0,0)$ over any finite sequence of coefficient
boxes $(A_t,U_t)$, $t=1,\dots,n$, periodic or not, with $\ahi_t<1$, and put $q_t:=\arad_t\abs{c_{t-1}}+\urad_t$ and
$Q_t:=q_t/(1-\ahi_t)$ along the resulting centre trajectory. Then
\[
r_n=\sum_{t=1}^nw_t\,(1-\ahi_t)\,Q_t\ \le\ \Big(1-\prod_{t\le n}\ahi_t\Big)\max_{t\le n}Q_t,\qquad w_t:=\prod_{t<m\le n}\ahi_m,
\]
and the radius of the endpoint scan obeys the same bound. The centre satisfies
$\abs{c_n}\le\big(1-\prod_{t\le n}\acen_t\big)\max_{t\le n}\abs{\ucen_t}/(1-\acen_t)$.
\end{lemma}

\begin{proof}
Unrolling $r_t=\ahi_tr_{t-1}+q_t$ from $r_0=0$ gives $r_n=\sum_tw_tq_t=\sum_tw_t(1-\ahi_t)Q_t$. With $w_0:=\prod_{t\le n}\ahi_t$
we have $w_{t-1}=\ahi_tw_t$, so the nonnegative weights $w_t(1-\ahi_t)=w_t-w_{t-1}$ telescope to $1-\prod_t\ahi_t$. The endpoint
scan's box lies in the \eqref{eq:cr} box at every step (\cref{lem:cr}(b), (c), induction). The centre is the same computation with
$\acen_t$ and $\ucen_t$: $c_n=\sum_tw^{\rm c}_t(1-\acen_t)\,\ucen_t/(1-\acen_t)$ with $w^{\rm c}_t:=\prod_{t<m\le n}\acen_m$, and
these weights telescope to $1-\prod_t\acen_t$ (note $\acen_t\le\ahi_t<1$).
\end{proof}

\begin{proof}[Proof of \cref{prop:horizon}]
(a)~With $s\mapsto s-\omega$, $\Delta^{\rm lo/hi}_j=\softplus(s^{\rm lo/hi}_j-\omega)\to0$, so $1-\ahi_j\to0$ and
$\pih=\exp(-\alpha\sum_j\Delta^{\rm lo}_j)\to1$. By \cref{lem:dc}(b),
$Q_j(c)\le\tfrac12(e^{\eta_j}-1)\,\abs{c_{j-1}(c)}+e^{\eta_j}\varphi(\Delta^{\rm lo}_j)M_j$, with $M_j$ the product of the largest
magnitudes in the boxes of $x$ and of $B$ at phase $j$ and $\varphi(\Delta^{\rm lo}_j)\to1/\alpha$. The centres stay bounded:
$c^*=\cfree/(1-\pic)$ is a convex combination of the $\ucen_j/(1-\acen_j)$, because $\cfree=\sum_jw^{\rm c}_j\ucen_j$ with
$w^{\rm c}_j=\prod_{i>j}\acen_i$ and the weights $w^{\rm c}_j(1-\acen_j)$ telescope to $1-\pic$, and \cref{lem:dc}(b) bounds these
quotients as well (note $1-\acen_j\ge1-\ahi_j$). So $\bar Q$ stays bounded, and $r_K\le\bar QK\denh\to0$ by part (b). For the
limit of $r^*$: $\softplus(s-\omega)=e^{s-\omega}(1+o(1))$, so $\Delta^{\rm hi}_j/\Delta^{\rm lo}_j\to e^{\eta_j}$ and
$(1-\ahi_j)/(\alpha\Delta^{\rm lo}_j)\to1$. Interval products are positively homogeneous in the positive factor, so the injection
box equals $\Delta^{\rm lo}_j$ times a box converging to $[1,e^{\eta_j}]\cdot x_j\cdot B_j$, and $\urad_j/(1-\ahi_j)$ converges to
the radius of that box divided by $\alpha$. Likewise $\arad_j/(1-\ahi_j)\to\frac12(e^{\eta_j}-1)$, $c^*$ converges, and the weights
converge, $\mu_j=w_j(1-\ahi_j)/\denh\to e^{s^{\rm lo}_j}/\sum_ie^{s^{\rm lo}_i}>0$. Hence $r^*$ converges to a convex combination,
with positive weights, of the limiting DC levels, which is positive under the stated condition. For the minGRU-type layer use
\cref{lem:dc}(c) with $s\mapsto s-\omega$. For the S4D-type layer the decay is a point and $u=xB$ (\cref{lem:dc}(d)), so the DC level
is $Q_t=\abs{B}\xrad_t/(1-a)$, and the document radius $r_n=\sum_ta^{n-t}\abs{B}\xrad_t$ tends to $\sum_t\abs{B}\xrad_t$ as
$a\to1$: slowing an S4D-type channel at fixed $B$ does not hide it (\cref{rem:otherterms}).

(b)~From $(0,0)$ the start centres of the copies satisfy $c_k=\pic c_{k-1}+\cfree$, so $c_k=(1-\pic^k)c^*$ by induction,
$\pic(1-\pic^{k-1})c^*+(1-\pic)c^*=(1-\pic^k)c^*$; every start centre lies on the segment between $0$ and $c^*$. The radius obeys
$r_k=\pih r_{k-1}+\beta(c_{k-1})$, so $r_K=\sum_{k=0}^{K-1}\pih^{K-1-k}\beta(c_k)\le\max_k\beta(c_k)\,(1-\pih^K)/\denh$. The map $\beta$
is convex, being a nonnegative combination of the $q_j$, each the absolute value of an affine function of $c$ (slope
$\Gc_{j-1}\ge0$) times $\arad_j\ge0$ plus a constant; so $\beta(c_k)\le\max\{\beta(0),\beta(c^*)\}$. With $w_{j-1}=\ahi_jw_j$, $w_0=\pih$
and $w_P=1$ the weights telescope, $\sum_jw_j(1-\ahi_j)=1-\pih=\denh$, so for every $c$
$\beta(c)=\sum_jw_j(1-\ahi_j)\,Q_j(c)\le\denh\max_jQ_j(c)$. Hence $r_K\le\bar Q(1-\pih^K)$, and $(1-\denh)^K\ge1-K\denh$
(Bernoulli) gives the second bound. The expression for $r^*$ is \cref{lem:period}, and $r^*\ge\min_jQ_j(c^*)$ because it is a convex
combination of the $Q_j(c^*)$. In the scalar case ($P=1$, point decay $a=1-\delta$, injection radius $\urad$, $Q=\urad/\delta$)
$\beta\equiv\urad$ and the bounds are equalities: $r_K=\urad(1-\pi^K)/(1-\pi)=Q\,(1-(1-\delta)^K)\le QK\delta$ and $r_\infty=Q$.

(c)~For a fixed horizon $K$ and fixed DC levels, (b) gives $r_K\to0$ as $\denh\to0$, while $r_\infty$ does not change. A training
term that depends on the channel only through its radius at horizon $K$, and is non-decreasing in that radius, is therefore driven to
its value at radius $0$ by slowing the channel down at fixed DC levels, and the infinite-context box does not shrink at all. For the
document horizon, \cref{lem:dochorizon} applies: as $\prod_{t\le n}\alo_t\to1$ at bounded DC levels (as along the move of (a), where
$\Delta^{\rm lo}$ and $\Delta^{\rm hi}\to0$), both the radius and the centre of the channel vanish at every position of a document of
at most $n$ tokens, since $\prod_t\ahi_t\ge\prod_t\acen_t\ge\prod_t\alo_t$. The channel's box tends to $\{0\}$, and by continuity
(\cref{cor:stack}(iii)) a standard IBP term on such documents tends to its value with the channel removed. The channel's
infinite-context box keeps its DC-level width ((a) and \cref{cor:fixed}). $\prod_t\ahi_t\to1$ alone is not enough: the centre bound
carries $1-\prod_t\acen_t$, and a wide decay box keeps $\acen_t$ away from $1$. With $n=2$, $A_1=[1/2,1-\delta]$, $U_1=\{1\}$,
$A_2=\{1-\delta\}$, $U_2=\{0\}$ and the start $(0,0)$, every radius DC level is $0$ and the centre DC levels are at most $4$,
$\prod_t\ahi_t\to1$ as $\delta\to0$, yet the box at position $2$ is the point $1-\delta\to1$.
\end{proof}

\begin{proof}[Proof of \cref{cor:fixed}]
Suppose the robust term is computed from the $L=\infty$ fixed point in the DC form of \cref{lem:period} (\cref{rem:dcform}), as in
fixed-point fine-tuning.
Along the move of \cref{prop:horizon}(a), its radius $r^*=\sum_j\mu_jQ_j(c^*)$ converges to the positive limit given there (under the
condition stated there) instead of to $0$, so slowing a channel down cannot reduce the term to its zero-radius value.
\end{proof}
The DC form is designed to stay stable as $\denh\to0$ (\cref{rem:impl}): it never divides two vanishing quantities, so the term can be
evaluated and differentiated for arbitrarily slow channels. A training objective meant to control the infinite-context box should
therefore be evaluated at $K=\infty$, or through a K-uniform bound such as \cref{thm:hull}, and not at a finite horizon, unless every
time constant is capped well below that horizon. The blind spot of \cref{prop:horizon} needs the factor $1-\pih^K$ of its part (b), or
$1-\prod_t\ahi_t$ of \cref{lem:dochorizon} on the bare document, to tend to $0$ at bounded DC levels; a cap well below the horizon
keeps that factor away from $0$, and the boxes after $K$ copies approach the fixed point geometrically, at any rate above $\pih$ per
copy (proof of \cref{lem:period}). This is a recommendation, not a theorem. In the second training study the cap at
$\tau_{\max}=128$, a quarter of the 512-token training documents, trained with the standard finite-horizon term, kept at least 80\% of
the width-only $L=\infty$ certificates and at least 70\% of the K-uniform ones on both selective models, at clean-accuracy differences
of 0.5 [$-$0.8, 1.8] (minGRU-type) and $-$0.9 [$-$8.8, 6.9] points (Mamba) from the linear-ramp recipe (\cref{app:remedy}).

\Cref{cor:fixed} is a statement about objectives, not a training method. What trained models did is reported in
\cref{app:gate2,app:byte,app:remedy}: the first training study and the byte-level fine-tunes, and a second study in which the
fixed-point term from the start kept at least 80\% of the certificates of both selective models at $L=\infty$, and the K-uniform
hull from the start kept at least 80\% for every number of copies (for Mamba at $-$9.9 [$-$28.6, 8.8] and $-$8.8 [$-$28.5, 10.9]
points of clean accuracy with and without the cap, two of three seeds below 78\%), while under its gentler schedule standard IBP
already kept most of Mamba's at $L=\infty$; \cref{prop:horizon} states what a finite-horizon objective permits, not what every
training run does.

\begin{remark}[Other finite-horizon terms, and the S4D-type layer]\label{rem:otherterms}
The finite-$K$ training term of our implementation multiplies the injection of the last copy by $(1-\pih^{K-1})/\denh$; it is a
training surrogate, not a sound bound for $p^Kx$ (it is exact when $\arad=0$). It obeys the same estimate $\le\bar Q(1-\pih^{K-1})$,
so \cref{prop:horizon} applies to it. In the S4D-type layer the injection $u=xB$ has no structural factor $1-a$: slowing a channel
at fixed $B$ makes its DC level grow like $1/(1-a)$, and $r_K$ tends to $K$ times the per-copy injection rather than to $0$. Freezing a
channel and shrinking its injection are then two separate moves.
A zero-order-hold discretisation of a real-diagonal S4D layer would carry the factor: with the decay $a=e^{-\alpha\Delta}$ of
\cref{tab:layers}, its injection $\bar Bx$ with $\bar B=(1-a)B/\alpha$ hides a channel slowed through $\Delta\to0$ at fixed $\alpha$
and $B$, as Mamba's does (\cref{rem:injnorm}).
Predicted before the slow-channel attribution ran:
standard IBP training does not create slow S4D-type channels. The attribution found none: no channel of a non-degenerate
standard-IBP S4D-type seed was slower than 512 tokens (largest time constants 343 and 425 tokens, against 541.9 on one seed of the
rerun in \cref{app:remedy}), and its $L=\infty$ certified accuracy under synonym substitution was 39.8\% against 42.5\% at $L=0$. This is
consistent with the remark; it does not show that S4D-type training cannot exhibit the loophole, for instance through slow channels
with small injections.
\end{remark}

\begin{remark}[Injection normalisation]\label{rem:injnorm}
How a layer scales its injection decides whether slowing a channel hides it from a finite-horizon objective. Take one channel with a
point decay $a\in(0,1)$, the same at every position, and an injection box $(1-a)^pG$, with a fixed box $G$ of radius $g>0$ and an
exponent $p\ge0$. The scan \eqref{eq:cr} is exact for a point decay (\cref{lem:cr}(b)), and from the zero state the radius after $K$ positions
and the infinite-context radius, the channel's DC level, are
\[
r_K=g\,(1-a)^p\sum_{k<K}a^k=g\,(1-a)^{p-1}(1-a^K)\le K\,g\,(1-a)^p,\qquad r_\infty=g\,(1-a)^{p-1}.
\]
As $a\to1$ at fixed $G$: for $p>0$, $r_K\to0$ at every fixed horizon $K$, so a finite-horizon robust term can be lowered by slowing the
channel, as in \cref{prop:horizon}; and $r_\infty\to0$ for $p>1$, $r_\infty=g$ for $p=1$, $r_\infty\to\infty$ for $p<1$. For $p=0$,
$r_K\to Kg$, and slowing does not hide the channel. So $p=1$ gives the blind spot of \cref{prop:horizon} with a bounded DC level:
Mamba, where $\Delta=(1-a)\varphi(\Delta)$ with $\varphi(\Delta)\to1/\alpha$ (\cref{lem:dc}(b)), the minGRU-type layer, where
$z=1-a$, and a zero-order-hold real-diagonal S4D layer slowed through $\Delta\to0$ (\cref{rem:otherterms}). Our S4D-type layer, with
$u=xB$, has $p=0$. An exponent $0<p<1$ gives a stronger blind spot: the radius at every finite horizon vanishes while the
infinite-context box diverges. The normalisation $\sqrt{1-a^2}$ of the LRU and of RG-LRU \citep{orvieto2023,de2024} is $p=\frac12$ up
to the factor $\sqrt{1+a}\in[1,\sqrt2]$: $r_K\le Kg\sqrt{2(1-a)}\to0$, while $r_\infty=g\sqrt{(1+a)/(1-a)}\to\infty$. For a complex
mode, the disk recursion of \cref{lem:disks} with $a=\abs{\lambda}$ is the same computation. This remark concerns point decays;
\cref{lem:dc,prop:horizon} treat the interval decays of the three families of \cref{tab:layers}, and other normalisations of interval
decays are not analysed here.
\end{remark}

\subsection{IBP over a context in the linear regime}\label{app:proofs-loss}

\Cref{prop:exact} compares boxes with the true spread of the states. The context-length probe measures something else: on naturally
trained models, how much margin plain IBP loses over a context of $L$ tokens as $\eps\to0$, and how this loss compares with the true
first-order sensitivity of the margin. This subsection gives the theory of the measured quantities. Its part (a) is the first-order
behaviour of the natural interval extension, classical in interval analysis \citep{moore1966,neumaier1990}; for feed-forward
networks $D/G$ is the propagation tightness of \citet{mao2024}. What is specific to recurrences is its form over a context: parts
(a$_\infty$), (b1)--(b3) and (c1)--(c3). Parts (b) and (c) describe the first-order gain of finite contexts. For nonnegative diagonal
stacks they give exact temporal transport and convergence in $L$, asymptotically and with no rate below the slowest time constant;
for one-layer gated cells on stationary stretches they give geometric growth. They do not establish a separation at the context
lengths of the probe; that is measured (\cref{rem:gate3reading}).

\paragraph{Setting.}
Fix a model, a token sequence with clean embeddings $e^0_t$, a class $y$ (in the context-length probe the clean prediction $y_0$ of
the bare document) and a set $S$ of perturbed positions. The document occupies positions $0,\dots,n-1$ and the context the negative
positions, as in \cref{sec:setting}; the head reads position $n-1$. The probe boxes are
\[
E_t(\eps)=[\,e^0_t-\eps\one,\ e^0_t+\eps\one\,]\quad(t\in S),\qquad E_t=\{e^0_t\}\quad(t\notin S).
\]
In the context-length probe, $S$ is every position (for $W$, $D$, $R$) or one band of distances before the document (for $W_b$,
$D_b$, $R_b$). Its sequence families are \famname{txt}\,$L$, the document after $L$ tokens of contiguous neutral text;
\famname{per}\,$K$, the document after $K$ copies of a period $p$ of $P=1024$ neutral tokens, from the zero state with zero
convolution padding; \famname{inf}, the $L=\infty$ certificate of \cref{thm:inf}; and \famname{hull}, the K-uniform certificate of
\cref{thm:hull}; with $\eps\in\{10^{-10},10^{-8},10^{-6},10^{-4}\}$, in float64. On \famname{txt} and \famname{per} the probe runs the
endpoint IBP pass from the zero state with zero convolution padding; on \famname{inf} it runs the centre-radius pass with the
closed-form fixed point for the nonnegative diagonal models, the invariant-box search of \cref{rem:invbox} for LSTM and GRU, and the
exact max-pool box for the CNN.

An IBP pass is a finite directed acyclic graph. Its input nodes are the coordinates $e_{t,j}$; every other node $v$ is the output of
one interval primitive $\Phi_v$ applied to the boxes of its parents $\mathrm{pa}(v)$, and $\Phi_v$ encloses a real function
$\varphi_v$, the operation of the model at that node. A primitive with several outputs (for example a linear map) is a group of nodes
with the same parents. Evaluating the $\varphi_v$ at $e^0$ gives the clean values $v^0$. The margin bound is
$b(\eps)=\ell^{\rm lo}_y-\max_{k\ne y}\ell^{\rm hi}_k$, the lower end of the exact-range interval extension of
$m=\ell_y-\max_{k\ne y}\ell_k$ applied to the logit boxes, and $m^0$ is the clean margin. As in \cref{sec:theory-loss},
\[
W(\eps):=b(0)-b(\eps),\qquad D:=\sum_{t\in S}\norm{\nabla_{e_t}m(e^0)}_1,\qquad R(\eps):=\frac{W(\eps)}{\eps D};
\]
the probe reports $W/\eps$, $D/\eps$ and their ratio.

\begin{definition}[Absolute-value graph]\label{def:absgraph}
Suppose every $\varphi_v$ is differentiable at its clean argument. The \emph{absolute-value graph} of the pass is the same graph in
which the edge $u\to v$ carries the weight $\abs{\partial\varphi_v/\partial u}$, evaluated at the clean values. For an input coordinate
$(t,j)$ and a node $v$ let $\bar g_v(t,j)$ be the sum, over all directed paths from $e_{t,j}$ to $v$, of the product of the edge
weights, and $g_v(t,j)$ the same sum with the signed partial derivatives; by the chain rule, $g_v(t,j)=\partial v/\partial e_{t,j}$
at $e^0$. For the margin node write $\bar g_{t,j}$ and $g_{t,j}$, and put $G:=\sum_{t\in S}\sum_j\bar g_{t,j}=\sum_{t\in S}\norm{\bar
g_t}_1$.
\end{definition}

Every signed path product is bounded in absolute value by the corresponding product of absolute values, so
\begin{equation}\label{eq:domination}
\abs{g_{t,j}}\le\bar g_{t,j}\quad\text{for every }(t,j)\qquad\text{(domination); in particular }D\le G.
\end{equation}
The absolute-value graph depends on how the computation is split into primitives, that is, on the code. Two primitives that read the
same box are two separate paths, and any cancellation between them is lost. Three instances matter below: RMSNorm
(\cref{lem:rmsnorm}); the decay and the injection of a scan step, which share the gate (\cref{rem:gatepath}); and the margin, which is
formed from two separate logit bounds: the edge weights from the last hidden vector $u$ to $m$ are
$\abs{W^{\rm head}_{y,\cdot}}+\abs{W^{\rm head}_{k,\cdot}}$, while the true gradient is $W^{\rm head}_{y,\cdot}-W^{\rm head}_{k,\cdot}$
(\cref{rem:head}).

\begin{definition}[First-order property]\label{def:firstorder}
Let $\Phi$ be an interval extension of $\varphi:\reals^k\to\reals^q$, acting on boxes given by centre and radius. $\Phi$ is
\emph{first order at} $x^0$ if $\varphi$ is differentiable at $x^0$ and, for every $\rho\in[0,\infty)^k$ and every family of input
boxes with centres $\xcen(\eps)=x^0+o(\eps)$ and radii $\xrad(\eps)=\eps\rho+o(\eps)$ ($\eps\downarrow0$), the output boxes have
\[
y^{\rm c}(\eps)=\varphi(x^0)+o(\eps),\qquad y^{\rm r}(\eps)=\eps\,\abs{D\varphi(x^0)}\,\rho+o(\eps),
\]
where $\abs{D\varphi(x^0)}$ is the entrywise absolute Jacobian. We write $\dot r:=\lim_{\eps\downarrow0}r(\eps)/\eps$ for the
first-order radius of a box.
\end{definition}

\begin{lemma}[Exact-range extensions are first order]\label{lem:exactrange}
If $\varphi$ is differentiable at $x^0$, its exact-range extension, which returns for each output coordinate $i$ the interval
$[\min_B\varphi_i,\max_B\varphi_i]$ over the input box $B$, is first order at $x^0$.
\end{lemma}

\begin{proof}
Fix $i$ and a family $B_\eps$ as in \cref{def:firstorder}. For $x\in B_\eps$,
$\norm{x-x^0}_\infty\le\norm{\xcen-x^0}_\infty+\norm{\xrad}_\infty=O(\eps)$, so differentiability gives
$\varphi_i(x)=\varphi_i(x^0)+\nabla\varphi_i(x^0)\cdot(x-x^0)+o(\eps)$ uniformly on $B_\eps$. Write $x-x^0=(\xcen-x^0)+\delta$ with
$\abs{\delta}\le\xrad$ entrywise. The linear part is maximised at $\delta=\xrad\odot\mathrm{sign}\,\nabla\varphi_i(x^0)$ and minimised
at the opposite corner, so
\[
\max_{B_\eps}\varphi_i=\varphi_i(x^0)+\abs{\nabla\varphi_i(x^0)}\cdot\xrad+o(\eps),\qquad
\min_{B_\eps}\varphi_i=\varphi_i(x^0)-\abs{\nabla\varphi_i(x^0)}\cdot\xrad+o(\eps),
\]
using $\xcen-x^0=o(\eps)$. Half the sum and half the difference give the centre $\varphi_i(x^0)+o(\eps)$ and the radius
$\eps\abs{\nabla\varphi_i(x^0)}\cdot\rho+o(\eps)$.
\end{proof}

\begin{lemma}[The primitives of the implementation]\label{lem:primitives}
In real arithmetic, with the threshold of the implemented softplus and the SiLU slack as implemented, every interval primitive of our
IBP passes is first order (\cref{def:firstorder}) at the clean arguments listed below.
\begin{center}\upshape\small
\begin{tabular}{@{}>{\raggedright\arraybackslash}p{0.42\textwidth}>{\raggedright\arraybackslash}p{0.3\textwidth}>{\raggedright\arraybackslash}p{0.2\textwidth}@{}}
\toprule
interval primitive & operation & first order at\\
\midrule
linear map and causal depthwise convolution (centre $W\mu+b$, radius $\abs{W}\,\mathrm{rad}$) & affine map, convolution & every point\\
sum of boxes (residual, $ah+u$, readout sums, LSTM $c'=fc+ig$) & addition & every point\\
constant times a box & scaling & every point\\
product of two boxes (minimum and maximum of the four endpoint products) & product $ab$ & every point\\
product with a nonnegative factor (corner rule) & product $ab$ with $a^{\rm lo}\ge0$ ($\Delta$, decay, sigmoid, rsqrt) & every point\\
square (lower end $0$ when the box contains $0$) & $x^2$ & every point, $x^0=0$ included\\
mean over coordinates inside RMSNorm & positive linear map & every point\\
endpoints of $\exp$, sigmoid, $\tanh$, rsqrt (argument $\ge\epsilon_{\rm n}>0$), decay $e^{-\alpha\Delta}$ & monotone $C^\infty$ maps & every point\\
GRU update (exact range of the convex combination) & $(1-z)\,n+z\,h$, $z\in[0,1]$ & every point\\
endpoint scan step (corner rule, then sum) & $ah+u$, $a\ge0$ & every point\\
centre-radius step \eqref{eq:cr} & $ah+u$, $a\ge0$ & every point\\
RMSNorm (square, mean, rsqrt, product, scale) & RMSNorm & every point; first-order map of \cref{lem:rmsnorm}\\
implemented softplus (threshold $20$) & softplus & every $s^0\ne20$\\
implemented SiLU (lowered minimum) & SiLU & every $x^0\ne x^*$\\
endpoints of ReLU (CNN, bag of words) & ReLU & every $x^0\ne0$\\
max-pool over positions (CNN), $\max_{k\ne y}$ in the margin & maximum & points with a unique maximiser\\
\bottomrule
\end{tabular}
\end{center}
\end{lemma}

At $x^0=x^*$, the minimiser of SiLU, the exact SiLU extension is still first order (with derivative $0$), but the implemented one is
not: it returns the lower end $\mathrm{silu}(x^*)-\delta_{\rm s}$ for every box that contains $x^*$, with $\delta_{\rm s}=10^{-15}$
in float64 and $10^{-7}$ in float32, so its centre and radius carry the $\eps$-independent term $\delta_{\rm s}/2$. At $s^0=20$ the
implemented softplus is itself discontinuous (it drops by $2.1\times10^{-9}$), so $\varphi$ is not differentiable there. ReLU at $0$
and ties of a maximum are treated in \cref{rem:nonsmooth}. In every case the pass on point boxes returns the clean values, with two
exceptions: a SiLU input exactly equal to $x^*$, and, in the minGRU-type $L=\infty$ and hull passes, a period gate logit above $20$,
where the period part takes $\log\bar a=-s$ instead of $\log(1-\sigma(s))$ (\cref{rem:impl}). So $b(0)=m^0$ on \famname{txt} and
\famname{per} whenever no clean SiLU input equals $x^*$.

\begin{proof}[Proof of \cref{lem:primitives}]
By \cref{lem:exactrange} it suffices that each primitive returns the exact range of a function that is differentiable at the given
points. An affine map has, coordinate by coordinate, the range centre $W\mu+b$ and radius $\abs{W}\,\mathrm{rad}$; a sum of independent
boxes and a constant multiple are exact. A bilinear map attains its extremes over a box at corners, which is the product of two boxes;
the corner rule for $a\ge0$ takes the lower end $a^{\rm lo}b^{\rm lo}$ if $b^{\rm lo}\ge0$, else $a^{\rm hi}b^{\rm lo}$, and the upper
end $a^{\rm hi}b^{\rm hi}$ if $b^{\rm hi}\ge0$, else $a^{\rm lo}b^{\rm hi}$. The square returns the range of $x^2$, and $x^2$ is
differentiable everywhere. A monotone map attains its extremes at the endpoints. The GRU update is exact because $(1-z)n+zh$ is
increasing in $n$ and $h$ for $z\in[0,1]$ and affine in $z$. The endpoint scan step is a composition of the corner rule and a sum.
For \eqref{eq:cr} the claim is read off the formula: with $\acen=a^0+o(\eps)$, $\arad=\eps\dot r_a+o(\eps)$, $c=h^0+o(\eps)$,
$r=\eps\dot r_h+o(\eps)$ and the same for $u$,
\[
c'=\acen c+\ucen=a^0h^0+u^0+o(\eps),\qquad
r'=\ahi r+\arad\abs{c}+\urad=\eps\big(a^0\dot r_h+\abs{h^0}\,\dot r_a+\dot r_u\big)+o(\eps),
\]
because $\abs{c}\to\abs{h^0}$ (continuity of $\abs{\cdot}$ suffices, also at $h^0=0$), and $a^0\dot r_h+\abs{h^0}\dot r_a+\dot r_u$
is $\abs{\partial(ah+u)/\partial(h,a,u)}$ applied to the radii. (Equivalently: by \cref{lem:cr}(b) the excess of \eqref{eq:cr} over
the exact range is $\arad\min(r,\abs{c})=O(\eps^2)$.) RMSNorm is a composition of rows of the table (\cref{lem:rmsnorm}). The
implemented softplus box is the pair of endpoint images unless the box straddles $20$; the softplus is smooth and increasing on each
side of $20$, and a box of radius $O(\eps)$ around $s^0\ne20$ does not straddle $20$ for small $\eps$. SiLU has exactly one critical
point: $\mathrm{silu}'(x)=\sigma(x)\,k(x)$ with $k(x)=1+x\,\sigma(-x)$, which is at least $1$ for $x\ge0$ and on $(-\infty,0)$ strictly
increasing ($k'(x)=\sigma(-x)(1-x\sigma(x))>0$) from $-\infty$ to $1$; its zero is $x^*$. So SiLU is monotone on every box that avoids
$x^*$, where the implemented SiLU returns the endpoint images, the exact range; a box of radius $O(\eps)$ around $x^0\ne x^*$ avoids
$x^*$ for small $\eps$. The maximum of boxes is $[\max_k\ell^{\rm lo}_k,\max_k\ell^{\rm hi}_k]$, the exact range of a function that is
increasing in each argument, and it is differentiable at $x^0$ iff the maximiser is unique; ReLU is exact by monotonicity and
differentiable iff $x^0\ne0$. On a point box every exact-range primitive returns the point, the implemented softplus cannot straddle,
and the implemented SiLU returns a point unless $x^0=x^*$.
\end{proof}

\begin{lemma}[RMSNorm as implemented]\label{lem:rmsnorm}
Let $d=d_m$, $x^0\in\reals^d$, weight $w$, $\epsilon_{\rm n}>0$ ($10^{-5}$ in our models) and
$\varsigma:=(\norm{x^0}_2^2/d+\epsilon_{\rm n})^{-1/2}$. The first-order radius map of the interval RMSNorm at $x^0$ is
\[
\dot r_{\rm out}=M_{\rm N}\,\dot r_{\rm in},\qquad
M_{\rm N}=\varsigma\,\abs{\diag(w)}\Big(I+\frac{\varsigma^2}{d}\,\abs{x^0}\,\abs{x^0}^{\top}\Big),
\qquad\text{whereas}\qquad
J_{\rm N}=\varsigma\,\diag(w)\Big(I-\frac{\varsigma^2}{d}\,x^0x^{0\top}\Big)
\]
is the Jacobian of RMSNorm. $M_{\rm N}$ and $\abs{J_{\rm N}}$ agree off the diagonal, and
$(M_{\rm N})_{ii}-\abs{J_{\rm N}}_{ii}=2\varsigma^3\abs{w_i}(x^0_i)^2/d$. In the radial direction
$J_{\rm N}x^0=\epsilon_{\rm n}\varsigma^3\,(w\odot x^0)$, while
$M_{\rm N}\abs{x^0}=\varsigma\,(1+\varsigma^2\norm{x^0}_2^2/d)\,(\abs{w}\odot\abs{x^0})$ with $1\le1+\varsigma^2\norm{x^0}_2^2/d<2$. So
the true map almost annihilates the radial direction (the factor $\epsilon_{\rm n}\varsigma^2=\epsilon_{\rm n}/(\norm{x^0}_2^2/d+
\epsilon_{\rm n})$), and the first-order IBP map does not; this is a token-wise loss, of type (i) in \cref{prop:loss}(b2).
\end{lemma}

\begin{proof}
The interval RMSNorm computes the square of each coordinate (first-order radius $2\abs{x^0_j}\dot r_j$), the mean of the squares
($\frac2d\sum_j\abs{x^0_j}\dot r_j$), rsqrt of the mean plus $\epsilon_{\rm n}$ at the endpoints (derivative of magnitude
$\frac12\varsigma^3$, so radius $\frac{\varsigma^3}{d}\sum_j\abs{x^0_j}\dot r_j$ around $\varsigma$), the product of that box with the
box of $x_i$ by the corner rule (radius $\varsigma\dot r_i+\abs{x^0_i}\frac{\varsigma^3}{d}\sum_j\abs{x^0_j}\dot r_j$), and the scaling
by $w_i$ (factor $\abs{w_i}$); every step is first order (\cref{lem:primitives}) and the composition is \cref{lem:composition}. This is
$M_{\rm N}$. Differentiating $w_ix_i\,\varsigma(x)$ with $\partial\varsigma/\partial x_j=-\varsigma^3x_j/d$ gives $J_{\rm N}$. Off the
diagonal both entries are $\varsigma^3\abs{w_i}\abs{x^0_i}\abs{x^0_j}/d$. On the diagonal
$\varsigma^2(x^0_i)^2/d\le\varsigma^2\norm{x^0}^2_2/d<1$, so $\abs{J_{\rm N}}_{ii}=\varsigma\abs{w_i}(1-\varsigma^2(x^0_i)^2/d)$.
Finally $1-\varsigma^2\norm{x^0}_2^2/d=\epsilon_{\rm n}\varsigma^2$.
\end{proof}

\begin{lemma}[Composition]\label{lem:composition}
If every primitive of the pass is first order at its clean argument, then for the boxes $E_t(\eps)$ every node $v$ has a box with
centre $v^0+o(\eps)$ and first-order radius
\[
\dot r_v=\sum_{u\in\mathrm{pa}(v)}\Big|\frac{\partial\varphi_v}{\partial u}\Big|\,\dot r_u,\qquad \dot r_{e_{t,j}}=\one[t\in S],
\qquad\text{hence}\qquad \dot r_v=\sum_{t\in S}\sum_j\bar g_v(t,j).
\]
\end{lemma}

\begin{proof}
Induction in a topological order. The input boxes have centre $e^0_t$ and radius $\eps\one$ or $0$. If the parents of $v$ satisfy the
claim, the boxes handed to $\Phi_v$ are a family as in \cref{def:firstorder} with $\rho=(\dot r_u)_{u\in\mathrm{pa}(v)}$, so the box
of $v$ satisfies it too. Unrolling the linear recursion for $\dot r$ over the graph gives the path sums.
\end{proof}

\begin{proposition}[IBP over a context in the linear regime]\label{prop:loss}
Work in real arithmetic, with a model, a sequence, a class $y$, a set $S$ and the boxes $E_t(\eps)$ as above, and let $b$ be the margin
bound of the plain IBP pass, endpoint or centre-radius.

\emph{(a) Linear regime.} Assume that every primitive of the pass is first order at its clean argument. By \cref{lem:primitives} this
excludes, for the Mamba, S4D-type, minGRU-type, LSTM and GRU models, only clean gate logits equal to $20$, clean SiLU inputs equal to
$x^*$ and ties for the runner-up class (none with two classes). Then $b(0)=m^0$ and
\[
W(\eps)=\eps\,G+o(\eps),\qquad G=\sum_{t\in S}\norm{\bar g_t}_1\ \ge\ D=\sum_{t\in S}\norm{g_t}_1 .
\]
Hence $R(\eps)\to G/D\ge1$ as $\eps\downarrow0$ whenever $D>0$. The endpoint pass and the centre-radius pass have the same $G$.

\emph{(a$_\infty$) The infinite context.} Let the model be a stack of nonnegative diagonal layers whose clean periodic cascade
contracts ($\pih<1$ at $\eps=0$ in every channel of every layer, automatic in exact arithmetic), and let every primitive of the
cascade of \cref{def:cascade} be first order at its clean argument; for the minGRU-type layer, assume also that no clean gate logit of
the period exceeds $20$ (\cref{rem:impl}). Then the $L=\infty$ bound $b_\infty$ of \cref{thm:inf} satisfies $b_\infty(0)=m^0_\infty$,
the clean margin of $p^\infty x$, and $m^0_\infty-b_\infty(\eps)=\eps G_\infty+o(\eps)$. Here $G_\infty$ is given by the recursion of
\cref{lem:composition} on the clean cascade, in which every channel contributes over a period the first-order fixed-point radius
\[
\dot r^*=\frac{1}{1-\pi^0}\sum_{j=1}^Pw^0_j\big(\abs{h^0_{j-1}}\,\dot r_{a,j}+\dot r_{u,j}\big)
=\sum_{k\ge0}(\pi^0)^k\sum_{j=1}^Pw^0_j\big(\abs{h^0_{j-1}}\,\dot r_{a,j}+\dot r_{u,j}\big),
\]
with the clean decays $a^0_j$, $w^0_j=\prod_{i>j}a^0_i$, $\pi^0=\prod_ja^0_j$, the clean periodic states $h^0_j$ (the state of
$p^\infty$ after phase $j$, $h^0_0=h^0_P$; not the $c^0_j$ of \cref{sec:theory-periodic}) and the first-order radii
$\dot r_{a,j},\dot r_{u,j}$ of the coefficient boxes. So $G_\infty$ is the absolute-value gain of the infinite unrolled context, summed
copy by copy. If in addition the hypotheses of (b3) hold for the past $p^\infty$, then
$\lim_{K\to\infty}\lim_{\eps\downarrow0}R(\famname{per}\,K)(\eps)=G_\infty/D^{\rm per}_\infty$, with
$D^{\rm per}_\infty:=\lim_KD(\famname{per}\,K)>0$ by (A3).

\emph{(b) Nonnegative diagonal layers.}
\emph{(b1) Exact transport.} In every state channel $i$ of a nonnegative diagonal layer, the edge $h_{t-1,i}\to h_{t,i}$ of the
absolute-value graph carries the weight $a^0_{t,i}=\partial h_{t,i}/\partial h_{t-1,i}$. If the coefficient boxes have first-order
radii $\dot r_{a,t},\dot r_{u,t}$, the state box of the endpoint or the centre-radius scan has
\[
\dot r_{h,t}=\sum_{s\le t}\Pi^0_{s,t}\big(\abs{h^0_{s-1}}\,\dot r_{a,s}+\dot r_{u,s}\big),\qquad \Pi^0_{s,t}:=\prod_{s<m\le t}a^0_m,
\]
and first-order coefficient perturbations $(\delta a_s,\delta u_s)$ move the true state by
$\delta h_t=\sum_{s\le t}\Pi^0_{s,t}\,(h^0_{s-1}\,\delta a_s+\delta u_s)$: the same kernel. The absolute-value sensitivity of the margin to
a token at lag $d$ is therefore a nonnegative combination of the true transport factors $\Pi^0$ of the channels (for $Q\ge2$, of
products of such factors along the layers).

\emph{(b2) What is lost.} The difference between $G$ and $D$ comes from three kinds of lost cancellation and from nothing else:
(i)~\emph{local}, inside the $k_c$-window of the token, including the merges at the scan nodes $h_t$ of that window between the
temporal path and the paths through $a_t$, $u_t$: at the token-wise maps (linear maps, the convolution, RMSNorm as in
\cref{lem:rmsnorm}, the gate $y\odot\mathrm{SiLU}(z)$, the head and the margin) and between the decay path and the injection path of a
scan step, which enter as $\abs{h^0_{s-1}}\,\abs{\partial a_s}+\abs{\partial u_s}$ instead of $\abs{h^0_{s-1}\,\partial a_s+\partial u_s}$
(\cref{rem:gatepath}); (ii)~\emph{across state channels} at the readout $y=\sum_\nu C_\nu h_\nu+D^{\rm sk}x$ ($D^{\rm sk}$ the skip
parameter) and the output map; (iii)~for $Q\ge2$, \emph{across the intermediate positions} $t\in[s,n-1]$ at which the response of
layer 1 to the token $s$ enters layer 2. For one layer ($Q=1$) and a token at distance $n-1-s\ge k_c$ from the readout, with
$\norm{g_s}_1>0$,
\begin{equation}\label{eq:zetachi}
\frac{\norm{\bar g_s}_1}{\norm{g_s}_1}=\zeta(s)\,\chi(s),\qquad
\zeta(s)=\frac{\sum_iA_i(s)}{\sum_i\norm{v_i(s)}_1}\ge1,\qquad
\chi(s)=\frac{\sum_i\norm{v_i(s)}_1}{\big\|\sum_iv_i(s)\big\|_1}\ge1,
\end{equation}
where $v_i(s)=\gamma_i\,\Pi_i(s)\,\kappa_i(s)$ is the signed contribution of channel $i$ and
$A_i(s)=\bar\gamma_i\,\Pi_i(s)\,\norm{\bar\kappa_i(s)}_1\ge\norm{v_i(s)}_1$ its absolute counterpart (definitions in the proof). The
transport factor $\Pi_i(s)$ cancels in each per-channel ratio $A_i/\norm{v_i}_1=\bar\gamma_i\norm{\bar\kappa_i(s)}_1/(\abs{\gamma_i}\,
\norm{\kappa_i(s)}_1)$; the lag enters $\zeta$ and $\chi$ only through the mix of channels, which shifts towards the slow channels as the
lag grows. Identity \eqref{eq:zetachi} covers no model measured in this paper, all of which have two layers; for $Q\ge2$ the ratio can
grow with the lag (\cref{rem:gate3reading}, item 6).

\emph{(b3) Convergence in $L$.} Fix an infinite past (a token at every negative position); \famname{txt}\,$L$ keeps its last $L$
tokens. Assume (A1) every $\varphi_v$ is $C^1$ with a locally Lipschitz derivative on a neighbourhood of the closure of the set of clean
values over all $L$, and every primitive is first order at every clean argument (for the three nonnegative diagonal families this holds
when softplus is the exact function $\log(1+e^s)$, which the implemented softplus approximates within $2.1\times10^{-9}$ above $20$, and
no clean SiLU input equals $x^*$); (A3) $D_\infty:=\lim_{L\to\infty}D(L)>0$ (the limit exists by the proof). Then the uniform
contraction (A2) holds automatically: there is $\bar a<1$ with $a^0_{t,i}\le\bar a$ in every layer, channel and position, for every
$L$ (\cref{lem:uniform}; here $\bar a$ is that uniform bound, not a decay of \cref{sec:setting}). The limits
$G^{\rm txt}_\infty=\lim_LG(L)$ and $D_\infty$ exist, with
\[
\abs{G(L)-G^{\rm txt}_\infty}+\abs{D(L)-D_\infty}\le C\,(n+L+1)^{3Q}\,\bar a^{\,L},
\]
so $R_1(L):=G(L)/D(L)\to G^{\rm txt}_\infty/D_\infty$; moreover $\sup_LG(L)<\infty$ and $R_1(L)$ is bounded in $L$. The constant $C$
depends on $n$, $Q$, $\bar a$ and on bounds of the clean values and of the local derivatives and their Lipschitz constants; it grows
without bound as $\bar a\to1$. Nothing is claimed about $R_1(L)$ for $L$ below the slowest time constant $\tau:=1/(1-\bar a)$.

\emph{(c) Gated cells.} For a one-layer model built from the interval LSTM or GRU cell of our implementation, the path sums of the
absolute-value graph from the state at $t-1$ ($(h,c)$, resp.\ $h$) to the state at $t$ form a nonnegative matrix $M_t$ with
$M_t\ge\abs{J_t}$ entrywise, $J_t$ being the true state Jacobian. Parts (c), (c1) and (c3) are stated for one layer. Every gated model
of our experiments has two layers (\cref{app:scope}, item 9 of what is not proved), and for those only (c2) applies verbatim. The
contribution of the token at $s$ is $\norm{\bar g_s}_1=\bar\ell^{\top}M_{n-1}\cdots M_{s+1}\bar B_s\one$, against
$\norm{g_s}_1=\norm{\ell^{\top}J_{n-1}\cdots J_{s+1}B_s}_1$, with readout vectors $\bar\ell\ge\abs{\ell}$ and injection matrices
$\bar B_s\ge\abs{B_s}$ (defined in the proof); $\abs{J_{n-1}\cdots J_{s+1}}\le M_{n-1}\cdots M_{s+1}$.

\emph{(c1)} If on a stretch of the context that ends at the readout the clean trajectory is stationary ($M_t=M$, $J_t=J$,
$\bar B_t=\bar B$), $M$ is primitive with $\rho(M)>1$, and $\bar\ell\ne0$ (any head), $\bar B\ne0$, then the contribution of a token at
lag $d$ inside the stretch is at least $c\,\rho(M)^d$ for some $c>0$ and all large $d$, while the true one is at most $C_\kappa\kappa^d$
for every $\kappa>\rho(J)$. With $\rho(J)<1$ the distance-resolved ratio grows geometrically. The LSTM of \cref{prop:exact}(b) on the
zero input, with any head, is such a case: $M=\abs{J_0}$, $\rho(M)=1-\gamma+1.2\gamma^2>1$, while its true state Jacobians contract at
the rate $\kappa<1$ of \cref{prop:lstmfull}.

\emph{(c2) Finite $\eps$.} At every finite $L$, $b$ is continuous at $\eps=0$ (the argument of \cref{cor:stack}(iii), which applies to
any composition of exact-range primitives) and, under the hypothesis of (a), $W(\eps)/\eps\to G(L)<\infty$: plain IBP is first order at
every finite context length. The LSTM and GRU $h$ boxes stay in $[-1,1]^H$ (the LSTM's $c$ box does not), so $W(\eps;L)\le W_{\rm sat}$
for a constant $W_{\rm sat}$ independent of $\eps$ and $L$, and the bound can be $\eta$-linear at $\eps$, that is
$\abs{W(\eps)-\eps G(L)}\le\eta\,\eps G(L)$, only if $\eps\le W_{\rm sat}/((1-\eta)\,G(L))$. Under (c1), with the stretch equal to the
whole context, this range shrinks like $\rho(M)^{-L}$: at fixed $\eps$ the box leaves the linear regime once $\eps G(L)$ exceeds
$W_{\rm sat}/(1-\eta)$. Part (c2) holds for gated models of any depth: the head reads the last layer's $h$ box.

\emph{(c3) Non-commuting limits.} For the LSTM of \cref{prop:exact}(b) on the zero input, with $\rho_A:=1-\gamma+1.2\gamma^2$, the
radius $\nu_t$ of the $h$ box after $t$ steps satisfies, for every $\eps>0$ and $t\ge0$,
\[
\nu_t\le\bar\nu_t:=\tfrac12\gamma^2\eps\,\frac{\rho_A^{\,t}-1}{\rho_A-1},\qquad \lim_{\eps\downarrow0}\nu_t/\eps=\bar\nu_t/\eps,
\qquad \lim_{t\to\infty}\nu_t\ge\nu^* ,
\]
so the number of steps before $\nu_t$ reaches $\nu^*/2$ is at least $\log\big(1+(\rho_A-1)\nu^*/(\gamma^2\eps)\big)/\log\rho_A$, which
grows like $\log(1/\eps)$. At every finite $t$ the box vanishes as $\eps\to0$; at $t=\infty$ it does not (\cref{prop:exact}(b)).
\Cref{prop:exact}(b$''$) adds the matching asymptotic upper bound.
\end{proposition}

Part~(a) is the classical first-order behaviour of the natural interval extension \citep{moore1966,neumaier1990}; for feed-forward
networks $D/G$ is the propagation tightness of \citet{mao2024}, so $R$ is the reciprocal of a margin-level tightness over the context.
$R\ge1$ also follows from soundness at the box corner (\cref{cor:observables}(iii)); it is a check, not a finding. Parts~(b) and~(c)
predict no profile at the lengths we measure.

\begin{proof}[Proof of \cref{prop:loss}(a)]
By \cref{lem:primitives} the pass on point boxes returns the clean values, so $b(0)=m^0$. By \cref{lem:composition} the logit boxes
have centres $\ell^0_k+o(\eps)$ and first-order radii $\dot r_{\ell_k}=\sum_{t\in S}\sum_j\bar g_{\ell_k}(t,j)$. The bound $b(\eps)$
is the lower end of the exact-range extension of $m=\ell_y-\max_{k\ne y}\ell_k$ applied to these boxes. With a unique runner-up
$k^*$, $m$ is differentiable at $\ell^0$ with partial derivatives $+1$, $-1$ and $0$, so by \cref{lem:exactrange} the margin box has
centre $m^0+o(\eps)$ and radius $\eps(\dot r_{\ell_y}+\dot r_{\ell_{k^*}})+o(\eps)$, and $\dot r_{\ell_y}+\dot r_{\ell_{k^*}}=G$ is
the path sum of \cref{def:absgraph} for the margin node (its two incoming edges have weight $1$). Hence $b(\eps)=m^0-\eps G+o(\eps)$.
$G\ge D$ is \eqref{eq:domination}. The endpoint and the centre-radius pass differ only in the scan step, and the two scan steps have
the same first-order map (\cref{lem:primitives}), so they have the same $\dot r$ at every node.
\end{proof}

\begin{proof}[Proof of \cref{prop:loss}(a$_\infty$)]
\emph{One channel.} Suppose the coefficient boxes of the period satisfy $\acen_j=a^0_j+o(\eps)$, $\arad_j=\eps\dot r_{a,j}+o(\eps)$,
$\ucen_j=u^0_j+o(\eps)$, $\urad_j=\eps\dot r_{u,j}+o(\eps)$; then $\ahi_j=a^0_j+O(\eps)$. In exact arithmetic every $1-\ahi_j$ is
positive, and the DC-form fixed point of the implementation is $c^*=\cfree/(1-\pic)$, $r^*=\beta(c^*)/(1-\pih)$ (\cref{lem:period}).
$\cfree$ and $\pic$ are polynomials in the $\acen_j,\ucen_j$, and $1-\pi^0>0$, so $c^*=h^0_P+o(\eps)$, where $h^0_P$, the value of
$\cfree/(1-\pic)$ at $\eps=0$, is the clean periodic state (\cref{lem:period} at $\eps=0$). The phase centres
$\Gc_jc^*+\sum_{i\le j}(\prod_{i<m\le j}\acen_m)\ucen_i$ are then $h^0_j+o(\eps)$. Next,
$q_j=\arad_j\abs{c_{j-1}}+\urad_j=\eps(\abs{h^0_{j-1}}\dot r_{a,j}+\dot r_{u,j})+o(\eps)$, $w_j=w^0_j+O(\eps)$ and
$\denh=1-\pi^0+O(\eps)$, which gives $r^*=\eps\dot r^*+o(\eps)$ with $\dot r^*$ as stated, and the phase radii
\[
\Gh_jr^*+\beta_j(c^*)=\eps\Big(\Gamma^0_j\dot r^*+\sum_{i\le j}\Big(\prod_{i<m\le j}a^0_m\Big)\big(\abs{h^0_{i-1}}\dot r_{a,i}+\dot r_{u,i}\big)\Big)+o(\eps),
\qquad\Gamma^0_j=\prod_{i\le j}a^0_i .
\]
Writing $1/(1-\pi^0)=\sum_{k\ge0}(\pi^0)^k$ shows that this first-order radius is the sum, over all positions of all copies before the
phase, of the clean transport factor times the first-order injection, that is, the formula of (b1) on the infinite periodic past.

\emph{Layers and document.} Layer 1's period coefficient boxes come from the boxes $E^p_j$ of the period through token-wise primitives
and the periodic convolution, so they satisfy the hypothesis above (\cref{lem:primitives,lem:composition}). If layer $\ell$'s phase
boxes have centres $h^0+o(\eps)$ and radii $\eps\dot r+o(\eps)$, so do its output boxes and the period coefficient boxes of layer
$\ell+1$ (\cref{lem:composition}). The document part (convolution reading the period's tail, \eqref{eq:cr} from the phase-$P$ box,
readout, final RMSNorm, head) is a finite pass of first-order primitives. At $\eps=0$ all boxes are points and equal the clean values
of $p^\infty x$ (\cref{thm:inf}(ii) with zero radii; for the minGRU-type layer this uses the assumption on the gate logits), so
$b_\infty(0)=m^0_\infty$, and the margin is handled as in (a).

\emph{Identification with $\lim_KG(\famname{per}\,K)$.} The proof of (b3) shows that $G(L)$ converges to the sum, over all positions of
the infinite context, of the absolute-value gains computed on the limiting clean trajectory. For the past $p^\infty$ this trajectory
is the periodic one of \cref{thm:inf}, and the expansion of $1/(1-\pi^0)$ above shows that the recursion defining $G_\infty$ computes
the same sum (nonnegative terms, so the order of summation does not matter). \famname{per}\,$K$ is the truncation at $L=KP$, a
subsequence, so $G(\famname{per}\,K)\to G_\infty$, and $D(\famname{per}\,K)$ converges by (b3). The limit in $\eps$ is taken first, at
each $K$ (part (a)), and the limit in $K$ second.
\end{proof}

\begin{proof}[Proof of \cref{prop:loss}(b1)]
By \cref{lem:primitives}, both scan steps propagate first-order radii by
$\dot r_{h,t}=a^0_t\,\dot r_{h,t-1}+\abs{h^0_{t-1}}\,\dot r_{a,t}+\dot r_{u,t}$, with $\dot r_h=0$ at the zero start (or $\dot r^*$ of
(a$_\infty$) at the fixed point), and the true linearisation is $\delta h_t=a^0_t\,\delta h_{t-1}+h^0_{t-1}\,\delta a_t+\delta u_t$.
Unrolling both gives the formulas; the weight of the temporal edge is $a^0_t\ge0$ in both. For $Q\ge2$, a path from $e_s$ to the margin
enters each layer at some position through the coefficient or the input boxes and leaves it at a later position through the readout;
inside the layer its only edges between different positions are the convolution (a window of $k_c$ positions) and the temporal edges
of one channel, whose product is a factor $\Pi^0$.
\end{proof}

\begin{proof}[Proof of \cref{prop:loss}(b2)]
$\bar g_{t,j}-\abs{g_{t,j}}$ is positive only if two paths from $e_{t,j}$ to $m$ with products of different sign exist. Two such
paths separate at some node and meet again at a node with two parents on them. By (b1) the temporal edges have the same nonnegative
weight in both graphs, so the sign difference is created on the other edges: token-wise maps and the convolution window, together with
the merges at the scan nodes $h_t$ of the window between the temporal path and the paths through $a_t$ and $u_t$ (type (i)); the
readout sums over state channels and the output map (type (ii)); and, for $Q\ge2$, the inputs of a deeper layer at different
positions $t$, where the paths through different intermediate positions meet in that layer's state (type (iii)). For
\eqref{eq:zetachi}, let $Q=1$ and $n-1-s\ge k_c$. The readout position $n-1$ reads $h_{n-1}$ and quantities ($x_{n-1}$, $z_{n-1}$,
$C_{n-1}$, the residual) that depend on the positions $n-k_c,\dots,n-1$ only, so every path from $e_s$ to $m$ passes through exactly
one final state $h_{n-1,i}$. Put $\gamma_i=\partial m/\partial h_{n-1,i}$, let $\bar\gamma_i\ge\abs{\gamma_i}$ be the absolute-value
gain from $h_{n-1,i}$ to $m$, and by (b1)
\[
\frac{\partial h_{n-1,i}}{\partial e_s}=\Pi_i(s)\,\kappa_i(s),\qquad
\Pi_i(s):=\Pi^0_{s+k_c-1,\,n-1},\qquad
\kappa_i(s):=\sum_{t=s}^{s+k_c-1}\Pi^0_{t,\,s+k_c-1}\big(h^0_{t-1,i}\nabla_{e_s}a_{t,i}+\nabla_{e_s}u_{t,i}\big)\in\reals^{d_m},
\]
and $\bar\kappa_i(s)$ the same sum with $\abs{h^0_{t-1,i}}$ and the absolute-value gains of $a_{t,i}$ and $u_{t,i}$ with respect to
$e_s$; $\bar\kappa_i(s)\ge\abs{\kappa_i(s)}$ entrywise by \eqref{eq:domination} on the subgraphs. Then $g_s=\sum_iv_i(s)$ and
$\bar g_s=\sum_i\bar\gamma_i\Pi_i(s)\bar\kappa_i(s)$, so $\norm{\bar g_s}_1=\sum_iA_i(s)$ with $A_i(s)\ge\norm{v_i(s)}_1$, and
\eqref{eq:zetachi} is the identity
$\frac{\sum_iA_i}{\norm{\sum_iv_i}_1}=\frac{\sum_iA_i}{\sum_i\norm{v_i}_1}\cdot\frac{\sum_i\norm{v_i}_1}{\norm{\sum_iv_i}_1}$ together
with the triangle inequality.
\end{proof}

\begin{lemma}[Fading memory]\label{lem:fading}
Under the hypotheses of \cref{prop:loss}(b3), with $\bar a$ and the compact set $\mathcal K$ of \cref{lem:uniform}, there is
$C<\infty$ such that for all $L\le L'$:
(i)~$\norm{\bar g^{(L)}_s}_1\le C\,(n-s)^{Q-1}\,\bar a^{\,n-1-s}$ for $-L\le s\le n-1$, and the same bound holds for $\norm{g^{(L)}_s}_1$
and, with $n-1$ replaced by $t$, for the absolute-value gain from $e_s$ to any node at position $t\ge s$;
(ii)~every clean value at a position $t\ge-L$ differs between the truncations at $L$ and at $L'$ by at most
$C\,(t+L+1)^{Q-1}\,\bar a^{\,t+L}$;
(iii)~$\big|\norm{\bar g^{(L')}_s}_1-\norm{\bar g^{(L)}_s}_1\big|+\norm{g^{(L')}_s-g^{(L)}_s}_1\le C\,(n+L+1)^{3Q-3}\,\bar a^{\,n+L}$
for $-L\le s\le n-1$.
Here a superscript $(L)$ marks the context truncated at $-L$.
\end{lemma}

\begin{proof}
Let $\Lambda\ge1$ bound every local absolute partial derivative on $\mathcal K$, $\Lambda_2$ the Lipschitz constants of the local
derivatives on a neighbourhood of $\mathcal K$ (assumption (A1)), and $M_h$ a bound of the clean states. Constants $C$ may change from
line to line; they depend only on $n,Q,\bar a,k_c,\Lambda,\Lambda_2,M_h$ and the widths.

(i)~Propagate first-order radii from position $s$ alone ($\dot r=1$ on the coordinates of $e_s$, $0$ elsewhere); by
\cref{lem:composition} the first-order radius of a node is its absolute-value gain from $e_s$. In layer 1 the coefficient boxes have
nonzero first-order radius only at the $k_c$ positions $s,\dots,s+k_c-1$, at most $C$, so by (b1) $\dot r_{h,t}\le C\bar a^{\,t-s}$ for
$t\ge s$, and the layer's outputs inherit this bound. If the input of layer $\ell$ has $\dot r\le C(t-s+1)^{\ell-2}\bar a^{\,t-s}$, its
coefficient boxes obey the same bound with another $C$ (a window of $k_c$ positions), and by (b1)
$\dot r_{h,t}\le\sum_{m\le t}\bar a^{\,t-m}C(m-s+1)^{\ell-2}\bar a^{\,m-s}\le C(t-s+1)^{\ell-1}\bar a^{\,t-s}$. After $Q$ layers the head
gives (i); the signed gradient is dominated by \eqref{eq:domination}.

(ii)~In layer 1 the inputs of the two contexts agree on $[-L,n)$, and the convolution outputs agree from $-L+k_c-1$ on. There the state
difference is at most $2M_h$, and afterwards it obeys $\delta h_t=a^0_t\delta h_{t-1}$ with the common decay, so
$\abs{\delta h_t}\le C\bar a^{\,t+L}$. If the inputs of layer $\ell$ differ by at most $C(t+L+1)^{\ell-2}\bar a^{\,t+L}$, its coefficients
differ by at most $\Lambda$ times this (over a window), and $\delta h_t=a'_t\delta h_{t-1}+(a'_t-a_t)h_{t-1}+(u'_t-u_t)$ gives
$\abs{\delta h_t}\le\bar a\abs{\delta h_{t-1}}+C(t+L+1)^{\ell-2}\bar a^{\,t+L}$, hence $\abs{\delta h_t}\le C(t+L+1)^{\ell-1}\bar a^{\,t+L}$
(as in the proof of \cref{lem:limit}). The outputs are Lipschitz in the states and the window.

(iii)~Fix $s\ge-L$. Both propagations from $e_s$ live on the positions $t\ge s$, which exist in both contexts, on the same graph, with
edge weights $w$ (truncation at $L$) and $w'$ (at $L'$). An edge at position $t$ has $\abs{w'-w}\le\Lambda_2C(t+L+1)^{Q-1}\bar a^{\,t+L}$
by (ii). Let $\rho,\rho'$ be the two propagated first-order radii; $\delta=\rho'-\rho$ vanishes at the inputs and satisfies
$\delta_v=\sum_uw'_{uv}\delta_u+\sum_u(w'_{uv}-w_{uv})\rho_u$. So $\abs{\delta}$ at the margin is at most the propagation, with the
nonnegative weights $w'$, of the forcing $\sum_u\abs{w'_{uv}-w_{uv}}\rho_u$. At position $t$ the forcing is at most
$C(t+L+1)^{Q-1}\bar a^{\,t+L}\cdot(t-s+1)^{Q-1}\bar a^{\,t-s}$ (by (i) for $\rho$), and its propagation to the margin multiplies it by at
most $C(n-t)^{Q-1}\bar a^{\,n-1-t}$ (by (i) for $w'$, started at position $t$). The exponents add up to $(n-1+L)+(t-s)$, so the sum over
$t\ge s$ is at most $C(n+L+1)^{3Q-3}\bar a^{\,n+L}/(1-\bar a)$. The signed sensitivities obey the same identity with
$\abs{\rho}\le\bar\rho$, which gives the bound for $\norm{g^{(L')}_s-g^{(L)}_s}_1$.
\end{proof}

\begin{proof}[Proof of \cref{prop:loss}(b3)]
\Cref{lem:uniform} gives (A2). For $L<L'$,
\[
G(L')-G(L)=\sum_{s<-L}\norm{\bar g^{(L')}_s}_1+\sum_{s=-L}^{n-1}\big(\norm{\bar g^{(L')}_s}_1-\norm{\bar g^{(L)}_s}_1\big).
\]
By \cref{lem:fading}(i) the first sum is a tail of $\sum_{d\ge0}C(d+1)^{Q-1}\bar a^{\,d}$ beyond $d=n+L$, at most
$C(n+L+1)^{Q-1}\bar a^{\,n+L}$; by \cref{lem:fading}(iii) the second has $n+L$ terms of size at most $C(n+L+1)^{3Q-3}\bar a^{\,n+L}$.
Both bounds are uniform in $L'$ and at most $C(n+L+1)^{3Q}\bar a^{\,L}$ together (with $\bar a^{\,n}\le1$), so $G(L)$ is Cauchy, and
letting $L'\to\infty$ gives the stated rate. The same argument applies to $D(L)$. For each fixed $s$, $\norm{\bar g^{(L)}_s}_1$
converges by \cref{lem:fading}(iii) to the gain computed on the limiting clean trajectory (which exists by \cref{lem:fading}(ii)), and
by the summable bound (i) and dominated convergence $G^{\rm txt}_\infty$ is the sum of these limits over all $s\le n-1$; this is the
identification used in (a$_\infty$). By (i), $\sup_LG(L)\le\sum_{d\ge0}C(d+1)^{Q-1}\bar a^{\,d}<\infty$. With (A3),
$D(L)\ge D_\infty/2$ for $L\ge L_0$, so $R_1(L)\to G^{\rm txt}_\infty/D_\infty$ and $R_1$ is bounded on $L\ge L_0$; the finitely many
$L<L_0$ with $D(L)>0$ add finitely many values.
\end{proof}

\begin{proof}[Proof of \cref{prop:loss}(c)--(c3)]
\emph{The matrices $M_t$.} The interval LSTM cell computes the pre-activations with two linear maps and a sum (edges of weight
$\abs{W_{hh}}$ and $\abs{W_{ih}}$), the gates at the endpoints of $\sigma$ and $\tanh$ (weights $\sigma',\tanh'\ge0$), $fc$ and $ig$ by
the corner rule for a nonnegative factor (weights $\abs{c^0}$ and $\sigma(f^0)$, resp.\ $\abs{\tanh g^0}$ and $\sigma(i^0)$),
$c'=fc+ig$, and $h'$ by the corner rule for $\sigma(o)$ and $\tanh(c')$ (weights $\abs{\tanh c'^0}$ and $\sigma(o^0)$). The true partial
derivatives are the same path sums with the signed weights $W_{hh}$, $c^0$, $\tanh g^0$, $\tanh c'^0$. The interval GRU cell uses
linear maps, $\sigma$, the corner rule for $r\,(W_{hn}h+b_{hn})$, $\tanh$ and the exact convex combination of \cref{lem:primitives},
whose first-order map is the exact absolute Jacobian $(\abs{h^0-n^0},\,1-z^0,\,z^0)$ of $(1-z)n+zh$ (\cref{lem:exactrange}). In both
cells each entry of $M_t$ is a sum of absolute path products whose signed counterpart is the entry of $J_t$, so $M_t\ge\abs{J_t}$. For
nonnegative matrices $A\le A'$ and $B\le B'$ imply $AB\le A'B'$, and $\abs{XY}\le\abs{X}\abs{Y}$, so
$\abs{J_{n-1}\cdots J_{s+1}}\le M_{n-1}\cdots M_{s+1}$. The token $e_s$ enters only the cell at step $s$ (there is no convolution);
let $B_s$ be the Jacobian of the state at $s$ with respect to $e_s$ and $\bar B_s\ge\abs{B_s}$ its absolute-value counterpart. The head
reads $h_{n-1}$: $\ell=W^{\rm head}_y-W^{\rm head}_k$ and $\bar\ell=\abs{W^{\rm head}_y}+\abs{W^{\rm head}_k}$ on the $h$-coordinates
($0$ on $c$). Every path from $e_s$ to $m$ goes through the state at $s$, the temporal edges and the head, which gives both expressions.

(c1)~$M$ primitive implies $\rho(M)^{-d}M^d\to vw^{\top}$ with $v,w>0$ entrywise and $w^{\top}v=1$ (Perron-Frobenius), so
$\rho(M)^{-d}\,\bar\ell^{\top}M^d\bar B\one\to(\bar\ell^{\top}v)(w^{\top}\bar B\one)>0$, because $\bar\ell$ and $\bar B\one$ are
nonnegative and nonzero. For the true contribution, $\norm{\ell^{\top}J^dB}_1\le C\norm{J^d}$ and $\norm{J^d}^{1/d}\to\rho(J)$
(Gelfand). For the LSTM of \cref{prop:exact}(b) with $x\equiv0$ the clean trajectory stays at $h=c=0$, where the gate edges carry
$\abs{c^0}=\abs{\tanh g^0}=\abs{\tanh c'^0}=0$ and the input, forget and output gates have no recurrent weights; so $M_t=\abs{J_0}$ of
\cref{prop:lstmfull}. Its $(h,h)$ block $\gamma^2\abs{U}$ and $(c,h)$ block $\gamma\abs{U}$ are positive and its $(h,c)$ and $(c,c)$
blocks have positive diagonals, so its graph is strongly connected with self-loops and $\abs{J_0}$ is primitive;
$\rho(\abs{J_0})=1-\gamma+1.2\gamma^2>1$, and the true products contract at rate $\kappa<1$ (\cref{prop:lstmfull}). The injection of
$x$ is $\tfrac12\gamma$ into each $c$-coordinate and $\tfrac12\gamma^2$ into each $h$-coordinate.

(c2)~Continuity at $\eps=0$ follows from the argument of \cref{cor:stack}(iii), which applies to any composition of exact-range
primitives, and the limit of $W/\eps$ is (a). The LSTM's $h$ box is the corner-rule product of a box in $[0,1]$ and a box in $[-1,1]$,
and the GRU's lies in the hull of the boxes of $n$ and $h$ (the exact convex combination), so every $h$ box, in every layer, lies in
$[-1,1]^H$ (\cref{prop:exactac}(c)). The head is a linear map, so $\ell^{\rm lo}_y\ge\beta_y-\norm{W^{\rm head}_y}_1$ and
$\ell^{\rm hi}_k\le\beta_k+\norm{W^{\rm head}_k}_1$ for every $\eps$ and $L$ ($\beta$ the head bias), while
$m^0\le\beta_y-\beta_k+\norm{W^{\rm head}_y-W^{\rm head}_k}_1$. Hence
$W(\eps;L)\le W_{\rm sat}:=2\big(\norm{W^{\rm head}_y}_1+\norm{W^{\rm head}_k}_1\big)$ (two classes; otherwise maximise over $k$). If
$\abs{W(\eps)-\eps G(L)}\le\eta\eps G(L)$, then $(1-\eta)\eps G(L)\le W_{\rm sat}$. If the stretch of (c1) is the whole context,
$G(L)\ge c\,\rho(M)^L$ for large $L$.

(c3)~Let $(C_t,\nu_t)$ be the iterates of $T_\eps$ of \cref{prop:lstmfull} from $(0,0)$ and $(\bar C_t,\bar\nu_t)$ those of its linear
part, $\bar C_t=(1-\gamma)\bar C_{t-1}+\gamma(\tfrac12\eps+1.2\bar\nu_{t-1})$, $\bar\nu_t=\gamma\bar C_t$. Since $0\le\tanh u\le u$ for
$u\ge0$ and $T_\eps$ is non-decreasing in both arguments, induction gives
$C_t=(1-\gamma)C_{t-1}+\gamma\tanh(\tfrac12\eps+1.2\nu_{t-1})\le\bar C_t$ and $\nu_t=\gamma\tanh(C_t)\le\gamma C_t\le\bar\nu_t$.
Substituting $\bar\nu_{t-1}=\gamma\bar C_{t-1}$ gives $\bar C_t=\rho_A\bar C_{t-1}+\tfrac12\gamma\eps$, hence
$\bar C_t=\tfrac12\gamma\eps(\rho_A^{\,t}-1)/(\rho_A-1)$ and the formula for $\bar\nu_t$. As $\tanh u=u+O(u^3)$ and the iterates are
$O(\eps)$ at fixed $t$, induction gives $\nu_t=\bar\nu_t+O(\eps^3)$. The limit in $t$ is \cref{prop:exact}(b). Finally
$\nu_t\le\bar\nu_t<\nu^*/2$ as long as $\rho_A^{\,t}-1<(\rho_A-1)\nu^*/(\gamma^2\eps)$.
\end{proof}

The first-order coefficient $\bar\nu_t/\eps$ in (c3) agrees with (a): it is the absolute-value gain of the $h$ box, built from
$M=\abs{J_0}$ and the injection above.

\begin{corollary}[The observables of the context-length probe]\label{cor:observables}
Assume the hypotheses of \cref{prop:loss}(a) for every family and set of perturbed positions involved, and $D>0$. Write
$\varpi(\eps):=\tfrac12\log_{10}\big(W(100\eps)/W(\eps)\big)$; the width exponent of the context-length probe,
$1+\tfrac12\log_{10}\big((W/\eps)(10^{-8})/(W/\eps)(10^{-10})\big)$, is $\varpi(10^{-10})$. Call the bound \emph{$\eta$-linear on} an
interval $I$ if $(1-\eta)\eps G\le W(\eps)\le(1+\eta)\eps G$ for $\eps\in I$.

(i)~\emph{Width exponent.} $\varpi(\eps)\to1$ as $\eps\downarrow0$. If the bound is $\eta$-linear on $[\eps,100\eps]$, then
$\abs{\varpi(\eps)-1}\le\tfrac12\log_{10}\frac{1+\eta}{1-\eta}$. Hence a document has $\abs{\varpi-1}\le0.05$ on a family whenever the
bound is $\eta$-linear on $[10^{-10},10^{-8}]$ with $(1+\eta)/(1-\eta)\le10^{0.1}$; the median clause of the amended exponent criterion
of the probe (median $\abs{\varpi-1}\le0.05$) holds if this is true for more than half of the documents on each of the five families it
is checked on. A document with $\varpi\le0.5$ is not $\eta$-linear on $[10^{-10},10^{-8}]$ for any $\eta<9/11$, so the amended
saturation criterion for gated models (median width exponent at most $0.5$) means that at least half of the documents are not. If
$W(\eps)=\eps G+\eps^2G_2+o(\eps^2)$, then $\varpi(\eps)-1=\frac{99}{2\ln10}\,\eps\,G_2/G+o(\eps)$, and the linearity statistic of the
original criterion, $\frac{W(\eps)/\eps}{W(100\eps)/(100\eps)}-1=-99\,\eps\,G_2/G+o(\eps)$, measures the second-order term, as the
amendment of the probe states.

(ii)~\emph{Distance bands.} For disjoint sets $S_b$ (distance bands, the document) with $D_b:=\sum_{t\in S_b}\norm{g_t}_1>0$,
$W_b(\eps)=\eps G_b+o(\eps)$ with $G_b:=\sum_{t\in S_b}\norm{\bar g_t}_1$, so $R_b\to G_b/D_b\ge1$. The vectors $\bar g_t$ and $g_t$
are those of the full pass. If the $S_b$ partition $S$, the widths are additive at first order, $W(\eps)-\sum_bW_b(\eps)=o(\eps)$, and
$G/D=\sum_b(D_b/D)(G_b/D_b)$. For one nonnegative diagonal layer and a band of the context (all its positions are at distance at least
$n\ge k_c$ from the readout), \eqref{eq:zetachi} gives $G_b/D_b\ge\sum_{s\in S_b}\zeta(s)\chi(s)\norm{g_s}_1/\sum_{s\in S_b}\norm{g_s}_1$,
with equality when $\norm{g_s}_1>0$ for every $s\in S_b$. This covers none of the models of the probe, which all have two layers. For
$Q\ge2$ the type (iii) loss can make $G_b/D_b$ grow with the distance of the band (\cref{rem:gate3reading}, item 6). For a one-layer
gated cell the profile of $G_b$ is governed by the products $M_{n-1}\cdots M_{s+1}$ (\cref{prop:loss}(c)).

(iii)~\emph{Soundness at finite $\eps$, without linearisation.} At every $\eps$, $b(\eps)\le m(e^0-\eps\,\mathrm{sign}\,\nabla m)$, the
margin at the box corner, which the probe checks. If $m$ is $C^2$ on the box, $\abs{\delta^{\top}\nabla^2m\,\delta}\le H$ there for
$\norm{\delta}_\infty\le1$, and $b(0)=m^0$ (no clean SiLU input at $x^*$), then $R(\eps)\ge1-H\eps/(2D)$. So a value
$R<1-10^{-3}$ at $\eps=10^{-10}$, which the soundness check of the probe flags, requires $H\eps/(2D)>10^{-3}$ at that $\eps$.

(iv)~\emph{Certified radius.} Let $m^0>0$, $r_{\rm cert}:=\sup\{\eps\ge0:b(\eps)>0\}$, $r_1:=m^0/G$ and $r_D:=m^0/D$, the radius at
which the linearised margin $m^0-\eps D$ vanishes (the first-order $\ell_\infty$ distance to the decision boundary under perturbations
of every position in $S$). Then $r_1=r_D/R_1$ with $R_1=G/D$. If the bound is $\eta$-linear on $[0,r_1/(1-\eta)]$, then
$r_1/(1+\eta)\le r_{\rm cert}\le r_1/(1-\eta)$. Under (b3) with $m^0_\infty>0$, $r_1(L)\to m^0_\infty/G^{\rm txt}_\infty$; under (c1)
on the whole context, $r_1(L)\le m^0(L)/(c\,\rho(M)^L)$ for large $L$. A document is certified at $\eps$ (the certified share the probe
reports) iff $\eps<r_{\rm cert}$.
\end{corollary}

\begin{proof}
(i)~$W(100\eps)/W(\eps)=100\,(G+o(1))/(G+o(1))\to100$. Under $\eta$-linearity the ratio lies in
$[100\frac{1-\eta}{1+\eta},100\frac{1+\eta}{1-\eta}]$, which gives the bound on $\abs{\varpi-1}$; $\varpi\le\frac12$ means
$W(100\eps)\le10\,W(\eps)$, incompatible with $100\frac{1-\eta}{1+\eta}>10$, that is with $\eta<9/11$. With the second-order expansion,
$W(100\eps)/W(\eps)=100\,(1+99\eps G_2/G)+o(\eps)$, and $\log_{10}(1+x)=x/\ln10+O(x^2)$; the second statistic is
$(G+\eps G_2)/(G+100\eps G_2)-1$. (ii)~Apply (a) with $S$ replaced by $S_b$: by \cref{lem:composition} the first-order radius is the
path sum restricted to the inputs in $S_b$, on the same absolute-value graph, because the graph is evaluated at the clean point, which
does not depend on $S_b$. Additivity follows from $G=\sum_bG_b$. Positions of the band with $\norm{g_s}_1=0$ add $\norm{\bar g_s}_1\ge0$
to the numerator only, hence the inequality. (iii)~Soundness (\cref{cor:stack}(i)) at the corner, and Taylor's formula:
$m(e^0-\eps\,\mathrm{sign}\,\nabla m)\le m^0-\eps D+\tfrac12H\eps^2$. (iv)~$b$ is non-increasing (\cref{cor:stack}(ii)) and continuous
(\cref{cor:stack}(iii)), so $\{\eps:b(\eps)>0\}=[0,r_{\rm cert})$. For $\eps<r_1/(1+\eta)$, $b(\eps)\ge m^0-(1+\eta)\eps G>0$; at
$\eps_1=r_1/(1-\eta)$, $b(\eps_1)\le m^0-(1-\eta)\eps_1G=0$. The limits follow from (b3), from \cref{lem:fading}(ii) for $m^0(L)$, and
from (c2).
\end{proof}

For the family \famname{inf} the probe uses $D(\famname{per}\,4)$ in place of the gradient norm of the infinite-context model, and it
marks $R_{\rm inf}$ as a proxy unless $D(\famname{per}\,4)/D(\famname{per}\,2)$ and $W(\famname{per}\,4)/W(\famname{inf})$ are within 5\%
of $1$. By (a$_\infty$) this proxy converges to $G_\infty/D^{\rm per}_\infty$ as the number of copies grows, at the rate of (b3); with
$K\le4$ copies of $P=1024$ tokens the check is the empirical part. \famname{per}\,$K$ uses the endpoint scan and \famname{inf} the
centre-radius scan; the two agree to first order (\cref{lem:primitives}), so the comparison $W(\famname{per}\,K)/W(\famname{inf})$ is
between first-order quantities only.

\begin{remark}[Non-smooth points]\label{rem:nonsmooth}
1.~\emph{Max-pool ties (CNN).} In \famname{per}\,$K$ with $K\ge2$ and in \famname{inf}, copies of the period produce identical
convolution windows, so the maximum over positions is tied at the clean point by construction and $m$ is not differentiable there. The
pass is still first-order homogeneous: maxima, ReLU and the smooth primitives are locally Lipschitz and directionally differentiable,
the exact range of such a function over $x^0+\eps(\mu+[-\rho,\rho])+o(\eps)$ is $f(x^0)+\eps\{f'(x^0;\delta):\delta\in\mu+[-\rho,\rho]\}
+o(\eps)$, and composing gives $W(\eps)=\eps G^\sharp+o(\eps)$ (a sketch; \cref{app:scope}, item 6 of what is not proved), so
$\varpi(\eps)\to1$ still. But $G^\sharp$ is not an absolute-value gain, and the gradient that defines $D$ is not the first-order
worst-case deviation: the gradient of the implemented maximum splits evenly between tied entries, while a perturbation of one of two
tied windows moves the maximum only upwards. Soundness gives only $G^\sharp\ge D^\sharp:=\max_{\norm{\delta}_\infty\le1}\big(-m'(e^0;
\delta)\big)$, and $D^\sharp$ can be smaller than $D$. If, for instance, $W^{\rm head}_{y,c}=-W^{\rm head}_{k,c}>0$ for the tied feature
$c$, perturbing only one of the two tied windows leaves $b$ unchanged (only the upper end of the maximum moves), while the split gradient
gives $D_b>0$, so $R_b=0$. This mechanism is consistent with the band looseness below $1-10^{-3}$ that the smoke run of the probe
flagged for the CNN; it was not checked that the flagged bands contain such a tie, and the amended band check excludes the CNN.

2.~\emph{ReLU at $0$} (CNN, bag of words): a clean pre-activation exactly $0$; generic inputs avoid it, and the same directional
statement applies.

3.~\emph{The SiLU minimum.} With the slack $\delta_{\rm s}$, every $\eps>0$ at which some SiLU box contains $x^*$ adds to $W$ an
$\eps$-independent term of order $\delta_{\rm s}$ times the absolute-value gain from that node to the margin. In the limit $\eps\to0$
this happens only if a clean input equals $x^*$; at a finite $\eps$ it happens when a clean input lies within about $\eps\dot r$ of
$x^*$. The float32 slack $10^{-7}$ made $W/\eps$ grow like $1/\eps$ on such elements, which is why the amendment of the probe set
$\delta_{\rm s}=10^{-15}$ in float64.

4.~\emph{The softplus threshold.} A gate-logit box that straddles $20$ is widened by at most the drop $2.1\times10^{-9}$, again
independent of $\eps$; this requires a clean logit within the box radius of $20$.

5.~\emph{Not an issue:} ties among the four products of an interval product, the switch of the corner rule at $b^0=0$, $\abs{c}$ at
$c=0$ in \eqref{eq:cr}, and the square at $0$ (\cref{lem:primitives}).

6.~\emph{K-uniform bounds.} For \famname{hull} and for the gated \famname{inf} (the invariant box of \cref{rem:invbox}), $b(0)$ is
already a wide bound (it covers every number of copies, or the transient from the zero state), not $m^0$, so \cref{prop:loss} does not
apply to them; the probe reports them by $b(0)$, vacuity and the gap to the clean margins at $\eps=0$, never as a tightness ratio.
\end{remark}

\begin{remark}[The gate path of a slow channel]\label{rem:gatepath}
In a Mamba channel the decay $a=e^{-\alpha\Delta}$ and the injection $u=\Delta xB$ read the same box of $\Delta$ through two primitives,
so the absolute-value graph gives $\partial(ah+u)/\partial\Delta$ the weight $\alpha a\abs{h}+\abs{xB}$, while the true partial
derivative is $-\alpha ah+xB$. At the DC level $h=u/(1-a)$ of constant coefficients, with $\phi(z):=z/(e^z-1)\in(0,1)$ for $z>0$,
\[
-\alpha ah+xB=xB\,\big(1-\phi(\alpha\Delta)\big),\qquad \alpha a\abs{h}+\abs{xB}=\abs{xB}\,\big(1+\phi(\alpha\Delta)\big),
\]
a ratio $(1+\phi)/(1-\phi)$ that tends to $\infty$ as the channel slows down ($1-\phi(z)=z/2+O(z^2)$). The derivative with respect to the
gate logit multiplies both by $\sigma(s)$. In the minGRU-type layer the true partial derivative of $(1-z)h+zg$ with respect to $z$ is
$g-h$, which vanishes at the DC level $h=g$, while our implementation, which bounds $(1-z)h$ and $zg$ as two separate products, gives
$\abs{h}+\abs{g}$. In both layers the true sensitivity of a channel at its DC level to its own gate cancels, because the DC level does
not depend on the gate to leading order (the forgetting-factor structure of \cref{lem:dc}), and IBP adds the two paths. The GRU's
exact convex combination treats $(1-z)n+zh$ as one primitive and keeps $\abs{h-n}$ (\cref{lem:primitives}), and the S4D-type layer
has no gate path ($\dot r_a=0$). This is a type (i) loss that grows as a channel slows down; how much it contributes in trained models
is not known. It is one candidate for the difference between the measured looseness of the S4D-type model ($R$ between 734 and 858
for seed 0, 508--906 over its three seeds) and that of the selective models ($R$ between $2.07\times10^{4}$ and $1.78\times10^{5}$ over
all seeds) in the context-length probe and its replication; this attribution is not tested.
\end{remark}

\begin{remark}[What \cref{prop:loss} says about the context-length probe, and what it does not]\label{rem:gate3reading}
Here we only say which outcomes of the probe (\cref{sec:exp-gate3}, \cref{app:gate3}) the theory predicts.

1.~\emph{The growth criterion} (the looseness after 4096 tokens of contiguous text at most 10 times that after 256 tokens, and after
four periods at most 10 times that after one). (b3) gives convergence and boundedness of $R_1(L)$ with no rate below $\tau$, and
\cref{prop:exact}(a$'$) gives $W(\eps)\le4C'\eps$ uniformly in $L$ with a constant that can be astronomically large; neither predicts a
factor at $L=256$--$4096$. The measured ratios changed by factors $0.69$--$1.06$ (contiguous text) and $0.95$--$1.06$ (periods) over the
ten nonnegative diagonal models of the probe and its replication: no geometric accumulation, all ratios within $[0.6,1.1]$; an
empirical finding, not a prediction. Failing the criterion would not have contradicted (b3) if $\tau\gg4096$.

2.~\emph{The amended exponent criterion} (median $\abs{\varpi-1}\le0.05$). By \cref{cor:observables}(i) it holds when
$[10^{-10},10^{-8}]$ is in the linear regime of the model, which depends on the size of its second-order terms, not on
\cref{prop:loss}. Measured medians: $0.997$--$1.000$. The pre-registered linearity statistic of the original criterion (at most
$0.01$) measures $-99\eps G_2/G$ at $\eps=10^{-8}$; the byte-level Mamba had $0.0162$ and failed it, the only failed clause of that
criterion.

3.~\emph{A finite $L=\infty$ certificate.} In exact arithmetic the \famname{inf} box is finite (\cref{thm:inf}, \cref{lem:period}); a
non-finite one comes from floating point (underflow of $1-\bar a$, \cref{rem:impl}, or overflow). Measured: finite on every document of
the six nonnegative diagonal models of the first run and of the four of the replication.

4.~\emph{The saturation criterion for gated models.} (c2) gives saturation of bounded interval states for the two-layer gated models
of the probe. (c) and (c1) give a mechanism, a first-order gain built from $M_t\ge\abs{J_t}$ that grows geometrically on stationary
stretches, only for one-layer cells. Measured: the boxes were saturated at $\eps=10^{-10}$ inside the bare document (median width
exponent at most $7\times10^{-14}$ on the bare document), so the pre-registered growth criterion for gated models, which asked for
growth with $L$, could not pass (growth $0.44$--$1.03$). The reading of the probe's specification, ``saturation at $\eps=10^{-10}$ means
a first-order gain above about $10^{10}$ times the margin scale'', is heuristic: (c2) bounds the linear range from above by
$W_{\rm sat}/((1-\eta)G)$, so saturation at $\eps$ is what one expects once $\eps G$ is of order $W_{\rm sat}$, but a non-linear $W$ at
$10^{-10}$ does not by itself bound $G$ from below. The interval-map growth rates measure the growth rate directly
(\cref{sec:exp-r4}, \cref{app:gate3}).

5.~\emph{Distance bands.} For the two-layer models measured, \cref{prop:loss} predicts no band profile: \eqref{eq:zetachi} is a
one-layer identity, (c1) is stated for one-layer cells, and for $Q\ge2$ the band ratio can grow with the distance (item 6). Measured:
the looseness for perturbations 1024--4096 positions before the document was $0.28$--$1.01$ times that for perturbations within 64
positions of it (after 4096 tokens of contiguous text) on the ten models of the probe and its replication, and the single-token
state-level ratios of the growth-rate study at lags 64, 256 and 1024 stayed within $0.82$--$2.64$ times their value at lag 8 in both
layers (seed 0; within $0.31$--$1.93$ on seeds 1 and 2 of the word-level models). A flat or falling band profile of a two-layer model is
an empirical finding.

6.~\emph{Band ratios can grow for $Q\ge2$} (first order, in the absolute-value graph). Take two layers with one channel each, kernel
size 1 and no residual: layer 1 is $h^1_t=a\,h^1_{t-1}+e_t$, read out as $v^2_t=h^1_t$; layer 2 is $h^2_t=a\,h^2_{t-1}+u^2_t$ with
$\partial u^2_t/\partial v^2_t=c_t$ along the clean trajectory; the margin is $m=h^2_{n-1}$. Both layers are LTI and nonnegative diagonal.
Suppose the clean sequence gives $c_t=(-1)^t$, which can happen in an S4D-type layer, where $c_t$ contains the factor $\mathrm{silu}'$
of the layer-2 pre-activation, and that factor changes sign at $x^*$. Then $\bar g_s=a^{n-1-s}(n-s)$ and
$g_s=a^{n-1-s}\sum_{t=s}^{n-1}(-1)^t\in\{0,\pm a^{n-1-s}\}$, so $\bar g_s/\abs{g_s}\ge n-s$, and a band of distances $[d_0,d_1)$ before
the document (position $s=-1-d$) has $G_b/D_b\ge n+1+d_0$: the ratio grows linearly with the distance. $G(L)$ and $D(L)$ still converge,
as (b3) states, and (b1) holds: $\bar g_s$ is a nonnegative combination of transport products, but the number of terms grows with the
lag. With decays $a\ne b$ in the two layers the ratio stays bounded, so growth needs near-equal time constants in consecutive layers. A
flat band profile of a two-layer model is therefore not a consequence of the nonnegative diagonal structure.

7.~\emph{The $L=\infty$ proxy.} $R_{\rm inf}$ was flagged as converged for the five word-level models and not for the byte-level Mamba,
whose $D(\famname{per}\,4)/D(\famname{per}\,2)$ was $1.05$.
\end{remark}

\begin{remark}[The head]\label{rem:head}
The margin bound is formed from separate logit bounds, so the edge weights at the head are $\abs{W^{\rm head}_y}+\abs{W^{\rm head}_k}$
rather than $\abs{W^{\rm head}_y-W^{\rm head}_k}$. Bounding the row $W^{\rm head}_y-W^{\rm head}_k$ on the box of the last hidden vector,
the elision of \citet{gowal2018}, would remove this factor. The reported $R$ measures this implementation, not IBP with elision.
\end{remark}

\subsection{Assumptions, scope and what is not proved}\label{app:scope}

\begin{remark}[What the results do not claim]\label{rem:notclaimed}
1.~\emph{One period at a time, or one token set.} \Cref{thm:inf,thm:hull} certify $p^Kx$, $K\le\infty$, for the given period $p$ and
its per-position boxes, and say nothing about other contexts. \Cref{cor:anyctx} certifies every context over a token set $S$, at the
price of the box hull $E_S$ at every position. Over a single repeated token ($\abs{S}=1$, the most frequent neutral token, period
$P=8$) it certified 53--94\% of the documents for the certified-trained S4D-type models with a clean document (27.5--61.5\% with the
document's own substitution box, leaving out the near-constant standard-IBP seed~1, clean accuracy 0.54, which reached 86\% and which the
degeneracy rule of the slow-channel attribution excludes), at most $5.3\pm9.2\%$ (mean $\pm$ sd over seeds) for the certified-trained
selective models (the first study's, with a clean document; with the cap and the hull term, \cref{app:remedy}, 85.5\% [73.7, 97.3]
and 75.7\% [63.2, 88.2] of the documents of Mamba and the minGRU-type model are certified with the document in its substitution
box), and nothing for any natural nonnegative diagonal or gated model; for LSTM and GRU the construction is the invariant
box of \cref{rem:invbox}, used in place of the hull, so their ``nothing'' is a statement about that construction and not an instance of
\cref{cor:anyctx}. With the 64 most frequent neutral tokens every non-degenerate stateful model was at or near $0$. With $P=64$, Mamba after
fixed-point fine-tuning reached $23.2\pm38.8\%$ at $\abs{S}=1$. These sets are frequent neutral tokens, not a model of real contexts.

2.~\emph{The limit only.} \Cref{thm:inf} is about $p^\infty x$ in the sense of \cref{def:limit} (\cref{ex:finite}). Finite contexts
need \cref{thm:hull} or \cref{thm:hullK}, which cost width (\cref{rem:price}).

3.~\emph{Incompleteness.} IBP certificates lower-bound robust accuracy; substitution attacks and PGD give upper bounds. The gap is an
empirical quantity.

4.~\emph{Non-vacuity.} The theorems hold for any weights. Small boxes are an empirical property of trained models;
\cref{prop:exactac}(a) and \cref{prop:exact}(a$'$) give $O(\eps)$ only with constants that contain $1/\denh$ or $1/(1-\bar a')$ per
layer (\cref{rem:notsay}).

5.~\emph{Real arithmetic} (\cref{rem:float}).

6.~\emph{Stateless baselines} have their own exact semantics, outside \cref{thm:inf,thm:hull}. The CNN's max-pool box over the window
set of $p^Kx$ is exact for each $K$ but not nested in $K$, because a maximum over more windows is larger, not wider; its K-uniform box is
the hull over $K=0$, $K=1$, $K\ge2$ and $K=\infty$. For bag-of-words the document's weight $R/(KP+R)$ in the mean vanishes as
$K\to\infty$, so the $L=\infty$ certificate says nothing about the document.

7.~\emph{Gated RNN baselines.} A bounded invariant box exists (\cref{prop:exactac}(c)), but neither its non-vacuity nor its discovery
by the bounded search of \cref{rem:invbox} is guaranteed.

8.~\emph{Threat model.} Every position is perturbed independently inside its box; insertions and deletions are not modelled. The
substitution boxes use the counter-fitted vectors also as model inputs, a deviation from \citet{jia2019}, who feed GloVe vectors and
filter the substitution sets with a language model.

9.~\emph{\Cref{prop:loss} is about $\eps\to0$.} Whether $\eps\in[10^{-10},10^{-8}]$ is in the linear regime of a given model, and how
$R(L)$ behaves for $L$ below the slowest time constant, are empirical (\cref{rem:gate3reading}).

10.~\emph{Depth.} \Cref{prop:exactac}(a), \cref{prop:exact}(a$'$), \cref{prop:loss}(b1), (b3) and (c2) hold for any number of
layers; identity \eqref{eq:zetachi}, \cref{prop:loss}(c), (c1) and (c3) and \cref{prop:exact}(b) and (b$''$) are one-layer statements.
Every trained model of this paper has two layers, so its band profile and the mechanism of its gated saturation are measured, not
proved.

11.~\emph{The boundary.} \Cref{lem:boxes} gives sufficient conditions for exactness in time in boxes, and for point transitions a
characterisation (\cref{lem:boxes}(ii)). \Cref{lem:disks} gives a bound that does not compound, not exactness, for disks under
complex-diagonal transitions with input-independent $\lambda$. The complex-diagonal control tests one complex-diagonal LTI
architecture with two seeds. Selective complex-diagonal layers, S5 with dense input and output maps, and delta-rule layers are not
tested.

12.~\emph{Gated transitions and fixed-size domains.} Boxes and the budgeted zonotopes we tried give no time-uniform certificate for the
gated models (the distance probe and the budget sweep, \cref{app:gate5}); whether some fixed-size domain does is open. Nothing here
says that gated transitions preserve no fixed-size domain, and nothing here says that a relational domain fixes gated certification.

13.~\emph{Reduced zonotopes never lose against boxes on linear point recursions} (\cref{lem:zono}). The paper does not claim that a
budgeted zonotope loses, or compounds a reduction loss, in time on the box-exact transitions of \cref{lem:boxes}(ii), nor that it
dominates the box: where a zonotope lost to the box in the distance probe (S4D-type model, $d=512$), the loss came at a large $\eps$ and
is of higher order (\cref{rem:archs}, \cref{app:gate5}). Nor does it claim that a reduced zonotope cannot do better than boxes in time:
with budget 0 it is the box, which diverges on \cref{prop:exact}(b$'$), where an unreduced zonotope is exact.
\end{remark}

\paragraph{Assumptions.}
\begin{itemize}
\item[(A0)] \emph{Real arithmetic.} No theorem covers rounding. Every evaluation from the first training study under synonym
  substitution on runs in float64 without directed rounding, apart from the outward-rounded re-verification of \cref{rem:float}; the
  earlier byte-level evaluations ran a float32 model around a float64 period part,
  and their float64 re-evaluation is in \cref{app:byte} (\cref{rem:float}). \Cref{prop:loss}(a) is about the implemented softplus away
  from $20$; part (b3) uses the exact softplus.
\item[(A1)] \emph{Differentiability} of every primitive at its clean argument: no clean gate logit equal to $20$, no clean SiLU input
  equal to $x^*$, no ReLU pre-activation equal to $0$, no tie in a maximum, a unique runner-up class (automatic for two classes). For
  (b3): $C^1$ with locally Lipschitz derivatives on a neighbourhood of the closure of the clean values, and the first-order property at
  every $L$.
\item[(A2)] \emph{Uniform contraction} $a^0\le\bar a<1$: automatic in exact arithmetic for the three nonnegative diagonal families
  (\cref{lem:uniform}), with $\bar a$ possibly very close to $1$.
\item[(A3)] $D_\infty>0$ and $D(L)>0$ for the $L$ considered.
\item[(A4)] For \cref{prop:loss}(a$_\infty$): the clean periodic cascade contracts (automatic in exact arithmetic) and, for the
  minGRU-type layer, no clean period gate logit above $20$; for $G_\infty=\lim_KG(\famname{per}\,K)$: the hypotheses of (b3) for the
  past $p^\infty$.
\item[(A5)] For \cref{prop:loss}(c), (c1), (c3): a one-layer LSTM or GRU cell (none of the models of the context-length probe). For
  (c1): a stationary stretch of the clean trajectory with $M$ primitive and $\rho(M)>1$ (and $\rho(J)<1$ for the geometric growth of the
  ratio). For (c2): LSTM or GRU cells with any number of layers (bounded $h$ boxes).
\item[(A6)] \emph{The probe threat model}: an $\ell_\infty$ box of radius $\eps$ on every perturbed embedding, as in the context-length
  probe, not the substitution boxes of the training studies. \Cref{prop:loss} holds for boxes of radius $\eps\rho_t$ with any fixed
  $\rho_t\ge0$ (\cref{lem:composition}), but the probe's quantities use $\rho_t=\one$.
\item[(A7)] For \cref{prop:exact}(a$'$): embeddings in a compact set, input boxes that contain the clean embedding and have radius at
  most $C_0\eps$, and exact interval extensions (the SiLU slack and the softplus widening of \cref{rem:silu} add $\eps$-independent
  terms).
\item[(A8)] For \cref{lem:boxes,lem:disks}: the coefficients range over their boxes independently across positions and independently of
  the state. This is the semantics of IBP given its coefficient boxes. The true model correlates the coefficients of nearby positions
  through the shared input; in the first layer that loss is local (\cref{prop:loss}(b2)(i)). From the second layer on, it correlates
  the coefficients of all positions through the states of the layers below. That is the type (iii) loss of \cref{prop:loss}(b2), which
  is not local and can grow with the distance (\cref{rem:gate3reading}, item 6). Neither is part of ``exactness in time''.
\end{itemize}

\paragraph{What is not proved, or only sketched.}
\begin{enumerate}
\item No rate or bound for $R_1(L)$ at the lengths of the context-length probe: (b3) is asymptotic, with constants that blow up as
  $\bar a\to1$, and the constant of \cref{prop:exact}(a$'$) is not explicit. The growth criterion of the probe is not a consequence of
  either.
\item No a-priori bound on the second-order term $G_2$, hence none on whether $[10^{-10},10^{-8}]$ is in the linear regime (the amended
  exponent criterion). The expansion $W=\eps G+\eps^2G_2+o(\eps^2)$ in \cref{cor:observables}(i) is assumed, not derived.
\item No limit theorem for the band ratios $R_b$ as the distance grows. \eqref{eq:zetachi} is an exact decomposition for one layer; for
  $Q\ge2$ the band ratio can grow with the distance (\cref{rem:gate3reading}, item 6), and the measured flat profiles are empirical.
\item For gated cells on natural contexts there is no theorem that $G(L)$ grows: (c1) needs a stationary stretch, and in general the
  growth rate is the top Lyapunov exponent of the products $M_{n-1}\cdots M_{s+1}$, which is not controlled. The interval-map growth
  rates measure it for the two-layer natural models (\cref{app:gate3}).
\item \Cref{prop:exact}(b$''$) is proved for the construction of (b) only. $\nu^*=0.4716676$ and $\rho_A=1.1995306$ were evaluated
  numerically (\cref{rem:lstmnum}); the upper bound of (b$''$) has no explicit constant.
\item The directional (B-derivative) version for max-pool ties and ReLU at $0$ (\cref{rem:nonsmooth}, item 1) is a sketch that relies
  on the chain rule for directional derivatives of locally Lipschitz maps; it is not written out.
\item No first-order theory for the K-uniform bounds (\famname{hull}, the gated invariant box), whose $b(0)$ is not the clean margin.
\item The $L=\infty$ proxy $W_{\rm inf}/D(\famname{per}\,4)$ converges to $G_\infty/D^{\rm per}_\infty$ only as the number of copies
  grows; nothing guarantees convergence at $K=4$, which is why the probe checks it empirically.
\item The extension of \cref{prop:loss}(c) and (c1) to stacked gated layers is not written out. It is needed for every gated model of the
  context-length probe: the word-level LSTM and GRU and the byte-level LSTM were all trained with two layers. For them $M_t\ge\abs{J_t}$
  holds in each layer and (c2) holds verbatim. The lag-resolved formula becomes a sum over the position at which a path crosses to
  layer 2, as in (b2)(iii), and (c1) is not stated for this case.
\item The domain-relative principle is a set of sufficient conditions. ``Boxes for nonnegative transitions'' is not an equivalence: real
  diagonal transitions of either sign (\cref{lem:boxes}(i)) and the sign-pattern-nonnegative transitions $DN_tD$ and $D_tN_tD_{t-1}$
  (\cref{rem:provisos}) are also exact in boxes, and only for point transitions does \cref{lem:boxes}(ii) give a necessary and
  sufficient condition, on the injection radii; $D_tN_tD_{t-1}$ is the exact class only for transitions without zero entries and with
  positive injection radii (\cref{rem:provisos}). For disks there is no exactness statement, only the non-compounding bound of
  \cref{lem:disks}.
\item Floating-point soundness rests on a measured assumption: the outward-rounded re-verification of \cref{rem:float} assumes that
  the platform's $\exp$, $\mathrm{expm1}$, $\mathrm{log1p}$, $\tanh$ and square root err by at most two ulps, which was measured on
  the platform and not proved, and it leaves out the certificates listed there.
\item The attribution of the gap between the S4D-type and the selective models' looseness to the gate path (\cref{rem:gatepath}) is
  untested.
\item No result on fixed-size domains for gated transitions. Boxes and the budgeted zonotopes of the distance probe and the budget sweep
  give no time-uniform certificate for the gated models; whether some fixed-size domain does is open (\cref{rem:notclaimed}, item 12).
\item No theory of budgeted relational domains for gated cells: nothing bounds the budget needed to keep a certified radius over $d$
  steps. The budget sweep measured seven budgets on one seed, with a reach that grows with the budget up to the end of its grid
  ($d=1024$); that is a measurement, not a bound. On linear point-transition recursions a reduced zonotope is never wider than boxes
  (\cref{lem:zono}). The higher-order relaxation loss that let boxes beat the $B=256$ zonotope on the S4D-type model at $d=512$ is not
  bounded here.
\item Selective complex-diagonal layers are not covered, and this is the open boundary of the disk result. Under IBP their $\lambda_t$
  are set-valued and \cref{lem:disks} does not apply: with $\lambda\in\{1,i\}$ the ratio of the composed disk radius to the hull radius
  tends to $(2+\sqrt2)/2>\pi/2$, and with $\lambda\in\{1,e^{i\vartheta}\}$, or $\{q,qe^{i\vartheta}\}$ with $q<1$, it tends to
  $(1+\cot(\vartheta/4))/2$, which is unbounded as $\vartheta\to0$ (\cref{rem:pihalf}). Whether some domain follows input-dependent
  complex-diagonal transitions uniformly in time is open.
\item No result says that \cref{prop:exact}(a$'$) or \cref{prop:loss}(b3) predicts the flat profiles of the context-length probe and
  its replication, or that (b$''$) explains the measured saturation of the two-layer gated models; both are measured.
\end{enumerate}

\clearpage
\section{Notation, implementation and numerics}\label{app:numerics}

\paragraph{Notation.}
\Cref{tab:notation,tab:notation2} list the symbols of the paper. Section~3 works with one state channel at a time; in a stack,
every quantity of \cref{tab:notation} carries a layer and a channel index, which we drop.

\begin{table}[h]
\caption{Notation (1): models, contexts, and one state channel over a period.}
\label{tab:notation}
\centering\small
\begin{tabular}{@{}>{\raggedright\arraybackslash}p{0.25\textwidth}>{\raggedright\arraybackslash}p{0.71\textwidth}@{}}
\toprule
symbol & meaning\\
\midrule
$Q$, $\ell$ & number of layers ($Q=2$ in every trained model), layer index\\
$d_m$, $d_i$, $N$ & model width ($d_m=64$), inner width ($d_i=2d_m$), states per inner channel ($N=16$ for Mamba and the S4D-type
layer; the minGRU-type layer has one)\\
$k_c$ & kernel size of the causal depthwise convolution ($k_c=4$)\\
$e_t$, $E_t$, $\tilde e_t$ & embedding at position $t$, its input box, an admissible perturbation $\tilde e_t\in E_t$\\
$S$, $E_S$ & a token set, and the box hull of the embeddings of its tokens (\cref{cor:anyctx}); in \cref{sec:theory-loss} $S$ is the
set of perturbed positions\\
$V$, $p\in V^P$, $x\in V^n$ & vocabulary; a context period of $P$ tokens; a document of $n$ tokens\\
$p^Kx$, $K$, $L$ & $K\in\{0,1,\dots,\infty\}$ copies of $p$ followed by $x$; the context length $L=KP$ ($L=0$: the bare document)\\
$P_0$ & convolution-context positions placed before the period in the implementation's layout ($P_0=8$)\\
$j$ & phase: the position inside the period, $j=1,\dots,P$\\
$F_K(\tilde e)$, $F_\infty(\tilde e)$ & logits on $p^Kx$ from the zero state, and their limit $K\to\infty$ (\cref{def:limit})\\
$a_t$, $u_t$ & decay and injection of one state channel at position $t$, as in \eqref{eq:scan}\\
$A=[\alo,\ahi]$, $\acen$, $\arad$ & decay box of one channel at one position, its centre and its radius\\
$U$, $\ucen$, $\urad$ & injection box, its centre and its radius\\
$H=[c-r,c+r]$ & a state box with centre $c$ and radius $r$\\
$\alpha$, $s$, $\Delta$, $\bar a$ & Mamba: decay rate $\alpha>0$, gate logit $s$, step $\Delta=\softplus(s)$, decay $\bar a=e^{-\alpha\Delta}$
(\cref{lem:gate,lem:dc}); from \cref{lem:uniform} on, $\bar a$ is the uniform bound on the clean decays\\
$\eta$, $\varphi(\Delta)$ & width $s^{\rm hi}-s^{\rm lo}$ of a gate-logit box; $\varphi(\Delta)=\Delta/(1-e^{-\alpha\Delta})$, so that
$\Delta=(1-\bar a)\,\varphi(\Delta)$\\
\midrule
$\Gc_j$, $\Gh_j$ & prefix products $\prod_{i\le j}\acen_i$ and $\prod_{i\le j}\ahi_i$ over a period ($\Gamma_0=1$)\\
$w_j$, $w^{\rm c}_j$ & suffix products $\prod_{i=j+1}^P\ahi_i$ and $\prod_{i=j+1}^P\acen_i$ ($w_P=w^{\rm c}_P=1$)\\
$\pic$, $\pih$ & contraction of the centre and of the radius over one period, $\Gc_P$ and $\Gh_P$\\
$\denh$ & the period leak $1-\pih$\\
$\tau_{\rm h}$ & time constant of a channel, $P/(-\ln\pih)$ tokens, about $1/(1-\ahi)$ for a constant decay; a channel is
\emph{slow} when $\tau_{\rm h}$ exceeds the length of the training documents\\
$c^0_j$, $\cfree$ & centre after phase $j$ of one period started at centre $0$; $\cfree=c^0_P$\\
$q_j(c)$ & radius injected at phase $j$ when the period starts at centre $c$: $\arad_j\abs{\Gc_{j-1}c+c^0_{j-1}}+\urad_j$\\
$\beta_j(c)$, $\beta(c)$ & radius accumulated over phases $1,\dots,j$, and over the whole period ($\beta(c)=\beta_P(c)$), from $(c,0)$\\
$\Psi$ & period map $(c,r)\mapsto(\pic c+\cfree,\ \pih r+\beta(c))$ (\cref{lem:period})\\
$(c^*,r^*)$, $H^*$, $H^*_j$ & fixed point of $\Psi$, its box $[c^*-r^*,c^*+r^*]$, and the phase boxes of the fixed point\\
$\mu_j$ & DC weights $w_j(1-\ahi_j)/\denh$: nonnegative, summing to one (\cref{rem:dcform})\\
$Q_j$, $Q_j(c)$, $\bar Q$ & DC level $q_j(c^*)/(1-\ahi_j)$, the steady-state level of the radius of phase $j$ under constant
coefficients; the same from start centre $c$; $\bar Q=\max_j\max\{Q_j(0),Q_j(c^*)\}$ (\cref{prop:horizon})\\
$\bar r$, $d_j$, $\varrho_j$, $\hat H_j$ & hull construction of \cref{thm:hull}: $\bar r=\max\{r^*,\beta(0)/\denh\}$, $d_j=\Gc_jc^*$,
$\varrho_j=\Gh_j\bar r+\max\{\beta_j(0),\beta_j(c^*)\}$, and the hull phase boxes $\hat H_j$\\
$\Binf(p,x)$, $\Bcup$ & logit box of the periodic cascade (\cref{thm:inf}) and of the hull construction (\cref{thm:hull})\\
$r_K$, $\omega$ & radius after $K$ copies from the zero state; shift of all gate logits of a period (\cref{prop:horizon})\\
\bottomrule
\end{tabular}
\end{table}

\begin{table}[h]
\caption{Notation (2): domains, looseness, gated cells and the experiments.}
\label{tab:notation2}
\centering\small
\begin{tabular}{@{}>{\raggedright\arraybackslash}p{0.25\textwidth}>{\raggedright\arraybackslash}p{0.71\textwidth}@{}}
\toprule
symbol & meaning\\
\midrule
$T_t$, $\Phi(t,s)$ & transition of a recursion $h_t=T_th_{t-1}+u_t$, and the product $T_t\cdots T_{s+1}$\\
$\mathcal C_t$, $\Reach_t$ & set of admissible coefficients, reachable set (\cref{def:exact})\\
$\mathcal D$, $\hullD$, $\mathcal F_t$ & abstract domain, its hull map, its best step\\
$D_t$, $N_t$ & sign matrices $\diag(\pm1)$ and entrywise nonnegative matrices of the class $T_t=D_tN_tD_{t-1}$ (\cref{lem:boxes})\\
$\disk(c,\rho)$, $\rho_t$, $\Omega_t$ & closed disk of $\complexes$; disk radius of \cref{lem:disks}; circumradius
$\max_{h\in\Reach_t}\abs{h-c_t}$ of the reachable set\\
$\lambda=\abs{\lambda}e^{i\theta}$, $\gbox$ & eigenvalue of a complex mode; box growth factor $\abs{\lambda}(\abs{\cos\theta}+\abs{\sin\theta})$\\
$y$, $y_0$, $m$, $m^0$ & class; clean prediction on the bare document; margin $\ell_y-\max_{k\ne y}\ell_k$; clean margin\\
$b(\eps)$, $W(\eps)$ & IBP lower bound of the margin; margin loss $b(0)-b(\eps)$\\
$D$, $G$, $R(\eps)$ & first-order true sensitivity $\sum_t\norm{\nabla_{e_t}m}_1$; path sum of the absolute-value graph, the bound IBP
would give if every operation were linear (\cref{def:absgraph}); looseness $W(\eps)/(\eps D)$\\
$g_t$, $\bar g_t$, $\dot r$ & gradient of the margin with respect to $e_t$, its absolute-value counterpart; first-order radius of a box
(\cref{def:firstorder})\\
$\varpi(\eps)$ & width exponent (\cref{cor:observables}): 1 for a width linear in $\eps$, 0 for a width independent of $\eps$\\
$M_t$, $J_t$, $\lambda_M$, $\lambda_J$ & first-order radius map of a gated cell's interval step ($M_t\ge\abs{J_t}$) and its true state
Jacobian; their top Lyapunov exponents per token\\
$\gamma$, $\kappa$, $\rho_A$, $\nu^*$, $\nu_t$, $\tsat$ & the contracting LSTM of \cref{prop:lstmfull}: $\gamma=\sigma(8)$, the contraction
rate $\kappa$ of its true dynamics, the spectral radius $\rho_A$ of its absolute Jacobian at the zero state, the positive root $\nu^*$,
the radius $\nu_t$ of its $h$ box after $t$ steps, and the time to saturation\\
$\varsigma$ & RMSNorm scale at the clean argument (\cref{lem:rmsnorm})\\
$r(d)$ & certified radius of one token at distance $d$ before the readout (distance probe): the largest $\eps$ at which a method
certifies the clean prediction; a radius of the input ball, not of a state box\\
$r_{\rm hi}$ & upper end of the final bisection bracket of a certified radius, an upper bound of the radius (\cref{app:lirpa})\\
$b(0)$ & a method's lower bound of the margin at a zero-width input, which equals $m^0$ in exact arithmetic; for IBP the
$b(\eps)$ above at $\eps=0$, in \cref{app:lirpa} for any method (\cref{tab:lirpafwd})\\
$T$ & in \cref{app:lirpa}, the length of the whole sequence [$L$ neutral tokens $\mid$ document]; elsewhere the truncation length of
the documents\\
$B$, $d^*_B$ & zonotope budget (promoted symbols kept per step); reach: the largest grid distance up to which at least half of the
documents keep a tenth of their adjacent-token certified radius\\
\bottomrule
\end{tabular}
\end{table}

\paragraph{Symbols with two meanings.}
A few letters keep two meanings, each local to one part of the paper. $\lambda$ is the eigenvalue of a complex mode, and with the
subscripts $J$ and $M$ a Lyapunov exponent, the usual notation in both fields. $\rho$ is the radius of a disk ($\rho_t$) and a spectral
radius ($\rho(\cdot)$, $\rho_A$). $B$ is the zonotope budget, as in the figures, and $B_t$ Mamba's input matrix (\cref{tab:layers},
\cref{prop:horizon}(a), \cref{lem:dc}). $D$ alone is the first-order sensitivity; $D_t$ are sign matrices, $\mathcal D$ is a domain,
$\disk$ a disk and $D^{\rm sk}$ the skip parameter of the readout below. $W$ alone is the margin loss; with a subscript it is a weight
matrix ($W_h$, $W_{\rm in}$, $W_{\rm out}$). $Q$ is the number of layers, $Q_j$ and $\bar Q$ are DC levels. $S$ is a set of perturbed
positions in \cref{sec:theory-loss} and a token set in \cref{cor:anyctx}. $\bar a$ is Mamba's decay as a function of the gate logit in
\cref{lem:gate,lem:dc} and a uniform bound on the clean decays from \cref{lem:uniform} on. $T$ is the truncation length of the
documents ($T=512$ tokens, \cref{sec:exp-setup}; 1024 bytes at byte level) and, in \cref{app:lirpa}, the length of the whole
sequence [$L$ neutral tokens $\mid$ document]; $T_t$ is the transition of a recursion.

\paragraph{The layer instances.}
\Cref{tab:instances} gives the three nonnegative diagonal layers and the complex-diagonal control in full. All have $d_i=2d_m$ inner
channels, $k_c=4$, and the output map $o_t=W_{\rm out}\big(y_t\odot\mathrm{SiLU}(z_t)\big)$ with the residual $v^{\ell+1}_t=v^\ell_t+o_t$;
before the scan, layer $\ell$ computes $(\tilde x_t,z_t)=W_{\rm in}\,\mathrm{RMSNorm}(v^\ell_t)$,
$\hat x_t=b+\sum_{k=0}^{k_c-1}w_k\odot\tilde x_{t-k}$ (zero padding before the first position) and $x_t=\mathrm{SiLU}(\hat x_t)$. The
S4D-type and minGRU-type layers are the Mamba block of our code with the selective scan replaced; they are not reimplementations of
the original architectures in every detail. The minGRU-type decay is the Mamba decay with $\alpha=1$, and in all three nonnegative
diagonal families $a_{t,i}<1$ in exact arithmetic.

\begin{table}[h]
\caption{The layer instances, per inner channel $d$ (vectors over the $N$ states of the channel, operations elementwise). $D^{\rm sk}$:
skip parameter. The complex-diagonal S4D is a control and not a nonnegative diagonal layer.}
\label{tab:instances}
\centering\small
\begin{tabular}{@{}>{\raggedright\arraybackslash}p{0.12\textwidth}>{\raggedright\arraybackslash}p{0.10\textwidth}>{\raggedright\arraybackslash}p{0.32\textwidth}>{\raggedright\arraybackslash}p{0.16\textwidth}>{\raggedright\arraybackslash}p{0.18\textwidth}@{}}
\toprule
layer & state $h_{t,d}$ & decay & injection & readout $y_{t,d}$\\
\midrule
Mamba \citep{gu2023} & $\reals^N$ & $\exp(-\alpha_d\Delta_{t,d})$, $\alpha_d\in(0,\infty)^N$, $\Delta_t=\softplus(s_t)$, $s_t=W_{\rm dt}s^{\rm
low}_t+b_{\rm dt}$, $(s^{\rm low}_t,B_t,C_t)=W_xx_t$ & $\Delta_{t,d}\,x_{t,d}\,B_t$ & $C_t^\top h_{t,d}+D^{\rm sk}_dx_{t,d}$\\
S4D-type, real diagonal \citep{gu2022b} & $\reals^N$ & $\exp\big(-e^{A_{\log,d}}\,\softplus(\delta_d)\big)$ with parameters $A_{\log,d}\in\reals^N$,
$\delta_d\in\reals$; input-independent & $x_{t,d}\,B_d$ & $C_d^\top h_{t,d}+D^{\rm sk}_dx_{t,d}$\\
minGRU-type \citep{feng2024} & $\reals$ & $1-\sigma(s_{t,d})=\exp(-\softplus(s_{t,d}))$, $s_t=W_zx_t+b_z$ & $\sigma(s_{t,d})\,(W_hx_t)_d$ &
$C_dh_{t,d}+D^{\rm sk}_dx_{t,d}$\\
complex-diagonal S4D (control) & $\complexes^{N/2}$ & $\lambda_d=\exp(\Delta_dA_d)$, $A_d=-e^{A_{{\rm re},\log,d}}+iA_{{\rm im},d}$,
$\Delta_d=\softplus(\delta_d)$; input-independent & $\bar B_d\,x_{t,d}$, $\bar B=(\lambda-1)B/A$ (zero-order hold) &
$2\,\mathrm{Re}(C_d^\top h_{t,d})+D^{\rm sk}_dx_{t,d}$\\
\bottomrule
\end{tabular}
\end{table}

\paragraph{Interval rules of the implementation.}
Every token-wise map and the convolution are propagated by their natural interval extension, the composition of the exact-range
interval versions of the elementary operations \citep{moore1966,gowal2018}: linear maps and the convolution in centre-radius form
($c=W\mu+b$, $r=\abs{W}\,\mathrm{rad}$), products by their four endpoint products, monotone scalar functions (softplus, $\exp$,
sigmoid, $\tanh$, rsqrt) at the endpoints, and SiLU at the endpoints together with its global minimum when the minimiser lies inside
the box. This gives, per position and state channel, a decay box $A_t\subset[0,1]$ and an injection box $U_t$ that depend only on the
boxes of the convolution window $t-k_c+1,\dots,t$ of the layer input. For Mamba, $\alo_t=\exp(-\alpha\Delta^{\rm hi}_t)$ and
$\ahi_t=\exp(-\alpha\Delta^{\rm lo}_t)$ with $\Delta^{\rm lo/hi}_t=\softplus(s^{\rm lo/hi}_t)$, because $\bar a$ is decreasing in $s$.
The GRU's update $(1-z)n+zh$ is bounded by the exact range of the convex combination over the boxes of $z$, $n$ and $h$; the LSTM's
gate products by the endpoint rule for a factor in $[0,1]$.

\paragraph{Two scans.}
The scan is implemented twice. The centre-radius scan \eqref{eq:cr} (\texttt{ibp\_mixer\_cr} in \texttt{ibp\_mamba.py} and
\texttt{zoo\_ssm.py}) is used by the $L=\infty$ certificate and by the hull. The endpoint scan (\texttt{ibp\_mixer}) bounds $ah$ by its
exact range and is used for the plain certificates of finite contexts; it is tighter by $\arad\min(r,\abs{c})$ per step
(\cref{lem:cr}(b)). Both are exact to first order and have the same first-order radius at every node (\cref{prop:loss}(a)), so their
numbers are comparable at first order for the nonnegative diagonal models, and only there. In the context-length probe, the finite
contexts use the endpoint scan and $L=\infty$ the centre-radius scan. The period is laid out as
$[\,p_{P-P_0+1},\dots,p_P\mid p\mid x\,]$, which computes the periodic cascade whenever $P\ge P_0\ge(k_c-1)Q$
(\cref{lem:implcascade}); the code asserts this inequality, and every experiment uses $Q=2$ and $P_0=8$.
Both unbounded-context certificates run \eqref{eq:cr}, which is sound but not the best box step (\cref{lem:cr}(b)). Its
per-step excess is of second order when the boxes have radius $O(\eps)$, but with wide coefficient boxes it accumulates over the time
constant. For one channel with $a\in[0.98,0.999]$ and $u\in[0.9,1.1]$ at every position, \eqref{eq:cr} settles at
$c^*=2000/21\approx95.2$ and $r^*=21100/21\approx1004.8$, the box $[-19100/21,\,1100]\approx[-910,1100]$, while the hull of the limit
set, the limit of the endpoint scan, is $[0.9/0.02,\ 1.1/0.001]=[45,1100]$: the upper end is exact, and the excess, all at the lower
end, makes the box about 1.9 times as wide.

\begin{remark}[SiLU and softplus as implemented]\label{rem:silu}
The SiLU primitive takes as lower end the global minimum of SiLU lowered by a slack whenever the minimiser lies in the box:
$10^{-15}$ in float64 and $10^{-7}$ in float32, a margin against rounding of the constant. The softplus primitive widens a box that
straddles the threshold $20$ of the library softplus by the drop $2.1\times10^{-9}$ of that function at its threshold. Both enlarge
boxes, keep the soundness and the inclusion monotonicity of \cref{cor:stack}, and break its continuity statement by at most the slack,
respectively the drop. The proofs of \cref{app:proofs} use the exact extension; \cref{rem:enclosure} transfers their conclusions to any
enclosure that contains it. Statements about $\eps\to0$ (\cref{prop:exact,prop:loss}) hold for the code only where no clean SiLU input
lies within the box radius of the minimiser and no clean gate logit within the box radius of $20$ (\cref{rem:nonsmooth}).
\end{remark}

\begin{remark}[Computing the fixed point and the hull]\label{rem:impl}
The period part (\texttt{\_ss\_period} in \texttt{ibp\_mamba.py} and \texttt{zoo\_ssm.py}) computes $(c^*,r^*)$ in float64 in the DC form
of \cref{lem:period}. For Mamba, $\log\bar a=-\alpha\Delta$ is taken exactly; the weights $\mu_j$ are a softmax of
$\log w_j+\log(1-\ahi_j)$; and the phase trajectory needs one centre scan from $0$ and one radius scan from $c^*$. The DC form is
designed never to divide two vanishing quantities: for a slow channel both $\denh$ and $\beta(c^*)$ tend to $0$, while their quotient
is a convex combination of DC levels, which \cref{lem:dc} keeps bounded. No rounding-error bound is proved for this float64
computation; the re-verification of \cref{rem:float} encloses the same closed form with outward rounding. Phases
with $1-\ahi_j=0$ arise only from floating-point underflow; their injections are added with the full denominator, and $\denh=0$
returns $r^*=+\infty$, so the certificate is then vacuous (as in exact arithmetic, \cref{rem:denzero}).

\emph{Two numerical issues of the minGRU-type decay.} (i)~\emph{The softplus threshold.} The period part takes the logarithm of the
minGRU-type decay as $-\softplus(s)$ through the library softplus, which returns $s$ above its threshold $20$, while the model
computes $1-\sigma(s)$. So the period part uses the decay $e^{-s}$ instead of $1-\sigma(s)=e^{-s}/(1+e^{-s})$: a relative overestimate
by the factor $1+e^{-s}$, of the order of the drop of the library softplus at $20$, and this path is not exactly sound, even in real
arithmetic, in channels with a period gate logit above $20$. It changed no certificate: on every minGRU-type checkpoint and setting
of the earlier studies that \cref{rem:float} re-verified, this path, the direct path below (in float64) and the outward-rounded evaluation, which encloses
the exact decay, certify the same documents, and the margin bounds of the two float64 paths differ by at most $9.6\times10^{-13}$
relative to $1+\abs{b}$. Mamba is not affected, because its model uses the same softplus. (ii)~\emph{The complement.} The model computes
the decay as $1-\sigma(s)$, which evaluates to $0$ once $\sigma(s)$ rounds to $1$: in float32 for $s>17.3$, and in float64 only at
much larger logits, where the true decay $\sigma(-s)$ is still positive. The code now has a decay path that computes $\sigma(-s)$ and
$\log\sigma(-s)$ directly, without the threshold and without the subtraction, which removes both issues; every minGRU-type number of
the earlier studies was computed with the earlier path; the training remedy of \cref{app:remedy} used the direct path throughout, and
re-evaluated on it the three standard-IBP minGRU-type models of \cref{app:gate2} give the same clean and certified shares as
\cref{tab:trainseeds}, to 0.1 point. With the cap of \cref{app:remedy}, the centre-radius scan of the minGRU-type layer takes the
radius coefficient of the decay box $\rho\,\sigma(-s)$ from the injection weight $\sigma(s)+\sigma(-s)/\tau_{\max}$ (half its spread
over the gate box). That is the decay box's half-width exactly when $1-\rho=1/\tau_{\max}$, which holds in float64 for the caps
used (128, 512 and 2048, powers of two), and the outward-rounded evaluator refuses a cap for which it does not hold.
\end{remark}

\begin{remark}[Floating point]\label{rem:float}
All results of this paper are in real arithmetic. The implementation computes the period part of both certificates (fixed point
and hull) in float64, with exact logarithms of the decays (for the minGRU-type layer above gate logit $20$, see \cref{rem:impl}), and
casts the result to the working precision by rounding to nearest, without outward rounding. The word-level experiments run the
whole model and all boxes in float64: IBP training under synonym substitution, the context-length probe, the complex-diagonal control,
the interval-map growth rates, the slow-channel attribution, the token-set certificate, the distance probe, the budget sweep and the
training remedy for the blind spot (on GPUs, \cref{app:remedy}). The
early byte-level length evaluations ran a float32 model around the float64 period part; their float64 re-evaluation on 1000 reviews
reproduced the float32-period numbers (\cref{app:byte}).

\emph{What float32 does.} On a trained byte-level Mamba checkpoint, in the float32 document scan, $\ahi=1-(1-\bar a)$ rounds to exactly
$1.0$ for 11.6--12.7\% of the layer-1 (position, channel) pairs, which is conservative for the box, and lies below
$\exp(-\alpha\Delta^{\rm lo})$ for 45--51\% of them, which is anti-conservative by up to $2^{-25}$ per step. The decays that round to
$1.0$ lie within $2^{-25}$ of $1$, so $1.0$ is their correctly rounded float32 value: no formula can represent them below $1$ in
float32, and in a float32 network those channels do not forget at those positions. This concerns Mamba; the minGRU-type issues of
\cref{rem:impl} are separate. In the capped models of \cref{app:remedy} no decay rounds to 1 in float32: every decay is at most
$1-1/\tau_{\max}$, so every complement is at least $1/\tau_{\max}\ge1/2048>4.8\times10^{-4}$. The float32 logit box misses the float64 box by
$1.4\times10^{-6}$ at $\eps=0.002$ ($7.7\times10^{-7}$ of the width), but by 1--2\% of the width at $\eps=10^{-6}$, both without a prefix and
at $L=\infty$. The float32 synonym boxes lie within $1.7\times10^{-6}$ of a word-by-word brute force (\cref{app:audits}). For the
context-length probe, the rounding floor of $W$ at $\eps=10^{-10}$ is about $10^{-6}$ to $10^{-5}$ relative, with at least two orders of
magnitude of margin in every validity check. The box arm of the complex-diagonal control overflowed float64 at long contexts; this is
real growth of the box (\cref{prop:exact}(b$'$)), not rounding, and the pre-registration exempted that arm from the finiteness check.

\emph{Re-verification with outward rounding.} An evaluator that mirrors the certificate code operation by operation in float64
encloses every exact intermediate value \citep{rump2010}. It steps the result of each IEEE operation one float outward and that of
$\exp$, $\mathrm{expm1}$, $\mathrm{log1p}$, $\tanh$ and the square root four floats outward, and it widens sums and matrix products,
in any summation order, by $\gamma_n\abs{x}^\top\abs{y}$, $\gamma_n=nu/(1-nu)$ with $u$ the unit roundoff, plus an underflow term.
Four steps enclose the exact value if the platform's implementation of each function errs by at most two ulps.
This is the library assumption, measured and not proved: over $10^7$ random arguments per function, adversarial ones and between $5.4\times10^4$ and
$10^6$ recorded from the certificate computations, the largest errors were 0.52, 1.00, 1.00, 0.75 and 0.53 ulp, in the order above.
An error is measured against a 128-bit reference (mpmath) and counted in ulps of the exact value, the spacing $2^{e-52}$ of the floats
in the binade $[2^e,2^{e+1})$ that contains it (with $e$ clamped at $-1022$, the subnormal spacing), on the tensor layouts of the
evaluator's calls; in this convention a
correctly rounded result errs by at most 0.5 ulp. The library square root that the evaluator calls, which returned the same values on
every layout, is therefore not correctly rounded on this platform, and the evaluator does not treat it as an IEEE operation, one float
outward, but as a library function, four floats outward.
With every decay carried
as an enclosure of its complement $1-a$, the $L=\infty$ fixed point and the K-uniform hull are enclosed from the DC closed form of
\cref{lem:period}, in which the period leak $1-\pih$ is a sum of nonnegative terms rather than a difference, so the enclosure has
no $1/(1-\pih)$ amplification and no limit in the time constant.
Certificate checking runs beside it as a cross-check, and it is an inductive-invariant argument. Layer by layer in cascade order, the
float64 box of every channel is inflated slightly to a box $S$, and one outward-rounded period map, which encloses the exact one, is
run from a start box. At $L=\infty$ the start box is $S$, and the map must end inside $S$; the iterates of the exact period map from
$S$ are then nested (\cref{lem:cr}(c)) and converge to $H^*$ (\cref{lem:period}), so $H^*\subset S$. For the hull, $S$ is the hull's
end-of-period box, the start box is $\hull(S,0)$, with the convolution history hulled with $0$, and the map must end inside $S$; then,
by induction over the copies, the box at the start of every copy lies in $\hull(S,0)$, which contains the zero state, so the phase
boxes of this one period map contain those of every copy and of the cascade, and they take the place of \cref{lem:hull1} in the proof
of \cref{thm:hull}. The certified boxes are those of the check,
not the closed form's: for a point decay, as in the S4D-type layer, the period map from $\hull(S,0)$ does not widen the trajectory from
the zero state by the radius bound $\bar r$, as \cref{def:hullcons} does, so this route can certify a document that the closed form
does not certify. A document counts as certified if either route certifies it. In float64 the check at $L=\infty$ passed on all of 15,400 pairs of
document and layer but one, on every pair whose slowest channel has a time constant up to $2.5\times10^9$ tokens, and failed on the
one at $1.6\times10^{14}$ tokens: the inflation that absorbs the rounding of one period map grows like $u\,\tau_{\rm h}$. The closed
form has no such limit.

\emph{Which network is certified.} The certificates concern the real-arithmetic network, not a deployed float32 network. On the
byte-level Mamba checkpoint above, the deployed float32 network differs from it: 11\% of its clean layer-1 decays equal $1.0$ exactly,
and quantisation changes $1-\prod\bar a$ of its 55 slow channels by up to 0.5\%, which moves its own $L\to\infty$ state by up to
$1.3\times10^{-3}$. A statement about the deployed float32 network would have to treat its rounding as an additional perturbation.
The re-verified network keeps the float64 constants of the code and the library's softplus, which returns its argument above 20
(\cref{rem:silu}), and its embedding lookup returns the rows of an archived float64
table $E$ of input embeddings, computed once per checkpoint by the code's own embedding head on one CPU thread (sha256 of the
manifest of these tables: \texttt{376f135f6e6ec0296eef16d7bfa8c38b\allowbreak 50f5db2aa9b7a79ac1cda555863c9d63}; the hash of each
table is in the code release).
The float64 code recomputes these embeddings at every position, and their rounding can differ from $E$ in some entries,
since it depends on the processor's linear-algebra kernel; substituting $E$ changed none of the 8549 certificate verdicts on which
both were evaluated, and their margin bounds by at most $7.7\times10^{-10}$ relative. The checkpoints of \cref{app:remedy} have
their own archive of such tables, made in the same way (sha256 of its manifest:
\texttt{92039b280de8ae3acb79081d00e53861\allowbreak a770f6e6401ee650705db93be8e20cb6}); on them, too, substituting $E$ changed
none of the 6374 verdicts on which both were evaluated. Their re-verified network also keeps the float64 constants of the cap,
recorded and checked per layer: Mamba adds
$\log\rho$ to every log decay, and the minGRU-type layer multiplies its decay by $\rho=1-1/\tau_{\max}$ and adds $\sigma(-s)/\tau_{\max}$
to its injection weight (its float64 period part adds $\log\rho$ to $\log\sigma(-s)$, which differs from $\log(\rho\,\sigma(-s))$
only by the rounding of $\log\rho$). For $\tau_{\max}=512$, $\rho=1-2^{-9}$ and $1/\tau_{\max}=2^{-9}$
exactly, and $\log\rho$ is the float64 number nearest to $\log(1-2^{-9})$ (\texttt{-0x1.0040155d5889ep-9}). This is not the
deployed float32 capped network.

\emph{Trusted base.} A re-verified certificate rests on the evaluator's code, which mirrors the certificate code operation by
operation and was tested against exact and high-precision arithmetic and by mutation (\cref{app:audits}), not proved correct; on
correctly rounded basic operations with rounding to nearest, no flush to zero and an exact nextafter, which were checked on the
production node type (rounding mode, subnormals and nextafter again at the start of every job) and are not proved; on the library
assumption above, measured and not proved; and on what defines the certified network: the code's float64 weights, the archived
embedding tables, the library softplus and, for the capped models, the cap constants.

\begin{figure}[h]
\centering
\includegraphics[width=\textwidth]{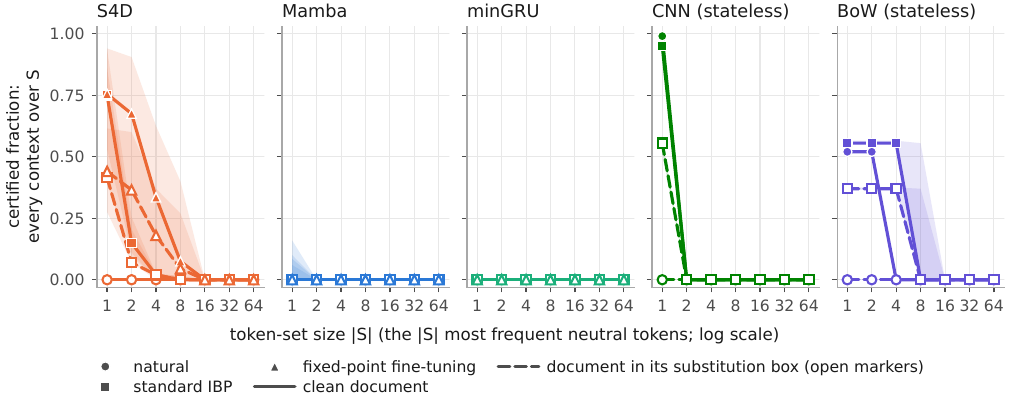}
\caption{The token-set certificate: share of the test documents certified for every context of every length over the token set
$S$ (\cref{cor:anyctx}: the box hull of the embeddings of $S$ at every context position, one K-uniform certificate per residue
modulo the period $P=8$), against $|S|$; the class is the clean prediction of the bare document. Solid: clean document; dashed:
the document also in its own synonym substitution box. Line: median over the non-degenerate seeds (degenerate: more than 80\% of
the predictions in one class), band: seed minimum to maximum; the share certified at $L=0$ with the substitution box, an upper
bound for the dashed curves, is the last column of \cref{tab:anyctx} for the models listed there. Among the recurrent models of this test only the certified-trained real-diagonal
S4D-type model keeps a substantial share at $|S|=1$; the stateless CNN and bag-of-words models also certify at $|S|=1$; every model falls to
zero or near zero as $|S|$ grows. The plotted shares are the float64 ones; their re-verification with outward rounding lost no
certificate and added one document to one share (\cref{tab:anyctx}).}
\label{fig:anyctx}
\end{figure}

\emph{What was re-verified.} The certificates of \cref{tab:train,tab:trainseeds} (every context length, $L=\infty$ and the hull,
1000 documents), the token-set certificates of \cref{tab:anyctx,fig:anyctx} with the share certified at $L=0$, the certificates at
$\eps=10^{-8}$ of the context-length probe (\cref{sec:exp-gate3}, \cref{tab:gate3,tab:stateless}) and the box radii of the Mamba and
S4D-type models in the distance probe (\cref{sec:exp-gate5}, \cref{app:gate5}), on the documents of the runs before the extension of
\cref{app:cleanctx} and on 48 of the 54 checkpoints evaluated (the other
six, the natural GRU and LSTM models, only for the box radii below).
Every one of these certificates
was kept: none was lost in any of the 363 combinations of checkpoint and setting that hold one. One was gained: in \cref{tab:anyctx}
the fixed-point fine-tuned S4D-type seed~0 with $|S|=1$ and a clean document is certified on 152 instead of 151 of 200 documents
(certificate checking certified a document whose closed-form float64 margin bound was $-6.9\times10^{-7}$: it is the token-set
certificate of \cref{cor:anyctx} with the boxes of the invariant above in place of those of \cref{def:hullcons}), and the table prints
that share.
Relative to $1+\abs{b}$, the outward-rounded margin bound lay below the float64 one by at most $6.2\times10^{-12}$ at finite
contexts and $1.6\times10^{-11}$ at $L=\infty$ and for the hull in the first training study, $9.4\times10^{-8}$ for the token sets
(more than the $10^{-8}$ we had predicted, at $|S|=8$ for the fixed-point fine-tuned S4D-type seed~1, which kept every certificate
there) and $3.8\times10^{-7}$ for the probe certificates at $\eps=10^{-8}$ and the box radii. The threshold issue (i) of \cref{rem:impl} changed no certificate.
A second pre-registered run, with its own platform record (the same largest library errors, to two decimals; $\tanh$, which these
certificates do not evaluate, without recorded arguments), re-verified the certificates of the training remedy (\cref{app:remedy})
on 39 of its checkpoints: the linear-ramp, cap-128, cap-512, fixed-point-term, cap + hull and hull rows of Mamba and the minGRU-type
model and the linear-ramp row of the S4D-type model, at every context length, after one and four copies of the period, at
$L=\infty$ and for the hull (1000 documents), also at box scale 0 on the cap-128 rows and the minGRU-type cap + hull and hull rows,
and the token-set certificates of the six cap + hull checkpoints over one token (200 documents). Its float64 reference, run on CPUs,
reproduced the archived outcome of every document, computed on GPUs. None changed: all 208,020 archived certificates were kept and
none was gained, in all 309 combinations of checkpoint and setting, and the summary of \cref{app:remedy}, recomputed from the
re-verified outcomes, equals the published one in every number and verdict. Relative to $1+\abs{b}$, the outward-rounded margin
bound lay below the float64 one by at most $1.5\times10^{-11}$ at finite contexts, $2.5\times10^{-11}$ at $L=\infty$ and for the
hull, $2.1\times10^{-11}$ on the capped models, $3.1\times10^{-12}$ for the token sets, and $2.1\times10^{-11}$ (capped) and
$4.7\times10^{-12}$ (uncapped) at box scale 0.
On the capped models certificate checking passed at an inflation of at most $10^{-9}$ on 61,232 of the 83,364 pairs of document and
layer where it ran (73.5\%; every time constant there is at most 1184 tokens); the rest needed a larger inflation or did not pass,
and a document counts if either route certifies it. Of the 31,794 such pairs of the uncapped models at $L=\infty$, at both box
scales, none whose slowest channel has a time constant of at most $10^8$ tokens failed, and the inflation needed grew with the time
constant (rank correlation 0.74 over the 2752 pairs above $10^5$ tokens).
Not re-verified are the GRU and LSTM box radii of the distance probe, by design: an interval enclosure of a long unperturbed gated
trajectory cannot be tight, because the gated interval growth amplifies the rounding-error bounds of the about 1460 unperturbed
positions around the perturbed token, so every such certificate is lost under outward rounding; a rigorous version would need a
contraction-based enclosure of the point trajectory. Not re-verified either are the zonotope radii of the distance probe and the
budget sweep, the comparison with auto\_LiRPA (\cref{app:lirpa}), the complex-diagonal disks of \cref{sec:exp-r2}, the certified
shares at $L=0$ and $L=\infty$ of the slow-channel attribution (\cref{tab:attrib}), the byte-level table of \cref{app:byte}, the
earlier character-level study (\cref{app:charlm}), the certificates on the documents added by the extension and those at box scale
0 in its clean-accuracy tables (\cref{app:cleanctx}), the certificates of the zero-order-hold control of \cref{app:remedy} and every
certificate of the memory task of \cref{app:memory} (its main stage and its phase 2) and, of the training remedy, the first-recipe and cap-2048 rows, the
first-study rows after one and four copies and at box scale 0, the development-split shares, the box-scale-0 certificates and the width-only and flip shares that use them
outside the cap-128 rows and the minGRU-type cap + hull and hull rows, the token sets outside the cap + hull rows and the
diagnostics, named per row and column in \cref{app:remedy}.
In the re-verification of the earlier studies, under the pre-registered implementation one validity check failed, on a single
document without a certificate, so by the pre-registered rule nothing would have been read; under an amendment written after the
production run every validity check passed, and this remark reports that reading (\cref{app:prereg-amend}). The re-verification of
the training remedy was registered before it ran and was not amended, and every validity check passed; its ten tasks that reached their time
budget were resumed by resubmission under the same frozen identity, as registered (a summary run before that failed only the check
that every planned document is present; the paper uses none of its results), and its cap-128 part, submitted later under that
identity, completed in one submission. No statement here concerns a float32 network.
\end{remark}

\section{Soundness checks against sampled inputs, and other checks}\label{app:audits}

Every certificate in the paper is a real-arithmetic statement, and the reported ones were re-verified with outward rounding except
those listed in \cref{rem:float}; the checks below test the implementation and that re-verification, and none of them is used in a
proof. They differ by experiment. The certificates of the distance probe and of the budget sweep were checked
against sampled inputs. Those of the context-length probe and of the complex-diagonal control were checked at the box corner along
the gradient sign, by containment of each K-uniform box in the boxes of the members it covers, and by the point limit; the
growth-rate measurements by first-order soundness and the point limit. The token-set certificates were checked against concrete
contexts. In the comparison with auto\_LiRPA's linear relaxations (\cref{app:lirpa}) ours and auto\_LiRPA's bounds were checked
against sampled inputs and by the point limit, and auto\_LiRPA's IBP against an interval evaluation of the same traced graph. The
closed forms of the period part were checked against brute-force repetition, and the implementation against randomized and
integration tests. The outward-rounded evaluator was tested against exact and high-precision arithmetic and by mutation before its
production run, and checked in that run against the float64 code and against sampled inputs. Every soundness check of our
certificates passed in every production run. auto\_LiRPA's bounds passed the
sampled-input checks and, except those of its forward mode at $d>1$, the point limit; there its zero-width bounds missed the clean
margin at larger distances, and 23 of them lay above it (\cref{tab:lirpafwd}): an outcome of that verifier, not of our
certificates. Two document-count checks needed completion jobs, because runs had stopped at their time budget,
and no criterion was changed (\cref{app:prereg}); in the re-verification of the training remedy, ten tasks that had stopped at their
time budget were resumed, as registered (\cref{rem:float}); one check of the re-verification of the earlier studies, against
over-widening, passed only under an amendment written after its production run (\cref{app:prereg-amend}). \Cref{tab:checks}
summarises the checks.

\begin{table}[h]
\caption{Checks of the implementation and their outcomes (details in the paragraphs below).}
\label{tab:checks}
\centering\small
\begin{tabular}{@{}>{\raggedright\arraybackslash}p{0.22\textwidth}>{\raggedright\arraybackslash}p{0.43\textwidth}>{\raggedright\arraybackslash}p{0.29\textwidth}@{}}
\toprule
certificates & check & outcome\\
\midrule
distance probe, three seeds (box and zonotopes) & sampled inputs (random points, random corners, the gradient corner) at two radii per
method, distance and document; point limit; first-order ratio & no violation in 66,300--84,983 checks per model on seed~0 and
61,200--85,000 on seeds 1 and 2\\
budget sweep & the same checks; reproduction of the distance probe & no violation in 76,398 and 78,982 checks; 500 of 500 shared cells
reproduced per model\\
context-length probe & margin at the box corner, containment of the K-uniform boxes, point limit, finiteness, band ratios & passed on
every model of both runs and of the two completed runs\\
complex-diagonal control & the same per arm, and identity of every domain-independent quantity across the arms & passed on both
seeds\\
interval-map growth rates & first-order soundness at every resolved lag, point limit, $M_t\ge\abs{J_t}$ entrywise & passed on every model
and seed\\
slow-channel attribution & re-implementation of the period part against the certificate code, and its $K\to\infty$ limit & no mismatch;
within $10^{-9}$ on every checkpoint\\
token-set certificate & every residue certificate against concrete contexts and against the $L=0$ and $L=\infty$ margins & no violation
on any checkpoint\\
closed forms and hull & brute-force repetition of the period; containment of finite numbers of copies in the hull & agreement to
$10^{-6}$ or better; no containment violation\\
comparison with auto\_LiRPA (ours and its bounds; \cref{app:lirpa}) & rewritten models against the clean margin; auto\_LiRPA's IBP
against an interval evaluation of the same graph, ours never looser; sampled inputs; point limit; chord correction on every CROWN-type
bound & no violation and no error in any group; margins within $2.4\times10^{-14}(1+\abs{m^0})$; forward mode's zero-width
bounds: \cref{tab:lirpafwd}\\
training remedy (\cref{app:remedy}) & the cap on every capped checkpoint; per-document implications between the certificates of one
document (K-uniform implies $L=0$, one and four copies and $L=\infty$; certified implies classified correctly); the checks of the
token-set certificate and of the attribution; reproduction of the first study & no violation: the implications on 87 checkpoints,
the cap on every capped checkpoint, the token-set and attribution checks wherever they ran; the first study's Mamba and S4D-type
counts reproduced within two documents of 1000\\
floating-point re-verification (\cref{rem:float}) & unit tests of every primitive against exact rational and high-precision
values, under cancellation, underflow and overflow; mutation tests, each error term removed in turn; an end-to-end high-precision
oracle on tiny random models, also in every production job; the library-function errors in every job; the float64 reference against
the archived counts; an unwidened mirror against the float64 code; sampled inputs; for the training remedy also its own tests, the
float64 reference against the archived outcome of every document, the cap and the constants of the re-verified network, and the
summary of \cref{app:remedy} recomputed from the re-verified outcomes & earlier studies: every test passed and every mutation was
caught; every archived count reproduced; the mirror's reproduction check failed on none of 8489 documents (278 of them not
comparable, mostly with a float64 bound of $-\infty$); no sampled input outside an outward-rounded box (64 per document on the same
documents, none at the $L=\infty$ fixed point, which bounds a limit and no finite input); the check against over-widening passed
only under an amendment (\cref{app:prereg-amend}); training remedy: every test and every check passed, the archived outcome of
every document reproduced, the cap kept in every capped layer\\
\bottomrule
\end{tabular}
\end{table}

\paragraph{Sampled inputs: the distance probe and the budget sweep.}
In the distance probe every certificate, of the box and of the zonotopes, was checked against 17 sampled inputs per method, distance
and check point: random points, random corners and the corner along the sign of the margin gradient, inside the $\ell_\infty$ ball of
the perturbed embedding, at the fixed probe radius and at the certified radius. A violation is a sample whose logits leave the method's
logit box or whose margin lies below its margin bound (tolerance $10^{-9}(1+\abs{\cdot})$). There was no violation, on 200 documents
per seed (the $B=2048$ rows of the nonnegative diagonal models on seed~0: 50): in 81,617 (GRU) and 84,983 (LSTM) checks on seed~0 and in 81,617 and 81,685 (GRU) and 85,000 and 85,000 (LSTM) checks on
seeds 1 and 2, and in 66,300 checks per nonnegative diagonal model on seed~0 and 61,200 on each of seeds 1 and 2. The point evaluation
reproduced the clean margin to within $1.4\times10^{-14}$ for every method, distance and document, and no resolved first-order ratio
$R$ fell below 1. The budget sweep reused these
checks with 8 sampled inputs per check point, on 50 documents: no sample outside a box or below a bound in 76,398 (GRU) and 78,982
(LSTM) checks, the point limit within $10^{-13}$ of the clean margin, and no resolved first-order ratio below $1-10^{-3}$. It also
recomputed the distance probe's certified radius in all 500 cells that the two runs share for each model: every value lay inside the
sweep's final bisection bracket.

\paragraph{Box corners and containment: the context-length probe and the complex-diagonal control.}
The context-length probe applied four validity checks to every model before any other number was read: (i) at $\eps=10^{-10}$ no
document has $R<1-10^{-3}$ or $W<0$ on any finite context, which tests soundness at the box corner $e-\eps\,\sgn\nabla m$ to first
order (\cref{cor:observables}(iii)); (ii) no K-uniform certificate (the hull, and the invariant box of the gated cells) exceeds any
member it covers, at any $\eps$ including $0$ (tolerance $10^{-9}(1+\abs{b})$); (iii) the point pass reproduces the clean margin to
within $10^{-8}$; (iv) at least 150 documents completed. The amended version added: no non-finite bound at $10^{-10}$; no bound above the
margin at the box corner at any $\eps$, which is soundness without linearisation; no resolved distance band with a looseness below
$1-10^{-3}$; clean accuracy on the probe documents of at least 0.70 with the predicted class not above 80\% one class; and the
pre-registered grid. Both versions passed on every model, in the first run and in the replication on seeds 1 and 2. A static review
of the code before the production run found that point boxes have width exactly $0$ in every primitive of the nonnegative diagonal
models and in the closed-form fixed point, and estimated the rounding floor of $W$ at $\eps=10^{-10}$ at about $10^{-6}$ to $10^{-5}$
relative, at least two orders of magnitude below every threshold. The complex-diagonal control applied the amended checks to each arm
(the box arm exempt from finiteness, since a wrapping box may overflow float64) and checked that every quantity that does not depend on
the domain (clean margins, sensitivities $D$, margins at the box corner and the bound at $\eps=0$) agreed between the box arm and the
disk arm within $10^{-9}(1+\abs{\cdot})$ on every family and document; both checks passed on both seeds.

\paragraph{Interval-map growth rates.}
Before reading any exponent, the growth-rate measurement checked on every model that no resolved lag had an IBP-to-true ratio below
$1-10^{-3}$ at any $\eps\le10^{-8}$ (first-order soundness), that the point row reproduced the clean margin to within $10^{-8}(1+\abs{m})$
with a state radius of exactly $0$, and, for the gated cells, that the first-order radius map dominates the true Jacobian entrywise,
$M_t\ge\abs{J_t}$ up to a floating-point floor, at every entry, position, layer and document (\cref{prop:loss}(c)). These checks passed
on every model of seed~0 and of the replication on seeds 1 and 2.

\paragraph{Slow-channel attribution and the token-set certificate.}
The slow-channel attribution replaces the period part of the certificate by its own re-implementation, which also reports time
constants and DC levels per channel. Before any criterion was read it checked that the re-implementation returns the same boxes, bit for
bit (no mismatch), that its limit $K\to\infty$ agrees with the closed-form certificate within $10^{-9}$ relative with no mismatch in
finiteness, that its boxes are monotone and contain the clean infinite-context state, and that every checkpoint has at least 150
documents; this passed on all 38 checkpoints. The token-set certificate compared every residue certificate with the clean margins of
concrete contexts of its residue class (four random contexts over a subset of $S$, of lengths 3 and 1030), with the $L=\infty$ bound and the
clean $L=\infty$ margin of every single token of $S$, the residue $0$ with the $L=0$ margin, and the bound with the document's
substitution box against the bound for the clean document, at tolerance $10^{-9}(1+\abs{b})$; there was no violation on any of its 61
checkpoints.

\paragraph{Closed forms against brute-force repetition, and the hull.}
The closed-form fixed point was compared with brute-force repetition of the period: for moderate time constants the relative
difference after 64 copies was $10^{-6}$; the closed form for 16 and 64 copies matched brute force with a width ratio of $1.0000$; at
$K=10^9$ copies it matched the fixed point to $10^{-17}$; and for the S4D-type and minGRU-type layers the fixed point matched brute
force at $K=64$ to within $3\times10^{-12}$. For the K-uniform hull, the integration tests checked that its boxes contain those of every
number of copies $K=0,\dots,64$, and that the invariant boxes of the GRU and the LSTM contain the brute-force boxes of $K=1,\dots,64$
copies (and are correctly reported as vacuous for large weights). On a byte-level checkpoint after fixed-point fine-tuning (32 reviews,
$\eps=0.01$), the plain centre-radius boxes of $K=0,1,2,4$ and $\infty$ all lay inside the hull boxes, with a worst float64 slack of
$-6\times10^{-16}$ against a test tolerance of $-10^{-12}$ relative. The context-length probe repeated the containment check in
production at every $\eps$ (check (ii) above). No K-uniform certificate exceeded a member it covers in any test (\cref{rem:price}).

\paragraph{Randomized and integration tests of the implementation.}
A randomized test of the centre-radius step (\cref{lem:cr}(a) and (c)) on $2\times10^6$ random boxes, 35\% of them containing $0$, with corner
and interior points, found no violation and a monotonicity slack of at least $1.9\times10^{-5}$. Every model's interval pass on point boxes
reproduces its clean forward pass (to $10^{-16}$ for the centre-radius scan of Mamba; for the other models within the test tolerances
of $10^{-5}$ relative in float32 and $10^{-12}$ in float64), and random interior points and random corners of random input boxes stayed
inside the computed boxes (no violation). The layout boundary of \cref{lem:implcascade} was measured: with two layers, $P_0=5$ changes the state boxes by
$3.1\times10^{-4}$ and $P_0=6$ by $2.8\times10^{-17}$; with three layers, $P_0=8$ changes the layer-2 boxes by $9.6\times10^{-4}$ and the logit
box by $1.9\times10^{-4}$, and $P_0=9$ by $3\times10^{-16}$. For the first training study under synonym substitution, the synonym boxes were
compared with a word-by-word brute force over the substitutions (within $1.7\times10^{-6}$ in float32, from the rounding of the table
product; the evaluations use float64), and the document certificates of seven architectures, at random initialisation and after
training, were tested against 128 random substitution patterns and the attack's outputs (60 tests, no violation). When the
substitution attack was built, its outputs satisfied certified $\le$ attacked $\le$ clean document by document: every certified document
resisted the attack, and every document that resisted it was classified correctly. The audit of the training pipeline found one real
error, since corrected: the closed-form fixed point dropped the injection of
phases whose $1-\bar a$ underflows to exactly $0$, which underestimates $r^*$; such injections are now added with the full denominator
(\cref{rem:impl}), and the float64 re-evaluation of the earlier byte-level runs with the corrected code reproduced their numbers
(\cref{app:byte}). A test checks that a resumed training run gives the same weights, bit for bit on CPU, as an uninterrupted one.

\paragraph{What is not checked.}
The re-verification with outward rounding rests on the trusted base listed in \cref{rem:float}: the evaluator's code and the
platform's basic operations, tested and checked but not proved; what defines the certified network (its float64 constants, the
archived embedding tables, the library softplus and the cap constants); and one assumption, that the platform's $\exp$,
$\mathrm{expm1}$, $\mathrm{log1p}$, $\tanh$ and square root err by at most two ulps, measured and not proved. It covers
certificates only:
the measurements of the probes (looseness, margin losses, sensitivities, width exponents, band ratios and growth rates) carry an
estimated rounding floor, not a bounded one, and neither the zonotopes nor the other certificates that \cref{rem:float} lists as not
re-verified are covered. The sampled-input checks bound the implementation's errors only at the sampled points; the guarantees are
the proofs of \cref{app:proofs}, which concern the real-arithmetic network.

\section{The context-length probe, the complex-diagonal control and the interval-map growth rates in full}\label{app:gate3}

\subsection{The context-length probe}\label{app:gate3-ctx}

\paragraph{Setting.}
Each model reads the first 200 test documents (two runs, of Mamba seed~1 and of the standard-IBP Mamba seed~2 of
\cref{tab:stateless}, had stopped at their time budget after 190 and were completed by the extension of \cref{app:cleanctx}) after a
neutral context
of licence text; the class is $y_0$, the clean prediction on the bare document. The sequence families are: $L$ tokens of contiguous
neutral text before the document, $L\in\{0,64,256,1024,4096\}$ ($L=0$ is the bare document); $K\in\{1,2,4\}$ copies of a period, the
first 1024 tokens of the same neutral chunk, before the document, read from the zero state; the $L=\infty$ certificate of
\cref{thm:inf} on the layout of \cref{lem:implcascade} (for LSTM and GRU the bounded invariant-box search of \cref{rem:invbox}); and
the K-uniform hull of \cref{thm:hull}. Every embedding of every position gets an $\ell_\infty$ box of radius
$\eps\in\{10^{-10},10^{-8},10^{-6},10^{-4}\}$; the probe measures how a box propagates over the context and is not a threat model. Per
family, radius and document it records the IBP lower bound $b(\eps)$ of the margin of $y_0$, the margin loss $W=b(0)-b(\eps)$, the
first-order true sensitivity $D$ and the looseness $R=W/(\eps D)$ (\cref{sec:theory-loss}); the width exponent
$1+\frac12\log_{10}\big(W(10^{-8})/W(10^{-10})\big)$, which is 1 for a width linear in $\eps$ and 0 for a width independent of it; the
same ratio $R$ for perturbations restricted to the distance bands 0--64, 64--256, 256--1024 and 1024--4096 before the document, and to
the document itself; and soundness checks (\cref{app:audits}). Medians are over the documents, a vacuous document counting as
$+\infty$; everything runs in float64 without directed rounding (its certificates at $\eps=10^{-8}$ were re-verified with outward
rounding, except those on the ten documents that the completion added to each of the two runs, \cref{rem:float}). K-uniform bounds
(the hull, and the invariant box of LSTM and GRU)
are reported by their bound at $\eps=0$, their vacuity and their share of certified documents, never by $R$, because their bound at
$\eps=0$ is not the clean margin. $R$ measures our implementation, which bounds the two logits of the margin separately
(\cref{rem:head}). At $L=\infty$ the sensitivity $D$ is that of four periods, a proxy, marked as such where the four-period width
has not converged (\cref{cor:observables}).

\paragraph{Predictions.}
The probe was pre-registered with two predictions (record and dates: \cref{tab:prereg}). For the nonnegative diagonal models, $R$ grows
by at most a factor of 10 from $L=256$ to $L=4096$ and from one to four periods, the $L=\infty$ certificate is finite on at least 95\%
of the documents at $\eps=10^{-8}$, and a linearity statistic, the median relative difference between the $L=\infty$ width per unit
$\eps$ at $10^{-10}$ and at $10^{-8}$, is at most 0.01. For LSTM and GRU, $R$ grows by at least a factor of 10 from $L=256$ to $4096$, or
the bound becomes non-finite on at least half of the documents. The first prediction held for every nonnegative diagonal model except
one clause, the linearity statistic of the byte-level Mamba (0.0162). The second failed for every gated model (growth 0.44--1.03),
because their bounded interval states were already saturated inside the bare document. An amendment, written after a smoke run on
2--4 documents and before the production run, kept both predictions and their computation, replaced the linearity statistic by a
median width exponent within 0.05 of 1, and added for the gated models the alternative of a median width exponent of at most 0.5 (a
width almost independent of $\eps$). Both amended criteria held; the gated one describes the regime that the smoke run had shown and is
not a confirmed prediction. The replication on seeds 1 and 2 (eight new models) kept the amended criteria unchanged; there the original
prediction for the nonnegative diagonal models held in full, and the original growth criterion for the gated models failed again
(growth 0.60--0.88).

\begin{table}[htbp]
\caption{Context-length probe, nonnegative diagonal models after natural training (200 documents), medians at
$\eps=10^{-8}$. (a)~Clean accuracy on these documents and the looseness $R$ after $L$ tokens of neutral text, after $K$ copies of
the 1024-token period, and at $L=\infty$; its growth from $L=256$ to $4096$ and from $K=1$ to $4$. (b)~Median width exponent after
4096 tokens; percentage of documents certified after 0 and 4096 tokens, at $L=\infty$ (\cref{thm:inf}) and by the K-uniform hull
(\cref{thm:hull}), which also covers every finite number of copies, with its 95\% Wilson interval per checkpoint
(\cref{app:cleanctx-method}); ratio of $R$ for perturbations 1024--4096 positions before the
document to $R$ for perturbations within 64 positions of it (band far/near). $^\dagger$Proxy: the four-period width has not
converged.}
\label{tab:gate3}
\centering\footnotesize
\setlength{\tabcolsep}{4pt}
\textit{(a)}\par\smallskip
\begin{tabular}{@{}lcccccccccc@{}}
\toprule
 & & & \multicolumn{6}{c}{looseness $R$} & \multicolumn{2}{c}{growth of $R$}\\
\cmidrule(lr){4-9}\cmidrule(l){10-11}
model & seed & clean & $L=0$ & $L=256$ & $L=4096$ & $K=1$ & $K=4$ & $L=\infty$ & $L$: 256$\to$4096 & $K$: 1$\to$4\\
\midrule
Mamba & 0 & 0.87 & $1.17{\times}10^{5}$ & $8.6{\times}10^{4}$ & $8.35{\times}10^{4}$ & $9.66{\times}10^{4}$ & $9.86{\times}10^{4}$ & $9.91{\times}10^{4}$ & 0.971 & 1.02\\
 & 1 & 0.85 & $6.36{\times}10^{4}$ & $4.26{\times}10^{4}$ & $2.95{\times}10^{4}$ & $3.22{\times}10^{4}$ & $3.06{\times}10^{4}$ & $3.07{\times}10^{4}$ & 0.693 & 0.953\\
 & 2 & 0.885 & $1.78{\times}10^{5}$ & $1.39{\times}10^{5}$ & $1.3{\times}10^{5}$ & $1.39{\times}10^{5}$ & $1.4{\times}10^{5}$ & $1.4{\times}10^{5}$ & 0.937 & 1.01\\
S4D-type & 0 & 0.86 & 847 & 858 & 734 & 774 & 790 & 790 & 0.856 & 1.02\\
 & 1 & 0.875 & 722 & 630 & 520 & 508 & 509 & 509 & 0.826 & 1.00\\
 & 2 & 0.86 & 864 & 858 & 906 & 756 & 793 & 794 & 1.06 & 1.05\\
minGRU-type & 0 & 0.86 & $2.86{\times}10^{4}$ & $2.31{\times}10^{4}$ & $2.25{\times}10^{4}$ & $2.2{\times}10^{4}$ & $2.31{\times}10^{4}$ & $2.32{\times}10^{4}$ & 0.978 & 1.05\\
 & 1 & 0.885 & $2.99{\times}10^{4}$ & $2.59{\times}10^{4}$ & $2.45{\times}10^{4}$ & $2.53{\times}10^{4}$ & $2.69{\times}10^{4}$ & $2.72{\times}10^{4}$ & 0.944 & 1.06\\
 & 2 & 0.895 & $2.75{\times}10^{4}$ & $2.23{\times}10^{4}$ & $2.07{\times}10^{4}$ & $2.24{\times}10^{4}$ & $2.24{\times}10^{4}$ & $2.24{\times}10^{4}$ & 0.927 & 0.999\\
byte Mamba & 0 & 0.81 & $8.84{\times}10^{4}$ & $8.24{\times}10^{4}$ & $6.62{\times}10^{4}$ & $7.28{\times}10^{4}$ & $6.88{\times}10^{4}$ & $6.95{\times}10^{4}$\,$^\dagger$ & 0.804 & 0.946\\
\bottomrule
\end{tabular}

\bigskip
\textit{(b)}\par\smallskip
\setlength{\tabcolsep}{3pt}
\begin{tabular}{@{}lccccccc@{}}
\toprule
 & & & \multicolumn{4}{c}{documents certified (\%)} & \\
\cmidrule(lr){4-7}
model & seed & width exp.\ ($L=4096$) & $L=0$ & $L=4096$ & $L=\infty$ & hull (every $K$) & band far/near\\
\midrule
Mamba & 0 & 0.999 & 99.5 [97.2, 99.9] & 93.5 [89.2, 96.2] & 93.5 [89.2, 96.2] & 0 [0, 1.9] & 0.802\\
 & 1 & 0.999 & 100 [98.1, 100] & 86.5 [81.1, 90.6] & 87.0 [81.6, 91.0] & 0 [0, 1.9] & 0.284\\
 & 2 & 0.998 & 99.5 [97.2, 99.9] & 93.5 [89.2, 96.2] & 91.0 [86.2, 94.2] & 0 [0, 1.9] & 0.764\\
S4D-type & 0 & 1 & 100 [98.1, 100] & 91.0 [86.2, 94.2] & 87.5 [82.2, 91.4] & 0 [0, 1.9] & 0.941\\
 & 1 & 1 & 100 [98.1, 100] & 92.5 [88.0, 95.4] & 91.5 [86.8, 94.6] & 0.5 [0.1, 2.8] & 0.643\\
 & 2 & 1 & 100 [98.1, 100] & 91.0 [86.2, 94.2] & 89.0 [83.9, 92.6] & 0.5 [0.1, 2.8] & 1.01\\
minGRU-type & 0 & 1 & 100 [98.1, 100] & 89.5 [84.5, 93.0] & 87.5 [82.2, 91.4] & 0 [0, 1.9] & 0.886\\
 & 1 & 1 & 100 [98.1, 100] & 90.0 [85.1, 93.4] & 88.5 [83.3, 92.2] & 0 [0, 1.9] & 0.978\\
 & 2 & 1 & 99.5 [97.2, 99.9] & 90.5 [85.6, 93.8] & 91.5 [86.8, 94.6] & 0 [0, 1.9] & 0.617\\
byte Mamba & 0 & 0.997 & 99.5 [97.2, 99.9] & 86.0 [80.5, 90.1] & 84.0 [78.3, 88.4] & 0 [0, 1.9] & 0.77\\
\bottomrule
\end{tabular}
\end{table}

\begin{table}[htbp]
\caption{Context-length probe, gated models after natural training (200 documents). Their boxes are saturated, so $R$ is not a
tightness ratio; the table gives instead the margin loss per unit radius $W/\eps$ at $\eps=10^{-8}$ on the bare document and after
4096 tokens, the median width exponent there, the growth of $R$ from $L=256$ to 4096, the share of documents certified at any
context length, and the invariant box of \cref{rem:invbox} at $\eps=0$.}
\label{tab:gated}
\centering\footnotesize
\setlength{\tabcolsep}{4pt}
\begin{tabular}{@{}lccccccl@{}}
\toprule
 & & & \multicolumn{2}{c}{$W/\eps$} & width exponent & growth & invariant box\\
\cmidrule(lr){4-5}
model & seed & clean & $L=0$ & $L=4096$ & $L=0$ / 4096 & of $R$ & at $\eps=0$\\
\midrule
LSTM & 0 & 0.88 & $1.86{\times}10^{9}$ & $1.82{\times}10^{9}$ & $6.84{\times}10^{-14}$ / $6.96{\times}10^{-14}$ & 0.903 & vacuous on every document\\
 & 1 & 0.895 & $1.81{\times}10^{9}$ & $1.79{\times}10^{9}$ & $7.49{\times}10^{-14}$ / $7.52{\times}10^{-14}$ & 0.881 & vacuous on every document\\
 & 2 & 0.88 & $1.81{\times}10^{9}$ & $1.81{\times}10^{9}$ & $5.12{\times}10^{-14}$ / $5.13{\times}10^{-14}$ & 0.714 & vacuous on every document\\
GRU & 0 & 0.88 & $2.02{\times}10^{9}$ & $1.98{\times}10^{9}$ & $2.2{\times}10^{-16}$ / 0 & 0.441 & finite, median $b(0)=-14.9$\\
 & 1 & 0.895 & $2.2{\times}10^{9}$ & $2.14{\times}10^{9}$ & 0 / 0 & 0.604 & finite, median $b(0)=-15.9$\\
 & 2 & 0.88 & $2.02{\times}10^{9}$ & $1.95{\times}10^{9}$ & $2.2{\times}10^{-16}$ / 0 & 0.612 & finite, median $b(0)=-14.5$\\
byte LSTM & 0 & 0.81 & $1.72{\times}10^{9}$ & $1.73{\times}10^{9}$ & $3.3{\times}10^{-16}$ / $3.3{\times}10^{-16}$ & 1.03 & vacuous on every document\\
\bottomrule
\end{tabular}
\par\smallskip
No document was certified at any context length by any gated model.
\end{table}

\begin{table}[htbp]
\caption{Context-length probe, stateless references and certified-trained models (seed as listed; descriptive; medians at
$\eps=10^{-8}$ over 200 documents): clean accuracy, looseness $R$ on the bare document, after 4096
tokens and at $L=\infty$, and the share of documents certified at $L=\infty$ and by the K-uniform hull. The certified-trained models
were trained under synonym substitution (\cref{app:gate2}) and are probed here with the $\ell_\infty$ boxes of the probe.
$^\dagger$Proxy, not converged.}
\label{tab:stateless}
\centering\footnotesize
\begin{tabular}{@{}lcccccc@{}}
\toprule
 & & \multicolumn{3}{c}{looseness $R$} & \multicolumn{2}{c}{share certified}\\
\cmidrule(lr){3-5}\cmidrule(l){6-7}
model & clean & $L=0$ & $L=4096$ & $L=\infty$ & $L=\infty$ & hull (every $K$)\\
\midrule
CNN, natural & 0.90 & 1.56 & 1.08 & 1.17 & 0.86 & 0.855\\
bag of words, natural & 0.88 & 1.49 & 1.48 & n/a & n/a & n/a\\
Mamba, standard IBP, seed 2 & 0.765 & 1.68 & 1.80 & 22\,$^\dagger$ & 0.45 & 0.285\\
Mamba, fixed-point fine-tuning, seed 2 & 0.785 & 1.76 & 1.65 & 0.575\,$^\dagger$ & 0.80 & 0.685\\
S4D-type, standard IBP, seed 0 & 0.785 & 1.71 & 1.69 & 1.70 & 0.96 & 0.855\\
minGRU-type, fixed-point fine-tuning, seed 0 & 0.80 & 1.88 & 2.01 & 1.77\,$^\dagger$ & 0.86 & 0.18\\
\bottomrule
\end{tabular}
\end{table}

\paragraph{Reading.}
\emph{No geometric accumulation over the context, on three seeds.} For all ten nonnegative diagonal models the looseness changed by
factors of 0.69--1.06 from $L=256$ to $L=4096$ and of 0.95--1.06 from one to four periods, all within $[0.6,1.1]$; the width was
linear in $\eps$ (median width exponent 0.997--1.000), and the $L=\infty$ certificate was finite on every document and certified
84--93.5\% of them at $\eps=10^{-8}$. The looseness is large but set by the network, not by the context: $R$ is 508--906 for the
S4D-type seeds, $2.07\times10^{4}$--$2.99\times10^{4}$ for the minGRU-type seeds and $2.95\times10^{4}$--$1.78\times10^{5}$ for the
Mamba seeds, at every length. The band ratios (far/near) were 0.28--1.01: the looseness for perturbations 1024--4096 positions before
the document was at most about that for perturbations within 64 positions of it. \Cref{prop:exact}(a$'$) allows growth up to a
constant that can be astronomically large, and no result of \cref{sec:theory} predicts these flat profiles (\cref{rem:gate3reading});
what the measurements show is no geometric accumulation at these lengths, not the absence of any growth.

\emph{The K-uniform hull certifies almost nothing on the natural models at the probe radius.} At $\eps=10^{-8}$ the hull of
\cref{thm:hull}, which also covers every finite number of copies of the period, certified no document of the natural Mamba,
minGRU-type and byte-level Mamba models and 0--0.5\% of the S4D-type documents, against 84--93.5\% for the $L=\infty$ certificate,
which covers the infinite context only. For Mamba and the minGRU-type model the hull was vacuous already at $\eps=0$ on 87.5--99.5\% of
the documents: it contains the clean trajectories from the zero state and from the fixed point, and slow channels make that segment
of start states long (\cref{rem:price}). For the S4D-type model it was finite but far below the clean margin (median bound at
$\eps=0$ between $-3.39\times10^{4}$ and $-1.61\times10^{4}$ over the seeds). The certified-trained models of \cref{tab:stateless}
keep part of it: 18--85.5\% under the hull against 45--96\% at $L=\infty$.

\emph{Gated models saturate inside the document, on three seeds.} Every LSTM and GRU was saturated already on the bare 512-token
document at $\eps=10^{-10}$ (median width exponent at most $7.52\times10^{-14}$), no document was certified at any context length, and
the LSTMs' invariant boxes were vacuous. This is the finite-horizon regime of \cref{prop:exact}(b$''$), whose mechanism
\cref{app:gate3-r4} measures. It is not evidence for the infinite-horizon statement \cref{prop:exact}(b).

\emph{Certified training changes the constant, not the shape.} The certified-trained models had $R$ between 1.65 and 2.01 at every
finite context length (natural models: 508 to $1.78\times10^{5}$), with equally flat profiles. For standard-IBP Mamba seed~2 the
$L=\infty$ width was about fifteen times the four-period width, and the model certified 45\% of the documents at $L=\infty$ against
54\% after 4096 tokens: the slow-channel effect of \cref{prop:horizon} (\cref{app:gate2}).

\begin{figure}[htbp]
\centering
\includegraphics[width=\textwidth]{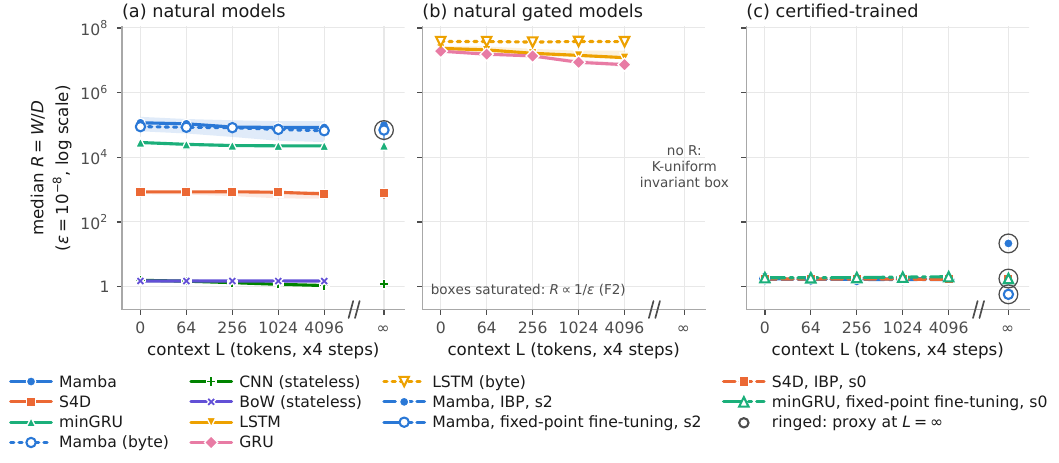}
\caption{Context-length probe: IBP looseness $R=W/(\eps D)$ against the context length; median over the documents at
$\eps=10^{-8}$, for the bare 512-token document ($L=0$), after $L$ tokens of contiguous neutral text, and at $L=\infty$ (the
certificate of \cref{thm:inf}; its $D$ is that of four periods, a proxy where the four-period width has not converged: ringed).
Lines: median over seeds (three for the word-level natural models); bands: seed minimum to maximum. (a)~Nonnegative diagonal models
and the stateless references: $R$ is large but shows no geometric accumulation with $L$ (for each nonnegative diagonal model,
$R(4096)/R(256)$ lies within $[0.6,1.1]$). (b)~Gated models: their boxes are already saturated inside the document, so $R$ scales
like $1/\eps$ and is not a tightness ratio (the plot's ``(F2)'' refers to \cref{fig:eps}). (c)~Certified-trained models.}
\label{fig:gate3}
\end{figure}

\begin{figure}[htbp]
\centering
\includegraphics[width=\textwidth]{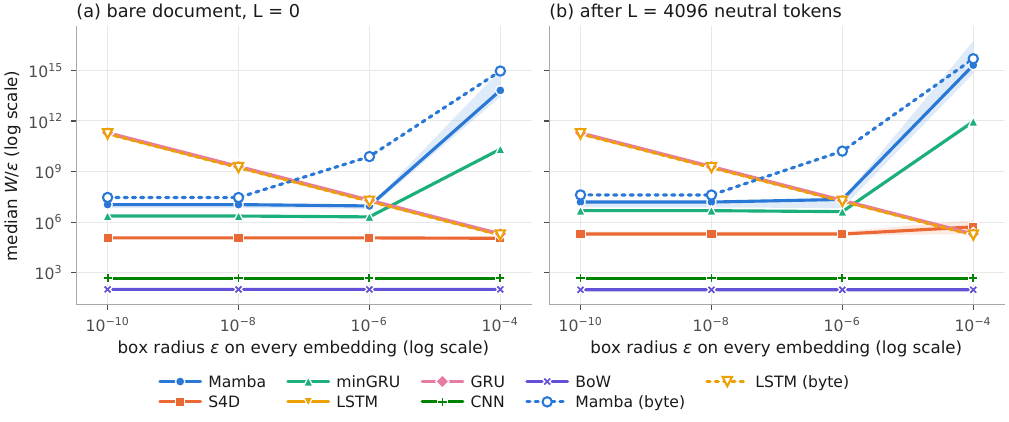}
\caption{Context-length probe, the $\eps$ signature of the IBP width: median over the documents of the margin loss per unit radius,
$W/\eps=(b(0)-b(\eps))/\eps$, at the four probe radii, on the bare document (a) and after 4096 neutral tokens (b); a vacuous
document counts as $+\infty$. Lines: median over seeds; bands: seed minimum to maximum. A flat line means a width proportional to
$\eps$, slope $-1$ a width independent of $\eps$. For the word-level nonnegative diagonal models and the stateless references
$W/\eps$ hardly depends on $\eps$ up to $10^{-6}$: the width is linear in $\eps$ (width exponent close to 1); the byte-level Mamba
leaves the linear regime already at $10^{-6}$. For LSTM and GRU $W/\eps$ falls like
$1/\eps$: the width does not depend on $\eps$, and the box is saturated already at $\eps=10^{-10}$ on the bare document (width
exponent close to 0). At $\eps=10^{-4}$ the Mamba and minGRU-type widths leave the linear regime (median $W/\eps$ of
$6.53\times10^{13}$ and $2.13\times10^{10}$ on the bare document, against $1.09\times10^{7}$ and $2.3\times10^{6}$ at $10^{-8}$); the
S4D-type width does not ($1.1\times10^{5}$ against $1.17\times10^{5}$).}
\label{fig:eps}
\end{figure}

\begin{figure}[htbp]
\centering
\includegraphics[width=\textwidth]{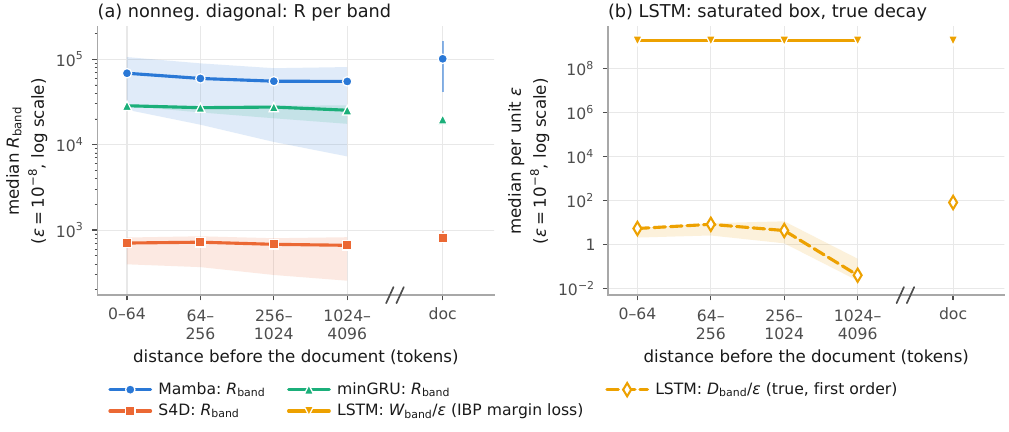}
\caption{Context-length probe, distance bands ($\eps=10^{-8}$, on the sequences with 4096 neutral tokens): only the positions in one
band of distances before the document, or only the document, are perturbed. (a)~Median over the documents of
$R_{\rm band}=W_{\rm band}/D_{\rm band}$ for the natural nonnegative diagonal models (a vacuous document counts as $+\infty$; documents
whose change of the bound in a band is below the rounding floor would be dropped and counted, and there were none); line: median over
the three seeds, band: seed minimum to maximum. Where $R_{\rm band}$ does not grow with the distance, the median IBP sensitivity to a
distant band decays at least as fast as the median true one. The far/near ratios of the medians over seeds are 0.802 (Mamba), 0.941
(S4D-type) and 0.886 (minGRU-type); per seed they span 0.28--1.01, so this is a statement about these medians, not about every seed or
document. (b)~Natural LSTM (three seeds): the median margin loss per unit $\eps$ of each band is the same saturated value
($1.81\times10^{9}$) in every band, while the true first-order sensitivity falls from 5.39 (distance 0--64) to 0.0406 (1024--4096), so
their ratio is not a tightness ratio for this model. The document is set apart on the right.}
\label{fig:bands}
\end{figure}

\subsection{The complex-diagonal control}\label{app:gate3-r2}

\paragraph{Setting.}
Two complex-diagonal LTI S4D models in the block of our other models (\cref{tab:layers}; 8 complex modes per inner channel, S4D-Lin
initialisation, zero-order hold) were trained naturally (seeds 0 and 1) and probed twice with the context-length probe, once with the
state propagated in real boxes and once in complex disks (\cref{lem:disks}), with the same weights, documents and forward pass. A
cross-arm check confirmed that every quantity that does not depend on the domain (clean margins, $D$, the margins at the box corner,
$b(0)$) agreed between the arms on every document. The pre-registered prediction was that the box arm fails and the disk arm meets
the amended regime-1 criterion of the context-length probe: growth of $R$ by at most a factor of 10, an $L=\infty$ certificate finite on
at least 95\% of the documents, and a median width exponent within 0.05 of 1. In every layer of both models 86--87\% of the modes have
a box growth factor $\gbox=\abs{\lambda}(\abs{\cos\theta}+\abs{\sin\theta})\ge1$ (largest 1.38--1.39), while every disk factor
$\abs{\lambda}$ is at most 0.999 (slowest disk time constant 612--804 tokens).

\begin{table}[htbp]
\caption{Complex-diagonal control (200 documents per arm; medians at $\eps=10^{-8}$): the same naturally trained complex-diagonal S4D
in real boxes and in complex disks. Clean accuracy; looseness $R$ after 0, 256 and 4096 neutral tokens, after four periods and at
$L=\infty$; its growth from 256 to 4096 tokens and from one to four periods; median width exponent on the bare document and after 4096
tokens; share of documents with a finite $L=\infty$ bound; share certified after 0 and 4096 tokens and at $L=\infty$. Overflow:
non-finite in float64 (the box grows, \cref{prop:exact}(b$'$)); vacuous: no finite invariant box. Per checkpoint and arm the 95\%
Wilson interval of a share has a half-width of at most 0.069 at $n=200$ (\cref{app:cleanctx-method}; those of the shares the text
cites: \cref{tab:ci}).}
\label{tab:r2}
\centering\scriptsize
\setlength{\tabcolsep}{3pt}
\begin{tabular}{@{}lccccccccccc@{}}
\toprule
 & & & \multicolumn{5}{c}{looseness $R$} & growth & width exp. & $L=\infty$ & certified\\
\cmidrule(lr){4-8}
seed & arm & clean & $L=0$ & $L=256$ & $L=4096$ & $K=4$ & $L=\infty$ & $L$ / $K$ & $L=0$ / 4096 & finite & $L=0$ / 4096 / $\infty$\\
\midrule
0 & box & 0.88 & $2.14{\times}10^{65}$ & $1.05{\times}10^{137}$ & overflow & overflow & vacuous & inf / nan & 2.01 / 0 & 0 & 0 / 0 / 0\\
0 & disk & 0.88 & $5.96{\times}10^{3}$ & $6.19{\times}10^{3}$ & $5.62{\times}10^{3}$ & $5.2{\times}10^{3}$ & $5.21{\times}10^{3}$ & 0.908 / 0.924 & 1 / 1 & 1 & 1 / 0.845 / 0.89\\
1 & box & 0.845 & $7.49{\times}10^{65}$ & $4.44{\times}10^{137}$ & overflow & overflow & vacuous & inf / nan & 2.01 / 0 & 0 & 0 / 0 / 0\\
1 & disk & 0.845 & $5.48{\times}10^{3}$ & $5.38{\times}10^{3}$ & $5.91{\times}10^{3}$ & $5.51{\times}10^{3}$ & $5.51{\times}10^{3}$ & 1.1 / 0.988 & 1 / 1 & 1 & 1 / 0.935 / 0.935\\
\bottomrule
\end{tabular}
\end{table}

\paragraph{Reading.}
The prediction held on both seeds (\cref{fig:regimes}b of the main text draws the same data). In boxes the $L=\infty$ certificate was
vacuous on every document, as \cref{prop:exact}(b$'$) requires for any mode with $\gbox\ge1$; already on the bare document the bound
was $2\times10^{65}$ to $7\times10^{65}$ times the true sensitivity and grew like $\eps^2$ (width exponent about 2: the second layer's
gate multiplies two wide intervals); at $L=1024$ the bound was non-finite in float64 on 98--99\% of the documents and at $L=4096$ on
all; no document was certified at any length. In disks the same weights behaved like the real-diagonal models of \cref{tab:gate3}:
$R$ flat between $5.2\times10^{3}$ and $6.2\times10^{3}$ from the bare document to $L=\infty$, a width exponent of 1, and the $L=\infty$
certificate finite everywhere (certified in float64 on 89\% and 93.5\% of the documents; not re-verified with outward rounding,
\cref{rem:float}). Limits: one LTI architecture and two seeds. Other domains
would not wrap either: in per-mode co-rotating coordinates the transition is the positive scalar $\abs{\lambda}$ and boxes there are
exact in time, and an unreduced zonotope (one generator per input per step) is exact as well. A budgeted zonotope in the original
coordinates is never wider than the boxes (\cref{lem:zono}), but its dropped generators are boxed and boxes wrap, so nothing here shows
that it avoids the wrapping; it was not run on this model. The control therefore shows a mismatch between transition and domain, not
that no box-like domain works. The disk arm is looser than the real-diagonal S4D-type models (508--906), which \cref{lem:disks}
allows: the injection segments are replaced by disks, and the factor $\pi/2$ bounds only the circumradius, not the looseness in the
readout direction (\cref{rem:pihalf}); these $R$ measure the whole model and do not isolate the directional looseness of disks.

\begin{figure}[htbp]
\centering
\includegraphics[width=\textwidth]{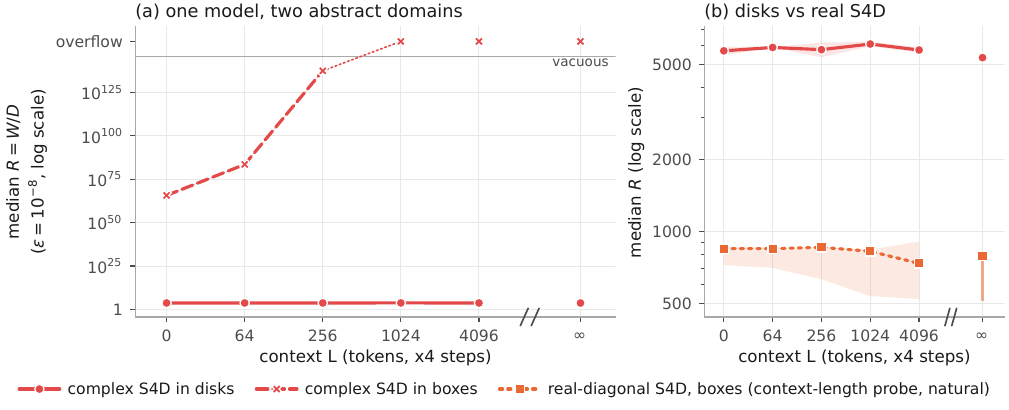}
\caption{Complex-diagonal control: the same weights of a naturally trained complex-diagonal LTI S4D (two seeds), the same documents,
prefixes and forward pass; only the abstract domain of the state differs: real boxes (each $2\times2$ rotation-scaling block widens
a box by $\abs{\lambda}(\abs{\cos\theta}+\abs{\sin\theta})$, which is at least 1 for most modes) or complex disks (a constant-factor
loss, at most $\pi/2$ in circumradius, that does not compound in time). Median over the documents of $R=W/(\eps D)$ at
$\eps=10^{-8}$ against the context length; line: mean of the two seeds' medians, band: their range. (a)~In boxes $R$ is
astronomically large already on the bare document and overflows float64 at longer contexts (non-finite medians drawn in the
overflow strip, never dropped; at $L=\infty$ the box certificate is vacuous: no finite invariant box exists); in disks $R$ is flat
from $L=0$ to $L=\infty$. (b)~The disk arm on a narrower axis, with the real-diagonal S4D-type models of the context-length probe
(three seeds) for reference.}
\label{fig:twoarm}
\end{figure}

\subsection{The interval-map growth rates}\label{app:gate3-r4}

\paragraph{Setting.}
For each model one token 1024 positions before the document gets an $\ell_\infty$ box of radius $\eps=10^{-10}$ (nonnegative diagonal
models: $10^{-6}$), and the probe follows its state box against the true first-order width of the same state, with finite differences
in float64. Along the clean trajectories of the bare documents it measures top Lyapunov exponents per token of the true state
Jacobians ($\lambda_J$) and of the first-order radius map $M_t\ge\abs{J_t}$ of the interval cell, read off the implementation
($\lambda_M$; \cref{prop:loss}(c)). For one-layer cells, \cref{prop:exact}(b$''$) and \cref{prop:loss}(c) predict
$\lambda_M>0>\lambda_J$ and saturation after about $\ln(1/\eps)/\lambda_M$ steps; our models have two layers, so this is a
measurement, not a check of a theorem. Every model has 200 documents per seed: the pre-registered runs had 100 on seed~0 and 100 on
each of seeds 1 and 2 (the replication on the five word-level natural models), and the extension of \cref{app:cleanctx} added 100
more to each. The replication kept the criteria unchanged, with the rule that every criterion met on seed~0 must be met on both
seeds; the byte-level models have seed~0 only. Every pre-registered criterion was met on every seed (\cref{tab:prereg}), on the 200
documents and on the 100 added ones alone.

\begin{figure}[htbp]
\centering
\includegraphics[width=\textwidth]{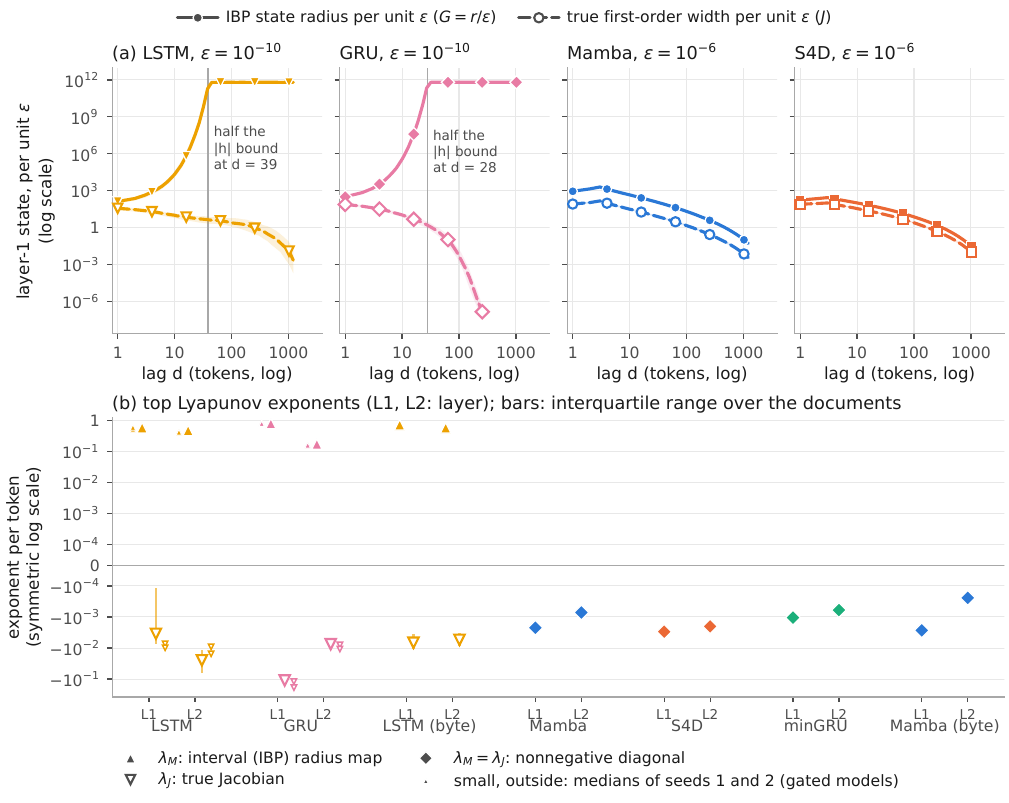}
\caption{Interval-map growth rates. (a)~One token 1024 positions before the document gets an $\ell_\infty$ box; per unit $\eps$, the
IBP radius of its layer-1 state box summed over the units (solid) and the true first-order width of the same state (dashed) against
the lag; seed~0, medians over the documents, bands: interquartile range. Gated models at $\eps=10^{-10}$ (vertical line: the median lag
at which the mean unit radius of $h$ reaches half its a-priori bound $\abs{h}\le1$), nonnegative diagonal models at $\eps=10^{-6}$. The
gated box grows geometrically while the true width decays, until it saturates; for Mamba and the S4D-type model both decay together.
(b)~Top Lyapunov exponents per token along the clean trajectories of the bare documents, of the interval radius map $M_t\ge\abs{J_t}$
($\lambda_M$) and of the true Jacobian ($\lambda_J$): medians over the documents (200 per model and seed), bars: their interquartile
range on seed~0, small markers: the medians of seeds 1 and 2 of the gated models. The median gated cell contracts ($\lambda_J<0$) while its interval
map expands ($\lambda_M>0$); the long $\lambda_J$ bar of the LSTM's first layer ($-0.0035$, interquartile range
$[-0.0073,\,-1.1\times10^{-4}]$) ends just below zero, and 24\% of those documents have $\lambda_J>0$. For the nonnegative diagonal
models the maps coincide and the top exponent, the slowest channel's, is negative.}
\label{fig:r4}
\end{figure}

\Cref{tab:r4} gives the per-seed values behind \cref{sec:exp-r4}, and \cref{fig:r4} the profiles. The criterion $\lambda_J<0$ holds for
the median and the pooled value in every layer of every seed, and the 95\% interval of every median $\lambda_J$ lies below 0 (that
of every median $\lambda_M$ above 0), but not on every document: on seed~0, 24.1\% [18.7, 30.5] of the documents of the LSTM's
first layer, about a quarter, have a slightly positive top exponent of the true Jacobian (interquartile range of $\lambda_J$
$[-0.0073,\,-1.1\times10^{-4}]$), and 10.1\% and 5.0\% on seeds 1 and 2; in the LSTM's second layer 0\%, 4.0\% and 10.1\%, in
the GRU's layers at most 0.6\%.

\begin{table}[htbp]
\caption{Interval-map growth rates, gated models (200 documents per seed; medians per layer, with the 95\% order-statistic interval
of the median for $\lambda_M$ and $\lambda_J$, \cref{app:cleanctx-method}): top
Lyapunov exponents per token of the interval radius map ($\lambda_M$) and of the true Jacobian ($\lambda_J$); lag at which one token's
$h$ box reaches half its a-priori bound $\abs{h}\le1$; growth of the IBP-to-true ratio before saturation; measured saturation rate over
the largest $\lambda_M$ of the layers up to the one measured.}
\label{tab:r4}
\centering\footnotesize
\setlength{\tabcolsep}{4pt}
\begin{tabular}{@{}lccrrrrr@{}}
\toprule
model & seed & layer & $\lambda_M$ & $\lambda_J$ & lag & growth & rate\,/\,$\max\lambda_M$\\
\midrule
LSTM & 0 & 1 & 0.575 [0.570, 0.578] & $-0.0035$ [$-0.00464$, $-0.0026$] & 39 & $4.0{\times}10^{10}$ & 1.001\\
 & 0 & 2 & 0.469 [0.457, 0.480] & $-0.025$ [$-0.0423$, $-0.0164$] & 33 & $3.0{\times}10^{9}$ & 1.03\\
 & 1 & 1 & 0.512 [0.507, 0.517] & $-0.0073$ [$-0.0081$, $-0.00615$] & 44 & $6.4{\times}10^{10}$ & 0.994\\
 & 1 & 2 & 0.426 [0.413, 0.450] & $-0.0153$ [$-0.0178$, $-0.0136$] & 36 & $2.5{\times}10^{9}$ & 0.999\\
 & 2 & 1 & 0.583 [0.579, 0.588] & $-0.0099$ [$-0.0108$, $-0.00887$] & 39 & $4.35{\times}10^{10}$ & 0.9974\\
 & 2 & 2 & 0.409 [0.389, 0.430] & $-0.0091$ [$-0.0114$, $-0.00801$] & 34 & $3.50{\times}10^{9}$ & 1.0015\\
GRU & 0 & 1 & 0.780 [0.778, 0.782] & $-0.111$ [$-0.115$, $-0.108$] & 28 & $1.08{\times}10^{11}$ & 1.003\\
 & 0 & 2 & 0.171 [0.162, 0.180] & $-0.0075$ [$-0.00852$, $-0.00673$] & 26 & $1.28{\times}10^{9}$ & 0.998\\
 & 1 & 1 & 0.837 [0.834, 0.839] & $-0.189$ [$-0.192$, $-0.183$] & 26 & $5.02{\times}10^{11}$ & 0.998\\
 & 1 & 2 & 0.160 [0.152, 0.163] & $-0.0079$ [$-0.00901$, $-0.00729$] & 25 & $1.67{\times}10^{9}$ & 0.993\\
 & 2 & 1 & 0.776 [0.772, 0.780] & $-0.120$ [$-0.125$, $-0.117$] & 29 & $2.39{\times}10^{11}$ & 1.004\\
 & 2 & 2 & 0.156 [0.146, 0.168] & $-0.0108$ [$-0.0119$, $-0.00967$] & 27 & $1.54{\times}10^{9}$ & 1.008\\
byte LSTM & 0 & 1 & 0.712 [0.710, 0.714] & $-0.0067$ [$-0.00857$, $-0.0056$] & 31 & $4.66{\times}10^{10}$ & 1.0013\\
 & 0 & 2 & 0.568 [0.564, 0.573] & $-0.0055$ [$-0.00634$, $-0.00488$] & 27 & $2.11{\times}10^{9}$ & 1.012\\
\bottomrule
\end{tabular}
\end{table}

\paragraph{Reading.}
On every seed the median document's true dynamics contract along clean text in both layers of the three gated models, while their
interval maps expand at 0.16--0.84 nats per token (200 documents per seed). One token's box at $\eps=10^{-10}$ grows geometrically
against the true width by a
factor of $1.3\times10^{9}$ to $5.0\times10^{11}$ and reaches half the a-priori bound after 25--44 tokens (seed~0: 26--39). The measured
saturation rate matched the largest $\lambda_M$ of the layers up to the one measured (ratio 0.99--1.03 over the three seeds), that
is, saturation after about $\ln(1/\eps)/\max\lambda_M$ steps. The maximum matters: the GRU's second layer has $\lambda_M$ of
0.156--0.171 on its own, for which $\ln(10^{10})/\lambda_M$ would be about 135 tokens at the seed-0 rate, yet it saturated after 25--27
tokens, driven by its first layer ($\lambda_M$ of 0.776--0.837) through the layer input. This is the mechanism behind the saturation
inside the document of \cref{app:gate3-ctx} and behind the collapse of the box's certified radius within 16--128 steps
(\cref{app:gate5}); for two-layer models it is measured, not proved, since \cref{prop:exact}(b$''$) and \cref{prop:loss}(c) are
one-layer statements.

For the nonnegative diagonal models ($M_t=\abs{J_t}=J_t$) the slowest channel's mean log decay was negative on every document and
segment on every seed (200 documents per seed; largest value $-1.5\times10^{-4}$ per token, byte-level Mamba on seed~0; at most
$-1.9\times10^{-4}$ on seeds 1 and 2), the single-token box was linear at $\eps=10^{-6}$, and the state-level IBP-to-true ratio at
lags 64, 256 and 1024 stayed within 0.82--2.64 (seed~0), 0.31--1.93 (seed~1) and 0.76--1.92 (seed~2) times its value at lag 8 in both
layers of every model. On Mamba seed~1 the median exponent ($\lambda_M=\lambda_J$) of the first and the second layer,
$-0.00197$ [$-0.00200$, $-0.00192$] and $-0.00267$ [$-0.00276$, $-0.00255$], carries four of the five statistics on which the added
documents differ from the archived ones: on the 100 added documents alone it is $-0.00192$ [$-0.00198$, $-0.00182$] and
$-0.00256$ [$-0.00270$, $-0.00242$] (\cref{tab:ci-b}), negative in either set.

\clearpage

\section{IBP training under synonym substitution and the slow-channel attribution in full}\label{app:gate2}

\paragraph{Setting.}
Seven architectures (Mamba, the S4D-type and minGRU-type layers, LSTM, GRU, CNN and bag of words) were trained on word-level IMDB
under synonym substitution at box scale 1 (\cref{sec:setting}): naturally (one seed; Mamba three seeds), with standard IBP (three seeds)
and with fixed-point fine-tuning, a fine-tuning with the robust term at the $L=\infty$ fixed point of \cref{cor:fixed} (three seeds;
training recipes
in \cref{app:compute}). Each checkpoint was evaluated in float64 on the first 1000 test documents, with every position of the context
and of the document perturbed: certified accuracy on the bare document, after 1024 and 4096 tokens of contiguous neutral text (endpoint
scan), at $L=\infty$ (the repeated 1024-token period, \cref{thm:inf}; for LSTM and GRU the invariant box of \cref{rem:invbox}) and under
the K-uniform hull of \cref{thm:hull}, which covers every number of copies of the period; and accuracy under a substitution attack on
the first 100 documents at $L=0$. The study was pre-registered as a test of whether certified training of Mamba transfers to unbounded
contexts. Four of its five criteria failed, and its own rule stopped the certified-training line (\cref{tab:prereg}): the attack did not
bring the natural Mamba to at most 40\% accuracy, the level set to show that the threat is real (50.3\%); standard IBP did not reach
60\% certified at $L=0$ with 78\% clean accuracy (47.6\% with 61.7\%); fixed-point fine-tuning did not keep at least 0.8 of its
$L=0$ certificates at
$L=\infty$ within 3 points of the standard-IBP clean accuracy on two of three seeds (per-seed ratios 0.89, 0.52 and 0.35, the first at
51.9\% clean accuracy); and the ratios predicted for the other architectures held for the S4D-type model and the LSTM but not for the
minGRU-type model and the GRU. The fifth criterion, which would have called for a long-range task had the stateless CNN certified at
least as much as every stateful model, did not trigger. The numbers below therefore serve as measurements of what the certificates say
and as tests of \cref{prop:horizon}, not as a training result. A second pre-registered study, with the linear-ramp recipe (after
\citealp{jia2019}) and remedies for the blind spot, is reported in \cref{app:remedy}.

\begin{table}[htbp]
\caption{IBP training under synonym substitution (box scale 1, float64): clean and certified accuracy in \% on the first 1000 test
documents at growing context length, mean $\pm$ sd over three seeds unless marked; hull: the K-uniform certificate of \cref{thm:hull};
attack: accuracy under the substitution attack on the first 100 documents at $L=0$. Finite contexts are contiguous neutral text,
$L=\infty$ is a repeated 1024-token period (for LSTM and GRU the invariant box). Per-seed values, which matter because three seeds are
close to constant classifiers, are in \cref{tab:trainseeds}. The S4D-type seed~1 after fixed-point fine-tuning was evaluated
separately and is listed alone; LSTM and GRU after standard IBP, and LSTM after fixed-point fine-tuning, predict one class (47.4\%).
Every certificate in this table was re-verified with outward rounding, and none changed (\cref{rem:float}). For one checkpoint the
95\% Wilson interval of a share has a half-width of at most 3.1 points at $n=1000$ and 9.7 at $n=100$ (the attack); it does not
apply to the means over seeds (\cref{app:cleanctx-method}).}
\label{tab:train}
\centering\scriptsize
\setlength{\tabcolsep}{3.5pt}
\begin{tabular}{@{}lcccccccc@{}}
\toprule
 & & \multicolumn{5}{c}{certified} & & \\
\cmidrule(lr){3-7}
model & clean & $L=0$ & $L=1024$ & $L=4096$ & $L=\infty$ & hull & $L=\infty$\,/\,$L=0$ & attack\\
\midrule
\multicolumn{9}{@{}l}{\emph{standard IBP}}\\
Mamba & $61.7\pm12.7$ & $47.6\pm4.5$ & $39.0\pm9.4$ & $18.0\pm3.9$ & $3.8\pm4.7$ & $2.8\pm4.7$ & 0.08 & $60.0\pm10.4$\\
minGRU-type & $67.6\pm1.6$ & $52.3\pm6.6$ & $37.0\pm9.7$ & $22.7\pm5.0$ & $1.5\pm1.2$ & $0.9\pm0.9$ & 0.03 & $67.3\pm5.0$\\
S4D-type & $64.2\pm12.2$ & $42.5\pm4.6$ & $39.8\pm8.4$ & $39.9\pm8.3$ & $39.8\pm8.3$ & $33.2\pm10.9$ & 0.93 & $58.3\pm10.4$\\
CNN & $82.3\pm0.6$ & $52.8\pm1.0$ & $41.9\pm0.6$ & $34.5\pm5.5$ & $39.1\pm1.4$ & $30.7\pm1.0$ & 0.74 & $67.7\pm2.1$\\
bag of words & $83.4\pm0.5$ & $59.7\pm1.2$ & $35.5\pm2.1$ & $36.8\pm3.7$ & n/a & n/a & n/a & $66.0\pm2.0$\\
LSTM & 47.4 & 47.4 & 47.4 & 47.4 & 0.0 & n/a & 0.00 & 46.0\\
GRU & 47.4 & 47.4 & 47.4 & 47.4 & 47.4 & n/a & 1.00 & 46.0\\
\midrule
\multicolumn{9}{@{}l}{\emph{fixed-point fine-tuning}}\\
Mamba & $68.2\pm14.6$ & $53.0\pm2.3$ & $44.7\pm5.8$ & $35.9\pm13.0$ & $30.7\pm13.1$ & $19.9\pm19.4$ & 0.59 & $67.3\pm11.7$\\
minGRU-type & $70.4\pm6.1$ & $59.6\pm2.9$ & $57.6\pm5.2$ & $54.8\pm8.1$ & $43.7\pm12.1$ & $10.6\pm7.5$ & 0.74 & $69.0\pm6.2$\\
S4D-type, seeds 0, 2 & $77.6\pm0.9$ & $52.8\pm4.3$ & $46.2\pm5.7$ & $46.4\pm5.8$ & $46.0\pm5.4$ & $39.2\pm5.2$ & 0.87 & $68.5\pm3.5$\\
S4D-type, seed 1 & 72.9 & 53.3 & 52.6 & 52.7 & 52.7 & 48.9 & & \\
LSTM & 47.4 & 47.4 & 47.4 & 47.4 & 0.0 & n/a & 0.00 & 46.0\\
GRU & $63.1\pm22.9$ & $33.3\pm29.0$ & $33.3\pm29.0$ & $33.3\pm29.0$ & $33.3\pm29.0$ & n/a & 0.67 & $53.7\pm7.5$\\
\midrule
\multicolumn{9}{@{}l}{\emph{natural training} (one seed; Mamba three)}\\
Mamba & $87.9\pm0.8$ & 0.0 & 0.0 & 0.0 & 0.0 & 0.0 & & $50.3\pm5.7$\\
S4D-type & 86.8 & 0.0 & 0.0 & 0.0 & 0.0 & 0.0 & & 45.0\\
minGRU-type & 86.0 & 0.0 & 0.0 & 0.0 & 0.0 & 0.0 & & 57.0\\
LSTM & 89.4 & 0.0 & 0.0 & 0.0 & 0.0 & n/a & & 63.0\\
GRU & 90.7 & 0.0 & 0.0 & 0.0 & 0.0 & n/a & & 60.0\\
CNN & 89.8 & 0.0 & 0.0 & 0.0 & 0.0 & 0.0 & & 48.0\\
bag of words & 85.8 & 0.3 & 0.0 & 0.0 & n/a & n/a & & 49.0\\
\bottomrule
\end{tabular}
\end{table}

\begin{table}[htbp]
\caption{IBP training under synonym substitution, per seed (1000 test documents, \%; seeds 0 / 1 / 2): clean accuracy, certified
accuracy at growing context length, and the K-uniform hull. Close to constant classifiers: the standard-IBP Mamba seed~1 (51.9\%
clean), the standard-IBP S4D-type seed~1 (51.4\% clean; on the 200 documents of \cref{tab:attrib} clean accuracy 0.54 with 80\% of its
predictions in one class) and the Mamba seed~0 after fixed-point fine-tuning (51.9\% clean); their certified shares stay near their
clean accuracy at every length. After fixed-point fine-tuning the GRU seeds 0 and 1 are constant classifiers (certified share equal
to the clean accuracy) and its seed~2 certifies nothing, and the minGRU-type seed~2 predicts one class on 81\% of the documents of
the token-set certificate (\cref{app:anyctx}). Fixed-point: fixed-point fine-tuning. Every certificate in this table was
re-verified with outward rounding, and none changed (\cref{rem:float}). Per checkpoint the 95\% Wilson interval of a share has a
half-width of at most 3.1 points at $n=1000$ (\cref{app:cleanctx-method}; clean accuracy at every context length:
\cref{tab:cleanlen-ct}).}
\label{tab:trainseeds}
\centering\scriptsize
\setlength{\tabcolsep}{3pt}
\begin{tabular}{@{}llcccccc@{}}
\toprule
model & training & clean & $L=0$ & $L=1024$ & $L=4096$ & $L=\infty$ & hull\\
\midrule
Mamba & standard IBP & 57.2 / 51.9 / 76.1 & 42.4 / 50.0 / 50.3 & 35.4 / 49.6 / 31.9 & 14.4 / 22.2 / 17.4 & 0.7 / 9.2 / 1.4 & 0.1 / 8.2 / 0.0\\
minGRU-type & standard IBP & 66.7 / 66.7 / 69.4 & 58.9 / 52.2 / 45.7 & 46.0 / 38.3 / 26.8 & 26.6 / 24.4 / 17.1 & 2.0 / 2.4 / 0.1 & 0.9 / 1.8 / 0.1\\
S4D-type & standard IBP & 75.7 / 51.4 / 65.6 & 40.2 / 47.8 / 39.4 & 32.1 / 48.8 / 38.5 & 32.1 / 48.7 / 38.9 & 32.3 / 48.7 / 38.5 & 27.7 / 45.7 / 26.2\\
Mamba & fixed-point & 51.9 / 72.6 / 80.2 & 50.4 / 54.0 / 54.6 & 50.6 / 44.4 / 39.0 & 50.6 / 31.1 / 25.9 & 44.9 / 28.2 / 19.1 & 41.9 / 12.5 / 5.4\\
minGRU-type & fixed-point & 76.8 / 69.8 / 64.7 & 59.4 / 56.8 / 62.5 & 58.6 / 52.0 / 62.3 & 59.7 / 45.5 / 59.2 & 57.7 / 37.7 / 35.8 & 2.1 / 16.4 / 13.4\\
S4D-type & fixed-point & 78.3 / 72.9 / 77.0 & 55.8 / 53.3 / 49.7 & 50.2 / 52.6 / 42.1 & 50.5 / 52.7 / 42.3 & 49.8 / 52.7 / 42.1 & 42.9 / 48.9 / 35.6\\
GRU & fixed-point & 52.6 / 47.4 / 89.4 & 52.6 / 47.4 / 0.0 & 52.6 / 47.4 / 0.0 & 52.6 / 47.4 / 0.0 & 52.6 / 47.4 / 0.0 & n/a\\
\bottomrule
\end{tabular}
\end{table}

\paragraph{Attacks.}
A certified share is a lower bound on the robust accuracy under the threat model, and an attack gives an upper bound. The
pre-registered attack is greedy and best-first: the substitutable positions are visited in the order of the gradient saliency of the
margin, and each is replaced by the substitute that lowers the margin most (two sweeps); a beam search of width 5 then runs over the 20
most salient positions, with random restarts. A success is counted only when the model mispredicts the substituted words. On the first
100 documents at $L=0$ it left the natural Mamba at $50.3\pm5.7\%$ accuracy (\cref{tab:train}). A run that adds the genetic attack of
\citet{alzantot2018} (population 60, 40 generations) on the documents that the greedy phase did not flip, registered afterwards as a
separate test, on 200 documents at $L=0$, found every success in the greedy phase (one from a random restart) and none in the genetic
search. Under it the natural Mamba kept 50.5, 51.5 and 53.5\% accuracy (clean 87, 85 and 88.5\% on these documents), the natural
S4D-type model 45.5\%, the CNN 50\%, the bag of words 52\%, the minGRU-type model 58.5\%, the GRU 58.5\% and the LSTM 61.5\%; the
certified-trained Mamba models lost little (fixed-point fine-tuning, seed~2: 72\% against 78.5\% clean). The robust accuracy of the
natural Mamba under
this threat is therefore only known to lie between 0\% (IBP certificates) and about 50\% (attacks).

\paragraph{The slow-channel attribution.}
\Cref{prop:horizon} predicts that a robust term on documents of at most 512 tokens is blind to slow Mamba and minGRU-type channels,
whose injections carry the forgetting factor, and not to S4D-type ones (\cref{rem:otherterms}). The attribution tests this on the $L=\infty$
failures of the standard-IBP models (\cref{tab:attrib}): 200 test documents with the periods of the context-length probe, the true
label as the class, the time constant $\tau_{\rm h}=P/(-\ln\pih)$ of every state channel at the threat, and a knockout that gives every
channel slower than the 512-token training document its radius after one 1024-token period at its own DC levels (the $r_K$ of
\cref{prop:horizon}(b)), a diagnostic and not a certificate. A checkpoint counts as degenerate, and enters no criterion, if its clean
accuracy is below 0.55 or more than 80\% of its predictions are one class.

\begin{table}[htbp]
\caption{Slow-channel attribution of the $L=\infty$ failures of the standard-IBP models (200 test documents with the periods of the
context-length probe; the class is the true label). Clean accuracy and certified share at $L=0$ and $L=\infty$ on these documents;
width failures (certified at $L=0$ but not at $L=\infty$, with a correct clean $L=\infty$ prediction) and flip failures (the clean
$L=\infty$ prediction is wrong, which no box can certify); share of the width failures restored by the knockout, with its 95\%
percentile-bootstrap interval over the documents of the checkpoint (\cref{app:cleanctx-method}; a point where every resample gives
the same share); share of the
(document, channel) pairs with a time constant above 4096 tokens at the threat and in the clean dynamics, maximised over the layers.
The S4D-type seed~1 (clean accuracy 0.54, 80\% of its predictions in one class) is degenerate and left out. $^\S$No channel is slower
than 512 tokens (largest time constant 343 and 425 tokens), so the knockout replaces nothing.}
\label{tab:attrib}
\centering\footnotesize
\begin{tabular}{@{}lcccccccc@{}}
\toprule
 & & & \multicolumn{2}{c}{certified} & & & \multicolumn{2}{c}{slower than 4096 tokens}\\
\cmidrule(lr){4-5}\cmidrule(l){8-9}
model & seed & clean & $L=0$ & $L=\infty$ & width / flip failures & restored & at the threat & clean\\
\midrule
Mamba & 0 & 0.58 & 0.46 & 0.03 & 57 / 31 & 0.912 [0.830, 0.981] & 0.217 & 0.207\\
 & 1 & 0.565 & 0.55 & 0.08 & 91 / 3 & 1.0 [1.0, 1.0] & 0.0024 & 0.0010\\
 & 2 & 0.765 & 0.51 & 0.06 & 49 / 46 & 0.796 [0.675, 0.905] & 0.319 & 0.298\\
minGRU-type & 0 & 0.67 & 0.59 & 0.02 & 76 / 38 & 0.711 [0.606, 0.810] & 0.695 & 0.611\\
 & 1 & 0.685 & 0.565 & 0.055 & 39 / 67 & 0.769 [0.629, 0.897] & 0.460 & 0.397\\
 & 2 & 0.67 & 0.46 & 0 & 59 / 33 & 0.661 [0.537, 0.780] & 0.417 & 0.280\\
S4D-type & 0 & 0.785 & 0.44 & 0.35 & 18 / 0 & 0 [0, 0]\,$^\S$ & 0 & 0\\
 & 2 & 0.66 & 0.385 & 0.395 & 4 / 2 & 0 [0, 0]\,$^\S$ & 0 & 0\\
\bottomrule
\end{tabular}
\end{table}

\begin{figure}[htbp]
\centering
\includegraphics[width=\textwidth]{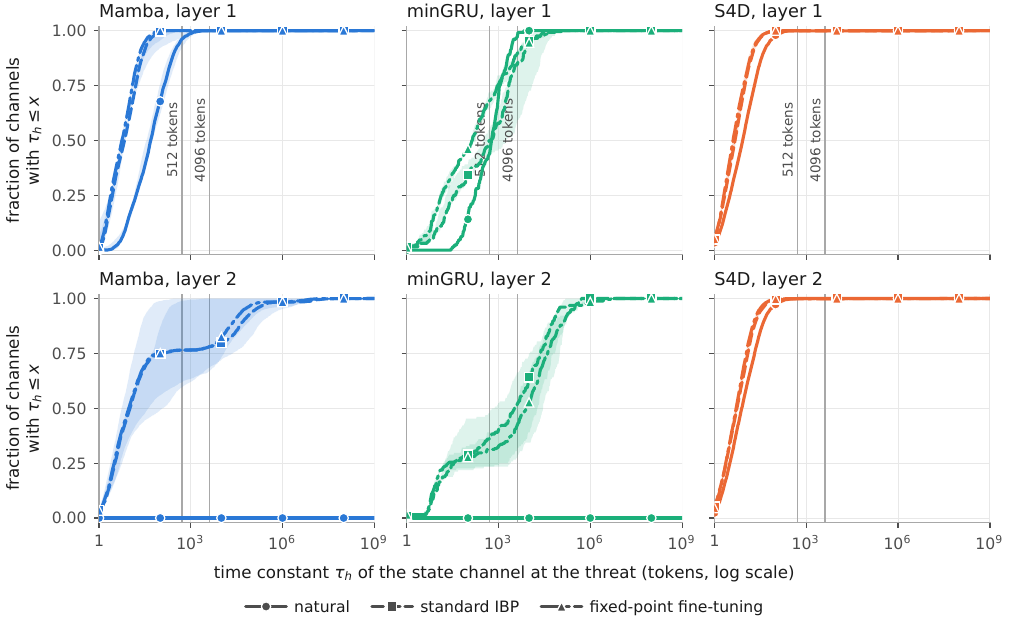}
\caption{Slow-channel attribution: empirical distribution over the state channels of one layer (Mamba and S4D-type: $d_i\times N$
channels, minGRU-type: $d_i$) of the time constant $\tau_{\rm h}=P/(-\ln\pih)$, with $\pih$ the product over a period of $P=1024$
neutral tokens of the upper decay bound $\ahi$ in the synonym threat box (box scale 1; $\tau_{\rm h}=\infty$ where $\pih=1$). Per channel
the median over the 200 documents; line: median over the non-degenerate seeds of the distribution function, band: seed minimum to
maximum. Guides at 512 tokens (the training document length) and 4096 tokens. A curve that ends below 1 has channels slower than the
axis ($\tau_{\rm h}=\infty$ included; a curve at 0 everywhere: no channel contracts over the period, as in the second layer of the natural
Mamba and minGRU-type models, whose boxes are too wide at this threat). After standard IBP and fixed-point fine-tuning a sizeable
fraction of the Mamba
and minGRU-type channels is slower than the training document (for example the minGRU-type model after standard IBP, layer 2: 63\% of
the channels above 512 tokens and 48\% above 4096, medians over seeds), while the real-diagonal S4D-type models have at most 0.1\% above
512 tokens. The pooled (document, channel) fractions per layer agree with \cref{tab:attrib}, whose columns are their maximum over the
layers.}
\label{fig:tau}
\end{figure}

\paragraph{Reading.}
With this recipe, standard IBP keeps the certificates of the real-diagonal S4D-type model at every context length (39.8\% at
$L=\infty$ against 42.5\% at $L=0$; the seed~1 after fixed-point fine-tuning: 52.7\% against 53.3\%, flat from $L=1024$ on), while
those of Mamba and the minGRU-type model decay with the
length and vanish at $L=\infty$ (under the linear-ramp recipe of \cref{app:remedy}, Mamba keeps most of them at $L=\infty$). The
S4D-type means include seed~1, which is close to a constant classifier; its other two seeds certify
32.3\% and 38.5\% at $L=\infty$ against 40.2\% and 39.4\% at $L=0$ (\cref{tab:trainseeds}). The attribution separates two causes of the
selective models' losses. Part of the collapse is a clean effect: on 3--48\% (Mamba) and 33--63\% (minGRU-type) of the failing documents
the infinite neutral context changes the clean prediction itself, and no box can certify such a flip; on the first 1000 test
documents the infinite clean context alone costs the standard-IBP Mamba seed~2 and the three standard-IBP minGRU-type seeds 9.4--27.1
points of clean accuracy (\cref{tab:cleanlen-ct}; Mamba seeds 0 and 1, marked degenerate there, left out); for the minGRU-type model slow
channels therefore explain well under the whole loss, and what follows concerns width failures only. Of the width failures, the
knockout restores 92\% (Mamba) and 71\% (minGRU-type) pooled, per seed 91, 100 and 80\% (Mamba) and 71, 77 and 66\% (minGRU-type);
restricting it to channels slower than 4096 tokens restores 92\% and 70\%, and a conservative variant, whose boxes provably contain the
clean infinite-context state, restores 90\% and 67\%. The pooled Mamba number leans on seed~1, close to a constant classifier (clean
accuracy 0.565 on these documents) but above the degeneracy line, with 91 width failures, all restored, against 57 and 49 for seeds 0
and 2. Standard IBP training with this recipe created these channels: in the clean dynamics 0.10--30\% (Mamba) and 28--61\% (minGRU-type) of the
(document, channel) pairs are slower than 4096 tokens, against none in the natural models, while the two non-degenerate S4D-type seeds
have no channel slower than 512 tokens and lose little, as \cref{rem:otherterms} predicted before the attribution ran. Two
pre-registered predictions failed on one seed each: that every selective standard-IBP seed has at least 1\% of its channels slower than
4096 tokens at the threat (Mamba seed~1: 0.24\%), and that the byte-level fine-tunes slow their channels at unchanged DC levels
(\cref{app:byte}). The time constants at the threat are drawn in \cref{fig:tau} and discussed in \cref{rem:timescales}. The knockout is a
diagnostic, not a certificate.

\clearpage

\section{The byte-level study}\label{app:byte}

\paragraph{Setting.}
Before the word-level experiments, a byte-level study trained the same small models (two layers, width 64) on byte-level IMDB with
documents of $T=1024$ bytes, under an $\ell_\infty$ box of radius 0.01 on every byte embedding, context included. An audit found that
this radius is not a threat: under a strong PGD attack the natural model's median worst-case margin was 6.05 against 6.3 clean, and
the ball is about 200 times smaller than the distance between any two byte embeddings. The study therefore appears here only as a
diagnostic of \cref{prop:horizon}, not as a robustness result. Its natural Mamba and LSTM are also the byte-level rows of the
context-length probe and of the interval-map growth rates (\cref{app:gate3}). Its evaluation reads 1000 test reviews after 4096 bytes
of contiguous licence text, at $L=\infty$ (a repeated 1024-byte period, \cref{thm:inf}) and under the K-uniform hull (\cref{thm:hull}).

\paragraph{Pre-registered tests and their outcomes.}
Two byte-level tests were fixed in writing before their runs. (i)~Fine-tuning three standard-IBP Mamba checkpoints with a finite-horizon
curriculum, a robust term on up to $10^{6}$ copies of a 256-byte period (a horizon of $2.56\times10^{8}$ tokens), was to keep at least
0.8 of the $L=0$ certificates at $L=\infty$ with at least 70\% clean accuracy, while a control trained further with standard IBP stays
at or below 10\% at $L=\infty$; a clean accuracy below 65\% or an $L=\infty$ share below 30\% was to stop the line. All three
fine-tuned seeds stayed below 30\% at $L=\infty$ (1.0, 27.3 and 22.3\% against 70.0, 70.9 and 66.4\% at $L=0$; control 2.0\%), the
stop condition. One run trained from scratch with the same curriculum met the main condition (48.5\% at $L=\infty$ against 56.7\% at
$L=0$, 73.5\% clean). (ii)~A final round, fixed after (i) and before its own runs: fixed-point fine-tuning, with the robust term at the
$L=\infty$ fixed point in the DC form of \cref{lem:period} (the curriculum now on the radius of the prefix boxes), on three seeds, and a
repetition of the from-scratch curriculum on seeds 1 and 2. A run passed with an $L=\infty$ share of at least 0.8 times its $L=0$ share,
at least 70\% clean accuracy and at least 55\% certified at $L=0$; fixed-point fine-tuning passed if two of its three fine-tuned seeds
did, the curriculum if one of its two new seeds did. Fixed-point fine-tuning passed: its seeds reached clean 74.6, 74.2 and 70.5\%,
certified 68.4, 68.5 and 65.5\% at $L=0$ and 59.6, 60.0 and 49.7\% at $L=\infty$ (ratios 0.87, 0.88 and 0.76), and a run trained from
scratch with the fixed-point term met the condition at the threshold (ratio 0.80). The from-scratch curriculum failed on both new seeds
(ratios 0.16 and 0.71), so its pass on seed~0 did not replicate. These runs evaluated a float32 model around a float64 period part. The
curriculum test met its stop condition, and round (ii) was fixed and run after it; we report both as they happened. Fixed-point
fine-tuning did not carry over to the word-level threat (\cref{app:gate2}).

\begin{table}[htbp]
\caption{Byte-level study, certificates in float64 (1000 reviews; $\ell_\infty$ radius 0.01 on every byte embedding, not a threat;
fractions of the reviews; seeds 0 / 1 / 2). The 4096-byte prefix is contiguous licence text, $L=\infty$ the repeated 1024-byte period;
the hull covers every number of copies of the period.}
\label{tab:byte}
\centering\scriptsize
\setlength{\tabcolsep}{3pt}
\begin{tabular}{@{}>{\raggedright\arraybackslash}p{3.2cm}ccccc@{}}
\toprule
training & clean, $L=0$ & certified, $L=0$ & certified, $L=4096$ & certified, $L=\infty$ & hull (every $K$)\\
\midrule
standard IBP from scratch & 0.745 / 0.754 / 0.716 & 0.62 / 0.635 / 0.588 & 0.437 / 0.476 / 0.386 & 0.0 / 0.009 / 0.019 & 0.0 / 0.0 / 0.005\\
control: more standard IBP steps (one model) & 0.753 & 0.713 & 0.664 & 0.02 & 0.001\\
finite-horizon curriculum fine-tuning & 0.731 / 0.744 / 0.696 & 0.7 / 0.709 / 0.664 & 0.7 / 0.686 / 0.665 & 0.01 / 0.273 / 0.223 & 0.003 / 0.111 / 0.102\\
fixed-point fine-tuning & 0.746 / 0.742 / 0.705 & 0.684 / 0.685 / 0.655 & 0.652 / 0.652 / 0.606 & 0.596 / 0.6 / 0.497 & 0.511 / 0.561 / 0.399\\
\bottomrule
\end{tabular}
\end{table}

\paragraph{Float64 re-evaluation.}
The byte-level checkpoints were re-evaluated in float64 on the same 1000 reviews (\cref{tab:byte}). At $L=0$, 4096 and $\infty$ this
reproduced the numbers obtained with the float32 model around a float64 period part, and it added the K-uniform hull, for which no
float32 value exists. Under the hull the fixed-point fine-tunes kept 51.1, 56.1 and 39.9\%, the curriculum fine-tunes 0.3, 11.1 and
10.2\%, the control
0.1\% and standard IBP from scratch 0, 0 and 0.5\%. A second float64 evaluation, on 200 documents with the periods of the context-length
probe, found 66.5, 66 and 65.5\% at $L=0$ and 60, 58.5 and 49\% at $L=\infty$ for the fixed-point fine-tunes, and 68, 69 and 65.5\%
at $L=0$ against 1.5,
26.5 and 14\% at $L=\infty$ for the curriculum fine-tunes.

\paragraph{Slow channels at byte level.}
In the slow-channel attribution (\cref{app:gate2}), giving the curriculum fine-tunes' channels slower than $10^{8}$ tokens their radius at
the curriculum horizon ($K=250000$ periods of 1024 tokens) restored 89, 96 and 89\% of their width failures. These seeds had 20.6\%,
12.7\% (layer 1) and 25.8\% (layer 0) of their (document, channel) pairs above $10^{8}$ tokens, with largest time constants of
$1.38\times10^{17}$, $1.24\times10^{18}$ and $3.45\times10^{14}$ tokens. The fine-tuning slowed channels at roughly unchanged DC levels in
two of the three seeds (58\% and 24\% of the layer-1 channels slowed at least tenfold, at median DC-level ratios of 0.65 and 1.11), but
not in the third (38\% of the layer-0 channels slowed, at DC levels 10.3 times higher): the pre-registered prediction of unchanged DC
levels failed on that seed. The control, trained further with standard IBP, slowed 9.3\% of its layer-1 channels as well. So "the
fine-tuning slowed channels at unchanged DC levels" holds for two of three seeds only. Two further byte-level measurements are stated
with the theory: on a trained checkpoint the infinite-context box does not contain the boxes of finite contexts (\cref{rem:covers}),
and a standard-IBP checkpoint has a channel whose time constant is about $6\times10^{7}$ tokens (\cref{rem:timescales}).

\section{Pre-registration record}\label{app:prereg}

Every experiment of \cref{sec:exp} was pre-registered: its criteria were written into a specification before the production run,
and a summary script applied them mechanically to the outputs. Where a criterion was amended or a validity check corrected, the
amendment is dated, both versions are computed, and both are printed; \cref{app:prereg-amend} lists every such change. The
specifications, the summary scripts and their outputs are in the code release (\cref{app:compute}). They use internal labels
for the tests and short codes for the criteria. This appendix is the only place where those labels and codes appear:
\cref{tab:prereg} maps them to the names used in the rest of the paper, \cref{app:prereg-criteria} states every criterion in words with
its outcome, and \cref{app:prereg-notshown} lists what the experiments do not show. The main text summarises the tests in
\cref{tab:tests}.
Dates are those of the entries of our laboratory log that fixed each specification; the log, in the code release, gives their times of
day.

\begin{table}[htbp]
\caption{The pre-registered tests: the name used in the paper, the internal label of the specification and of its outputs, the
date on which the specification was fixed (before the production run), the criteria, and the outcome. Criteria are stated in words
in \cref{app:prereg-criteria}.}
\label{tab:prereg}
\centering\scriptsize
\setlength{\tabcolsep}{4pt}
\begin{tabular}{@{}>{\raggedright\arraybackslash}p{0.19\textwidth}>{\raggedright\arraybackslash}p{0.09\textwidth}>{\raggedright\arraybackslash}p{0.125\textwidth}>{\raggedright\arraybackslash}p{0.19\textwidth}>{\raggedright\arraybackslash}p{0.315\textwidth}@{}}
\toprule
name in the paper & internal label & fixed on & criteria & outcome\\
\midrule
byte-level curriculum fine-tuning (\cref{app:byte}) & pilot, round of 2026-10-09 & \mbox{2026-10-09} & per-run rule; stop rule & stop condition met: $L=\infty$ below 30\% on all three fine-tuned seeds\\
byte-level test of fixed-point fine-tuning (\cref{app:byte}) & pilot, round ss4 & \mbox{2026-10-10} & per-run rule; fixed-point fine-tuning: 2 of 3 seeds; from-scratch curriculum: 1 of 2 & fixed-point fine-tuning PASS (2 of 3 seeds); from-scratch curriculum FAIL\\
IBP training under synonym substitution (\cref{app:gate2}) & Gate 2 & \mbox{2026-10-10} & G2-a, G2-b, G2-c, G2-d, G2-e & G2-a, G2-b, G2-c, G2-d FAIL; G2-e no fork; the certified-training line stopped\\
genetic attack (\cref{app:gate2}) & follow-up of Gate 2 & \mbox{2026-10-21} & separate test; cannot revise G2-a & no success beyond the greedy phase; G2-a stands\\
context-length probe (\cref{app:gate3}) & Gate 3 & \mbox{2026-10-21}; Amendment 1 the same day, before the production run & C0, C1, C2; amended C0$'$, C1$'$, C2$'$ & C0 PASS; C1 FAIL (one clause, byte-level Mamba); C2 FAIL; C0$'$, C1$'$ (6 models), C2$'$ (3 gated models) PASS\\
its replication on seeds 1 and 2 & Gate 3R & \mbox{2026-10-22} & the same, Amendment 1 unchanged & C0, C0$'$, C1 PASS; C2 FAIL; C1$'$, C2$'$ PASS\\
complex-diagonal control (\cref{app:gate3}) & R2 (Gate 4) & \mbox{2026-10-21} & T, M, V, X, B (B1, B2), Dk & CONFIRMED on both seeds\\
interval-map growth rates (\cref{app:gate3}) & R4 (Gate 4) & \mbox{2026-10-21} & V1--V4, G1--G6, J1--J3 & all PASS (3 gated, 4 nonnegative diagonal models)\\
their replication on seeds 1 and 2 & Gate 6C & \mbox{2026-10-22} & every seed-0 PASS must replicate & every criterion replicates\\
slow-channel attribution (\cref{app:gate2}) & R1 (Gate 4) & \mbox{2026-10-21} & C0, P1--P6 & C0, P1, P2, P5, P6 PASS; P3, P4 FAIL (one seed each)\\
token-set certificate (\cref{app:anyctx}) & R9 (Gate 4) & \mbox{2026-10-21} & V, A1, A2, B--F, G, H1, H2, I & V PASS (after a completion job); A1, A2, B, C, D, F, I PASS; E, G, H1, H2 FAIL; on the arms completed by the extension (200 documents) G holds (the seed that failed: 61.5\% [54.6, 68.0] at both periods; \cref{app:cleanctx-record})\\
certified radius against distance (\cref{app:gate5}) & Gate 5 & \mbox{2026-10-22} & V1--V4, Z1a, Z1b, Z2a, Z2b, Z3 & V PASS (after a corrected document count); GRU, LSTM: Z1a PASS, Z1b FAIL, Z3 PASS; Mamba: Z2a, Z2b PASS; S4D-type: Z2a PASS, Z2b FAIL\\
its replication on seeds 1 and 2 & Gate 6B & \mbox{2026-10-22} & Gate 5's criteria; replication rule; P1--P4 & every outcome replicates; P1--P4 hold on every seed\\
budget sweep (\cref{app:sweep}) & Gate 6A & \mbox{2026-10-22}; Amendment 1 before any output & V (V1--V5), S1, S2, S3 & all PASS (GRU, LSTM)\\
linear relaxations at finite context (\cref{app:lirpa}) & Gate 7 & \mbox{2026-10-27} (criteria frozen with the code, before the production run) & V1--V9; A1--A5, A4L, A4E, N1; B1--B5; C1 & V PASS; A1 PASS (standard-IBP Mamba, S4D-type, minGRU-type); A2 PASS (standard-IBP S4D-type, minGRU-type), n/a (standard-IBP Mamba); A3 FAIL (standard-IBP Mamba, S4D-type, minGRU-type); A4 PASS (standard-IBP S4D-type, minGRU-type; natural GRU, LSTM, S4D-type, minGRU-type), FAIL (standard-IBP Mamba), n/a (natural Mamba); A4L PASS (standard-IBP Mamba, S4D-type, minGRU-type; natural LSTM and Mamba); A4E PASS (standard-IBP S4D-type, minGRU-type; natural LSTM), n/a (standard-IBP and natural Mamba); A5 n/a; N1 PASS; B1, B2 PASS (GRU, LSTM); B3, B4 PASS (S4D-type, minGRU-type), n/a (Mamba); B5 PASS (GRU, LSTM), FAIL (Mamba, S4D-type, minGRU-type); C1 PASS\\
training remedy for the blind spot (\cref{app:remedy}) & Gate 8 & \mbox{2026-10-27}, frozen by hash before the first training run & V1--V7; P1--P10, C1, C2 & V PASS; minGRU-type: P1--P7, C1, P8a PASS; C2, P8b, P8c, P10 FAIL; Mamba: P1, P2, P4a, P4b, P6, C1, P8a PASS; P3a, P3b, P5, P7 (its accuracy clauses), C2, P8b, P8c FAIL; P10 n/a; S4D-type: P9 PASS, P10 n/a\\
floating-point re-verification (\cref{rem:float}) & floating-point track; amendment 8q & \mbox{2026-10-28}, before the production run; one amendment after it & validity checks; six predictions & as registered: the check against over-widening failed on one document, so nothing read; amended: every validity check passed; five predictions held, the one for token sets failed\\
floating-point re-verification of the training remedy (\cref{rem:float}) & floating-point track, tier T5 & \mbox{2026-10-31}, before any of its runs; inputs frozen before the production runs & validity checks; eight predictions & every validity check passed; no certificate changed; seven predictions held, the one for certificate checking on the capped models failed\\
extension to 200 documents and intervals (\cref{app:cleanctx}) & W15; amendments W15-A1, W15-A2 & \mbox{2026-10-31}, before any of its runs; two amendments of its summary script & validity checks; six predictions & as registered, two checks failed (a reworded text line; 148 task logs read from their first line), so nothing read; amended: every check that can stop a reading passed; all six predictions held: archived outcomes of the predicted criteria kept, 159 of 164 statistics agreeing between new and archived documents, the universal statements, the median gated growth rates, the clean prediction of the natural nonnegative diagonal models settling within 16 periods, no clean prediction of the natural Mamba and minGRU-type models certified by the hull\\
\bottomrule
\end{tabular}
\end{table}

\subsection{Criteria and outcomes}\label{app:prereg-criteria}

\paragraph{Byte-level curriculum fine-tuning (pilot, round of 2026-10-09; fixed before the run was submitted).}
Three standard-IBP byte-level Mamba checkpoints fine-tuned with a finite-horizon curriculum (a robust term on up to $10^6$ copies of a
256-byte period), a fourth model trained from scratch with the curriculum, and a control trained further with standard IBP; $\ell_\infty$
radius 0.01 on every byte embedding (not a threat, see below), 1000 reviews. \emph{Pass:} the fine-tuned models keep at $L=\infty$ at
least 0.8 times their certified share at $L=0$ with at least 70\% clean accuracy, while the control stays at or below 10\% at
$L=\infty$. \emph{Stop:} a clean accuracy below 65\% or an $L=\infty$ share below 30\%. \emph{Outcome:} the stop condition was met:
all three fine-tuned seeds stayed below 30\% at $L=\infty$ (1.0, 27.3 and 22.3\%, against 70.0, 70.9 and 66.4\% at $L=0$; control
2.0\%); the run from scratch met the main condition (48.5\% at $L=\infty$ against 56.7\% at $L=0$, 73.5\% clean). The next round,
below, was fixed after this outcome and before its own runs (\cref{app:byte}).

\paragraph{Byte-level test of fixed-point fine-tuning (pilot, round ss4; fixed 2026-10-10, before the round was submitted).}
Byte-level IMDB at $T=1024$, a two-layer Mamba of width 64, an $\ell_\infty$ box of radius 0.01 on every byte embedding, 1000
reviews with a licence-text prefix, every position perturbed. An audit before this round had found that the radius 0.01 is not a
threat (the certified ball is about 200 times smaller than the distance between two byte embeddings), so the test bears only on the
mechanism of \cref{prop:horizon,cor:fixed}. A run passes if its certified share at $L=\infty$ is at least 0.8 times its share at
$L=0$, its clean accuracy at $L=0$ is at least 70\% and its certified share at $L=0$ at least 55\%. Fixed-point fine-tuning
(fine-tuning with the robust term at the fixed point, three seeds) passes if at least two of its fine-tuned seeds pass; the
from-scratch curriculum (training from scratch with a finite-horizon curriculum up to $K=10^6$ copies, the only earlier recipe that
had passed, on seed~0) passes if at least one of its repetitions on seeds 1 and 2 passes. \emph{Outcome.} Fixed-point fine-tuning
PASS: seed~0 had clean accuracy 74.6\% and certified 68.4\% at $L=0$ and 59.6\% at $L=\infty$ (ratio 0.87), seed~1 74.2\%, 68.5\%
and 60.0\% (0.88), seed~2 70.5\%, 65.5\% and 49.7\% (0.76, not passed). From-scratch curriculum FAIL (ratios 0.16 and 0.71). A run
trained from scratch with the fixed-point term (seed~0, outside the rule) met the per-run rule at a ratio of 0.80. These numbers come from a float32 model around a float64 period part; the float64 re-evaluation reproduced them
(\cref{app:byte}).

\paragraph{IBP training under synonym substitution (Gate 2; fixed 2026-10-10).}
Word-level IMDB at $T=512$, synonym substitution at box scale 1 at every position of the context and the document, seven
architectures (Mamba, S4D-type, minGRU-type, LSTM, GRU, CNN, bag of words), natural training, standard IBP and fixed-point
fine-tuning, 1000 test
documents, float64. \emph{G2-a} (the threat is real): natural Mamba keeps at most 40\% accuracy under the greedy attack with beam at
$L=0$, with clean accuracy at least 85\%. FAIL: 50.3\% under attack, clean 87.9\%. \emph{G2-b} (certified training works):
standard-IBP Mamba is certified at $L=0$ on at least 60\% of the documents with clean accuracy at least 78\%. FAIL: 47.6\% at a
clean accuracy of 61.7\%. \emph{G2-c} (unbounded context): Mamba after fixed-point fine-tuning is certified at $L=\infty$ on at
least 0.8 times its own
$L=0$ share, with clean accuracy within 3 points of standard-IBP Mamba of the same seed, on two of three seeds. FAIL: the seed
ratios were 0.89 (at a clean accuracy of 51.9\%, against 57.2\% for standard IBP), 0.52 and 0.35. \emph{G2-d} (the class, not the
model): the S4D-type and minGRU-type models after fixed-point fine-tuning reach the same ratio of at least 0.8 on two of three
seeds, and the LSTM
and GRU trained with the period term reach at most 0.3 at $L=\infty$ with their invariant box. FAIL: the minGRU-type ratios were 0.97,
0.66 and 0.57 and the GRU's 1.0, 1.0 and 0.0; the S4D-type clause held (0.89 and 0.85; the third seed was evaluated later) and so
did the LSTM clause (0, 0, 0). Every standard-IBP LSTM and GRU collapsed to the majority class (47.4\% clean accuracy).
\emph{G2-e} (relevance; a fork, not a stop): if the CNN's certified share is at least every stateful model's at $L=\infty$ and in the
evidence probe, a long-range task is required. No fork: the CNN certified 39.1\% at $L=\infty$, against 47.4\% for the best stateful
row, an IBP-trained GRU that had collapsed to the majority class. \emph{Rule:} a failure of G2-a, G2-b or G2-c stops the
certified-training line. It stopped on 2026-10-20, and its numbers serve only as measurements of what the certificates say
(\cref{sec:exp-train}, \cref{app:gate2}). A second pre-registered study, with another schedule and remedies for the blind spot,
followed (training remedy for the blind spot, below). The specification was amended during the build of the code, on the day the criteria were
fixed, in data preparation and implementation only (vocabulary, embedding scale, the exact range of the GRU update, the certificates
of the stateless models); the criteria were not changed.

\paragraph{Genetic attack (follow-up of Gate 2; registered 2026-10-21).}
A stronger attack, the genetic substitution attack of \citet{alzantot2018}, registered as a separate test that cannot revise G2-a:
200 documents at $L=0$, the natural models of every architecture and the certified-trained Mamba models. \emph{Outcome:} every
success came from the greedy phase (one from a random restart), none from the genetic search; natural Mamba kept 50.5, 51.5 and
53.5\% accuracy under attack (clean 87, 85 and 88.5\%), so G2-a stands.

\paragraph{Context-length probe (Gate 3; fixed 2026-10-21, before any run of its evaluation script).}
The natural word-level Mamba (seeds 0--2), S4D-type, minGRU-type, LSTM and GRU (seed~0) and the byte-level Mamba and LSTM, 200
documents; the CNN, the bag-of-words model and certified-trained models are descriptive rows. With $R$ the median looseness at
$\eps=10^{-8}$, growth\_txt $=R(\text{4096 neutral tokens})/R(\text{256 neutral tokens})$ and growth\_per $=R(\text{4 periods})/
R(\text{1 period})$, the criteria as pre-registered were: \emph{C0} (validity): (i) at $\eps=10^{-10}$ no document has $R<1-10^{-3}$
or $W<0$ on any finite-context family; (ii) no K-uniform certificate (the hull, the gated invariant box) exceeds any member it covers,
at any $\eps$ including 0, to a tolerance of $10^{-9}(1+|b|)$; (iii) $|b(0)-m^0|\le10^{-8}$ on every finite-context family; (iv) at
least 150 documents. \emph{C1} (nonnegative diagonal models): growth\_txt $\le10$, growth\_per $\le10$, a finite $L=\infty$ certificate
for at least 95\% of the documents at $10^{-8}$, and median $|W_{\infty}(10^{-10})/W_\infty(10^{-8})-1|\le0.01$. \emph{C2} (gated
models): growth\_txt $\ge10$, or a non-finite $W/\eps$ after 4096 neutral tokens for at least 50\% of the documents. Amendment~1
(\cref{app:prereg-amend}) replaced C2 by C2$'$ and the last clause of C1 by C1$'$, and added the checks C0$'$. \emph{Outcome:} C0
PASS; C1 FAIL, on one clause only: the linearity statistic of the byte-level Mamba was 0.0162; C2 FAIL (the gated growth was
0.44--1.03, because the boxes were saturated inside the document); C0$'$, C1$'$ (six models) and C2$'$ (three gated models) PASS.
C2$'$ was written after the smoke run had shown the saturation: it describes the observed regime and is not a confirmed prediction.

\paragraph{Replication of the context-length probe on seeds 1 and 2 (Gate 3R; fixed 2026-10-22).}
Eight new models (seeds 1 and 2 of the S4D-type, minGRU-type, LSTM and GRU models, trained with the natural recipe), with the
criteria of Amendment~1 unchanged. \emph{Outcome:} C0, C0$'$ and C1 PASS (growth\_txt 0.83, 1.06, 0.94 and 0.93; linearity statistic
at most 0.0014); C2 FAIL (growth 0.60--0.88, saturated inside the document); C1$'$ and C2$'$ PASS.

\paragraph{Complex-diagonal control (R2, one of four tests registered as Gate 4; fixed 2026-10-21).}
Two complex-diagonal LTI S4D models (seeds 0 and 1) trained with the natural recipe; the context-length probe ran twice per model,
in real boxes and in complex disks, with the same weights, documents and forward pass. In the order of reading: \emph{T}
(descriptive): validation accuracy, expected at least 0.80. \emph{M}: from the parameters alone, a model is testable if some layer
has modes with $\gbox\ge1$ (predicted: at least half of the modes of every layer). \emph{V} (validity per arm): C0$'$, with the box
arm exempt from ``no non-finite bound at $10^{-10}$'', because a wrapping box may overflow float64 (the exempted items are
recorded), and the domain of the output equal to the arm. \emph{X} (cross-arm identity): the clean margins, $D$, the corner margins
and $b(0)$ agree between the arms within $10^{-9}(1+|v|)$ on every document. \emph{B} (box arm): fails C1$'$; \emph{B1}: the
$L=\infty$ certificate is vacuous on at least 95\% of the documents; \emph{B2}: the finite-context box blows up (median width
exponent after 4096 neutral tokens at most 0.5, or growth\_txt $\ge10$). \emph{Dk} (disk arm): passes C1$'$. A model is CONFIRMED
if V and X pass on both arms, it is testable, the box arm fails C1$'$ and the disk arm passes; the control is confirmed if both models
are. \emph{Outcome:} CONFIRMED on both models (validation accuracy 0.876 and 0.858; 86--87\% of the modes of every layer have
$\gbox\ge1$; B1 and B2 true).

\paragraph{Interval-map growth rates (R4, Gate 4; fixed 2026-10-21).}
Gated models: the natural word-level LSTM and GRU and the byte-level LSTM; nonnegative diagonal models: the natural word-level
Mamba, S4D-type and minGRU-type and the byte-level Mamba; seed~0, 100 documents, both layers. \emph{V1--V4} (validity): first-order
soundness (no resolved lag with an IBP-to-true ratio below $1-10^{-3}$ at $\eps\le10^{-8}$); the point row reproduces the clean
margin within $10^{-8}(1+|m|)$ with zero state radius; for gated cells $M_t\ge|J_t|$ entrywise up to a floating-point floor, the
power-iteration log norms ordered $J\le|J|\le M$, a linear first-order radius map and a reproducible clean step; at least 80 of the
100 documents. \emph{G1}: $\lambda_M>0$ and \emph{G2}: $\lambda_J<0$, for the median and the pooled value over the documents, in every
layer. \emph{G3}: a median gap $\lambda_M-\lambda_J$ of at least 0.02 per token in every layer, and a largest per-layer median
$\lambda_M$ of at least $\ln(10^{10})/512=0.045$. \emph{G4}: one token's box at $\eps=10^{-10}$ reaches half its a-priori bound within
the $1024+n$ lags for at least 80\% of the documents, in every layer. \emph{G5}: the IBP-to-true ratio grows at least tenfold before
saturation (median over the documents), in every layer. \emph{G6}: the median measured saturation rate over the largest $\lambda_M$
of the layers up to the one measured lies in $[1/2,2]$, in every layer. \emph{J1}: the slowest channel's mean log decay is negative
on every document, layer, setting and segment. \emph{J2}: linear regime at $\eps=10^{-6}$ (the IBP radius per unit $\eps$ changes by
at most 5\% between $10^{-8}$ and $10^{-6}$, and no exit from the linear regime on at least 95\% of the documents). \emph{J3}: the
median IBP-to-true ratio at lags 64, 256 and 1024 is at most 10 times its value at lag 8, in every layer. \emph{Outcome:} every
criterion PASS (three gated and four nonnegative diagonal models). G2 holds for the median and the pooled value, not on every
document: on seed~0 about a quarter of the LSTM's layer-1 documents have $\lambda_J$ slightly above 0.

\paragraph{Replication of the growth rates on seeds 1 and 2 (Gate 6C; fixed on 2026-10-22 with the budget sweep).}
The same script and criteria on seeds 1 and 2 of the five word-level natural models, 100 documents each, with the rule that every
criterion that passed on seed~0 must pass on both seeds. \emph{Outcome:} V and G1--G6 PASS for GRU and LSTM on both seeds, J1--J3
PASS for Mamba, the S4D-type and the minGRU-type model on both seeds.

\paragraph{Slow-channel attribution (R1, Gate 4; fixed 2026-10-21).}
38 checkpoints (the word-level nonnegative diagonal models after natural training, standard IBP and fixed-point fine-tuning, and
the byte-level
Mamba models), 200 documents each, the periods of the context-length probe, the class the true label, at each model's own threat
(synonym substitution at box scale 1; at byte level the $\ell_\infty$ radius 0.01) and at $\eps=0$. Seeds that are degenerate (clean
accuracy below 0.55, or more than 80\% of the predictions in one class) enter no criterion. \emph{C0} (validity): the instrumented
computation reproduces the closed form bit for bit; the closed form agrees with $K\to\infty$ within $10^{-9}$, with no finiteness
mismatch; no monotonicity or clean-containment violation; at least 150 documents; and the $L=0$ certified share within three binomial
standard deviations of the full evaluation of the same model (a row failing one of the last two is reported and not read).
\emph{P1}: for standard-IBP Mamba and, separately, the minGRU-type model, replacing the radius of every channel with $\tau_{\rm h}>512$
tokens (slower than the training document) by its radius after one period of 1024 tokens restores at least 50\% of the width failures
at $L=\infty$, pooled, and at least 25\% on every non-degenerate seed (per-seed clauses need at least five width failures).
\emph{P2}: the effect is concentrated on channels slower than 4096 tokens (restoring only those gives at least 0.70 times the pooled
restoration). \emph{P3}: (i) at least 1\% of the (document, channel) pairs have $\tau_{\rm h}>4096$ at the threat (maximum over the
layers) for every non-degenerate selective IBP seed, and (ii) every non-degenerate S4D-type IBP seed has fewer than the smallest of
them. \emph{P4}: every byte-level finite-horizon fine-tune slowed, in some layer, at least 1\% of its channels at least tenfold at a
median DC-level ratio in $[1/3,3]$ among them, against its initial checkpoint, and the control (more standard IBP steps) slowed a
smaller fraction. \emph{P5}: for those fine-tunes, replacing the channels with $\tau_{\rm h}>10^8$ by their radius at the curriculum
horizon ($K=250000$ periods of 1024 tokens) restores at least 50\% of the width failures. \emph{P6} (secondary): standard IBP training
slows channels in the clean dynamics (the smallest share of channels with a clean $\tau_{\rm h}>4096$ over the non-degenerate IBP
seeds exceeds the largest over the natural seeds, Mamba and the minGRU-type model separately). \emph{Outcome:} C0 PASS; P1 PASS
(92\% pooled for Mamba, per seed 91, 100 and 80\%; 71\% for the minGRU-type model, per seed 71, 77 and 66\%); P2 PASS (92\% and 70\%
from the channels slower than 4096 tokens); P3 FAIL at (i) on Mamba seed~1 (0.24\%); P4 FAIL (two of three fine-tunes slowed their
layer-2 channels at median DC-level ratios of 0.65 and 1.11, the third its layer-1 channels at a 10.3 times higher DC level); P5
PASS (89, 96 and 89\%); P6 PASS.

\paragraph{Token-set certificate (R9, Gate 4; fixed 2026-10-21).}
61 checkpoints, 150 to 200 documents per arm, the setting of \cref{app:anyctx}; ``share'' is the certified share at $P=8$ over the
most frequent neutral token unless stated. \emph{V} (validity): every planned checkpoint, at least the minimum number of documents in
each arm, zero check violations, no merge issue or duplicate document, the registered grid. \emph{A1}: natural nonnegative diagonal
models have a share of at most 0.05 with a clean document. \emph{A2}: for the natural selective ones the bound is vacuous on at least
half of the documents. \emph{B}: the certified-trained S4D-type models, each training kind separately, have a share of at least 0.2
with a clean document and at least 0.1 with the substitution box on two of three seeds. \emph{C}: every non-degenerate stateful model
has a share of at most 0.05 at $|S|=64$, and at most half its $|S|=1$ share where that is at least 0.1, in both modes. \emph{D}: the
standard-IBP S4D-type model has a larger share than both selective standard-IBP models of the same seed, on two of three seeds.
\emph{E}: fixed-point fine-tuning gives a larger share than standard IBP for the selective models, same seed, on two of three pairs,
Mamba and the minGRU-type model each. \emph{F}: every non-degenerate gated model has a share of at most 0.05 in both modes. \emph{G}: a
longer period is not worse: the share at $P=64$ is at least the share at $P=8$ minus 0.02 for $|S|=1$, 2 and 4, in both modes, for
every non-degenerate model. \emph{H1}: the byte-level fixed-point fine-tunes have a share of at least 0.1 with their $\ell_\infty$
document box on two of three seeds. \emph{H2}: at byte level, the fixed-point fine-tune has a larger share than the finite-horizon
fine-tune of the same seed on two of three pairs. \emph{I} (a stateless reference, outside the reading): the natural CNN has a share
of at least 0.5. \emph{Outcome:} V PASS after the completion job of \cref{app:prereg-amend}; A1, A2, B, C, D, F and I PASS (the CNN
at 0.99); E FAIL (fixed-point fine-tuning was better on one of three Mamba seeds and on neither minGRU-type pair); G FAIL (one
S4D-type seed after fixed-point fine-tuning, with the substitution box: 0.581 at $P=64$ against 0.615 at $P=8$; on the arms that the
extension completed to 200 documents both periods certify the same 123 documents, 61.5\% [54.6, 68.0], and G holds there,
\cref{app:cleanctx-record}); H1 FAIL (no seed); H2 FAIL (no pair).

\paragraph{Certified radius against distance (Gate 5; fixed 2026-10-22, before any run; the smoke test changed no criterion).}
The natural seed-0 GRU, LSTM, Mamba and S4D-type models, 50 documents, the distance probe of \cref{app:gate5}. \emph{V1}: no sampled
input (random point, random corner, gradient corner) outside a method's logit hull or below its margin bound, to a tolerance of
$10^{-9}(1+|\cdot|)$, and the clean logits inside every hull. \emph{V2}: the point limit reproduces the clean margin within
$10^{-8}(1+|m^0|)$ for every method, distance and document. \emph{V3}: no resolved first-order ratio below $1-10^{-3}$ where
$D\ge10^{-4}(1+|m^0|)$. \emph{V4}: at least 40 documents with every method of the run plan. \emph{Z1a} (gated): median
$r_{\rm box}(128)/r_{\rm box}(1)\le10^{-3}$. \emph{Z1b} (gated): $r_{2048}(1024)/r_{2048}(1)\ge0.1$ for at least 80\% of the documents,
where $r_{2048}$ is the certified radius of the $B=2048$ zonotope. \emph{Z2a} (nonnegative diagonal): median
$r_{\rm box}(1024)/r_{\rm box}(1)\ge0.1$. \emph{Z2b} (nonnegative diagonal): the $B=256$ zonotope certifies at least 0.99 times the
box's certified radius on at least 80\% of the documents, at every distance where it ran. \emph{Z3} (cost; a prediction for the gated
models, descriptive for the others): one $B=2048$ evaluation at $d=1024$ costs at least ten box evaluations. \emph{Outcome:} V PASS
on all four models (after the corrected document count of \cref{app:prereg-amend}); GRU: Z1a PASS, Z1b FAIL (0\%), Z3 PASS (19
times); LSTM: Z1a PASS, Z1b FAIL (24\%), Z3 PASS (35 times); Mamba: Z2a PASS (median ratio 7.7), Z2b PASS (100\% of the documents at
every distance); S4D-type: Z2a PASS (27.2), Z2b FAIL (54\% at $d=512$).

\paragraph{Replication on seeds 1 and 2 (Gate 6B; fixed 2026-10-22).}
Seeds 1 and 2 of the four models, 20 documents each, Gate~5's criteria unchanged except that V4 requires 18 of the 20 documents; the
nonnegative diagonal models without the $B=2048$ rows. \emph{Replication rule:} a criterion replicates when its outcome on seeds 1 and
2 equals the seed-0 outcome. \emph{P1--P4}, profile statements of the paper checked per seed (a different set from P1--P6 of the
slow-channel attribution): \emph{P1}: the GRU's $B=2048$ zonotope keeps a tenth of its adjacent-token radius at $d=128$ on at least
half of the documents; \emph{P2}: the LSTM's does at $d=512$; \emph{P3}: the box's median certified radius is 0 at $d=128$ for both gated
models; \emph{P4}: the median $r_{\rm box}(1024)/r_{\rm box}(1)$ is at least 1 for Mamba and the S4D-type model. \emph{Outcome:} V PASS
on the eight models; every outcome replicates (GRU and LSTM: Z1a PASS, Z1b FAIL with 0\% and 0\% and 25\% and 65\% (the 65\% not
decisive at $n=20$; decisive on the extended set, 58.5\% [51.6, 65.1]), Z3 PASS with 33
and 32 and 47 and 45 times; Mamba: Z2a, Z2b PASS; S4D-type: Z2a PASS, Z2b FAIL with 25\% and 45\% at $d=512$, and on seed~1 already
60\% at $d=128$); P1--P4 hold on every seed.

\paragraph{Budget sweep (Gate 6A; fixed 2026-10-22; Amendment 1 the same day, before the production run).}
The zonotope alone on the natural seed-0 GRU and LSTM, documents 0--15, seven budgets and seven distances (\cref{app:sweep}).
\emph{V}: V1--V3 of Gate~5 at every (budget, distance, document); \emph{V4}: at least 14 of the 16 documents carry every budget at every
distance; \emph{V5} (reproduction): at the cells Gate~5 also ran, Gate~5's certified radius lies in this run's final bracket up to a
factor 1.001 for at least 95\% of the cells. \emph{S1} (monotone in $B$): for every distance the median certified radius at a larger
budget is at least 0.93 times the one at a smaller budget, and the reach does not decrease. \emph{S2} (budget buys distance):
$d^*_{16384}\ge\min(4d^*_{2048},1024)$. \emph{S3} (price at most linear in $B$): the median time of one bound at $d=1024$ grows at most
8 times from $B=2048$ to $B=16384$. \emph{Outcome:} V (V5: 160 of 160 shared cells per model), S1 (no failing pair among 147 pairs of
medians and 21 pairs of reaches), S2 ($d^*_{16384}=1024$ for both models, against 512 and 1024 required) and S3 (5.03 and 6.54) PASS
for the GRU (15 complete documents) and the LSTM (16); the reading over every document that carries a cell gives the same verdicts.

\paragraph{Linear relaxations at finite context (Gate 7; criteria frozen with the code on 2026-10-27, before the production run).}
auto\_LiRPA's IBP, CROWN-IBP, CROWN, $\alpha$-CROWN and forward mode against our certificates (\cref{app:lirpa}): the distance probe
of \cref{app:gate5} on the natural GRU and LSTM (seeds 0--2, 20 documents each; the GRU also in a one-product form of its update),
on the natural Mamba, S4D-type and minGRU-type models (seed~0, 10 documents) and on the two models of the complex-diagonal control
(10 documents each); and synonym substitution on the standard-IBP Mamba, S4D-type and minGRU-type models (three seeds each;
documents 0--199 at $L=0$ and 0--49 up to $L=4096$, auto\_LiRPA up to $L=1024$; full CROWN on seed~0) and on the natural seed-0
models. The thresholds were written into
the specification and into the code, which was frozen by its hash before the production run; the criteria section of the
specification was last changed earlier that day. They were set after correctness and timing runs on a few documents: the
timing runs fixed the largest distance of full CROWN (64, and 128 on the complex-diagonal control), the largest context of
auto\_LiRPA under synonym substitution (1024 neutral tokens) and the rule for forward mode at a zero-width input below, and the cost
prediction under synonym substitution was kept as written although the timing runs had made its failure likely. Statistics are over
the planned cells; a criterion holds for an architecture when at least two of its three seeds pass, except the cost exponent of the
distance probe, which is read on seed~0; a criterion is n/a when its statistics cover fewer than half of the planned cells.
\emph{V1--V9} (validity, every group; any failure voids the group): the rewritten networks reproduce the clean margin within
$10^{-9}(1+|m^0|)$; auto\_LiRPA's IBP equals an interval evaluation of the same traced graph within $10^{-9}(1+|b|)$, and ours is
never looser than that evaluation (equal to auto\_LiRPA's IBP on the LSTM); no sampled input has a margin below any method's bound,
to $10^{-9}(1+|m|)$; the certified fractions of the standard-IBP models reproduce those of the first training study within 0.0025;
the point limit holds within $10^{-8}(1+|m^0|)$ for every method other than forward mode at every distance and for forward mode at
$d=1$, while at $d>1$ a forward-mode cell that is more than that off gets no radius, an outcome reported with the number of its
bounds above the clean margin; errors on at most 5\% of the auto\_LiRPA runs; ours on at least 90\% of the planned cells, every
planned output complete and one frozen code version; the box radii of \cref{app:gate5} inside this run's final bisection bracket,
up to a small tolerance, on at least 95\% of the complete cells both runs share; and every CROWN-type bound computed with the chord
correction of \cref{app:lirpa}, added when the correctness runs had found the two soundness gaps described there.
\emph{B1} (GRU, LSTM, each seed): CROWN-IBP collapses like the box, median $r_{\rm hi}(128)/r(1)\le10^{-3}$, with $r_{\rm hi}$ the
upper end of the final bisection bracket. \emph{B2}: full CROWN keeps $r(64)/r(1)\ge0.1$ on at least 80\% of the documents with
complete radii at both distances (Wilson 95\% interval reported); for the GRU, B1 and B2 use the better update form per document.
\emph{B3} (Mamba, S4D-type, minGRU-type): CROWN-IBP, forward mode and CROWN each keep a median $r(64)/r(1)\ge0.1$ over complete
pairs. \emph{B4}: the median ratio of CROWN's certified radius to the box's at $d=1$ is at least 10 for Mamba and 2 for the other
two, and its median change from $d=1$ to 64 at most threefold. \emph{B5} (seed~0): the exponent of the time of one evaluation in $d$
over $d\ge16$ is at least 1.5 for CROWN and at most 1.3 for CROWN-IBP and forward mode. \emph{C1} (complex-diagonal control, each
seed; six parts): boxes and CROWN-IBP have median $r_{\rm hi}(128)/r(1)\le10^{-3}$; disks keep a median $r(1024)/r(1)\ge0.1$;
CROWN keeps $r(128)/r(1)\ge0.1$ on at least 80\% of the complete pairs; the median time of one disk bound at $d=128$ is at most a
tenth of that of one CROWN bound; and the disk's time grows with exponent at most 1.3 in $d$. \emph{A1} ($L=0$, documents
0--199): CROWN-IBP, and IBP and CROWN-IBP together, certify at most 3 points more than IBP; n/a where IBP is within 3 points of the
clean accuracy. \emph{A2} (documents 0--49 that every method bounded at $L=0$ and 1024 and the model classified correctly at both):
for Mamba and the minGRU-type model, where IBP's certified accuracy falls by at least 10 points from $L=0$ to 1024, the best
auto\_LiRPA certificate falls by at least half as much; for the S4D-type model, the best auto\_LiRPA certificate at $L=1024$ keeps
at least 85\% of its value at $L=0$ (a 90\% document-bootstrap interval is reported). \emph{A3}: at $L=1024$ the median time of one CROWN-IBP bound is at
least five times that of one bound of auto\_LiRPA's IBP, and its exponent in $T$ is larger by at least 0.5. \emph{A4} (seed~0; the
first 30 test documents of at most 64 tokens, $L=0$): full CROWN completes on at least 80\% within 30 minutes per bound. \emph{A4L} (a
doubling sweep on two documents): full CROWN does not finish on at least 80\% of the cells with $T\ge512$. \emph{A4E}: its time
grows with exponent at least 1.5 in $T$ over the completed sweep cells, at least three lengths. \emph{A5}: $\alpha$-CROWN adds in
median at most a fifth of CROWN's median gain over IBP. \emph{N1} (natural models, $L=0$): CROWN-IBP certifies at most 1\% and
CROWN at most 5\% of the documents.
\emph{Outcome:} V PASS on every group (the standard-IBP models reproduced the certified fractions of the first training study
exactly, Mamba and the S4D-type model in a separate re-evaluation of their checkpoints with the same code).
B1 PASS on every seed (median $r_{\rm hi}(128)/r(1)$ at most $4.7\times10^{-11}$ in the better GRU form and
$2.1\times10^{-10}$ in the one-product form, upper bounds set by the end of the search range); B2 PASS on every seed and in both
GRU forms (20 of 20 documents, Wilson interval $[0.84,1.00]$); B3 and B4 PASS on the S4D-type and minGRU-type models (CROWN 19.8
and 44.0 times the box at $d=1$, changes 0.23 and 1.44), n/a on Mamba, whose CROWN bound became non-finite in 39 of 40 cells; B5
PASS on GRU and LSTM (CROWN 1.96 and 1.95; CROWN-IBP and forward mode 0.86--0.98), FAIL on Mamba, the S4D-type and the minGRU-type
model (CROWN 1.48, 1.38 and 1.43); C1 PASS, all six parts, on both seeds. A1 PASS on the three standard-IBP architectures: on
the three minGRU-type seeds (CROWN-IBP 17.5, 28.0 and 33.5\% against IBP's 59.0, 56.5 and 46.0\%), on the three S4D-type seeds
(21.5, 48.0 and 25.0\% against 44.0, 50.5 and 38.5\%) and on Mamba's seeds 0 and 2 (9.0 and 9.0\% against 46.0 and 51.0\%; seed~1
n/a, with 1.5 points of headroom); IBP and CROWN-IBP together certified exactly IBP's documents on every seed. A2 PASS on the
minGRU-type model (IBP fell 27.3, 25.0 and 25.8 points, the best auto\_LiRPA certificate 21.2, 38.9 and 32.3; the interval of
seed~0 includes 0) and on the S4D-type model, where the best auto\_LiRPA certificate kept at least 85\% of its $L=0$ certificates
at $L=1024$ on seeds 1 and 2 (87.0 to 100\% and 63.6 to 59.1\%; seed~1 is close to a constant classifier and the interval of
seed~2 includes 0) but not on seed~0 (27.8 to 16.7\%); n/a on Mamba, where IBP fell by less than 10 points on seeds 0 and 1 (7.7
and 0.0) and the prediction held on seed~2 (IBP fell 20.0 points, the best auto\_LiRPA certificate 28.6). A3 FAIL on every seed
(ratios 1.69, 1.84--1.86 and 1.52--1.54 on the minGRU-type, Mamba and S4D-type models; exponents 1.03--1.04 against 1.01--1.02,
1.12--1.14 against 1.11--1.12 and 0.99--1.01 against 0.98--0.99). A4 PASS on the standard-IBP S4D-type and minGRU-type models
and the natural GRU, LSTM and S4D-type models (30 of 30) and the natural minGRU-type model (29 of 30; the 30th returned no finite
bound, which is not assessable), FAIL on the standard-IBP Mamba (5 of 30; the other 25 timed out), n/a on the natural Mamba (no
finite bound); A4L PASS on the natural LSTM and Mamba and the standard-IBP minGRU-type model (a 30-minute time-out at $T=250$),
the standard-IBP S4D-type model (at $T=506$) and the standard-IBP Mamba (at $T=58$), every longer cell skipped, none attempted;
A4E PASS on the natural LSTM (2.04) and the standard-IBP minGRU-type model (2.06), over $T=58$--128: four bounds at two sweep
lengths, ``three lengths'' having been implemented as three distinct values of $T$, and on the standard-IBP S4D-type model (1.92
over $T=58$--256, six bounds at three lengths); n/a on the natural and the standard-IBP Mamba (no completed sweep cell); A5 n/a
on every model ($\alpha$-CROWN returned no bound); N1 PASS (no verifier certified any document of a natural model). Forward mode missed the clean
margin at a zero-width input on the gated models from $d=32$ and on the complex-diagonal control from $d=64$, and 23 of its
zero-width bounds lay above it (\cref{tab:lirpafwd}).

\paragraph{Training remedy for the blind spot (Gate 8; fixed 2026-10-27, before the first training run; frozen by its hash).}
Word-level IMDB under the synonym threat of the first training study, two-layer width-64 Mamba, minGRU-type and S4D-type models
(\cref{app:remedy}). The last 2500 training documents were held out as a development split; the test documents were read only by the
evaluation. A selection stage chose the recipe of the main stage among four schedules on the development split (seed~0) by
a rule fixed in advance: Mamba and the minGRU-type model each at least 82\% clean, then the highest mean certified share; it chose
the linear-ramp recipe (after \citealp{jia2019}), to which the criteria below refer. The main stage trained seeds 1--3 of
every arm and evaluated them on the first 1000 test documents. A checkpoint is degenerate if its clean accuracy is below 60\% or it
predicts one class for more than 80\% of the documents. Clean and $L=0$ shares average every seed, everything else the
non-degenerate seeds, with 95\% t-intervals over the seeds. The specification, the summary script and the run plan were frozen by
their hash before the first selection-stage job, together with the outcome of the smoke test's rule for a weight ramp of the hull term (not
triggered). \emph{V1--V7} (validity): the smoke test passed; every planned run is complete with its planned arguments; the cap holds
on every capped checkpoint; no per-document implication is violated on 87 checkpoints (a document certified K-uniformly is certified
at $L=0$, after one and four copies and at $L=\infty$; a certified document is classified correctly at that context); the checks of
the token-set certificate and of the attribution are clean; training health is reported (no run flagged); and the re-evaluated
Mamba and S4D-type checkpoints of the first training study reproduce its counts within two documents. \emph{P1}: the linear-ramp
recipe has a mean clean accuracy of at least 80\% and no degenerate seed. \emph{P2}: its mean certified share at $L=0$ is at least
50\%. \emph{P3a}: it keeps at most half of its $L=0$ certificates at $L=\infty$ (mean ratio). \emph{P3b}: resetting the channels
slower than 512 tokens restores at least half of its width failures (pooled, at least five). \emph{P4a, P4b}: the cap at
$\tau_{\max}=128$ keeps at least 0.8 of the width-only $L=\infty$ certificates and at least 0.7 of the K-uniform ones, on two of
three seeds. \emph{P5}: it raises the mean $L=\infty$ ratio over the linear-ramp recipe by at least $\max(0.10, 2\,\mathrm{SE})$.
\emph{P6}: the fixed-point term from the start (the $L=\infty$ term from the first robust step) keeps at least 0.8 at $L=\infty$ on
two of three seeds, at a mean clean accuracy at most 3 points below the linear-ramp recipe's. \emph{P7} (headline): the cap at
$\tau_{\max}=512$ with the hull term keeps at least 0.8 for every number of copies, with a clean accuracy of at least 78\% and an
$L=0$ share at most 5 points below the linear-ramp recipe's, on two of three seeds. \emph{C1, C2}: paired by seed, this combination
exceeds the cap alone (C1) and the hull term without the cap (C2) by at least 0.10 in that ratio on two of three pairs.
\emph{P8a--c}: the combination is certified for every context of every length over the 1, 8 and 64 most frequent neutral tokens,
with the document in its substitution box, on at least 25, 10 and 5\% of the documents (the last also more than the linear-ramp
recipe). \emph{P9}: the S4D-type model keeps at least 0.8 at $L=\infty$ under the linear-ramp recipe. \emph{P10}: the first recipe,
rerun, reproduces the clean, $L=0$ and $L=\infty$ shares of the first training study within 5 points (n/a with fewer than two
non-degenerate seeds on a side). A reading table, fixed with the criteria, states which sentence each outcome allows;
\cref{sec:exp-train,app:remedy} follow it. \emph{Outcome:} V PASS. minGRU-type: P1--P7, C1 and P8a PASS (the headline at 83.7\%
[80.3, 87.1] clean, 72.2\% [69.6, 74.9] K-uniform, one copy 73.8\% [69.9, 77.7], 1024 tokens 73.8\% [70.9, 76.6]; with the cap at
$\tau_{\max}=512$ this is close to a one-period certificate); C2 FAIL (the hull term without the cap met P7's rule as well); P8b,
P8c FAIL (0\%); P10 FAIL (the rerun first recipe's $L=0$ share differs by $-$8.8 points, Welch interval [$-$22.7, 5.1]). Mamba: P1,
P2, P4a, P4b, P6 and P8a PASS; P3a FAIL (the linear-ramp recipe keeps 78.5\% [27.3, 129.6] of its certificates at $L=\infty$, per
seed 54.8, 92.5 and 88.1\%); P3b FAIL (36.2\% of 47 width failures restored); P5 FAIL ($+$20.1 points, Welch interval [$-$31.1,
71.3]); P7 FAIL on its accuracy clauses (only one seed of three meets them, while every seed keeps at least 0.98 for every number of
copies); C1 PASS (paired differences: K-uniform ratio $+$14.8 [0.8, 28.8] points, K-uniform share $+$1.7 [$-$23.1, 26.4],
clean accuracy $-$10.5 [$-$28.4, 7.3]; the reading table reads C1 and C2 only next to a passing P7); C2, P8b, P8c FAIL; P10 n/a.
S4D-type: P9
PASS (93.8\% [89.9, 97.6]); P10 n/a.

\paragraph{Floating-point re-verification (floating-point track; fixed 2026-10-28, before the production run).}
The outward-rounded evaluator of \cref{rem:float} on 54 checkpoints: the first training study (1000 documents; every context length,
$L=\infty$ and the hull), the token-set certificate (the documents of \cref{app:anyctx}, 200 in almost every arm; $|S|=1$, 2, 4 and 8
at $P=8$ and $|S|=1$ at $P=64$, in both modes, and the $L=0$ ceiling), the certificates at $\eps=10^{-8}$ of the context-length
probe (its 190 or 200 documents) and the box radii of the distance probe (on its archived documents), each at the reported radius. The criteria, a reading table, the thresholds of the summary script, the code (by
its hash, after a smoke test that passed) and the hashes of the archived embedding tables were frozen before the production run;
the GRU and LSTM box radii had been placed outside the claim before that, when the smoke tests showed their outward enclosures
saturating. The outward evaluator runs on the documents whose float64 margin bound exceeds $-10^{-6}(1+|b|)$ (the skip rule); the
others count as not certified. \emph{Validity checks} (any failure: nothing is read): the platform has correctly rounded basic
operations, no flush to zero and library functions whose measured errors are at most 2 ulps (four outward steps where the measured
error is at most 1 ulp, eight where it is at most 2: twice the steps that enclose that error), on the production node type and in
every job, with one platform fingerprint; the float64 reference reproduces the archived certified counts of every checkpoint and
setting, up to documents with $|b|\le10^{-9}$; an unwidened mirror of the evaluator reproduces the float64 margin bound within
$10^{-10}(1+|b|)$ or within four times the outward slack, on 20 documents per checkpoint and setting (within $10^{-6}(1+|b|)$,
without the slack, on the gated box-radius rows outside the claim; not applicable where the float64 bound is $-\infty$ or where the
invariant-box search of a gated model flips at rounding level); on every evaluated document the outward bound is never above the
mirror's by more than $10^{-12}(1+|b|)$ and, as a check against over-widening, never below it by more than $10^{-6}(1+|b|)$; a
high-precision oracle on tiny models passes in every job; no sampled input lies outside an outward box, 64 per document on the same
20 documents (not applicable to the fixed point at $L=\infty$, which bounds a limit); every planned document is present; where both
routes give boxes, they meet. \emph{Predictions:} at finite contexts every certificate is kept, apart from documents within
$10^{-6}(1+|b|)$ of zero, which are counted, and the largest margin change is at most $10^{-8}(1+|b|)$; the same at $L=\infty$ and
for the hull; certificate checking passes where the slowest channel's time constant is at most $10^8$ tokens, may fail above
$10^{11}$, and needs more inflation the slower the channel (rank correlation at least 0.5 over the pairs of document and layer with
a time constant of at least $10^5$ tokens); for the minGRU-type layer, the earlier decay path, the direct path in float64 and the
outward evaluation certify the same documents, and the two float64 bounds agree within $10^{-9}(1+|b|)$; for token sets, the same as
at finite contexts; for the probes, every certificate at $\eps=10^{-8}$ and every reported box radius inside the claim is kept. The
reading table states which sentence each outcome allows (for example, a cell that changes by more than 0.1 point prints its
re-verified value); \cref{rem:float} follows it. \emph{Outcome:} every validity check passed under the amendment of
\cref{app:prereg-amend}; under the registered implementation the check against over-widening failed on one document, so nothing
would have been read. Five predictions held: at finite contexts (largest margin change $6.2\times10^{-12}$), at $L=\infty$ and for
the hull ($1.6\times10^{-11}$), for certificate checking (every pair of document and layer up to $2.5\times10^9$ tokens passed, one
failure, at $1.6\times10^{14}$; rank correlation 0.78 over 3548 pairs), for the minGRU-type layer ($9.5\times10^{-13}$) and for the
probes ($3.7\times10^{-7}$). The prediction for token sets failed ($9.3\times10^{-8}$, at $|S|=8$ in both modes for the fixed-point
fine-tuned S4D-type seed~1, which kept all its certificates there). No certificate inside the claim was lost, and one was gained
(\cref{rem:float}).

\paragraph{Floating-point re-verification of the training remedy (floating-point track; fixed 2026-10-31, before any of its runs; inputs frozen before the production runs).}
The outward-rounded evaluator of \cref{rem:float} on 39 checkpoints of \cref{app:remedy}: Mamba and the minGRU-type model with the
linear-ramp recipe, the cap at 128 and at 512, the fixed-point term, the cap with the hull term and the hull term, and the S4D-type
model with the linear-ramp recipe, three seeds each, at every context length, after one and four copies of the period, at
$L=\infty$ and for the hull (1000 documents), also at box scale 0 for the minGRU-type models with the hull term and for both models
with the cap at 128, and the token-set certificates of the six models with the cap and the hull term over one token (200
documents, both modes); the cap at 128 forms a separate part, read by the same rules. The specification with its decisions, the thresholds of the summary script and the code were frozen by their hashes
before the smoke test, the embedding tables, the platform record, the manifest of the archived outcomes and the shard limits after
it, before the production runs; the code of the earlier record stayed frozen. The float64 reference runs on CPUs; the archived
outcomes were computed on GPUs. \emph{Validity checks} (any failure: nothing is read): those of the earlier record, with a
document-by-document comparison of the reference with the archived outcome (agreement except on documents with $|b|\le10^{-9}$ or
with $|b|$ within four times the outward slack and within $10^{-6}(1+|b|)$), and five more: in the outward evaluation of every capped
model every time constant is at most $P/(1-\rho^P)$ (within $10^{-9}$ relative); the constants of the certified network are those of
each checkpoint; the data and the archived outcomes are the recorded ones; the code of the earlier record, with its frozen
specification and the hashes of its embedding tables, is unchanged at every submission and in the summary; and the summary of
\cref{app:remedy}, recomputed from the re-verified outcomes, passes its inputs check, with every violated per-document implication
explained by rounding.
\emph{Predictions:} at finite contexts, at $L=\infty$ and for the hull, for the token sets and at box scale 0 every certificate is
kept, apart from documents within $10^{-6}(1+|b|)$ of zero, which are counted, and the largest margin change is at most
$10^{-8}(1+|b|)$; on the capped models it is at most $10^{-10}(1+|b|)$ (secondary); on the capped models certificate checking passes
at an inflation of at most $10^{-9}$ on every pair of document and layer, and on the uncapped ones as in the earlier record
(secondary); no document of the comparison is tolerated and the embedding substitution flips no verdict; and every number of
these rows whose inputs were all re-verified, and the verdict of every criterion that reads only such inputs, is unchanged when
recomputed. The reading table states which sentence each outcome
allows; \cref{sec:exp-train,rem:float,app:remedy} follow it. \emph{Outcome:} every validity check passed, after ten tasks that had
stopped at their time budget were resumed as registered (a summary run before that failed only the completeness check; the paper
uses none of its results); the cap-128 part, submitted later with the same frozen hashes, completed in one submission; nothing was
amended. No certificate changed: 208,020 were kept and none was gained. Seven predictions held
(largest margin changes at most $1.5\times10^{-11}$ at finite contexts, $2.5\times10^{-11}$ at $L=\infty$ and for the hull,
$2.1\times10^{-11}$ on the capped models, $3.1\times10^{-12}$ for the token sets and $2.1\times10^{-11}$ at box scale 0; no
tolerated document and no flip; every printed number and verdict unchanged). The one for certificate checking failed on its capped
part: 61,232 of the 83,364 pairs of document and layer passed at an inflation of at most $10^{-9}$; no claim depends on it. The
certificates entering P2, P3a, P4a, P4b, P5, P6, P7, C1, C2, P8a and P9 were re-verified; the counts of \cref{tab:tests} also
include P3b and P10, which read a diagnostic and the first recipe.

\paragraph{Extension to 200 documents and intervals (fixed 2026-10-31, before any of its runs; code frozen after its smoke test, and again after an amendment).}
The extension of \cref{app:cleanctx}: clean accuracy at every context on the 1000 documents of the first training study (53
checkpoints) and on the 200 of the context-length probe (25 checkpoints); the distance probe (four natural architectures, three seeds)
and the growth rates (17 pairs of model and seed) to 200 documents per seed (the $B=2048$ zonotope of the nonnegative diagonal models
on seed~0: 50), the budget sweep to 50, and the completion of the two context-length-probe rows and the twelve token-set arms that had
stopped short of 200; and, descriptive, the growth rates of two
certified-trained gated checkpoints. Every new document continues the archived order and draw with the archived options, so the
archived documents are a prefix of each extended set, and the original summary scripts, run with their archived invocations, read the
extended sets and, separately, the new documents. \emph{Validity checks} (a failure of one of the first three stops every reading, of
any other the reading of its experiment): every task checks the code, the data, the checkpoints and the archived outputs against a
hash manifest; the draws and the options equal the archived ones; the summary reproduces every value of the archived summaries that it
recomputes; archived documents recomputed by each extension match field by field (discrete fields exactly, bounds and margins within
$10^{-12}$ relative on the archived type of CPU, certified radii inside the new bisection bracket; floating-point differences on the
other type are reported); every planned document is present exactly once; the experiments' own validity checks pass on the new
documents; the clean runs reproduce the archived clean counts and agreement flags, up to documents with a clean margin within
$10^{-9}$ of zero; the per-document implications between a certificate and the clean prediction hold; and a second resampling seed
checks the bootstrap (reported, never a stop). \emph{Predictions:} (i) every criterion of the extended tests except criterion G of
the token-set certificate, which is left without a prediction and reported paired on the common documents, keeps its archived
outcome on the extended sets and, where the new documents alone meet the criterion's minimum, on them; (ii) the new documents agree
with the archived ones: of 164 listed statistics of the distance probe, the growth rates and the budget sweep, at most 16 have a 95\%
interval of the difference between the new and the archived documents that excludes 0 (about 8 are expected under exchangeable
documents); (iii) the universal statements of the paper hold on every new document (the gated box certifies no radius from $d=128$ on,
the $B=2048$ zonotope certifies a positive radius at $d=1024$, the slowest nonnegative diagonal channel's mean log decay is negative on
every segment, the $L=\infty$ box of the completed probe row is finite at $\eps=10^{-8}$, every byte-level token-set row stays 0); (iv)
the median gated document has $\lambda_J<0<\lambda_M$, with 95\% intervals excluding 0, in every layer of every seed, on the extended
sets and on the new documents alone, and the Newcombe 95\% interval of the difference between the new and the archived documents in
the share of the seed-0 LSTM's first-layer documents with $\lambda_J>0$ contains 0; (v) at most 10 of the 1000 documents change their clean prediction between $K=16$ and
$L=\infty$ on each natural Mamba, S4D-type and minGRU-type checkpoint; (vi) the K-uniform hull at box scale 0 certifies the clean
prediction of at most 10 of the 1000 documents of each natural Mamba and minGRU-type checkpoint. A reading table fixed with the
predictions states which sentence each outcome allows and keeps the archived verdicts as the pre-registered outcomes; \cref{app:cleanctx}
and the extended numbers of the paper follow it. \emph{Outcome:} every validity check that can stop a reading passed under the two
amendments of \cref{app:prereg-amend}; as registered, two checks failed, so nothing would have been read. All six predictions held:
(i) every criterion that it predicted kept its archived outcome on the extended sets (200 documents per cell, 50 for the budget sweep, the archived documents
included; 162 criterion outcomes) and on the new documents alone (145 outcomes of the distance probe, the growth rates and the budget
sweep); (ii) on 159 of the 164 statistics (the five exceptions: \cref{app:cleanctx-record}); (iii) and (iv) without exception; (v)
and (vi) with 0 documents on every checkpoint. Criterion G of the token-set certificate, which the extension did not predict and
reported paired on the common documents, holds on the completed arms (200 documents; 61.5\% [54.6, 68.0] at both periods for the seed
that failed); its failure above stays the pre-registered outcome. On the extended sets the S4D-type share of Z2b at $d=128$ is
not decisive at $n=200$ on seeds 1 and 2 (74.5\% [68.0, 80.0] and 85.5\% [80.0, 89.7], against 80\% required); the extended values
that the paper prints are in \cref{tab:ci,tab:ci-b}, with their intervals.

\paragraph{Where the information goes (evidence probe on the training remedy; fixed 2026-11-02, before any run of its driver; code and data frozen after its smoke test).}
Every checkpoint of the training remedy's main stage (\cref{app:remedy}; 78 checkpoints) and the four archived references of the
first training study, read on the 1000 test documents followed by $d\in\{64,128,256,512,1024,2048,4096\}$ trailing licence words, on
the document, on another document with the same words and on the words alone, in float64 (82 outputs). The quantities are clean
accuracies and their differences: the carry (accuracy minus the other-document rule's accuracy, with a paired bootstrap interval
over the documents), the agreement of the two predictions, the largest change of the margin, and the per-seed classes E, N, P and D
(\cref{app:memory}, ``Quantities''); means over the non-degenerate seeds with 95\% t-intervals, paired contrasts with paired
t-intervals, an unpaired one with its Welch interval. The specification's criteria section, the summary script and the run plan were
frozen by their hashes on 2026-11-02 (section 5: 4abf834f, summary script 7999cb04, plan 2ab20133), the code and data manifest after
the smoke test on 2026-11-03 (12539a02), with the evaluation's time limit of 10 minutes taken from the smoke test; the production
manifest equalled the frozen one. \emph{V1--V7} (validity; V2 reported; a failure of V3--V7 stops every reading): every planned output is complete; the
draw equals the training remedy's; the bare documents reproduce its counts within two documents; the four references reproduce
their archived CPU runs on every document; the configuration equals the manifest; the outputs are internally consistent.
\emph{F1}: on every cap-128 checkpoint (12), after 4096 words the two predictions agree on at least 99.9\% of the documents and the
largest change of the margin is at most $10^{-4}$. \emph{F2}: in each of ten cells (cap 128 at 1024 words; cap 512, cap + hull and
natural with cap 512 at 4096 words; both architectures), at least two of three non-degenerate seeds are class P. \emph{F3}: in each
of four units (cap 128 against the linear ramp, natural with cap 128 against natural; both architectures), applicable when the
uncapped arm's mean carry after 256 words has an interval above 0, the capped run's carry after 256 words is below its partner's in
at least two of three pairs and the mean paired difference is at most $-$2 points. \emph{F4}: in each of nine cells (the linear ramp
for Mamba, the S4D-type and the minGRU-type model; the fixed-point term, the hull term and natural training for Mamba and the
minGRU-type model), at least two of three non-degenerate seeds are class P or D at 4096 words. A FAIL is read by its class
majority; a reading table fixed with the criteria states which sentence each outcome allows, and \cref{sec:exp-train,app:remedy}
follow it. \emph{Outcome} (2026-11-03): V1--V7 PASS; F1 PASS (12 of 12; largest agreement deficit 0, largest change of the margin
$8.65\times10^{-12}$); F2 PASS (10 of 10 cells, 3 of 3 seeds each); F3 PASS (4 of 4 units, 3 of 3 pairs each); F4 PASS in 8 of 9
cells and FAIL, class E, for the linear-ramp minGRU-type models (2 of 3 seeds class E at 4096 words). No amendment. The same probe
then ran on the twelve checkpoints of the zero-order-hold control (2026-11-03; read by no criterion; its own summary and supplementary
manifest recorded).

\paragraph{Zero-order-hold S4D-type control (fixed 2026-11-02, before any run; code and data frozen after its smoke test).}
The S4D-type block of \cref{app:remedy} with the standard zero-order-hold injection $\bar B=\frac{1-a}{\alpha}B$ in place of our
unnormalised $xB$, matched initialisation and evaluation after conversion to our layer's form (\cref{app:remedy}, ``A
zero-order-hold S4D-type control''): the linear-ramp recipe alone (A1z, paired by seed with this study's S4D-type linear-ramp runs),
the hull term from the start on it (A5z), and natural training on the zero-order-hold layer (Nz) and on ours (Nu); seeds 1--3, the
training remedy's statistics and degeneracy rule, the hull term's weight-ramp rule applied to A5z (not triggered). The specification's
criteria section, the summary script and the run plan were frozen by their hashes on 2026-11-02 (section 7: cc209147, summary
script 0cee9898, plan 94cc11ad), the code, data and archive manifest after the second smoke test on 2026-11-03 (8518c980); the time
limits of the jobs were not pre-registered objects. \emph{V1--V9} (validity): the smoke test passed; completeness (reported); the
conversion reproduces $\bar B$ entry by entry on every checkpoint; no per-document implication is violated; the token-set and
diagnostic checks are clean; training health (reported); the reference rows equal this study's summary; the census equals the
diagnostic's time constants; the initialisation census was recomputed. \emph{Comparable accuracy}: no degenerate A1z seed and a
mean clean accuracy of at least 76.3\%. \emph{Z1}: in at least two of three non-degenerate seeds A1z develops a largest time constant
above $\max(1000,\text{its initial value})$ and a layer with at least 1\% of its channel--state pairs slower than 4096 words.
\emph{Z2}: our layer's $L=\infty$ ratio exceeds A1z's by at least 0.10 in at least two of three seed pairs. \emph{Z3}: resetting
A1z's channels slower than 512 words restores at least half of its width failures (pooled, at least five). \emph{Z4}: A5z keeps at
least 0.8 of its $L=0$ certificates K-uniformly at a clean accuracy of at least 78\% and an $L=0$ share within 5 points of A1z's mean,
in two of three seeds; its remedy reading needs A1z's ratio below 0.8 in those seeds. A reading table fixed with the criteria states
which sentence each outcome allows, with every contrast ``at fixed $B$''; \cref{app:remedy} follows it. \emph{Outcome}
(2026-11-03): V1--V9 PASS (V2 and V6 reported; every A5z run 20{,}999 steps with no excluded robust row); comparable accuracy PASS
(81.2\% [80.8, 81.6]); Z1 FAIL (0 of 3 seeds); Z2 FAIL (0 of 3 pairs); Z3 FAIL, vacuous (0.0 of 18 width failures restored: no A1z
channel is slower than 512 words); Z4 PASS (3 of 3 seeds) without its remedy reading (A1z's ratios 0.880, 0.853 and 0.850). No
amendment.

\paragraph{A task that needs memory (fixed 2026-11-03, before any smoke or production run; code and data frozen after its smoke test; phase 2 frozen separately on 2026-11-05).}
Word-level IMDB under the synonym threat with a gap of up to 1024 neutral words and a marker between the review and the prediction
(\cref{app:memory}): the natural arms (no cap; caps 64, 128, 512, 2048), plain IBP (the linear-ramp recipe), cap 128, cap 2048, the
fixed-point term, cap 512 + hull and hull, seeds 1--3, two architectures; the training remedy's recipes, statistics and degeneracy
rule; the development split read only by the stage rules; the test documents read only by the evaluation. The specification's
pre-registration section (definitions, stage and smoke rules, criteria, classification, validity, reading table), the summary
script and the run plan were frozen by their hashes on 2026-11-03 (section 9: e84b6eb4; summary script d7444132; plan 3dd7686f; the
criteria table's own hash 07175992 is printed by every summary). \emph{Stage rules, fixed with them}: a smoke test on an H200
node sets the gap range (1024 words unless a memory class exceeds 122 GB, then 512), the evaluation chunk, the node class per
memory class and the hull term's weight ramp by rules on its measurements; a natural stage reads the premise (M1, below) and, if it
fails at 4000 steps, forks to 21{,}000-step reruns and stops the architecture if it fails again; two pilots of plain IBP must pass a
health bar before the IBP arms train; the core tier (every arm) applies if the last natural-stage summary starts by 2026-11-20.
The smoke test of record (job 532418) passed on 2026-11-04 and froze the code and data manifest (a14f6822), its rules file (24cd63a8)
and the data hashes of both gap ranges (358b8572, 25179696): gap range 1024, evaluation chunk 64, the class with the largest peak on
H200 nodes, no weight ramp. Two earlier smoke jobs were superseded before any result existed: their memory probe measured a cache
pool inherited across evaluation calls (the allocator's reserved peak was reset to the pool the preceding evaluation had left, so a
training window of plain IBP reported 122.6 GB reserved against 36.9 GB allocated, which triggered the fallback rule; the second
job then failed on a naming error in the smoke's own driver step for the fallback range); the probe was corrected to empty the cache
before each reset (2026-11-03), no threshold or rule changed, and the three frozen objects kept their hashes. The production manifest
(aa0bb191) differs from the smoke's in one line, the submission script: its first stage-P submission had stopped at a wrong chain
dependency before any task ran, the one-line fix touched no file that a task executes, and the cluster rebuilt and verified the
manifest at submission. \emph{V0--V10} (validity; V2 and V6 reported): the driver reproduces the training remedy's archived
evaluations document by document on six checkpoints; the smoke test passed; completeness; the cap holds on every capped checkpoint;
no per-document implication is violated; the draws equal the registered ones; training health; the data identity (gap range,
variants, data seed, marker, data hashes); the configurations; the arguments; the output paths and flags of every task line.
\emph{M1} (premise): the natural models' mean development accuracy after the longest gap is at least 80\%. \emph{M2} (theorem
check): on every checkpoint with a cap of 64 or 128, after 4096 trailing words the two predictions agree on at least 99.9\% of the
documents and the largest change of the margin is at most $10^{-4}$. \emph{M3}: in each of three cells (natural with caps 64 and 128,
IBP with cap 128), at least two of three non-degenerate seeds are class P after 1024 words. \emph{M4a}: the cap 128 keeps a
K-uniform ratio of at least 0.7 at gap 0 in two of three seeds. \emph{M4b}: plain IBP's mean K-uniform ratio at gap 0 is at most
0.5. \emph{M4c}: plain IBP's mean accuracy after 1024 words is at least 70\%. \emph{M5} (headline; read for the fixed-point term
and for the hull term, n/a unless M4c passes): in at least two of three usable pairs with plain IBP of the same seed, the arm
meets the headline bar of the training remedy at gap 0 (ratio $\ge0.8$, clean accuracy $\ge78\%$, $L=0$ share within 5 points of
plain IBP's mean), reaches at least 70\% after 1024 words, is at most 5 points behind plain IBP in accuracy and in carry after
1024 words, and forgets at most 5 points more between gap 0 and gap 1024. \emph{M7}: natural accuracy after 1024 words is at least
15 points higher with the cap 2048 than with the cap 128. \emph{C1}: the drop from gap 0 to gap 1024 of cap 512 + hull exceeds the
hull term's by at least 10 points in two of three pairs. \emph{Classification} (per architecture, the first label that applies):
no stage; no IBP memory baseline (M4c fails; or a remedy alone reaches the bar); plain IBP keeps both; none certifiable; FREE LUNCH (an arm that keeps the headline bar
and remembers within 5 points of plain IBP in at least two pairs); TRADE-OFF (every keeping arm at least 10 points behind, every
free arm losing its certificates); MIXED. \emph{M6} and phase 2 are reported without a prediction. A reading table fixed with the
criteria states which sentence each outcome allows, with every K-uniform number at gap 0 next to its shares and every per-seed rule
``in $k$ of 3 seeds''; \cref{sec:exp-train,app:memory} follow it. \emph{Outcome}: natural stage (summary of 2026-11-04 08:09,
515d9fce): M1 PASS for both architectures, no fork, no stop, both pilots PASS, V0 PASS, core tier. Main stage (2026-11-05): V0--V10
PASS, readable; 68 of 68 runs; M2 PASS (both); M3 FAIL, class E, in all three Mamba cells; M3 PASS for the minGRU-type natural cap 64
and IBP cap 128, FAIL, class E, for its natural cap 128; M4a, M4b, M4c PASS (both); M5 FAIL for the fixed-point term (both
architectures) and for the hull term on the minGRU-type model, n/a on Mamba (one usable pair; two of its three hull seeds and one cap
512 + hull seed are degenerate); M7 FAIL (both; Mamba's Welch interval excludes 0, so the difference is reported ``below the
pre-registered bar''); C1 FAIL on the minGRU-type model (its paired interval excludes 0, reported likewise), n/a on Mamba;
classification FREE LUNCH for both architectures, through cap 2048, with cap 128, the fixed-point term and cap 512 + hull in memory
dead zones on Mamba and cap 512 + hull and the hull term on the minGRU-type model (summary 9d24846d). \emph{Phase 2} (the sandwich
certificate, evaluation only, no criterion): its manifest (f2adda63; the production manifest plus three phase-2 files) and driver
rule (007053e1) were frozen after its smoke test on 2026-11-05; three earlier phase-2 smoke jobs had failed only in a negative
control of the unit test (the control lacked the power to detect a wrongly started trailing hull; every check of the certificate
itself passed from the first job), the control was rewritten three times, the last two at the state level, and the certificate code gained a
recording hook, with no change to the production manifest or the pre-registered section. Outcome (2026-11-05): the containment
check and the reduction check passed on the 36 checkpoints (0 violations; 0 of 36{,}000 documents differ); the quantities are
reported in \cref{app:memory} (summary 665d41cf).

\subsection{Amendments and corrections}\label{app:prereg-amend}

\paragraph{Context-length probe, Amendment 1 (2026-10-21, before the production run).}
A smoke run on two to four documents and a static review of the code showed two design errors in the criteria. First, C2 asked the
gated models' looseness to grow tenfold from 256 to 4096 neutral tokens, or the bound to become non-finite; but the interval states
of LSTM and GRU are bounded ($|h|\le1$), so the bound never becomes non-finite, and the smoke run showed the gated boxes already
saturated on the bare document, where the growth ratio is about 1. C2 would therefore have failed exactly when the gated bound is
loosest. \emph{C2$'$} replaces it: growth\_txt $\ge10$, or a median width exponent of at most 0.5 after 4096 neutral tokens, where the
width exponent of a document is $1+\frac12\log_{10}(W(10^{-8})/W(10^{-10}))$ (1 for a width linear in $\eps$, 0 for a width
independent of $\eps$). Second, the linearity clause of C1 measured the second-order term at $10^{-8}$ (about $10^{-8}$ times the
ratio of the second- to the first-order coefficient), not linearity as $\eps\to0$; the smoke run's word-level Mamba already sat at
0.006--0.009. \emph{C1$'$} replaces that clause by a median $|\text{width exponent}-1|\le0.05$ on the $L=\infty$ box, after 256 and
4096 neutral tokens, and after one and four periods. \emph{C0$'$} adds: no non-finite bound at $10^{-10}$ on the finite-context
families; no bound above the margin at the box corner, at any $\eps$; no resolved distance band with a band looseness below
$1-10^{-3}$ among the bands with $D_{\rm band}>10^{-6}D$ (the CNN excluded, because perturbing one copy of a repeated period breaks an
exact tie of its max-pool, which legitimately fails this check); clean accuracy on the probe documents of at least 0.70 and at most
80\% of the predictions in one class; and the registered grid. The same amendment fixed two interval primitives shared by every float64
certificate (the SiLU slack and the softplus threshold; \cref{app:numerics}). The original criteria were kept verbatim and computed;
the verdict is read from C0$'$, C1$'$ and C2$'$, and the original verdict is printed next to it. Two to four of the 200 documents of the
production run had been seen in the smoke run.

\paragraph{Token-set certificate: a completion job.}
The validity check first failed because seven byte-level runs had stopped at their time budget below 150 documents. A completion
job brought each to at least 150 documents (150 to 200), the check's minimum, with the criteria unchanged.
The extension of \cref{app:cleanctx} later completed every arm to 200 documents (the four byte-level arms at $P=8$, both arms of three
GRU checkpoints and two S4D-type arms at $P=64$); on the completed arms every criterion kept its outcome except G, which holds there
(200 documents; 61.5\% [54.6, 68.0] at both periods for the seed that failed; \cref{app:prereg-criteria}, the extension's paragraph).

\paragraph{Certified radius against distance: a corrected document count.}
The criteria were fixed before any result, and the smoke test changed none. The first summary counted documents over the union of
the shards, whereas V4 requires 40 documents for every method; budget stops had left the Mamba and S4D-type shards of the box and of
the $B=256$ zonotope at 34 and 29 documents. The count was corrected to be per method, the missing documents were completed, and the
criteria were recomputed on 50 documents per method. The preliminary values on the incomplete set had been inspected and logged
before the completion; no verdict changed (Z2b of the S4D-type model at $d=512$: 45\% of the documents preliminary, 54\% final,
against 80\% required).

\paragraph{Budget sweep, Amendment 1 (2026-10-22, before the production run was submitted).}
A static review of the code found that the medians and shares of the sweep were not tied to one set of documents: if a shard
stopped early, a median over 15 documents and one over 16 could differ by more than the 0.93 tolerance of S1 through the change of
the set alone. The amendment computes $F_B(d)$, the reach $d^*_B$ and the medians of S1 and S3 over the documents complete in every
(budget, distance) cell, prints the reading over every document that carries a cell beside it, and makes the amended reading the
verdict. It was written into the specification and logged before the production run was submitted, so before any output of the
sweep or of the replications existed. One GRU document lacks the $B=16384$ cells, because its shard stopped at its time budget, so
the GRU verdicts rest on 15 documents and the LSTM verdicts on 16; the reading over every document gives the same verdicts and the
same reaches. In the extension to 50 documents (\cref{app:cleanctx}) the same document still lacks those cells, so the verdicts rest
on 49 (GRU) and 50 (LSTM) complete documents, and both readings again give the same verdicts and reaches. The replications on seeds 1
and 2 kept the criteria of the certified radius against distance and of the growth rates unchanged, except that the former requires
18 of its 20 documents.

\paragraph{Floating-point re-verification: an amendment after the production run (2026-10-28).}
A tightness (over-widening) check flagged one ill-conditioned, uncertified document; no certificate depends on it. As registered,
this check (the outward bound is not below the unwidened mirror's by more than $10^{-6}(1+|b|)$) applied to every evaluated document,
including the 20 per checkpoint and setting that are evaluated, whatever their float64 bound, for the mirror and sampled-input
checks. The flagged document (the standard-IBP minGRU-type seed~2 at $L=\infty$) had an outward bound 0.045 (relative) below the
mirror's, where the code and the mirror, two round-to-nearest evaluations of the same real bound, already differed by
$2.3\times10^{-4}$ relative; the other 20 documents evaluated for that checkpoint and setting passed, and its one certificate was
kept. Under the
registered rule nothing would have been read. The amendment, written after the production outputs had been read, judges the check
on the documents that the skip rule admits to the outward evaluation, whose certificates the re-verification decides, and reports
it on the others (3689 of them, one above the threshold: the failure above). Only the summary script changed; the evaluator and
its outputs did not. Under the amendment every validity check passes; both summaries are in the code release, and \cref{rem:float}
reports the amended reading.

\paragraph{Extension to 200 documents and intervals: an amendment before its code was frozen (2026-10-31).}
As registered, the check that the summary reproduces the archived summaries compared every text line exactly, and the first smoke test
found one difference, in no number: the context-length probe's archived verdict line says that ``the gated boxes blow up (C2)'', which
the summary script, reworded later with the same verdict, writes as ``saturate or grow''; the amendment, written before the code was
frozen, reports this listed pair of lines instead of failing and compares every number, every verdict and every other line as before,
and every summary prints both readings: under the registered rule the check fails on that line and stops every reading, under the
amendment it passes.

\paragraph{Extension to 200 documents and intervals: an amendment after the production run (2026-10-31).}
As registered, the identity check took the run line of each task log from its first line, which in the 148 logs written on the
cluster's virtual nodes is a notice of the batch system, so the check failed on their outputs, although every one of these logs ends
with exit status 0 and records no failed identity check of its own, and nothing would have been read; the amendment, written after
the production summary and before any of its results was read, takes as run line the first line in the job script's format and
looks for the host line in the eight lines after it, every other rule unchanged, and changes only the summary script and its tests,
frozen again by their hashes (under that freeze the five tasks that had
stopped at their time budget were resumed), and under it every validity check that can stop a reading passes; both summaries are in
the code release.

\subsection{What the experiments do not show}\label{app:prereg-notshown}

\begin{enumerate}
\item \emph{One certified-training result, for small models on one task.} The first training study failed its
pre-registered criteria; its numbers, and those of fixed-point fine-tuning, appear only as measurements of what the certificates
say and as tests of \cref{prop:horizon}. The training remedy (\cref{app:remedy}) met its headline criterion for the minGRU-type
model only; for Mamba the hull term's K-uniform certificates came with two of three seeds below 78\% clean accuracy, and the
capped models hold no information beyond about $\tau_{\max}$ tokens.
\item \emph{No threat-model claim from the probes.} The boxes of radius at most $10^{-4}$ of the context-length probe, the
complex-diagonal control and the growth rates, and the one-token balls of the distance probe and the budget sweep, are not threats;
neither is the byte-level radius 0.01. Under synonym substitution the natural models certify nothing, and their true robust accuracy is
only known to lie between 0\% and about 50\%.
\item \emph{No claim that $R$ is small.} For natural models the looseness is $10^2$--$10^5$ at every length; only its shape in $L$ is
claimed.
\item \emph{No theorem behind the flat profiles} of the two-layer models (\cref{rem:gate3reading} allows growth), and no claim of ``no
growth'': the ratios reached 1.06 and the band ratios 1.01; what is claimed is no geometric accumulation, with ratios within
$[0.6,1.1]$.
\item \emph{No claim that a relational domain fixes gated certification.} The reach of the zonotope grows with its budget, from 1--16
tokens at $B=256$ to at least 1024 tokens, the end of the grid, at $B=16384$, and the cost of one bound grows with the budget and with
the distance; at $B=2048$ the zonotope kept a tenth of its adjacent-token certified radius at $d=1024$ on at most 59\% of the
documents on any seed (200 per seed), against 80\% required. The exponents of the reach against the budget (2.1 and 1.8) describe a grid of seven
distances and are not a rate.
\item \emph{No claim about fixed-size domains for gated RNNs.} Boxes and the budgeted zonotopes we tried give no time-uniform
certificate for the gated models; whether some fixed-size domain does is open, and nothing here says that gated transitions preserve
none. A reach that grows with the budget is consistent with ``no fixed size suffices'' but does not prove it, and no reach beyond 1024
tokens was measured.
\item \emph{No claim that the zonotope dominates the box, none that a budgeted zonotope loses in time, and none that a relational
domain changes only the per-step constant outside the nonnegative diagonal transitions.} On the S4D-type model the $B=256$ zonotope
certified less than 0.99 times the box's certified radius on 48\% of the documents at $d=512$ on seed~0, and reached 0.99 times the
box's on only 28\% and 36\% of the documents on seeds 1 and 2 (200 per seed), a large-$\eps$ effect of its relaxations of the nonlinear operations;
on linear point-transition recursions a reduced zonotope of any budget is never wider than boxes (\cref{lem:zono}, a standard fact,
which does not cover Mamba's interval-decay scan or whole blocks);
on the gated cells the zonotope reached farther than the box (in float64; not re-verified with outward rounding, \cref{rem:float}).
\item \emph{No claim that disks are exact in time.} Their loss on complex-diagonal LTI transitions is at most $\pi/2$ in circumradius
and does not compound; it is not a bound in a readout direction (\cref{rem:pihalf}); and for selective (input-dependent)
complex-diagonal modes no constant bound holds: over the family of \cref{rem:pihalf} the limit ratio is unbounded as the angle tends
to 0, while for each fixed angle it stays bounded in $t$. Selective complex-diagonal layers are not tested.
\item \emph{No reading of $r(d)$ as a box radius.} The distance probe's $r(d)$ is a certified input radius; ``0 from $d=128$'' means that
the box certified no radius at all there, not that a state box shrank to a point.
\item \emph{One complex-diagonal architecture} in the complex-diagonal control, with two seeds.
\item \emph{The slow-channel counterfactual is not a certificate}, and flip failures are a property of the models. In the first
study, slow channels explain 92\% (Mamba) and 71\% (minGRU-type) of the width failures, not the $L=\infty$ losses as a whole:
3--48\% and 33--63\% of the failures are flips; under the linear-ramp recipe of \cref{app:remedy} the counterfactual restores 36\% of
47 width failures (Mamba) and 75\% of 141 (minGRU-type).
\item \emph{Token-set certificates only over frequent neutral tokens}, and substantial shares mainly over a single repeated token (at
the 8 most frequent tokens only the fixed-point fine-tuned S4D-type models keep up to 40\%, \cref{app:anyctx}); over the 64 most
frequent tokens no non-degenerate stateful model exceeds 5\%. The
gated rows of the token-set certificate use the invariant box (\cref{rem:invbox}) and are not instances of \cref{cor:anyctx}.
\item \emph{Seeds and scale.} Every natural word-level recurrent architecture of each probe, the complex-diagonal control aside, has
three seeds in the context-length probe (200 documents), in
the certified radius against distance and in the growth rates (200 documents per seed since the extension of \cref{app:cleanctx}; the
$B=2048$ zonotope of the nonnegative diagonal models on seed~0 has 50). Still one seed: the budget sweep (seed~0, 50 documents, 49
complete for the GRU, one perturbed token, a grid capped at $d=1024$) and the byte-level models of the growth rates; the
complex-diagonal control has two seeds. Below 200 documents remain the budget sweep; the comparison with auto\_LiRPA's linear
relaxations (\cref{app:lirpa}), where the gated models have 20 documents per seed and the nonnegative diagonal models of the distance
probe one seed with 10 documents, and full CROWN under synonym substitution ran on seed~0 only; and the attack columns of the two
training studies (100 documents; \cref{app:cleanctx-record}). All models have two layers of width 64 and none is pretrained; the
distance probe and the budget sweep use natural models only.
\item \emph{Floating point.} Every number is computed in float64, without directed rounding except in the re-verification of
\cref{rem:float} (\cref{app:numerics}).
That re-verification covers the certificates listed there, under a measured but unproved bound on the errors of five library
functions and the rest of the trusted base listed there; the others, among them the disks, the zonotopes, the comparison with
auto\_LiRPA and those of the training remedy named in \cref{app:remedy}, were not re-verified.
\Cref{prop:exact}(a$'$) is an
exact-arithmetic statement about small $\eps$ and says nothing about channels whose float64 time constant is infinite at the threat
(the slow-channel attribution).
\item \emph{Baselines.} No Transformer, no randomized smoothing, no complete verifier ($\alpha$-CROWN without branch and bound, no
$\beta$-CROWN), no POPQORN, Cert-RNN or Prover \citep{ko2019,du2021,ryou2021}; the linear relaxations that were run are those of
auto\_LiRPA (\cref{app:lirpa}); the zonotope of the distance probe and the budget sweep uses generic relaxations.
\item \emph{Cost.} The timings of \cref{app:wallclock} are one batch of 10 documents on one CPU node for the four seed-0 models; the
token-set certificate was not timed, and the plain-IBP length that costs as much as the $L=\infty$ certificate differs by model (about
1.1--2.0k tokens for Mamba and the minGRU-type model, 4.4--5.3k for the S4D-type model), so no single equivalent length is claimed.
\end{enumerate}

\section{Further architectures, and related work in full}\label{app:related}

This appendix places further recurrent layers in the framework of \cref{sec:theory-exact} (\cref{app:related-archs}) and gives
the related work of \cref{sec:related} in full (\cref{app:related-full}).

\subsection{Further architectures}\label{app:related-archs}

\begin{remark}[What \cref{def:exact} says about architectures]\label{rem:archs}
\leavevmode
\begin{enumerate}
\item \emph{State independence} makes $(T_t,u_t)$ independent of $h_{t-1}$. It is what makes the natural interval extension of a
  step, which bounds the decay and the state separately, coincide with the best step. If a decay depended on the state, the
  product $a(h)h$ would lose the dependency between its factors, the dependency problem of interval arithmetic
  \citep{moore1966,neumaier1990}, even in a scalar nonnegative channel. For $a(h)=\sigma(3+\ln9-3h)$ and
  $u\in[1/10-\eps,1/10+\eps]$, the fixed point $h^*=1$ has true derivative $a+a'h^*=63/100$, while the natural extension propagates
  radii with gain $a+\abs{a'h^*}=117/100$, so no invariant box of radius $O(\eps)$ contains $h^*$. The best step, the exact range
  of $a(h)h+u$ over an interval, would still be exact in this scalar case. SRU is an example of a state-dependent decay: its forget
  gate reads $c_{t-1}$ \citep{lei2018}. In a stack, the coefficients of layer $\ell+1$ depend on the states of layer $\ell$. That
  dependence is acyclic and creates no feedback, only, at first order, the loss of \cref{prop:loss}(b2): a loss of type (iii)
  across positions and of type (i) inside a window.
\item \emph{For boxes}, diagonality suffices (\cref{lem:boxes}(i), decays of either sign), and so does entrywise nonnegativity of a
  coupled transition (\cref{lem:boxes}(iii)); for point transitions \cref{lem:boxes}(ii) gives the exact condition. Neither
  property is necessary: the exact condition also covers the orthant-monotone transitions $DN_tD$ and, more generally,
  $D_tN_tD_{t-1}$ (\cref{rem:provisos}). The three nonnegative diagonal families have both properties. Nonnegativity is also what
  the centre-radius step (\cref{lem:cr}) and the hull (\cref{lem:hull1}) use; a real diagonal layer with negative decays would need
  a different \eqref{eq:cr} step and hull, which we do not treat.
\item \emph{Complex-diagonal LTI} transitions (S4D with complex eigenvalues, S5 and the LRU
  \citep{gu2022b,smith2023,orvieto2023}) preserve disks (\cref{lem:disks}, input-independent $\lambda$) and wrap in real boxes
  whenever $\theta\notin\frac\pi2\mathbb Z$. In a selective complex-diagonal layer $\lambda_t$ is set-valued under IBP,
  \cref{lem:disks} does not apply, and no constant disk bound holds: over the family $\lambda_t\in\{1,e^{i\vartheta}\}$ of
  \cref{rem:pihalf} the limit ratio of the composed disk radius to the hull radius is unbounded as $\vartheta\to0$. The
  complex-diagonal control tests the LTI case on one architecture with two seeds (\cref{sec:exp-r2}). In per-mode co-rotating
  coordinates the transition becomes the positive scalar $\abs{\lambda}$, and boxes are then exact in time by \cref{lem:boxes},
  which is stronger than disks. An unreduced zonotope (one generator per input per step) represents the reachable set of a linear
  point-transition recursion exactly, and order reduction cannot make it worse than boxes: by \cref{lem:zono} a reduced zonotope of
  any budget is no wider than the composed box steps in every coordinate and every linear readout, and its interval hull is exact
  wherever boxes are exact in time. So the control shows a mismatch between transition and domain, not that no box-like domain
  works.

  \emph{What the certified radius against distance shows about zonotopes on these transitions.} These are float64 results; the
  zonotope radii were not re-verified with outward rounding (\cref{rem:float}). The zonotopes of the distance
  probe never reduce the input symbols, and on the Mamba and S4D-type models their median looseness $R(10^{-6})$ was 1, to within
  1\%, at every distance they ran, on every seed (200 documents per seed; the $B=2048$ rows of seed~0: 50), against 829--1020 for the S4D-type box and
  $4.4\times10^{4}$ to $1.75\times10^{5}$ for the Mamba box on seed~0 (316--1050 and $1.08\times10^{4}$--$3.69\times10^{5}$ over the
  three seeds). On seed~0 their median
  certified radius ($B=256$) was about 110 times the box's at $d=1$ and about 30 times at $d=512$ on Mamba, and about 10 times at
  $d=1$ and 1.1 times at $d=512$ on the S4D-type model. On the S4D-type model the $B=256$ zonotope certified at least 0.99 times
  the box's certified radius on 94.5\% of the documents at $d=128$, but on only 52.5\% at $d=512$ (seed~0; on seeds 1 and 2, 74.5\%
  and 85.5\% at $d=128$, both not decisive at $n=200$ against the predicted 80\%, \cref{tab:gate5loose}, and 28\% and 36\% at
  $d=512$, with the zonotopes again exact to first order at every distance). The box's certified
  radius grows with $d$ (seed~0: median 0.00186 at $d=1$ and 0.034 at $d=512$, per document a median factor of 16.9; seeds 1 and 2:
  factors of 40.3 and 16.4), so these comparisons are made at a much larger $\eps$, where the higher-order terms of the zonotope's
  relaxations of SiLU, RMSNorm and the products bind (\cref{lem:zono} and the scope paragraph after it). The distance probe does
  not separate relaxation from reduction losses outside the linear part, and nothing in it shows a loss in time
  (\cref{app:gate5}).
\item \emph{Delta-rule} layers (DeltaNet \citep{yang2024}, Gated DeltaNet \citep{yang2025}, RWKV-7 \citep{peng2025}) have
  state-independent but non-diagonal, sign-indefinite transitions; neither \cref{lem:boxes} nor \cref{lem:disks} applies, and
  \cref{lem:boxes}(ii), which concerns point transitions, predicts wrapping in boxes only at first order (under IBP their
  transitions are set-valued). Plain DeltaNet's $I-\beta kk^\top$ has eigenvalue 1 on the complement of $k$ and does not contract
  there; the scalar gate of Gated DeltaNet restores contraction. RG-LRU, Mamba-2, minLSTM and QRNN
  \citep{de2024,dao2024,feng2024,bradbury2017} are nonnegative diagonal. Their injections differ (RG-LRU scales its injection by
  $\sqrt{1-a_t^2}$; \citealp{de2024}, Eq.~(4)), so \cref{lem:dc} and \cref{prop:horizon}(a) do not transfer automatically.
\item \emph{Gated cells} (LSTM, GRU) depend on the state through their gates and couple their coordinates through sign-indefinite
  recurrent matrices; interval arithmetic propagates radii with a matrix $M_t\ge\abs{J_t}$ built from the primitives of the cell
  (\cref{prop:loss}(c)).
\end{enumerate}
\end{remark}

\paragraph{Where the layers in use fall.}
S4 uses structured, input-independent state matrices \citep{gu2022a}; S4D, S5 and the LRU make them diagonal
\citep{gu2022b,smith2023,orvieto2023}. Mamba \citep{gu2023} and Mamba-2 \citep{dao2024} compute a diagonal or scalar decay from the
input, and RG-LRU in Griffin \citep{de2024}, minGRU and minLSTM \citep{feng2024} and QRNN \citep{bradbury2017} compute elementwise
gates from the input alone. Except in S4, the transition is diagonal and independent of the state, which these papers exploit for
convolutional or parallel-scan training; in the real-valued layers it is also nonnegative, since their decays lie in $(0,1)$ by
construction (item 4 of \cref{rem:archs}; their injections differ, so not every result transfers). Complex-diagonal layers act on
real coordinates by rotation-scaling blocks, so real boxes wrap \citep{moore1966} and can grow while every mode contracts
(\cref{lem:disks}(iii), \cref{prop:exact}(b$'$)); in disks, the domain of circular arithmetic \citep{gargantini1971}, an LTI mode
loses at most a factor $\pi/2$ in circumradius at every step, without compounding (\cref{lem:disks}), but for input-dependent
eigenvalues no constant disk bound holds (\cref{rem:pihalf}). LSTM and GRU \citep{hochreiter1997,cho2014} have state-dependent,
sign-indefinite Jacobians. Weight conditions for the stability of GRUs exist \citep{bonassi2021}, and the saturating LSTM and GRU
cells we use always have bounded invariant boxes (\cref{prop:exactac}), but stability does not make the box shrink with $\eps$, as
the contracting LSTM of \cref{prop:exact}(b) shows. The dividing line is thus whether the abstract domain can follow the transition,
not SSM against RNN, nor the sign of the Jacobian alone. \Cref{tab:archs} places the layers in use and the further ones above; only
the rows marked ``tested'' were measured in \cref{sec:exp}.

\begin{table}[h]
\caption{Recurrent layers and where the theory of \cref{sec:theory} places them, given their coefficient boxes. Regime~1: boxes are
exact in time (\cref{lem:boxes}); regime~2: boxes wrap, and disks lose at most a factor $\pi/2$ in circumradius without compounding
(\cref{lem:disks}, input-independent eigenvalues only); regime~3: the interval map can expand while the cell contracts
(\cref{prop:exact}). Last column: whether every injection carries the forgetting factor $1-a$ (\cref{lem:dc}), the condition under
which slowing a channel hides it from a finite-horizon robust objective (\cref{prop:horizon}); ``not analysed'': the injection
differs from the three we treat, and the lemma does not transfer automatically.}
\label{tab:archs}
\centering\small
\begin{tabular}{@{}>{\raggedright\arraybackslash}p{0.17\textwidth}>{\raggedright\arraybackslash}p{0.29\textwidth}>{\raggedright\arraybackslash}p{0.32\textwidth}>{\raggedright\arraybackslash}p{0.13\textwidth}@{}}
\toprule
layer & transition under IBP & regime and domain & injection carries $1-a$\\
\midrule
Mamba \citep{gu2023} & input-dependent, nonnegative, diagonal ($e^{-\alpha\Delta_t}$) & regime 1, boxes; tested & yes\\
S4D-type, real diagonal \citep{gu2022b} & input-independent, nonnegative, diagonal & regime 1, boxes; tested & no ($u_t=x_tB$)\\
minGRU-type \citep{feng2024} & input-dependent, nonnegative, diagonal ($1-\sigma(s_t)$) & regime 1, boxes; tested & yes\\
Mamba-2 \citep{dao2024} & input-dependent, nonnegative, a scalar decay & nonnegative diagonal, boxes (\cref{lem:boxes}(i)); not tested &
not analysed\\
RG-LRU \citep{de2024} & input-dependent, nonnegative, diagonal & nonnegative diagonal, boxes; not tested & not analysed (scaled by
$\sqrt{1-a_t^2}$)\\
minLSTM \citep{feng2024}, QRNN \citep{bradbury2017} & input-dependent, nonnegative, diagonal (elementwise gates) & nonnegative
diagonal, boxes; not tested & not analysed\\
S4D, complex \citep{gu2022b} & input-independent, complex diagonal: a rotation-scaling block per mode in real coordinates & regime 2:
boxes wrap for $\theta\notin\frac\pi2\mathbb Z$; disks within $\pi/2$ (\cref{lem:disks}); tested & --\\
S5 \citep{smith2023}, LRU \citep{orvieto2023} & input-independent, complex diagonal, dense input maps & regime 2: disks are sound and
keep the factor $\pi/2$, without the best-step property (\cref{rem:pihalf}); not tested & --\\
DeltaNet \citep{yang2024}, Gated DeltaNet \citep{yang2025}, RWKV-7 \citep{peng2025} & state-independent, non-diagonal,
sign-indefinite; set-valued under IBP & neither \cref{lem:boxes} nor \cref{lem:disks} applies; \cref{lem:boxes}(ii) predicts
wrapping at first order; not tested & not analysed\\
SRU \citep{lei2018} & diagonal, but the decay reads the state ($c_{t-1}$) & outside both: the natural extension loses the
dependency (item 1 of \cref{rem:archs}); not tested & not analysed\\
LSTM \citep{hochreiter1997}, GRU \citep{cho2014} & gates read $h$; sign-indefinite recurrent matrices & regime 3
(\cref{prop:exact}(b), \cref{prop:loss}(c)); tested & --\\
\bottomrule
\end{tabular}
\end{table}

A selective complex-diagonal layer, whose eigenvalues would be input-dependent, falls outside every row: its eigenvalue sets are
not points under IBP, and over the family of \cref{rem:pihalf} no constant disk bound holds (item 3 of \cref{rem:archs}).

\subsection{Related work in full}\label{app:related-full}

\paragraph{Interval bound propagation and certified training.}
IBP propagates a box through each layer by the natural interval extension of its operations \citep{gowal2018,mirman2018}; it is
cheap and differentiable and underlies most certified training. Linear relaxations are tighter \citep{wong2018,zhang2018}, CROWN-IBP
combines them with IBP \citep{zhang2020}, \citet{shi2021} shorten the warm-up of IBP training, and SABR, TAPS and MTL-IBP reduce its
over-regularization with small boxes or adversarial terms \citep{muller2023,mao2023,depalma2024}. Zonotopes, polyhedra and automatic
linear bound propagation are tighter and costlier \citep{gehr2018,singh2018,singh2019,xu2020}, and set-based training propagates
such sets during training \citep{koller2025}. \citet{mao2024} define the tightness of IBP as the ratio of the optimal box radius to
the IBP radius (in a deep linear network, the absolute value of the product of the weight matrices against the product of their
absolute values), and show that it decreases with depth and width at initialization and increases under IBP training once the
network is wide enough; \citet{krukowski2024} attribute the looseness of IBP on trained networks to the wrapping effect. Both
analyses concern depth. We study the time direction: in a nonnegative diagonal layer every state channel is an interval whose
radius contracts at the rate $\ahi$, the largest admissible decay; given its coefficient boxes the endpoint scan returns the
reachable set exactly, so boxes are exact in time (\cref{lem:boxes}(i)), and the centre-radius step exceeds the exact image of its
own previous box by $\arad\min(r,\abs{c})$ (\cref{lem:cr}(b)). The looseness $R=W/(\eps D)$ of our context-length probe is the
reciprocal of a margin-level version of the tightness of \citet{mao2024}, taken over the context length instead of depth; as
$\eps\to0$ it tends to $G/D\ge1$, with $G$ the gain of the absolute-value network (\cref{prop:loss}(a)), the context analogue of
comparing $\abs{W_L\cdots W_1}$ with $\abs{W_L}\cdots\abs{W_1}$. For the natural nonnegative diagonal models $R$ was between 508
and $1.78\times10^5$, with no geometric accumulation in $L$ (\cref{sec:exp-gate3}). \Cref{prop:horizon} concerns the horizon at
which a robust term is evaluated, whatever the relaxation.

\paragraph{Limits of interval certification.}
Networks certified with interval analysis are universal approximators \citep{baader2020}, interval-certified robustness has
fundamental limits \citep{mirman2022}, and tighter relaxations do not necessarily give better certified training
\citep{jovanovic2022}. These results concern the size of interval bounds and how training shapes it; ours concern how the bounds
evolve with the context length.

\paragraph{Stronger verifiers.}
$\alpha$-CROWN and $\beta$-CROWN \citep{xu2021,wang2021} build on CROWN and automatic linear bound propagation
\citep{zhang2018,xu2020}. They bound the network unrolled over a context of given length, so their cost grows with that length.
\Cref{app:lirpa} compares CROWN-IBP, CROWN and $\alpha$-CROWN with our certificates at finite context: on the gated cells full
CROWN keeps a tenth of its adjacent-token certified radius at distance 64, the largest distance at which it ran, at a time per
bound that grows about as the square of the distance over distances 16 to 64 on seed~0, while $\alpha$-CROWN (20 iterations, no
branch and bound) exceeded its memory limit at distance 16 there (seed~0, five documents) and gave no complete radius on the other
models; no complete verifier was run.

\paragraph{Verification of recurrent and sequence models.}
RNN verifiers unroll the sequence and relax every step: POPQORN \citep{ko2019}, Cert-RNN with abstract transformers built from
bounding planes \citep{du2021}, Prover with polyhedra \citep{ryou2021}, RNN-Guard against perturbations of many frames
\citep{zhang2023}, differential verification \citep{mohammadinejad2021}, star sets \citep{tran2023} and abstraction refinement
\citep{lin2026}. \citet{jacoby2020} replace unrolling for ReLU RNNs by inferred invariants that bound each memory unit as a function
of time up to a given horizon, and \citet{tran2023} note that the over-approximation error of invariant-based methods accumulates
over time. For word substitutions, the literature covers IBP training \citep{jia2019,huang2019}, certification of LSTMs against
programmable string transformations \citep{zhang2021}, randomized smoothing \citep{ye2020}, and growth-bound regularization,
evaluated under attack for LSTM, S4 and CNN classifiers \citep{bouri2025} and certified for LSTMs and CNNs \citep{bouri2026}.
Transformer verifiers bound the cross-position dependencies of attention \citep{shi2020,bonaert2021,wei2023} and now reach
GPT-2-sized models \citep{duong2026}. All of these address inputs of a given finite length: their cost grows with the length or,
for the invariant-based methods, the horizon is fixed in advance. Our certificates cover a context of any number
$K\in\{0,1,\dots,\infty\}$ of copies of a given period, every position perturbed (\cref{thm:inf,thm:hull}; every $K\le K_{\max}$ in
\cref{thm:hullK}), and every context of every length over a token set at the price of the box hull of its embeddings
(\cref{cor:anyctx}), at a cost independent of the context length: on our models the $L=\infty$ certificate and the hull took
0.29--2.12\,s per batch of 10 documents, 10--31\% of one plain IBP pass at $L=16384$, or about as much as a plain pass over
1.1--2.0k (Mamba, minGRU-type) or 4.4--5.3k (S4D-type) context tokens (\cref{app:wallclock}). They say nothing about other contexts
(\cref{rem:notclaimed}).
\Cref{sec:theory-exact} explains why they can be informative for nonnegative diagonal layers and need not
be for LSTMs and GRUs: given their coefficient boxes, boxes are exact in time for the former (\cref{lem:boxes}), while for a gated cell interval propagation
follows a matrix $M_t\ge\abs{J_t}$ that can expand while the cell contracts (\cref{prop:exact}(b), \cref{prop:loss}(c)). We reuse
the substitution boxes of \citet{jia2019} and apply them at every position of the context as well (\cref{sec:setting}).
Other analyses of recurrent networks extract automata from them \citep{weiss2018}, which read inputs of any length, verify them and
certify their robustness by property-directed methods \citep{khmelnitsky2021}, or verify agent-environment systems with RNN-based
agents \citep{akintunde2019}. Randomized smoothing certifies text classifiers against word substitutions \citep{ye2020} and other
text attacks \citep{zeng2023}; its certificates are probabilistic and need sampling at every input, while ours are deterministic and, for a given period or token set,
cover every context length at once.

\paragraph{Robustness, stability and verification of state-space models.}
Pure SSMs benefit little from adversarial training \citep{qi2024}, and short triggers can erase much of a Mamba model's state
\citep{lemercier2026}. Control-theoretic analyses give Lyapunov stability of Mamba's recurrence \citep{halloran2025},
input-to-state stability of selective SSMs \citep{zubic2026}, and BIBO stability of the selective scan with robustness bounds whose
terms grow linearly with the sequence length \citep{cao2026}; L2RU, LipSSM and recurrent equilibrium networks bound gains or
Lipschitz constants by construction \citep{massai2025,yoshino2026,revay2024}. The closest work is \citet{thapliyal2026}: an interval
certificate for an LTI SSM classifier under $\ell_\infty$ perturbations of every embedding, with the contraction condition
$\norm{A}_\infty<1$ enforced by a hinge penalty and a steady-state half-width bound $\delta\norm{B}_\infty/(1-\norm{A}_\infty)$ for
the state box; selection mechanisms are out of its scope, and certificates are checked at a finite length. We treat selective
scans, where the decay is itself an interval computed from the layer input, and multi-layer stacks. The contraction comes from the
architecture, since every decay is below one in exact arithmetic, not from a constraint on the weights, although it can be
arbitrarily slow (\cref{rem:timescales}, \cref{lem:uniform}). The closed-form fixed point of the whole cascade certifies the
infinite-context limit (\cref{thm:inf}), and the hull certifies every number of copies at once (\cref{thm:hull}). Unlike the bounds
of \citet{cao2026}, our boxes are bounded uniformly in the number of copies; their constants contain $1/\denh$ instead
(\cref{rem:notsay}). In our searches we found no earlier work that trains or verifies a selective SSM with interval bounds;
\citet{thapliyal2026} trains with a contraction penalty, not an IBP loss. \citet{benkish2025} explore the length extrapolation of
Mamba; \cref{prop:horizon} concerns a related effect of the training horizon, on robust objectives instead of clean predictions.

\paragraph{Interval analysis, monotone systems and invariant sets.}
IBP is the natural interval extension of interval analysis \citep{moore1966,neumaier1990}: sound and inclusion monotone, but it
loses dependencies and re-wraps boxes that the dynamics rotates \citep{moore1966,neumaier1993}, which zonotope-based methods reduce
\citep{kuhn1998,girard2005,kochdumper2021}. In box-type order reduction the dropped generators are enclosed by their interval hull
\citep{kuhn1998} after a ranking of the generators \citep{girard2005} (attributions as in \citealp{kopetzki2017}); the enclosure
keeps every coordinate radius, and a linear map widens interval radii by at most $\abs{T}$, so on linear recursions with point
transitions a reduced zonotope of any budget is never wider than boxes. We state this elementary consequence as \cref{lem:zono},
because it decides the reading of the certified radius against distance. For monotone systems interval reasoning is tight: by
M\"uller's theorem, trajectories from the corners of the box, with the inputs at their bounds, bound all others
\citep{muller1927,smith1995}, and for positive linear systems \citep{farina2000} with interval inputs interval arithmetic returns the
interval hull of the reachable set, because every coordinate attains its extremes at the same corner; the same holds, for point
transitions, for systems that are monotone with respect to an orthant order \citep{smith1995} and for the sign-pattern-nonnegative
transitions $D_tN_tD_{t-1}$ of \cref{lem:boxes}(ii) and \cref{rem:provisos}. This underlies interval observers
\citep{gouze2000,efimov2013}, also for periodic systems \citep{efimov2012}, and the embedding systems of mixed-monotone reachability
\citep{coogan2015,coogan2020,meyer2021}, whose equilibria and periodic orbits certify robustly forward-invariant sets
\citep{abate2022}. Robust positively invariant (RPI) sets and the minimal RPI set are classical
\citep{kolmanovsky1998,blanchini1999,rakovic2005}, also for linear difference inclusions \citep{kouramas2005}, as are componentwise
ultimate bounds \citep{kofman2007}. Abstract domains designed for digital filters handle linear time-invariant recursions
\citep{feret2004,monniaux2005,roux2012}, and disk (circular) arithmetic is classical \citep{gargantini1971}. For rotations,
ellipsoids \citep{kurzhanski1997} and support functions \citep{leguernic2010} are further set representations for linear
reachability; a linear map sends an ellipsoid onto a (possibly degenerate) ellipsoid, as a rotation-scaling sends a disk onto a
disk. We did not test them. Our certificates are
real-arithmetic statements computed in float64; verified numerics gives the tools for rigorous results in floating-point arithmetic
\citep{rump2010}, and with them we re-verified the reported certificates with outward rounding, except those listed in
\cref{rem:float}, under a measured bound on the errors of five library functions.

We use these facts and do not claim them. A nonnegative diagonal channel is a positive scalar recursion with interval coefficients,
\cref{lem:cr} is its interval step, \cref{lem:boxes}(iii) is the comparison principle in the form we use, and $H^*$ is an RPI box of
the period map whose phase boxes form a periodic orbit of the box dynamics (\cref{lem:period}). For point decays (the S4D-type layer)
$H^*$ is exactly the minimal RPI interval of the channel for its injection boxes: with a point decay one period maps an interval $H$
exactly onto $\pih H+[\cfree-\beta(0),\cfree+\beta(0)]$ (\cref{lem:cr}(b) with $\arad=0$), so
$H^*=\sum_{i\ge0}\pih^{\,i}\,[\cfree-\beta(0),\cfree+\beta(0)]$, the minimal RPI set of that map \citep{kolmanovsky1998}; for selective
channels $H^*$ is RPI (\cref{lem:period}) but, because of the excess of \cref{lem:cr}(b), it need not be minimal. \Cref{prop:exact} is
the mixed-monotone caveat in IBP terms: interval propagation follows a matrix $M_t\ge\abs{J_t}$ built from the primitives, not the
Jacobian (\cref{prop:loss}(c)), and our LSTM contracts while the width of its infinite-horizon IBP box stays bounded away from zero
as $\eps\to0$, at $2\nu^*=0.9433$ (\cref{prop:lstmfull}, \cref{rem:lstmnum}). \citet{harapanahalli2026} certify the invariant set
itself during training; \cref{prop:horizon} shows what goes wrong when a recurrent network is trained against a finite-horizon
surrogate of its invariant box instead.

\paragraph{Completeness in abstract interpretation.}
\Cref{def:exact} is the completeness of abstract interpretation \citep{cousot1977,giacobazzi2000}, restricted to the iterates from
the zero state: composing the best one-step abstractions loses nothing against the best abstraction of the reachable set.
\Cref{lem:boxes} identifies transitions for which the box domain is complete in this sense, given the coefficient boxes, and
\cref{lem:disks} bounds the loss of disks where they are not exact. \citet{giacobazzi2000} make abstract interpretations complete by
refining or simplifying the abstract domain; in these terms, the open question of regime~3 asks whether boxes have a refinement of
fixed size that certifies gated transitions uniformly in time.

\paragraph{Fixed points of implicit models.}
IBP certificates for deep equilibrium models bound the interval fixed point of an implicit layer, made well defined by a
parameterization \citep{wei2022}; certifiable deep equilibrium models are designed for such certificates \citep{li2022}; and
non-Euclidean and mixed-monotone contraction give robust implicit networks with certificates \citep{jafarpour2021,jafarpour2022}.
Our $L=\infty$ certificate is an interval fixed point too, but over time instead of depth: a cascade of period maps whose
coefficient boxes the network computes from the boxes of the layer below (\cref{def:cascade}), solved in closed form layer by layer,
in a form that stays stable as channels slow down (\cref{lem:period}, \cref{rem:impl}). It exists for all weights and bounded input
boxes, because every decay is below one in exact arithmetic, so $\denh>0$ (\cref{sec:setting}, \cref{lem:period}). Here the
deployed model reads finite contexts and the fixed point is their limit, so the $L=\infty$ certificate covers no finite context
(\cref{ex:finite}); the K-uniform hull (\cref{thm:hull}) covers both.

\paragraph{Networks in the loop.}
Reachability analysis of neural feedback loops \citep{everett2021}, interval reachability of nonlinear systems with neural network
controllers \citep{jafarpour2023} and forward invariance in neural network controlled systems \citep{harapanahalli2023} bound the
states of a dynamical system driven by a network, and controllers have been trained so that a polytope is certifiably robustly
invariant \citep{harapanahalli2026}; robustly forward-invariant sets of mixed-monotone systems follow from the equilibria and
periodic orbits of their embedding systems \citep{abate2022}. The comparison principle behind \cref{lem:boxes}(iii) also underlies
monotone control systems \citep{angeli2003} and orthant-monotone systems \citep{smith1995}. These works bound a system driven by a
network; we bound a network driven by an unbounded context, and the invariant box is computed inside the network, layer by layer.

\paragraph{Long contexts and prefixes.}
Universal adversarial triggers \citep{wallace2019}, many-shot jailbreaking \citep{anil2024} and hidden-state poisoning of Mamba
\citep{lemercier2026} act through the context, and \citet{thapliyal2026} studies SSM heads for jailbreak detection. Our threat model
is weaker at each position, a box around a fixed token with no insertions or deletions (\cref{rem:notclaimed}), but it covers every
position of an arbitrarily long periodic context, or every context over a token set (\cref{cor:anyctx}); inserted triggers lie
outside it.

\paragraph{Positioning.}
Relative to IBP and certified training, we move the analysis of IBP from depth to context length; for nonnegative diagonal layers
boxes are exact in time given their coefficient boxes (\cref{lem:boxes}), which yields closed-form certificates for the
infinite-context limit and for every
number of copies of the period (\cref{thm:inf,thm:hull}). The finite-horizon blind spot (\cref{prop:horizon}) belongs to the
training objective, not to the relaxation, and we found no earlier description of it; \cref{cor:fixed} gives an objective that
slowing a channel cannot reduce. We do not claim a better certified-training method in general: our first pre-registered IBP
training under synonym substitution missed four of its five criteria and was stopped by its own rule, and the second
(\cref{app:remedy}) used the linear-ramp recipe (after \citealp{jia2019}), under which the cap with the hull term met its headline
criterion for the minGRU-type model only; for Mamba the cap with the hull term kept at least 80\% of the certificates for every
number of copies, but at $-$9.9 [$-$28.6, 8.8] points of clean accuracy, with two of three seeds below 78\%. Relative to RNN and transformer verification, the new
axis is an unbounded perturbed context (any number of copies of a given period, or every context over a token set), certified at a
cost independent of its length (\cref{app:wallclock}), with a domain-relative criterion (\cref{sec:theory-exact}: boxes exact in time
for real diagonal and sign-pattern-nonnegative point transitions, disks with a constant, non-compounding loss for complex-diagonal
LTI layers, gated cells open) and an LSTM that shows how a gated cell can fail in boxes (\cref{prop:exact}(b)). Relative to work on
SSMs, we found no earlier interval analysis of selective scans; ours needs no contraction constraint and covers depth and the
infinite-context limit, at the price of constants in $1/\denh$. Relative to long-context attacks, our threat model is weak at each
position and silent on inserted triggers; what it adds is the length of the perturbed context.

\paragraph{What is classical and what is new.}
Classical, and used without claim: the comparison principle for monotone systems (\cref{lem:boxes}(iii)), the wrapping effect of
interval arithmetic (\cref{lem:boxes}(ii), with Moore's rotation example in \cref{prop:exact}(b$'$)), disk arithmetic (\cref{lem:disks}), contraction and robust positively invariant sets
(\cref{lem:period}), box-type order reduction of zonotopes (\cref{lem:zono}), the first-order behaviour of interval extensions
(\cref{prop:loss}(a)), and interval fixed points of contracting maps, as in the certification of implicit models. New: the
composition of these facts through deep networks whose input-dependent coefficient boxes the network computes itself; the limit
semantics with the K-uniform hull and the token-set certificate; the domain-relative boundary of exactness in time for the recurrent
layers in use, with the contracting LSTM of \cref{prop:exact}(b) and (b$''$); and the finite-horizon blind spot of robust
objectives (\cref{prop:horizon}).

\section{Code, compute and reproducibility}\label{app:compute}

\paragraph{What the code release contains.} The code release will be made publicly available.
(1)~The code, in Python and PyTorch, with one line per file:
\begin{itemize}
\item \emph{Models and interval arithmetic.} \texttt{small.py}: the small Mamba block that the three nonnegative diagonal layers are
built on. \texttt{ibp\_mamba.py}: differentiable IBP for the Mamba models and an explicit LSTM baseline, with the interval primitives,
the endpoint and centre-radius scans, the closed-form fixed point of a period and the K-uniform hull. \texttt{zoo\_ssm.py}: the
S4D-type and minGRU-type layers (the same block with the selective scan replaced) with endpoint IBP, the fixed-point certificate and
the hull. \texttt{zoo\_rnn.py}: GRU and LSTM with IBP through the gates and the invariant-box search for the infinite context.
\texttt{zoo\_s4dc.py}: the complex-diagonal S4D of the control, with IBP in real boxes and in complex disks. \texttt{zoo\_pool.py},
\texttt{ibp\_zoo.py}: the embedding head on counter-fitted vectors, the CNN and bag-of-words baselines with their unbounded-context
certificates, and the model zoo.
\item \emph{Data and threat model.} \texttt{wordsub.py}: tokenisation, vocabulary, synonym sets and the exact embedding box of a
position under synonym substitution. \texttt{prep\_wordsub.py}: builds the word-level IMDB data, the vocabulary and the neutral texts as
token ids. \texttt{data.py}: the byte-level data.
\item \emph{Training.} \texttt{train\_ibp.py}: natural training, standard IBP and fixed-point fine-tuning under either threat model,
with resumable checkpoints, and for the training remedy (\cref{app:remedy}) the time-constant cap, the $L=\infty$ and K-uniform
robust terms from the start, and the linear-ramp recipe (after \citealp{jia2019}).
\item \emph{Evaluation, one script per experiment.} \texttt{eval\_len.py}: certified accuracy with finite neutral prefixes, at $L=\infty$
and under the K-uniform hull, with the substitution attack of \texttt{attack\_sub.py} (greedy search by gradient saliency, random
restarts, beam search and a genetic phase, every success re-checked by a forward pass). \texttt{eval\_cert.py}: clean, certified and
attacked accuracy and certified radii of a checkpoint at several radii. \texttt{eval\_ctx.py}: the context-length probe and the complex-diagonal control.
\texttt{zono\_ctx.py}: the distance probe (box against time-direction zonotope, certified radius by bisection, sampled-input checks).
\texttt{diag\_growth.py}: the interval-map growth rates. \texttt{diag\_tau.py}: time constants, DC levels and the slow-channel
attribution. \texttt{eval\_anyctx.py}: the token-set certificate with its checks against concrete contexts. \texttt{paper\_numbers.py}:
the wall-clock timings and the constants of \cref{prop:lstmfull}. \texttt{lirpa\_ctx.py}: the comparison with auto\_LiRPA's linear
relaxations (\cref{app:lirpa}; the models rewritten in operators that auto\_LiRPA bounds, the interval evaluation of the traced
graph, the distance probe and synonym substitution with their checks against sampled inputs and at zero width, and the summary),
on auto\_LiRPA~0.7.0 \citep{xu2020}, whose chord slopes of exp, sigmoid and tanh it corrects on narrow intervals, with the square
of RMSNorm written as a product.
\item \emph{Re-verification with outward rounding} (\cref{rem:float}). \texttt{fpround.py}: the directed interval primitives
(outward steps from round-to-nearest, the library-function steps, the bounds of sums and matrix products, the centre-radius step
and the DC closed forms in interval arithmetic). \texttt{fp\_libm.py}: the platform's fingerprint, the checks of its basic
operations and the measurement of its library functions against high-precision values. \texttt{fp\_ibp.py}: the outward-rounded
mirrors of the certificate code of every architecture, with certificate checking at $L=\infty$ and for the hull. \texttt{fp\_eval.py}:
the driver, which archives the embedding tables, evaluates every reported certificate with its float64 reference and runs the
validity checks per document. A driver for the checkpoints of the training remedy (\cref{app:remedy}), built on
\texttt{fp\_eval.py}, with their archive format, the certificates after one and four copies of the period and at box scale 0, the
token sets, and the comparison of its float64 reference with the archived outcome of every document.
\item \emph{Clean accuracy and the extension to 200 documents} (\cref{app:cleanctx}). An evaluator of the clean prediction at every
context, per document (the neutral contexts, the copies of the period and the limit, on the documents and chunks of the first
training study and of the context-length probe), and the summary script of the extension, which builds the extended and the new-only
views of the stored outputs, runs the summary scripts below on them unchanged, and computes the intervals of
\cref{app:cleanctx-method}, the tables of \cref{app:cleanctx} and the checks of its record.
\item \emph{Summaries and figures.} \texttt{summarize\_ctx.py}, \texttt{summarize\_fp.py}, the summary script of the training
remedy's re-verification (which also recomputes that study's summary from the re-verified outcomes), the summary scripts of the budget sweep
and of the training remedy, and the summary modes of the evaluation scripts apply the pre-registered criteria mechanically and write
the tables that the paper quotes. \texttt{make\_figures.py} draws
every figure from the stored outputs; it computes nothing beyond medians, quartiles and fractions of stored per-document values.
\item \emph{Tests.} \texttt{test\_soundness.py}, \texttt{test\_ss.py} and a test directory with acceptance tests of the hull, of every
model of the zoo, of the complex-diagonal model in both domains, of the substitution threat model and the attack, of the distance
probe and the sweep summary, of the rewritten models and the checks of the comparison with auto\_LiRPA, of the time-constant cap and
the direct minGRU-type decay path, of the outward-rounded evaluator (its primitives against exact and high-precision values, its
mirrors, an end-to-end oracle and mutation tests) and of its driver and summary for the training remedy,
of the extension's evaluator and summary (its intervals against hand-computed values, its draws against those of the archived runs,
its views of the stored outputs),
of training resumption, and integration tests of training and evaluation (\cref{app:audits}).
\end{itemize}
(2)~One pre-registration specification per test, identified by the internal labels of \cref{tab:prereg}, each fixed before the
production run of its test (with every amendment marked as such), and the summary outputs (JSON and Markdown) that the summary scripts
wrote; every number of the paper is quoted from them. (3)~The neutral context texts: licence texts for evaluation, disjoint from the
GNU manuals that supply the training periods of fixed-point fine-tuning. The IMDB reviews \citep{maas2011} and the counter-fitted vectors
\citep{mrksic2016} are public and are read by \texttt{prep\_wordsub.py} from their original distributions. (4)~For the
re-verification of \cref{rem:float}: the archived embedding tables with their hash manifest, the record of the platform checks and of
the library-function measurement, and its summary under the registered implementation and under the amendment (\cref{app:prereg-amend});
and, for the training remedy, its own archived embedding tables with their manifest, its platform record and its summaries, of the
part read first, of the rest of the re-verified scope and of the cap-128 part.
(5)~For the extension of \cref{app:cleanctx}: the hash manifests of its frozen code and inputs, its job record and task logs, and its
summaries under the registered and the amended rules (\cref{app:prereg-amend}).

\paragraph{Data.}
Word-level IMDB \citep{maas2011}: reviews lowercased, HTML line breaks replaced by spaces, contractions split in the style of the
Penn Treebank, and
tokens matched against the counter-fitted vectors \citep{mrksic2016} (65,713 words, 300 dimensions, every vector normalised). The
vocabulary holds the training-set tokens that have a vector, most frequent first, 43,566 entries; the 64 most frequent tokens without a
vector (punctuation, possessives, numbers) receive fixed random unit vectors and are neither substitutable nor substitutes, and the
remaining tokens without a vector map to an unknown-word token (1.4\% of the training tokens, 1.6\% of the test tokens). Documents are
truncated at $T=512$ tokens. A word's synonym set holds its up to 8 nearest vocabulary words within distance $0.5$ in the counter-fitted
space; 32.9\% of the tokens have at least one substitute, 2.93 on average, and every substitutable position of the context and of the
document may be replaced independently. Unlike \citet{jia2019}, we feed the counter-fitted vectors also as model inputs, through a
trainable linear map to $d_m=64$ dimensions; the box of a position is then exact for the embedding layer. The training and test splits
are shuffled with seed 0, and each experiment takes the first $n$ test documents: 1000 for IBP training under synonym substitution
and for the clean accuracy over the context (and 200, on the documents of the context-length probe); 200 for the context-length probe,
the complex-diagonal control, the slow-channel
attribution and the token-set certificate; 200 per seed for the interval-map growth rates and for the distance probe (50 for its
$B=2048$ zonotope on the nonnegative diagonal models of seed~0); 50 for the budget sweep; the comparison with auto\_LiRPA's linear
relaxations uses the documents listed in \cref{app:lirpa}. The byte-level study reads byte-level IMDB at $T=1024$ (\cref{app:byte}).

\paragraph{Training.}
The recurrent models of the earlier studies have two layers of width $d_m=64$ ($N=16$ states per inner channel for Mamba and the
S4D-type layer, hidden size 64 for LSTM and GRU) and are trained on batches of 64 documents with AdamW (weight decay 0.05) and a
one-cycle learning-rate schedule. The
loss is $(1-\kappa)\,\mathrm{CE}(\text{clean logits})+\kappa\,\mathrm{CE}(\text{worst-case logits})$, the worst-case logits being the
lower bound of the true-class logit and the upper bounds of the others. \emph{Natural} training: 4000 steps at learning rate
$2\times10^{-3}$, $\kappa=0$. \emph{Standard IBP}: 8000 steps at $2\times10^{-3}$; after 2000 clean warm-up steps the substitution box
scale ramps geometrically from $10^{-3}$ to 1 over two thirds of the remaining steps, and $\kappa$ from 0 to 0.5 over the same window.
\emph{Fixed-point fine-tuning}: 3000 steps of fine-tuning from the standard-IBP checkpoint at $5\times10^{-4}$ with $\kappa=0.5$, where half of the
robust batch carries the $L=\infty$ fixed-point term of \cref{cor:fixed} with training periods of 256 tokens from the GNU manuals, the
radius of the period positions ramping geometrically from $10^{-4}$ times the box to the full box over 2000 steps; for LSTM and GRU the
prefix term is a training surrogate of their invariant box, four brute-force periods. Every natural word-level recurrent architecture
has three seeds (seeds 1 and 2 of the S4D-type, minGRU-type, LSTM and GRU models were trained for the replication of the
context-length probe, with validation accuracy 0.857--0.887); standard IBP and fixed-point fine-tuning have three seeds each; the CNN and the
bag-of-words model have one natural seed; the complex-diagonal S4D of the control was trained with the natural recipe on two seeds.
Training jobs stop at 280 minutes with a resumable checkpoint and continue in a new job; the final in-training evaluation uses 2000
documents, and the evaluations of the paper are run separately on the test documents above. The training remedy's recipes, arms,
seeds and development split are given in \cref{app:remedy}.

\paragraph{How to run.}
\begin{enumerate}
\item Prepare the data with \texttt{prep\_wordsub.py} from the raw reviews and the counter-fitted vectors ($T=512$); the byte-level data
come through \texttt{data.py}.
\item Train each model with one call of \texttt{train\_ibp.py} with the settings above; the run matrix in the code release lists every
call.
\item Evaluate each experiment with its script (list above). Every script writes one JSON file per model, in float64, records the
settings and job of the run, and can be split over documents and merged.
\item Summarise: the summary scripts apply the pre-registered criteria and print both the criteria and the verdicts, together with
every amended reading next to the original one.
\item Draw the figures with \texttt{make\_figures.py}, and run the tests (\cref{app:audits}).
\end{enumerate}

\paragraph{Compute.}
Training ran as GPU jobs (one GPU and 4 CPU threads per job) on a shared cluster with NVIDIA H100 and H200 GPUs, under a 5-hour limit
per job. Natural training took about 2.5 minutes for the S4D-type and minGRU-type models and about 18 minutes for LSTM and GRU (the
replication models of seeds 1 and 2). Every certificate of the earlier studies was evaluated on CPU in float64 (those of the
training remedy on GPUs in float64, \cref{app:remedy}): the first training study with 16
threads per job; the context-length probe with 8 threads (Mamba and S4D-type rows) or 2 (the others), the replication on seeds 1 and 2
taking 49, 34, 35 and 12 minutes per model for the S4D-type, minGRU-type, LSTM and GRU models; the distance probe in shards of 18
minutes to 2.2 hours (GRU 38 minutes, LSTM 35--56 minutes, Mamba 2.2 hours with $B=256$ and 40 minutes with $B=2048$, S4D-type 1.6--1.9
hours and 18 minutes); the budget sweep with 4 threads, its 17 jobs and those of the replications on seeds 1 and 2 finishing within
about one and a half hours of wall-clock time; and the comparison with auto\_LiRPA with one thread for ours and four for every
auto\_LiRPA process (\cref{app:lirpa}). The re-verification of \cref{rem:float} ran on CPU, in float64 with outward rounding, that
of the training remedy included. The
training remedy's main stage took about 75 GPU-hours of training by the runs' own timers, and its selection stage about 9
(\cref{app:remedy}). The extension of \cref{app:cleanctx} ran on CPU in float64, with 2 to 16 threads per task, and used 1,012
core-hours by its task logs (\cref{app:cleanctx-record}). \Cref{app:wallclock} gives the cost of one certificate.

\paragraph{Determinism.}
The budget sweep recomputed the certified radii of the distance probe in all 500 cells the two runs share for each model, and every
value lay inside the sweep's final bisection bracket (\cref{app:audits}). A test checks that a resumed training run gives the same
weights, bit for bit on CPU, as an uninterrupted one.

\section{An earlier study: character-level language models}\label{app:charlm}

In an earlier phase of this work, before the experiments of \cref{sec:exp}, we compared certified radii at growing distance for two
trained character-level language models, a GRU and a Mamba (validation losses of 2.08 and 1.96 bits per character), with the verifier
of that time. It differs from the IBP of the paper: one token at distance $d$ before the prediction was perturbed inside a random
16-dimensional subspace of its embedding, and the ``box'' method was a degree-2 polynomial zonotope on the 16 input symbols whose
remainders are carried as intervals, not IBP. For the GRU, a time-direction zonotope promoted the remainders to fresh symbols at every
step and kept a budget of 2048.

\paragraph{Results.}
On six samples, the GRU's box certificates were zero from $d=128$ on, while its time-direction zonotope kept certified radii of
0.042--0.115 at $d=1024$, at a cost of 135\,s against 3\,s for the box. In a first run of the comparison, projected gradient descent
found no change of a GRU prediction within radius 10 at $d\ge8$. The Mamba's box certificates did not collapse (0.024--0.057 at $d=1024$), and the promoted zonotope gave 0.060--0.080 at $d=1024$, at a median cost of 1661\,s against about
280\,s for the box.

\paragraph{Agreement and differences with the distance probe.}
For the box the direction agrees with the distance probe of \cref{sec:exp-gate5}: the GRU's box certificate collapses with the
distance and the Mamba's does not. The zonotope result does not carry over at the same budget: the character-level GRU held its
certified radius at $d=1024$ with a budget of 2048, while the word-level GRU of the distance probe did not (median ratio $r(1024)/r(1)$
of at most $2.3\times10^{-6}$ on seed~0 and $3.6\times10^{-5}$ on seeds 1 and 2, 200 documents per seed); with budgets of 8192 and
16384 the word-level GRU kept a tenth of its adjacent-token certified radius at $d=1024$ on 51\% and 76\% of the documents (budget
sweep, 49 documents, \cref{app:gate5}). The two
studies differ in four ways: the perturbation (16 of the embedding's directions against all 64), the verifier (degree 2 with interval
remainders against degree 1 with Girard-type reduction), the model and the task (character-level language modelling against
word-level sentiment classification), and the sample size (6 against 200 documents per seed). The distance probe is the evidence the
paper relies on; this study is reported for completeness.

\section{Certified radius against distance on three seeds, and the budget sweep}\label{app:gate5}

This appendix gives the full record of the certified radius against distance (\cref{sec:exp-gate5}) on its three seeds, and of
the budget sweep on the gated models.

\paragraph{Probe.}
The sequence is [1024 neutral tokens $\mid$ document], and the class $y_0$ is its clean prediction. Only the token at distance
$d\in\{1,16,128,512,1024\}$ before the readout is perturbed, inside the $\ell_\infty$ ball of its 64 embedding coordinates. The
\emph{certified radius} $r(d)$ of a method is the largest $\eps\in[10^{-12},10^2]$ at which the method certifies $y_0$, found by
bisection in log space with 12 halvings; $r=0$ when $10^{-12}$ is not certified. It is a radius of the input ball, not of a state
box: ``the box's $r(d)$'' is the largest ball that the box method certifies. We also report $R(10^{-6})$, the looseness of a
method at $\eps=10^{-6}$ (\cref{sec:theory-loss}), which is 1 for a method that is exact to first order.

\paragraph{Methods and run plan.}
The \emph{box} is plain endpoint IBP. The \emph{zonotope with budget $B$} uses affine forms over the 64 input symbols, which are
never reduced, so the zonotope is exact to first order; affine maps are exact, products use the product rule of affine forms,
the nonlinearities sound linear relaxations, and in-flight values are promoted to fresh symbols; at the end of every step a
Girard-type reduction, ranked by $\ell_1$ weight, keeps $B$ promoted symbols and boxes the rest. For GRU and LSTM every method ran
at every distance. For Mamba and the S4D-type model, for cost, the $B=256$ zonotope ran up to $d=512$ and the $B=2048$ zonotope up
to $d=128$; the box ran at every distance. The models are the natural word-level models of the context-length probe, on the
first 200 test documents of each seed: the pre-registered runs had the first 50 on seed~0 and the first 20 on seeds 1 and 2 (at
least 18 required), and the extension of \cref{app:cleanctx} added the others with the same configuration. The nonnegative diagonal
models ran without the $B=2048$ rows on seeds 1 and 2; on seed~0 these rows, which were descriptive, cover the first 50 documents.
Everything is in float64, and every bound was checked against sampled inputs (\cref{app:audits}); the box radii of Mamba and the
S4D-type model on the documents of the pre-registered runs were re-verified with outward rounding at the radius itself, those on
the documents added by the extension were not, nor those of GRU and LSTM, nor the zonotopes (\cref{rem:float}).

\paragraph{What was predicted, and what happened.}
For GRU and LSTM the test predicted that the box's certified radius collapses (median $r_{\rm box}(128)/r_{\rm box}(1)\le10^{-3}$),
that the $B=2048$ zonotope keeps $r(1024)/r(1)\ge0.1$ on at least 80\% of the documents, and that one zonotope bound at $d=1024$
costs at least ten box bounds. For Mamba and the S4D-type model it predicted that the box's certified radius does not collapse
(median $r_{\rm box}(1024)/r_{\rm box}(1)\ge0.1$) and that the $B=256$ zonotope certifies at least 0.99 times the box's certified
radius at every distance it ran, on at least 80\% of the documents. In the pre-registered run on seed~0 (50 documents) the box
collapsed on both gated models, the zonotope
failed its prediction at $d=1024$ (it kept a tenth of its adjacent-token radius on 0\% of the GRU and 24\% of the LSTM documents),
the cost prediction held (19 and 35 box evaluations), the box did not collapse on either nonnegative diagonal model, and the
zonotope reached 0.99 times the box on every Mamba document at every distance but on only 54\% of the S4D-type documents at
$d=512$. The replication on seeds 1 and 2 (20 documents each) applied the same criteria, with the rule that a criterion replicates
when its outcome there equals the seed-0 one: every outcome replicated. The criteria and their codes are in \cref{app:prereg}. On
the extended sets of 200 documents per seed (the archived documents included) every criterion kept its outcome on every
seed, and it did so on the added documents alone (150 on seed~0, 180 on seeds 1 and 2), so these outcomes replicated on the new
documents (\cref{app:cleanctx-record}). There the zonotope kept a tenth of its adjacent-token radius at $d=1024$ on 0\% of the GRU
documents of every seed and on 26.0\% [20.4, 32.5], 14.0\% [9.9, 19.5] and 58.5\% [51.6, 65.1] of the LSTM documents (seeds 0, 1
and 2; on seed~2 the interval of the 20 archived documents, 65\% [43.3, 81.9], contained the predicted 80\%, that of the 200
does not), and the S4D-type zonotope reached 0.99 times the box at $d=512$ on 52.5\% [45.6, 59.3],
28.0\% [22.2, 34.6] and 36.0\% [29.7, 42.9] of the documents (95\% intervals per checkpoint, \cref{app:cleanctx-method}).

\begin{table}[h]
\caption{Certified radius against distance, seed~0: median certified radius $r(d)$ of one token at distance $d$ (200 documents per
method; the $B=2048$ rows of Mamba and the S4D-type model: 50; $r=0$: no radius certified; --: not run, for cost). 95\% intervals
of the medians that the text cites: \cref{tab:ci-b}. Seeds 1 and 2 (200 documents each): \cref{tab:gate5seeds}.}
\label{tab:gate5}
\centering\small
\begin{tabular}{@{}llrrrrr@{}}
\toprule
model & method & $r(1)$ & $r(16)$ & $r(128)$ & $r(512)$ & $r(1024)$\\
\midrule
GRU & box & 0.0219 & $1.36{\times}10^{-7}$ & 0 & 0 & 0\\
 & $B=256$ & 0.0461 & 0.00302 & 0 & 0 & 0\\
 & $B=2048$ & 0.0461 & 0.019 & 0.00813 & $1.18{\times}10^{-4}$ & $9.72{\times}10^{-8}$\\
LSTM & box & 0.0296 & $1.52{\times}10^{-6}$ & 0 & 0 & 0\\
 & $B=256$ & 0.0777 & 0.0148 & $1.48{\times}10^{-9}$ & 0 & 0\\
 & $B=2048$ & 0.0777 & 0.0507 & 0.0375 & 0.0124 & 0.00238\\
Mamba & box & $7.45{\times}10^{-5}$ & $5.87{\times}10^{-5}$ & $2.45{\times}10^{-4}$ & $5.26{\times}10^{-4}$ & $6.75{\times}10^{-4}$\\
 & $B=256$ & 0.00822 & 0.00809 & 0.0127 & 0.0157 & --\\
 & $B=2048$ & 0.00933 & 0.0111 & 0.0147 & -- & --\\
S4D-type & box & 0.00186 & 0.00728 & 0.0155 & 0.034 & 0.056\\
 & $B=256$ & 0.0185 & 0.0292 & 0.0323 & 0.0364 & --\\
 & $B=2048$ & 0.019 & 0.0296 & 0.0327 & -- & --\\
\bottomrule
\end{tabular}
\end{table}

\begin{table}[h]
\caption{Certified radius against distance on seeds 1 and 2 (200 documents each): median certified radius $r(d)$ ($r=0$: no radius
certified; --: not run). 95\% intervals of these medians: the extension's summary in the code release (\cref{app:compute}).
Seed~0 is in \cref{tab:gate5}.}
\label{tab:gate5seeds}
\centering\small
\begin{tabular}{@{}llcrrrrr@{}}
\toprule
model & method & seed & $r(1)$ & $r(16)$ & $r(128)$ & $r(512)$ & $r(1024)$\\
\midrule
GRU & box & 1 & 0.0235 & $6.15{\times}10^{-8}$ & 0 & 0 & 0\\
 & & 2 & 0.0207 & $1.45{\times}10^{-7}$ & 0 & 0 & 0\\
 & $B=256$ & 1 & 0.047 & 0.00308 & 0 & 0 & 0\\
 & & 2 & 0.045 & 0.00332 & 0 & 0 & 0\\
 & $B=2048$ & 1 & 0.047 & 0.0192 & 0.01 & $3.63{\times}10^{-4}$ & $1.39{\times}10^{-6}$\\
 & & 2 & 0.045 & 0.0191 & 0.01 & $3.68{\times}10^{-4}$ & $1.57{\times}10^{-6}$\\
LSTM & box & 1 & 0.035 & $3.73{\times}10^{-6}$ & 0 & 0 & 0\\
 & & 2 & 0.0376 & $2.41{\times}10^{-6}$ & 0 & 0 & 0\\
 & $B=256$ & 1 & 0.0819 & 0.0186 & $8.66{\times}10^{-9}$ & 0 & 0\\
 & & 2 & 0.0824 & 0.0167 & $2.63{\times}10^{-9}$ & 0 & 0\\
 & $B=2048$ & 1 & 0.0819 & 0.0554 & 0.04 & 0.0107 & 0.00103\\
 & & 2 & 0.0824 & 0.0528 & 0.0398 & 0.0201 & 0.00924\\
Mamba & box & 1 & $7.01{\times}10^{-5}$ & $6.66{\times}10^{-5}$ & $3.61{\times}10^{-4}$ & $7.28{\times}10^{-4}$ & $1.06{\times}10^{-3}$\\
 & & 2 & $7.14{\times}10^{-5}$ & $6.75{\times}10^{-5}$ & $2.58{\times}10^{-4}$ & $4.65{\times}10^{-4}$ & $6.19{\times}10^{-4}$\\
 & $B=256$ & 1 & 0.00737 & 0.00866 & 0.0135 & 0.0151 & --\\
 & & 2 & 0.00781 & 0.00847 & 0.0124 & 0.0155 & --\\
S4D-type & box & 1 & 0.00134 & 0.00672 & 0.0236 & 0.0523 & 0.0696\\
 & & 2 & 0.00247 & 0.0101 & 0.0232 & 0.0433 & 0.0615\\
 & $B=256$ & 1 & 0.0176 & 0.0298 & 0.0342 & 0.0401 & --\\
 & & 2 & 0.0199 & 0.03 & 0.0336 & 0.0392 & --\\
\bottomrule
\end{tabular}
\end{table}

\begin{table}[h]
\caption{Gated models: percentage of the documents whose certified radius keeps at least a tenth of its adjacent-token value,
$r(d)/r(1)\ge0.1$, with its 95\% Wilson interval per checkpoint (\cref{app:cleanctx-method}; 200 documents per seed), and the
median time of one evaluation at $d=1024$ on the archived documents (seed~0: 50; seeds 1 and 2: 20 each).}
\label{tab:gate5shares}
\centering\footnotesize
\setlength{\tabcolsep}{3pt}
\begin{tabular}{@{}llcccccr@{}}
\toprule
 & & & \multicolumn{4}{c}{documents with $r(d)/r(1)\ge0.1$ (\%) at $d=$} & one evaluation\\
\cmidrule(lr){4-7}
model & seed & method & 16 & 128 & 512 & 1024 & at $d=1024$ (s)\\
\midrule
GRU & 0 & box & 0 [0, 1.9] & 0 [0, 1.9] & 0 [0, 1.9] & 0 [0, 1.9] & 0.418\\
 & & $B=256$ & 13.5 [9.4, 18.9] & 0 [0, 1.9] & 0 [0, 1.9] & 0 [0, 1.9] & 4.03\\
 & & $B=2048$ & 100 [98.1, 100] & 90.5 [85.6, 93.8] & 0.5 [0.1, 2.8] & 0 [0, 1.9] & 8.31\\
 & 1 & box & 0 [0, 1.9] & 0 [0, 1.9] & 0 [0, 1.9] & 0 [0, 1.9] & 0.264\\
 & & $B=256$ & 10.0 [6.6, 14.9] & 0 [0, 1.9] & 0 [0, 1.9] & 0 [0, 1.9] & 4.18\\
 & & $B=2048$ & 100 [98.1, 100] & 96.0 [92.3, 98.0] & 0.5 [0.1, 2.8] & 0 [0, 1.9] & 8.63\\
 & 2 & box & 0 [0, 1.9] & 0 [0, 1.9] & 0 [0, 1.9] & 0 [0, 1.9] & 0.281\\
 & & $B=256$ & 19.0 [14.2, 25.0] & 0 [0, 1.9] & 0 [0, 1.9] & 0 [0, 1.9] & 4.09\\
 & & $B=2048$ & 100 [98.1, 100] & 95.5 [91.7, 97.6] & 0.5 [0.1, 2.8] & 0 [0, 1.9] & 8.15\\
LSTM & 0 & box & 0 [0, 1.9] & 0 [0, 1.9] & 0 [0, 1.9] & 0 [0, 1.9] & 0.356\\
 & & $B=256$ & 93.5 [89.2, 96.2] & 0 [0, 1.9] & 0 [0, 1.9] & 0 [0, 1.9] & 5.29\\
 & & $B=2048$ & 100 [98.1, 100] & 100 [98.1, 100] & 69.0 [62.3, 75.0] & 26.0 [20.4, 32.5] & 11.6\\
 & 1 & box & 0 [0, 1.9] & 0 [0, 1.9] & 0 [0, 1.9] & 0 [0, 1.9] & 0.243\\
 & & $B=256$ & 93.0 [88.6, 95.8] & 0 [0, 1.9] & 0 [0, 1.9] & 0 [0, 1.9] & 5.22\\
 & & $B=2048$ & 100 [98.1, 100] & 99.0 [96.4, 99.7] & 61.5 [54.6, 68.0] & 14.0 [9.9, 19.5] & 11.4\\
 & 2 & box & 0 [0, 1.9] & 0 [0, 1.9] & 0 [0, 1.9] & 0 [0, 1.9] & 0.24\\
 & & $B=256$ & 96.5 [93.0, 98.3] & 0 [0, 1.9] & 0 [0, 1.9] & 0 [0, 1.9] & 5.18\\
 & & $B=2048$ & 100 [98.1, 100] & 100 [98.1, 100] & 92.5 [88.0, 95.4] & 58.5 [51.6, 65.1] & 10.4\\
\bottomrule
\end{tabular}
\end{table}

\begin{table}[h]
\caption{Nonnegative diagonal models (200 documents per seed; the $B=2048$ rows: seed~0, 50 documents; --: not run). (a)~Median
looseness $R(10^{-6})$ of each method per distance (1: exact to first order). (b)~Percentage of the documents on which the $B=256$
zonotope certifies at least 0.99 times the box's certified radius, with its 95\% Wilson interval per checkpoint
(\cref{app:cleanctx-method}); $^\ddagger$not decisive at $n=200$: the interval contains the predicted 80\%.}
\label{tab:gate5loose}
\centering\footnotesize
\setlength{\tabcolsep}{3pt}
\textit{(a)}\par\smallskip
\begin{tabular}{@{}llcccccc@{}}
\toprule
 & & & \multicolumn{5}{c}{median $R(10^{-6})$ at $d=$}\\
\cmidrule(l){4-8}
model & seed & method & 1 & 16 & 128 & 512 & 1024\\
\midrule
Mamba & 0 & box & $4.41{\times}10^{4}$ & $1.75{\times}10^{5}$ & $1.22{\times}10^{5}$ & $7.56{\times}10^{4}$ & $6.52{\times}10^{4}$\\
 & & $B=256$ & 1 & 1.01 & 1.01 & 1 & --\\
 & & $B=2048$ & 1 & 1 & 1 & -- & --\\
 & 1 & box & $1.5{\times}10^{4}$ & $8.33{\times}10^{4}$ & $4.74{\times}10^{4}$ & $1.91{\times}10^{4}$ & $1.08{\times}10^{4}$\\
 & & $B=256$ & 1 & 1 & 1 & 1 & --\\
 & 2 & box & $7.3{\times}10^{4}$ & $3.69{\times}10^{5}$ & $1.68{\times}10^{5}$ & $1.08{\times}10^{5}$ & $8.38{\times}10^{4}$\\
 & & $B=256$ & 1 & 1.01 & 1.01 & 1 & --\\
S4D-type & 0 & box & 874 & $1.02{\times}10^{3}$ & 829 & 834 & 855\\
 & & $B=256$ & 1 & 1 & 1 & 1 & --\\
 & & $B=2048$ & 1 & 1 & 1 & -- & --\\
 & 1 & box & 824 & 978 & 528 & 372 & 316\\
 & & $B=256$ & 1 & 1 & 1 & 1 & --\\
 & 2 & box & 929 & $1.05{\times}10^{3}$ & 840 & 847 & 758\\
 & & $B=256$ & 1 & 1 & 1 & 1 & --\\
\bottomrule
\end{tabular}

\bigskip
\textit{(b)}\par\smallskip
\begin{tabular}{@{}llcccc@{}}
\toprule
 & & \multicolumn{4}{c}{documents with $r_{B=256}(d)\ge0.99\,r_{\rm box}(d)$ (\%) at $d=$}\\
\cmidrule(l){3-6}
model & seed & 1 & 16 & 128 & 512\\
\midrule
Mamba & 0 & 100 [98.1, 100] & 100 [98.1, 100] & 100 [98.1, 100] & 100 [98.1, 100]\\
 & 1 & 100 [98.1, 100] & 100 [98.1, 100] & 100 [98.1, 100] & 100 [98.1, 100]\\
 & 2 & 100 [98.1, 100] & 100 [98.1, 100] & 100 [98.1, 100] & 100 [98.1, 100]\\
S4D-type & 0 & 100 [98.1, 100] & 100 [98.1, 100] & 94.5 [90.4, 96.9] & 52.5 [45.6, 59.3]\\
 & 1 & 100 [98.1, 100] & 100 [98.1, 100] & 74.5 [68.0, 80.0]$^\ddagger$ & 28.0 [22.2, 34.6]\\
 & 2 & 100 [98.1, 100] & 100 [98.1, 100] & 85.5 [80.0, 89.7]$^\ddagger$ & 36.0 [29.7, 42.9]\\
\bottomrule
\end{tabular}
\end{table}

\paragraph{Further measurements.}
\emph{Gated models.} Within 16 steps the box's certified radius fell to a median $6.34\times10^{-6}$, $2.96\times10^{-6}$ and
$7.01\times10^{-6}$ (GRU) and $5.18\times10^{-5}$, $1.13\times10^{-4}$ and $6.58\times10^{-5}$ (LSTM) of its value for the
adjacent token (seeds 0, 1, 2; 95\% intervals in \cref{tab:ci}), and from $d=128$ on it certified no radius on any document of any
seed. The median $r(1024)/r(1)$
of the $B=2048$ zonotope was $2.2\times10^{-6}$, $2.9\times10^{-5}$ and $3.6\times10^{-5}$ (GRU) and 0.033, 0.013 and 0.121
(LSTM), although it certified a positive radius at $d=1024$ on every document. Times and cost factors here are those of the
archived documents (50 on seed~0, 20 on seeds 1 and 2): one $B=2048$ evaluation at $d=1024$ cost 19, 33 and
32 (GRU) and 35, 47 and 45 (LSTM) box evaluations; over all evaluations of a document the factors were 11.6, 18.2 and 18.6 and
20.0, 25.0 and 25.1. The zonotope evaluation runs only over the $d$ steps after the token, so its time grows with $d$ (GRU, $B=2048$,
seed~0: 0.090\,s at $d=16$, 0.95\,s at $d=128$, 8.3\,s at $d=1024$), while a box evaluation runs over the whole sequence and took
0.23--0.44\,s at every $d$. Four profile statements of the paper were checked on every seed: the GRU's $B=2048$ zonotope keeps a
tenth of its adjacent-token radius at $d=128$ on at least half of the documents (90.5\%, 96.0\% and 95.5\%), the LSTM's does so at
$d=512$ (69.0\%, 61.5\% and 92.5\%), and the box's median certified radius is 0 at $d=128$ for both gated models; all held, as they
did in the pre-registered runs (0.88, 1.00, 1.00 and 0.76, 0.75, 0.90 on 50, 20 and 20 documents).
\emph{Nonnegative diagonal models.} Per document, the box's certified radius at $d=512$ was a median 7.76, 12.5 and 7.32 (Mamba) and
16.9, 40.3 and 16.4 (S4D-type) times its value at $d=1$, and at $d=128$ 3.6, 5.21 and 3.63 and 7.55, 15.5 and 8.11 times; from $d=1$
to $d=1024$ the median ratio was 9.73, 17.5 and 10.1 (Mamba) and 29.2, 58.0 and 23.6 (S4D-type), so the fourth profile statement
(a median $r_{\rm box}(1024)/r_{\rm box}(1)$ of at least 1) held on every seed. Seed~0's Mamba ratio, 9.73 [8.24, 11.0], is one
of the five statistics on which the added documents differ from the archived ones (the 95\% interval of the difference
excludes 0): on the 150 added documents alone it is 10.4 [8.78, 12.1] (\cref{tab:ci-b}), so the box's radius grows with the
distance in both sets. On seed~0 the Mamba ratio dips to 0.715 at $d=16$
before it grows (0.943 and 0.964 at $d=16$ on seeds 1 and 2). On the archived documents one $B=256$ evaluation at $d=512$
cost 93, 37 and 34 (Mamba) and 110, 49 and 44 (S4D-type) box evaluations, a timing and not a criterion. On seed~0 the median certified radius of the $B=256$
zonotope was about 110 times the box's at $d=1$ and 30 times at $d=512$ on Mamba (95\% intervals of these ratios of medians:
[99.8, 132] and [27.9, 31.4]), and about 10 times and 1.1 times on the S4D-type model ([8.92, 11.1] and [0.98, 1.17]).
\emph{Clean accuracy} on these sequences was 0.83, 0.86 and 0.855 (GRU), 0.785, 0.865 and 0.845 (LSTM), 0.845, 0.80 and 0.865
(Mamba) and 0.84, 0.85 and 0.895 (S4D-type) on seeds 0, 1 and 2, on 200 documents each (\cref{tab:cleanctx}, column
$L=1024$; Wilson half-width at most 0.069 at $n=200$).

\paragraph{Reading, as pre-registered.}
\emph{Gated cells.} The box's certificates collapse with the distance, as the expanding interval maps measured by the growth rates
predict (\cref{sec:exp-r4}). The $B=2048$ zonotope keeps a tenth of its adjacent-token certified radius up to $d=128$ on most GRU
documents and up to $d=512$ on most LSTM documents, and then decays strongly; with $B=256$ the certified radius is lost by $d=128$
on every seed. With these two budgets, the budget needed to verify a gated RNN up to distance $d$ grows with $d$, and so does the cost
of one bound; the budget sweep below measures the relation on seven budgets. Boxes and the budgeted zonotopes we tried give no
time-uniform certificate for the gated models; whether some fixed-size domain does is open. We do not say that changing the domain
fixes gated certification, nor that gated transitions preserve no fixed-size domain.
\emph{Nonnegative transitions.} The box's certified radius does not collapse with the distance, because a distant token matters
less to the prediction and boxes are exact in time for these transitions (\cref{lem:boxes}). The zonotopes were exact to first
order at every distance on every seed (median $R(10^{-6})$ of 1 to within 1\%, against $1.08\times10^{4}$ to $3.69\times10^{5}$
for the Mamba box and 316 to 1050 for the S4D-type box), so they remove the box's loss within each step. Their advantage shrinks
with $d$ because the comparison moves to larger $\eps$: the box's certified radius grows with $d$, and at a larger $\eps$ the
higher-order terms of the zonotope's relaxations of SiLU, RMSNorm and the gate products bind. On the S4D-type model this made the
$B=256$ zonotope certify less than 0.99 times the box's certified radius on 47.5\% of the documents at $d=512$ on seed~0, at an
$\eps$ a median 17 times larger than at $d=1$; on seeds 1 and 2 it reached 0.99 times the box's certified radius at $d=512$ on only
28\% and 36\% of the documents, at a factor of 40 and 16, and on seed~1 the shortfall starts at $d=128$ (74.5\% [68.0, 80.0], not
decisive at $n=200$: the interval contains the predicted 80\%), where the factor is already 16. This is consistent with a
large-$\eps$ effect of the relaxations, and nothing in these runs shows a loss in time (they did not
measure how the higher-order loss at a fixed large $\eps$ depends on $d$): on the S4D-type scan, a linear point recursion, a
reduced zonotope of any budget is never wider than boxes for the same injection sets (\cref{lem:zono}; in the model each domain
computes its own injections through the nonlinear maps, and Mamba's interval-decay scan is not covered), and the runs do not
separate relaxation from reduction losses outside the scan. On these transitions a relational domain tightens the per-step
constant, and its large-$\eps$ relaxations can be looser than boxes; we do not say that the zonotope dominates the box, nor that a
budgeted zonotope loses in time.

\paragraph{Limits.}
Natural models only (200 documents per seed, the $B=2048$ rows of Mamba and the S4D-type model on seed~0 with 50); one perturbed
token, a probe and not a threat model; the
zonotope uses generic relaxations, a conservative stand-in for the best zonotope, in particular for Mamba; for the SSMs most state
remainders stay boxes, because the carried state is larger than either budget; and the box evaluation runs over the whole sequence
while the zonotope's runs only over the $d$ steps after the token, which favours the zonotope in the cost ratio.

\begin{figure}[h]
\centering
\includegraphics[width=\textwidth]{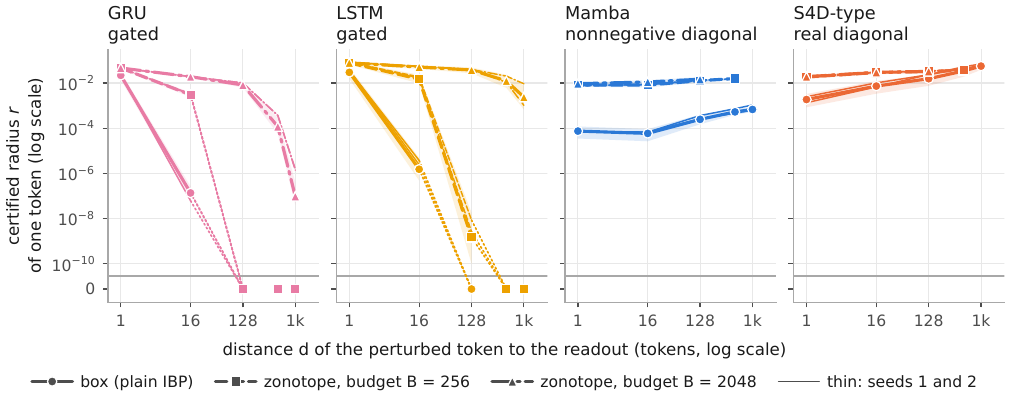}
\caption{Certified radius against distance on three seeds: certified radius $r(d)$ of one token at distance $d$ before the readout,
for the box (plain IBP) and the zonotopes with $B=256$ and $2048$ (for Mamba and the S4D-type model $B=256$ up to $d=512$ and
$B=2048$ up to $d=128$, as run; on their seeds 1 and 2 only $B=256$); seed~0 (200 documents; the $B=2048$ rows of Mamba and the
S4D-type model: 50): median over the documents, band: interquartile range; thin lines: the medians of seeds 1 and 2 (200 documents
each); $r=0$ in the floor strip. For GRU and LSTM the
box's certified radius collapses to 0 by $d=128$ on every seed, and the zonotope keeps a positive certified radius longer the larger
its budget; at $d=1024$ the $B=2048$ zonotope keeps a tenth of its adjacent-token radius on 0\%, 0\% and 0\% (GRU) and 26\%, 14\%
and 58.5\% (LSTM) of the documents (seeds 0, 1, 2). For Mamba and the S4D-type model the box's certified radius does not collapse: its
median ratio $r(1024)/r(1)$ is 9.73 [8.24, 11.0], 17.5 and 10.1 (Mamba; seed~0 on the added documents alone 10.4 [8.78, 12.1],
\cref{tab:ci-b}) and 29.2, 58.0 and 23.6 (S4D-type).}
\label{fig:gate5}
\end{figure}

\subsection{The budget sweep}\label{app:sweep}

\paragraph{Question and configuration.}
The certified radius against distance ran two budgets on the gated models: with $B=256$ the certified radius was lost by $d=128$,
with $B=2048$ it was kept to $d=128$ (GRU) and $d=512$ (LSTM) on most documents. Two budgets cannot say how the distance a zonotope
covers scales with its budget. The budget sweep asked whether budget buys distance, and at what rate and price, or whether the
reach saturates, beyond some budget, because of the relaxations of the gate nonlinearities rather than the reduction. It ran the
zonotope alone, at seven budgets $B\in\{256,512,1024,2048,4096,8192,16384\}$ and seven distances $d\in\{1,16,64,128,256,512,1024\}$,
on documents 0--49 of the natural seed-0 GRU and LSTM (the pre-registered run had documents 0--15, and the extension of
\cref{app:cleanctx} added 16--49 with the same configuration), with 8 bisection halvings (final bracket ratio at most 1.075) and
every other option as in the distance probe; the box was not rerun, and the box rows of the distance probe for the same documents
are the comparison.

\paragraph{Derived quantity and criteria.}
Per budget, document and distance, the ratio $r(d)/r(1)$, and $F_B(d)$, the share of the documents with a ratio of at least 0.1:
those that keep a tenth of their adjacent-token certified radius. The \emph{reach} $d^*_B$ is the largest grid distance $d$ with
$F_B(d')\ge0.5$ for every grid distance $d'\le d$; 1024 is the end of the grid, so a reach of 1024 means at least 1024. The test
predicted that the median certified radius does not fall, beyond the bisection tolerance, as the budget grows, and that the reach
does not decrease; that budget buys distance, $d^*_{16384}\ge\min(4d^*_{2048},1024)$; and that one bound at $d=1024$ costs at most 8
times more at $B=16384$ than at $2048$. Medians, shares and reaches are taken over the documents complete in every (budget,
distance) cell, 49 of the 50 for the GRU (document~3 lacks its $B=16384$ cells: a shard of the pre-registered run stopped at its
time budget before them) and 50 for the LSTM; this rule was
added to the specification before any output existed (\cref{app:prereg}), and the reading over every document that carries a cell
gives the same verdicts and the same reaches.

\paragraph{Verdicts.}
Every pre-registered criterion passed for both models, in the pre-registered run (16 documents), on the 50 documents and on the
added documents 16--49 alone (\cref{app:cleanctx-record}). The sweep reproduced the distance probe's certified radius in all 500
shared cells of each model (it lay inside this run's final bracket). None of the 147 pairs of medians and 21 pairs of reaches
compared across budgets failed the monotonicity criterion. The reach at $B=16384$ was 1024 for both models, against the required
512 (GRU: four times $d^*_{2048}=128$) and 1024 (LSTM: the grid cap, since $4d^*_{2048}=2048$). On the archived documents
one bound at $d=1024$ cost 5.03 (GRU) and 6.54 (LSTM) times more at $B=16384$ than at $2048$. Validity: no sample outside a hull or below a bound in 76,398 (GRU)
and 78,982 (LSTM) checks, the point limit within $10^{-13}$ of the clean margin, and no resolved first-order ratio below
$1-10^{-3}$ (\cref{app:audits}). Clean accuracy on these documents was 0.74 (GRU) and 0.80 (LSTM).

\begin{table}[h]
\caption{The budget sweep at the longest distance, $d=1024$ (seed~0; 49 complete documents for the GRU, 50 for the LSTM): median
certified radius, percentage of documents that keep a tenth of their adjacent-token certified radius with its 95\% Wilson interval
per checkpoint (\cref{app:cleanctx-method}), median looseness $R(10^{-6})$, and median time of one bound on the archived
documents (15 complete for the GRU, 16 for the LSTM); and the reach $d^*_B$ (1024: the end of the grid; its bootstrap interval:
\cref{tab:ci}).}
\label{tab:sweepfull}
\centering\footnotesize
\setlength{\tabcolsep}{3pt}
\begin{tabular}{@{}llccccccc@{}}
\toprule
model & quantity & $B=256$ & 512 & 1024 & 2048 & 4096 & 8192 & 16384\\
\midrule
GRU & $d^*_B$ & 1 & 16 & 64 & 128 & 512 & 1024 & 1024\\
 & median $r(1024)$ & 0 & 0 & 0 & $8.82{\times}10^{-8}$ & $5.62{\times}10^{-4}$ & 0.00453 & 0.00562\\
 & share (\%) & 0 & 0 & 0 & 0 & 8.2 & 51.0 & 75.5\\
 & \quad 95\% interval & [0, 7.3] & [0, 7.3] & [0, 7.3] & [0, 7.3] & [3.2, 19.2] & [37.5, 64.4] & [61.9, 85.4]\\
 & $R(10^{-6})$ & $4.0{\times}10^{11}$ & $3.73{\times}10^{11}$ & $3.03{\times}10^{11}$ & $3.39{\times}10^{10}$ & 3.06 & 1 & 1\\
 & time (s) & 4.87 & 5.29 & 6.74 & 9.54 & 17.9 & 41.2 & 48.0\\
LSTM & $d^*_B$ & 16 & 16 & 128 & 512 & 1024 & 1024 & 1024\\
 & median $r(1024)$ & 0 & 0 & $6.77{\times}10^{-11}$ & 0.00274 & 0.0255 & 0.0316 & 0.0328\\
 & share (\%) & 0 & 0 & 0 & 24.0 & 92.0 & 100 & 100\\
 & \quad 95\% interval & [0, 7.1] & [0, 7.1] & [0, 7.1] & [14.3, 37.4] & [81.2, 96.8] & [92.9, 100] & [92.9, 100]\\
 & $R(10^{-6})$ & $1.67{\times}10^{10}$ & $1.67{\times}10^{10}$ & $9.91{\times}10^{9}$ & 5.11 & 1 & 1 & 1\\
 & time (s) & 5.51 & 5.81 & 7.5 & 10.8 & 19.5 & 37.2 & 70.8\\
\bottomrule
\end{tabular}
\end{table}

\paragraph{Reading, as pre-registered.}
The reach of the time-direction zonotope with per-step promotion on these gated cells grows with its budget: $d^*_B$ rose from 1
(GRU) and 16 (LSTM) at $B=256$ to at least 1024 at $B=16384$, a value the GRU reached at $B=8192$ and the LSTM at $B=4096$. A
least-squares fit of $\log d^*_B$ against $\log B$ over the budgets below the cap gives exponents 2.1 (GRU, five budgets) and 1.8
(LSTM, four budgets); on a grid of seven distances the reach is quantized, so these exponents describe the grid and are not a
measured rate. The price of one bound, measured on the archived documents, grew less than in proportion to the budget: over
the seven budgets the least-squares line $t=a+bB$ has $a=5.44$\,s and $b=2.94$\,ms per symbol for the GRU and $a=3.39$\,s and
$b=4.1$\,ms for the LSTM, with log-log slopes 0.615 and 0.635. As in the distance probe the price grows with the distance (GRU,
$B=16384$: 0.075\,s at $d=16$, 3.3\,s at $d=128$, 48\,s at $d=1024$). The median $r(1)$ is 0.0453 (GRU) and 0.075 (LSTM) at every
budget. On the same documents the box had a median $r(1)$ of 0.0258 and 0.0319, certified no radius from $d=128$ on, and took about
0.4\,s per bound at every $d$, there and on the archived documents. At $d=1024$ and
$B=16384$ the zonotope kept a tenth of its adjacent-token certified radius on 75.5\% [61.9, 85.4] (GRU) and 100\% [92.9, 100]
(LSTM) of the documents (median ratios 0.143 and 0.422) and was exact to first order (median $R(10^{-6})=1$). So the collapse of the
box certificates on these gated
cells is a property of the domain, and a budgeted relational domain trades budget for distance. We do not say that some fixed-size
domain certifies gated cells uniformly in time: that stays open, and a reach that grows with the budget is consistent with ``no
fixed size suffices'' without proving it.

\paragraph{Descriptive.}
Where the largest budget reduced nothing ($B=16384$ at $d\le128$ for the GRU and $d\le64$ for the LSTM: no symbol dropped and no
value left unpromoted), the zonotope is unreduced and its only loss is the relaxation of the nonlinearities; there the median ratio
$r(d)/r(1)$ was 0.294 (GRU, $d=128$) and 0.649 (LSTM, $d=64$). With $B\ge8192$ the LSTM's median ratio stayed between 0.422 and
0.649 at every distance from 16 to 1024, while the GRU's fell from 0.392 at $d=16$ to 0.100 ($B=8192$) and 0.143 ($B=16384$) at
$d=1024$. The median $R(10^{-6})$ at $d=1024$ shows where the reduction breaks first-order exactness: it is at least
$9.9\times10^{9}$ up to $B=2048$ for the GRU and up to $B=1024$ for the LSTM, 3.06 (GRU, $B=4096$) and 5.11 (LSTM, $B=2048$) at the
next budget, and 1 from there on.

\paragraph{Limits.}
One seed (seed~0); 50 documents (49 complete for the GRU); one perturbed token, a probe and not a threat model; natural models with
two layers of 64 units; seven distances, capped at $d=1024$, so the reach at the largest budgets is a lower bound and nothing was
measured beyond 1024 tokens; 8 halvings (bracket ratio at most 1.075); the generic relaxations of the distance probe.

\begin{figure}[h]
\centering
\includegraphics[width=\textwidth]{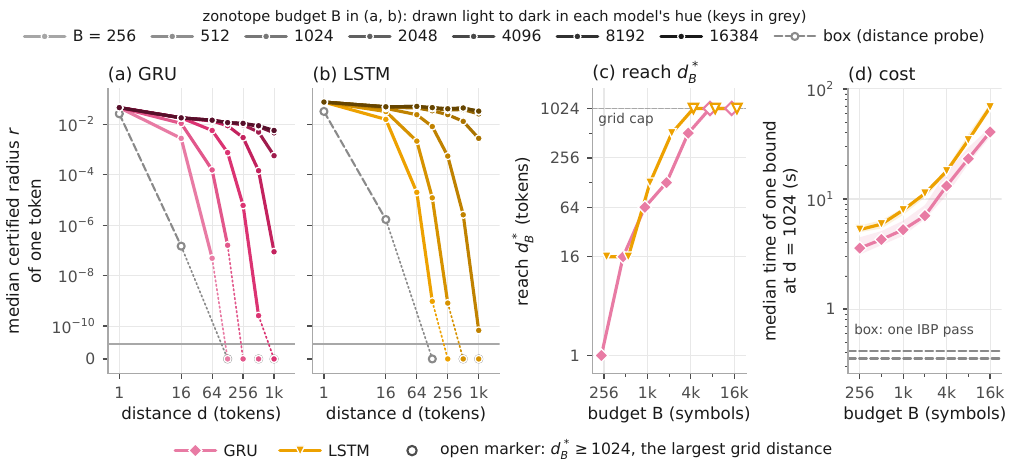}
\caption{The budget sweep of the zonotope on the naturally trained GRU and LSTM (seed~0). (a,~b)~Median certified radius $r(d)$
over the complete documents (GRU 49 of 50, LSTM 50 of 50) against the distance, one line per budget $B=256,\dots,16384$ from light
to dark; grey dashed: the box of the distance probe on the same documents; $r=0$ in the floor strip. (c)~Reach $d^*_B$ against $B$;
an open marker is the end of the grid,
$d^*_B\ge1024$; a fit of $\log d^*_B$ against $\log B$ below the cap gives exponents 2.1 (GRU) and 1.8 (LSTM), a description of the
quantized grid. (d)~Median time of one zonotope bound at $d=1024$ (band: interquartile range), from 3.58 to 40.6\,s (GRU) and from
5.27 to 68.2\,s (LSTM); grey dashed: one box bound on the same documents (0.418 and 0.356\,s); these times include the added
documents and come from machines of two CPU types, while the text and \cref{tab:sweepfull} give the times of the pre-registered
documents. Every pre-registered criterion passed for both models. A reach that grows with the budget does not show that some fixed-size domain certifies gated cells
uniformly in time.}
\label{fig:sweep}
\end{figure}

\clearpage
\section{The token-set certificate in full}\label{app:anyctx}

\paragraph{Setting.}
\Cref{cor:anyctx} certifies every context of every length over a token set $S$ by putting the box hull $E_S$ of the embeddings of
$S$ at every context position and running the K-uniform certificate of \cref{thm:hull} once per residue modulo the period. The test
used the period $P=8$ and a second arm with $P=64$, $P_0=8$ convolution-context positions and the layout
[$E_S^{\,P_0}\mid E_S^{\,P}\mid E_S^{\,r}\mid x$]. The token set $S$ is the $|S|$ most frequent tokens of the neutral evaluation text
(padding and unknown tokens excluded), $|S|\in\{1,2,4,8,16,32,64\}$; $|S|=1$ is the most frequent neutral token, repeated. The
document is either clean or in its own substitution box (box scale 1), and the class is the clean prediction of the bare document.
The test covered 61 checkpoints with 200 documents per arm (150 to 200 in the pre-registered reading; the extension of
\cref{app:cleanctx} completed the twelve shorter arms). For LSTM and GRU the invariant-box search (\cref{rem:invbox})
replaces the hull, so their rows are not instances of \cref{cor:anyctx}; their uniformity in the context is that of the invariant
box. For the CNN a window-set hull and for the bag-of-words model an exact closed form replace it. For the minGRU-type layer the
period part's threshold issue (\cref{rem:impl}) changed no certificate: the re-verification with outward rounding, which uses the
exact decay, certified the same documents (\cref{rem:float}). These token sets are frequent neutral tokens, not a model of real
contexts.

\begin{table}[h]
\caption{The token-set certificate: share of documents certified for every context of every length over $S$ ($P=8$ unless
marked; the column $P=64$ has a clean document), next to the clean accuracy of the model on the same 200 documents. Last column: the share certified at $L=0$ with the substitution box, an upper bound for the column ``subst.\ box''.
$^*$Near-constant classifier (clean accuracy 0.54, 80\% of its predictions in one class). $^\dagger$Degenerate under the test's own
rule (81\% of its predictions in one class). $^\ddagger$LSTM and GRU with their invariant box, not the token-set certificate.
Every certified share was re-verified with outward rounding (\cref{rem:float}), except on the 20 and 40 documents that the extension
of \cref{app:cleanctx} added to the two cells marked $^\P$, and none lost a document; $^\S$one document more
was certified with outward rounding than in float64 (0.755). Per checkpoint the 95\% Wilson interval of a share has a half-width of
at most 0.069 at $n=200$ (\cref{app:cleanctx-method}; those of the shares the text cites: \cref{tab:ci-b}). Fixed-point: fixed-point
fine-tuning.}
\label{tab:anyctx}
\centering\footnotesize
\setlength{\tabcolsep}{3pt}
\begin{tabular}{@{}lllccccccc@{}}
\toprule
 & & & clean & \multicolumn{3}{c}{$|S|=1$} & $|S|=8$ & ceiling\\
\cmidrule(lr){5-7}
model & training & seed & acc. & clean & subst.\ box & $P=64$ & clean & $L=0$, subst.\\
\midrule
S4D-type & std.\ IBP & 0 & 0.785 & 0.75 & 0.275 & 0.765$^\P$ & 0 & 0.495\\
 & & 1$^*$ & 0.54 & 0.935 & 0.86 & 0.94 & 0.015 & 0.925\\
 & & 2 & 0.66 & 0.53 & 0.415 & 0.535 & 0 & 0.605\\
 & fixed-point & 0 & 0.815 & 0.76$^\S$ & 0.44 & 0.765 & 0.075 & 0.665\\
 & & 1 & 0.765 & 0.94 & 0.615 & 0.94$^\P$ & 0.4 & 0.675\\
 & & 2 & 0.77 & 0.72 & 0.39 & 0.76 & 0 & 0.645\\
Mamba & std.\ IBP & 0 & 0.58 & 0 & 0 & 0 & 0 & 0.65\\
 & & 1 & 0.565 & 0.085 & 0.06 & 0.155 & 0 & 0.965\\
 & & 2 & 0.765 & 0 & 0 & 0 & 0 & 0.6\\
 & fixed-point & 0 & 0.565 & 0 & 0 & 0.005 & 0 & 0.985\\
 & & 1 & 0.745 & 0.16 & 0.1 & 0.68 & 0 & 0.645\\
 & & 2 & 0.785 & 0 & 0 & 0.01 & 0 & 0.66\\
minGRU-type & std.\ IBP & 0 & 0.67 & 0 & 0 & 0 & 0 & 0.805\\
 & & 1 & 0.685 & 0 & 0 & 0 & 0 & 0.705\\
 & & 2 & 0.67 & 0 & 0 & 0.025 & 0 & 0.59\\
 & fixed-point & 0 & 0.8 & 0 & 0 & 0 & 0 & 0.775\\
 & & 1 & 0.72 & 0 & 0 & 0.025 & 0 & 0.78\\
 & & 2$^\dagger$ & 0.63 & 0.25 & 0.225 & 0.395 & 0 & 0.96\\
all recurrent$^\ddagger$ & natural & all & 0.85--0.885 & 0 & 0 & 0 & 0 & 0\\
CNN & natural & 0 & 0.9 & 0.99 & 0 & 0.99 & 0 & 0\\
 & std.\ IBP & 0--2 & 0.795--0.825 & $0.960\pm0.017$ & $0.548\pm0.031$ & 0.960 & 0 & 0.558\\
byte Mamba (15) & all & all & & 0 & 0 & 0 & 0 & \\
\bottomrule
\end{tabular}
\end{table}

\paragraph{Group means.}
At $P=8$, $|S|=1$ and a clean document, over the non-degenerate seeds, the certified share was $0.738\pm0.203$ for the standard-IBP
and $0.807\pm0.117$ for the fixed-point fine-tuned S4D-type models (the near-constant standard-IBP seed~1 included, as the test's rule
admits it), $0.028\pm0.049$ and $0.053\pm0.092$ for the standard-IBP and fixed-point fine-tuned Mamba models ($0.232\pm0.388$ for the
fixed-point fine-tuned Mamba with $P=64$),
and 0 for the minGRU-type models. Every non-degenerate stateful model was certified on at most 5\% of the documents at $|S|=64$. The
stateless references certify at $|S|=1$ too (natural CNN 0.99, natural bag of words 0.52).

\paragraph{Reading.}
Over one repeated token, IBP certifies every context of every length for the certified-trained real-diagonal S4D-type models: on
53--94\% of the documents with a clean document and on 27.5--61.5\% with the document's own substitution box. The highest
standard-IBP values (93.5\% and 86\%) come from the near-constant classifier, which the degeneracy rule of the slow-channel
attribution excludes and the rule of this test admits; the substitution-box range above leaves its value out, and without it the
prediction for the S4D-type models still holds (seed~0: 75\% and 27.5\%; seed~2: 53\% and 41.5\%). For the certified-trained
selective models the share is 5.3\% or less on average at $P=8$, with a large spread between seeds: the second fixed-point fine-tuned
Mamba seed
reaches 16\% at $P=8$ and 68\% at $P=64$, the others at most 1\%. Natural nonnegative diagonal models certify nothing, the natural
selective ones vacuously (bound $-\infty$). For the selective models this is the slow-channel effect again: the hull contains the
clean trajectories from $0$ and from the fixed point, and slow channels with large DC levels make that segment long
(\cref{rem:price}). The natural gated models certify nothing with their invariant box, which is a statement about that
construction, not about \cref{cor:anyctx}. The price of ``any context'' is steep in $|S|$: at the 8 most frequent tokens only the
fixed-point fine-tuned S4D-type models keep up to 40\%, and at 64 every non-degenerate stateful model is at most 5\%. The models trained with the
cap and the hull term (\cref{app:remedy}), not part of this test, are certified over one repeated token ($|S|=1$) on 85.5\% [73.7,
97.3] (Mamba) and 75.7\% [63.2, 88.2] (minGRU-type) of 200 documents with the document in its substitution box (the 5.3\% above
is with a clean document), and on none at $|S|=8$ or 64.

\paragraph{What was predicted, and what failed.}
The test held: natural nonnegative diagonal models certify at most 5\% over one token, and the natural selective ones vacuously; the
certified-trained S4D-type models certify at least 20\% with a clean document and 10\% with the substitution box on two of three
seeds of each training kind; the share falls to at most 5\% at $|S|=64$ for every non-degenerate stateful model, and to at most half
its $|S|=1$ value where that is at least 0.1; the standard-IBP S4D-type model certifies more than both selective standard-IBP models
on every seed; the gated models certify at most 5\% in both modes; and the natural CNN, a stateless reference outside the reading,
certifies at least half of the documents (99\%). Four predictions failed: fixed-point fine-tuning was predicted to certify more than
standard IBP for the selective models on two of three seeds, and did so on one of three Mamba seeds and on neither of the two
minGRU-type pairs; a longer period was worse beyond the
tolerance of 2 points on one fixed-point fine-tuned S4D-type seed with the substitution box (58.1\% at $P=64$ on the first 160
documents against 61.5\% at $P=8$ on 200; on the arm completed by the extension (200 documents) the criterion held: both periods
certified the same 123 documents, 61.5\% [54.6, 68.0]);
and the byte-level fixed-point fine-tunes certified nothing with their $\ell_\infty$ box of radius 0.01 on the document, so they neither
reached the predicted 10\% nor beat the finite-horizon fine-tunes. The validity check first failed because seven byte-level runs
stopped at their time budget below 150 documents; a completion job brought each to at least 150 documents (150 to 200), the
check's minimum, with the criteria unchanged (\cref{app:prereg}), and the extension later brought every arm to 200
(\cref{app:cleanctx-record}).

\section{Cost}\label{app:wallclock}

\paragraph{Protocol.}
Natural word-level seed-0 models, float64 on a CPU with 8 threads, one batch of 10 documents with $\ell_\infty$ boxes of radius
$10^{-8}$ on every position (the cost does not depend on the radius), median of 3 timed repetitions after a warm-up; the period has
$P=1024$ tokens and $P_0=8$ convolution-context positions (\cref{tab:wallclock}).

\begin{table}[h]
\caption{Wall-clock time per batch of 10 documents, in seconds. $^*$LSTM: the invariant-box search (\cref{rem:invbox}).}
\label{tab:wallclock}
\centering\small
\begin{tabular}{@{}lcccccc@{}}
\toprule
 & \multicolumn{4}{c}{plain IBP} & $L=\infty$ certificate & K-uniform hull\\
\cmidrule(lr){2-5}
model & $L=0$ & $L=1024$ & $L=4096$ & $L=16384$ & (\cref{thm:inf}) & (\cref{thm:hull})\\
\midrule
Mamba & 0.47 & 1.30 & 3.84 & 16.29 & 1.73 & 2.12\\
S4D-type & 0.15 & 0.45 & 1.35 & 5.75 & 1.46 & 1.77\\
minGRU-type & 0.09 & 0.28 & 0.84 & 3.06 & 0.30 & 0.29\\
LSTM & 0.15 & 0.46 & 1.43 & 5.28 & 20.4$^*$ & n/a\\
\bottomrule
\end{tabular}
\end{table}

\paragraph{Reading.}
Plain IBP grows about linearly in $L$, and brute-force repetition of $K$ copies of a 1024-token period has the length $L=1024K$, so
it never reaches $L=\infty$. The $L=\infty$ certificate and the hull take a fixed time, whatever context length they cover: 10--31\%
of one plain pass at $L=16384$, or the cost of a plain pass over about 1.1--2.0k (Mamba, minGRU-type) and 4.4--5.3k (S4D-type)
context tokens, so no single equivalent context length is claimed. The hull costs about 1.2 times the $L=\infty$ certificate (0.96
times for the minGRU-type model, within timing noise). The LSTM's invariant-box search took 20.4\,s, consistent with its cap of 64
periods (\cref{rem:invbox}). The token-set certificate (one hull per residue modulo the period) was not timed.

\paragraph{Cost of the certificates, analysed.}
Brute-force repetition costs $O(KP)$ steps per layer and never reaches $K=\infty$; in trained models the relevant $K$ is of order
$1/\denh$, and radius time constants up to $1.24\times10^{18}$ tokens (finite) and infinite ones (at the synonym threat, for the
natural models) were measured (\cref{rem:timescales}, \cref{app:gate2,app:byte}). The $L=\infty$ certificate costs, per layer and channel,
two linear scans over the period (the centre from $0$, then the radius from $c^*$), $O(P)$ for each DC-normalised fixed point
(cumulative log-products and one softmax-weighted sum), and one (CR) scan over the document: $O\big(Q\,(P+n)\,d_iN\big)$ time,
independent of $K$, with $d_i$ inner channels of $N$ states each. The hull certificate adds one radius scan and one fixed point;
\cref{cor:anyctx} runs it once per residue, $P$ times. The period part depends only on $p$ and its boxes, not on the document, so it
can be computed once per period and reused across documents. The document part is a streaming recursion with $O(1)$ state per
channel and layer (the pair $(c,r)$ and the last $k_c-1$ convolution inputs), like the model itself. The implementation keeps the
period trajectories in memory ($O(P)$ per channel), because the next layer reads the phase boxes. In training, the $L=\infty$ term
costs these extra scans on the part of the batch that carries the prefix.

\section{Linear relaxations at finite context}\label{app:lirpa}

This appendix gives the full record of the comparison summarised in \cref{sec:exp-gate5} (``Linear relaxations at finite
context''): how much the linear-relaxation verifiers of auto\_LiRPA~0.7.0 \citep{xu2020} add to our interval certificates as
the context grows, at what time and memory, and up to which length or distance they run, on models, documents and perturbation
sets drawn from \cref{app:gate5,app:gate2} (listed below) and on the two models of the complex-diagonal control
(\cref{app:gate3-r2}).

\paragraph{Verifiers.}
Ours are the box (plain IBP), on the complex-diagonal control also the disk, and, under synonym substitution, the $L=\infty$
certificate and the K-uniform hull. From auto\_LiRPA we ran its IBP; CROWN-IBP \citep{zhang2020}, which bounds the
intermediate nodes by intervals and the margin by one backward linear pass; CROWN \citep{zhang2018}, which bounds every
intermediate node by its own backward linear pass; $\alpha$-CROWN \citep{xu2021} with 20 iterations at learning rate 0.1;
forward-mode linear relaxation \citep{xu2020}; and forward mode followed by a backward pass. auto\_LiRPA cannot trace our
models as written, so every model was rewritten as the same function of the same weights in operators that auto\_LiRPA bounds
(linear maps, products, sigmoid, tanh, exp, log, square root, reciprocal, ReLU, sums, slices and concatenations); on every
cell the rewritten network reproduced the model's clean margin $m^0$ to within $2.4\times10^{-14}(1+\abs{m^0})$. auto\_LiRPA sees
SiLU as a product with a sigmoid and the GRU update as products of independent intervals, where our box uses the exact ranges,
so its IBP is looser than ours wherever the two differ, and equal on the LSTM. The GRU ran in its own two-product form
$(1-z)n+zh$ and in the one-product form $n+z(h-n)$. Every verifier ran in float64 on CPU, on cluster nodes whose type was not
controlled; each auto\_LiRPA graph and its bounds ran in a separate process on four threads, with a memory limit and
caps of 30 minutes per graph build and per CROWN,
CROWN-IBP or $\alpha$-CROWN bound (15 minutes for the other methods); ours ran on one thread, so times of ours against
auto\_LiRPA's are descriptive.

\paragraph{Probes and run plan.}
The distance probe is that of \cref{app:gate5}: [1024 neutral tokens $\mid$ document], one token at distance
$d\in\{1,16,32,64,128,512,1024\}$ before the readout in the $\ell_\infty$ ball of its embedding, and the clean prediction as
the class. The certified radius is found by bisection in log space on $[10^{-12},10^2]$ after one evaluation at
$\eps=10^{-6}$, with 8 halvings (final bracket ratio at most 1.075), and with 6 for CROWN (ratio 1.33, so its radii lie on a
grid of ratio $10^{1/8}$). A bisection stopped by a non-finite bound, the memory limit or a time limit keeps
its last verified radius and is marked censored; the tests use complete radii, or the end $r_{\rm hi}$ of the final bracket,
an upper bound of the radius. The gated models ran on documents 0--19 on each of seeds 0, 1 and 2, the nonnegative diagonal
models on documents 0--9 on seed~0, and the complex-diagonal control on documents 0--9 on each of its two seeds. For the models
of \cref{app:gate5} these are the first of its 200 documents per seed, so medians over them differ from the medians there, while
the box radii agree document by document (``Validity'' below). CROWN ran up to $d=64$, on the
complex-diagonal control up to $d=128$: in a timing run one CROWN bound at $d=128$ took 8.4 minutes on the GRU (one document)
and did not finish within 15 minutes on Mamba. $\alpha$-CROWN ran at $d\le16$ on five documents (on Mamba at $d=16$ on one);
auto\_LiRPA's IBP and the forward--backward mode at $d\in\{1,16,128\}$, without bisection; on the gated models auto\_LiRPA's
IBP, forward and forward--backward mode on ten documents of seed~0, and CROWN-IBP and forward mode at $d\ge512$ on two (LSTM: up
to $d=512$); forward mode was run only up to $d=128$ on Mamba and on the complex-diagonal control.

Under synonym substitution the setting is that of \cref{app:gate2}: every perturbable position of [$L$ neutral tokens $\mid$
document] in its exact embedding box, and the true label. In this appendix $T$ is the length of that whole sequence, $L$ plus
the length of the document (elsewhere $T=512$ is the truncation length of the documents). The standard-IBP models (in this
appendix those of the first training study, \cref{app:gate2}: Mamba, S4D-type and minGRU-type, three seeds each) ran on documents
0--49 at $L\in\{0,256,1024,4096\}$ and on documents 50--199 at
$L=0$, auto\_LiRPA only at $L\le1024$: in a timing run its graph at $L=4096$ did not build within 30 minutes for the standard-IBP
Mamba and was still being built after more than 12 minutes for the S4D-type and minGRU-type models. Full CROWN ran on the first
30 test documents of at most 64 tokens at $L=0$ (seed~0 of each standard-IBP model, and the natural models), $\alpha$-CROWN on
those of the first ten with at most 48 tokens (three per model), and a doubling sweep of CROWN at $L\in\{0,64,192,448,960\}$ on
two documents of 58 and 64 tokens (the seed-0 standard-IBP models, the natural Mamba and LSTM) skipped every longer cell after
the first failure. The natural seed-0 models also ran at $L=0$ on documents 0--99.

\paragraph{Two soundness gaps of auto\_LiRPA 0.7.0.}
Both came out of one violation of the sampled-input check (\cref{app:audits}) by CROWN on the natural Mamba, one token
perturbed at distance 16, and both were explained from auto\_LiRPA's source before a correction was adopted.
(i)~\emph{The relaxation of $x^2$.} For an even power, auto\_LiRPA first sets the correct convex relaxation of $x^2$, with the
chord as upper line, but on every input interval $[l,u]$ that does not contain 0 a smoothness enhancement that is on by default
replaces the upper line by the tangent at the midpoint $m$, which lies below the function by $(x-m)^2$, up to $(u-l)^2/4$. On
$x\in[0.5,1.5]$ CROWN's upper bound of $x^2$ is 2.0 against a maximum of 2.25, and on 60 random boxes with half-widths from
$10^{-6}$ to 1 the bound
was unsound on 36, by up to 1.0. auto\_LiRPA switches to its sound square operator only when the exponent is a constant node,
which \texttt{torch.pow(x, 2)} does not trace to; we write the square in RMSNorm as $u\cdot u$, which auto\_LiRPA bounds with
that operator (exact interval, chord as upper line).
(ii)~\emph{Clamped chord slopes of exp, sigmoid and tanh.} auto\_LiRPA divides the chord slope of these functions by
$\max(u-l,10^{-8})$, so on an input interval narrower than $10^{-8}$ the upper line of exp, and of sigmoid and tanh on their
convex part below 0, is too flat by the factor $(u-l)/10^{-8}$ and misses $f(u)$ by about $f'(u)(u-l)$: on $[0,10^{-9}]$ the
upper line of exp has slope 0.1 and stays $9\times10^{-10}$ below $e^{u}$ at $u=10^{-9}$; its forward mode misses $e^u$ by
about $e^l\,10^{-12}$ at any width. In long-context recurrent models this matters: the perturbation of one token reaches the
later gate pre-activations attenuated, so CROWN relaxes thousands of exponentials on intervals far narrower than $10^{-8}$
(38{,}071 of the 45{,}696 exp inputs of the Mamba probe graph at $\eps=10^{-8}$), and there its certified lower bound of the
margin exceeded the margin of a sampled input in the box by $1.3\times10^{-8}$ to $2.7\times10^{-8}$ at $\eps$ from
$10^{-8}$ to $3\times10^{-6}$, at $\eps=10^{-8}$ 62\% of the margin's first-order variation. Wherever the input interval is
narrower than $10^{-8}$ we compute these relaxations on $[\min(l,u-10^{-8}),u]$ if $u\le0$ and on $[l,\max(u,l+10^{-8})]$
otherwise: the interval contains $[l,u]$, so every line remains a bound; it is wide enough that the clamp never applies; and
it lies on the same side of 0, so auto\_LiRPA's choice among its convex, concave and crossing cases does not change. With
this correction no sample violated a bound on the cell of the violation at any $\eps$ from $10^{-8}$ to $10^{-5}$, while
auto\_LiRPA as distributed still did (by $2.4\times10^{-8}$ at $\eps=10^{-7}$), and a diagnosis on every architecture with
CROWN, CROWN-IBP, forward and forward--backward mode at $d\in\{1,4,8,16\}$ gave 176 verdicts without a violation. Every
CROWN-type bound in this appendix carries the correction, so our CROWN-type numbers are those of auto\_LiRPA~0.7.0 with it and
with the square written as a product. The
other relaxations in our graphs are trusted on the evidence of the sampled-input and zero-width checks, which do not prove
soundness.

\paragraph{Validity.}
Every group passed every validity check. The rewritten networks reproduced the margins (above); auto\_LiRPA's
IBP equalled an interval evaluation of the same traced graph with auto\_LiRPA's rules to within $2\times10^{-11}(1+\abs{b})$, and
our box and IBP were never looser than that evaluation (on the LSTM equal to auto\_LiRPA's IBP). No sampled input had a margin
below any certified bound: under synonym substitution random substitutions, corners and points of the box, its gradient corner and
the clean input; on the probe random points and corners of the ball and its gradient corner at $\eps=10^{-6}$ and at the
certified radius. Every completed zero-width check (the point limit of \cref{app:audits}) returned the clean margin to within
$10^{-8}(1+\abs{m^0})$, for every method other than forward mode at every distance and for forward mode at $d=1$; for forward
mode at $d>1$ see the paragraph ``Forward mode at a zero-width input'' below. No auto\_LiRPA run ended in an error; time-outs, the memory limit and non-finite
bounds are outcomes, reported per method. Every CROWN-type bound used the chord correction. On the documents that both runs
share, the box radii reproduced those of \cref{app:gate5} on every complete cell (80 of 80 on seed~0 of the GRU and of the
LSTM, 60 of 60 on each of their seeds 1 and 2, 50 of 50 on Mamba and on the S4D-type model; no earlier records exist for the
minGRU-type model and the complex-diagonal control, and the GRU's one-product form had no box run here), and our certified
fractions on the standard-IBP models reproduced those of \cref{app:gate2} exactly at $L=0$, 1024, 4096 and $\infty$ and for the
hull (for Mamba and the S4D-type model in a separate re-evaluation of their checkpoints with the same code). One code version,
fixed before the production run, ran every task.

\paragraph{What was predicted, and what happened.}
The criteria were frozen with the code before the production run (\cref{app:prereg}).
\emph{Gated models} (\cref{tab:lirpagated,tab:lirpagatedcrit,tab:lirparatio}). The test predicted that CROWN-IBP collapses
like the box (median $r_{\rm hi}(128)/r(1)\le10^{-3}$); that CROWN keeps $r(64)/r(1)\ge0.1$ on at least 80\% of the documents,
for the GRU in the better of its two forms per document; and, on seed~0, that the time of one CROWN evaluation grows with
exponent at least 1.5 in $d$ and those of CROWN-IBP and forward mode with at most 1.3. The first two held on every seed,
the third on seed~0, where it is read.
\emph{Nonnegative diagonal models} (\cref{tab:lirpanonneg,tab:lirpanonnegratio,tab:lirpacost}). It predicted that CROWN-IBP,
forward mode and CROWN keep a median $r(64)/r(1)\ge0.1$; that the median ratio of CROWN's certified radius to the box's at $d=1$
is at least 10 (Mamba) or 2 (S4D-type, minGRU-type), and that the median change of that ratio from $d=1$ to $d=64$ is at most
threefold; and the same cost exponents. The radius and tightness
predictions held on the S4D-type and minGRU-type models and could not be assessed on Mamba, whose CROWN bound became
non-finite during the bisection in 39 of 40 cells; the CROWN cost exponent failed on all three models (Mamba 1.48,
minGRU-type 1.43, S4D-type 1.38).
\emph{Complex-diagonal control} (\cref{tab:lirpas4dc,tab:lirpacost}). It predicted that boxes and CROWN-IBP lose the radius by
$d=128$ (median $r_{\rm hi}(128)/r(1)\le10^{-3}$), that disks keep a median $r(1024)/r(1)\ge0.1$ and CROWN $r(128)/r(1)\ge0.1$
on at least 80\% of the documents, that the median time of one disk bound at $d=128$ is at most a tenth of that of one CROWN
bound, and that the disk's time grows at most as $d^{1.3}$. All six parts held on both seeds.
\emph{Synonym substitution} (\cref{tab:lirpasyn,tab:lirpasyndecl,tab:lirpasyncost,tab:lirpacrown}). For the standard-IBP
models it predicted that at $L=0$ neither CROWN-IBP nor IBP and CROWN-IBP together certify more than 3 points more than IBP,
where IBP leaves more than 3 points of headroom; that on documents classified correctly at both lengths, where IBP's certified
accuracy declines by at least 10 points from $L=0$ to $1024$, the best auto\_LiRPA certificate declines at least half as much
(Mamba, minGRU-type), and that the best auto\_LiRPA certificate at $L=1024$ keeps at least 85\% of its value at $L=0$
(S4D-type); and that at $L=1024$ the median time of one CROWN-IBP bound is at least five times that of one bound of
auto\_LiRPA's IBP, with an exponent in $T$ larger by at least 0.5. It predicted that full CROWN completes on at least 80\% of
the short documents within 30 minutes per bound, does not finish on at least 80\% of the cells at $T\ge512$, and grows with
exponent at least 1.5 in $T$; that $\alpha$-CROWN adds in median at most a fifth of CROWN's median gain over IBP; and that on
natural models CROWN-IBP certifies at most 1\% and CROWN at most 5\% of the documents. The first prediction held on all three
architectures: on every seed of the minGRU-type and S4D-type models, and on Mamba's seeds 0 and 2, where CROWN-IBP certified 9.0
and 9.0\% against IBP's 46.0 and 51.0\% (on seed~1 IBP leaves only 1.5 points of headroom); on the S4D-type seeds CROWN-IBP
certified 21.5, 48.0 and 25.0\% against 44.0, 50.5 and 38.5\%. The decline prediction held on the three minGRU-type seeds and on
Mamba's seed~2 (IBP fell 20.0 points, the best auto\_LiRPA certificate 28.6); on Mamba's seeds 0 and 1 IBP fell by less than 10
points (7.7 and 0.0), so the prediction did not apply there. On the S4D-type model the best auto\_LiRPA certificate kept at
least 85\% of its $L=0$ certificates at $L=1024$, on seeds 1 and 2 but not on seed~0 (\cref{tab:lirpasyndecl}). The cost
prediction failed on every seed of the three architectures. Full CROWN completed on every short document of the standard-IBP
S4D-type and minGRU-type models and of the natural GRU, LSTM and S4D-type models, on all but one of the natural minGRU-type model
(no finite bound there) and on only 5 of the 30 of the standard-IBP Mamba (median 1268\,s per bound, interquartile range
1013--1693\,s; the other 25 timed out), the one model on which that prediction failed; it returned no finite bound on the natural
Mamba. In the sweep it stopped at $T=250$ on the standard-IBP minGRU-type model and the natural LSTM and Mamba, at $T=506$
on the standard-IBP S4D-type model and at $T=58$ on the standard-IBP Mamba, and grew as $T^{2.04}$ (LSTM) and $T^{2.06}$
(minGRU-type) over $T=58$--128 (four bounds at two sweep lengths) and as $T^{1.92}$ (S4D-type) over $T=58$--256 (six bounds at
three lengths). $\alpha$-CROWN returned no bound, so its prediction could not be assessed, and no verifier certified any document
of a natural model.

\begin{table}[htbp]
\caption{Gated cells: median certified radius $r(d)$ of one token at distance $d$ (20 documents per seed; $\alpha$-CROWN on
five; forward mode on ten, two at $d\ge512$; CROWN-IBP at $d\ge512$ on two; the box at $d\ge512$ on ten). 0: no radius
certified on any document; --: not run; off: forward mode's zero-width bound missed the clean margin (\cref{tab:lirpafwd}) on every document, or on the
number given, and no forward-mode radius is given at that distance; nf: no finite bound; OOM: out of memory (about 25.5\,GB).
Zonotopes as in \cref{app:gate5} and, for $B=16384$, the budget sweep (\cref{app:sweep}) as it was before its extension to 50
documents, on the same documents (at $d=64$ on
16; $B=16384$ on 15 for the GRU and 16 for the LSTM). $^*$On 5 of 20 documents the box certified about $10^{-12}$, on the others nothing.
$^\dagger$One of the cell's ten zero-width bounds lay above the clean margin, by less than the off threshold (\cref{tab:lirpafwd}); a
lower bound above the margin at the clean input is unsound in floating point, so the cell gives no radius, although the registered
rule gives it one.}
\label{tab:lirpagated}
\centering\footnotesize
\setlength{\tabcolsep}{3pt}
\begin{tabular}{@{}llcrrrrrrr@{}}
\toprule
model & method & seed & $r(1)$ & $r(16)$ & $r(32)$ & $r(64)$ & $r(128)$ & $r(512)$ & $r(1024)$\\
\midrule
GRU & box & 0 & 0.0255 & $1.22{\times}10^{-7}$ & 0$^*$ & 0 & 0 & 0 & 0\\
 & CROWN-IBP & 0 & 0.0238 & $5.48{\times}10^{-9}$ & 0 & 0 & 0 & 0 & 0\\
 & \quad one-product form & 0 & 0.0047 & 0 & -- & 0 & 0 & -- & --\\
 & CROWN & 0 & 0.100 & 0.0422 & 0.0422 & 0.0422 & -- & -- & --\\
 & \quad one-product form & 0 & 0.075 & 0.0369 & -- & 0.0422 & -- & -- & --\\
 & $\alpha$-CROWN & 0 & 0.115 & OOM & -- & -- & -- & -- & --\\
 & forward & 0 & 0.0523 & no radius$^{\dagger}$ & off & off & off & off & off\\
 & zonotope $B=2048$ & 0 & 0.0488 & 0.0190 & -- & 0.0133 & 0.00847 & $1.59{\times}10^{-4}$ & $1.51{\times}10^{-7}$\\
 & zonotope $B=16384$ & 0 & 0.0453 & 0.0178 & -- & 0.0143 & 0.0115 & 0.00866 & 0.00649\\
 & box & 1 & 0.0264 & $6.05{\times}10^{-8}$ & -- & 0 & 0 & -- & --\\
 & CROWN-IBP & 1 & 0.0255 & $2.95{\times}10^{-9}$ & -- & 0 & 0 & -- & --\\
 & CROWN & 1 & 0.100 & 0.0422 & -- & 0.0422 & -- & -- & --\\
 & box & 2 & 0.0205 & $1.10{\times}10^{-7}$ & -- & 0 & 0 & -- & --\\
 & CROWN-IBP & 2 & 0.0213 & $6.65{\times}10^{-9}$ & -- & 0 & 0 & -- & --\\
 & CROWN & 2 & 0.100 & 0.0422 & -- & 0.0422 & -- & -- & --\\
LSTM & box & 0 & 0.0340 & $2.64{\times}10^{-6}$ & $1.78{\times}10^{-10}$ & 0 & 0 & 0 & 0\\
 & CROWN-IBP & 0 & 0.0470 & $3.79{\times}10^{-6}$ & $2.53{\times}10^{-10}$ & 0 & 0 & 0 & --\\
 & CROWN & 0 & 0.133 & 0.0656 & 0.075 & 0.0656 & -- & -- & --\\
 & $\alpha$-CROWN & 0 & 0.221 & OOM & -- & -- & -- & -- & --\\
 & forward & 0 & 0.0836 & $7.52{\times}10^{-4}$ & off (5 of 10) & off & nf & nf & --\\
 & zonotope $B=2048$ & 0 & 0.0783 & 0.0482 & -- & 0.0453 & 0.0383 & 0.0174 & 0.00345\\
 & zonotope $B=16384$ & 0 & 0.075 & 0.0437 & -- & 0.0523 & 0.0453 & 0.0437 & 0.0407\\
 & box & 1 & 0.0379 & $5.25{\times}10^{-6}$ & -- & 0 & 0 & -- & --\\
 & CROWN-IBP & 1 & 0.0523 & $7.52{\times}10^{-6}$ & -- & 0 & 0 & -- & --\\
 & CROWN & 1 & 0.133 & 0.100 & -- & 0.075 & -- & -- & --\\
 & box & 2 & 0.0407 & $2.55{\times}10^{-6}$ & -- & 0 & 0 & -- & --\\
 & CROWN-IBP & 2 & 0.0523 & $3.92{\times}10^{-6}$ & -- & 0 & 0 & -- & --\\
 & CROWN & 2 & 0.133 & 0.075 & -- & 0.0656 & -- & -- & --\\
\bottomrule
\end{tabular}
\end{table}

\begin{table}[htbp]
\caption{Gated cells, the predicted quantities. CROWN-IBP: median $r_{\rm hi}(128)/r(1)$ (predicted at most $10^{-3}$); since
CROWN-IBP certified no radius at $d=128$ on any document, $r_{\rm hi}=10^{-12}$, the end of the search range, and the value is
an upper bound. CROWN: share of the documents with $r(64)/r(1)\ge0.1$ (predicted at least 80\%; Wilson 95\% interval) and the
medians of $r(64)/r(1)$ and of $r_{\rm hi}(64)/r(1)$. Exponent: least-squares slope of the logarithm of the time of one
evaluation against $\log d$ over the single evaluations at $d\ge16$ (predicted: CROWN at least 1.5, CROWN-IBP and forward mode at
most 1.3; seed~0 only). For the GRU the test took the better form per document; its values equal those of the two-product
form.}
\label{tab:lirpagatedcrit}
\centering\small
\setlength{\tabcolsep}{4pt}
\begin{tabular}{@{}llcccccc@{}}
\toprule
 & & CROWN-IBP & \multicolumn{3}{c}{CROWN} & \multicolumn{2}{c}{exponent in $d$}\\
\cmidrule(lr){3-3}\cmidrule(lr){4-6}\cmidrule(l){7-8}
model & seed & $r_{\rm hi}(128)/r(1)$ & share $\ge0.1$ & $r(64)/r(1)$ & $r_{\rm hi}(64)/r(1)$ & CROWN & CROWN-IBP; forward\\
\midrule
GRU & 0 & $4.23{\times}10^{-11}$ & 1.00 [0.84, 1.00] & 0.492 & 0.656 & 1.96 & 0.86; 0.98\\
 & 0, one product & $2.13{\times}10^{-10}$ & 1.00 [0.84, 1.00] & 0.562 & 0.750 & -- & --\\
 & 1 & $3.92{\times}10^{-11}$ & 1.00 [0.84, 1.00] & 0.422 & 0.562 & & \\
 & 2 & $4.70{\times}10^{-11}$ & 1.00 [0.84, 1.00] & 0.492 & 0.656 & & \\
LSTM & 0 & $2.13{\times}10^{-11}$ & 1.00 [0.84, 1.00] & 0.562 & 0.750 & 1.95 & 0.96; 0.96\\
 & 1 & $1.91{\times}10^{-11}$ & 1.00 [0.84, 1.00] & 0.562 & 0.750 & & \\
 & 2 & $1.91{\times}10^{-11}$ & 1.00 [0.84, 1.00] & 0.492 & 0.656 & & \\
\bottomrule
\end{tabular}
\end{table}

\begin{table}[htbp]
\caption{Gated cells: median per-document ratio of certified radii (over the documents where both methods certify) and the
median looseness $R(10^{-6})$ of the box at $d=1$. For the GRU's one-product form, which has no box run, the box is that of
\cref{app:gate5}. CROWN's $R(10^{-6})$ was 1 at every distance; forward mode certified 2.75 (GRU) and 2.37 (LSTM) times the
box's radius at $d=1$ and, on the LSTM, 246 times at $d=16$ (on the GRU that cell gives no radius, \cref{tab:lirpagated}), and
$\alpha$-CROWN 5.62 and 4.22 times at $d=1$ (seed~0).}
\label{tab:lirparatio}
\centering\footnotesize
\setlength{\tabcolsep}{3pt}
\begin{tabular}{@{}llccccccc@{}}
\toprule
 & & \multicolumn{2}{c}{CROWN / box} & \multicolumn{2}{c}{CROWN-IBP / box} & \multicolumn{2}{c}{CROWN / zonotope at $d=1$; 16; 64} & box $R(10^{-6})$\\
\cmidrule(lr){3-4}\cmidrule(lr){5-6}\cmidrule(lr){7-8}
model & seed & $d=1$ & $d=16$ & $d=1$ & $d=16$ & $B=2048$ & $B=16384$ & at $d=1$\\
\midrule
GRU & 0 & 3.92 & $3.43{\times}10^{5}$ & 1.00 & 0.0488 & 1.98; 2.46; 3.16 & 2.05; 2.55; 2.94 & 34.2\\
 & 0, one product & 2.88 & $3.26{\times}10^{5}$ & 0.173 & 0 & 1.45; 2.04; 2.94 & 1.54; 2.05; 2.74 & --\\
 & 1 & 3.79 & $7.05{\times}10^{5}$ & 1.00 & 0.0487 & 2.03; 2.33; -- & -- & 42.0\\
 & 2 & 4.37 & $3.86{\times}10^{5}$ & 1.07 & 0.0673 & 2.06; 2.31; -- & -- & 47.4\\
LSTM & 0 & 3.92 & $2.94{\times}10^{4}$ & 1.49 & 1.49 & 1.70; 1.42; 1.65 & 1.78; 1.54; 1.43 & 41.1\\
 & 1 & 3.65 & $1.73{\times}10^{4}$ & 1.43 & 1.43 & 1.74; 1.56; -- & -- & 37.9\\
 & 2 & 3.40 & $2.94{\times}10^{4}$ & 1.43 & 1.38 & 1.64; 1.49; -- & -- & 35.4\\
\bottomrule
\end{tabular}
\end{table}

\begin{table}[htbp]
\caption{Nonnegative diagonal models (seed~0, 10 documents; Mamba's forward mode at $d=1$ on 9): median certified radius
$r(d)$ and the median of $r(64)/r(1)$ over the documents. n/a: not assessable (on Mamba CROWN's bound became non-finite during
the bisection in 39 of its 40 cells, which leaves one complete radius of ten, at $d=1$, and none at $d\ge16$). --: not run.}
\label{tab:lirpanonneg}
\centering\footnotesize
\setlength{\tabcolsep}{2pt}
\begin{tabular}{@{}llrrrrrrrc@{}}
\toprule
model & method & $r(1)$ & $r(16)$ & $r(32)$ & $r(64)$ & $r(128)$ & $r(512)$ & $r(1024)$ & $r(64)/r(1)$\\
\midrule
Mamba & box & $5.45{\times}10^{-5}$ & $5.05{\times}10^{-5}$ & $7.78{\times}10^{-5}$ & $1.54{\times}10^{-4}$ & $2.13{\times}10^{-4}$ & $5.43{\times}10^{-4}$ & $6.04{\times}10^{-4}$ & 2.46\\
 & CROWN-IBP & $4.85{\times}10^{-5}$ & $4.07{\times}10^{-5}$ & $6.04{\times}10^{-5}$ & $1.08{\times}10^{-4}$ & $1.54{\times}10^{-4}$ & $3.40{\times}10^{-4}$ & $3.52{\times}10^{-4}$ & 2.37\\
 & forward & $2.74{\times}10^{-3}$ & $2.46{\times}10^{-3}$ & $3.16{\times}10^{-3}$ & $4.23{\times}10^{-3}$ & $4.87{\times}10^{-3}$ & -- & -- & 1.54\\
 & CROWN & n/a & n/a & n/a & n/a & -- & -- & -- & n/a\\
minGRU-type & box & $8.66{\times}10^{-4}$ & $4.58{\times}10^{-4}$ & $5.68{\times}10^{-4}$ & $6.98{\times}10^{-4}$ & $8.02{\times}10^{-4}$ & $1.98{\times}10^{-3}$ & $2.64{\times}10^{-3}$ & 0.778\\
 & CROWN-IBP & $5.83{\times}10^{-4}$ & $3.41{\times}10^{-4}$ & $4.07{\times}10^{-4}$ & $4.54{\times}10^{-4}$ & $5.45{\times}10^{-4}$ & $1.11{\times}10^{-3}$ & $1.43{\times}10^{-3}$ & 0.778\\
 & forward & $8.98{\times}10^{-3}$ & $7.52{\times}10^{-3}$ & $8.06{\times}10^{-3}$ & $8.66{\times}10^{-3}$ & $9.70{\times}10^{-3}$ & 0.0154 & 0.0165 & 1.00\\
 & CROWN & 0.0422 & 0.0422 & 0.0422 & 0.0422 & -- & -- & -- & 1.00\\
S4D-type & box & 0.00250 & 0.00840 & 0.0112 & 0.0149 & 0.0201 & 0.0422 & 0.0649 & 7.00\\
 & CROWN-IBP & 0.00172 & 0.00525 & 0.00649 & 0.00849 & 0.0107 & 0.0223 & 0.0352 & 4.88\\
 & forward & 0.0138 & 0.0237 & 0.0275 & 0.0307 & 0.0328 & 0.0470 & OOM & 2.05\\
 & CROWN & 0.0422 & 0.0562 & 0.0656 & 0.0656 & -- & -- & -- & 1.33\\
\bottomrule
\end{tabular}
\end{table}

\begin{table}[htbp]
\caption{Nonnegative diagonal models: median per-document ratio of CROWN's certified radius to the box's at each distance
(interquartile range in brackets), the median change of that ratio from $d=1$ to $d=64$ (predicted at most 3), the range over
$d$ of the median ratios of CROWN-IBP and of forward mode to the box, and the box's own $r(1024)/r(1)$. On Mamba only one of
the ten documents has a complete CROWN radius, too few to assess the ratio (n/a).}
\label{tab:lirpanonnegratio}
\centering\footnotesize
\setlength{\tabcolsep}{3pt}
\begin{tabular}{@{}lcccccccc@{}}
\toprule
 & \multicolumn{4}{c}{CROWN / box at $d=$} & & \multicolumn{2}{c}{range over $d$} & \\
\cmidrule(lr){2-5}\cmidrule(lr){7-8}
model & 1 & 16 & 32 & 64 & change & CROWN-IBP / box & forward / box & box $r(1024)/r(1)$\\
\midrule
Mamba & n/a & -- & -- & -- & -- & 0.60--0.84 & 25.5--48.8 & 11.4\\
minGRU-type & 44.0 [34.6, 60.9] & 94.0 & 77.8 & 60.6 [45.3, 81.2] & 1.44 & 0.58--0.70 & 6.49--16.0 & 2.94\\
S4D-type & 19.8 [14.9, 47.3] & 6.77 & 5.83 & 4.37 [2.70, 12.1] & 0.230 & 0.58--0.70 & 1.15--5.83 & 26.4\\
\bottomrule
\end{tabular}
\end{table}

\begin{table}[htbp]
\caption{Complex-diagonal control (two seeds, 10 documents each): median certified radius $r(d)$, with the median $r(d)/r(1)$ in
brackets; seed~0 above seed~1. $r=0$: no radius certified on any document; off: forward mode's zero-width bound missed the
clean margin on every document (on 9 of 10 at $d=64$ on seed~1), and no forward-mode radius is given at that distance
(\cref{tab:lirpafwd}); nf: no finite bound on any document; --: not run. $\alpha$-CROWN ($d\le16$, five documents) gave no complete radius: its bisection ended non-finite in every cell.}
\label{tab:lirpas4dc}
\centering\footnotesize
\setlength{\tabcolsep}{3pt}
\begin{tabular}{@{}lcccccc@{}}
\toprule
$d$ & seed & box & disk & CROWN-IBP & forward & CROWN\\
\midrule
1 & 0 & $1.92{\times}10^{-3}$ & $1.84{\times}10^{-3}$ & $1.38{\times}10^{-3}$ & 0.0133 & 0.0422\\
 & 1 & $1.29{\times}10^{-3}$ & $1.24{\times}10^{-3}$ & $9.31{\times}10^{-4}$ & 0.0104 & 0.0422\\
16 & 0 & $7.05{\times}10^{-5}$ [0.038] & $1.62{\times}10^{-3}$ [0.97] & $5.13{\times}10^{-5}$ [0.041] & $2.86{\times}10^{-3}$ [0.24] & 0.0562 [1.33]\\
 & 1 & $3.93{\times}10^{-5}$ [0.034] & $9.33{\times}10^{-4}$ [0.75] & $3.40{\times}10^{-5}$ [0.037] & $2.21{\times}10^{-3}$ [0.22] & 0.0562 [1.33]\\
32 & 0 & $7.23{\times}10^{-7}$ [$3.5{\times}10^{-4}$] & $1.65{\times}10^{-3}$ [0.93] & $5.10{\times}10^{-7}$ [$3.7{\times}10^{-4}$] & $2.74{\times}10^{-4}$ [0.021] & 0.0562 [1.39]\\
 & 1 & $4.97{\times}10^{-7}$ [$3.9{\times}10^{-4}$] & $1.20{\times}10^{-3}$ [0.97] & $4.11{\times}10^{-7}$ [$4.1{\times}10^{-4}$] & $2.55{\times}10^{-4}$ [0.024] & 0.0562 [1.56]\\
64 & 0 & $3.25{\times}10^{-11}$ [$2.1{\times}10^{-8}$] & $1.73{\times}10^{-3}$ [1.20] & $2.35{\times}10^{-11}$ [$2.3{\times}10^{-8}$] & off & 0.0562 [1.33]\\
 & 1 & $2.84{\times}10^{-11}$ [$1.7{\times}10^{-8}$] & $1.44{\times}10^{-3}$ [1.04] & $2.29{\times}10^{-11}$ [$2.0{\times}10^{-8}$] & off & 0.0562 [1.56]\\
128 & 0 & 0 & $3.41{\times}10^{-3}$ [1.78] & 0 & off & 0.075 [1.56]\\
 & 1 & 0 & $2.94{\times}10^{-3}$ [2.13] & 0 & off & 0.075 [2.37]\\
512 & 0 & 0 & 0.0154 [8.98] & 0 & -- & --\\
 & 1 & 0 & 0.0154 [14.3] & 0 & -- & --\\
1024 & 0 & 0 & 0.0317 [24.0] & nf & -- & --\\
 & 1 & 0 & 0.0422 [28.4] & nf & -- & --\\
\bottomrule
\end{tabular}
\end{table}

\begin{table}[htbp]
\caption{Time of one evaluation at $\eps=10^{-6}$ (median over the documents, seconds; seed~0) and the least-squares exponent of
that time in $d$ over all single evaluations at $d\ge16$. Ours (box, disk) on one thread, auto\_LiRPA on four. $^*$Includes
cells whose zero-width bound missed the clean margin (at $d=32$ on the LSTM half of them); their evaluation at $10^{-6}$ is
kept. $^\dagger$The bisection ended non-finite after this evaluation (Mamba's CROWN: in 39 of its 40 cells). nf: no finite
bound; OOM: out of memory; --: not run.
Predicted on the gated and nonnegative diagonal models: CROWN at least 1.5, CROWN-IBP and forward mode at most 1.3; on the
complex-diagonal control: the disk at most 1.3.}
\label{tab:lirpacost}
\centering\footnotesize
\setlength{\tabcolsep}{3pt}
\begin{tabular}{@{}llrrrrrrrc@{}}
\toprule
model & method & $d=1$ & 16 & 32 & 64 & 128 & 512 & 1024 & exponent\\
\midrule
GRU & box & 0.235 & 0.235 & 0.234 & 0.234 & 0.233 & 0.237 & 0.231 & 0.004\\
 & CROWN-IBP & 0.0248 & 0.161 & 0.313 & 0.462 & 0.899 & 4.26 & 8.79 & 0.86\\
 & forward & 0.0308 & 0.256 & 0.490$^*$ & 1.01$^*$ & 1.89$^*$ & 7.43$^*$ & 16.7$^*$ & 0.98\\
 & CROWN & 0.144 & 10.8 & 41.0 & 160 & -- & -- & -- & 1.96\\
 & $\alpha$-CROWN & 6.47 & OOM & -- & -- & -- & -- & -- & --\\
LSTM & box & 0.220 & 0.220 & 0.224 & 0.221 & 0.221 & 0.222 & 0.223 & $-0.009$\\
 & CROWN-IBP & 0.0305 & 0.207 & 0.396 & 0.793 & 1.49 & 7.32 & -- & 0.96\\
 & forward & 0.0361 & 0.301 & 0.581$^*$ & 1.14$^*$ & nf & nf & -- & 0.96\\
 & CROWN & 0.189 & 12.2 & 45.8 & 181 & -- & -- & -- & 1.95\\
 & $\alpha$-CROWN & 8.72 & OOM & -- & -- & -- & -- & -- & --\\
Mamba & box & 0.253 & 0.252 & 0.253 & 0.254 & 0.254 & 0.257 & 0.269 & 0.011\\
 & CROWN-IBP & 0.0383 & 0.118 & 0.186 & 0.352 & 0.675 & 2.83 & 5.64 & 0.94\\
 & forward & 0.0645 & 0.236 & 0.461 & 0.902 & 1.61 & -- & -- & 0.96\\
 & CROWN & 3.07$^\dagger$ & 71.3$^\dagger$ & 187$^\dagger$ & 570$^\dagger$ & -- & -- & -- & 1.48\\
minGRU-type & box & 0.100 & 0.101 & 0.101 & 0.102 & 0.103 & 0.103 & 0.103 & 0.008\\
 & CROWN-IBP & 0.0291 & 0.0667 & 0.110 & 0.183 & 0.342 & 1.37 & 2.99 & 0.91\\
 & forward & 0.0371 & 0.105 & 0.187 & 0.338 & 0.627 & 2.33 & 4.89 & 0.92\\
 & CROWN & 0.406 & 6.08 & 14.8 & 44.7 & -- & -- & -- & 1.43\\
 & $\alpha$-CROWN & 24.5$^\dagger$ & OOM & -- & -- & -- & -- & -- & --\\
S4D-type & box & 0.102 & 0.103 & 0.104 & 0.103 & 0.104 & 0.106 & 0.108 & 0.014\\
 & CROWN-IBP & 0.0240 & 0.0465 & 0.0667 & 0.117 & 0.212 & 0.852 & 1.78 & 0.90\\
 & forward & 0.0344 & 0.0888 & 0.160 & 0.291 & 0.588 & 2.23 & OOM & 0.93\\
 & CROWN & 0.223 & 2.11 & 5.16 & 14.4 & -- & -- & -- & 1.38\\
 & $\alpha$-CROWN & 9.37$^\dagger$ & 75.9$^\dagger$ & -- & -- & -- & -- & -- & --\\
complex-diag. & box & 0.121 & 0.120 & 0.125 & 0.126 & 0.126 & 0.126 & 0.131 & 0.005\\
 & disk & 0.0946 & 0.0939 & 0.0929 & 0.0969 & 0.0975 & 0.102 & 0.105 & 0.014\\
 & CROWN-IBP & 0.0268 & 0.0666 & 0.111 & 0.194 & 0.366 & 1.53 & nf & 0.91\\
 & forward & 0.0414 & 0.148 & 0.265 & 0.552$^*$ & 1.02$^*$ & -- & -- & 0.97\\
 & CROWN & 0.262 & 2.59 & 7.22 & 23.7 & 66.8 & -- & -- & 1.58\\
 & $\alpha$-CROWN & 10.3$^\dagger$ & 100$^\dagger$ & -- & -- & -- & -- & -- & --\\
\bottomrule
\end{tabular}
\end{table}

\begin{table}[htbp]
\caption{Forward mode at a zero-width input: its zero-width bound minus the clean margin, $b(0)-m^0$ in margin units,
quartiles over the cells with a finite $b(0)$ (seed~0 unless marked); off: more than $10^{-8}(1+\abs{m^0})$ away (no radius);
above: $b(0)>m^0+10^{-9}(1+\abs{m^0})$. No sampled input at $\eps=10^{-6}$ fell below these cells' bounds at that $\eps$. The rows
not shown (the GRU at $d=1$, the LSTM at $d\le16$, the complex-diagonal control at $d\le32$, forward--backward at $d\le16$) had
no cell off or above. On the nonnegative diagonal models no cell was off or above at any distance (gaps at most
$3.5\times10^{-12}(1+\abs{m^0})$). $^*$The other 8 cells had no finite zero-width bound; all 10 ended non-finite.}
\label{tab:lirpafwd}
\centering\footnotesize
\setlength{\tabcolsep}{4pt}
\begin{tabular}{@{}lllrrcr@{}}
\toprule
model & method & $d$ & cells & off & $b(0)-m^0$: q25 / median / q75 & above\\
\midrule
GRU & forward & 16 & 10 & 0 & $-1.48{\times}10^{-9}$ / $-3.96{\times}10^{-10}$ / $1.95{\times}10^{-9}$ & 1\\
 & & 32 & 10 & 10 & $-0.0232$ / $-0.0043$ / $-0.00152$ & 1\\
 & & 64 & 10 & 10 & $-2.10{\times}10^{3}$ / $-1.77{\times}10^{3}$ / $-327$ & 1\\
 & & 128 & 10 & 10 & $-2.75{\times}10^{3}$ / $-2.53{\times}10^{3}$ / $-636$ & 1\\
 & & 512 & 2 & 2 & $-8.91{\times}10^{3}$ / $-8.81{\times}10^{3}$ / $-8.70{\times}10^{3}$ & 0\\
 & & 1024 & 2 & 2 & $-1.62{\times}10^{4}$ / $-1.62{\times}10^{4}$ / $-1.61{\times}10^{4}$ & 0\\
 & forward--backward & 128 & 10 & 10 & $-2.55{\times}10^{3}$ / $-2.38{\times}10^{3}$ / $-591$ & 0\\
LSTM & forward & 32 & 10 & 5 & $-1.34{\times}10^{-7}$ / $-5.74{\times}10^{-8}$ / $-3.12{\times}10^{-8}$ & 0\\
 & & 64 & 10 & 10 & $-55.5$ / $-42.2$ / $-0.11$ & 1\\
 & & 128 & 2$^*$ & 2 & $-2.98$ / $-2.52$ / $-2.06$ & 0\\
 & forward--backward & 128 & 2$^*$ & 2 & $-7.10$ / $-5.26$ / $-3.43$ & 0\\
complex-diag., seed 0 & forward & 64 & 10 & 10 & $-3.15{\times}10^{-7}$ / $2.35{\times}10^{-7}$ / $7.76{\times}10^{-7}$ & 6\\
 & & 128 & 10 & 10 & $-4.61{\times}10^{5}$ / $-7.78$ / $2.2$ & 3\\
 & forward--backward & 128 & 10 & 10 & $-2.33{\times}10^{5}$ / $-7.82$ / $-2.37$ & 2\\
complex-diag., seed 1 & forward & 64 & 10 & 9 & $-6.04{\times}10^{-8}$ / $1.96{\times}10^{-7}$ / $1.27{\times}10^{-6}$ & 7\\
 & & 128 & 10 & 10 & $-7.70$ / $-7.30$ / $-6.80$ & 0\\
 & forward--backward & 128 & 10 & 10 & $-7.92$ / $-7.30$ / $-6.97$ & 0\\
\bottomrule
\end{tabular}
\end{table}

\paragraph{Forward mode at a zero-width input.}
At a zero-width input (the point limit of \cref{app:audits}) these relaxations are exact in exact arithmetic, so each should
return the clean margin $m^0$ up to rounding, and a sound lower bound can never exceed it. On the traced graphs of the gated
models and of the complex-diagonal control, auto\_LiRPA's IBP and CROWN-IBP did so to within $10^{-13}$ (quartiles) at every
distance up to 1024, and CROWN's zero-width bound to within $2.4\times10^{-14}$; forward mode did not (\cref{tab:lirpafwd}). On
the GRU its zero-width bound missed $m^0$ on every document from $d=32$ on; on the LSTM on half of the documents at $d=32$ and
on all at $d=64$, and from $d=128$ on every cell ended non-finite (at $d=128$ only 2 of the 10 zero-width bounds of each mode
were finite, and all were off); on the complex-diagonal control on 59 of its 60 cells at $d\ge64$. Most of these bounds lie
below $m^0$, a loss of precision, but 23 lay above $m^0$: one GRU document at each of $d=16$, 32, 64 and 128, one LSTM document
at $d=64$, and on the complex-diagonal control 13 of the 20 documents at $d=64$ and five cells at $d=128$, three of them by about
2.9 or more. A lower bound above the margin at the clean input is a floating-point soundness failure of forward mode at that
input. A cell has no radius when its zero-width bound is more than $10^{-8}(1+\abs{m^0})$ off, and we give no forward-mode
radius at a distance with such a cell; the GRU cell at $d=16$ lay above $m^0$ by less than that, so the registered rule gives it a
radius, but a lower bound above the margin at the clean input is unsound, and we give it none (\cref{tab:lirpagated}).
A consistent reading, not a proof: forward mode computes the lower and the upper linear bound of each node by
different floating-point expressions, so at a zero-width input their concretisations differ by rounding, and that width grows
along the recursion with the per-step factor of the interval map, as a box does; on the complex-diagonal control the
interquartile range of $b(0)-m^0$ grew by a factor of 1.28 to 1.39 per token from $d=16$ to 64, close to the largest per-step
box factor of its modes, 1.39. The error is two-sided, not only a widening. We do not claim that forward mode is unsound in
exact arithmetic, nor that the gap lies in the graph, on which auto\_LiRPA's IBP and CROWN-IBP are exact.

\begin{table}[htbp]
\caption{Standard-IBP models under synonym substitution (Mamba, minGRU-type and S4D-type, three seeds each): certified accuracy
(\%) of our IBP, auto\_LiRPA's IBP and CROWN-IBP, and the better of the two auto\_LiRPA certificates, with the clean accuracy.
Top: $L=0$, documents 0--199, with the share of documents on which CROWN-IBP's bound exceeds ours and our $L=\infty$ certificate
and K-uniform hull. Bottom: documents 0--49 at every $L$. auto\_LiRPA did not run at $L=4096$. At every $L$ the documents
certified by ours or by any auto\_LiRPA certificate are exactly those certified by ours. Seed~1 of Mamba and of the S4D-type model
is close to a constant classifier (\cref{tab:trainseeds}).}
\label{tab:lirpasyn}
\centering\footnotesize
\setlength{\tabcolsep}{4pt}
\begin{tabular}{@{}lllrrrrrrrr@{}}
\toprule
model & seed & & clean & ours & auto\_LiRPA IBP & CROWN-IBP & best & CROWN-IBP above & $L=\infty$ & hull\\
\midrule
Mamba & 0 & $L=0$ & 58.0 & 46.0 & 10.5 & 9.0 & 11.0 & 0.0 & 0.5 & 0.0\\
 & 1 & $L=0$ & 56.5 & 55.0 & 18.0 & 15.0 & 18.5 & 1.0 & 6.5 & 6.0\\
 & 2 & $L=0$ & 76.5 & 51.0 & 12.5 & 9.0 & 13.5 & 0.0 & 3.0 & 0.0\\
minGRU-type & 0 & $L=0$ & 67.0 & 59.0 & 16.5 & 17.5 & 17.5 & 1.0 & 2.5 & 1.0\\
 & 1 & $L=0$ & 68.5 & 56.5 & 25.5 & 28.0 & 28.5 & 2.0 & 2.0 & 1.5\\
 & 2 & $L=0$ & 67.0 & 46.0 & 29.5 & 33.5 & 33.5 & 9.0 & 0.0 & 0.0\\
S4D-type & 0 & $L=0$ & 78.5 & 44.0 & 20.0 & 21.5 & 23.0 & 1.5 & 33.5 & 30.0\\
 & 1 & $L=0$ & 54.0 & 50.5 & 48.0 & 48.0 & 48.5 & 4.5 & 51.0 & 48.5\\
 & 2 & $L=0$ & 66.0 & 38.5 & 22.5 & 25.0 & 25.0 & 2.0 & 39.0 & 26.5\\
\midrule
Mamba & 0 & $L=0$ & 56 & 50 & 6 & 8 & 8 & & & \\
 & & $L=256$ & 54 & 48 & 2 & 2 & 2 & & & \\
 & & $L=1024$ & 56 & 42 & 0 & 0 & 0 & & & \\
 & & $L=4096$ & 56 & 18 & -- & -- & -- & & & \\
 & 1 & $L=0$ & 54 & 52 & 18 & 12 & 18 & & & \\
 & & $L=256$ & 54 & 52 & 10 & 8 & 10 & & & \\
 & & $L=1024$ & 54 & 52 & 8 & 6 & 8 & & & \\
 & & $L=4096$ & 54 & 22 & -- & -- & -- & & & \\
 & 2 & $L=0$ & 80 & 52 & 16 & 10 & 20 & & & \\
 & & $L=256$ & 76 & 50 & 6 & 4 & 6 & & & \\
 & & $L=1024$ & 74 & 34 & 0 & 0 & 0 & & & \\
 & & $L=4096$ & 56 & 18 & -- & -- & -- & & & \\
minGRU-type & 0 & $L=0$ & 68 & 62 & 14 & 16 & 16 & & & \\
 & & $L=256$ & 68 & 58 & 0 & 0 & 0 & & & \\
 & & $L=1024$ & 66 & 42 & 0 & 0 & 0 & & & \\
 & & $L=4096$ & 64 & 26 & -- & -- & -- & & & \\
 & 1 & $L=0$ & 78 & 70 & 32 & 34 & 34 & & & \\
 & & $L=256$ & 74 & 58 & 18 & 20 & 20 & & & \\
 & & $L=1024$ & 72 & 48 & 4 & 6 & 6 & & & \\
 & & $L=4096$ & 70 & 38 & -- & -- & -- & & & \\
 & 2 & $L=0$ & 72 & 48 & 32 & 38 & 38 & & & \\
 & & $L=256$ & 70 & 34 & 22 & 24 & 24 & & & \\
 & & $L=1024$ & 68 & 28 & 14 & 14 & 14 & & & \\
 & & $L=4096$ & 62 & 16 & -- & -- & -- & & & \\
S4D-type & 0 & $L=0$ & 78 & 30 & 16 & 20 & 20 & & & \\
 & & $L=256$ & 76 & 26 & 14 & 12 & 14 & & & \\
 & & $L=1024$ & 74 & 26 & 12 & 12 & 12 & & & \\
 & & $L=4096$ & 72 & 26 & -- & -- & -- & & & \\
 & 1 & $L=0$ & 52 & 46 & 42 & 42 & 42 & & & \\
 & & $L=256$ & 48 & 48 & 44 & 46 & 46 & & & \\
 & & $L=1024$ & 46 & 46 & 44 & 46 & 46 & & & \\
 & & $L=4096$ & 46 & 46 & -- & -- & -- & & & \\
 & 2 & $L=0$ & 58 & 34 & 26 & 28 & 28 & & & \\
 & & $L=256$ & 54 & 36 & 24 & 28 & 28 & & & \\
 & & $L=1024$ & 48 & 38 & 24 & 26 & 26 & & & \\
 & & $L=4096$ & 48 & 40 & -- & -- & -- & & & \\
\bottomrule
\end{tabular}
\end{table}

\begin{table}[htbp]
\caption{From $L=0$ to $L=1024$ on the standard-IBP models, on the documents of 0--49 that every method bounded at both lengths
and the model classified correctly at both. Top (Mamba, minGRU-type): number of documents, decline of our certified accuracy and
of the best auto\_LiRPA certificate (points; predicted: at least half of ours where ours is at least 10; n/a: ours declined by
less than 10, so the prediction does not apply), and a 90\% document-bootstrap interval of the best decline minus half of ours.
Bottom (S4D-type): our certified accuracy and the best auto\_LiRPA certificate at $L=0$ and at $L=1024$ (\%; predicted: the
best keeps at least 85\% of its $L=0$ value), and a 90\% interval of the best at $L=1024$ minus 0.85 times the best at $L=0$
(points). Seed~1 of Mamba and of the S4D-type model is close to a constant classifier.}
\label{tab:lirpasyndecl}
\centering\small
\begin{tabular}{@{}llcccc@{}}
\toprule
model & seed & documents & ours & best auto\_LiRPA & 90\% interval\\
\midrule
Mamba & 0 & 26 & 7.7 (n/a) & 15.4 & [$-1.9$, 25.0]\\
 & 1 & 27 & 0.0 (n/a) & 18.5 & [7.4, 29.6]\\
 & 2 & 35 & 20.0 & 28.6 & [7.1, 31.4]\\
minGRU-type & 0 & 33 & 27.3 & 21.2 & [$-4.5$, 21.2]\\
 & 1 & 36 & 25.0 & 38.9 & [11.1, 41.7]\\
 & 2 & 31 & 25.8 & 32.3 & [4.8, 35.5]\\
\midrule
 & & & ours, $L=0\to1024$ & best, $L=0\to1024$ & \\
\cmidrule(lr){4-5}
S4D-type & 0 & 36 & 38.9 $\to$ 36.1 & 27.8 $\to$ 16.7 & [$-15.3$, 0.1]\\
 & 1 & 23 & 95.7 $\to$ 100 & 87.0 $\to$ 100 & [18.7, 37.2]\\
 & 2 & 22 & 77.3 $\to$ 86.4 & 63.6 $\to$ 59.1 & [$-3.4$, 10.9]\\
\bottomrule
\end{tabular}
\end{table}

\begin{table}[htbp]
\caption{Standard-IBP models: median time per document (s) of one bound against the context length $L$ (median length $T$ of
[$L$ neutral tokens $\mid$ document] in brackets, the same for every model), seeds 0 / 1 / 2: our IBP (one thread),
auto\_LiRPA's IBP and CROWN-IBP (four threads), the build of auto\_LiRPA's graph, and CROWN-IBP's peak memory (GB; above the
process's resident memory in brackets). Below each model: the median ratio of one CROWN-IBP bound to one bound of auto\_LiRPA's
IBP at $L=1024$ (predicted at least 5) and the exponents of the bound times in $T$ (predicted: CROWN-IBP's at least 0.5 above
that of auto\_LiRPA's IBP). auto\_LiRPA ran without a failure up to $T=1536$ on every model.}
\label{tab:lirpasyncost}
\centering\footnotesize
\setlength{\tabcolsep}{3pt}
\begin{tabular}{@{}lccccc@{}}
\toprule
$L$ ($T$) & ours & auto\_LiRPA IBP & CROWN-IBP & graph build & CROWN-IBP memory\\
\midrule
\multicolumn{6}{@{}l}{\emph{Mamba}}\\
0 (231) & 0.052 / 0.053 / 0.052 & 0.77 / 0.77 / 0.75 & 1.34 / 1.35 / 1.32 & 11.9 / 11.8 / 11.6 & 1.09 / 1.08 / 1.08 (0.21 / 0.20 / 0.20)\\
256 (480) & 0.18 / 0.20 / 0.18 & 1.99 / 1.95 / 1.89 & 3.59 / 3.60 / 3.32 & 41.3 / 41.5 / 42.4 & 1.68 / 1.61 / 1.66 (0.46 / 0.48 / 0.48)\\
1024 (1248) & 0.53 / 0.54 / 0.52 & 5.54 / 5.46 / 5.33 & 10.2 / 10.2 / 9.68 & 387 / 395 / 412 & 3.96 / 4.37 / 3.97 (1.96 / 2.43 / 1.97)\\
4096 (4320) & 1.9 / 2.0 / 2.1 & -- & -- & -- & --\\
\multicolumn{6}{@{}l}{CROWN-IBP / auto\_LiRPA IBP at $L=1024$: 1.86 / 1.86 / 1.84}\\
\multicolumn{6}{@{}l}{exponents in $T$: ours 1.18--1.21, auto\_LiRPA IBP 1.11--1.12, CROWN-IBP 1.12--1.14}\\
\midrule
\multicolumn{6}{@{}l}{\emph{minGRU-type}}\\
0 (231) & 0.018 / 0.017 / 0.017 & 0.37 / 0.37 / 0.38 & 0.57 / 0.59 / 0.61 & 5.3 / 5.3 / 5.4 & 0.81 / 0.82 / 0.83 (0.07)\\
256 (480) & 0.034 / 0.033 / 0.034 & 0.79 / 0.76 / 0.79 & 1.29 / 1.23 / 1.26 & 16.8 / 15.5 / 17.1 & 1.09 / 0.94 / 1.07 (0.26 / 0.13 / 0.21)\\
1024 (1248) & 0.085 / 0.084 / 0.087 & 2.12 / 2.10 / 2.12 & 3.61 / 3.49 / 3.55 & 219 / 203 / 219 & 3.27 / 3.95 / 3.66 (2.12 / 2.85 / 2.49)\\
4096 (4320) & 0.31 / 0.31 / 0.31 & -- & -- & -- & --\\
\multicolumn{6}{@{}l}{CROWN-IBP / auto\_LiRPA IBP at $L=1024$: 1.69 / 1.69 / 1.69}\\
\multicolumn{6}{@{}l}{exponents in $T$: ours 0.96, auto\_LiRPA IBP 1.01--1.02, CROWN-IBP 1.03--1.04}\\
\midrule
\multicolumn{6}{@{}l}{\emph{S4D-type}}\\
0 (231) & 0.018 / 0.019 / 0.018 & 0.26 / 0.27 / 0.26 & 0.37 / 0.39 / 0.38 & 2.9 / 2.9 / 2.8 & 0.86 / 0.86 / 0.87 (0.10)\\
256 (480) & 0.035 / 0.038 / 0.037 & 0.51 / 0.55 / 0.55 & 0.76 / 0.79 / 0.80 & 7.9 / 8.4 / 8.4 & 1.28 / 1.23 / 1.28 (0.38 / 0.36 / 0.39)\\
1024 (1248) & 0.096 / 0.095 / 0.095 & 1.41 / 1.44 / 1.43 & 2.12 / 2.17 / 2.18 & 45.1 / 66.0 / 53.2 & 4.16 / 3.92 / 4.16 (2.76 / 2.65 / 2.80)\\
4096 (4320) & 0.34 / 0.35 / 0.34 & -- & -- & -- & --\\
\multicolumn{6}{@{}l}{CROWN-IBP / auto\_LiRPA IBP at $L=1024$: 1.52 / 1.52 / 1.54}\\
\multicolumn{6}{@{}l}{exponents in $T$: ours 0.98--1.00, auto\_LiRPA IBP 0.98--0.99, CROWN-IBP 0.99--1.01}\\
\bottomrule
\end{tabular}
\end{table}

\begin{table}[htbp]
\caption{Full CROWN under synonym substitution (seed~0 of every model). Top: the first 30 test documents of at most 64 tokens at
$L=0$ (median 56 tokens), time limit 30 minutes per bound: clean accuracy, documents on which CROWN completed, CROWN's certified
accuracy (\%), the
median time of one CROWN bound, its exponent in $T$ over these documents, and $\alpha$-CROWN on three documents. Bottom: the
doubling sweep on two documents of 58 and 64 tokens: status per length $T$ (a timeout skips every longer cell), CROWN's time per
bound, and its exponent in $T$ over the completed cells (two lengths; on the S4D-type model three). nf: no finite bound; --: none
(no exponent is given for the five completions of the standard-IBP Mamba, since the 30-minute limit censors its other 25
documents).}
\label{tab:lirpacrown}
\centering\footnotesize
\setlength{\tabcolsep}{3pt}
\begin{tabular}{@{}lccccccc@{}}
\toprule
model & training & clean & completed & certified & time per bound (s) & exponent & $\alpha$-CROWN\\
\midrule
Mamba & standard IBP & 63.3 & 5 (timeout 25) & 6.7 (IBP 50.0) & 1268 & -- & timeout 1, OOM 2\\
minGRU-type & standard IBP & 63.3 & 30 of 30 & 43.3 (IBP 56.7) & 155 & 2.21 & OOM 3\\
S4D-type & standard IBP & 76.7 & 30 of 30 & 36.7 (IBP 60.0) & 49.3 & 1.71 & OOM 3\\
GRU & natural & 76.7 & 30 of 30 & 0 & 143 & 2.16 & OOM 3\\
LSTM & natural & 80.0 & 30 of 30 & 0 & 127 & 1.94 & OOM 3\\
Mamba & natural & 76.7 & 0 (nf 30) & 0 & -- & -- & OOM 1, nf 2\\
minGRU-type & natural & 73.3 & 29 (nf 1) & 0 & 153 & 1.86 & OOM 3\\
S4D-type & natural & 80.0 & 30 of 30 & 0 & 57.5 & 1.61 & OOM 2, nf 1\\
\midrule
 & & $T\approx64$ & $T\approx128$ & $T\approx256$ & $T\approx512$, 1024 & time per bound (s) & exponent\\
\cmidrule(l){3-8}
Mamba & standard IBP & timeout & skipped & skipped & skipped & -- & --\\
minGRU-type & standard IBP & ok & ok & timeout & skipped & 164; 721 & 2.06\\
S4D-type & standard IBP & ok & ok & ok & timeout ($T=506$) & 73.8; 327; 1145 & 1.92\\
LSTM & natural & ok & ok & timeout & skipped & 167; 720 & 2.04\\
Mamba & natural & nf & nf & timeout & skipped & -- & --\\
\bottomrule
\end{tabular}
\end{table}

\paragraph{Reading, as pre-registered.}
\emph{Gated cells.} The collapse of the box with the distance is a failure of interval intermediate bounds: CROWN-IBP, a linear
final bound over box intermediate bounds, inherits it (no radius from $d=64$ on, on any document of any seed), while full CROWN
keeps a tenth of its adjacent-token radius at $d=64$ on every document, in both forms of the GRU, at a cost that grows
superlinearly in $d$ (exponents 1.96 and 1.95 on seed~0; one bound at $d=128$ took 8.4 minutes in a timing run on one GRU
document). Whether CROWN keeps the radius at $d=128$ or 1024 was not measured. The model's own two-product form of the GRU was
the better form for CROWN-IBP and for CROWN; the one-product form was faster for CROWN (118 against 160\,s per bound at
$d=64$ on seed~0), by a margin within the spread between runs (see Limits). On the GRU, CROWN-IBP certified 5--7\% of the box's radius
at $d=16$, consistent with auto\_LiRPA's intermediate boxes treating the products of the update as independent intervals,
where our box uses their exact range (we did not test the cause): a linear final bound over interval intermediate bounds can
lose to IBP. CROWN certified 1.45 to 2.1 times the radius of the zonotopes of \cref{app:gate5} at $d=1$ (1.6 to 2.1 in the
GRU's own two-product form and on the LSTM), where both are exact to first order, so the difference comes from higher-order
terms at the radius, and 2.9 (GRU) and 1.4 (LSTM) times that of the $B=16384$ zonotope at $d=64$.
\emph{Nonnegative transitions.} On the S4D-type and minGRU-type models the box, CROWN-IBP, forward mode and CROWN kept their
radii to $d=64$, and CROWN was tighter than the box by a factor whose median change from $d=1$ to 64 was at most 1.44-fold
(minGRU-type; limit 3); on the S4D-type model it fell, from 19.8 at $d=1$ to 4.37 at $d=64$, since the box's radius grows with
$d$ and CROWN's hardly does. CROWN-IBP certified 0.58--0.84 of the box's radius in median on all three models although it is
far less loose at $\eps=10^{-6}$: first-order tightness at a small $\eps$ did not carry over to the radius, consistent with
auto\_LiRPA's interval bounds, looser than ours, dominating there (not tested); per document it was sometimes the tighter one.
On Mamba CROWN could not be assessed (one complete radius of ten). CROWN's time grew with exponent 1.38--1.48 in $d$ over
$d=16$--64, below the predicted 1.5, consistent with a part of one CROWN evaluation that does not grow with $d$ still weighing
at these distances (we did not test this).
\emph{Complex-diagonal control.} Boxes and CROWN-IBP lose the radius within 128 tokens (wrapping); disks keep it to $d=1024$ and
CROWN to $d=128$, the largest distance it ran, with a median radius of 0.075 against 0.0034 and 0.0029 for the disks. At
$d=128$ one disk bound took 0.098 and 0.092\,s on one thread and one CROWN bound 66.8 and 72.8\,s on four, and the disk's time
did not grow with $d$, since it runs over the whole sequence at every $d$. Disks are not as tight as CROWN; their advantage is
their cost.
\emph{Synonym substitution.} On every seed of the standard-IBP Mamba, S4D-type and minGRU-type models CROWN-IBP certified fewer
documents than IBP at $L=0$, and none that IBP did not; since its intermediate boxes are looser than ours, this compares IBP with
one linear relaxation, not with the best one. On the minGRU-type models, on the documents classified correctly at both lengths,
the decline of the certified accuracy from $L=0$ to $1024$ is not removed by a linear final bound over interval intermediate
bounds (CROWN-IBP); whether full CROWN, which completed only up to $T=128$ there, would remove it is open. On the S4D-type model
the best auto\_LiRPA certificate kept at least 85\% of its $L=0$ certificates at $L=1024$, on two of three seeds: trivially on
seed~1, which is close to a constant classifier, and on seed~2 with a 90\% interval that includes 0; on seed~0 it fell from 27.8
to 16.7\% while ours went from 38.9 to 36.1\%. On Mamba, per seed: on seed~2 IBP fell 20.0 points and the best auto\_LiRPA
certificate 28.6, as predicted; on seeds 0 and 1 IBP fell 7.7 and 0.0 points, too little for the prediction to apply, and the
best auto\_LiRPA certificate 15.4 and 18.5. Within auto\_LiRPA one CROWN-IBP bound cost 1.69 (minGRU-type), 1.84--1.86 (Mamba)
and 1.52--1.54 (S4D-type) times one bound of its IBP at $L=1024$, with the same growth in $T$, and the graph build dominated both.
The K-uniform hull, valid for every number of copies of the period, took 0.099\,s per document on one thread on the minGRU-type
model, against 1.23--1.29\,s for one CROWN-IBP bound at $L=256$ after a build of 15.5--17.1\,s on four; on Mamba 0.64--0.65\,s,
against 3.32--3.60\,s after 41.3--42.4\,s, and on the S4D-type model 0.56--0.57\,s, against 0.76--0.80\,s after 7.9--8.4\,s. Full
CROWN completed on the short documents of the minGRU-type and S4D-type models and stopped at $T=250$ and $T=506$ in the sweep; on
Mamba it completed on only 5 of the 30 short documents within 30 minutes per bound, not in order of length (it timed out on a
document of 39 tokens and completed on one of 58), and stopped at $T=58$ in the sweep. Beyond these lengths only IBP-type bounds
(ours to $T\approx4600$, auto\_LiRPA's to $T=1536$) and our length-independent certificates were obtained. We do not say that
CROWN cannot be engineered to run longer.

\paragraph{Limits.}
auto\_LiRPA~0.7.0 as distributed on the rewritten models, with the chord correction and the square written as a product; SiLU
as a product with a sigmoid (an exact SiLU relaxation could be tighter); float64 without directed rounding for every method;
$\alpha$-CROWN with 20 iterations and no branch and bound, so no complete verifier; full CROWN only on short documents at $L=0$
and in the sweep on two documents of 58 and 64 tokens with neutral prefixes, a biased sample of reviews; the nonnegative
diagonal models on one seed with 10 documents, the gated models on 20 documents per seed, the complex-diagonal control on 10;
CROWN only up to $d=64$ on the gated and nonnegative diagonal models (128 on the complex-diagonal control); times that depend on
the node, its load and the thread count (tasks shared nodes: CROWN at $d=64$ took 160--229\,s per bound in the GRU's own form
across its three seeds, against 127\,s alone in the timing run, and 118\,s in the one-product form), on CPU nodes whose type was
not controlled, ours on one thread and auto\_LiRPA on four; the union of certificates is a combination, not one verifier; under
synonym substitution, of the standard-IBP models the Mamba, S4D-type and minGRU-type ones are reported (the LSTM and GRU are
constant classifiers, \cref{app:gate2}); and POPQORN, Cert-RNN and Prover \citep{ko2019,du2021,ryou2021} were not run.

\section{The training remedy for the blind spot in full}\label{app:remedy}

This appendix gives the full record of the second pre-registered training study summarised in \cref{sec:exp-train}: whether IBP
training of the selective models reaches a competitive clean accuracy under synonym substitution, whether the blind spot of
\cref{prop:horizon} survives such a recipe, and which remedy removes it, at what price. Criteria and outcomes:
\cref{app:prereg-criteria}.

\paragraph{Setting.}
The threat, data, models and certificates are those of the first training study (\cref{app:gate2}): word-level IMDB truncated at
512 tokens, synonym substitution at box scale 1 at every position of the context and of the document, two-layer models of width 64
($N=16$), and float64 certificates (their re-verification with outward rounding is stated at the end of this appendix). Two
things differ. The last 2500 training documents form a
development split that is never trained on, so training uses 22{,}500 documents; and every model uses the direct decay path of
\cref{rem:impl}. A selection stage chose the recipe of the main stage on the development split with seed~0 among four schedules of
21{,}000 steps (batch 64), by a rule fixed in advance: eligible if Mamba and the minGRU-type model each reach at least 82\% clean
accuracy on the development split, then the highest mean certified share.
\begin{enumerate}
\item 5250 clean steps, then a geometric ramp of the box scale from $10^{-3}$ to 1 over two thirds of the remaining steps; $\kappa$ to
0.5; one-cycle learning rate $2\times10^{-3}$, gradient clip 1.
\item 1750 clean steps, then a linear ramp over 14{,}000 steps; the rest as in 1.
\item No clean steps; linear ramps of the box scale and of $\kappa$ (to 0.8) over 14{,}000 steps, then 7000 steps at full scale;
constant learning rate $10^{-3}$, gradient clip 0.25: the recipe of \citet{jia2019}.
\item As 3 with $\kappa$ to 0.5.
\end{enumerate}
All four were eligible: mean clean accuracy 84.3, 86.6, 83.6 and 84.7\%, mean certified share at $L=0$ 65.7, 72.1, 75.2 and 72.6\%.
The rule selected the third, the \emph{linear-ramp recipe} (Mamba 84.3 / 75.6\%, minGRU-type 83.0 / 74.8\%; one seed, development
split). We kept AdamW with weight decay 0.05, no dropout, no early stopping and the substitution sets of \cref{sec:setting}. These
differ from those of \citet{jia2019}, whose threat is smaller, so their numbers are not a baseline here.

\paragraph{Arms.}
Seeds 1--3 of each arm, evaluated on the first 1000 test documents:
\begin{itemize}
\item \emph{first recipe}: standard IBP as in the first training study (\cref{app:gate2}; 8000 steps, 2000 clean steps, a geometric
ramp, $\kappa$ to 0.5), for Mamba, the minGRU-type and the S4D-type models;
\item \emph{linear ramp}: standard IBP with the linear-ramp recipe, the same three models;
\item and, for Mamba and the minGRU-type model only:
  \begin{itemize}
  \item \emph{cap 128, 512, 2048}: the linear-ramp recipe with every decay multiplied by $\rho=1-1/\tau_{\max}$;
  \item \emph{fixed-point term} (the fixed-point term from the start): the linear-ramp recipe with the $L=\infty$ term of
  \cref{cor:fixed} on half of every robust batch from the first robust step (training periods of 256 tokens from the GNU manuals, at
  the document's box scale); it trains from scratch and is not the fixed-point fine-tuning of \cref{app:gate2};
  \item \emph{cap + hull}: the cap at $\tau_{\max}=512$ with the K-uniform hull of \cref{thm:hull} in place of the fixed-point term;
  \item \emph{hull}: the hull term without the cap;
  \item \emph{natural}, with and without the cap.
  \end{itemize}
\end{itemize}
For a given seed, ``cap + hull'', ``cap 512'' and ``hull'' start from the same weights: the cap is a fixed reparametrisation and leaves
the initialisation unchanged. ``Cap + hull'' and ``hull'' also see the same batches. A smoke test at the production slope fixed in
advance whether the hull term needs a weight ramp: a ramp would have been used if the median robust loss of its rows over the first
50 robust steps had exceeded 5. It did not, so the term entered at full weight. ``First study'' denotes the standard-IBP
checkpoints of the first training study (\cref{app:gate2}), re-evaluated (the minGRU-type ones on the direct decay path).

\paragraph{The cap.}
Multiplying every decay by $\rho$ shifts the log decay by the constant $\log(1-1/\tau_{\max})$. This is a fixed monotone
reparametrisation of the gate, so the interval of a capped decay over a gate box is still the image of the box's endpoints, and IBP
stays the exact-range interval extension. The injection is unchanged ($\Delta\,x\,B$ for Mamba, $x\,B$ for the S4D-type layer), and the
minGRU-type layer keeps its convex form, with complement $\sigma(s)+\sigma(-s)/\tau_{\max}$. Every decay is at most $1-1/\tau_{\max}$,
so every channel keeps at most the fraction $1-1/\tau_{\max}$ of its state per token, and each channel's state retains at most
$(1-1/\tau_{\max})^d$ of any contribution after $d$ tokens;
information goes''). It is the certified network. Its float32 complements $1-\bar a$ are at least
$1/\tau_{\max}\ge1/2048>4.8\times10^{-4}$, far above the float32 resolution near 1, so none of its decays rounds to 1 in float32
(\cref{rem:float}). The time constant of the per-period contraction is at most $-1/\log(1-1/\tau_{\max})<\tau_{\max}$; over a
period of $P=1024$ tokens, $\pih\le(1-1/512)^{1024}\approx e^{-2}$ at $\tau_{\max}=512$.

\paragraph{Evaluation and statistics.}
Each checkpoint was evaluated as in \cref{app:gate2}, on GPUs in float64, with the neutral prefixes and the 1024-token period of
that study. In addition we record:
\begin{itemize}
\item the certified share after $K=1$ and $K=4$ copies of the period read from the zero state;
\item the clean accuracy at every context;
\item the attack on the first 100 documents at $L=0$;
\item the time constants and the counterfactual of \cref{app:gate2} (200 documents);
\item the token-set certificate of \cref{app:anyctx} (200 documents; $P=8$; the 1, 8 and 64 most frequent neutral tokens; a clean
document and the document in its substitution box).
\end{itemize}
A checkpoint is \emph{degenerate} if its clean accuracy is below 60\% or more than 80\% of its predictions fall in one class. This
rule, fixed for this study on the 1000 test documents, is stricter than that of the attribution in \cref{app:gate2} (clean accuracy
below 55\% or more than 80\% of its predictions in one class, on its 200 documents): the first study's Mamba seed~0 (57.2\% clean)
is degenerate here and not there. Clean and $L=0$
shares average every seed; the context shares, $Y$ (the $L=0$ share minus the K-uniform share, per seed), the ratios, the flip
shares, the token sets and the counterfactual average the non-degenerate seeds. Every mean comes with its 95\% t-interval over the
seeds (and is marked when fewer than three seeds enter); a three-seed t-interval of a share can extend beyond $[0,100]$, and the
per-seed values are in \cref{tab:remedy-seeds}.
Differences carry Welch intervals: an arm against the linear-ramp recipe,
and capped against uncapped natural training. The seed-paired contrasts carry paired t-intervals. Per checkpoint, the intervals are
95\% bootstrap intervals over the documents (2000 resamples). A ratio is the certified share at $L=\infty$ (or K-uniformly) over the
share at $L=0$; its \emph{width-only} version counts only documents whose clean prediction after the context is still correct. A
\emph{flip} is a document whose clean prediction after the context is wrong; no box certifies it.

\paragraph{Validity and training health.}
Every check passed. The smoke test passed every step. Every planned run (78) finished with its planned arguments, and every
evaluation is complete. The observed time constants of every capped model are within the cap's bound (for the natural capped models
at the bound itself, 127.5, 511.5 and 2047.5 tokens on every seed). No per-document implication is violated on 87 checkpoints. The
internal checks of the token-set certificate and of the attribution are clean; the token-set summary also prints the validity check
and the criteria of the earlier token-set test (\cref{app:anyctx}), which refer to that test's checkpoints and do not apply here, and
it counts a bound that is not a number as not certified, which can only lower a share (such bounds occur only for uncapped
checkpoints). The re-evaluated Mamba and S4D-type checkpoints of the first study reproduce its counts within two documents of 1000. No
run skipped a step or overflowed in its gradients. One first-recipe Mamba run (seed~2) excluded 112 batch rows from its robust term
in 5 steps because their logit bounds overflowed, and one Mamba run with the fixed-point term (seed~1) one row in one step. Each run
with a prefix term had 671{,}968 prefix rows; one of them was excluded, in that same Mamba run, and none in any other. By the runs' own timers, training
took 25.5 / 22.2 / 13.6 minutes per run for the first recipe (Mamba / minGRU-type / S4D-type), 72--76 / 57--72 / 43 minutes for the
linear-ramp recipe and the caps, 123--149 minutes with a prefix term, and 2--2.5 minutes for natural training
(\cref{tab:remedy-clean}); the main stage took about 75 GPU-hours of training in all, and the selection stage about 9 more.

\begin{table}[htbp]
\caption{IBP training under synonym substitution with the linear-ramp recipe and the remedies (\cref{sec:exp-train}): clean accuracy
and certified accuracy in \% on 1000 test documents (box scale 1, real-arithmetic certificates in float64), mean over three seeds
with its 95\% t-interval over the seeds. Clean and $L=0$ average every seed, the other columns the non-degenerate ones. $L=1024$:
after 1024 tokens of contiguous neutral text; $K=1$: after one copy of the 1024-token period; K-uniform: for every number of copies
(\cref{thm:hull}); $Y$: $L=0$ minus K-uniform, in points. Which certificates of this appendix were re-verified with outward
rounding: its last paragraph.}
\label{tab:remedy}
\centering\scriptsize\setlength{\tabcolsep}{2pt}
\begin{tabular}{@{}lccccccc@{}}
\toprule
training & clean & $L=0$ & $L=1024$ & $K=1$ & $L=\infty$ & K-uniform & $Y$\\
\midrule
\multicolumn{8}{@{}l}{\emph{Mamba}}\\
first recipe & 60.0 [42.5, 77.6] & 48.0 [29.5, 66.4] & \multicolumn{5}{c}{two non-degenerate seeds (\cref{tab:remedy-fin,tab:remedy-inf})}\\
linear ramp & 82.1 [74.6, 89.6] & 76.1 [70.5, 81.7] & 69.5 [64.7, 74.4] & 68.7 [62.3, 75.1] & 59.4 [23.5, 95.4] & 46.1 [$-$9.5, 101.6] & 30.0 [$-$30.1, 90.1]\\
cap 128 & 81.2 [78.7, 83.7] & 74.9 [72.5, 77.3] & 74.2 [70.7, 77.7] & 73.9 [70.0, 77.7] & 73.9 [70.0, 77.7] & 72.1 [65.6, 78.7] & 2.8 [$-$1.3, 6.9]\\
cap 512 & 82.8 [78.8, 86.7] & 75.2 [73.4, 76.9] & 70.1 [63.4, 76.8] & 69.6 [61.0, 78.2] & 69.0 [60.0, 78.0] & 63.3 [53.7, 72.9] & 11.9 [1.0, 22.7]\\
fixed-point term & 79.9 [77.0, 82.8] & 74.2 [70.7, 77.7] & 73.4 [71.8, 75.1] & 73.5 [71.3, 75.7] & 73.5 [71.1, 75.9] & 70.6 [68.6, 72.6] & 3.6 [0.5, 6.7]\\
cap + hull & 72.2 [55.1, 89.4] & 65.7 [41.5, 89.8] & 65.3 [41.7, 88.8] & 65.4 [42.0, 88.7] & 65.4 [42.0, 88.7] & 65.0 [42.5, 87.5] & 0.7 [$-$0.9, 2.3]\\
hull & 73.3 [55.0, 91.5] & 66.9 [45.3, 88.5] & 66.7 [45.3, 88.2] & 66.6 [45.5, 87.6] & 66.5 [45.6, 87.5] & 66.2 [45.6, 86.8] & 0.7 [$-$1.1, 2.4]\\
\multicolumn{8}{@{}l}{\emph{minGRU-type}}\\
first recipe & 66.7 [51.0, 82.4] & 43.5 [34.3, 52.7] & 30.4 [24.0, 36.8] & 28.7 [21.7, 35.6] & 1.0 [$-$2.1, 4.1] & 0.8 [$-$1.9, 3.6] & 42.6 [31.3, 53.9]\\
linear ramp & 84.0 [83.0, 85.1] & 75.7 [75.2, 76.3] & 61.1 [50.0, 72.3] & 61.1 [50.0, 72.3] & 30.3 [$-$1.9, 62.5] & 11.1 [$-$1.7, 23.9] & 64.6 [52.3, 76.9]\\
cap 128 & 84.5 [83.1, 86.0] & 75.4 [74.2, 76.5] & 72.4 [69.9, 74.9] & 71.0 [69.7, 72.4] & 70.8 [69.3, 72.3] & 67.3 [63.6, 71.1] & 8.0 [4.6, 11.5]\\
cap 512 & 83.7 [83.0, 84.5] & 75.0 [73.2, 76.9] & 69.4 [65.9, 73.0] & 69.0 [67.9, 70.0] & 67.9 [66.5, 69.4] & 62.5 [56.7, 68.3] & 12.5 [7.1, 17.9]\\
fixed-point term & 82.9 [80.5, 85.3] & 74.6 [73.0, 76.2] & 72.9 [71.1, 74.7] & 73.0 [72.0, 74.0] & 72.6 [71.3, 73.9] & 69.6 [68.5, 70.7] & 5.0 [2.4, 7.6]\\
cap + hull & 83.7 [80.3, 87.1] & 75.0 [71.0, 79.0] & 73.8 [70.9, 76.6] & 73.8 [69.9, 77.7] & 73.2 [69.5, 76.9] & 72.2 [69.6, 74.9] & 2.8 [1.4, 4.2]\\
hull & 82.3 [78.8, 85.8] & 74.3 [69.5, 79.1] & 73.2 [68.4, 78.1] & 73.3 [69.0, 77.7] & 72.6 [68.3, 76.8] & 71.9 [68.6, 75.2] & 2.4 [0.5, 4.3]\\
\multicolumn{8}{@{}l}{\emph{S4D-type}}\\
linear ramp & 81.3 [75.7, 86.9] & 72.4 [69.9, 74.9] & 68.8 [64.1, 73.5] & 68.2 [64.2, 72.1] & 67.9 [65.1, 70.6] & 59.7 [54.9, 64.5] & 12.7 [9.1, 16.2]\\
\bottomrule
\end{tabular}
\end{table}

\begin{table}[htbp]
\caption{Training remedy: clean accuracy and certified accuracy at $L=0$ in \% on the first 1000 test documents (mean over three seeds
with its 95\% t-interval; every seed included), the same on the 2500 development documents at the end of training, the number of
degenerate seeds (clean accuracy below 60\% or more than 80\% of the predictions in one class) and the mean training time per run in
minutes. Natural models were evaluated on the test documents for clean accuracy only (their development $L=0$ share is that of the
end-of-training evaluation). ``First study'': the standard-IBP checkpoints of the first training study (\cref{app:gate2}),
re-evaluated, shown for the reproduction check only and not as a baseline of the other rows.}
\label{tab:remedy-clean}
\centering\scriptsize\setlength{\tabcolsep}{3pt}
\begin{tabular}{@{}llcccccc@{}}
\toprule
 & & & \multicolumn{2}{c}{test} & \multicolumn{2}{c}{development} & \\
\cmidrule(lr){4-5}\cmidrule(lr){6-7}
model & training & degenerate & clean & $L=0$ & clean & $L=0$ & minutes\\
\midrule
Mamba & first recipe & 1 of 3 & 60.0 [42.5, 77.6] & 48.0 [29.5, 66.4] & 59.5 [42.3, 76.6] & 48.5 [28.0, 68.9] & 25.5\\
 & linear ramp & 0 & 82.1 [74.6, 89.6] & 76.1 [70.5, 81.7] & 81.2 [74.3, 88.2] & 75.0 [70.6, 79.5] & 72.5\\
 & cap 128 & 0 & 81.2 [78.7, 83.7] & 74.9 [72.5, 77.3] & 80.5 [74.7, 86.3] & 74.5 [71.1, 78.0] & 71.7\\
 & cap 512 & 0 & 82.8 [78.8, 86.7] & 75.2 [73.4, 76.9] & 81.4 [74.9, 87.9] & 74.7 [71.0, 78.3] & 75.5\\
 & cap 2048 & 0 & 82.4 [74.6, 90.2] & 75.9 [71.5, 80.4] & 81.2 [73.7, 88.8] & 74.7 [69.5, 79.9] & 72.2\\
 & fixed-point term & 0 & 79.9 [77.0, 82.8] & 74.2 [70.7, 77.7] & 79.3 [76.7, 81.9] & 73.8 [72.7, 75.0] & 145.5\\
 & cap + hull & 0 & 72.2 [55.1, 89.4] & 65.7 [41.5, 89.8] & 72.4 [55.3, 89.4] & 66.0 [44.9, 87.1] & 145.8\\
 & hull & 0 & 73.3 [55.0, 91.5] & 66.9 [45.3, 88.5] & 72.4 [53.5, 91.3] & 65.9 [41.2, 90.7] & 149.2\\
 & natural, cap 128 & 0 & 86.5 [83.9, 89.1] & -- & 86.5 [84.0, 88.9] & 0.0 [0.0, 0.0] & 2.5\\
 & natural, cap 512 & 0 & 88.2 [86.8, 89.5] & -- & 87.1 [86.0, 88.3] & 0.0 [0.0, 0.0] & 2.5\\
 & natural, cap 2048 & 0 & 87.3 [86.2, 88.4] & -- & 86.9 [85.6, 88.3] & 0.0 [0.0, 0.0] & 2.5\\
 & natural & 0 & 88.2 [86.9, 89.5] & -- & 87.2 [86.0, 88.4] & 0.0 [0.0, 0.0] & 2.5\\
 & first study & 2 of 3 & 61.7 [30.1, 93.3] & 47.6 [36.4, 58.7] & -- & -- & --\\
minGRU-type & first recipe & 0 & 66.7 [51.0, 82.4] & 43.5 [34.3, 52.7] & 65.2 [53.8, 76.6] & 42.4 [31.7, 53.2] & 22.2\\
 & linear ramp & 0 & 84.0 [83.0, 85.1] & 75.7 [75.2, 76.3] & 83.6 [82.0, 85.2] & 75.5 [73.3, 77.6] & 72.0\\
 & cap 128 & 0 & 84.5 [83.1, 86.0] & 75.4 [74.2, 76.5] & 83.6 [82.2, 85.0] & 74.6 [72.5, 76.8] & 58.0\\
 & cap 512 & 0 & 83.7 [83.0, 84.5] & 75.0 [73.2, 76.9] & 83.5 [81.8, 85.2] & 74.1 [71.3, 77.0] & 70.4\\
 & cap 2048 & 0 & 84.0 [82.1, 85.8] & 75.4 [74.2, 76.7] & 83.6 [82.5, 84.6] & 74.4 [71.6, 77.2] & 57.5\\
 & fixed-point term & 0 & 82.9 [80.5, 85.3] & 74.6 [73.0, 76.2] & 82.0 [78.4, 85.5] & 74.0 [70.5, 77.6] & 133.4\\
 & cap + hull & 0 & 83.7 [80.3, 87.1] & 75.0 [71.0, 79.0] & 83.1 [80.4, 85.8] & 74.9 [71.1, 78.7] & 122.6\\
 & hull & 0 & 82.3 [78.8, 85.8] & 74.3 [69.5, 79.1] & 82.0 [77.4, 86.6] & 74.4 [70.1, 78.6] & 124.0\\
 & natural, cap 128 & 0 & 86.7 [83.6, 89.8] & -- & 86.4 [85.4, 87.4] & 0.0 [0.0, 0.0] & 2.0\\
 & natural, cap 512 & 0 & 88.0 [87.3, 88.6] & -- & 87.5 [85.8, 89.1] & 0.0 [0.0, 0.0] & 2.0\\
 & natural, cap 2048 & 0 & 87.8 [86.6, 88.9] & -- & 87.6 [86.3, 89.0] & 0.0 [0.0, 0.0] & 2.0\\
 & natural & 0 & 87.7 [86.2, 89.2] & -- & 87.7 [86.4, 88.9] & 0.0 [0.0, 0.0] & 2.1\\
 & first study & 0 & 67.6 [63.7, 71.5] & 52.3 [35.9, 68.7] & -- & -- & --\\
S4D-type & first recipe & 2 of 3 & 58.9 [42.7, 75.0] & 43.0 [30.9, 55.1] & 59.8 [44.4, 75.3] & 45.2 [32.5, 57.9] & 13.6\\
 & linear ramp & 0 & 81.3 [75.7, 86.9] & 72.4 [69.9, 74.9] & 80.2 [78.9, 81.6] & 72.0 [67.3, 76.7] & 42.7\\
 & first study & 1 of 3 & 64.2 [33.9, 94.6] & 42.5 [30.9, 54.0] & -- & -- & --\\
\bottomrule
\end{tabular}
\end{table}

\begin{table}[htbp]
\caption{Training remedy: certified accuracy in \% at finite context lengths (non-degenerate seeds; mean with its 95\% t-interval
over the seeds; $^{(2)}$, $^{(1)}$: two seeds, one seed): on the bare document, after 1024 and 4096 tokens of contiguous neutral text,
and after $K=1$ and $K=4$ copies of the 1024-token period read from the zero state; and accuracy under the substitution attack on the
first 100 documents at $L=0$, an upper bound on the robust accuracy of those documents (the other columns use 1000 documents, so the
attack share can fall below the $L=0$ share). First study: for the reproduction check only.}
\label{tab:remedy-fin}
\centering\scriptsize\setlength{\tabcolsep}{2pt}
\begin{tabular}{@{}lcccccc@{}}
\toprule
training & $L=0$ & $L=1024$ & $L=4096$ & $K=1$ & $K=4$ & attack\\
\midrule
\multicolumn{7}{@{}l}{\emph{Mamba}}\\
first recipe & 48.4 [$-$45.6, 142.4]$^{(2)}$ & 45.7 [$-$19.1, 110.5]$^{(2)}$ & 34.5 [$-$145.3, 214.2]$^{(2)}$ & 46.3 [$-$28.7, 121.3]$^{(2)}$ & 36.0 [$-$134.3, 206.3]$^{(2)}$ & 62.0 [49.3, 74.7]$^{(2)}$\\
linear ramp & 76.1 [70.5, 81.7] & 69.5 [64.7, 74.4] & 66.6 [55.1, 78.2] & 68.7 [62.3, 75.1] & 64.4 [46.3, 82.5] & 74.0 [66.5, 81.5]\\
cap 128 & 74.9 [72.5, 77.3] & 74.2 [70.7, 77.7] & 74.2 [70.7, 77.7] & 73.9 [70.0, 77.7] & 73.9 [70.0, 77.7] & 76.0 [62.2, 89.8]\\
cap 512 & 75.2 [73.4, 76.9] & 70.1 [63.4, 76.8] & 69.6 [62.0, 77.3] & 69.6 [61.0, 78.2] & 69.2 [60.7, 77.7] & 74.3 [69.2, 79.5]\\
cap 2048 & 75.9 [71.5, 80.4] & 69.7 [66.2, 73.2] & 68.3 [61.3, 75.3] & 68.9 [63.5, 74.2] & 66.3 [56.2, 76.4] & 75.0 [66.0, 84.0]\\
fixed-point term & 74.2 [70.7, 77.7] & 73.4 [71.8, 75.1] & 73.5 [71.8, 75.2] & 73.5 [71.3, 75.7] & 73.5 [71.2, 75.8] & 73.3 [69.5, 77.1]\\
cap + hull & 65.7 [41.5, 89.8] & 65.3 [41.7, 88.8] & 65.3 [41.7, 88.8] & 65.4 [42.0, 88.7] & 65.4 [42.0, 88.7] & 69.3 [53.4, 85.3]\\
hull & 66.9 [45.3, 88.5] & 66.7 [45.3, 88.2] & 66.7 [45.3, 88.2] & 66.6 [45.5, 87.6] & 66.6 [45.5, 87.6] & 71.0 [68.5, 73.5]\\
first study & 50.3$^{(1)}$ & 31.9$^{(1)}$ & 17.4$^{(1)}$ & 28.5$^{(1)}$ & 20.0$^{(1)}$ & --\\
\multicolumn{7}{@{}l}{\emph{minGRU-type}}\\
first recipe & 43.5 [34.3, 52.7] & 30.4 [24.0, 36.8] & 16.9 [0.2, 33.7] & 28.7 [21.7, 35.6] & 17.2 [2.4, 32.0] & 66.3 [40.0, 92.7]\\
linear ramp & 75.7 [75.2, 76.3] & 61.1 [50.0, 72.3] & 45.9 [34.7, 57.1] & 61.1 [50.0, 72.3] & 47.2 [37.1, 57.3] & 76.0 [71.0, 81.0]\\
cap 128 & 75.4 [74.2, 76.5] & 72.4 [69.9, 74.9] & 72.4 [69.9, 74.9] & 71.0 [69.7, 72.4] & 71.0 [69.7, 72.4] & 78.3 [60.4, 96.3]\\
cap 512 & 75.0 [73.2, 76.9] & 69.4 [65.9, 73.0] & 68.7 [65.6, 71.9] & 69.0 [67.9, 70.0] & 68.4 [67.2, 69.7] & 78.3 [68.3, 88.4]\\
cap 2048 & 75.4 [74.2, 76.7] & 65.1 [64.9, 65.3] & 60.4 [59.3, 61.5] & 64.3 [62.9, 65.6] & 58.4 [55.3, 61.5] & 76.7 [67.3, 86.1]\\
fixed-point term & 74.6 [73.0, 76.2] & 72.9 [71.1, 74.7] & 72.9 [71.3, 74.5] & 73.0 [72.0, 74.0] & 72.8 [71.5, 74.0] & 77.7 [72.5, 82.8]\\
cap + hull & 75.0 [71.0, 79.0] & 73.8 [70.9, 76.6] & 73.6 [70.6, 76.6] & 73.8 [69.9, 77.7] & 73.7 [69.9, 77.5] & 79.3 [69.9, 88.7]\\
hull & 74.3 [69.5, 79.1] & 73.2 [68.4, 78.1] & 73.2 [68.6, 77.9] & 73.3 [69.0, 77.7] & 73.1 [68.8, 77.4] & 78.7 [72.9, 84.4]\\
first study & 52.3 [35.9, 68.7] & 37.0 [13.0, 61.0] & 22.7 [10.3, 35.1] & 36.2 [13.8, 58.5] & 23.0 [11.2, 34.8] & 67.3 [54.8, 79.8]\\
\multicolumn{7}{@{}l}{\emph{S4D-type}}\\
first recipe & 37.4$^{(1)}$ & 42.6$^{(1)}$ & 44.0$^{(1)}$ & 42.7$^{(1)}$ & 44.5$^{(1)}$ & 55.0$^{(1)}$\\
linear ramp & 72.4 [69.9, 74.9] & 68.8 [64.1, 73.5] & 68.5 [64.3, 72.8] & 68.2 [64.2, 72.1] & 67.9 [65.1, 70.6] & 74.7 [62.2, 87.2]\\
first study & 39.8 [34.7, 44.9]$^{(2)}$ & 35.3 [$-$5.4, 76.0]$^{(2)}$ & 35.5 [$-$7.7, 78.7]$^{(2)}$ & 35.3 [$-$1.5, 72.1]$^{(2)}$ & 35.4 [$-$4.0, 74.8]$^{(2)}$ & --\\
\bottomrule
\end{tabular}
\end{table}

\begin{table}[htbp]
\caption{Training remedy: clean and certified accuracy in \% for unbounded contexts (non-degenerate seeds; mean with its 95\%
t-interval). Clean: after four copies of the period, at $L=\infty$, and certified correct for every number of copies (the K-uniform
hull of the clean trajectories, a box statement; 0.0 on every seed of the natural models). Certified: at $L=\infty$
(\cref{thm:inf}) and K-uniform (\cref{thm:hull}); $Y$ is the $L=0$ share minus the K-uniform share, per seed, in points.
$^{(2)}$, $^{(1)}$: mean over two seeds, one seed. First study: for the reproduction check only.}
\label{tab:remedy-inf}
\centering\scriptsize\setlength{\tabcolsep}{2pt}
\begin{tabular}{@{}lcccccc@{}}
\toprule
 & \multicolumn{3}{c}{clean} & \multicolumn{2}{c}{certified} & \\
\cmidrule(lr){2-4}\cmidrule(lr){5-6}
training & $K=4$ & $L=\infty$ & every $K$ & $L=\infty$ & K-uniform & $Y$\\
\midrule
\multicolumn{7}{@{}l}{\emph{Mamba}}\\
first recipe & 62.5 [20.0, 105.1]$^{(2)}$ & 59.1 [$-$28.0, 146.1]$^{(2)}$ & 37.8 [$-$311.0, 386.5]$^{(2)}$ & 23.2 [$-$265.2, 311.6]$^{(2)}$ & 22.7 [$-$265.1, 310.4]$^{(2)}$ & 25.8 [$-$168.0, 219.5]$^{(2)}$\\
linear ramp & 73.6 [68.9, 78.3] & 71.3 [58.2, 84.4] & 58.7 [23.6, 93.9] & 59.4 [23.5, 95.4] & 46.1 [$-$9.5, 101.6] & 30.0 [$-$30.1, 90.1]\\
cap 128 & 81.1 [78.0, 84.2] & 81.1 [78.0, 84.2] & 79.2 [76.7, 81.7] & 73.9 [70.0, 77.7] & 72.1 [65.6, 78.7] & 2.8 [$-$1.3, 6.9]\\
cap 512 & 77.6 [76.5, 78.6] & 77.6 [76.5, 78.6] & 72.8 [69.4, 76.3] & 69.0 [60.0, 78.0] & 63.3 [53.7, 72.9] & 11.9 [1.0, 22.7]\\
cap 2048 & 74.4 [74.1, 74.7] & 74.4 [74.1, 74.7] & 67.2 [64.7, 69.8] & 65.9 [54.3, 77.5] & 58.1 [43.0, 73.2] & 17.9 [$-$0.9, 36.7]\\
fixed-point term & 78.7 [75.7, 81.8] & 78.7 [75.7, 81.8] & 76.3 [74.4, 78.2] & 73.5 [71.1, 75.9] & 70.6 [68.6, 72.6] & 3.6 [0.5, 6.7]\\
cap + hull & 72.3 [55.1, 89.5] & 72.3 [55.1, 89.5] & 72.0 [55.3, 88.8] & 65.4 [42.0, 88.7] & 65.0 [42.5, 87.5] & 0.7 [$-$0.9, 2.3]\\
hull & 73.0 [55.4, 90.6] & 73.0 [55.4, 90.6] & 72.5 [55.7, 89.4] & 66.5 [45.6, 87.5] & 66.2 [45.6, 86.8] & 0.7 [$-$1.1, 2.4]\\
natural, cap 128 & 86.0 [83.7, 88.3] & 86.0 [83.7, 88.3] & 0.0 [0.0, 0.0] & -- & -- & --\\
natural, cap 512 & 86.2 [85.1, 87.3] & 86.2 [85.1, 87.3] & 0.0 [0.0, 0.0] & -- & -- & --\\
natural, cap 2048 & 85.8 [84.4, 87.2] & 85.8 [84.4, 87.2] & 0.0 [0.0, 0.0] & -- & -- & --\\
natural & 85.7 [83.9, 87.5] & 85.7 [84.0, 87.4] & 0.0 [0.0, 0.0] & -- & -- & --\\
first study & 60.7$^{(1)}$ & 49.0$^{(1)}$ & 21.5$^{(1)}$ & 1.4$^{(1)}$ & 0.0$^{(1)}$ & 50.3$^{(1)}$\\
\multicolumn{7}{@{}l}{\emph{minGRU-type}}\\
first recipe & 60.0 [47.7, 72.3] & 52.7 [45.2, 60.3] & 28.4 [20.5, 36.3] & 1.0 [$-$2.1, 4.1] & 0.8 [$-$1.9, 3.6] & 42.6 [31.3, 53.9]\\
linear ramp & 69.6 [52.6, 86.6] & 64.0 [41.6, 86.4] & 44.4 [11.8, 76.9] & 30.3 [$-$1.9, 62.5] & 11.1 [$-$1.7, 23.9] & 64.6 [52.3, 76.9]\\
cap 128 & 83.0 [81.5, 84.5] & 83.0 [81.5, 84.5] & 79.8 [76.3, 83.3] & 70.8 [69.3, 72.3] & 67.3 [63.6, 71.1] & 8.0 [4.6, 11.5]\\
cap 512 & 82.3 [82.2, 82.5] & 82.3 [82.2, 82.5] & 76.6 [74.5, 78.7] & 67.9 [66.5, 69.4] & 62.5 [56.7, 68.3] & 12.5 [7.1, 17.9]\\
cap 2048 & 76.3 [66.0, 86.6] & 75.9 [65.3, 86.4] & 56.6 [37.9, 75.3] & 56.5 [53.1, 60.0] & 35.6 [18.6, 52.6] & 39.8 [24.0, 55.6]\\
fixed-point term & 81.8 [78.0, 85.7] & 81.8 [78.0, 85.6] & 79.3 [76.7, 82.0] & 72.6 [71.3, 73.9] & 69.6 [68.5, 70.7] & 5.0 [2.4, 7.6]\\
cap + hull & 83.3 [79.6, 86.9] & 83.3 [79.6, 86.9] & 82.1 [79.0, 85.2] & 73.2 [69.5, 76.9] & 72.2 [69.6, 74.9] & 2.8 [1.4, 4.2]\\
hull & 82.1 [78.4, 85.8] & 82.1 [78.4, 85.8] & 81.1 [78.3, 83.9] & 72.6 [68.3, 76.8] & 71.9 [68.6, 75.2] & 2.4 [0.5, 4.3]\\
natural, cap 128 & 85.6 [84.2, 87.0] & 85.6 [84.2, 87.0] & 0.0 [0.0, 0.0] & -- & -- & --\\
natural, cap 512 & 85.6 [84.3, 86.8] & 85.6 [84.3, 86.8] & 0.0 [0.0, 0.0] & -- & -- & --\\
natural, cap 2048 & 84.8 [83.6, 85.9] & 84.7 [83.6, 85.9] & 0.0 [0.0, 0.0] & -- & -- & --\\
natural & 84.7 [81.7, 87.7] & 84.7 [81.7, 87.7] & 0.0 [0.0, 0.0] & -- & -- & --\\
first study & 65.8 [62.5, 69.1] & 54.5 [47.2, 61.9] & 26.4 [10.8, 42.0] & 1.5 [$-$1.6, 4.6] & 0.9 [$-$1.2, 3.0] & 51.3 [35.8, 66.9]\\
\multicolumn{7}{@{}l}{\emph{S4D-type}}\\
first recipe & 52.0$^{(1)}$ & 52.0$^{(1)}$ & 41.6$^{(1)}$ & 44.5$^{(1)}$ & 27.7$^{(1)}$ & 9.7$^{(1)}$\\
linear ramp & 76.0 [71.4, 80.6] & 76.0 [71.4, 80.6] & 70.0 [64.2, 75.8] & 67.9 [65.1, 70.6] & 59.7 [54.9, 64.5] & 12.7 [9.1, 16.2]\\
first study & 65.8 [$-$53.0, 184.7]$^{(2)}$ & 65.8 [$-$53.0, 184.7]$^{(2)}$ & 53.5 [$-$92.6, 199.6]$^{(2)}$ & 35.4 [$-$4.0, 74.8]$^{(2)}$ & 27.0 [17.4, 36.5]$^{(2)}$ & 12.8 [8.4, 17.3]$^{(2)}$\\
\bottomrule
\end{tabular}
\end{table}

\begin{table}[htbp]
\caption{Training remedy: ratios and flip shares in \% (non-degenerate seeds; mean with its 95\% t-interval). $L=\infty$ and
K-uniform: certified share over the $L=0$ share. Width-only: among the documents certified at $L=0$ whose clean prediction after the
context is still correct. Flips: the share of the failures (certified at $L=0$, not after the context) whose clean prediction after
the context is wrong, which no box can certify. A ratio exceeds 100 when more documents are certified after the context than at
$L=0$, and intervals of ratios can exceed 100: they are t-intervals of a mean. $L=\infty$ ratio of the linear-ramp recipe per seed:
Mamba 54.8, 92.5 and 88.1, minGRU-type 53.2, 46.1 and 20.6 (other per-seed values: \cref{tab:remedy-seeds}). $^{(2)}$, $^{(1)}$:
mean over two seeds, one seed; in the flip columns of the Mamba models trained with the hull term, $^{(2)}$ because seed~3 has no
$L=\infty$ failure, so its flip share is undefined (no reading uses these two means). First study: for the reproduction check
only.}
\label{tab:remedy-ratio}
\centering\scriptsize\setlength{\tabcolsep}{2pt}
\begin{tabular}{@{}lcccccc@{}}
\toprule
 & \multicolumn{2}{c}{$L=\infty$} & \multicolumn{2}{c}{K-uniform} & \multicolumn{2}{c}{flips} \\
\cmidrule(lr){2-3}\cmidrule(lr){4-5}\cmidrule(lr){6-7}
training & ratio & width-only & ratio & width-only & $L=\infty$ & K-uniform\\
\midrule
\multicolumn{7}{@{}l}{\emph{Mamba}}\\
first recipe & 41.7 [$-$473.1, 556.6]$^{(2)}$ & 41.5 [$-$475.8, 558.9]$^{(2)}$ & 40.6 [$-$475.2, 556.3]$^{(2)}$ & 40.7 [$-$476.0, 557.3]$^{(2)}$ & 20.7 [$-$242.4, 283.9]$^{(2)}$ & 38.3 [$-$436.0, 512.6]$^{(2)}$\\
linear ramp & 78.5 [27.3, 129.6] & 85.6 [50.0, 121.2] & 61.1 [$-$14.3, 136.4] & 77.7 [17.4, 137.9] & 67.4 [40.7, 94.0] & 66.5 [53.9, 79.0]\\
cap 128 & 98.6 [96.5, 100.6] & 98.0 [95.1, 100.8] & 96.2 [90.6, 101.9] & 97.6 [94.2, 101.0] & 37.2 [3.1, 71.3] & 36.5 [0.8, 72.3]\\
cap 512 & 91.8 [78.5, 105.2] & 93.2 [86.3, 100.1] & 84.2 [70.1, 98.3] & 92.1 [84.5, 99.8] & 56.2 [54.2, 58.2] & 54.7 [47.6, 61.7]\\
cap 2048 & 86.9 [67.7, 106.0] & 91.7 [78.1, 105.2] & 76.6 [53.3, 99.9] & 90.0 [70.3, 109.8] & 67.8 [40.2, 95.4] & 67.5 [39.1, 95.9]\\
fixed-point term & 99.1 [97.4, 100.7] & 97.9 [95.9, 99.8] & 95.2 [91.2, 99.1] & 96.8 [93.8, 99.7] & 33.7 [6.8, 60.6] & 34.7 [14.4, 55.0]\\
cap + hull & 99.6 [98.4, 100.8] & 99.2 [97.1, 101.2] & 99.0 [96.8, 101.2] & 99.2 [97.1, 101.2] & 16.3 [$-$93.6, 126.3]$^{(2)}$ & 16.3 [$-$93.6, 126.3]$^{(2)}$\\
hull & 99.5 [98.2, 100.7] & 99.3 [97.7, 101.0] & 99.1 [96.5, 101.6] & 99.2 [97.4, 101.1] & 15.4 [$-$180.1, 210.9]$^{(2)}$ & 14.3 [$-$167.2, 195.8]$^{(2)}$\\
first study & 2.8$^{(1)}$ & 4.9$^{(1)}$ & 0.0$^{(1)}$ & 0.0$^{(1)}$ & 60.9$^{(1)}$ & 67.2$^{(1)}$\\
\multicolumn{7}{@{}l}{\emph{minGRU-type}}\\
first recipe & 2.3 [$-$5.5, 10.2] & 4.4 [$-$11.0, 19.7] & 2.0 [$-$4.9, 8.9] & 4.1 [$-$10.1, 18.4] & 35.6 [$-$1.4, 72.6] & 47.7 [24.2, 71.2]\\
linear ramp & 40.0 [$-$2.5, 82.5] & 54.5 [$-$12.7, 121.8] & 14.6 [$-$2.2, 31.4] & 28.5 [$-$8.5, 65.6] & 50.7 [$-$15.0, 116.4] & 52.2 [1.9, 102.4]\\
cap 128 & 93.9 [92.0, 95.9] & 94.2 [92.5, 95.9] & 89.3 [84.8, 93.9] & 92.1 [88.1, 96.0] & 24.5 [3.1, 45.9] & 27.9 [19.1, 36.7]\\
cap 512 & 90.5 [88.1, 93.0] & 90.8 [87.5, 94.1] & 83.3 [76.1, 90.5] & 86.9 [76.6, 97.2] & 22.3 [$-$3.6, 48.3] & 25.9 [$-$5.9, 57.8]\\
cap 2048 & 74.9 [71.0, 78.8] & 79.9 [77.4, 82.4] & 47.2 [25.4, 68.9] & 65.2 [51.0, 79.3] & 38.7 [6.9, 70.6] & 51.9 [24.3, 79.5]\\
fixed-point term & 97.3 [94.1, 100.5] & 96.7 [93.8, 99.7] & 93.3 [90.0, 96.7] & 94.8 [92.8, 96.7] & 27.8 [$-$3.7, 59.3] & 21.9 [$-$11.8, 55.5]\\
cap + hull & 97.6 [96.2, 98.9] & 97.1 [96.7, 97.5] & 96.3 [94.6, 97.9] & 96.8 [95.7, 97.8] & 15.1 [0.6, 29.5] & 13.7 [2.4, 25.0]\\
hull & 97.7 [96.4, 98.9] & 97.5 [95.6, 99.4] & 96.8 [94.4, 99.1] & 97.3 [95.2, 99.3] & 14.5 [0.9, 28.0] & 15.4 [9.7, 21.2]\\
first study & 2.7 [$-$2.9, 8.4] & 5.4 [$-$7.6, 18.4] & 1.7 [$-$2.3, 5.8] & 4.3 [$-$7.5, 16.1] & 43.3 [3.3, 83.4] & 55.7 [31.6, 79.8]\\
\multicolumn{7}{@{}l}{\emph{S4D-type}}\\
first recipe & 119.0$^{(1)}$ & 97.9$^{(1)}$ & 74.1$^{(1)}$ & 85.0$^{(1)}$ & 84.1$^{(1)}$ & 49.5$^{(1)}$\\
linear ramp & 93.8 [89.9, 97.6] & 93.6 [90.5, 96.6] & 82.5 [77.4, 87.6] & 92.0 [89.5, 94.5] & 61.6 [50.1, 73.1] & 58.9 [52.6, 65.2]\\
first study & 89.0 [$-$21.3, 199.4]$^{(2)}$ & 87.8 [$-$10.0, 185.6]$^{(2)}$ & 67.7 [52.4, 83.0]$^{(2)}$ & 73.6 [57.5, 89.7]$^{(2)}$ & 28.0 [$-$251.9, 307.8]$^{(2)}$ & 24.3 [$-$90.9, 139.5]$^{(2)}$\\
\bottomrule
\end{tabular}
\end{table}

\begin{table}[htbp]
\caption{Training remedy, non-degenerate seeds, \% (mean with its 95\% t-interval): the width-only $L=\infty$ ratio on documents
shorter than 128 tokens and of at least 256 tokens; the share of 200 documents certified for every context of every length over the
most frequent neutral token, with the document in its substitution box (\cref{cor:anyctx}, $P=8$; over the 8 and the 64 most frequent
tokens every share is 0.0 in both modes); the share of the width failures at $L=\infty$ that the counterfactual of \cref{app:gate2}
restores (channels slower than 512 tokens reset to their one-period radius; a diagnostic; 0.0 where the reset restores none of the
failures, which includes every model with no channel slower than 512 tokens, where it replaces nothing); and the share of (document,
channel) pairs slower than 4096 tokens in the clean dynamics (maximum over the layers). The last three columns use 200 documents.
$^{(2)}$, $^{(1)}$: mean over two seeds, one seed. First study: for the reproduction check only.}
\label{tab:remedy-attr}
\centering\scriptsize\setlength{\tabcolsep}{2pt}
\begin{tabular}{@{}lccccc@{}}
\toprule
 & \multicolumn{2}{c}{width-only $L=\infty$ ratio} & token set & restored & slow\\
\cmidrule(lr){2-3}
training & $<128$ tokens & $\ge256$ tokens & 1 token & by the reset & pairs\\
\midrule
\multicolumn{6}{@{}l}{\emph{Mamba}}\\
first recipe & 42.9 [$-$501.7, 587.4]$^{(2)}$ & 42.9 [$-$486.6, 572.5]$^{(2)}$ & 32.8 [$-$383.4, 448.9]$^{(2)}$ & 65.9 [$-$178.3, 310.1]$^{(2)}$ & 14.8 [$-$171.9, 201.5]$^{(2)}$\\
linear ramp & 85.9 [39.3, 132.6] & 84.9 [49.3, 120.5] & 32.8 [$-$39.5, 105.2] & 22.7 [$-$40.1, 85.6] & 36.8 [$-$38.1, 111.7]\\
cap 128 & 98.2 [94.4, 102.0] & 99.2 [98.2, 100.3] & 86.3 [82.7, 89.9] & 0.0 [0.0, 0.0] & 0.0 [0.0, 0.0]\\
cap 512 & 94.1 [87.5, 100.7] & 92.8 [84.9, 100.7] & 66.2 [34.9, 97.4] & 0.0 [0.0, 0.0] & 0.0 [0.0, 0.0]\\
cap 2048 & 91.8 [74.6, 108.9] & 91.8 [78.7, 105.0] & 48.3 [22.7, 74.0] & 19.1 [$-$2.5, 40.6] & 0.0 [0.0, 0.0]\\
fixed-point term & 97.4 [95.4, 99.3] & 98.5 [97.6, 99.4] & 83.5 [79.0, 88.0] & 0.0 [0.0, 0.0] & 8.5 [$-$16.2, 33.3]\\
cap + hull & 97.4 [91.3, 103.6] & 99.9 [99.3, 100.4] & 85.5 [73.7, 97.3] & 0.0$^{(1)}$ & 0.0 [0.0, 0.0]\\
hull & 97.9 [93.4, 102.5] & 99.9 [99.4, 100.4] & 86.5 [77.5, 95.5] & 0.0$^{(1)}$ & 0.0 [0.0, 0.0]\\
first study & 0.0$^{(1)}$ & 7.6$^{(1)}$ & -- & -- & --\\
\multicolumn{6}{@{}l}{\emph{minGRU-type}}\\
first recipe & 3.7 [$-$12.2, 19.6] & 5.0 [$-$11.4, 21.4] & 0.0 [0.0, 0.0] & 77.5 [47.6, 107.5] & 42.4 [11.2, 73.6]\\
linear ramp & 54.6 [$-$16.7, 126.0] & 55.4 [$-$11.0, 121.9] & 0.0 [0.0, 0.0] & 68.6 [18.5, 118.6] & 7.8 [$-$2.0, 17.5]\\
cap 128 & 89.2 [86.3, 92.1] & 97.4 [94.7, 100.2] & 47.0 [12.2, 81.8] & 0.0 [0.0, 0.0] & 0.0 [0.0, 0.0]\\
cap 512 & 89.1 [80.9, 97.3] & 92.9 [89.2, 96.6] & 60.0 [26.6, 93.4] & 0.0 [0.0, 0.0] & 0.0 [0.0, 0.0]\\
cap 2048 & 78.8 [68.8, 88.8] & 84.7 [83.1, 86.2] & 20.0 [$-$1.5, 41.5] & 27.1 [7.0, 47.3] & 0.0 [0.0, 0.0]\\
fixed-point term & 97.7 [96.2, 99.1] & 97.4 [95.6, 99.2] & 14.3 [$-$45.2, 73.9] & 0.0 [0.0, 0.0] & 0.3 [$-$0.9, 1.4]\\
cap + hull & 95.2 [90.8, 99.6] & 97.9 [96.9, 98.9] & 75.7 [63.2, 88.2] & 0.0 [0.0, 0.0] & 0.0 [0.0, 0.0]\\
hull & 95.6 [87.9, 103.3] & 98.2 [96.8, 99.7] & 0.2 [$-$0.6, 0.9] & 0.0 [0.0, 0.0] & 0.0 [0.0, 0.0]\\
first study & 5.1 [$-$8.2, 18.5] & 6.0 [$-$8.3, 20.3] & 0.0 [0.0, 0.0] & 71.4 [57.9, 84.8] & 42.9 [1.1, 84.7]\\
\multicolumn{6}{@{}l}{\emph{S4D-type}}\\
first recipe & 100.0$^{(1)}$ & 97.8$^{(1)}$ & 42.5$^{(1)}$ & 0.0$^{(1)}$ & 0.0$^{(1)}$\\
linear ramp & 93.8 [89.0, 98.6] & 94.3 [90.1, 98.5] & 65.2 [44.0, 86.3] & 0.0 [0.0, 0.0] & 0.0 [0.0, 0.0]\\
first study & 73.0 [$-$152.4, 298.4]$^{(2)}$ & 94.5 [42.0, 147.0]$^{(2)}$ & -- & -- & --\\
\bottomrule
\end{tabular}
\end{table}

\begin{table}[htbp]
\caption{Training remedy, per checkpoint (1000 test documents, \%): clean accuracy; certified accuracy at $L=0$, at $L=\infty$ and
K-uniform, and the K-uniform ratio, each with its 95\% bootstrap interval over the documents (the ratio paired); and the largest time
constant observed at the threat and in the clean dynamics, over every layer (tokens, on the 200 documents of the attribution; for a
capped model at most $-1/\log(1-1/\tau_{\max})$, which is 127.5, 511.5 and 2047.5). $^\ast$Degenerate (clean accuracy below 60\% or
more than 80\% of the predictions in one class). First study: seeds 0--2 of \cref{app:gate2}, re-evaluated for the reproduction
check (its time constants were not recomputed for Mamba and the S4D-type model). Natural models: \cref{tab:remedy-clean} and the
descriptive observation (v) below.}
\label{tab:remedy-seeds}
\centering\scriptsize\setlength{\tabcolsep}{2.5pt}
\begin{tabular}{@{}llccccccc@{}}
\toprule
model & training & seed & clean & $L=0$ & $L=\infty$ & K-uniform & K-uniform ratio & largest $\tau_{\rm h}$\\
\midrule
Mamba & first recipe & 1$^\ast$ & 52.7 & 47.1 [44.1, 50.3] & 0.8 [0.3, 1.4] & 0.0 [0.0, 0.0] & 0.0 [0.0, 0.0] & $1.3\cdot10^{7}$\\
 & & 2 & 60.6 & 41.0 [37.9, 44.3] & 0.5 [0.1, 1.0] & 0.0 [0.0, 0.0] & 0.0 [0.0, 0.0] & $1.3\cdot10^{10}$\\
 & & 3 & 66.8 & 55.8 [52.8, 58.9] & 45.9 [42.9, 49.1] & 45.3 [42.3, 48.4] & 81.2 [78.0, 84.4] & $1.0\cdot10^{4}$\\
 & linear ramp & 1 & 85.5 & 78.4 [75.9, 80.7] & 43.0 [39.8, 46.0] & 20.7 [18.1, 23.2] & 26.4 [23.1, 29.5] & $6.0\cdot10^{5}$\\
 & & 2 & 81.0 & 75.9 [73.2, 78.4] & 70.2 [67.3, 72.9] & 62.9 [59.8, 65.7] & 82.9 [80.2, 85.5] & $7.3\cdot10^{6}$\\
 & & 3 & 79.8 & 73.9 [71.2, 76.5] & 65.1 [62.1, 68.0] & 54.6 [51.5, 57.6] & 73.9 [70.6, 77.0] & $3.7\cdot10^{8}$\\
 & cap 128 & 1 & 81.8 & 73.9 [71.2, 76.6] & 72.3 [69.5, 74.9] & 69.3 [66.3, 72.2] & 93.8 [92.0, 95.4] & 127.5\\
 & & 2 & 80.0 & 75.1 [72.4, 77.7] & 73.9 [71.2, 76.6] & 72.6 [69.9, 75.2] & 96.7 [95.4, 97.9] & 127.5\\
 & & 3 & 81.7 & 75.8 [72.9, 78.4] & 75.4 [72.7, 78.0] & 74.5 [71.8, 77.2] & 98.3 [97.3, 99.1] & 127.5\\
 & cap 512 & 1 & 84.6 & 75.8 [73.2, 78.3] & 64.9 [61.9, 67.8] & 58.9 [55.9, 61.9] & 77.7 [74.7, 80.7] & 511.0\\
 & & 2 & 81.7 & 75.3 [72.6, 77.8] & 71.7 [68.9, 74.3] & 66.2 [63.3, 69.0] & 87.9 [85.5, 90.2] & 511.4\\
 & & 3 & 82.0 & 74.4 [71.7, 77.1] & 70.4 [67.6, 73.2] & 64.8 [61.7, 67.7] & 87.1 [84.7, 89.4] & 511.5\\
 & cap 2048 & 1 & 85.9 & 77.7 [75.1, 80.1] & 60.6 [57.6, 63.7] & 51.1 [48.0, 54.2] & 65.8 [62.2, 69.2] & 2033.4\\
 & & 2 & 81.6 & 76.0 [73.4, 78.4] & 69.5 [66.5, 72.2] & 62.3 [59.3, 65.1] & 82.0 [79.2, 84.7] & 2047.2\\
 & & 3 & 79.8 & 74.1 [71.4, 76.7] & 67.6 [64.7, 70.4] & 60.8 [57.8, 63.8] & 82.1 [79.3, 84.8] & 2047.5\\
 & fixed-point term & 1 & 81.1 & 73.2 [70.5, 75.8] & 73.0 [70.3, 75.7] & 69.7 [66.9, 72.5] & 95.2 [93.6, 96.7] & $2.2\cdot10^{4}$\\
 & & 2 & 79.8 & 75.8 [73.3, 78.3] & 74.6 [71.9, 77.2] & 70.9 [68.1, 73.6] & 93.5 [91.8, 95.2] & $3.1\cdot10^{4}$\\
 & & 3 & 78.8 & 73.6 [70.9, 76.3] & 72.9 [70.2, 75.6] & 71.2 [68.5, 73.9] & 96.7 [95.4, 97.9] & $1.1\cdot10^{7}$\\
 & cap + hull & 1 & 72.5 & 68.0 [65.2, 70.8] & 67.8 [65.0, 70.6] & 67.2 [64.3, 70.0] & 98.8 [97.9, 99.6] & 511.4\\
 & & 2 & 79.0 & 74.0 [71.3, 76.6] & 73.3 [70.5, 75.9] & 72.7 [69.9, 75.3] & 98.2 [97.2, 99.1] & 362.3\\
 & & 3 & 65.2 & 55.0 [51.8, 58.1] & 55.0 [51.8, 58.1] & 55.0 [51.8, 58.1] & 100.0 [100.0, 100.0] & 41.5\\
 & hull & 1 & 72.4 & 68.6 [65.9, 71.4] & 67.9 [65.3, 70.7] & 67.2 [64.5, 70.0] & 98.0 [96.8, 99.0] & 173.8\\
 & & 2 & 81.0 & 74.6 [71.8, 77.2] & 74.2 [71.5, 76.9] & 74.0 [71.2, 76.6] & 99.2 [98.5, 99.7] & 236.2\\
 & & 3 & 66.4 & 57.5 [54.6, 60.6] & 57.5 [54.6, 60.6] & 57.5 [54.6, 60.6] & 100.0 [100.0, 100.0] & 21.4\\
 & first study & 0$^\ast$ & 57.2 & 42.4 [39.5, 45.6] & 0.7 [0.2, 1.2] & 0.1 [0.0, 0.3] & 0.2 [0.0, 0.7] & --\\
 & & 1$^\ast$ & 51.9 & 50.0 [46.9, 53.2] & 9.2 [7.5, 11.0] & 8.2 [6.6, 9.9] & 16.4 [13.2, 19.6] & --\\
 & & 2 & 76.1 & 50.3 [47.4, 53.3] & 1.4 [0.7, 2.2] & 0.0 [0.0, 0.0] & 0.0 [0.0, 0.0] & --\\
\midrule
minGRU-type & first recipe & 1 & 72.9 & 40.2 [37.3, 43.2] & 2.4 [1.5, 3.4] & 2.1 [1.3, 3.0] & 5.2 [3.2, 7.5] & $7.4\cdot10^{5}$\\
 & & 2 & 60.3 & 42.7 [39.7, 45.9] & 0.1 [0.0, 0.3] & 0.1 [0.0, 0.3] & 0.2 [0.0, 0.7] & $1.5\cdot10^{6}$\\
 & & 3 & 66.9 & 47.5 [44.5, 50.5] & 0.4 [0.1, 0.8] & 0.3 [0.0, 0.7] & 0.6 [0.0, 1.5] & $7.2\cdot10^{12}$\\
 & linear ramp & 1 & 83.9 & 75.6 [72.9, 78.2] & 40.2 [37.1, 43.3] & 10.8 [8.9, 12.8] & 14.3 [11.8, 16.9] & $6.2\cdot10^{5}$\\
 & & 2 & 83.7 & 76.0 [73.3, 78.6] & 35.0 [32.1, 37.9] & 16.4 [14.2, 18.7] & 21.6 [18.8, 24.5] & $4.4\cdot10^{4}$\\
 & & 3 & 84.5 & 75.6 [72.9, 78.2] & 15.6 [13.4, 17.9] & 6.1 [4.7, 7.6] & 8.1 [6.2, 10.1] & $2.2\cdot10^{5}$\\
 & cap 128 & 1 & 85.2 & 75.9 [73.3, 78.5] & 71.2 [68.5, 73.9] & 68.4 [65.6, 71.2] & 90.1 [87.8, 92.2] & 127.5\\
 & & 2 & 84.3 & 75.0 [72.5, 77.5] & 71.1 [68.4, 73.8] & 68.0 [65.2, 70.8] & 90.7 [88.5, 92.7] & 127.4\\
 & & 3 & 84.1 & 75.2 [72.4, 77.8] & 70.1 [67.3, 72.8] & 65.6 [62.6, 68.4] & 87.2 [84.7, 89.5] & 127.5\\
 & cap 512 & 1 & 83.9 & 75.3 [72.6, 77.9] & 68.6 [65.5, 71.5] & 65.2 [62.1, 68.1] & 86.6 [84.1, 89.1] & 511.5\\
 & & 2 & 83.4 & 74.2 [71.5, 77.0] & 67.6 [64.8, 70.6] & 61.0 [58.1, 64.2] & 82.2 [79.4, 85.0] & 511.5\\
 & & 3 & 83.9 & 75.6 [73.0, 78.2] & 67.6 [64.7, 70.4] & 61.3 [58.2, 64.3] & 81.1 [78.2, 83.9] & 511.5\\
 & cap 2048 & 1 & 83.7 & 75.3 [72.6, 77.8] & 55.1 [52.0, 58.3] & 35.7 [32.7, 38.6] & 47.4 [43.8, 51.0] & 2047.5\\
 & & 2 & 83.4 & 75.0 [72.4, 77.7] & 56.6 [53.5, 59.6] & 28.7 [26.0, 31.6] & 38.3 [34.8, 41.7] & 2047.5\\
 & & 3 & 84.8 & 76.0 [73.4, 78.5] & 57.9 [54.9, 60.9] & 42.4 [39.2, 45.4] & 55.8 [52.2, 59.4] & 2047.5\\
 & fixed-point term & 1 & 83.6 & 75.2 [72.6, 77.8] & 72.8 [70.1, 75.4] & 69.5 [66.7, 72.2] & 92.4 [90.5, 94.2] & 2047.2\\
 & & 2 & 81.8 & 73.9 [71.2, 76.6] & 73.0 [70.3, 75.6] & 70.1 [67.3, 72.8] & 94.9 [93.3, 96.4] & 442.6\\
 & & 3 & 83.3 & 74.7 [72.0, 77.4] & 72.0 [69.2, 74.7] & 69.2 [66.3, 72.1] & 92.6 [90.7, 94.4] & $5.5\cdot10^{4}$\\
 & cap + hull & 1 & 84.5 & 75.7 [73.1, 78.2] & 74.2 [71.4, 76.8] & 72.8 [70.0, 75.5] & 96.2 [94.8, 97.5] & 472.3\\
 & & 2 & 84.5 & 76.2 [73.6, 78.8] & 73.9 [71.2, 76.5] & 72.9 [70.2, 75.6] & 95.7 [94.2, 97.0] & 446.1\\
 & & 3 & 82.1 & 73.2 [70.4, 75.8] & 71.5 [68.7, 74.3] & 71.0 [68.2, 73.8] & 97.0 [95.7, 98.1] & 480.9\\
 & hull & 1 & 83.2 & 75.2 [72.5, 77.8] & 73.7 [70.8, 76.2] & 72.9 [70.0, 75.4] & 96.9 [95.7, 98.1] & 1535.3\\
 & & 2 & 83.1 & 75.6 [73.1, 78.1] & 73.4 [70.7, 76.1] & 72.4 [69.6, 75.1] & 95.8 [94.3, 97.2] & 1816.3\\
 & & 3 & 80.7 & 72.1 [69.4, 74.9] & 70.6 [67.9, 73.5] & 70.4 [67.7, 73.2] & 97.6 [96.4, 98.7] & 1454.3\\
 & first study & 0 & 66.7 & 58.9 [55.9, 62.0] & 2.0 [1.2, 2.9] & 0.9 [0.4, 1.5] & 1.5 [0.7, 2.6] & $6.0\cdot10^{6}$\\
 & & 1 & 66.7 & 52.2 [49.1, 55.3] & 2.4 [1.5, 3.4] & 1.8 [1.0, 2.6] & 3.4 [2.0, 5.0] & $1.7\cdot10^{6}$\\
 & & 2 & 69.4 & 45.7 [42.6, 48.6] & 0.1 [0.0, 0.3] & 0.1 [0.0, 0.3] & 0.2 [0.0, 0.7] & $1.2\cdot10^{12}$\\
\midrule
S4D-type & first recipe & 1$^\ast$ & 52.4 & 46.3 [43.2, 49.5] & 48.3 [45.3, 51.6] & 42.7 [39.6, 45.9] & 92.2 [89.6, 94.5] & 419.0\\
 & & 2 & 65.4 & 37.4 [34.5, 40.5] & 44.5 [41.4, 47.8] & 27.7 [25.0, 30.6] & 74.1 [69.7, 78.6] & 541.9\\
 & & 3$^\ast$ & 58.8 & 45.3 [42.3, 48.5] & 43.8 [40.9, 47.1] & 26.6 [23.9, 29.5] & 58.7 [54.0, 63.4] & 242.0\\
 & linear ramp & 1 & 78.8 & 72.3 [69.4, 75.0] & 69.0 [66.1, 71.8] & 61.2 [58.1, 64.0] & 84.6 [81.9, 87.1] & 401.9\\
 & & 2 & 83.2 & 73.4 [70.8, 76.1] & 67.8 [64.9, 70.7] & 60.4 [57.4, 63.4] & 82.3 [79.5, 85.0] & 353.5\\
 & & 3 & 81.8 & 71.4 [68.7, 74.0] & 66.8 [63.8, 69.6] & 57.5 [54.5, 60.5] & 80.5 [77.5, 83.4] & 279.4\\
 & first study & 0 & 75.7 & 40.2 [37.1, 43.3] & 32.3 [29.3, 35.2] & 27.7 [24.9, 30.5] & 68.9 [64.3, 73.6] & --\\
 & & 1$^\ast$ & 51.4 & 47.8 [44.8, 51.0] & 48.7 [45.7, 51.9] & 45.7 [42.7, 48.8] & 95.6 [93.7, 97.4] & --\\
 & & 2 & 65.6 & 39.4 [36.6, 42.6] & 38.5 [35.6, 41.7] & 26.2 [23.5, 29.0] & 66.5 [62.0, 71.2] & --\\
\bottomrule
\end{tabular}
\end{table}

\begin{table}[htbp]
\caption{Training remedy, token-set certificate (\cref{cor:anyctx}; $P=8$; 200 test documents): share in \% certified for every
context of every length over the most frequent neutral token, for the clean prediction of the bare document; mean over the
non-degenerate seeds with its 95\% t-interval (document in its substitution box), and per seed 1 / 2 / 3 with the document in its
substitution box and with a clean document, next to the clean accuracy on the same documents. Over the 8 and the 64 most frequent
tokens no non-degenerate checkpoint of any arm is certified on any document. $^\ast$Degenerate seed (rule of this appendix, on the
1000 test documents; the token-set test's own rule, more than 80\% of the predictions in one class on its 200 documents, flags
none of these checkpoints). $^{(2)}$, $^{(1)}$: mean over two seeds, one seed.}
\label{tab:remedy-tokens}
\centering\scriptsize\setlength{\tabcolsep}{3pt}
\begin{tabular}{@{}llcccc@{}}
\toprule
 & & mean & \multicolumn{2}{c}{per seed} & clean accuracy\\
\cmidrule(lr){4-5}
model & training & substitution box & substitution box & clean document & per seed\\
\midrule
Mamba & first recipe & 32.8 [$-$383.4, 448.9]$^{(2)}$ & 0.0$^\ast$ / 0.0 / 65.5 & 0.0$^\ast$ / 0.0 / 92.0 & 59.0 / 67.5 / 64.5\\
 & linear ramp & 32.8 [$-$39.5, 105.2] & 0.0 / 55.5 / 43.0 & 0.0 / 62.5 / 52.0 & 80.0 / 79.0 / 75.0\\
 & cap 128 & 86.3 [82.7, 89.9] & 85.5 / 85.5 / 88.0 & 99.5 / 95.5 / 98.5 & 80.0 / 77.5 / 79.5\\
 & cap 512 & 66.2 [34.9, 97.4] & 79.5 / 64.5 / 54.5 & 95.5 / 73.0 / 66.5 & 83.0 / 81.0 / 79.0\\
 & cap 2048 & 48.3 [22.7, 74.0] & 36.5 / 55.5 / 53.0 & 44.5 / 62.0 / 62.5 & 80.5 / 80.5 / 75.5\\
 & fixed-point term & 83.5 [79.0, 88.0] & 81.5 / 84.0 / 85.0 & 93.5 / 92.0 / 94.0 & 78.0 / 79.5 / 76.5\\
 & cap + hull & 85.5 [73.7, 97.3] & 88.5 / 88.0 / 80.0 & 99.0 / 99.5 / 100.0 & 76.5 / 78.5 / 66.0\\
 & hull & 86.5 [77.5, 95.5] & 89.5 / 87.5 / 82.5 & 99.0 / 99.5 / 100.0 & 72.5 / 78.0 / 68.0\\
minGRU-type & first recipe & 0.0 [0.0, 0.0] & 0.0 / 0.0 / 0.0 & 0.0 / 0.0 / 0.0 & 74.5 / 61.0 / 67.5\\
 & linear ramp & 0.0 [0.0, 0.0] & 0.0 / 0.0 / 0.0 & 0.0 / 0.0 / 0.0 & 81.5 / 82.0 / 82.5\\
 & cap 128 & 47.0 [12.2, 81.8] & 63.0 / 41.0 / 37.0 & 77.5 / 57.0 / 50.0 & 86.0 / 83.0 / 81.0\\
 & cap 512 & 60.0 [26.6, 93.4] & 62.5 / 72.0 / 45.5 & 75.5 / 83.5 / 61.0 & 84.0 / 83.0 / 83.0\\
 & cap 2048 & 20.0 [$-$1.5, 41.5] & 15.0 / 30.0 / 15.0 & 28.5 / 44.0 / 28.0 & 82.0 / 82.0 / 81.5\\
 & fixed-point term & 14.3 [$-$45.2, 73.9] & 1.0 / 42.0 / 0.0 & 3.0 / 52.5 / 0.0 & 82.0 / 81.5 / 79.5\\
 & cap + hull & 75.7 [63.2, 88.2] & 75.0 / 71.0 / 81.0 & 88.0 / 86.5 / 95.5 & 81.0 / 82.5 / 80.5\\
 & hull & 0.2 [$-$0.6, 0.9] & 0.5 / 0.0 / 0.0 & 2.5 / 0.0 / 0.0 & 80.0 / 83.5 / 80.0\\
S4D-type & first recipe & 42.5$^{(1)}$ & 80.0$^\ast$ / 42.5 / 37.5$^\ast$ & 90.5$^\ast$ / 60.5 / 58.0$^\ast$ & 53.5 / 66.0 / 59.5\\
 & linear ramp & 65.2 [44.0, 86.3] & 64.5 / 57.0 / 74.0 & 74.5 / 71.5 / 88.5 & 76.0 / 82.0 / 77.5\\
\bottomrule
\end{tabular}
\end{table}

\begin{table}[htbp]
\caption{Training remedy: differences in points. Top: arm minus the linear-ramp recipe (clean accuracy and certified accuracy at
$L=0$; every seed), and natural training with the cap minus natural training without it, with 95\% Welch intervals. Bottom: the
seed-paired contrasts of the cap with the hull term against the cap alone ($\tau_{\max}=512$) and against the hull term without the
cap, with 95\% paired t-intervals; ``pairs'': seed pairs in which the K-uniform ratio of the cap with the hull term exceeds the
other's by at least 10 points. $-$0.0 marks a negative difference that rounds to 0.0.}
\label{tab:remedy-cost}
\centering\scriptsize\setlength{\tabcolsep}{3pt}
\begin{tabular}{@{}lccccc@{}}
\toprule
 & \multicolumn{2}{c}{Mamba} & \multicolumn{2}{c}{minGRU-type} & \\
\cmidrule(lr){2-3}\cmidrule(lr){4-5}
against the linear-ramp recipe & clean & $L=0$ & clean & $L=0$ & \\
\midrule
cap 128 & $-$0.9 [$-$8.8, 6.9] & $-$1.1 [$-$7.2, 5.0] & 0.5 [$-$0.8, 1.8] & $-$0.4 [$-$1.7, 0.9] & \\
cap 512 & 0.7 [$-$5.6, 6.9] & $-$0.9 [$-$6.8, 5.0] & $-$0.3 [$-$1.2, 0.6] & $-$0.7 [$-$2.6, 1.2] & \\
cap 2048 & 0.3 [$-$7.6, 8.3] & $-$0.1 [$-$5.4, 5.2] & $-$0.1 [$-$1.6, 1.5] & $-$0.3 [$-$1.7, 1.1] & \\
fixed-point term & $-$2.2 [$-$10.2, 5.8] & $-$1.9 [$-$6.7, 3.0] & $-$1.1 [$-$3.7, 1.5] & $-$1.1 [$-$2.9, 0.6] & \\
cap + hull & $-$9.9 [$-$28.6, 8.8] & $-$10.4 [$-$35.2, 14.4] & $-$0.3 [$-$3.9, 3.3] & $-$0.7 [$-$4.7, 3.3] & \\
hull & $-$8.8 [$-$28.5, 10.9] & $-$9.2 [$-$31.4, 13.1] & $-$1.7 [$-$5.4, 2.0] & $-$1.4 [$-$6.2, 3.4] & \\
\midrule
natural: cap minus none & clean & & clean & & \\
cap 128 & $-$1.7 [$-$4.6, 1.2] & & $-$1.0 [$-$4.5, 2.4] & & \\
cap 512 & $-$0.0 [$-$1.4, 1.4] & & 0.2 [$-$1.4, 1.9] & & \\
cap 2048 & $-$0.9 [$-$2.1, 0.3] & & 0.0 [$-$1.4, 1.4] & & \\
\midrule
seed-paired & \multicolumn{5}{c}{K-uniform ratio / K-uniform share / clean / $L=0$ / pairs}\\
cap + hull $-$ cap 512, Mamba & \multicolumn{5}{c}{14.8 [0.8, 28.8] / 1.7 [$-$23.1, 26.4] / $-$10.5 [$-$28.4, 7.3] / $-$9.5 [$-$32.3, 13.3] / 3 of 3}\\
cap + hull $-$ cap 512, minGRU-type & \multicolumn{5}{c}{13.0 [5.1, 20.9] / 9.7 [4.4, 15.1] / $-$0.0 [$-$3.9, 3.8] / 0.0 [$-$5.5, 5.5] / 2 of 3}\\
cap + hull $-$ hull, Mamba & \multicolumn{5}{c}{$-$0.0 [$-$2.3, 2.2] / $-$1.3 [$-$4.4, 1.8] / $-$1.0 [$-$3.7, 1.6] / $-$1.2 [$-$4.0, 1.5] / 0 of 3}\\
cap + hull $-$ hull, minGRU-type & \multicolumn{5}{c}{$-$0.5 [$-$1.4, 0.4] / 0.3 [$-$0.6, 1.3] / 1.4 [1.2, 1.5] / 0.7 [$-$0.1, 1.5] / 0 of 3}\\
\bottomrule
\end{tabular}
\end{table}

\paragraph{Reading: the linear-ramp recipe.}
Standard IBP with the linear-ramp recipe reached a mean clean accuracy of 82.1\% [74.6, 89.6] (Mamba), 84.0\% [83.0, 85.1]
(minGRU-type) and 81.3\% [75.7, 86.9] (S4D-type), with no degenerate seed and 76.1\% [70.5, 81.7], 75.7\% [75.2, 76.3] and 72.4\%
[69.9, 74.9] certified at $L=0$ (\cref{tab:remedy,tab:remedy-clean}). The first recipe, rerun on the same data, reached 60.0\% [42.5, 77.6],
66.7\% [51.0, 82.4] and 58.9\% [42.7, 75.0] clean, with one degenerate Mamba seed and two degenerate S4D-type seeds. It did not
reproduce the first study's certified share at $L=0$ on the minGRU-type model within the registered 5 points: against the first study,
the rerun differs by $-$0.9 [$-$17.0, 15.2] points in clean accuracy, $-$8.8 [$-$22.7, 5.1] at $L=0$ and $-$0.5 [$-$3.8, 2.7] at
$L=\infty$ (Welch intervals). For Mamba and the S4D-type model one side has fewer than two non-degenerate seeds, so the comparison is
not made; the differences, without intervals, are $-$12.4, $-$1.9 and $+$21.8 points (Mamba) and $-$5.2, $-$2.4 and $+$9.1 points
(S4D-type). Where this appendix refers to the first recipe, it uses this rerun, not the first study's numbers; the remedies are
compared with the linear-ramp recipe.

\paragraph{Reading: the minGRU-type model.}
Without a remedy the blind spot persists: the linear-ramp recipe keeps 40.0\% [$-$2.5, 82.5] of its $L=0$ certificates at $L=\infty$
(per seed 53.2, 46.1 and 20.6\%; \cref{tab:remedy-ratio}). Resetting its channels slower than 512 tokens to their one-period radius
restores 75.2\% of the 141 width failures (pooled; a diagnostic, not a certificate; \cref{tab:remedy-attr}). The cap at
$\tau_{\max}=512$ with the hull term from the start gives 83.7\% [80.3, 87.1] clean accuracy, with certificates for every number of
copies of the context period on 72.2\% [69.6, 74.9] of the documents (one copy 73.8\% [69.9, 77.7]; 1024 contiguous tokens 73.8\%
[70.9, 76.6]; \cref{tab:remedy}). That is 2.8 [1.4, 4.2] points below $L=0$, at a
clean-accuracy difference of $-$0.3 [$-$3.9, 3.3] points from the linear-ramp recipe; natural training reaches 87.7\% [86.2, 89.2]
without the cap, and the same cap changes it by $+$0.2 [$-$1.4, 1.9] points. The hull term adds to the cap: with the cap alone, the
share certified for every number of copies was 9.7 [4.4, 15.1] points lower (paired by seed; in two of three pairs the combination's
ratio exceeds the cap's by at least 10 points). The hull term from the start gives the result without the cap as well (82.3\% [78.8,
85.8] clean; 71.9\% [68.6, 75.2] for every number of copies, 73.3\% [69.0, 77.7] for one copy, 73.2\% [68.4, 78.1] after 1024 tokens;
2.4 [0.5, 4.3] points below $L=0$), so the cap is not needed for it (\cref{tab:remedy-cost}). The cap at $\tau_{\max}=128$ alone keeps
94.2\% [92.5, 95.9] of the width-only $L=\infty$ certificates and 89.3\% [84.8, 93.9] of the K-uniform ones, at a clean-accuracy
difference of 0.5 [$-$0.8, 1.8] points, and raises the $L=\infty$ ratio over the linear-ramp recipe by 54.0 points (Welch interval
[11.4, 96.5]). The fixed-point term from the start keeps 97.3\% [94.1, 100.5] of the certificates at $L=\infty$ (72.6\% [71.3, 73.9]
certified; after one and four copies 73.0\% [72.0, 74.0] and 72.8\% [71.5, 74.0]), at $-$1.1 [$-$3.7, 1.5] points. Over the most
frequent neutral token, no context of any length changes the prediction of the cap + hull models, with the document in its
substitution box, for 75.7\% [63.2, 88.2] of the 200 documents; over the 8 and the 64 most frequent tokens, for none, so contexts of
a few dozen distinct tokens remain open (\cref{tab:remedy-tokens}).

\paragraph{Reading: Mamba.}
With a gentler schedule the standard objective already keeps most certificates at $L=\infty$: the linear-ramp recipe keeps 78.5\%
[27.3, 129.6] of its $L=0$ certificates there (per seed 54.8, 92.5 and 88.1\%; \cref{tab:remedy-ratio}), so the collapse in the
first study was partly its schedule. \Cref{prop:horizon} remains a statement about objectives, and the minGRU-type result shows that
slow channels can still cost the certificates. The K-uniform ratio of these models is lower, 61.1\% [$-$14.3, 136.4] (seed~1:
26.4\%; \cref{tab:remedy-seeds}). Of their remaining
$L=\infty$ width failures, resetting the slow channels restores 36.2\% of the 47 (pooled), so most of them are not attributed to slow
channels. The hull term, with or without the cap, keeps 99.0\% [96.8, 101.2] and 99.1\% [96.5, 101.6] of the $L=0$ certificates for
every number of copies, but at clean-accuracy differences of $-$9.9 [$-$28.6, 8.8] and $-$8.8 [$-$28.5, 10.9] points from the
linear-ramp recipe: two of three seeds stay below 78\% clean (72.5, 79.0 and 65.2\% with the cap; 72.4, 81.0 and 66.4\% without), so
this is not a result at competitive accuracy. The combination does not exceed the hull term without the cap by 10 ratio points in any
seed pair, and the paired differences are within seed noise (ratio $-$0.0 [$-$2.3, 2.2] points, K-uniform share $-$1.3 [$-$4.4, 1.8],
clean accuracy $-$1.0 [$-$3.7, 1.6], $L=0$ share $-$1.2 [$-$4.0, 1.5]; \cref{tab:remedy-cost}). Against the cap alone the
combination has a higher K-uniform
ratio in all three pairs ($+$14.8 [0.8, 28.8] points), with no measurable difference in the K-uniform share ($+$1.7 [$-$23.1, 26.4])
and a clean-accuracy difference of $-$10.5 [$-$28.4, 7.3]. The cap at $\tau_{\max}=128$ keeps 98.0\% [95.1, 100.8] of the width-only
$L=\infty$ certificates and 96.2\% [90.6, 101.9] of the K-uniform ones, at $-$0.9 [$-$8.8, 6.9] points of clean accuracy, but does not
raise the $L=\infty$ ratio beyond the seed noise of the linear-ramp recipe ($+$20.1 points, Welch interval [$-$31.1, 71.3]). The
fixed-point term from the start keeps 99.1\% [97.4, 100.7] of the certificates at $L=\infty$ (73.5\% [71.1, 75.9] certified; after
one and four copies 73.5\% [71.3, 75.7] and 73.5\% [71.2, 75.8]), at $-$2.2 [$-$10.2, 5.8] points. Over the most frequent neutral token
the cap + hull models are certified for every context of any length, with the document in its substitution box, on 85.5\% [73.7,
97.3] of the documents; over the 8 and the 64 most frequent tokens, on none (\cref{tab:remedy-tokens}).

\paragraph{Reading: the S4D-type model.}
Under the linear-ramp recipe the S4D-type model keeps 93.8\% [89.9, 97.6] of its $L=0$ certificates at $L=\infty$ (67.9\% [65.1,
70.6] certified; after one and four copies 68.2\% [64.2, 72.1] and 67.9\% [65.1, 70.6]); its slowest channel has a time constant of
279--402 tokens. Under this recipe the contrast with the selective models holds for the minGRU-type model and not for Mamba.

\paragraph{Descriptive observations (not read by any criterion).}
(i) The cap alone at $\tau_{\max}=512$ met the headline's per-seed rule on two of three Mamba seeds (seeds 2 and 3: K-uniform ratios
87.9 and 87.1\% at 81.7 and 82.0\% clean accuracy), with a K-uniform share of 63.3\% [53.7, 72.9] (one copy 69.6\% [61.0, 78.2]; 1024
tokens 70.1\% [63.4, 76.8]), 11.9 [1.0, 22.7] points below $L=0$, at 82.8\% [78.8, 86.7] clean. This is not a pre-registered result.
(ii) On seed~3, Mamba trained with the hull term converged, with and without the cap, to models whose slowest channel has a time
constant of 41.5 and 21.4 tokens: their certified share is the same at every context (55.0 and 57.5\%), at 65.2 and 66.4\% clean.
Both are non-degenerate by the rule of this study and enter every mean.
(iii) Over the most frequent neutral token, with the document in its substitution box and on 200 documents, the uncapped
minGRU-type models trained with the hull term or with the fixed-point term from the start are certified on 0.2\% [$-$0.6, 0.9] and
14.3\% [$-$45.2, 73.9] of the documents, and the uncapped Mamba models on 86.5\% [77.5, 95.5] and 83.5\% [79.0, 88.0]. No criterion
reads these shares, and no contrast with the capped models is pre-registered; the result that the cap is not needed concerns the
K-uniform certificate of the context period only.
(iv) Under the first recipe, one Mamba seed (seed~3) kept 81.2\% [78.0, 84.4] (95\% bootstrap interval over the documents) of its
$L=0$ certificates K-uniformly.
(v) The natural models reach 85.3--88.8\% (Mamba) and 85.5--88.3\% (minGRU-type) per seed with or without the cap; the cap costs at
most 1.7 points on average, and every interval contains 0 (\cref{tab:remedy-cost}).

\paragraph{Where the information goes.}
A pre-registered evidence probe (\cref{app:prereg-criteria}, the paragraph of the same name) measured how far the checkpoints of
this study carry a document's evidence past trailing neutral words. Every checkpoint of the main stage read the 1000 test documents
followed by $d\in\{64,128,256,512,1024,2048,4096\}$ words of licence text (the trailing part of the same draw as the period before
the document, so the two never overlap), on three inputs per document and $d$: the document and the words; another document (test
document $i+1$) and the same words; and the words alone. The quantities are those of \cref{app:memory}: the clean accuracy
$\mathrm{acc}(d)$, the accuracy of the other-document rule, their difference the carry (exactly 0 when the prediction does not
depend on which document precedes the words; paired bootstrap interval over the documents), the agreement of the two predictions,
the largest change of the margin between them, and the class E / N / P / D of each checkpoint at each $d$ (\cref{app:memory},
``Quantities''). Everything is a clean accuracy in float64, not a certificate; the intervals are per checkpoint, conditional on the
trained checkpoints and on the 56{,}880-word licence stream, and not simultaneous; every validity check passed (the draw equals
this study's, the bare documents reproduce this study's counts within two documents, and the four archived references of the first
study reproduce their archived CPU runs on every document). Four predictions were read on the non-degenerate seeds
(\cref{tab:remedy-evidence}). (i) \emph{Under the cap at 128 the models forget.} With every decay capped at $1-1/128$, after 4096
trailing neutral words the prediction no longer depends on the document: two different documents followed by the same words give
the same prediction on at least 99.9\% of the documents (100.0\% on every one of the 12 checkpoints: cap 128 with and without IBP,
both architectures, three seeds), and the document moves the margin by at most $8.7\times10^{-12}$. This is the fading memory the
cap implies (the argument of \cref{lem:fading}(ii) with $\bar a=\rho$); no certificate shows it. (ii) \emph{No detectable evidence
after $8\tau_{\max}$ words.} After $8\tau_{\max}$ trailing words, in 3 of 3 seeds, the capped models (cap 128 after 1024 words; cap
512, cap + hull and natural with cap 512 after 4096 words; both architectures, ten cells) give the same prediction for two different
documents on at least 98\% of the documents, and their predictions carry no detectable evidence of the document (carry 0.0 [0.0,
0.0] in eight cells, $-$0.0 [$-$0.7, 0.6] for the minGRU-type cap 128 and 0.0 [$-$0.1, 0.2] for the natural Mamba cap 128; accuracy
47.2--54.0\% per seed, at chance). At $d=\tau_{\max}$ the same models still carry evidence in most cells (\cref{tab:remedy-evidence}: 14.9 [11.0,
18.7] and 24.5 [22.2, 26.7] points for cap 128 after 128 words (Mamba, minGRU-type), 12.1 [1.5, 22.6] and 13.4 [9.5, 17.2] for cap 512
after 512 words, 10.7 [3.8, 17.6] for the minGRU-type cap + hull and 2.8 [$-$8.9, 14.6] for the Mamba cap + hull, classes P, E, P),
and at $d=2\tau_{\max}$ less (\cref{tab:remedy-evidence}). (iii) \emph{The cap at 128 costs memory.} After 256 trailing words the
capped IBP models carry less evidence than the uncapped models of the same seeds in 3 of 3 pairs: paired difference $-$16.2
[$-$24.0, $-$8.3] points for Mamba (the uncapped models' carry 21.8 [14.2, 29.4]) and $-$10.0 [$-$15.9, $-$4.1] for the minGRU-type
model (24.8 [20.5, 29.1]); the capped natural models $-$9.8 [$-$12.6, $-$6.9] (Mamba; 24.0 [21.8, 26.3]) and $-$13.6 [$-$16.4, $-$10.9]
(minGRU-type; 26.0 [22.6, 29.4]). Whether this cost matters is not measured here: on IMDB the cap costs at most 1.7 points of natural
accuracy (observation (v); \cref{app:memory} measures a task that does). (iv) \emph{The uncapped models after 4096
words.} In at least two of three seeds the uncapped models of this study carry no detectable evidence in the prediction after 4096
neutral words in eight of nine cells: the linear-ramp recipe for Mamba (2 of 3 seeds; seed~1 is class E with carry 4.8 [2.9, 6.7])
and for the S4D-type model (3 of 3), the fixed-point term, the hull term and natural training for both architectures (3 of 3 each),
with every accuracy in the chance band. The ninth cell fails: the linear-ramp minGRU-type models carry detectable evidence past
4096 words in 2 of 3 seeds (carry 6.4 [4.2, 8.5] and 6.8 [5.0, 8.8], agreement 87.6 and 90.2\%; the third seed 0.6 [$-$0.3, 1.4];
mean 4.6 [$-$4.0, 13.2]). Next to it their certificates: K-uniform share 11.1\% [$-$1.7, 23.9], ratio 14.6\% [$-$2.2, 31.4]
(\cref{tab:remedy}); for the Mamba linear ramp, K-uniform 46.1\% [$-$9.5, 101.6], ratio 61.1\% [$-$14.3, 136.4], where seed~1, the
seed with evidence at 4096 words and the largest carry after 1024 words (15.0), has the lowest ratio, 26.4\%. No certificate
certifies this memory: none covers a context after the document. For the fixed-point term this answers whether its slow channels
carry information: not past 4096 words in any seed, but at shorter lengths (last length with detectable evidence 256 words on
Mamba, 512 on the full grid; 1024 on the minGRU-type model, in every seed). The last length of the grid with detectable evidence per
seed: linear ramp, Mamba 4096, 1024, 1024; S4D-type 256, 256, 256 (512 on the full grid); minGRU-type 4096, 4096, 1024 (2048 on the
full grid). Reported contrasts beyond (iii) (carry, capped minus uncapped, paired by seed; cap + hull against the linear ramp
unpaired, Welch): cap 512 against the linear ramp after 1024 words $-$4.5 [$-$16.0, 7.0] (Mamba) and $-$12.7 [$-$14.7, $-$10.7]
(minGRU-type); cap + hull against the linear ramp $-$6.5 [$-$24.2, 11.2] and $-$14.0 [$-$15.2, $-$12.7]; natural with cap 512 against
natural $-$2.7 [$-$4.3, $-$1.1] and $-$4.2 [$-$6.5, $-$1.9]; after 4096 words, cap 2048 against the linear ramp $-$1.4 [$-$7.2, 4.4] and
$-$4.0 [$-$12.0, 4.0], natural with cap 2048 against natural 0.0 [$-$0.2, 0.2] and 0.0 [$-$0.1, 0.2]. No criterion reaches the cap at
2048 ($8\tau_{\max}>4096$); under it the carry after 2048 words is 2.1 [$-$6.8, 11.1] (Mamba IBP; classes E, P, P) and 2.7 [0.4, 4.9]
(minGRU-type; E, E, E), after 4096 words 0.2 [$-$0.8, 1.2] (P, P, P) and 0.6 [$-$1.1, 2.3] (seed~1 class E, 1.4 [0.7, 2.2]). The probe
says nothing about every model forgetting, nothing about the first recipe or the cap at 2048 beyond the values reported, nothing
about adversarial trailing text or contexts before the document, and nothing about a task that needs memory, which \cref{app:memory}
measures; no certificate covers a context after the document.

\begin{table}[htbp]
\caption{The evidence probe on the training remedy's checkpoints (1000 test documents; clean accuracy in float64, not a
certificate): the carry after $d$ trailing licence words (accuracy minus the other-document rule's, in points; mean over the
non-degenerate seeds with its 95\% t-interval over the seeds) with the class of each seed (E: the carry's bootstrap interval lies
above 0; P: it contains 0 and the two predictions agree on at least 98\% of the documents; D: it contains 0 with less agreement), at
$d=\tau_{\max}$ for the capped arms and at $d=256$, 1024 and 4096; and, next to them, the arm's certificates of \cref{tab:remedy}
(K-uniform share and ratio; the one-copy share next to each stands in \cref{tab:remedy}; for the fixed-point term the $L=\infty$
share and ratio, its K-uniform reading being outside its pre-registration). Natural models have no certificate. The first-recipe checkpoints are not read by any criterion and are left
out. $^{(w)}$The Mamba seed-3 models of cap + hull and hull (observation (ii)) are class P at every $d\ge64$ with 100\% agreement
(clean accuracy 65.2 and 66.4\%): non-degenerate by the rule of this study, their predictions barely depend on the document; they
enter the ``two of three seeds'' counts without changing any reading, since the other two seeds share the class.}
\label{tab:remedy-evidence}
\centering\scriptsize\setlength{\tabcolsep}{1.5pt}
\begin{tabular}{@{}lccccc@{}}
\toprule
training & $d=\tau_{\max}$ & $d=256$ & $d=1024$ & $d=4096$ & K-uniform; ratio (\cref{tab:remedy})\\
\midrule
\multicolumn{6}{@{}l}{\emph{Mamba}}\\
linear ramp & -- & 21.8 [14.2, 29.4] E E E & 6.8 [$-$10.9, 24.5] E E E & 1.6 [$-$5.2, 8.4] E P P & 46.1 [$-$9.5, 101.6]; 61.1 [$-$14.3, 136.4]\\
cap 128 & 14.9 [11.0, 18.7] E E E & 5.6 [4.5, 6.8] E E E & 0.0 [0.0, 0.0] P P P & 0.0 [0.0, 0.0] P P P & 72.1 [65.6, 78.7]; 96.2 [90.6, 101.9]\\
cap 512 & 12.1 [1.5, 22.6] E E E & 20.1 [14.4, 25.8] E E E & 2.3 [$-$4.3, 8.9] E E P & 0.0 [0.0, 0.0] P P P & 63.3 [53.7, 72.9]; 84.2 [70.1, 98.3]\\
cap 2048 & 2.1 [$-$6.8, 11.1] E P P & 21.9 [15.9, 27.9] E E E & 5.9 [$-$10.8, 22.5] E E D & 0.2 [$-$0.8, 1.2] P P P & 58.1 [43.0, 73.2]; 76.6 [53.3, 99.9]\\
fixed-point term & -- & 10.5 [4.1, 16.9] E E E & 0.1 [$-$0.3, 0.5] P P P & 0.0 [0.0, 0.0] P P P & $L=\infty$: 73.5 [71.1, 75.9]; 99.1 [97.4, 100.7]\\
cap + hull$^{(w)}$ & 2.8 [$-$8.9, 14.6] P E P & 7.2 [$-$13.0, 27.4] E E P & 0.3 [$-$1.0, 1.6] P D P & 0.0 [0.0, 0.0] P P P & 65.0 [42.5, 87.5]; 99.0 [96.8, 101.2]\\
hull$^{(w)}$ & -- & 5.8 [$-$11.0, 22.6] E E P & 0.1 [$-$0.2, 0.4] P P P & 0.0 [0.0, 0.0] P P P & 66.2 [45.6, 86.8]; 99.1 [96.5, 101.6]\\
natural & -- & 24.0 [21.8, 26.3] E E E & 4.0 [3.4, 4.5] E E E & $-$0.0 [$-$0.2, 0.1] P P P & --\\
natural, cap 128 & 25.9 [23.0, 28.8] E E E & 14.3 [9.6, 18.9] E E E & 0.0 [$-$0.1, 0.2] P P P & 0.0 [0.0, 0.0] P P P & --\\
natural, cap 512 & 10.7 [1.7, 19.8] E E E & 22.1 [13.9, 30.3] E E E & 1.3 [0.2, 2.4] E E E & 0.0 [0.0, 0.0] P P P & --\\
natural, cap 2048 & 0.2 [$-$0.4, 0.7] P P P & 25.3 [20.5, 30.2] E E E & 3.8 [0.2, 7.4] E E E & $-$0.0 [$-$0.2, 0.1] P P P & --\\
\multicolumn{6}{@{}l}{\emph{S4D-type}}\\
linear ramp & -- & 14.0 [4.0, 24.0] E E E & 0.2 [$-$0.5, 0.8] P P P & 0.0 [0.0, 0.0] P P P & 59.7 [54.9, 64.5]; 82.5 [77.4, 87.6]\\
\multicolumn{6}{@{}l}{\emph{minGRU-type}}\\
linear ramp & -- & 24.8 [20.5, 29.1] E E E & 15.4 [14.1, 16.6] E E E & 4.6 [$-$4.0, 13.2] E E P & 11.1 [$-$1.7, 23.9]; 14.6 [$-$2.2, 31.4]\\
cap 128 & 24.5 [22.2, 26.7] E E E & 14.7 [13.1, 16.4] E E E & $-$0.0 [$-$0.7, 0.6] P P P & 0.0 [0.0, 0.0] P P P & 67.3 [63.6, 71.1]; 89.3 [84.8, 93.9]\\
cap 512 & 13.4 [9.5, 17.2] E E E & 20.4 [15.3, 25.5] E E E & 2.7 [$-$0.3, 5.7] E E E & 0.0 [0.0, 0.0] P P P & 62.5 [56.7, 68.3]; 83.3 [76.1, 90.5]\\
cap 2048 & 2.7 [0.4, 4.9] E E E & 21.8 [17.7, 25.9] E E E & 10.1 [7.6, 12.5] E E E & 0.6 [$-$1.1, 2.3] E P P & 35.6 [18.6, 52.6]; 47.2 [25.4, 68.9]\\
fixed-point term & -- & 13.8 [12.5, 15.1] E E E & 1.1 [0.7, 1.5] E E E & 0.0 [0.0, 0.0] P P P & $L=\infty$: 72.6 [71.3, 73.9]; 97.3 [94.1, 100.5]\\
cap + hull & 10.7 [3.8, 17.6] E E E & 18.1 [14.6, 21.6] E E E & 1.4 [1.2, 1.6] E E E & 0.0 [0.0, 0.0] P P P & 72.2 [69.6, 74.9]; 96.3 [94.6, 97.9]\\
hull & -- & 19.3 [10.0, 28.7] E E E & 5.5 [0.0, 11.0] E E E & 0.0 [0.0, 0.0] P P P & 71.9 [68.6, 75.2]; 96.8 [94.4, 99.1]\\
natural & -- & 26.0 [22.6, 29.4] E E E & 6.5 [3.8, 9.1] E E E & 0.0 [0.0, 0.0] P P P & --\\
natural, cap 128 & 24.1 [22.9, 25.4] E E E & 12.3 [8.5, 16.2] E E E & 0.0 [0.0, 0.0] P P P & 0.0 [0.0, 0.0] P P P & --\\
natural, cap 512 & 12.4 [5.5, 19.4] E E E & 23.8 [21.9, 25.7] E E E & 2.2 [$-$0.9, 5.4] E E E & 0.0 [0.0, 0.0] P P P & --\\
natural, cap 2048 & 0.4 [$-$0.6, 1.5] E P P & 25.5 [24.7, 26.4] E E E & 5.0 [0.5, 9.4] E E E & 0.0 [$-$0.1, 0.2] P P P & --\\
\bottomrule
\end{tabular}
\end{table}

\paragraph{A zero-order-hold S4D-type control.}
Our S4D-type layer injects $x_tB$ with no forgetting factor (\cref{tab:layers}); the standard S4D discretisation injects $\bar Bx_t$
with $\bar B=(\Delta A)^{-1}(e^{\Delta A}-1)\Delta B$, which carries the factor $1-a_t$ and to which \cref{prop:horizon} applies at fixed
$B$ (\cref{rem:injnorm}). A pre-registered control (\cref{app:prereg-criteria}) trains the same block with this zero-order-hold
injection: per channel and state, $a=e^{-\alpha\Delta}$ and $\bar B=\frac{1-a}{\alpha}B$, the exact discretisation of
$\dot h=-\alpha h+Bx(t)$ with $x$ held over a step of length $\Delta$ ($\Delta$ a parameter, as in our layer), with the parameters,
shapes and initialisation of our layer (the same $A_{\log}$ and $\Delta$ per seed, so the time constants start equal), and every
checkpoint is evaluated after conversion to our layer's form, $B\leftarrow\frac{1-a}{\alpha}B$, the network the certificates refer to.
Arms, seeds 1--3: the linear-ramp recipe alone (A1z), paired with our layer's linear-ramp runs of the same seeds (the same initial
network and batches); the hull term from the start on it (A5z; the weight-ramp rule did not fire); and natural training on the
zero-order-hold layer (Nz) and on our layer (Nu), reported only. The predictions: the control develops slow channels (in two of
three seeds a largest time constant above $\max(1000,\text{its initial value})$ and a layer with at least 1\% of its channel--state
pairs slower than 4096 words); it loses at least 0.10 of its $L=\infty$ ratio against our layer in two of three seed pairs;
resetting its channels slower than 512 words restores at least half of its width failures; and the hull term meets the headline
bar on it. Every validity check passed (the conversion reproduces $\bar B$ entry by entry on the 12 checkpoints, no per-document
implication is violated, the reference rows equal this study's, the census equals the diagnostic's and the initialisation was
recomputed); the control is comparable (clean accuracy 81.2\% [80.8, 81.6], no degenerate seed, against a bar of 76.3\%). The first
three predictions failed and the fourth passed, so the reading table's rows for ``Z1 and Z2 fail'' and for ``Z4 passes while A1z's
ratio is already at least 0.8'' apply (\cref{tab:remedy-zoh}). Trained like ours, the zero-order-hold S4D-type layer developed no
channel slower than 512 words (largest time constant 274.9 / 259.8 / 270.9 words, initially at most 969.5 / 963.4 / 974.9; 0\% of
the largest layer's channels slower than 512 words in every seed, initially 1.07\% / 1.12\% / 1.27\%), and in none of the three
seed pairs did its $r_{\rm inf}$ fall 0.10 below our layer's (96.3\% [95.1, 97.5] against 93.8\% [89.9, 97.6]; paired difference
$+$0.025 [$-$0.022, $+$0.072], at fixed $B$). On this task and recipe the normalisation alone did not produce the $L=\infty$ loss.
\Cref{prop:horizon} still applies to its objectives at fixed $B$. Per seed, $r_{\rm inf}$ is 0.958, 0.963 and 0.967 against 0.954,
0.924 and 0.936 for our layer; the width-only ratio differs by $-$0.007 [$-$0.025, $+$0.012]; the intervals contain 0, so no
difference is detected in either direction, and the control's flip share (0.440 [0.176, 0.704]) is below the 0.5 at which the
flips reading would apply. The counterfactual that resets the channels slower than 512 words restored 0.0 of 18 width failures
(6, 5 and 7 per seed), mechanically: the control has no such channel, so the reset changes nothing, and nothing about the
mechanism of \cref{prop:horizon} is read from it. A5z keeps $r_{\rm hull}\ge0.8$ at 79.6\% [76.9, 82.2] clean, as A1z already did
without the hull term: no effect of the hull term is shown (the registered remedy reading needs A1z's ratio below 0.8 in the seeds
that support the prediction; A1z's K-uniform ratios are 88.0, 85.3 and 85.0\% per seed; next to it, the paired differences A5z
minus A1z: ratio $+$11.9 [6.4, 17.4] points, K-uniform share $+$6.7 [2.1, 11.3], clean accuracy $-$1.6 [$-$3.9, $+$0.7], $L=0$ share
$-$1.8 [$-$3.2, $-$0.4]; A5z's ratios 97.4, 98.0 and 98.7\% at 80.7, 79.4 and 78.6\% clean). The largest time constants of A5z are 81.4,
95.8 and 97.4 words. The natural arms, reported: clean accuracy 85.8\% [82.9, 88.7] (Nz) and 85.3\% [83.9, 86.7] (Nu); largest time
constants 562.3 / 471.1 / 331.3 (Nz) and 399.7 / 356.2 / 364.7 words (Nu), initially 969.5 / 963.4 / 974.9; 0\% slower than 4096 in
every seed, and slower than 512 only in Nz seed~1 (0.049\%); nothing about the robust term's role is inferred from them, whose
steps, schedule and weight-decay integral differ from the IBP arms'. This control compares two parametrisations, including their
optimiser geometry (Adam's relative step on $B$ differs between them by a factor of 10 to 2000), on two-layer width-64 models, one
task and three seeds, with the linear-ramp recipe, matched initialisation and the converted network; its certificates are float64,
not re-verified with outward rounding (\cref{rem:float}). Nothing is inferred about the normalisation's role in the training
contrast between our S4D-type layer and the selective models; the control does not test whether this layer can hide channels at
fixed $B$ (\cref{prop:horizon} says it can); no capped zero-order-hold arm was trained, so nothing is said about whether the cap is
needed for this layer.

\begin{table}[htbp]
\caption{The zero-order-hold S4D-type control (1000 test documents, \%; float64 certificates, not re-verified with outward
rounding; mean over the three seeds with its 95\% t-interval over the seeds): clean accuracy, the certified shares at $L=0$, after
1024 tokens of neutral text, after one copy of the 1024-token period, at $L=\infty$ and K-uniformly, and the ratios $L=\infty$ and
K-uniform over $L=0$; below, the largest time constant per seed (words; seeds 1--3; every arm of a seed starts from the same
initialisation, whose largest time constants are 969.5, 963.4 and 974.9). A1z: the zero-order-hold layer under the linear-ramp
recipe; A5z: with the hull term from the start; Nz: natural training on it; ``ours'': this study's linear-ramp S4D-type row
(\cref{tab:remedy}), the pair of A1z at fixed $B$; Nu: natural training on our layer. Natural arms: clean accuracy and the census
only.}
\label{tab:remedy-zoh}
\centering\scriptsize\setlength{\tabcolsep}{2pt}
\begin{tabular}{@{}lcccccccc@{}}
\toprule
arm & clean & $L=0$ & $L=1024$ & $K=1$ & $L=\infty$ & K-uniform & $L=\infty$ ratio & K-uniform ratio\\
\midrule
A1z & 81.2 [80.8, 81.6] & 71.0 [70.0, 72.0] & 69.1 [67.2, 71.1] & 68.5 [67.9, 69.0] & 68.3 [68.2, 68.5] & 61.1 [57.4, 64.8] & 96.3 [95.1, 97.5] & 86.1 [82.1, 90.1]\\
A5z & 79.6 [76.9, 82.2] & 69.1 [68.0, 70.3] & 68.9 [67.2, 70.6] & 69.3 [68.6, 69.9] & 69.3 [68.6, 69.9] & 67.8 [66.1, 69.4] & 100.2 [99.3, 101.1] & 98.0 [96.4, 99.6]\\
ours & 81.3 [75.7, 86.9] & 72.4 [69.9, 74.9] & 68.8 [64.1, 73.5] & 68.2 [64.2, 72.1] & 67.9 [65.1, 70.6] & 59.7 [54.9, 64.5] & 93.8 [89.9, 97.6] & 82.5 [77.4, 87.6]\\
Nz & 85.8 [82.9, 88.7] & -- & -- & -- & -- & -- & -- & --\\
Nu & 85.3 [83.9, 86.7] & -- & -- & -- & -- & -- & -- & --\\
\bottomrule
\end{tabular}

\medskip
\begin{tabular}{@{}lccccc@{}}
\toprule
largest $\tau_{\rm h}$ (words) & A1z & A5z & ours & Nz & Nu\\
\midrule
seeds 1 / 2 / 3 & 274.9 / 259.8 / 270.9 & 81.4 / 95.8 / 97.4 & 401.9 / 353.5 / 279.4 & 562.3 / 471.1 / 331.3 & 399.7 / 356.2 / 364.7\\
\bottomrule
\end{tabular}
\end{table}

\paragraph{The evidence probe on the control.}
The probe above also ran on the twelve converted checkpoints of the control (reported; no criterion reads them;
\cref{tab:remedy-zohprobe}). After 4096 trailing neutral words every one of the twelve is class P: no detectable evidence,
agreement 100.0\%, and clean accuracy in the chance band (A1z 51.7\% [49.7, 53.7], A5z 52.8\% [52.3, 53.3], Nz 49.2\% [47.4, 51.0],
Nu 48.4\% [46.9, 49.9]; our layer's linear ramp 50.6\% [47.8, 53.5]). A1z carries evidence up to 512 words (class E in every seed;
carry 18.4 [14.0, 22.9] after 64 words and 17.3 [15.5, 19.2] after 128, against 24.7 [18.3, 31.0] and 20.4 [13.1, 27.6] for our layer,
overlapping intervals) and none from 1024 words on, the same length as our layer, consistent with its census (largest time
constants 260--275 words; A5z 81--97, Nz 331--562, Nu 356--400). A5z, with the hull term from the start, carries less at every length
up to 1024 words (11.9 [8.0, 15.8] after 64 words) and is class P from 256 words on in two seeds.
over three seeds, two-layer width-64 models and one task; they are read as no verdict of the probe's criteria, as no certificate
and no immunity of the zero-order-hold layer, and they say nothing about the robust term's role.

\begin{table}[htbp]
\caption{The evidence probe on the zero-order-hold control's checkpoints (reported, not criteria; 1000 test documents; clean
accuracy in float64, not a certificate): the carry after $d$ trailing licence words (points; mean over the three seeds with its 95\%
t-interval over the seeds) with the class of each seed, as in \cref{tab:remedy-evidence}; ``ours'' is this study's linear-ramp
S4D-type row. At $d=4096$ every checkpoint is class P with 100\% agreement and a clean accuracy inside the chance band
$[42.2,57.8]\%$.}
\label{tab:remedy-zohprobe}
\centering\scriptsize\setlength{\tabcolsep}{2.5pt}
\begin{tabular}{@{}lccccc@{}}
\toprule
arm & $d=256$ & $d=512$ & $d=1024$ & $d=2048$ & $d=4096$\\
\midrule
ours (linear ramp) & 14.0 [4.0, 24.0] E E E & 4.1 [$-$2.9, 11.1] E E E & 0.2 [$-$0.5, 0.8] P P P & 0.0 [$-$0.1, 0.2] P P P & 0.0 [0.0, 0.0] P P P\\
A1z (zero-order hold, linear ramp) & 8.8 [3.8, 13.9] E E E & 2.5 [1.1, 3.8] E E E & 0.3 [$-$0.0, 0.7] P P P & 0.0 [0.0, 0.0] P P P & 0.0 [0.0, 0.0] P P P\\
A5z (zero-order hold, hull) & 0.9 [$-$1.0, 2.9] P P E & 0.1 [$-$0.3, 0.4] P P P & 0.0 [0.0, 0.0] P P P & 0.0 [0.0, 0.0] P P P & 0.0 [0.0, 0.0] P P P\\
Nz (zero-order hold, natural) & 14.0 [8.3, 19.8] E E E & 6.4 [3.1, 9.6] E E E & 1.4 [$-$1.6, 4.3] E E P & 0.2 [$-$0.4, 0.7] P P P & 0.0 [0.0, 0.0] P P P\\
Nu (ours, natural) & 16.9 [15.6, 18.1] E E E & 7.9 [4.4, 11.3] E E E & 1.6 [0.4, 2.9] E E E & 0.2 [0.0, 0.3] P P P & 0.0 [0.0, 0.0] P P P\\
\bottomrule
\end{tabular}
\end{table}

\paragraph{What this study does not show.}
These are real-arithmetic certificates, computed in float64 and, where the next paragraph says so, re-verified with outward
rounding, for two-layer width-64 models on one task, with three seeds per arm. The capped models are the certified networks: after
$8\tau_{\max}$ trailing neutral words they carry no detectable evidence of the document in 3 of 3 seeds (caps 128 and 512; ``Where the
information goes''), and
at $\tau_{\max}=512$ and the 1024-token period every channel contracts by at least $e^{-2}$ per period, so their K-uniform certificate is
close to a one-period certificate (a geometric-series bound puts its radius within about $1/(1-e^{-2})\approx1.16$ times the
one-period radius). Nothing is claimed about certified dependence beyond about $\tau_{\max}$ tokens (no certificate covers a context
after the document), about memory beyond what the evidence probe measured, other tasks, larger or
pretrained models, or contexts other than repeated copies of the evaluation period and contexts over the stated token sets, which are
frequent neutral tokens and not adversarial distractors. The threat differs from that of \citet{jia2019}, so no comparison with their
numbers is made. The training periods (256 tokens of GNU manuals) differ from the evaluation period (1024 tokens of licence text). The
development split shares movies with the training split; the test split does not.

\paragraph{Re-verification with outward rounding.}
The evaluator of \cref{rem:float} re-verified the certificates of these rows: linear ramp, cap 128, cap 512, fixed-point term, cap +
hull and hull for Mamba and the minGRU-type model, and linear ramp for the S4D-type model. On each of them it re-verified every
certificate at box scale 1 on the 1000 test documents (at $L=0$, 1024 and 4096 tokens, after one and four copies of the period, at
$L=\infty$ and K-uniformly) and, on the cap-128 rows and the minGRU-type cap + hull and hull rows, also those at box scale 0 (the
clean $L=\infty$ and every-$K$ columns of \cref{tab:remedy-inf}); it also re-verified the token-set certificates of the cap + hull
rows over the most frequent neutral token, on their 200 documents, in both modes. None changed: all 208{,}020 archived certificates
were kept and none was gained, and the outward-rounded margin bounds lay below the float64 ones by at most $2.5\times10^{-11}$
relative to $1+\abs{b}$. Every number and verdict of these rows, recomputed from the re-verified outcomes, is unchanged. So the
certificates entering the criteria on the linear-ramp recipe's certified share at $L=0$, the blind spot without a remedy, the cap at
$\tau_{\max}=128$, the fixed-point term, the headline, the contrasts with the cap alone and with the hull term alone, the token sets
over one token and the S4D-type model were re-verified; the counts of \cref{tab:tests} were not, because other criteria counted
there read the first recipe or a diagnostic (\cref{app:prereg-criteria}). Not re-verified
are: every certified number of the first-recipe and cap 2048 rows, in the tables and in the text; the first-study rows'
certificates after one and four copies of the period and at box scale 0, with the width-only ratios and flip shares that use them
(their other certificates re-evaluate the checkpoints of \cref{app:gate2} and agree, document by document, with the first study's
certificates re-verified in \cref{rem:float}); the development-split $L=0$ shares of \cref{tab:remedy-clean} and those of the
selection stage (Setting); the clean $L=\infty$ and every-$K$ columns of
\cref{tab:remedy-inf} on every other row, the natural ones included; the width-only ratios and the flip shares of
\cref{tab:remedy-ratio,tab:remedy-attr} on every other row, which combine re-verified certificates with the clean outcomes at
$L=\infty$ and for every $K$ (box scale 0), which were not re-verified; the token-set shares of every row but cap + hull
(\cref{tab:remedy-attr,tab:remedy-tokens} and
observation (iii)); and the diagnostics (the restored share and the slow pairs of \cref{tab:remedy-attr}, the largest time constants
of \cref{tab:remedy-seeds}). Clean accuracies and the attack are not certificates and were not re-verified. The re-verified network
is each trained network with the float64 constants of its cap, where it has one, and an archived embedding table
(\cref{rem:float}); the deployed float32 capped network is not certified. The record is in \cref{app:prereg}.

\clearpage
\section{Clean accuracy over the context, sample sizes and intervals}\label{app:cleanctx}

This appendix reports the clean accuracy of every model of the first training study and of the context-length probe at every
context length, and 95\% intervals per checkpoint for the statistics that the paper reports. Both come from an extension of the
experiments that was pre-registered before any of its runs and frozen before its production runs (\cref{app:prereg-criteria}). It
brought the distance probe and the interval-map growth rates to 200 documents per seed (the $B=2048$ zonotope of the nonnegative
diagonal models on seed~0: 50) and the budget sweep to 50, completed the two context-length-probe rows and
the twelve token-set arms that had stopped short of 200 documents, and evaluated the clean prediction of every document at every
context. Every new document continues the archived order, draws and configuration, so the archived documents are a prefix of each
extended set. The archived verdicts of \cref{tab:prereg} remain the pre-registered outcomes of their tests; the outcomes on the
extended sets stand beside them (\cref{app:cleanctx-record}).

\subsection{Document sets, contexts and definitions}

\paragraph{Document sets and contexts.}
Set G holds the first 1000 test documents of the first training study (word level, $T=512$), each with a 4096-token chunk of the
neutral licence stream (56,880 tokens) drawn once by a fixed generator; set Q holds the first 200 of these documents with the chunks
of the context-length probe, drawn by another fixed generator; set Q$_{\rm b}$ holds the first 200 byte-level reviews with chunks of
the byte-level licence stream. Their chunks differ, so clean accuracies at $L>0$ differ between G and Q by construction, and no
statistic combines two sets. A context of length $L\in\{0,256,1024,4096\}$ is the last $L$ tokens of the chunk, placed before the
document; $K\in\{1,4,16\}$ means $K$ copies of the period $p$, the first 1024 tokens of the chunk, read from the zero state, the
finite members of \cref{def:limit}; on set Q the probe's own contexts are $L\in\{0,64,256,1024,4096\}$ and $K\in\{1,2,4\}$.
$L=\infty$ is the limit $F_\infty=\lim_{K\to\infty}F_K$ of \cref{def:limit} at the clean input, computed per architecture. For the
nonnegative diagonal models (Mamba, S4D-type, minGRU-type) it is \cref{thm:inf} at box scale 0, where every box is a point; a
document whose clean cascade has a channel with $\denh=0$, or whose box is not finite, would be undefined and count as not correct,
and there was none. For the complex-diagonal control (set Q) it is the periodic steady state, every mode having modulus below 1. For
the CNN it is any $K\ge2$, which sees the same windows as every larger $K$. For the bag of words it is the model on the period alone,
a prediction that does not depend on the document. For LSTM and GRU no limit is computed (``n/a''); $K=16$, 16,384 tokens of
context, is the longest one reported, and their invariant box at box scale 0 is the box statement beside it.

\paragraph{Definitions.}
Clean accuracy at a context is the share of documents whose clean prediction there equals the label. A prediction \emph{change}
between $L=0$ and the limit (LSTM and GRU: $K=16$) is a document whose prediction differs. In this binary task such a document is
correct at one context and wrong at the other, so the change share is $(b+c)/n$, with $b$ documents lost (correct at $L=0$, wrong at
the limit) and $c$ gained, and the accuracy at the limit minus that at $L=0$ is $(c-b)/n$. \emph{Certified correct for every number of
copies} is the share of documents whose clean prediction the K-uniform certificate of \cref{thm:hull} at box scale 0 certifies as the
label for every $K\ge0$ and for $K=\infty$ (LSTM and GRU: the invariant box at box scale 0). It is a box statement, not a clean
accuracy: by \cref{thm:hull}(i) it is at most the share of documents correct at every $K$, and at $\eps=0$ the hull still contains the
clean trajectories from the zero state and from the fixed point (\cref{rem:price}), so it can lie far below. \emph{Correct at every
context} is the share of documents correct at every context evaluated and, where it is defined, at the limit: the empirical
counterpart of the box statement, extended to the contiguous contexts; since the box statement does not cover those, it can exceed
this share (the CNNs). The \emph{predicted-positive} share separates a collapse to one class from changes at a
near-chance accuracy. On set Q the \emph{agreement} at a context is the share of documents whose prediction there equals $y_0$, the
prediction on the bare document; it bounds the probe's certified share at that context. A checkpoint is \emph{degenerate} if its
clean accuracy at $L=0$ is below 60\% or more than 80\% of its predictions fall in one class (the rule of \cref{app:remedy}); degenerate
checkpoints are marked and left out of the readings.

\subsection{Clean accuracy at every context}

\Cref{tab:cleanlen,tab:cleanlen-ct} give the clean accuracy of 53 checkpoints at every context on set G, the documents of the
first training study (its 45 checkpoints and the natural replication seeds 1 and 2 of the S4D-type, minGRU-type, LSTM and GRU
models), with the change between $L=0$ and the limit, and \cref{tab:cleanctx} gives clean accuracy and agreement on sets Q and Q$_{\rm b}$ for the 25 checkpoints of the
context-length probe.

\begin{table}[!htbp]
\caption{Clean accuracy over the context on the 1000 test documents of the first training study (set G), natural word-level models
(\%). $L$: neutral licence text before the document; $K$: copies of the 1024-token period; $L=\infty$: the limit of
\cref{def:limit} at the clean input (CNN: any $K\ge2$; bag of words: the model on the period alone; LSTM and GRU: no limit computed).
Every $K$: certified correct for every number of copies, the K-uniform box of \cref{thm:hull} at box scale 0 (LSTM and GRU: the
invariant box), a box statement and not a clean accuracy. Changed: documents whose prediction differs between $L=0$ and the limit
(LSTM and GRU: $K=16$), with its Wilson 95\% interval and the counts lost and gained, $b/c$; $\Delta$: the limit minus $L=0$ in
points, with its paired bootstrap 95\% interval; all: correct at every context; positive: predicted-positive share. At $n=1000$ the
Wilson half-width of every share is at most 3.1 points. No document had an undefined limit. Neither the limits nor the box
statements at box scale 0 were re-verified with outward rounding (\cref{rem:float}).}
\label{tab:cleanlen}
\centering\scriptsize\setlength{\tabcolsep}{2pt}
\begin{tabular}{@{}llrrrrrrrrrlrlrrr@{}}
\toprule
 & & \multicolumn{4}{c}{neutral text, $L$} & \multicolumn{3}{c}{copies, $K$} & & every $K$ & \multicolumn{3}{c}{$L=0$ against the limit} & & \multicolumn{2}{c}{positive}\\
\cmidrule(lr){3-6}\cmidrule(lr){7-9}\cmidrule(lr){12-14}\cmidrule(l){16-17}
model & seed & 0 & 256 & 1024 & 4096 & 1 & 4 & 16 & $L=\infty$ & (box) & changed & $b/c$ & $\Delta$ (points) & all & $L=0$ & limit\\
\midrule
Mamba & 0 & 88.4 & 86.9 & 86.5 & 86.5 & 85.7 & 85.3 & 85.2 & 85.2 & 0.0 & 7.6 [6.1, 9.4] & 54/22 & $-$3.2 [$-$4.9, $-$1.5] & 79.7 & 52.4 & 55.0\\
 & 1 & 87.0 & 84.9 & 83.5 & 83.7 & 83.2 & 82.5 & 82.5 & 82.5 & 0.0 & 10.5 [8.7, 12.6] & 75/30 & $-$4.5 [$-$6.5, $-$2.5] & 76.1 & 53.4 & 45.3\\
 & 2 & 88.4 & 86.8 & 86.8 & 86.7 & 86.8 & 86.7 & 86.7 & 86.7 & 0.0 & 4.9 [3.7, 6.4] & 33/16 & $-$1.7 [$-$3.1, $-$0.3] & 82.4 & 53.0 & 51.1\\
S4D-type & 0 & 86.8 & 84.6 & 81.9 & 81.1 & 83.9 & 83.0 & 83.0 & 83.0 & 0.1 & 10.4 [8.7, 12.4] & 71/33 & $-$3.8 [$-$5.8, $-$1.9] & 74.7 & 51.4 & 51.4\\
 & 1 & 87.1 & 87.4 & 86.1 & 86.4 & 85.5 & 85.1 & 85.1 & 85.1 & 0.2 & 6.2 [4.9, 7.9] & 41/21 & $-$2.0 [$-$3.5, $-$0.5] & 81.4 & 51.3 & 47.7\\
 & 2 & 87.7 & 85.3 & 82.8 & 82.0 & 83.9 & 82.5 & 82.5 & 82.5 & 0.8 & 12.0 [10.1, 14.2] & 86/34 & $-$5.2 [$-$7.3, $-$3.1] & 74.1 & 51.5 & 60.9\\
minGRU-type & 0 & 86.0 & 83.6 & 82.9 & 82.6 & 82.5 & 82.2 & 82.2 & 82.2 & 0.0 & 9.0 [7.4, 10.9] & 64/26 & $-$3.8 [$-$5.7, $-$2.0] & 75.8 & 52.8 & 47.4\\
 & 1 & 88.4 & 84.7 & 83.5 & 83.4 & 84.1 & 82.9 & 82.9 & 82.9 & 0.0 & 9.5 [7.8, 11.5] & 75/20 & $-$5.5 [$-$7.4, $-$3.6] & 76.5 & 51.6 & 47.1\\
 & 2 & 89.0 & 86.4 & 85.0 & 84.9 & 86.3 & 85.5 & 85.5 & 85.5 & 0.0 & 6.9 [5.5, 8.6] & 52/17 & $-$3.5 [$-$5.1, $-$1.9] & 79.6 & 51.8 & 51.3\\
LSTM & 0 & 89.4 & 86.2 & 85.5 & 84.9 & 84.2 & 84.0 & 84.0 & n/a & 0.0 & 9.2 [7.6, 11.2] & 73/19 & $-$5.4 [$-$7.3, $-$3.6] & 77.9 & 52.2 & 47.4\\
 & 1 & 89.3 & 87.3 & 86.9 & 85.6 & 86.9 & 86.3 & 86.3 & n/a & 0.0 & 6.2 [4.9, 7.9] & 46/16 & $-$3.0 [$-$4.6, $-$1.5] & 80.3 & 51.9 & 49.5\\
 & 2 & 89.1 & 87.5 & 86.1 & 85.8 & 85.9 & 85.5 & 85.5 & n/a & 0.0 & 8.0 [6.5, 9.8] & 58/22 & $-$3.6 [$-$5.4, $-$1.9] & 79.1 & 53.5 & 52.3\\
GRU & 0 & 90.7 & 87.2 & 84.8 & 82.9 & 84.7 & 78.9 & 74.0 & n/a & 0.0 & 23.1 [20.6, 25.8] & 199/32 & $-$16.7 [$-$19.5, $-$13.9] & 62.8 & 52.5 & 58.4\\
 & 1 & 90.8 & 88.5 & 87.0 & 86.0 & 86.4 & 82.8 & 81.7 & n/a & 0.0 & 13.9 [11.9, 16.2] & 115/24 & $-$9.1 [$-$11.4, $-$6.9] & 73.6 & 51.8 & 53.7\\
 & 2 & 89.7 & 87.1 & 85.3 & 84.1 & 84.7 & 82.5 & 81.8 & n/a & 0.0 & 13.7 [11.7, 16.0] & 108/29 & $-$7.9 [$-$10.1, $-$5.7] & 72.5 & 51.1 & 57.6\\
CNN & 0 & 89.8 & 86.1 & 82.4 & 75.1 & 83.0 & 82.8 & 82.8 & 82.8$^{\dagger}$ & 78.7 & 14.4 [12.4, 16.7] & 107/37 & $-$7.0 [$-$9.3, $-$4.7] & 62.1 & 52.8 & 58.6\\
bag of words & 0 & 85.8 & 77.8 & 62.7 & 50.3 & 61.7 & 50.0 & 47.2 & 46.5 & n/a & 53.5 [50.4, 56.6] & 464/71 & $-$39.3 [$-$43.1, $-$35.5] & 34.7 & 53.0 & 13.5\\
\bottomrule
\end{tabular}
\par\smallskip
\raggedright\scriptsize $^{\dagger}$The code's own $L=\infty$ entry for the CNN, a max-pool over an infinite past without the
zero-padded start windows, gives the same clean accuracy on these documents: 82.8\% [80.3, 85.0] for the natural CNN, and 71.0, 75.4
and 71.3\% for the standard-IBP seeds of \cref{tab:cleanlen-ct}.
\end{table}

\begin{table}[!t]
\caption{As \cref{tab:cleanlen}, certified-trained models of the first training study: standard IBP and fixed-point fine-tuning,
seeds 0--2 (\cref{app:gate2}). $^\ast$Degenerate. The standard-IBP and fixed-point fine-tuned LSTM and GRU models marked so are
constant classifiers; a change share whose discordant documents are fewer than 10 gets the counts only ($\Delta$: --).}
\label{tab:cleanlen-ct}
\centering\scriptsize\setlength{\tabcolsep}{2pt}
\begin{tabular}{@{}llrrrrrrrrrlrlrrr@{}}
\toprule
 & & \multicolumn{4}{c}{neutral text, $L$} & \multicolumn{3}{c}{copies, $K$} & & every $K$ & \multicolumn{3}{c}{$L=0$ against the limit} & & \multicolumn{2}{c}{positive}\\
\cmidrule(lr){3-6}\cmidrule(lr){7-9}\cmidrule(lr){12-14}\cmidrule(l){16-17}
model & seed & 0 & 256 & 1024 & 4096 & 1 & 4 & 16 & $L=\infty$ & (box) & changed & $b/c$ & $\Delta$ (points) & all & $L=0$ & limit\\
\midrule
\multicolumn{17}{@{}l}{\emph{standard IBP}}\\
Mamba & 0$^\ast$ & 57.2 & 55.9 & 57.3 & 55.8 & 56.4 & 53.1 & 51.5 & 47.7 & 18.1 & 41.1 [38.1, 44.2] & 253/158 & $-$9.5 [$-$13.4, $-$5.6] & 29.3 & 28.2 & 37.1\\
 & 1$^\ast$ & 51.9 & 51.8 & 51.8 & 51.8 & 51.8 & 51.8 & 51.8 & 52.4 & 47.5 & 3.5 [2.5, 4.8] & 15/20 & 0.5 [$-$0.6, 1.7] & 50.4 & 25.7 & 29.0\\
 & 2 & 76.1 & 73.6 & 71.2 & 59.3 & 72.9 & 60.7 & 53.6 & 49.0 & 21.5 & 58.7 [55.6, 61.7] & 429/158 & $-$27.1 [$-$31.5, $-$22.6] & 27.7 & 66.3 & 17.2\\
S4D-type & 0 & 75.7 & 74.4 & 74.6 & 74.4 & 75.1 & 75.2 & 75.2 & 75.2 & 65.0 & 5.1 [3.9, 6.6] & 28/23 & $-$0.5 [$-$1.9, 0.9] & 71.2 & 54.7 & 57.2\\
 & 1$^\ast$ & 51.4 & 51.3 & 51.0 & 50.8 & 51.0 & 50.9 & 50.9 & 50.9 & 49.4 & 2.1 [1.4, 3.2] & 13/8 & $-$0.5 [$-$1.4, 0.4] & 50.0 & 19.6 & 17.5\\
 & 2 & 65.6 & 61.2 & 58.1 & 57.3 & 57.5 & 56.5 & 56.5 & 56.5 & 42.0 & 16.3 [14.1, 18.7] & 127/36 & $-$9.1 [$-$11.6, $-$6.7] & 51.4 & 38.4 & 22.1\\
minGRU-type & 0 & 66.7 & 66.5 & 66.4 & 65.1 & 66.1 & 66.0 & 62.8 & 57.3 & 33.6 & 35.4 [32.5, 38.4] & 224/130 & $-$9.4 [$-$13.0, $-$5.7] & 43.2 & 62.9 & 29.3\\
 & 1 & 66.7 & 66.3 & 66.1 & 63.1 & 65.8 & 64.4 & 60.3 & 51.4 & 22.1 & 65.3 [62.3, 68.2] & 403/250 & $-$15.3 [$-$20.3, $-$10.4] & 26.2 & 76.5 & 11.6\\
 & 2 & 69.4 & 70.3 & 69.9 & 66.1 & 70.1 & 67.0 & 59.1 & 54.9 & 23.5 & 42.1 [39.1, 45.2] & 283/138 & $-$14.5 [$-$18.4, $-$10.6] & 36.8 & 72.4 & 40.7\\
LSTM & 0$^\ast$ & 47.4 & 47.4 & 47.4 & 47.4 & 47.4 & 47.4 & 47.4 & n/a & 0.0 & 0.0 [0.0, 0.4] & 0/0 & -- & 47.4 & 0.0 & 0.0\\
 & 1$^\ast$ & 47.4 & 47.4 & 47.4 & 47.4 & 47.4 & 47.4 & 47.4 & n/a & 0.0 & 0.0 [0.0, 0.4] & 0/0 & -- & 47.4 & 0.0 & 0.0\\
 & 2$^\ast$ & 47.4 & 47.4 & 47.4 & 47.4 & 47.4 & 47.4 & 47.4 & n/a & 0.0 & 0.0 [0.0, 0.4] & 0/0 & -- & 47.4 & 0.0 & 0.0\\
GRU & 0$^\ast$ & 47.4 & 47.4 & 47.4 & 47.4 & 47.4 & 47.4 & 47.4 & n/a & 47.4 & 0.0 [0.0, 0.4] & 0/0 & -- & 47.4 & 0.0 & 0.0\\
 & 1$^\ast$ & 47.4 & 47.4 & 47.4 & 47.4 & 47.4 & 47.4 & 47.4 & n/a & 47.4 & 0.0 [0.0, 0.4] & 0/0 & -- & 47.4 & 0.0 & 0.0\\
 & 2$^\ast$ & 47.4 & 47.4 & 47.4 & 47.4 & 47.4 & 47.4 & 47.4 & n/a & 47.4 & 0.0 [0.0, 0.4] & 0/0 & -- & 47.4 & 0.0 & 0.0\\
CNN & 0 & 81.6 & 75.9 & 72.0 & 72.5 & 71.0 & 71.0 & 71.0 & 71.0$^{\dagger}$ & 62.9 & 24.6 [22.0, 27.4] & 176/70 & $-$10.6 [$-$13.6, $-$7.7] & 46.0 & 54.2 & 48.0\\
 & 1 & 82.7 & 80.6 & 76.5 & 71.0 & 75.5 & 75.4 & 75.4 & 75.4$^{\dagger}$ & 69.8 & 16.9 [14.7, 19.3] & 121/48 & $-$7.3 [$-$9.8, $-$4.8] & 55.9 & 54.7 & 60.2\\
 & 2 & 82.7 & 77.1 & 70.8 & 69.1 & 71.3 & 71.3 & 71.3 & 71.3$^{\dagger}$ & 65.8 & 19.8 [17.4, 22.4] & 156/42 & $-$11.4 [$-$14.1, $-$8.7] & 48.6 & 55.3 & 55.7\\
bag of words & 0 & 83.1 & 75.6 & 62.6 & 53.7 & 60.9 & 54.8 & 52.9 & 52.8 & n/a & 45.9 [42.8, 49.0] & 381/78 & $-$30.3 [$-$34.1, $-$26.6] & 41.6 & 53.9 & 95.4\\
 & 1 & 84.0 & 76.2 & 62.7 & 53.2 & 61.0 & 54.5 & 52.9 & 52.5 & n/a & 46.3 [43.2, 49.4] & 389/74 & $-$31.5 [$-$35.2, $-$27.8] & 41.8 & 54.4 & 95.7\\
 & 2 & 83.2 & 75.1 & 60.7 & 53.0 & 59.3 & 53.5 & 52.7 & 52.6 & n/a & 45.2 [42.1, 48.3] & 379/73 & $-$30.6 [$-$34.3, $-$26.9] & 42.1 & 55.0 & 97.0\\
\midrule
\multicolumn{17}{@{}l}{\emph{fixed-point fine-tuning}}\\
Mamba & 0$^\ast$ & 51.9 & 52.2 & 52.3 & 52.3 & 52.3 & 52.3 & 52.3 & 52.4 & 51.2 & 0.5 [0.2, 1.2] & 0/5 & -- & 51.9 & 25.9 & 25.4\\
 & 1 & 72.6 & 72.0 & 71.3 & 71.1 & 71.8 & 70.8 & 70.4 & 70.5 & 44.9 & 10.7 [8.9, 12.8] & 64/43 & $-$2.1 [$-$4.2, $-$0.1] & 63.8 & 65.2 & 68.3\\
 & 2 & 80.2 & 79.2 & 78.3 & 77.2 & 78.4 & 76.2 & 75.5 & 69.9 & 53.2 & 23.9 [21.4, 26.6] & 171/68 & $-$10.3 [$-$13.3, $-$7.3] & 59.3 & 55.4 & 77.9\\
S4D-type & 0 & 78.3 & 77.0 & 77.0 & 76.7 & 76.8 & 76.7 & 76.7 & 76.7 & 68.2 & 6.2 [4.9, 7.9] & 39/23 & $-$1.6 [$-$3.1, $-$0.1] & 72.9 & 55.7 & 56.9\\
 & 1 & 72.9 & 73.2 & 73.4 & 73.5 & 73.1 & 73.2 & 73.2 & 73.2 & 69.5 & 2.3 [1.5, 3.4] & 10/13 & 0.3 [$-$0.6, 1.2] & 71.7 & 57.5 & 55.6\\
 & 2 & 77.0 & 75.5 & 75.3 & 75.7 & 75.0 & 75.0 & 75.0 & 75.0 & 65.2 & 6.4 [5.0, 8.1] & 42/22 & $-$2.0 [$-$3.6, $-$0.4] & 71.0 & 53.8 & 52.4\\
minGRU-type & 0 & 76.8 & 74.7 & 70.8 & 68.0 & 70.6 & 68.2 & 67.6 & 66.3 & 14.8 & 16.7 [14.5, 19.1] & 136/31 & $-$10.5 [$-$13.0, $-$8.1] & 61.2 & 60.4 & 75.5\\
 & 1 & 69.8 & 70.1 & 69.0 & 67.9 & 69.4 & 68.2 & 67.4 & 66.5 & 59.1 & 11.7 [9.9, 13.8] & 75/42 & $-$3.3 [$-$5.4, $-$1.2] & 61.8 & 69.2 & 80.3\\
 & 2 & 64.7 & 64.6 & 64.6 & 63.1 & 63.9 & 63.7 & 64.3 & 64.3 & 40.8 & 0.4 [0.2, 1.0] & 4/0 & -- & 62.0 & 76.9 & 77.3\\
LSTM & 0$^\ast$ & 47.4 & 47.4 & 47.4 & 47.4 & 47.4 & 47.4 & 47.4 & n/a & 0.0 & 0.0 [0.0, 0.4] & 0/0 & -- & 47.4 & 0.0 & 0.0\\
 & 1$^\ast$ & 47.4 & 47.4 & 47.4 & 47.4 & 47.4 & 47.4 & 47.4 & n/a & 0.0 & 0.0 [0.0, 0.4] & 0/0 & -- & 47.4 & 0.0 & 0.0\\
 & 2$^\ast$ & 47.4 & 47.4 & 47.4 & 47.4 & 47.4 & 47.4 & 47.4 & n/a & 0.0 & 0.0 [0.0, 0.4] & 0/0 & -- & 47.4 & 0.0 & 0.0\\
GRU & 0$^\ast$ & 52.6 & 52.6 & 52.6 & 52.6 & 52.6 & 52.6 & 52.6 & n/a & 52.6 & 0.0 [0.0, 0.4] & 0/0 & -- & 52.6 & 100.0 & 100.0\\
 & 1$^\ast$ & 47.4 & 47.4 & 47.4 & 47.4 & 47.4 & 47.4 & 47.4 & n/a & 47.4 & 0.0 [0.0, 0.4] & 0/0 & -- & 47.4 & 0.0 & 0.0\\
 & 2 & 89.4 & 87.2 & 86.3 & 86.0 & 86.1 & 85.6 & 85.6 & n/a & 0.0 & 10.2 [8.5, 12.2] & 70/32 & $-$3.8 [$-$5.8, $-$1.9] & 77.9 & 51.6 & 51.0\\
\bottomrule
\end{tabular}
\end{table}

\begin{table}[!t]
\caption{Clean accuracy over the context on the documents of the context-length probe (\%; \cref{app:gate3-ctx}): set Q, 200
word-level documents, and set Q$_{\rm b}$, 200 byte-level reviews. Each cell gives clean accuracy and, after the slash, agreement with
$y_0$, the prediction on the bare document, which bounds the probe's certified share there (at $L=0$ the agreement is 100\% by
definition). $L=\infty$: as in \cref{tab:cleanlen} (complex-diagonal S4D: the periodic steady state; LSTM and GRU: no limit
computed). Every $K$: certified correct for every number of copies at box scale 0, a box statement (n/a: no K-uniform certificate for
that model); positive: predicted-positive share. At $n=200$ the Wilson half-width of every share is at most 6.9 points; no checkpoint
of this table is degenerate.}
\label{tab:cleanctx}
\centering\scriptsize\setlength{\tabcolsep}{2pt}
\begin{tabular}{@{}llrccccccccrrr@{}}
\toprule
 & & \multicolumn{5}{c}{neutral text, $L$} & \multicolumn{3}{c}{copies, $K$} & & every $K$ & \multicolumn{2}{c}{positive}\\
\cmidrule(lr){3-7}\cmidrule(lr){8-10}\cmidrule(l){13-14}
model & seed & 0 & 64 & 256 & 1024 & 4096 & 1 & 2 & 4 & $L=\infty$ & (box) & $L=0$ & limit\\
\midrule
\multicolumn{14}{@{}l}{\emph{natural word-level models (set Q)}}\\
Mamba & 0 & 87.0 & 86.0/98.0 & 85.0/95.0 & 84.5/94.5 & 83.0/94.0 & 84.0/95.0 & 84.0/95.0 & 84.0/95.0 & 83.5 & 0.0 & 55.0 & 54.5\\
 & 1 & 85.0 & 86.0/97.0 & 81.0/91.0 & 80.0/88.0 & 79.0/87.0 & 80.5/88.5 & 79.0/87.0 & 79.0/87.0 & 79.0 & 0.0 & 56.0 & 44.0\\
 & 2 & 88.5 & 88.5/97.0 & 88.0/95.5 & 86.5/95.0 & 85.5/94.0 & 85.0/92.5 & 85.5/92.0 & 85.5/92.0 & 85.5 & 0.0 & 52.5 & 52.5\\
S4D-type & 0 & 86.0 & 86.5/96.5 & 83.5/91.5 & 84.0/90.0 & 83.0/91.0 & 80.5/89.5 & 79.0/88.0 & 78.5/87.5 & 78.5 & 0.0 & 52.0 & 50.5\\
 & 1 & 87.5 & 88.0/98.5 & 87.5/95.0 & 85.0/92.5 & 85.0/92.5 & 86.0/93.5 & 85.5/92.0 & 85.0/91.5 & 85.0 & 0.5 & 47.5 & 44.0\\
 & 2 & 86.0 & 88.5/96.5 & 86.5/93.5 & 89.5/92.5 & 89.0/91.0 & 84.0/91.0 & 82.5/89.5 & 82.0/89.0 & 82.0 & 0.5 & 52.0 & 59.0\\
minGRU-type & 0 & 86.0 & 85.5/98.5 & 83.0/94.0 & 80.5/90.5 & 79.5/89.5 & 80.5/88.5 & 81.5/87.5 & 81.5/87.5 & 81.5 & 0.0 & 53.0 & 45.5\\
 & 1 & 88.5 & 87.5/97.0 & 85.0/92.5 & 84.5/91.0 & 84.0/90.5 & 83.0/90.5 & 81.5/89.0 & 82.0/88.5 & 82.0 & 0.0 & 53.5 & 45.0\\
 & 2 & 89.5 & 89.5/98.0 & 87.5/95.0 & 86.0/92.5 & 85.0/90.5 & 83.5/93.0 & 83.0/92.5 & 83.0/92.5 & 83.0 & 0.0 & 49.5 & 48.0\\
LSTM & 0 & 88.0 & 86.5/97.5 & 83.0/93.0 & 78.5/88.5 & 78.0/88.0 & 83.0/91.0 & 82.5/90.5 & 82.5/90.5 & n/a & 0.0 & 52.0 & 45.5\\
 & 1 & 89.5 & 89.0/97.5 & 86.5/96.0 & 86.5/95.0 & 85.5/93.0 & 86.0/92.5 & 86.0/92.5 & 85.5/92.0 & n/a & 0.0 & 50.5 & 46.5\\
 & 2 & 88.0 & 89.0/95.0 & 88.5/93.5 & 84.5/90.5 & 83.5/89.5 & 85.0/94.0 & 85.0/94.0 & 85.0/94.0 & n/a & 0.0 & 54.0 & 51.0\\
GRU & 0 & 88.0 & 88.5/97.5 & 86.5/93.5 & 83.0/88.0 & 78.5/84.5 & 82.0/87.0 & 79.0/83.0 & 77.0/80.0 & n/a & 0.0 & 51.0 & 54.0\\
 & 1 & 89.5 & 90.0/97.5 & 86.0/94.5 & 86.0/90.5 & 81.0/89.5 & 87.5/95.0 & 86.0/93.5 & 85.0/92.5 & n/a & 0.0 & 53.5 & 53.0\\
 & 2 & 88.0 & 88.5/96.5 & 86.0/93.0 & 85.5/91.5 & 82.0/89.0 & 81.5/90.5 & 82.0/90.0 & 82.0/89.0 & n/a & 0.0 & 52.0 & 53.0\\
CNN & 0 & 90.0 & 90.5/94.5 & 87.0/92.0 & 81.5/87.5 & 77.5/77.5 & 84.0/86.0 & 84.0/86.0 & 84.0/86.0 & 84.0 & 80.0 & 52.0 & 56.0\\
bag of words & 0 & 88.0 & 85.5/92.5 & 74.5/79.5 & 58.0/61.0 & 49.5/51.5 & 61.5/64.5 & 56.0/58.0 & 54.0/55.0 & 51.0 & n/a & 52.0 & 14.0\\
complex-diagonal S4D & 0 & 88.0 & 87.0/94.0 & 82.0/90.0 & 80.5/84.5 & 80.5/84.5 & 80.5/86.5 & 83.5/89.5 & 83.0/89.0 & 83.0 & n/a & 52.0 & 44.0\\
 & 1 & 84.5 & 85.5/96.0 & 83.0/94.5 & 82.0/93.5 & 82.0/93.5 & 81.5/93.0 & 81.5/94.0 & 81.0/93.5 & 81.0 & n/a & 52.5 & 50.0\\
\midrule
\multicolumn{14}{@{}l}{\emph{certified-trained word-level models (set Q)}}\\
Mamba, standard IBP & 2 & 76.5 & 75.5/98.0 & 74.0/92.5 & 69.5/81.0 & 60.5/54.0 & 72.5/87.0 & 67.0/76.5 & 62.0/67.5 & 57.5 & 26.5 & 67.5 & 19.5\\
S4D-type, standard IBP & 0 & 78.5 & 78.5/98.0 & 76.0/95.5 & 76.5/96.0 & 77.0/95.5 & 77.5/96.0 & 77.5/96.0 & 77.5/96.0 & 77.5 & 69.0 & 58.5 & 60.5\\
Mamba, fixed-point & 2 & 78.5 & 77.5/99.0 & 75.0/94.5 & 76.0/92.5 & 78.5/89.0 & 79.0/91.5 & 78.5/88.0 & 78.0/87.5 & 71.5 & 56.5 & 58.5 & 78.5\\
minGRU-type, fixed-point & 0 & 80.0 & 79.5/98.5 & 78.0/97.0 & 73.5/91.5 & 72.0/90.0 & 73.0/91.0 & 71.5/89.5 & 71.5/89.5 & 71.0 & 17.5 & 63.0 & 73.0\\
\midrule
\multicolumn{14}{@{}l}{\emph{natural byte-level models (set Q$_{\rm b}$)}}\\
Mamba & 0 & 81.0 & 81.5/97.5 & 76.5/94.5 & 77.5/93.5 & 73.0/89.0 & 77.5/93.5 & 74.5/89.5 & 73.5/87.5 & 72.5 & 0.0 & 51.0 & 50.5\\
LSTM & 0 & 81.0 & 79.0/97.0 & 80.0/96.0 & 77.0/96.0 & 78.5/95.5 & 78.5/95.5 & 77.0/94.0 & 77.0/94.0 & n/a & 0.0 & 53.0 & 51.0\\
\bottomrule
\end{tabular}
\end{table}

\paragraph{Reading.}
On our licence-text context the word-level natural models' clean accuracy falls from 85.8--90.8\% at $L=0$ to 50.3--86.7\% after
4096 tokens and 46.5--86.7\% at $L=\infty$ (where a limit is defined: not LSTM and GRU), and 4.9--53.5\% of their predictions change
between $L=0$ and $L=\infty$ (LSTM and GRU: $K=16$). The selective standard-IBP models lose 9.4--27.1 points to the infinite clean
context alone (Mamba seed~2 and the three minGRU-type seeds; Mamba seeds 0 and 1 are degenerate and left out). For the
certified-trained models the clean accuracy does not move in one direction: on some seeds it rises with the context (the
fixed-point fine-tuned S4D-type seed~1 from 72.9\% at $L=0$ to 73.5\% after 4096 tokens). The clean prediction of the natural
nonnegative diagonal models settles within 16 periods: on each of the nine checkpoints, 0 of the 1000 documents change between
$K=16$ and $L=\infty$. On the 1000 documents the K-uniform hull certified the clean prediction of none of the documents of each
natural Mamba and minGRU-type seed, a box statement and not a clean accuracy. A certified share at a context is read against the
clean accuracy at the same context in these tables. On set Q, after 4096 tokens the word-level natural models agree with their
bare-document prediction on 51.5--94.0\% of the documents, which bounds the certified shares of the context-length probe there.

\subsection{Intervals}\label{app:cleanctx-method}

The intervals of this appendix, and the bracketed intervals that the paper prints for the experiments that the extension covers,
are 95\% intervals per checkpoint with the test document as the unit. They are conditional on the trained checkpoint and on the
56,880-token licence stream of our contexts, they hold per statistic and not simultaneously, and none includes training variability.
A share (certified, clean, agreement, a prediction change, the predicted-positive share, the share of documents that keep a tenth of
$r(1)$ or that reach 0.99 times the box) gets the Wilson score interval. A median or quartile of a per-document quantity (looseness,
width exponent, certified radius or a per-document ratio of radii, $\lambda_M$, $\lambda_J$, a lag) gets the distribution-free
interval between two order statistics of the sample, with ranks from the binomial distribution, whose coverage is at least 95\% at
every $n$ (for the median of 200 documents, the 86th and the 115th value); a radius not certified at $10^{-12}$ counts as 0 and
enters by its rank. Every other statistic (a ratio of two medians on the same documents, a paired difference, a pooled rate, a reach,
a share over a data-dependent subset or over pairs of document and channel) gets the 95\% percentile bootstrap over the documents,
with 10,000 resamples drawn once per document set (generator seed 0), so that every comparison within a checkpoint is paired. A rerun
with generator seed 1 moved 124 of the 858 bootstrap intervals by more than one printed unit; these were recomputed with 50,000
resamples, whose values are printed. A difference between the new and the archived documents, two disjoint samples, gets Newcombe's
interval for shares and the bootstrap with the two sets resampled independently otherwise; a paired difference with fewer than 10
discordant documents is given as its two counts. Two studies keep their pre-registered intervals: the linear relaxations
(\cref{app:lirpa}) Wilson intervals and a 90\% document bootstrap, the training remedy (\cref{app:remedy}) 95\% t-intervals over seeds
and 95\% document-bootstrap intervals per checkpoint. Means over seeds keep their standard deviation over seeds (\cref{app:remedy}:
its pre-registered t-interval), and none of the intervals of the extension is placed on a quantity over seeds, training methods or
architectures. The marks of \cref{tab:ci,tab:ci-b} say whether the interval of
a statistic that a criterion reads lies on the verdict's side of the threshold (decisive) or contains it (not decisive at this $n$);
they are descriptive and unadjusted.
The disk certificates and the zonotope radii of these tables are float64 results, not re-verified with outward rounding
(\cref{rem:float}). For orientation, the half-width of the Wilson interval is at most 3.1, 6.9, 9.7, 13.4 and 20.1
points at $n=1000$, 200, 100, 50 and 20, and its upper end for 0 of $n$ documents is 0.38, 1.88, 3.70, 7.13 and 16.1\%.

\begin{table}[!htbp]
\caption{Per-checkpoint 95\% intervals of the statistics that the paper reports, part 1 (\cref{sec:exp-gate3,sec:exp-gate5}).
(a), (b): the context-length probe and the complex-diagonal control on set Q (200 documents; the probe ran again only for the two
rows that had stopped at 190 documents, whose 10 new documents are too few for a reading of their own; a superscript d marks a
decisive interval of a statistic that a criterion reads). (c), (d): the distance probe and the budget sweep. Archived: the documents of
the original run; extended: those and the new ones, the paper's numbers; new only: the new documents alone; new $-$ archived: the
interval of the difference between these two disjoint sets (shares in points); decisive: yes if the interval lies on the verdict's
side of the threshold of the criterion that reads the statistic, no if it contains it (not decisive at this $n$), -- if no criterion
reads it. Shares in \%. A cell that names several models or seeds stands for each of them, all with the same values. Not repeated
here: the first training study (\cref{tab:train,tab:trainseeds}; per seed a
Wilson half-width of at most 3.1 points), the clean accuracies (\cref{tab:cleanlen,tab:cleanlen-ct,tab:cleanctx}), and the linear
relaxations and the training remedy, which print their own pre-registered intervals (\cref{app:lirpa,app:remedy}).}
\label{tab:ci}
\centering\scriptsize\setlength{\tabcolsep}{2.5pt}
\begin{tabular}{@{}llrlllll@{}}
\multicolumn{8}{@{}l}{(a) Context-length probe, nonnegative diagonal models: median looseness ratios and certified shares at $\eps=10^{-8}$}\\
\toprule
model & seed & $n$ & $R(L{=}4096)/R(L{=}256)$ & $R(K{=}4)/R(K{=}1)$ & $L=\infty$ finite & $L=\infty$ certified & hull certified\\
\midrule
Mamba & 0 & 200 & 0.971 [0.823, 1.05]$^{\rm d}$ & 1.02 [0.928, 1.07]$^{\rm d}$ & 100.0$^{\rm d}$ & 93.5 [89.2, 96.2] & 0.0 [0.0, 1.9]\\
 & 1 & 200 & 0.693 [0.577, 0.779]$^{\rm d}$ & 0.953 [0.884, 1.01]$^{\rm d}$ & 100.0$^{\rm d}$ & 87.0 [81.6, 91.0] & 0.0 [0.0, 1.9]\\
 & 1, archived & 190 & 0.693 [0.566, 0.785]$^{\rm d}$ & 0.95 [0.879, 1.01]$^{\rm d}$ & 100.0$^{\rm d}$ & 86.8 [81.3, 90.9] & 0.0 [0.0, 2.0]\\
 & 2 & 200 & 0.937 [0.835, 1.06]$^{\rm d}$ & 1.01 [0.945, 1.06]$^{\rm d}$ & 100.0$^{\rm d}$ & 91.0 [86.2, 94.2] & 0.0 [0.0, 1.9]\\
S4D-type & 0 & 200 & 0.856 [0.74, 1.07]$^{\rm d}$ & 1.02 [0.947, 1.15]$^{\rm d}$ & 100.0$^{\rm d}$ & 87.5 [82.2, 91.4] & 0.0 [0.0, 1.9]\\
 & 1 & 200 & 0.826 [0.723, 0.917]$^{\rm d}$ & 1 [0.955, 1.02]$^{\rm d}$ & 100.0$^{\rm d}$ & 91.5 [86.8, 94.6] & 0.5 [0.1, 2.8]\\
 & 2 & 200 & 1.06 [0.805, 1.21]$^{\rm d}$ & 1.05 [0.902, 1.11]$^{\rm d}$ & 100.0$^{\rm d}$ & 89.0 [83.9, 92.6] & 0.5 [0.1, 2.8]\\
minGRU-type & 0 & 200 & 0.978 [0.806, 1.17]$^{\rm d}$ & 1.05 [0.96, 1.09]$^{\rm d}$ & 100.0$^{\rm d}$ & 87.5 [82.2, 91.4] & 0.0 [0.0, 1.9]\\
 & 1 & 200 & 0.944 [0.829, 1.11]$^{\rm d}$ & 1.06 [0.982, 1.18]$^{\rm d}$ & 100.0$^{\rm d}$ & 88.5 [83.3, 92.2] & 0.0 [0.0, 1.9]\\
 & 2 & 200 & 0.927 [0.802, 1.05]$^{\rm d}$ & 0.999 [0.912, 1.07]$^{\rm d}$ & 100.0$^{\rm d}$ & 91.5 [86.8, 94.6] & 0.0 [0.0, 1.9]\\
byte-level Mamba & 0 & 200 & 0.804 [0.717, 0.919]$^{\rm d}$ & 0.946 [0.853, 1.05]$^{\rm d}$ & 100.0$^{\rm d}$ & 84.0 [78.3, 88.4] & 0.0 [0.0, 1.9]\\
\bottomrule
\end{tabular}
\par\smallskip
\raggedright\scriptsize Finite at $L=\infty$: every document, Wilson interval [98.1, 100] (200 documents; [98.0, 100] for 190). Over
the ten models and every context the median width exponent lies between 0.997 [0.996, 0.997] (byte-level Mamba) and 1 [1, 1], and
the median looseness between 508 [426, 636] (S4D-type seed~1, one period) and $1.78\times10^5$ [$1.43\times10^5$,
$2.39\times10^5$] (Mamba seed~2, bare document).
\par\medskip
\centering\scriptsize
\begin{tabular}{@{}llllll@{}}
\multicolumn{6}{@{}l}{(b) Complex-diagonal control, both arms, $\eps=10^{-8}$}\\
\toprule
seed & clean & box: median $R$, $L=0$ & box: $L=\infty$ cert. & disk: median $R$ in $10^3$, range & disk: $L=\infty$ cert.\\
\midrule
0 & 88.0 [82.8, 91.8] & $2.14\cdot10^{65}$ [$4.23\cdot10^{53}$, $5.84\cdot10^{72}$] & 0.0 [0.0, 1.9] & 5.2 [4.31, 6.42] ($K{=}4$) to 6.24 [5.44, 7.21] ($L{=}1024$) & 89.0 [83.9, 92.6]\\
1 & 84.5 [78.8, 88.9] & $7.49\cdot10^{65}$ [$1.37\cdot10^{54}$, $1.56\cdot10^{73}$] & 0.0 [0.0, 1.9] & 5.38 [4.14, 6.85] ($L{=}256$) to 5.98 [4.2, 6.66] ($L{=}1024$) & 93.5 [89.2, 96.2]\\
\bottomrule
\end{tabular}
\par\medskip
\begin{tabular}{@{}llrlrlrlc@{}}
\multicolumn{9}{@{}l}{(c) Distance probe, gated models; (d) budget sweep, seed~0 (the complete documents of 0--49)}\\
\toprule
cell & archived & $n$ & extended & $n$ & new only & $n$ & new $-$ archived & decisive\\
\midrule
\multicolumn{9}{@{}l}{\emph{box: median $r(16)/r(1)$, GRU in units of $10^{-6}$, LSTM in units of $10^{-5}$}}\\
GRU 0 & 5.89 [4.8, 7.25] & 50 & 6.34 [5.68, 7.1] & 200 & 6.42 [5.73, 7.66] & 150 & 0.536 [$-$0.888, 2.14] & --\\
GRU 1 & 2.82 [2.27, 3.17] & 20 & 2.96 [2.64, 3.29] & 200 & 2.97 [2.64, 3.41] & 180 & 0.148 [$-$0.318, 0.975] & --\\
GRU 2 & 5.58 [4.05, 7.95] & 20 & 7.01 [6.29, 7.99] & 200 & 7.15 [6.53, 8.33] & 180 & 1.57 [$-$0.883, 3.21] & --\\
LSTM 0 & 5.67 [3.69, 9.69] & 50 & 5.18 [4.35, 6.47] & 200 & 5.12 [4.26, 6.38] & 150 & $-$0.552 [$-$4.59, 1.88] & --\\
LSTM 1 & 15 [5.88, 22.7] & 20 & 11.3 [9.96, 13.3] & 200 & 11.1 [9.31, 12.2] & 180 & $-$3.91 [$-$11.4, 3.34] & --\\
LSTM 2 & 5.84 [3.71, 11.9] & 20 & 6.58 [5.61, 7.5] & 200 & 6.65 [5.73, 7.5] & 180 & 0.812 [$-$4.18, 2.91] & --\\
\multicolumn{9}{@{}l}{\emph{box certifies a radius $r(d)>0$, at each of $d=128$, 512 and 1024}}\\
GRU, LSTM 0 & 0.0 [0.0, 7.1] & 50 & 0.0 [0.0, 1.9] & 200 & 0.0 [0.0, 2.5] & 150 & -- & --\\
GRU, LSTM 1, 2 & 0.0 [0.0, 16.1] & 20 & 0.0 [0.0, 1.9] & 200 & 0.0 [0.0, 2.1] & 180 & -- & --\\
\multicolumn{9}{@{}l}{\emph{$B=2048$ zonotope keeps $r(1)/10$, $d=128$, $F_{2048}(128)$}}\\
GRU 0 & 88.0 [76.2, 94.4] & 50 & 90.5 [85.6, 93.8] & 200 & 91.3 [85.7, 94.9] & 150 & 3.33 [$-$5.16, 15.7] & yes\\
GRU 1 & 100.0 [83.9, 100.0] & 20 & 96.0 [92.3, 98.0] & 200 & 95.6 [91.5, 97.7] & 180 & $-$4.44 [$-$8.52, 11.8] & yes\\
GRU 2 & 100.0 [83.9, 100.0] & 20 & 95.5 [91.7, 97.6] & 200 & 95.0 [90.8, 97.3] & 180 & $-$5 [$-$9.23, 11.3] & yes\\
\multicolumn{9}{@{}l}{\emph{the same at $d=512$, $F_{2048}(512)$}}\\
LSTM 0 & 76.0 [62.6, 85.7] & 50 & 69.0 [62.3, 75.0] & 200 & 66.7 [58.8, 73.7] & 150 & $-$9.33 [$-$21.8, 5.82] & yes\\
LSTM 1 & 75.0 [53.1, 88.8] & 20 & 61.5 [54.6, 68.0] & 200 & 60.0 [52.7, 66.9] & 180 & $-$15 [$-$30.6, 7.93] & yes\\
LSTM 2 & 90.0 [69.9, 97.2] & 20 & 92.5 [88.0, 95.4] & 200 & 92.8 [88.0, 95.7] & 180 & 2.78 [$-$5.85, 23.1] & yes\\
\multicolumn{9}{@{}l}{\emph{the same at $d=1024$, $F_{2048}(1024)$}}\\
GRU 0 & 0.0 [0.0, 7.1] & 50 & 0.0 [0.0, 1.9] & 200 & 0.0 [0.0, 2.5] & 150 & 0 [$-$7.14, 2.5] & yes\\
GRU 1, 2 & 0.0 [0.0, 16.1] & 20 & 0.0 [0.0, 1.9] & 200 & 0.0 [0.0, 2.1] & 180 & 0 [$-$16.1, 2.09] & yes\\
LSTM 0 & 24.0 [14.3, 37.4] & 50 & 26.0 [20.4, 32.5] & 200 & 26.7 [20.2, 34.3] & 150 & 2.67 [$-$12.2, 15] & yes\\
LSTM 1 & 25.0 [11.2, 46.9] & 20 & 14.0 [9.9, 19.5] & 200 & 12.8 [8.7, 18.4] & 180 & $-$12.2 [$-$34.5, 2.71] & yes\\
LSTM 2 & 65.0 [43.3, 81.9] & 20 & 58.5 [51.6, 65.1] & 200 & 57.8 [50.5, 64.8] & 180 & $-$7.22 [$-$25.6, 15.6] & yes\\
\midrule
\multicolumn{9}{@{}l}{\emph{reach $d^*_B$ at $B=256$ and at $B=16384$}}\\
GRU, $B=256$ & 1 [1, 1] & 15 & 1 [1, 1] & 49 & 1 [1, 1] & 34 & 0 [0, 0] & --\\
LSTM, $B=256$ & 16 [16, 16] & 16 & 16 [16, 16] & 50 & 16 [16, 16] & 34 & 0 [0, 0] & --\\
GRU, $B=16384$ & 1024 [512, 1024] & 15 & 1024 [1024, 1024] & 49 & 1024 [1024, 1024] & 34 & 0 [0, 512] & --\\
LSTM, $B=16384$ & 1024 [1024, 1024] & 16 & 1024 [1024, 1024] & 50 & 1024 [1024, 1024] & 34 & 0 [0, 0] & --\\
\multicolumn{9}{@{}l}{\emph{share that keeps $r(1)/10$ at $d=1024$, $F_B(1024)$}}\\
GRU, $B=8192$ & 60.0 [35.7, 80.2] & 15 & 51.0 [37.5, 64.4] & 49 & 47.1 [31.5, 63.3] & 34 & $-$12.9 [$-$38.4, 16.2] & --\\
GRU, $B=16384$ & 73.3 [48.0, 89.1] & 15 & 75.5 [61.9, 85.4] & 49 & 76.5 [60.0, 87.6] & 34 & 3.14 [$-$19.7, 30.7] & --\\
LSTM, $B=16384$ & 100.0 [80.6, 100.0] & 16 & 100.0 [92.9, 100.0] & 50 & 100.0 [89.8, 100.0] & 34 & 0 [$-$10.2, 19.4] & --\\
\bottomrule
\end{tabular}
\par\smallskip
\raggedright\scriptsize Reach $d^*_B$ at the other budgets, $B=512$, 1024, 2048, 4096 and 8192, extended (49 and 50 documents): GRU
16 [16, 16], 64 [64, 64], 128 [128, 128], 512 [256, 512] and 1024 [512, 1024]; LSTM 16 [16, 16], 128 [64, 128], 512 [512, 512],
1024 [1024, 1024] and 1024 [1024, 1024]. The archived and the new-only reaches have the same point values, except the GRU's at
$B=8192$ on the new documents alone (512 [512, 1024]); no difference interval excludes 0.
\end{table}

\begin{table}[!htbp]
\caption{Per-checkpoint 95\% intervals, part 2 (columns of (c) and the marks as in \cref{tab:ci}). (e) The interval-map growth rates
(\cref{sec:exp-r4,app:gate3-r4}): the cells at the ends of the ranges that the paper prints, and the two cells whose difference
interval excludes 0; every gated cell with its interval is in \cref{tab:r4}; layers as there; rates per token, $n$ the documents with
a rate. (f) The distance probe on the nonnegative diagonal models; $B=256$ over box: the ratio of the two median certified radii on the
same documents, and after the semicolon the median of the per-document ratios. (g) The slow-channel attribution and (h) the token-set
certificate (set Q, 200 documents; not extended except the token-set arms completed to 200). $^a$The interval of the difference excludes
0: these are the five statistics on which the new documents differ from the archived ones (each row of Mamba seed~1 is two, $\lambda_M$
and $\lambda_J$ being equal on Mamba). $^b$Newcombe interval of the growth-rate prediction for this share, which is not one of the
164 statistics of that comparison.}
\label{tab:ci-b}
\centering\scriptsize\setlength{\tabcolsep}{2.5pt}
\begin{tabular}{@{}llrlrlrlc@{}}
\multicolumn{9}{@{}l}{(e) Growth rates; (f) distance probe, nonnegative diagonal models}\\
\toprule
cell & archived & $n$ & extended & $n$ & new only & $n$ & new $-$ archived & decisive\\
\midrule
\multicolumn{9}{@{}l}{\emph{median $\lambda_M$, gated models}}\\
GRU 1, layer 1 & 0.835 [0.833, 0.839] & 99 & 0.837 [0.834, 0.839] & 199 & 0.838 [0.835, 0.84] & 100 & 0.00263 [$-$0.00225, 0.00581] & yes\\
GRU 1, layer 2 & 0.163 [0.152, 0.176] & 99 & 0.16 [0.152, 0.163] & 199 & 0.156 [0.146, 0.162] & 100 & $-$0.00719 [$-$0.0234, 0.00493] & yes\\
GRU 2, layer 2 & 0.164 [0.148, 0.179] & 99 & 0.156 [0.146, 0.168] & 199 & 0.151 [0.138, 0.167] & 100 & $-$0.0135 [$-$0.0335, 0.00538] & yes\\
\multicolumn{9}{@{}l}{\emph{median $\lambda_J$, gated models, in units of $10^{-3}$}}\\
LSTM 0, layer 1 & $-$3.75 [$-$5, $-$3.04] & 99 & $-$3.47 [$-$4.64, $-$2.6] & 199 & $-$3.04 [$-$5.63, $-$1.93] & 100 & 0.703 [$-$1.85, 2.16] & yes\\
\multicolumn{9}{@{}l}{\emph{share of the documents with $\lambda_J>0$}}\\
LSTM 0, layer 1 & 25.3 [17.7, 34.6] & 99 & 24.1 [18.7, 30.5] & 199 & 23.0 [15.8, 32.2] & 100 & [$-$14.0, 9.60]$^b$ & --\\
\multicolumn{9}{@{}l}{\emph{median lag to half the a-priori bound (tokens)}}\\
GRU 1, layer 2 & 25 [25, 25] & 100 & 25 [25, 25] & 200 & 25 [25, 25] & 100 & 0 [0, 0] & --\\
LSTM 1, layer 1 & 44 [43, 45] & 100 & 44 [44, 45] & 200 & 44 [44, 45] & 100 & 0 [$-$0.5, 2] & --\\
\multicolumn{9}{@{}l}{\emph{median saturation rate over the largest $\lambda_M$ of the layers up to the one measured}}\\
LSTM 1, layer 1 & 0.992 [0.986, 0.999] & 100 & 0.994 [0.989, 1] & 200 & 0.997 [0.989, 1.01] & 100 & -- & yes\\
GRU 1, layer 2 & 0.994 [0.979, 1.01] & 100 & 0.993 [0.985, 1] & 200 & 0.992 [0.98, 1.02] & 100 & -- & yes\\
LSTM 0, layer 2 & 1.03 [1.02, 1.06] & 100 & 1.03 [1.02, 1.04] & 200 & 1.03 [1.02, 1.04] & 100 & -- & yes\\
\multicolumn{9}{@{}l}{\emph{median $\lambda_M=\lambda_J$, natural Mamba, in units of $10^{-3}$}}\\
Mamba 1, layer 1 & $-$2.00 [$-$2.06, $-$1.94] & 100 & $-$1.97 [$-$2.00, $-$1.92] & 200 & $-$1.92 [$-$1.98, $-$1.82] & 100 & 0.082 [0.0033, 0.182]$^a$ & --\\
Mamba 1, layer 2 & $-$2.75 [$-$2.88, $-$2.64] & 100 & $-$2.67 [$-$2.76, $-$2.55] & 200 & $-$2.56 [$-$2.70, $-$2.42] & 100 & 0.191 [0.00315, 0.361]$^a$ & --\\
\midrule
\multicolumn{9}{@{}l}{\emph{box: median $r(1024)/r(1)$}}\\
Mamba 0 & 7.72 [6.73, 10.6] & 50 & 9.73 [8.24, 11] & 200 & 10.4 [8.78, 12.1] & 150 & 2.67 [0.408, 4.45]$^a$ & yes\\
Mamba 1 & 17.2 [11.4, 24.4] & 20 & 17.5 [15, 20.5] & 200 & 17.5 [15, 20.8] & 180 & 0.348 [$-$6.95, 6.47] & yes\\
Mamba 2 & 9.61 [6.95, 12.7] & 20 & 10.1 [8.78, 11.9] & 200 & 10.1 [8.78, 11.9] & 180 & 0.456 [$-$3.1, 3.75] & yes\\
S4D-type 0 & 27.2 [20.6, 39.8] & 50 & 29.2 [25.7, 33.4] & 200 & 29.7 [25, 35.5] & 150 & 2.49 [$-$10, 9.54] & yes\\
S4D-type 1 & 52.8 [28.8, 87.4] & 20 & 58 [49.1, 69.8] & 200 & 59 [49.4, 74.3] & 180 & 6.2 [$-$21.8, 34.4] & yes\\
S4D-type 2 & 21.7 [15.5, 44.1] & 20 & 23.6 [20.6, 26.7] & 200 & 23.8 [20.6, 26.9] & 180 & 2.04 [$-$14.2, 8.06] & yes\\
\multicolumn{9}{@{}l}{\emph{box: median certified radius $r(d)$, in units of $10^{-3}$}}\\
S4D-type 0, $d=1$ & 1.99 [1.22, 2.98] & 50 & 1.86 [1.63, 2.13] & 200 & 1.83 [1.62, 2.22] & 150 & -- & --\\
S4D-type 0, $d=512$ & 29.1 [21.4, 40.5] & 50 & 34 [31.1, 37.5] & 200 & 34.3 [31.3, 37.9] & 150 & -- & --\\
\multicolumn{9}{@{}l}{\emph{$B=256$ zonotope over box: ratio of the median certified radii; after the semicolon the median of the per-document ratios}}\\
Mamba 0, $d=1$ & 116 [90.8, 128]; 104 & 50 & 110 [99.8, 132]; 106 & 200 & 108 [93.3, 148]; 109 & 150 & -- & --\\
S4D-type 0, $d=1$ & 9.39 [7.28, 13.9]; 9.63 & 50 & 9.98 [8.92, 11.1]; 9.6 & 200 & 10.1 [8.74, 11.1]; 9.58 & 150 & -- & --\\
Mamba 0, $d=512$ & 30.1 [26.8, 32.9]; 29.2 & 50 & 30 [27.9, 31.4]; 28.8 & 200 & 30 [27.5, 31.6]; 28.7 & 150 & $-$0.134 [$-$3.8, 3.63] & --\\
S4D-type 0, $d=512$ & 1.27 [0.924, 1.68]; 1.12 & 50 & 1.07 [0.98, 1.17]; 1.05 & 200 & 1.06 [0.967, 1.14]; 1.04 & 150 & $-$0.213 [$-$0.618, 0.161] & --\\
\multicolumn{9}{@{}l}{\emph{share on which the $B=256$ zonotope reaches 0.99 times the box's certified radius}}\\
Mamba 0, $d=512$ & 100.0 [92.9, 100.0] & 50 & 100.0 [98.1, 100.0] & 200 & 100.0 [97.5, 100.0] & 150 & 0 [$-$2.5, 7.14] & yes\\
Mamba 1, 2, $d=512$ & 100.0 [83.9, 100.0] & 20 & 100.0 [98.1, 100.0] & 200 & 100.0 [97.9, 100.0] & 180 & 0 [$-$2.09, 16.1] & yes\\
S4D-type 0, $d=512$ & 54.0 [40.4, 67.0] & 50 & 52.5 [45.6, 59.3] & 200 & 52.0 [44.1, 59.8] & 150 & $-$2 [$-$17.3, 13.7] & yes\\
S4D-type 1, $d=512$ & 25.0 [11.2, 46.9] & 20 & 28.0 [22.2, 34.6] & 200 & 28.3 [22.3, 35.3] & 180 & 3.33 [$-$19.4, 18.8] & yes\\
S4D-type 2, $d=512$ & 45.0 [25.8, 65.8] & 20 & 36.0 [29.7, 42.9] & 200 & 35.0 [28.4, 42.2] & 180 & $-$10 [$-$31.8, 10.5] & yes\\
S4D-type 1, $d=128$ & 60.0 [38.7, 78.1] & 20 & 74.5 [68.0, 80.0] & 200 & 76.1 [69.4, 81.8] & 180 & 16.1 [$-$3.22, 38.2] & no\\
S4D-type 2, $d=128$ & 90.0 [69.9, 97.2] & 20 & 85.5 [80.0, 89.7] & 200 & 85.0 [79.1, 89.5] & 180 & $-$5 [$-$14.3, 15.6] & no\\
\bottomrule
\end{tabular}
\par\medskip
\begin{tabular}{@{}lllll@{}}
\multicolumn{5}{@{}l}{(g) Slow-channel attribution, standard-IBP selective models (archived, 200 documents; bootstrap per seed)}\\
\toprule
model & seed & restored of the width failures & failures that are flips & pairs slower than 4096 tokens, clean (largest layer)\\
\midrule
Mamba & 0 & 91.2 [83.0, 98.1] (52 of 57) & 35.2 [25.3, 45.3] (31 of 88) & 20.7 [20.5, 20.8] (layer 2)\\
 & 1 & 100.0 [100.0, 100.0] (91 of 91) & 3.2 [0.0, 7.2] (3 of 94) & 0.1 [0.1, 0.1] (layer 1)\\
 & 2 & 79.6 [67.5, 90.5] (39 of 49) & 48.4 [38.4, 58.5] (46 of 95) & 29.8 [29.4, 30.2] (layer 2)\\
minGRU-type & 0 & 71.1 [60.6, 81.0] (54 of 76) & 33.3 [24.8, 42.2] (38 of 114) & 61.1 [59.9, 62.3] (layer 2)\\
 & 1 & 76.9 [62.9, 89.7] (30 of 39) & 63.2 [53.9, 72.3] (67 of 106) & 39.7 [39.0, 40.4] (layer 2)\\
 & 2 & 66.1 [53.7, 78.0] (39 of 59) & 35.9 [26.3, 45.7] (33 of 92) & 28.0 [27.8, 28.2] (layer 1)\\
\bottomrule
\end{tabular}
\par\medskip
\begin{tabular}{@{}lllllll@{}}
\multicolumn{7}{@{}l}{(h) Token-set certificate over the most frequent neutral token, certified-trained S4D-type models ($P=8$ unless stated)}\\
\toprule
training & seed & clean document & substitution box & $P=64$, box: archived ($n$) & extended (200) & $P=64$ against $P=8$: lost/gained (200)\\
\midrule
standard IBP & 0 & 75.0 [68.6, 80.5] & 27.5 [21.8, 34.1] & 29.4 [23.3, 36.5] (180) & 29.0 [23.2, 35.6] & 0/3\\
 & 1 & 93.5 [89.2, 96.2] & 86.0 [80.5, 90.1] & 86.0 [80.5, 90.1] (200) & 86.0 [80.5, 90.1] & 0/0\\
 & 2 & 53.0 [46.1, 59.8] & 41.5 [34.9, 48.4] & 41.5 [34.9, 48.4] (200) & 41.5 [34.9, 48.4] & 0/0\\
fixed-point & 0 & 75.5 [69.1, 80.9] & 44.0 [37.3, 50.9] & 45.5 [38.7, 52.4] (200) & 45.5 [38.7, 52.4] & 0/3\\
 & 1 & 94.0 [89.8, 96.5] & 61.5 [54.6, 68.0] & 58.1 [50.4, 65.5] (160) & 61.5 [54.6, 68.0] & 0/0\\
 & 2 & 72.0 [65.4, 77.8] & 39.0 [32.5, 45.9] & 42.5 [35.9, 49.4] (200) & 42.5 [35.9, 49.4] & 0/7\\
\bottomrule
\end{tabular}
\par\smallskip
\raggedright\scriptsize Other token-set shares that the text cites (clean document, $P=8$ unless stated; 200 documents, not
extended): fixed-point fine-tuned Mamba seed~1 16.0 [11.6, 21.7], at $P=64$ 68.0 [61.2, 74.1]; its seeds 0 and 2 0.0 [0.0, 1.9],
at $P=64$ 0.5 [0.1, 2.8] and 1.0 [0.3, 3.6]; fixed-point fine-tuned S4D-type seed~1 at $|S|=8$ 40.0 [33.5, 46.9] (seeds 0 and 2:
7.5 [4.6, 12.0] and 0.0 [0.0, 1.9]); natural CNN 99.0 [96.4, 99.7] (decisive) and bag of words 52.0 [45.1, 58.8].
\end{table}

\subsection{The extension record}\label{app:cleanctx-record}

\paragraph{Runs.}
The extension ran on 2026-10-31 as arrays of CPU tasks, after its smoke test (run twice, \cref{app:prereg-amend}) and a summary of
the archive alone: 423 tasks and, before the final summary, the five tasks that had stopped at their time budget, resumed from
their last save. Per experiment, with its tasks and core-hours: the clean prediction at every
context for the 53 checkpoints of \cref{tab:cleanlen,tab:cleanlen-ct} on 1000 documents (53 tasks, 152 core-hours) and for the 25 of
\cref{tab:cleanctx} on 200 (25, 3); documents 190--199 of the two context-length-probe rows (2, 24); 10 to 50 documents for each of
the twelve token-set arms (13, 8); documents 100--199 of the growth rates for 17 pairs of model and seed (17, 52), and 200 documents for
each of two certified-trained gated checkpoints, described at the end of this appendix (2, 3); documents 50--199 on seed~0 and 20--199
on seeds 1 and 2 of the distance probe for the four architectures, 510 new documents each (204, 581); and documents 16--49 of the
budget sweep for the GRU and the LSTM (48, 146). The recomputed archived documents (59 tasks) took 28 core-hours, the two smoke tests
14, and the summaries and figures less than 1: 1,012 core-hours in all, against about 1,515 planned. Each number of core-hours is
counted from the task logs, which record the start, the host, the number of cores and the end of every task, as the time between the
run line and the exit line times the cores.

\paragraph{Reproduction and other checks.}
Every extension also recomputed archived documents with the archived configuration on the type of CPU that had produced them: two
documents per pair of model and seed of the growth rates and of the distance probe, the first document of each budget group of the
sweep, one whole batch of each completed token-set arm, and documents 0--189 of the two context-length-probe rows, 660 documents in
all; 26 more, in eight files, ran on the other type of CPU of the cluster to measure the node effect (686 in all). Every discrete field (a prediction,
a flag, certified or not) matched, every bound and margin lay within the registered tolerance, and every recomputed certified radius
lay inside the new bisection bracket, bitwise equal in all 474 radius cells; the only floating-point differences, on the cross-node
growth-rate document, were in stored six-digit values that interval maps amplify, which the registered rules report for another type
of CPU and do not count as failures. The budget sweep reproduced the distance probe's certified radii in all 500 cells per model
that the two share. The clean runs reproduced the archived clean counts of the first training study and the archived agreement flags
of the context-length probe within the registered tolerance, and the registered per-document implications between a certificate and
the clean prediction held on every checkpoint.

\paragraph{Verdicts.}
Every validity check that can stop a reading passed under the two amendments below, and all six predictions held. (i) Every
criterion that the extension predicted, all but the token-set certificate's criterion that a longer period is not worse (see
below), kept its archived outcome on the extended sets, archived documents included (162 criterion outcomes of the two completed
context-length-probe rows, the token-set certificate, the growth rates, the distance probe and the budget sweep), and on the new
documents alone where these meet a criterion's own minimum (the 145 outcomes of the growth rates, on 100 new documents per pair of
model and seed, of the distance probe, on 150 or 180, and of the budget sweep, on 34). (ii) The new documents agree with the archived
ones on 159 of 164 statistics (about 8 exclusions of 0 are expected under exchangeable documents, and at most 16 were allowed). The
five whose difference interval excludes 0 are the median ratio $r_{\rm box}(1024)/r_{\rm box}(1)$ of the natural Mamba seed~0 (on the
new documents alone 10.4 [8.78, 12.1], 150 documents) and the median $\lambda_M$ and $\lambda_J$, which are equal on Mamba, of both
layers of the natural Mamba seed~1 ($-0.00192$ [$-0.00198$, $-0.00182$] and $-0.00256$ [$-0.0027$, $-0.00242$], 100 documents); for
these as for the others the extended values with their intervals are the paper's numbers (\cref{tab:ci-b}). (iii) The universal
statements of the paper hold on every document with the extended $n$: the gated box certifies no radius at $d\ge128$ and the
$B=2048$ zonotope a positive one at $d=1024$ (200 documents per seed); the slowest nonnegative diagonal channel's mean log decay is
negative on every segment (200 documents per pair of model and seed); the $L=\infty$ box is finite at $\eps=10^{-8}$ on every document
of the completed natural Mamba row (200); and every byte-level token-set row stays 0 (200 documents per arm). (iv) $\lambda_J<0<
\lambda_M$ for the median document, with 95\% intervals excluding 0, in every layer of every seed of the gated models (200 documents
per seed), and the interval of the difference between the new and the archived documents in the share of the seed-0 LSTM's
first-layer documents with $\lambda_J>0$ contains 0. (v) On our licence-text context the clean prediction of the natural nonnegative
diagonal models settles within 16 periods: 0 of 1000 documents change between $K=16$ and $L=\infty$ on each checkpoint. (vi) On 1000
documents the K-uniform hull certified the clean prediction of none of the documents of each natural Mamba and minGRU-type seed. The
share of documents on which the $B=2048$ zonotope keeps a tenth of $r(1)$ at $d=1024$ on the LSTM seed~2, not decisive on its 20
archived documents, is decisive at $n=200$ (58.5\% [51.6, 65.1]). The share on which the $B=256$ zonotope reaches 0.99 times the box
at $d=128$ is not decisive at $n=200$ on the S4D-type seed~1 (74.5\% [68.0, 80.0]; decisive on its 20 archived documents) and seed~2
(85.5\% [80.0, 89.7], as on its 20); both intervals contain the threshold of 80\%. The token-set certificate's criterion that a longer
period is not worse, which the extension left without a prediction and reported paired on the common documents, failed as registered
on one fixed-point fine-tuned S4D-type seed, when 160 documents at $P=64$ were compared with 200 at $P=8$, and holds on the completed
arms: at $P=64$ that seed certifies the same 123 of its 200 documents as at $P=8$, 61.5\% [54.6, 68.0] (\cref{tab:ci-b}); the failure
stays the pre-registered outcome (\cref{app:prereg-criteria}).

\paragraph{Cells still below 200 documents, and why.}
The budget sweep has 50 documents (49 complete for the GRU), because 200 would have cost about 1,020 core-hours more, and so do the
$B=2048$ zonotope rows of the nonnegative diagonal models on seed~0 (about 47 more). The linear relaxations of \cref{app:lirpa} keep
their registered sizes (20 documents per seed on the gated models, 10 on the nonnegative diagonal models and on each model of the
complex-diagonal control, and up to 200 under synonym substitution), since full CROWN on 200 documents per seed would cost about
3,600 core-hours. The attack columns of the two training studies keep their registered first 100 documents (the genetic attack has
200 documents for 15 checkpoints, \cref{app:gate2}). The two completed context-length-probe rows and the completed token-set arms
reached 200 documents with 10 to 50 new documents each, too few for a reading on the new documents alone. Fewer than three seeds
remain for the budget sweep and the byte-level models of the growth rates (seed~0), for the complex-diagonal control (two), and in
\cref{app:lirpa} for the nonnegative diagonal models and for full CROWN under synonym substitution (seed~0).

\paragraph{Two amendments of the summary script.}
Both changed only the summary script and its tests, after the pre-registration, and both readings are reported
(\cref{app:prereg-amend}). The
first, made before the code was frozen, concerns the check that the summary reproduces every value of the archived summaries: as
registered it compared every text line exactly, and one archived verdict line of the context-length probe says that the gated boxes
``blow up'', where the summary script, reworded on 2026-10-22 with the same verdict, now writes ``saturate or grow''. As frozen this
check fails on that one line, which would have stopped every reading although no archived number differs; amended, a listed and
dated wording change of a summary script is reported, not failed, and every number, verdict and other line is compared as before.
The second, made after the production summary and before any of its results was read, concerns the identity check of the tasks: as
registered it read the run line of each task log from its first line, but on the cluster's virtual nodes the batch system writes two
notices before it, so the check failed on the 148 outputs whose logs begin with these notices, and nothing would have been read.
All 148 logs end with exit status 0 and contain no checksum warning. Amended, the run line is the first line in the format of the job
script and the host line is looked for in the eight lines after it; every other rule is unchanged. Under both amendments every
validity check that can stop a reading passes.

\paragraph{Floating point.}
Neither the certificates on the documents added by the extension nor the computations at box scale 0 of
\cref{tab:cleanlen,tab:cleanlen-ct,tab:cleanctx}, the limits and the box statements, were re-verified with outward rounding
(\cref{rem:float}).

\subsection{The evidence probe, and two certified-trained gated models}

\paragraph{The evidence probe.}
The first training study also stored, for each of its 45 checkpoints, the clean accuracy of a probe that moves the evidence away from
the readout: the document followed by 256, 1024 or 4096 neutral tokens and read at the end (1000 documents; aggregates only, so no
paired statistic; Wilson half-width at most 3.1 points). The natural recurrent models lose most of it: 70.5--82.7\% after 256
trailing tokens, 53.7--71.2\% after 1024 and 47.6--50.7\% after 4096 (natural Mamba seed~0: 77.6\% [74.9, 80.1], 56.3\% [53.2, 59.3]
and 49.8\% [46.7, 52.9]). After 4096 trailing tokens every checkpoint other than the CNNs lies within 47.0--56.5\%, near chance (a
constant classifier scores 47.4 or 52.6\% here). The CNN's max-pool keeps the evidence: 86.2\% [83.9, 88.2], 81.6\% [79.1, 83.9]
and 74.6\% [71.8, 77.2] for the natural CNN, and 68.9--73.0\% after 4096 tokens with standard IBP. The probe is descriptive; it
entered no criterion except, through its certified shares, the relevance fork of the first study (\cref{app:prereg-criteria}).
Because its context follows the document, it measures how far the document's evidence carries, not the effect of a context before
it.

\paragraph{Growth rates of two certified-trained gated models.}
The growth rates of \cref{app:gate3-r4} were also measured, without a prediction, on 200 documents of set Q for two certified-trained
gated checkpoints: the standard-IBP LSTM seed~0, a constant classifier (48.0\% [41.2, 54.9] clean on these documents), and the GRU
after fixed-point fine-tuning, seed~2, the one certified-trained gated checkpoint that is not a constant classifier (88.5\% [83.3,
92.2] clean here; 89.4\% clean and 0.0\% certified on the 1000 documents of \cref{tab:trainseeds}). Both keep the gated pattern of
the natural models: median $\lambda_M$ 0.713 [0.711, 0.716] and 0.0847 [0.0675, 0.0986] per token in the first and second layer of the
LSTM, 0.871 [0.869, 0.875] and 0.0803 [0.0658, 0.0963] in the GRU; median $\lambda_J$ $-0.0393$ [$-0.0414$, $-0.0358$] and $-0.0062$
[$-0.00722$, $-0.00569$] (LSTM), $-0.16$ [$-0.163$, $-0.155$] and $-0.0138$ [$-0.0163$, $-0.0125$] (GRU), over 199 documents with a
rate; and a median lag to half the a-priori bound of 32 and 30 tokens (LSTM), 26 and 24 (GRU). The validity checks of the growth rates
passed on both. One of the two is a constant classifier and the other certifies no document, so neither shows how the interval maps of
a gated model behave once IBP training gives it non-trivial certificates.

\section{A task that needs memory}\label{app:memory}

The training remedy of \cref{app:remedy} was measured on a task that does not charge for forgetting: a review followed by licence text,
where a capped model that carries no evidence of the review past $8\tau_{\max}$ words (\cref{app:remedy}, ``Where the information
goes'') loses nothing. This appendix reports a third pre-registered training study on a task that does: the review's sentiment has to
be read after a gap of up to 1024 neutral words. It asks whether a model that must remember still keeps its unbounded-context
certificates, and at what price in memory. Criteria, hashes and outcomes in the record's form: \cref{app:prereg-criteria}. The theory
behind the design is in \cref{app:proofs-periodic} (``Memory after the information'': no single box certifies trailing contexts of
every length after the information a prediction needs, \cref{prop:M1}; one box covers every context before the information and a
horizon after it, (M4) of \cref{rem:startboxM}); phase 2 below evaluates that construction.

\paragraph{Setting.}
Word-level IMDB truncated at 512 words, the synonym threat of \cref{app:gate2} at box scale 1 at every position (the review, the gap and
any period before the review), two-layer width-64 Mamba and minGRU-type models, the linear-ramp recipe and the certificates of
\cref{app:remedy}, all in float64 and, unlike that appendix's core rows, not re-verified with outward rounding
(\cref{rem:float}). A training row is a review, a gap of $G$ words of GNU-manual text and a marker token at the end; the prediction is
read at the marker. The gap is empty with probability $1/4$ and otherwise uniform on $1,\dots,1024$ words; four gap variants per
review, 90{,}000 training rows and 2500 development rows, one draw shared by every arm and seed (data seed 20261102). The marker is the
special token with the smallest count in both neutral streams (id 23): a point under the threat, never a substitute. The test inputs
are the first 1000 test documents (526 positive) with a gap of $G\in\{0,256,512,1024,2048,4096\}$ licence words drawn as in
\cref{app:remedy}'s probe (the trailing words of the same draw, disjoint from the 1024-word period that precedes the document), and,
for the K-uniform and one-copy certificates, the perturbed 1024-word licence period before the review at $G\in\{0,1024\}$. Gaps of
2048 and 4096 words lie beyond the training range.

\paragraph{Quantities.}
Per checkpoint and gap $G$, on the test split: the clean accuracy $\mathrm{acc}(G)$ (Wilson interval); the accuracy of the
\emph{other-document rule} $\mathrm{acc}'(G)$, which scores the prediction made after another review (test document $i+1$) followed
by the same gap and marker against document $i$'s label; their difference, the \emph{carry} $\mathrm{carry}(G)=\mathrm{acc}(G)-\mathrm{acc}'(G)$
in points, exactly 0 when the prediction does not depend on which review precedes the gap, with a paired bootstrap interval over the
documents ($B=2000$); the agreement $\mathrm{agrD}(G)$ of the two predictions; and the class of the checkpoint at $G$: E (``carries
evidence'') when carry's interval lies above 0, N below 0, otherwise P (``no detectable evidence'') if $\mathrm{agrD}\ge0.98$ and D if
not. The certificates at gap $G$ are the $L=0$ share $c_0(G)$ (the review and the gap in their boxes), the K-uniform share
$c_{\rm hull}(G)$ (every number of copies of the perturbed period before the review, \cref{thm:hull}), the one-copy share $c_{K1}(G)$
and the $L=\infty$ share $c_{\rm inf}(G)$, pooled and per class (the smaller class share is the ``worst class''), with the ratios
$r_{\rm hull}(G)=c_{\rm hull}(G)/c_0(G)$ and $r_{\rm inf}(G)=c_{\rm inf}(G)/c_0(G)$. A checkpoint is degenerate if its clean accuracy
at gap 0 is below 60\% or it predicts one class on more than 80\% of the documents; means are over the non-degenerate seeds with 95\%
t-intervals over the seeds, and are n/a with fewer than two. Seeds pair the arms (the same initial weights and rows): a paired
difference is a mean over the usable pairs (both members non-degenerate) with its paired t-interval. A rule that ignores the review
scores within the chance band $[42.2,57.8]\%$ with probability at least 0.99 at $n=1000$; ``at chance'' is written only when every
covered seed's accuracy lies inside it.

\paragraph{Arms.}
Seeds 1--3 of each arm; the natural arms train for 4000 steps, the IBP arms for 21{,}000 (batch 64), on the same rows:
\begin{itemize}
\item \emph{natural}, and \emph{natural with cap} $\tau_{\max}\in\{64,128,512,2048\}$ (every decay times $1-1/\tau_{\max}$): the
memory ceiling and the dose--response of the cap without IBP;
\item \emph{plain IBP}: the linear-ramp recipe of \cref{app:remedy} without a remedy, the reference of every paired comparison;
\item \emph{cap 128} and \emph{cap 2048}: plain IBP with the cap, at $\tau_{\max}=128$ (below the memory requirement) and at
$2048$ (twice it);
\item \emph{fixed-point term}: the $L=\infty$ term of \cref{cor:fixed} on half of every robust batch from the first robust step
(256-word training periods of GNU manuals at the document's box scale), as in \cref{app:remedy};
\item \emph{cap 512 + hull}: the paper's headline recipe, the cap at $\tau_{\max}=512$ with the K-uniform hull of \cref{thm:hull} as
the robust term; and \emph{hull}: the hull term without the cap. The smoke test's rule for a weight ramp of the hull term did not fire
(median robust loss over the first 50 robust steps: Mamba 0.749 and 0.750, minGRU-type 1.483 and 1.751, against 5), so the term entered
at full weight.
\end{itemize}
``Cap 512 + hull'' and ``hull'' of a seed share their batches; plain IBP, cap 128 and cap 2048 of a seed share theirs; the fixed-point
term draws prefix periods and diverges from plain IBP after the first robust step, paired by its initial weights and rows. A pilot of
plain IBP (seed 0) ran before the IBP arms. The second robust term of \cref{app:remedy} aside, every setting is Gate 8's: the recipe,
the prefix terms, the seeds, the degeneracy rule and the statistics.

\paragraph{Stages, rules and their outcomes.}
Three stages, each with rules fixed before it ran (\cref{app:prereg-criteria}). (i) A \emph{smoke test} (one H200 job) set the
production parameters by rules on its own measurements: the gap range stayed at 1024 words (no memory class above 122 GB; the
fallback to 512 words was not triggered), the evaluation runs in chunks of 64 documents (chunks of 256 peaked at 147.96 GB), the
IBP class with the largest training peak (86.54 GB: plain IBP and the two caps) runs on H200 nodes, the hull term enters without a
weight ramp (above), and the data
hashes of both gap ranges were recorded. The first smoke's memory probe measured a cache pool inherited across evaluation calls: the
allocator's reserved peak was reset to the pool left by the preceding evaluation, so the training window of plain IBP reported 122.6 GB
reserved against 36.9 GB allocated, which triggered the fallback rule; the probe was corrected (the cache is emptied before each
reset) before any production run, the two superseded smoke jobs produced no result data (120 training steps each), no threshold or
rule changed, and the pre-registered objects kept their hashes. (ii) The \emph{natural stage} trained the 30 natural runs and the two
pilots and read the premise: the natural models' mean development accuracy after 1024 words, 86.8\% [85.9, 87.7] (Mamba; seeds
86.4, 87.0, 87.0) and 87.3\% [85.0, 89.7] (minGRU-type; 88.4, 86.4, 87.2), passed the bar of 80\% at 4000 steps, so no fork to 21{,}000
steps and no stop fired; both pilots passed (no skipped step, no overflow row; development accuracy after 1024 words 81.9 and
81.3\%); the driver reproduced the archived Gate 8 evaluations document by document on six checkpoints (0 differences). Its summary
job started on 2026-11-04 at 08:09 (Berlin), before the deadline that selects the core tier (every arm above). (iii) The \emph{main
stage} trained and evaluated the 36 IBP checkpoints: 68 of 68 planned runs complete (the 18 long-class runs stopped at their 287-minute
budget and resumed once, as registered), the cap holds on every capped checkpoint, every per-document implication holds on the 68
checkpoints, the draws, data identity, configurations and arguments match the plan, and no training run was flagged. The summary is
readable. Three Mamba checkpoints are degenerate (a result, not an incompleteness): cap 512 + hull seed~2 and hull seeds 2 and 3
(clean accuracy at gap 0 58.8, 59.4 and 59.9\%; carry after 1024 words 2.9, 4.0 and 5.2 points); cap 512 + hull seed~3 is not
degenerate by the rule but reaches 60.5\% at gap 0 and 51.3\% after 1024 words. The minGRU-type model has no degenerate seed.
So the Mamba hull arm has one non-degenerate seed and is ``unread'': no paired criterion reads it, and every sentence that ranges over
the arms names it. The hull arms of \cref{app:remedy} (the same recipe, without gaps) had no degenerate seed; this is a new
phenomenon on this task and is described only.

\paragraph{Results per arm and per checkpoint.}
\Cref{tab:memory-mamba,tab:memory-mingru} give, per architecture, the clean accuracy at gaps 0, 512 and 1024, the carry and the class
after 1024 words, the certificates at gap 0 and after 1024 words with the worst-class share, and every checkpoint. The natural models
carry the review across 1024 words (clean accuracy after 1024 words 86.4\% [85.0, 87.7] (Mamba) and 86.8\% [83.4, 90.1] (minGRU-type);
carry 36.1 [34.5, 37.7] and 36.5 [35.1, 37.9] points) and, beyond the training range, after 4096 words still 80.6\% [76.5, 84.7] and
85.5\% [80.2, 90.8]. The natural dose--response of the cap after 1024 words: Mamba 70.2\% [40.0, 100.4] at $\tau_{\max}=64$, 80.8\% [76.9, 84.7]
at 128, 85.4\% [84.2, 86.6] at 512, 86.0\% [84.8, 87.3] at 2048; minGRU-type 51.2\% [49.5, 52.9], 84.1\% [80.9, 87.3], 86.4\% [83.2, 89.5],
86.8\% [86.0, 87.6]; after 4096 words the cap at 2048 gives 67.3\% [63.9, 70.6] and 81.0\% [74.5, 87.5], the cap at 512 50.9\% [48.4, 53.4]
and 50.7\% [46.5, 55.0]. The development-to-test differences lie within $\pm3$ points except the Mamba cap-64 models (seed~1
$+10.0$ points, development 66.2 against test 56.2; seed~2 $+4.5$). The largest observed time constant is infinite on every natural
model without a cap and sits at the cap on every capped arm (63.5, 127.5, 511.5 and 2047.5 words, to within a few words); plain IBP,
the fixed-point term and the hull term have finite but large ones (\cref{tab:memory-mamba,tab:memory-mingru}, per checkpoint).

\begin{table}[htbp]
\caption{The memory task, Mamba (1000 test documents; \%; mean over the non-degenerate seeds with its 95\% t-interval over the
seeds; an arm with fewer than three non-degenerate seeds prints its seeds' values, $^\ast$: degenerate). (a) Clean accuracy (float64) at a gap of $G$ licence words between the review and the marker,
the carry after 1024 words (accuracy minus the other-document rule's, in points) and the class of each seed at $G=1024$ (E: carries
detectable evidence; P: none, and the two predictions agree on at least 98\% of the documents; D: none, less agreement; $^\ast$:
degenerate). (b) Certificates of the IBP arms (float64, not re-verified with outward rounding): at gap 0 the $L=0$ share, the
K-uniform share (every number of copies of the perturbed 1024-word period before the review, \cref{thm:hull}), the one-copy share and
the ratio K-uniform$/L=0$; after 1024 trailing words the K-uniform share and its ratio; and the smaller of the two classes' K-uniform
shares at gap 0. (c) Every checkpoint: the carry with its paired bootstrap interval over the documents, the certificates at gap 0
(with the $L=\infty$ share and its ratio) and the largest time constant observed over the 150 documents of the diagnostic. Written by
the table script from the summary; the internal codes are mapped in \cref{app:prereg-criteria}.}
\label{tab:memory-mamba}
\centering\scriptsize\setlength{\tabcolsep}{2pt}
\begin{tabular}{@{}lccccc@{}}
\toprule
training & \multicolumn{3}{c}{clean accuracy at gap $G$ (\%)} & carry at $G=1024$ & class at $G=1024$\\
 & $G=0$ & $G=512$ & $G=1024$ & (points) & (per seed)\\
\midrule
natural & 87.1 [84.9, 89.4] & 86.9 [86.1, 87.6] & 86.4 [85.0, 87.7] & 36.1 [34.5, 37.7] & E E E\\
natural, cap 2048 & 87.7 [86.5, 89.0] & 87.0 [84.6, 89.5] & 86.0 [84.8, 87.3] & 35.6 [33.7, 37.5] & E E E\\
natural, cap 512 & 87.9 [85.3, 90.5] & 87.0 [85.1, 88.8] & 85.4 [84.2, 86.6] & 35.6 [33.0, 38.2] & E E E\\
natural, cap 128 & 86.0 [83.0, 89.0] & 85.8 [82.1, 89.5] & 80.8 [76.9, 84.7] & 29.9 [23.5, 36.3] & E E E\\
natural, cap 64 & 84.3 [79.0, 89.7] & 83.3 [75.9, 90.7] & 70.2 [40.0, 100.4] & 19.7 [$-$11.7, 51.1] & E E E\\
plain IBP & 81.0 [78.2, 83.7] & 80.7 [77.2, 84.1] & 80.5 [76.2, 84.7] & 29.2 [25.2, 33.2] & E E E\\
cap 128 & 75.2 [73.8, 76.6] & 74.1 [72.2, 75.9] & 71.1 [67.7, 74.4] & 20.0 [16.8, 23.2] & E E E\\
cap 2048 & 79.4 [77.6, 81.2] & 78.8 [77.8, 79.8] & 78.2 [77.3, 79.1] & 27.1 [24.7, 29.5] & E E E\\
fixed-point term & 74.2 [62.7, 85.7] & 72.3 [56.6, 88.0] & 68.8 [46.8, 90.9] & 18.6 [0.5, 36.8] & E E E\\
cap 512 + hull & 81.1 / 58.8$^\ast$ / 60.5 & 81.2 / 50.2$^\ast$ / 52.0 & 80.0 / 50.2$^\ast$ / 51.3 & 29.0 / 2.9$^\ast$ / 4.1 & E E$^\ast$ E\\
hull & 79.1 / 59.4$^\ast$ / 59.9$^\ast$ & 79.3 / 51.4$^\ast$ / 57.0$^\ast$ & 78.2 / 51.5$^\ast$ / 57.6$^\ast$ & 25.8 / 4.0$^\ast$ / 5.2$^\ast$ & E E$^\ast$ E$^\ast$\\
\bottomrule
\end{tabular}

\par\medskip
\resizebox{\textwidth}{!}{%
\begin{tabular}{@{}lccccccc@{}}
\toprule
training & \multicolumn{4}{c}{at gap 0} & \multicolumn{2}{c}{after 1024 words} & worst class\\
 & $L=0$ & K-uniform & $K=1$ & ratio & K-uniform & ratio & K-uniform, gap 0\\
\midrule
plain IBP & 75.1 [74.3, 75.9] & 2.2 [$-$6.1, 10.6] & 74.6 [74.4, 74.7] & 3.0 [$-$8.1, 14.0] & 2.1 [$-$5.8, 10.1] & 2.9 [$-$7.8, 13.6] & 2.0 [$-$5.6, 9.6]\\
cap 128 & 72.5 [70.1, 74.9] & 72.4 [70.0, 74.7] & 72.5 [70.1, 74.8] & 99.8 [99.6, 100.0] & 69.6 [66.8, 72.4] & 99.9 [99.7, 100.1] & 62.0 [58.6, 65.5]\\
cap 2048 & 74.6 [72.0, 77.2] & 68.8 [55.7, 81.9] & 73.6 [69.1, 78.0] & 92.2 [76.6, 107.8] & 68.9 [58.0, 79.7] & 94.4 [82.7, 106.1] & 63.2 [50.8, 75.7]\\
fixed-point term & 70.8 [56.0, 85.5] & 70.2 [54.0, 86.4] & 70.5 [54.8, 86.2] & 99.1 [96.8, 101.5] & 65.6 [41.7, 89.5] & 99.5 [98.7, 100.3] & 59.1 [43.3, 74.8]\\
cap 512 + hull & 74.3 / 57.4$^\ast$ / 58.3 & 74.3 / 57.4$^\ast$ / 58.3 & 74.4 / 57.4$^\ast$ / 58.3 & 100.0 / 100.0$^\ast$ / 100.0 & 73.9 / 50.2$^\ast$ / 50.4 & 100.0 / 100.0$^\ast$ / 100.0 & 68.6 / 44.1$^\ast$ / 48.7\\
hull & 75.9 / 57.7$^\ast$ / 57.2$^\ast$ & 73.9 / 57.6$^\ast$ / 57.1$^\ast$ & 75.4 / 57.7$^\ast$ / 57.2$^\ast$ & 97.4 / 99.8$^\ast$ / 99.8$^\ast$ & 72.6 / 51.1$^\ast$ / 56.3$^\ast$ & 96.3 / 99.8$^\ast$ / 98.6$^\ast$ & 61.6 / 46.8$^\ast$ / 31.2$^\ast$\\
\bottomrule
\end{tabular}
}

\par\medskip
\begin{tabular}{@{}llcccccccccc@{}}
\toprule
training & seed & clean, $G=0$ & clean, $G=1024$ & carry(1024) & class & $L=0$ & K-uniform & ratio & $L=\infty$ & $L=\infty$ ratio & largest $\tau_{\rm h}$\\
\midrule
natural & 1 & 88.0 & 86.4 & 36.1 [32.5, 39.7] & E & -- & -- & -- & -- & -- & $\infty$\\
 & 2 & 87.2 & 86.9 & 36.8 [33.2, 40.4] & E & -- & -- & -- & -- & -- & $\infty$\\
 & 3 & 86.2 & 85.8 & 35.5 [31.8, 39.3] & E & -- & -- & -- & -- & -- & $\infty$\\
natural, cap 2048 & 1 & 87.2 & 85.5 & 35.8 [32.1, 39.5] & E & -- & -- & -- & -- & -- & 2047.5\\
 & 2 & 87.8 & 86.5 & 36.3 [32.6, 40.0] & E & -- & -- & -- & -- & -- & 2047.5\\
 & 3 & 88.2 & 86.1 & 34.8 [31.1, 38.7] & E & -- & -- & -- & -- & -- & 2047.5\\
natural, cap 512 & 1 & 87.8 & 84.9 & 34.9 [31.2, 38.6] & E & -- & -- & -- & -- & -- & 511.5\\
 & 2 & 89.0 & 85.4 & 35.1 [31.6, 38.7] & E & -- & -- & -- & -- & -- & 511.5\\
 & 3 & 86.9 & 85.9 & 36.8 [33.0, 40.7] & E & -- & -- & -- & -- & -- & 511.5\\
natural, cap 128 & 1 & 84.9 & 79.4 & 27.0 [23.5, 30.7] & E & -- & -- & -- & -- & -- & 127.5\\
 & 2 & 87.3 & 82.5 & 31.8 [28.5, 35.4] & E & -- & -- & -- & -- & -- & 127.5\\
 & 3 & 85.9 & 80.5 & 31.0 [27.3, 34.8] & E & -- & -- & -- & -- & -- & 127.5\\
natural, cap 64 & 1 & 82.3 & 56.2 & 5.2 [3.3, 7.1] & E & -- & -- & -- & -- & -- & 63.5\\
 & 2 & 86.6 & 78.1 & 28.3 [25.0, 31.9] & E & -- & -- & -- & -- & -- & 63.5\\
 & 3 & 84.1 & 76.3 & 25.6 [21.9, 29.1] & E & -- & -- & -- & -- & -- & 63.5\\
plain IBP & 1 & 81.8 & 82.2 & 30.7 [27.2, 34.6] & E & 74.9 & 0.6 & 0.8 & 35.3 & 47.1 & $7.3\cdot10^{7}$\\
 & 2 & 79.7 & 78.8 & 27.5 [23.8, 31.3] & E & 75.0 & 0.0 & 0.0 & 19.0 & 25.3 & $3.2\cdot10^{8}$\\
 & 3 & 81.4 & 80.4 & 29.4 [25.6, 33.5] & E & 75.5 & 6.1 & 8.1 & 39.5 & 52.3 & $1.8\cdot10^{8}$\\
cap 128 & 1 & 75.3 & 72.4 & 21.1 [17.2, 24.9] & E & 72.6 & 72.4 & 99.7 & 72.4 & 99.7 & 127.5\\
 & 2 & 75.7 & 71.1 & 20.3 [16.7, 24.0] & E & 73.4 & 73.3 & 99.9 & 73.3 & 99.9 & 127.5\\
 & 3 & 74.6 & 69.7 & 18.6 [15.2, 22.1] & E & 71.5 & 71.4 & 99.9 & 71.4 & 99.9 & 127.5\\
cap 2048 & 1 & 80.0 & 77.8 & 27.6 [23.9, 31.4] & E & 74.5 & 63.8 & 85.6 & 70.4 & 94.5 & 2047.4\\
 & 2 & 78.6 & 78.4 & 26.0 [22.1, 29.7] & E & 75.7 & 74.3 & 98.2 & 74.9 & 98.9 & 2047.5\\
 & 3 & 79.6 & 78.4 & 27.7 [24.1, 31.6] & E & 73.6 & 68.3 & 92.8 & 72.1 & 98.0 & 2047.5\\
fixed-point term & 1 & 77.5 & 74.6 & 23.2 [19.5, 27.0] & E & 74.1 & 73.6 & 99.3 & 73.7 & 99.5 & 1890.1\\
 & 2 & 76.2 & 73.3 & 22.5 [19.1, 26.0] & E & 74.3 & 74.3 & 100.0 & 74.3 & 100.0 & $8.4\cdot10^{4}$\\
 & 3 & 68.9 & 58.6 & 10.2 [7.1, 13.3] & E & 63.9 & 62.7 & 98.1 & 63.0 & 98.6 & 1213.6\\
cap 512 + hull & 1 & 81.1 & 80.0 & 29.0 [25.2, 32.9] & E & 74.3 & 74.3 & 100.0 & 74.3 & 100.0 & 511.5\\
 & 2$^\ast$ & 58.8 & 50.2 & 2.9 [1.4, 4.6] & E & 57.4 & 57.4 & 100.0 & 57.4 & 100.0 & 511.5\\
 & 3 & 60.5 & 51.3 & 4.1 [2.2, 6.0] & E & 58.3 & 58.3 & 100.0 & 58.3 & 100.0 & 511.5\\
hull & 1 & 79.1 & 78.2 & 25.8 [22.1, 29.5] & E & 75.9 & 73.9 & 97.4 & 74.8 & 98.6 & 9532.0\\
 & 2$^\ast$ & 59.4 & 51.5 & 4.0 [2.2, 5.9] & E & 57.7 & 57.6 & 99.8 & 57.6 & 99.8 & $3.6\cdot10^{4}$\\
 & 3$^\ast$ & 59.9 & 57.6 & 5.2 [3.3, 7.1] & E & 57.2 & 57.1 & 99.8 & 57.1 & 99.8 & $2.6\cdot10^{4}$\\
\bottomrule
\end{tabular}

\end{table}

\begin{table}[htbp]
\caption{The memory task, minGRU-type model: as \cref{tab:memory-mamba}.}
\label{tab:memory-mingru}
\centering\scriptsize\setlength{\tabcolsep}{2pt}
\begin{tabular}{@{}lccccc@{}}
\toprule
training & \multicolumn{3}{c}{clean accuracy at gap $G$ (\%)} & carry at $G=1024$ & class at $G=1024$\\
 & $G=0$ & $G=512$ & $G=1024$ & (points) & (per seed)\\
\midrule
natural & 87.7 [86.3, 89.1] & 87.0 [84.2, 89.9] & 86.8 [83.4, 90.1] & 36.5 [35.1, 37.9] & E E E\\
natural, cap 2048 & 86.9 [84.9, 89.0] & 86.8 [84.8, 88.9] & 86.8 [86.0, 87.6] & 35.2 [33.4, 37.0] & E E E\\
natural, cap 512 & 86.6 [84.1, 89.1] & 86.7 [84.1, 89.3] & 86.4 [83.2, 89.5] & 35.3 [30.1, 40.6] & E E E\\
natural, cap 128 & 87.3 [87.0, 87.7] & 87.6 [86.9, 88.2] & 84.1 [80.9, 87.3] & 34.4 [28.7, 40.1] & E E E\\
natural, cap 64 & 85.8 [83.6, 88.0] & 76.7 [59.8, 93.7] & 51.2 [49.5, 52.9] & $-$0.1 [$-$0.4, 0.2] & P P P\\
plain IBP & 83.3 [81.0, 85.6] & 83.1 [80.5, 85.7] & 83.0 [81.4, 84.6] & 32.3 [29.7, 35.0] & E E E\\
cap 128 & 78.8 [74.6, 83.1] & 61.9 [55.1, 68.7] & 52.3 [50.3, 54.3] & 0.2 [0.2, 0.2] & P P P\\
cap 2048 & 82.1 [81.4, 82.8] & 81.4 [79.3, 83.5] & 81.0 [77.9, 84.1] & 29.8 [26.5, 33.0] & E E E\\
fixed-point term & 81.7 [79.9, 83.4] & 81.4 [79.9, 82.8] & 81.4 [80.9, 82.0] & 30.6 [28.5, 32.8] & E E E\\
cap 512 + hull & 80.2 [78.3, 82.1] & 78.0 [77.2, 78.7] & 72.8 [71.1, 74.4] & 20.6 [18.8, 22.5] & E E E\\
hull & 80.4 [79.6, 81.1] & 79.5 [78.6, 80.4] & 78.6 [76.6, 80.6] & 27.0 [26.2, 27.7] & E E E\\
\bottomrule
\end{tabular}

\par\medskip
\resizebox{\textwidth}{!}{%
\begin{tabular}{@{}lccccccc@{}}
\toprule
training & \multicolumn{4}{c}{at gap 0} & \multicolumn{2}{c}{after 1024 words} & worst class\\
 & $L=0$ & K-uniform & $K=1$ & ratio & K-uniform & ratio & K-uniform, gap 0\\
\midrule
plain IBP & 76.7 [74.7, 78.7] & 8.0 [2.7, 13.3] & 75.5 [69.8, 81.1] & 10.4 [3.4, 17.5] & 8.8 [3.1, 14.5] & 11.8 [3.7, 19.8] & 6.2 [3.1, 9.4]\\
cap 128 & 71.6 [67.6, 75.7] & 67.6 [59.1, 76.1] & 70.3 [64.0, 76.7] & 94.4 [87.6, 101.1] & 43.7 [41.4, 45.9] & 100.0 [100.0, 100.0] & 63.7 [52.9, 74.4]\\
cap 2048 & 76.3 [72.5, 80.1] & 68.1 [63.3, 72.9] & 74.6 [69.1, 80.0] & 89.3 [86.1, 92.4] & 66.0 [60.2, 71.9] & 91.5 [87.3, 95.6] & 64.4 [54.9, 73.9]\\
fixed-point term & 73.4 [71.2, 75.6] & 10.3 [$-$11.8, 32.4] & 72.6 [71.0, 74.3] & 14.0 [$-$16.2, 44.2] & 9.3 [$-$11.2, 29.9] & 13.0 [$-$15.6, 41.5] & 0.0 [0.0, 0.0]\\
cap 512 + hull & 75.4 [74.2, 76.5] & 74.1 [73.0, 75.3] & 74.7 [74.0, 75.4] & 98.4 [95.9, 100.8] & 67.3 [64.2, 70.4] & 99.2 [98.1, 100.3] & 65.8 [62.7, 69.0]\\
hull & 74.8 [73.4, 76.2] & 73.0 [70.7, 75.3] & 74.7 [73.2, 76.2] & 97.6 [94.2, 101.1] & 72.4 [71.3, 73.6] & 98.2 [95.2, 101.1] & 65.0 [62.1, 67.9]\\
\bottomrule
\end{tabular}
}

\par\medskip
\begin{tabular}{@{}llcccccccccc@{}}
\toprule
training & seed & clean, $G=0$ & clean, $G=1024$ & carry(1024) & class & $L=0$ & K-uniform & ratio & $L=\infty$ & $L=\infty$ ratio & largest $\tau_{\rm h}$\\
\midrule
natural & 1 & 88.3 & 88.2 & 37.0 [33.3, 40.7] & E & -- & -- & -- & -- & -- & $\infty$\\
 & 2 & 87.6 & 86.6 & 36.6 [33.1, 40.2] & E & -- & -- & -- & -- & -- & $\infty$\\
 & 3 & 87.2 & 85.5 & 35.9 [32.2, 39.7] & E & -- & -- & -- & -- & -- & $\infty$\\
natural, cap 2048 & 1 & 87.3 & 86.9 & 34.8 [31.4, 38.6] & E & -- & -- & -- & -- & -- & 2047.5\\
 & 2 & 87.5 & 87.0 & 36.0 [32.5, 39.8] & E & -- & -- & -- & -- & -- & 2047.5\\
 & 3 & 86.0 & 86.4 & 34.7 [31.0, 38.5] & E & -- & -- & -- & -- & -- & 2047.5\\
natural, cap 512 & 1 & 87.4 & 87.7 & 37.5 [34.0, 41.1] & E & -- & -- & -- & -- & -- & 511.5\\
 & 2 & 87.0 & 86.2 & 35.2 [31.6, 38.9] & E & -- & -- & -- & -- & -- & 511.5\\
 & 3 & 85.5 & 85.2 & 33.3 [29.7, 37.0] & E & -- & -- & -- & -- & -- & 511.5\\
natural, cap 128 & 1 & 87.3 & 83.2 & 32.8 [29.3, 36.2] & E & -- & -- & -- & -- & -- & 127.5\\
 & 2 & 87.5 & 83.6 & 33.3 [29.8, 36.9] & E & -- & -- & -- & -- & -- & 127.5\\
 & 3 & 87.2 & 85.6 & 37.0 [33.4, 40.6] & E & -- & -- & -- & -- & -- & 127.5\\
natural, cap 64 & 1 & 86.8 & 51.6 & $-$0.2 [$-$0.6, 0.2] & P & -- & -- & -- & -- & -- & 63.5\\
 & 2 & 85.1 & 50.4 & 0.0 [0.0, 0.0] & P & -- & -- & -- & -- & -- & 63.5\\
 & 3 & 85.5 & 51.6 & 0.0 [0.0, 0.0] & P & -- & -- & -- & -- & -- & 63.5\\
plain IBP & 1 & 84.4 & 83.6 & 32.1 [28.3, 35.9] & E & 77.5 & 5.6 & 7.2 & 25.1 & 32.4 & $5.0\cdot10^{7}$\\
 & 2 & 82.8 & 82.3 & 31.4 [27.8, 35.2] & E & 76.8 & 8.8 & 11.5 & 35.1 & 45.7 & $3.6\cdot10^{6}$\\
 & 3 & 82.8 & 83.1 & 33.5 [29.8, 37.4] & E & 75.9 & 9.6 & 12.6 & 41.9 & 55.2 & $1.4\cdot10^{6}$\\
cap 128 & 1 & 79.5 & 53.0 & 0.2 [$-$0.3, 0.7] & P & 72.9 & 70.8 & 97.1 & 72.4 & 99.3 & 127.5\\
 & 2 & 80.1 & 51.4 & 0.2 [0.0, 0.5] & P & 72.2 & 68.1 & 94.3 & 70.5 & 97.6 & 127.5\\
 & 3 & 76.9 & 52.4 & 0.2 [$-$0.2, 0.7] & P & 69.8 & 64.0 & 91.7 & 67.1 & 96.1 & 127.5\\
cap 2048 & 1 & 82.2 & 82.2 & 31.1 [27.4, 34.9] & E & 77.6 & 68.9 & 88.8 & 72.4 & 93.3 & 2047.5\\
 & 2 & 81.8 & 79.7 & 28.5 [24.9, 32.2] & E & 76.6 & 69.5 & 90.7 & 73.5 & 96.0 & 2047.5\\
 & 3 & 82.3 & 81.1 & 29.7 [25.9, 33.5] & E & 74.6 & 65.9 & 88.3 & 69.1 & 92.6 & 2045.7\\
fixed-point term & 1 & 80.9 & 81.5 & 31.4 [27.7, 35.3] & E & 72.4 & 15.4 & 21.3 & 47.4 & 65.5 & $3.9\cdot10^{6}$\\
 & 2 & 81.8 & 81.2 & 29.7 [26.1, 33.5] & E & 73.9 & 15.4 & 20.8 & 67.6 & 91.5 & $9.3\cdot10^{4}$\\
 & 3 & 82.3 & 81.6 & 30.8 [27.3, 34.6] & E & 73.9 & 0.0 & 0.0 & 52.6 & 71.2 & $6.8\cdot10^{5}$\\
cap 512 + hull & 1 & 81.0 & 73.1 & 20.9 [17.7, 24.3] & E & 75.8 & 73.7 & 97.2 & 74.1 & 97.8 & 511.1\\
 & 2 & 80.0 & 73.2 & 21.2 [17.9, 24.7] & E & 74.9 & 74.1 & 98.9 & 74.3 & 99.2 & 511.4\\
 & 3 & 79.5 & 72.0 & 19.8 [16.6, 23.2] & E & 75.4 & 74.6 & 98.9 & 74.7 & 99.1 & 511.3\\
hull & 1 & 80.3 & 78.8 & 26.9 [23.2, 30.7] & E & 74.6 & 74.0 & 99.2 & 74.2 & 99.5 & $2.6\cdot10^{4}$\\
 & 2 & 80.7 & 77.7 & 26.7 [22.9, 30.6] & E & 74.3 & 72.2 & 97.2 & 73.3 & 98.7 & $1.1\cdot10^{4}$\\
 & 3 & 80.1 & 79.3 & 27.3 [23.6, 31.2] & E & 75.4 & 72.8 & 96.6 & 73.0 & 96.8 & $2.7\cdot10^{4}$\\
\bottomrule
\end{tabular}

\end{table}

\paragraph{Criteria and classification.}
The criteria, fixed before any run, are read per architecture on the non-degenerate seeds; a pre-registered reading table states
which sentence each outcome allows, and the sentences below are its instantiations (\cref{tab:memory-criteria} lists the per-arm
flags, \cref{tab:memory-criteria-outcomes} every outcome with its quantity). The classification asks, per architecture, whether some remedy arm keeps its K-uniform
certificates at gap 0 (the headline bar of \cref{app:remedy}: ratio $\ge0.8$, clean accuracy $\ge78\%$ and an $L=0$ share within 5
points of plain IBP's, in a seed) \emph{and} remembers as well as plain IBP of the same seed (``free'': accuracy and carry after 1024
words at most 5 points below plain IBP's, a drop from gap 0 to gap 1024 at most 5 points larger, and at least 70\% after 1024 words),
in at least two of three usable pairs; an arm that keeps and is at least 10 points behind in accuracy, carry and drop is ``costly'';
a pair between the two bars lies in a dead zone, and an arm whose ratio lies in $[0.7,0.8)$ in a seed is in the ratio dead zone.

\emph{Mamba.} The premise holds: natural training reaches 86.8\% [85.9, 87.7] on the development split after 1024 words (above;
on the test split 86.4\% [85.0, 87.7] with carry 36.1 [34.5, 37.7]). The theorem check passes: with every decay capped at $1-1/128$ (and
$1-1/64$), after 4096 trailing words the review moves the margin by at most $7.0\times10^{-9}$ on every document, and the
prediction equals the one after another review on at least 99.9\% of them (on every one of the nine checkpoints), consistent with
the fading bound (\cref{lem:F}, a corollary of \cref{lem:fading}; under a time-constant cap every decay is at most
$\rho=1-1/\tau_{\max}$, so \cref{lem:F} applies with $\bar a\le\rho$). Plain IBP learns the memory: 80.5\%
[76.2, 84.7] clean accuracy after 1024 words (carry 29.2 [25.2, 33.2]), and on the memory task standard IBP training keeps at most
half of its K-uniform certificates at gap 0 ($r_{\rm hull}$ 3.0\% [$-$8.1, 14.0], per seed 0.8, 0.0 and 8.1; K-uniform share 2.2\%
[$-$6.1, 10.6], one copy 74.6\% [74.4, 74.7]; $r_{\rm inf}$ 41.6\% [6.0, 77.2]). On the memory task the cap at 128 keeps at least 70\% of
the bare-input certificates at gap 0 for every number of copies, in 3 of 3 seeds ($r_{\rm hull}$ 99.8\% [99.6, 100.0], per seed 99.7,
99.9 and 99.9; K-uniform 72.4\% [70.0, 74.7], one copy 72.5\% [70.1, 74.8]). But the Mamba models with a cap of 64 or 128 still carry
detectable evidence after $8\tau_{\max}$ words ($16\tau_{\max}$ at $\tau_{\max}=64$) in 3 of 3 seeds (carry after 1024 words: natural,
cap 64: 19.7 [$-$11.7, 51.1], per seed 5.2, 28.3 and 25.6, agreement 69.4\% [24.3, 114.6], accuracy 70.2\% [40.0, 100.4]; natural, cap
128: 29.9 [23.5, 36.3], agreement 56.7\% [46.4, 67.1], accuracy 80.8\% [76.9, 84.7]; IBP, cap 128: 20.0 [16.8, 23.2], agreement 60.8\%
[54.4, 67.2], accuracy 71.1\% [67.7, 74.4]): the cap attenuates each channel by $e^{-8}$ ($e^{-16}$), and the readout can still decode it.
Nothing is said about these arms holding no information beyond about $\tau_{\max}$ words. \emph{Classification: FREE LUNCH, through
the cap at 2048.} On a task that needs memory (a review's sentiment read after up to 1024 neutral words), IBP training with the
time-constant cap at $\tau_{\max}=2048$ (twice the memory requirement) keeps K-uniform certificates at the end of the review (at gap 0:
$r_{\rm hull}$ 92.2\% [76.6, 107.8], per seed 85.6, 98.2 and 92.8; K-uniform share 68.8\% [55.7, 81.9], one copy 73.6\% [69.1, 78.0];
clean accuracy 79.4\% [77.6, 81.2], per seed 80.0, 78.6 and 79.6) and carries the evidence across 1024 words (clean accuracy 78.2\%
[77.3, 79.1]; at most 5 points behind plain IBP of the same seed after 1024 words, paired differences in accuracy $-$2.3 [$-$7.3, 2.7]
and in carry $-$2.1 [$-$4.3, 0.1], and it forgets at most 5 points more between gap 0 and gap 1024, $\Delta$drop 0.7 [$-$3.5, 4.9]; at
gap 0 the paired accuracy difference is $-$1.6 [$-$2.6, $-$0.6]) in 3 of 3 seeds. After the 1024 words its K-uniform certified share is
68.9\% [58.0, 79.7] (ratio 94.4\% [82.7, 106.1]). The label's arm is the cap, not the hull term: the fixed-point term did not keep both
against plain IBP of the same seed (0 of 3 pairs) because it keeps the ratio but not the clean accuracy ($r_{\rm hull}$ at gap 0 99.1\%
[96.8, 101.5], K-uniform 70.2\% [54.0, 86.4], one copy 70.5\% [54.8, 86.2]; clean accuracy at gap 0 77.5, 76.2 and 68.9\%, each below
78\%; after 1024 words 68.8\% [46.8, 90.9], paired differences in accuracy $-$11.6 [$-$33.7, 10.4], in carry $-$10.6 [$-$29.4, 8.3],
$\Delta$drop 4.9 [$-$4.8, 14.5], at gap 0 $-$6.8 [$-$19.1, 5.6]; seed~3 the weakest, 68.9\% at gap 0 and 58.6\% after 1024 words), and
the hull term is unread (one usable pair; in that pair, seed~1, it keeps the ratio at 97.4\% with 79.1\% clean accuracy at gap 0 and
is 4.0 points behind in accuracy and 4.9 in carry, $\Delta$drop 1.3). Pairs in a memory dead zone: cap 128 (3 pairs), the fixed-point
term (3 pairs) and cap 512 + hull (pair 3); plain IBP's reference $L=0$ share is 75.1\%. Without IBP, natural accuracy after 1024
words is 5.2 points lower at $\tau_{\max}=128$ than at 2048 (80.8 against 86.0\%; Welch interval [1.1, 9.3], which excludes 0; at gap 0
the difference is 1.7 [$-$1.5, 4.9]), below the pre-registered bar of 15 points for a memory cost of the cap. Whether the cap at 512 of
the headline recipe costs memory against the hull term is n/a on Mamba (one usable pair: the drops from gap 0 to gap 1024 are 1.1 and
0.9 points, a difference of 0.2).

\emph{minGRU-type model.} The premise holds: 87.3\% [85.0, 89.7] on the development split after 1024 words (test split 86.8\%
[83.4, 90.1], carry 36.5 [35.1, 37.9]). The theorem check passes: under the caps 128 and 64, after 4096 trailing words the review moves
the margin by at most $9.5\times10^{-10}$ on every document, and the prediction equals the one after another review on at least 99.9\%
of them (every one of the nine checkpoints), consistent with the fading bound. Plain IBP learns the memory: 83.0\% [81.4, 84.6] after
1024 words (carry 32.3 [29.7, 35.0]), and keeps at most half of its K-uniform certificates at gap 0 ($r_{\rm hull}$ 10.4\% [3.4, 17.5],
per seed 7.2, 11.5 and 12.6; K-uniform 8.0\% [2.7, 13.3], one copy 75.5\% [69.8, 81.1]; $r_{\rm inf}$ 44.4\% [16.0, 72.9]). The cap at
128 keeps at least 70\% of the bare-input certificates at gap 0 for every number of copies, in 3 of 3 seeds ($r_{\rm hull}$ 94.4\%
[87.6, 101.1], per seed 97.1, 94.3 and 91.7; K-uniform 67.6\% [59.1, 76.1], one copy 70.3\% [64.0, 76.7]), at the price of memory:
after $8\tau_{\max}$ trailing words, in 3 of 3 seeds, the cap-128 IBP models give the same prediction after two different reviews on
at least 98\% of the documents, and their predictions carry no detectable evidence of the review (carry 0.2 [0.2, 0.2]; agreement 99.5\%
[99.0, 100.1]; accuracy 52.3, at chance); the same holds for the natural models with the cap at 64 after $16\tau_{\max}$ words (carry
$-$0.1 [$-$0.4, 0.2], agreement 99.9\% [99.3, 100.4]; accuracy 51.2, at chance), while the natural cap-128 models still carry detectable
evidence after $8\tau_{\max}$ words in 3 of 3 seeds (carry 34.4 [28.7, 40.1], agreement 54.0\% [48.6, 59.5], accuracy 84.1\% [80.9,
87.3]): the cap attenuates each channel by $e^{-8}$, and the readout can still decode it. \emph{Classification: FREE LUNCH, through the
cap at 2048.} On this task IBP training with the cap at $\tau_{\max}=2048$ keeps K-uniform certificates at the end of the review (at
gap 0: $r_{\rm hull}$ 89.3\% [86.1, 92.4], per seed 88.8, 90.7 and 88.3; K-uniform 68.1\% [63.3, 72.9], one copy 74.6\% [69.1, 80.0];
clean accuracy 82.1\% [81.4, 82.8], per seed 82.2, 81.8 and 82.3) and carries the evidence across 1024 words (clean accuracy 81.0\%
[77.9, 84.1]; at most 5 points behind plain IBP of the same seed after 1024 words, paired differences in accuracy $-$2.0 [$-$3.5,
$-$0.5] and in carry $-$2.6 [$-$6.1, 1.0], and it forgets at most 5 points more between gap 0 and gap 1024, $\Delta$drop 0.8
[$-$2.6, 4.1]; at gap 0 the paired accuracy difference is $-$1.2 [$-$3.4, 0.9]) in 3 of 3 seeds. After the 1024 words its K-uniform
certified share is 66.0\% [60.2, 71.9] (ratio 91.5\% [87.3, 95.6]). The time-constant cap is the trade-off: at $\tau_{\max}=128$ the
models carry no detectable evidence after 1024 words (accuracy 52.3, at chance). Here too the label's arm is the cap, not the hull
term. The fixed-point term did not keep both (0 of 3 pairs): it remembers as well as plain IBP (paired differences in accuracy
$-$1.6 [$-$2.8, $-$0.3], in carry $-$1.7 [$-$4.2, 0.8], $\Delta$drop $-$0.1 [$-$3.1, 2.9], at gap 0 $-$1.7 [$-$5.7, 2.3]; 81.4\% [80.9,
82.0] after 1024 words) but loses the K-uniform certificates at gap 0 ($r_{\rm hull}$ 14.0\% [$-$16.2, 44.2], per seed 21.3, 20.8 and
0.0; K-uniform 10.3\% [$-$11.8, 32.4], one copy 72.6\% [71.0, 74.3]). The hull term did not keep both either (1 of 3 pairs): it keeps
the certificates in 3 of 3 seeds ($r_{\rm hull}$ at gap 0 97.6\% [94.2, 101.1], K-uniform 73.0\% [70.7, 75.3], one copy 74.7\% [73.2,
76.2]) at a memory cost of 4--5 points (paired differences in accuracy $-$4.4 [$-$5.7, $-$3.1], in carry $-$5.4 [$-$7.3, $-$3.5],
$\Delta$drop 1.4 [$-$0.9, 3.8], at gap 0 $-$3.0 [$-$5.5, $-$0.4]; pairs 1 and 3 in the memory dead zone with carry differences of
$-$5.2 and $-$6.2; 78.6\% [76.6, 80.6] after 1024 words; after the 1024 words K-uniform 72.4\% [71.3, 73.6], ratio 98.2\% [95.2, 101.1]).
Cap 512 + hull keeps its certificates in 3 of 3 seeds ($r_{\rm hull}$ at gap 0 98.4\% [95.9, 100.8], K-uniform 74.1\% [73.0, 75.3],
one copy 74.7\% [74.0, 75.4]) with every pair in a memory dead zone ($\Delta$drop in $(5,10)$ points, or a gap-1024 difference in
$(-10,-5)$; paired differences in accuracy $-$10.2 [$-$12.8, $-$7.7], in carry $-$11.7 [$-$16.2, $-$7.2], $\Delta$drop 7.1 [5.2, 8.9], at
gap 0 $-$3.2 [$-$4.0, $-$2.4]; 72.8\% [71.1, 74.4] after 1024 words); plain IBP's reference $L=0$ share is 76.7\%. No memory cost of the
cap is measurable between 128 and 2048 in natural training (2.6 points, Welch interval [$-$0.7, 5.9]; 84.1 against 86.8\%; at gap 0
$-$0.4 [$-$2.5, 1.7]). At $2\tau_{\max}$ the capped hull models lose 5.6 [1.7, 9.6] points more accuracy between gap 0 and gap 1024 than
the uncapped hull models of the same seeds (per pair 6.4, 3.8 and 6.7 points; the interval excludes 0), below the pre-registered bar
of 10 points for a memory cost of the cap at 512 in every pair (0 of 3); after 1024 words the paired accuracy difference is $-$5.8
[$-$9.3, $-$2.3] and the carry difference $-$6.3 [$-$8.9, $-$3.7]; at gap 0 the paired difference is $-$0.2 [$-$2.1, 1.7].

\begin{table}[htbp]
\caption{The memory task: the per-arm flags of the classification and the paired differences from plain IBP (both architectures).
Flags (e1b\_spec 9.0): cert$_s$ counts the seeds that meet the headline bar at gap 0 ($r_{\rm hull}\ge0.8$, clean accuracy $\ge78\%$,
$L=0$ share within 5 points of plain IBP's mean); keeps: in at least two seeds; loses: $r_{\rm hull}(0)<0.7$ in at least two
non-degenerate seeds; free, cost, both: over the usable pairs with plain IBP of the same seed (free: accuracy and carry after 1024
words at most 5 points below plain IBP's, a drop from gap 0 to 1024 at most 5 points larger, at least 70\% after 1024 words; cost: at
least 10 points below in both and a drop at least 10 points larger; both: free and the headline bar in the same seed), each in at
least two pairs; a dead zone names the seeds whose pair falls between the bars. The paired differences (lower table) are means over
the usable pairs with paired t-intervals (points); an arm with fewer than three non-degenerate seeds prints its pairs' values per
seed ($^\ast$: a degenerate seed, no pair). Written by the table script from the summary.}
\label{tab:memory-criteria}
\centering\scriptsize\setlength{\tabcolsep}{2.5pt}
\begin{tabular}{@{}llcccccccl@{}}
\toprule
model & training & nd seeds & cert$_s$ & keeps & loses & free & cost & both & dead zone (seeds)\\
\midrule
Mamba & plain IBP (reference) & 3 & 0 of 3 & no & yes & -- & -- & -- & --\\
 & cap 128 & 3 & 0 of 3 & no & no & no & no & no & memory: 1, 2, 3\\
 & cap 2048 & 3 & 3 of 3 & yes & no & yes & no & yes & --\\
 & fixed-point term & 3 & 0 of 3 & no & no & no & no & no & memory: 1, 2, 3\\
 & cap 512 + hull & 2 & 1 of 3 & no & no & no & no & no & memory: 3\\
 & hull (unread) & 1 & 1 of 3 & no & no & no & no & no & --\\
\midrule
minGRU-type & plain IBP (reference) & 3 & 0 of 3 & no & yes & -- & -- & -- & --\\
 & cap 128 & 3 & 2 of 3 & yes & no & no & yes & no & --\\
 & cap 2048 & 3 & 3 of 3 & yes & no & yes & no & yes & --\\
 & fixed-point term & 3 & 0 of 3 & no & yes & yes & no & no & --\\
 & cap 512 + hull & 3 & 3 of 3 & yes & no & no & no & no & memory: 1, 2, 3\\
 & hull & 3 & 3 of 3 & yes & no & no & no & no & memory: 1, 3\\
\bottomrule
\end{tabular}

\par\medskip
\begin{tabular}{@{}llccccc@{}}
\toprule
model & training & usable pairs & $\Delta$acc(1024) & $\Delta$carry(1024) & $\Delta$drop & $\Delta$acc(0)\\
\midrule
Mamba & plain IBP (reference) & 3 & -- & -- & -- & --\\
 & cap 128 & 3 & $-$9.4 [$-$13.2, $-$5.6] & $-$9.2 [$-$13.8, $-$4.6] & 3.6 [2.9, 4.4] & $-$5.8 [$-$9.6, $-$1.9]\\
 & cap 2048 & 3 & $-$2.3 [$-$7.3, 2.7] & $-$2.1 [$-$4.3, 0.1] & 0.7 [$-$3.5, 4.9] & $-$1.6 [$-$2.6, $-$0.6]\\
 & fixed-point term & 3 & $-$11.6 [$-$33.7, 10.4] & $-$10.6 [$-$29.4, 8.3] & 4.9 [$-$4.8, 14.5] & $-$6.8 [$-$19.1, 5.6]\\
 & cap 512 + hull & 2 & $-$2.2 / $-$$-$$^\ast$ / $-$29.1 & $-$1.7 / $-$$-$$^\ast$ / $-$25.3 & 1.5 / $-$$-$$^\ast$ / 8.2 & $-$0.7 / $-$$-$$^\ast$ / $-$20.9\\
 & hull (unread) & 1 & -- & -- & -- & --\\
\midrule
minGRU-type & plain IBP (reference) & 3 & -- & -- & -- & --\\
 & cap 128 & 3 & $-$30.7 [$-$31.1, $-$30.4] & $-$32.1 [$-$34.8, $-$29.5] & 26.2 [21.9, 30.6] & $-$4.5 [$-$8.6, $-$0.4]\\
 & cap 2048 & 3 & $-$2.0 [$-$3.5, $-$0.5] & $-$2.6 [$-$6.1, 1.0] & 0.8 [$-$2.6, 4.1] & $-$1.2 [$-$3.4, 0.9]\\
 & fixed-point term & 3 & $-$1.6 [$-$2.8, $-$0.3] & $-$1.7 [$-$4.2, 0.8] & $-$0.1 [$-$3.1, 2.9] & $-$1.7 [$-$5.7, 2.3]\\
 & cap 512 + hull & 3 & $-$10.2 [$-$12.8, $-$7.7] & $-$11.7 [$-$16.2, $-$7.2] & 7.1 [5.2, 8.9] & $-$3.2 [$-$4.0, $-$2.4]\\
 & hull & 3 & $-$4.4 [$-$5.7, $-$3.1] & $-$5.4 [$-$7.3, $-$3.5] & 1.4 [$-$0.9, 3.8] & $-$3.0 [$-$5.5, $-$0.4]\\
\bottomrule
\end{tabular}

\end{table}

\begin{table}[htbp]
\caption{The memory task: the outcome of every criterion with the quantity it reads (both architectures; means over the
non-degenerate seeds with 95\% t-intervals, paired differences with paired t-intervals, the natural contrast with its Welch
interval); the internal codes are mapped in \cref{app:prereg-criteria}. Written by the table script from the summary.}
\label{tab:memory-criteria-outcomes}
\centering\scriptsize\setlength{\tabcolsep}{2.5pt}
\begin{tabular}{@{}p{0.24\textwidth}p{0.36\textwidth}p{0.36\textwidth}@{}}
\toprule
criterion & Mamba & minGRU-type\\
\midrule
the task permits memory (premise, development split) & PASS: mean development accuracy after 1024 words 86.8 [85.9, 87.7] (per seed 86.4 / 87.0 / 87.0), threshold 80.0 & PASS: mean development accuracy after 1024 words 87.3 [85.0, 89.7] (per seed 88.4 / 86.4 / 87.2), threshold 80.0\\
theorem check: no influence of the review after 4096 words under the caps 64 and 128 & PASS: 9 checkpoints read, 0 failing; largest influence of the review on the margin 6.9e-09, smallest agreement 100.0 & PASS: 9 checkpoints read, 0 failing; largest influence of the review on the margin 9.4e-10, smallest agreement 100.0\\
no detectable evidence after 16 $\tau_{\max}$: natural, cap 64 & FAIL (class E, per seed E E E): carry 19.7 [$-$11.7, 51.1], agreement after two different reviews 69.4 [24.3, 114.6], accuracy 70.2 [40.0, 100.4] & PASS (class P, per seed P P P): carry $-$0.1 [$-$0.4, 0.2], agreement after two different reviews 99.9 [99.3, 100.4], accuracy 51.2 [49.5, 52.9], every seed in the chance band\\
no detectable evidence after 8 $\tau_{\max}$: natural, cap 128 & FAIL (class E, per seed E E E): carry 29.9 [23.5, 36.3], agreement after two different reviews 56.7 [46.4, 67.1], accuracy 80.8 [76.9, 84.7] & FAIL (class E, per seed E E E): carry 34.4 [28.7, 40.1], agreement after two different reviews 54.0 [48.6, 59.5], accuracy 84.1 [80.9, 87.3]\\
no detectable evidence after 8 $\tau_{\max}$: IBP, cap 128 & FAIL (class E, per seed E E E): carry 20.0 [16.8, 23.2], agreement after two different reviews 60.8 [54.4, 67.2], accuracy 71.1 [67.7, 74.4] & PASS (class P, per seed P P P): carry 0.2 [0.2, 0.2], agreement after two different reviews 99.5 [99.0, 100.1], accuracy 52.3 [50.3, 54.3], every seed in the chance band\\
the cap 128 keeps the K-uniform certificates at gap 0 & PASS: ratio $\ge0.7$ in 3 of 3 seeds (per seed 99.7 / 99.9 / 99.9); K-uniform 72.4 [70.0, 74.7], one copy 72.5 [70.1, 74.8] & PASS: ratio $\ge0.7$ in 3 of 3 seeds (per seed 97.1 / 94.3 / 91.7); K-uniform 67.6 [59.1, 76.1], one copy 70.3 [64.0, 76.7]\\
plain IBP keeps at most half of its K-uniform certificates at gap 0 & PASS: mean ratio 3.0 [$-$8.1, 14.0] (K-uniform 2.2 [$-$6.1, 10.6], one copy 74.6 [74.4, 74.7]); $L=\infty$ ratio 41.6 [6.0, 77.2] & PASS: mean ratio 10.4 [3.4, 17.5] (K-uniform 8.0 [2.7, 13.3], one copy 75.5 [69.8, 81.1]); $L=\infty$ ratio 44.4 [16.0, 72.9]\\
plain IBP learns the memory & PASS: accuracy after 1024 words 80.5 [76.2, 84.7], carry 29.2 [25.2, 33.2] & PASS: accuracy after 1024 words 83.0 [81.4, 84.6], carry 32.3 [29.7, 35.0]\\
headline: the fixed-point term keeps certificates and memory & FAIL: certificates and memory kept together in 0 of 3 pairs; ratio 99.1 [96.8, 101.5] (K-uniform 70.2 [54.0, 86.4], one copy 70.5 [54.8, 86.2]); accuracy after 1024 words 68.8 [46.8, 90.9]; paired differences from plain IBP: accuracy $-$11.6 [$-$33.7, 10.4], carry $-$10.6 [$-$29.4, 8.3], drop 4.9 [$-$4.8, 14.5], at gap 0 $-$6.8 [$-$19.1, 5.6] & FAIL: certificates and memory kept together in 0 of 3 pairs; ratio 14.0 [$-$16.2, 44.2] (K-uniform 10.3 [$-$11.8, 32.4], one copy 72.6 [71.0, 74.3]); accuracy after 1024 words 81.4 [80.9, 82.0]; paired differences from plain IBP: accuracy $-$1.6 [$-$2.8, $-$0.3], carry $-$1.7 [$-$4.2, 0.8], drop $-$0.1 [$-$3.1, 2.9], at gap 0 $-$1.7 [$-$5.7, 2.3]\\
headline: the hull term keeps certificates and memory & n/a (one usable pair with plain IBP) & FAIL: certificates and memory kept together in 1 of 3 pairs; ratio 97.6 [94.2, 101.1] (K-uniform 73.0 [70.7, 75.3], one copy 74.7 [73.2, 76.2]); accuracy after 1024 words 78.6 [76.6, 80.6]; paired differences from plain IBP: accuracy $-$4.4 [$-$5.7, $-$3.1], carry $-$5.4 [$-$7.3, $-$3.5], drop 1.4 [$-$0.9, 3.8], at gap 0 $-$3.0 [$-$5.5, $-$0.4]\\
the cap costs memory without IBP (natural, 2048 against 128) & FAIL: 86.0 against 80.8 after 1024 words, difference 5.2 points, Welch [1.1, 9.3] (threshold 15); at gap 0 1.7 [$-$1.5, 4.9] & FAIL: 86.8 against 84.1 after 1024 words, difference 2.6 points, Welch [$-$0.7, 5.9] (threshold 15); at gap 0 $-$0.4 [$-$2.5, 1.7]\\
the headline recipe's cap costs memory (cap 512 + hull against hull) & n/a (one usable pair): drop difference 0.2 points & FAIL: drop difference $\ge10$ points in 0 of 3 pairs (per pair 6.4 / 3.8 / 6.7; mean 5.6 [1.7, 9.6]); accuracy after 1024 words $-$5.8 [$-$9.3, $-$2.3], at gap 0 $-$0.2 [$-$2.1, 1.7]\\
classification & FREE LUNCH (cap 2048); unread: hull & FREE LUNCH (cap 2048)\\
\bottomrule
\end{tabular}

\end{table}

\begin{figure}[htbp]
\centering
\includegraphics[width=\textwidth]{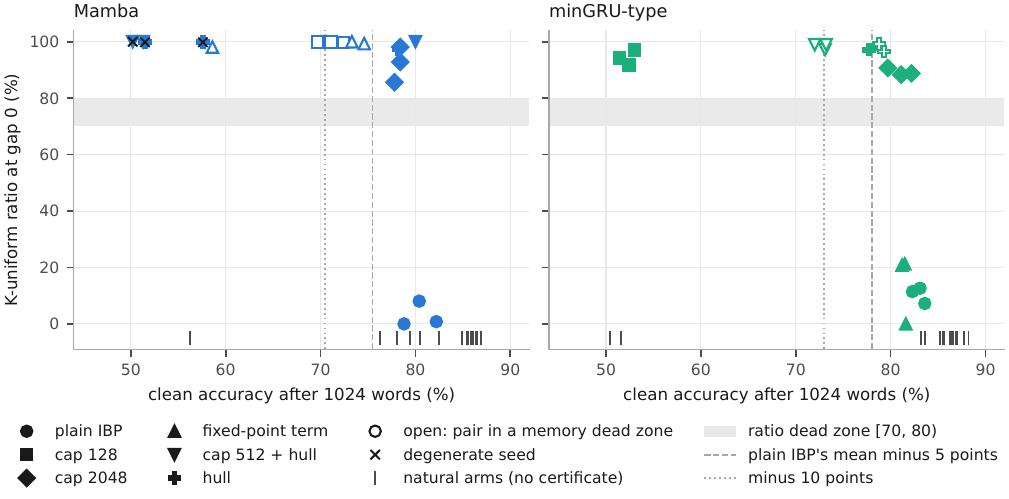}
\caption{The memory task: the trade-off the classification reads. K-uniform ratio at gap 0 (\%; float64, not re-verified) against the
clean accuracy after 1024 neutral words (\%; float64), one marker per IBP checkpoint (three seeds per arm; crosses: degenerate seeds),
one panel per architecture. Open markers: the checkpoint's pair with plain IBP of the same seed lies in a memory dead zone (a
difference in $(-10,-5)$ points after 1024 words, or a drop difference in $(5,10)$ points). The band is the ratio dead zone
$[0.7,0.8)$; the dashed and dotted rules stand 5 and 10 points below plain IBP's mean accuracy after 1024 words (the free and cost
bars); the ticks at the bottom mark the natural arms' accuracy after 1024 words (they have no certificate). Written by the table
script from the summary's scatter record.}
\label{fig:memory-scatter}
\end{figure}

\paragraph{Certified memory (reported, no prediction).}
For every number of copies of the preceding perturbed 1024-word period and every substitution in the review and in 1024 trailing
words, the K-uniform certificate after the 1024 words covers (measured; no prediction): Mamba, plain IBP 2.1\% [$-$5.8, 10.1] of the
documents (at least 2.0\% [$-$5.8, 9.9] of each class), cap 128 69.6\% [66.8, 72.4] (45.9\% [39.7, 52.2]), cap 2048 68.9\% [58.0, 79.7]
(58.2\% [44.2, 72.1]), fixed-point term 65.6\% [41.7, 89.5] (37.2\% [$-$3.1, 77.5]), cap 512 + hull 62.1\% (seeds 73.9 and 50.4;
worst class 36.1\%, seeds 61.2 and 11.0), hull n/a (one non-degenerate seed: 72.6\%, worst class 54.6\%); minGRU-type, plain IBP 8.8\% [3.1, 14.5] (7.4\% [2.7, 12.1]), cap 128
43.7\% [41.4, 45.9] (18.4\% [$-$0.1, 36.9]), cap 2048 66.0\% [60.2, 71.9] (55.1\% [43.5, 66.8]), fixed-point term 9.3\% [$-$11.2, 29.9]
(0.0\% [0.0, 0.0]), cap 512 + hull 67.3\% [64.2, 70.4] (36.5\% [30.3, 42.7]), hull 72.4\% [71.3, 73.6] (61.0\% [54.6, 67.3]). These
certificates cover the 1024 trailing words of the input and no longer trailing context (\cref{prop:M1}). \Cref{fig:memory-scatter}
plots the trade-off that the classification reads, per checkpoint.

\paragraph{Phase 2: one certificate for the trailing words as well.}
A single K-uniform box cannot cover trailing contexts of every length after the review and still certify reviews of both labels
(\cref{prop:M1}), but one certificate can cover every number of copies of a perturbed period before the review \emph{and} up to
$K'_{\max}$ copies of a perturbed period after it, by starting the trailing hull from the review's end box instead of from $0$
((M4) of \cref{rem:startboxM}, the ``sandwich''). Its evaluation ran after the main stage, with its own frozen manifest and no criterion: the
sandwich certificate $c_{\rm sand}(4)$ (every $K\ge0$ and $K=\infty$ copies of the preceding period, at most four copies of a 256-word
licence period after the review, the marker last) on the 36 IBP checkpoints and 1000 documents, next to the brute-force certificates
of its members ($K\in\{0,1\}$, $K'\in\{0,\dots,4\}$) and the clean accuracy after $K'=1$, 2, 4 and 16 copies. Two validity checks passed:
on every document every brute-force member box lies inside the sandwich box (36{,}000 documents, 10 members each, 0 violations at a
relative slack of $-10^{-9}$), and the reduction to $K'_{\max}=0$ certifies exactly the documents of the stage's K-uniform certificate
at gap 0 (0 documents differ on every checkpoint). The cost is independent of both numbers of copies: 63--66 s (minGRU-type) and
87--90 s (Mamba) per 1000 documents on one H200 at 25 documents per chunk, at a peak of 97.02 GB (Mamba). \Cref{tab:memory-phase2}
gives every arm. For every number of copies of a preceding period and up to four copies of a 256-word period after the review,
the cap-2048 models are certified on 53.6\% [31.3, 75.9] of the documents, at least 44.7\% [32.6, 56.7] of each class (Mamba), and on
49.1\% [45.4, 52.9], at least 40.7\% [20.3, 61.2] (minGRU-type), at a cost independent of both numbers (measured); the minGRU-type hull
models on 64.3\% [60.7, 67.9], at least 53.2\% [46.1, 60.3]. Against the K-uniform certificate after 1024 trailing words (68.9\%
[58.0, 79.7] and 66.0\% [60.2, 71.9] for the cap-2048 models, 72.4\% [71.3, 73.6] for the minGRU-type hull models) and the smallest
brute-force $K=1$ share over $K'\le4$ (72.6\% [70.0, 75.1], 71.2\% [66.0, 76.4] and 73.3\% [71.8, 74.8]), the hull over the trailing
copies costs these models about 15--19 and 9 points (a reading, not a criterion). The capped models with $\tau_{\max}\le512$ keep the
sandwich certificate in one class only (worst class: Mamba cap 128 0.2\% [$-$0.7, 1.1], minGRU-type cap 128 2.5\% [$-$0.4, 5.4], cap
512 + hull 1.7\% [0.8, 2.6], against K-uniform shares at gap 0 of 72.4, 67.6 and 74.1\%): a box that is uniform over $K'\le4$ must
contain the states after 0 and after 1024 trailing words, and after 1024 trailing words these models' clean carry has dropped
($K'=4$: cap 128 minGRU-type 0.3 and Mamba 20.0 points, minGRU-type cap 512 + hull 19.9, against 31.3 and 28.8 for plain IBP),
so the box covers
the prediction made with the review's evidence and the one made without it, and certifies only the class the latter agrees with.
Plain IBP's fixed-member brute-force certificates stand at 73--74\% (\cref{tab:memory-phase2}) while its K-uniform sandwich
certifies 1.6\% [$-$4.2, 7.4] (Mamba) and 6.4\% [2.5, 10.2] (minGRU-type): the sandwich is a K-uniform statement and inherits the
blind spot of plain IBP at gap 0 (above). After 16 trailing copies (4096 words, beyond the training range) plain IBP keeps 78--79\%
clean accuracy, the minGRU-type hull models 73\%, the cap-2048 models 62--66\% and the models with $\tau_{\max}\le512$ 52--53\%
(\cref{tab:memory-phase2}, with intervals). Nothing is claimed for trailing lengths beyond 1024 words or that are not multiples of
256.

\begin{table}[htbp]
\caption{The memory task, phase 2: the sandwich certificate (float64, not re-verified) per IBP arm. Upper table: the share of the
1000 documents certified for every number of copies of the perturbed 1024-word period before the review and at most four copies of
a perturbed 256-word period after it ($c_{\rm sand}(4)$), its worst class, the reduction to no trailing copy ($c_{\rm sand}(0)$, equal
to the K-uniform share at gap 0), the K-uniform share after 1024 trailing words and its worst class (main stage), and the smallest
brute-force share over $K'\le4$ at $K=1$. Lower table: clean accuracy (float64) after $K'$ trailing copies, the carry at $K'=4$ and
16, and the validity check of the reduction (the largest count difference between $c_{\rm sand}(0)$ and the K-uniform share at gap 0
over the seeds, and the documents on which they disagree, summed). Means over the non-degenerate seeds with 95\% t-intervals; an arm
with fewer than three non-degenerate seeds prints its seeds' values ($^\ast$: degenerate). Written by the table script from the
summary.}
\label{tab:memory-phase2}
\centering\scriptsize\setlength{\tabcolsep}{2.5pt}
\resizebox{\textwidth}{!}{%
\begin{tabular}{@{}llcccccc@{}}
\toprule
model & training & sandwich, $K'\le4$ & worst class & $K'=0$ ($=$ K-uniform, gap 0) & K-uniform, 1024 words & worst class, 1024 & brute force, $K=1$, $\min_{K'\le4}$\\
\midrule
Mamba & plain IBP & 1.6 [$-$4.2, 7.4] & 1.3 [$-$3.6, 6.3] & 2.2 [$-$6.1, 10.6] & 2.1 [$-$5.8, 10.1] & 2.0 [$-$5.8, 9.9] & 73.4 [72.5, 74.3]\\
 & cap 128 & 37.8 [34.4, 41.2] & 0.2 [$-$0.7, 1.1] & 72.4 [70.0, 74.7] & 69.6 [66.8, 72.4] & 45.9 [39.7, 52.2] & 69.9 [67.0, 72.7]\\
 & cap 2048 & 53.6 [31.3, 75.9] & 44.7 [32.6, 56.7] & 68.8 [55.7, 81.9] & 68.9 [58.0, 79.7] & 58.2 [44.2, 72.1] & 72.6 [70.0, 75.1]\\
 & fixed-point term & 52.2 [17.6, 86.9] & 26.4 [$-$18.4, 71.1] & 70.2 [54.0, 86.4] & 65.6 [41.7, 89.5] & 37.2 [$-$3.1, 77.5] & 65.7 [42.6, 88.9]\\
 & cap 512 + hull & 32.8 / 13.5$^\ast$ / 12.1 & 14.1 / 4.4$^\ast$ / 5.9 & 74.3 / 57.4$^\ast$ / 58.3 & 73.9 / 50.2$^\ast$ / 50.4 & 61.2 / 8.4$^\ast$ / 11.0 & 73.9 / 50.5$^\ast$ / 49.7\\
 & hull & 64.7 / 15.3$^\ast$ / 15.9$^\ast$ & 46.8 / 7.4$^\ast$ / 6.8$^\ast$ & 73.9 / 57.6$^\ast$ / 57.1$^\ast$ & 72.6 / 51.1$^\ast$ / 56.3$^\ast$ & 54.6 / 11.2$^\ast$ / 11.8$^\ast$ & 73.9 / 51.5$^\ast$ / 55.9$^\ast$\\
\midrule
minGRU-type & plain IBP & 6.4 [2.5, 10.2] & 4.4 [3.1, 5.6] & 8.0 [2.7, 13.3] & 8.8 [3.1, 14.5] & 7.4 [2.7, 12.1] & 73.6 [68.9, 78.3]\\
 & cap 128 & 14.2 [$-$3.1, 31.5] & 2.5 [$-$0.4, 5.4] & 67.6 [59.1, 76.1] & 43.7 [41.4, 45.9] & 18.4 [$-$0.1, 36.9] & 43.4 [40.9, 45.8]\\
 & cap 2048 & 49.1 [45.4, 52.9] & 40.7 [20.3, 61.2] & 68.1 [63.3, 72.9] & 66.0 [60.2, 71.9] & 55.1 [43.5, 66.8] & 71.2 [66.0, 76.4]\\
 & fixed-point term & 7.3 [$-$9.6, 24.1] & 0.0 [0.0, 0.0] & 10.3 [$-$11.8, 32.4] & 9.3 [$-$11.2, 29.9] & 0.0 [0.0, 0.0] & 70.2 [67.0, 73.4]\\
 & cap 512 + hull & 36.1 [33.7, 38.5] & 1.7 [0.8, 2.6] & 74.1 [73.0, 75.3] & 67.3 [64.2, 70.4] & 36.5 [30.3, 42.7] & 67.2 [64.6, 69.9]\\
 & hull & 64.3 [60.7, 67.9] & 53.2 [46.1, 60.3] & 73.0 [70.7, 75.3] & 72.4 [71.3, 73.6] & 61.0 [54.6, 67.3] & 73.3 [71.8, 74.8]\\
\bottomrule
\end{tabular}
}

\par\medskip
\resizebox{\textwidth}{!}{%
\begin{tabular}{@{}llccccccc@{}}
\toprule
model & training & $K'=1$ & $K'=2$ & $K'=4$ & $K'=16$ & carry, $K'=4$ & carry, $K'=16$ & V-C2: count / documents\\
\midrule
Mamba & plain IBP & 80.4 [77.2, 83.5] & 80.3 [77.0, 83.6] & 80.0 [76.8, 83.1] & 78.2 [75.3, 81.1] & 28.8 [26.0, 31.6] & 27.1 [20.9, 33.4] & 0 / 0\\
 & cap 128 & 75.2 [73.2, 77.2] & 74.2 [72.2, 76.3] & 71.2 [68.1, 74.3] & 52.7 [52.2, 53.3] & 20.0 [16.9, 23.1] & 0.0 [0.0, 0.0] & 0 / 0\\
 & cap 2048 & 79.2 [78.5, 79.9] & 79.0 [77.4, 80.6] & 78.8 [77.8, 79.8] & 66.4 [52.4, 80.5] & 27.5 [23.8, 31.2] & 15.6 [3.8, 27.3] & 0 / 0\\
 & fixed-point term & 74.1 [62.3, 85.9] & 72.4 [56.4, 88.4] & 68.9 [48.2, 89.7] & 53.6 [41.4, 65.8] & 18.6 [$-$0.1, 37.2] & 2.8 [$-$3.1, 8.6] & 0 / 0\\
 & cap 512 + hull & 80.7 / 50.6$^\ast$ / 51.8 & 81.3 / 50.6$^\ast$ / 51.7 & 80.3 / 50.6$^\ast$ / 51.7 & 52.6 / 47.5$^\ast$ / 51.8 & 29.1 / 2.9$^\ast$ / 4.0 & 0.0 / 0.0$^\ast$ / 4.2 & 0 / 0\\
 & hull & 79.2 / 51.9$^\ast$ / 56.6$^\ast$ & 79.2 / 51.9$^\ast$ / 56.6$^\ast$ & 78.4 / 51.9$^\ast$ / 56.6$^\ast$ & 69.8 / 51.9$^\ast$ / 56.5$^\ast$ & 25.4 / 4.1$^\ast$ / 5.3$^\ast$ & 16.6 / 4.1$^\ast$ / 5.2$^\ast$ & 0 / 0\\
\midrule
minGRU-type & plain IBP & 83.2 [81.1, 85.4] & 82.8 [79.8, 85.8] & 82.0 [79.5, 84.5] & 79.2 [75.8, 82.6] & 31.3 [29.1, 33.5] & 28.2 [25.2, 31.2] & 0 / 0\\
 & cap 128 & 74.0 [71.5, 76.5] & 62.0 [55.1, 69.0] & 52.2 [48.3, 56.1] & 52.0 [48.1, 55.9] & 0.3 [0.2, 0.5] & 0.0 [0.0, 0.0] & 0 / 0\\
 & cap 2048 & 81.8 [80.4, 83.2] & 81.5 [79.8, 83.1] & 80.5 [77.7, 83.3] & 62.2 [47.2, 77.2] & 29.7 [26.8, 32.6] & 11.8 [$-$2.6, 26.2] & 0 / 0\\
 & fixed-point term & 81.8 [81.2, 82.3] & 81.2 [79.4, 82.9] & 81.0 [80.4, 81.5] & 77.0 [69.0, 84.9] & 30.7 [29.5, 31.8] & 26.0 [16.1, 35.9] & 0 / 0\\
 & cap 512 + hull & 79.8 [77.7, 81.9] & 78.3 [77.1, 79.4] & 72.4 [71.1, 73.6] & 52.6 [52.6, 52.6] & 19.9 [18.4, 21.5] & 0.0 [0.0, 0.0] & 0 / 0\\
 & hull & 79.8 [79.3, 80.3] & 79.6 [78.7, 80.6] & 78.8 [78.5, 79.2] & 73.3 [65.2, 81.3] & 27.5 [26.6, 28.3] & 21.7 [13.6, 29.8] & 0 / 0\\
\bottomrule
\end{tabular}
}

\end{table}

\paragraph{What this study does not show.}
These are real-arithmetic certificates computed in float64 and not re-verified with outward rounding (\cref{rem:float}), for
two-layer width-64 models on one task, under the synonym threat at every position, with GNU-manual words in the training gaps and
the 256-word training periods, licence words in the test gaps and one 1024-word licence period before each test document, one draw
of training gaps shared by every seed (data seed 20261102) and three seeds per arm; the natural runs train for 4000 steps and the IBP
runs for 21{,}000. The IBP cap sweep is $\{128,2048,\text{none}\}$ plus the cap 512 with the hull term, while the natural sweep has
$\{64,128,512,2048,\text{none}\}$. The K-uniform ratio is read at gap 0; the memory requirement is bounded (1024 words), and the
certificates of \cref{tab:memory-phase2} cover trailing contexts of up to 1024 words in units of 256. No single certificate can cover
trailing contexts of every length for documents of different labels (\cref{prop:M1}: one box, real arithmetic, the nonnegative
diagonal class), and a single-box K-uniform certificate placed after the review certifies at most the accuracy of a review-blind
rule, which lies in $[42.2,57.8]\%$ with probability at least 0.99 under an i.i.d.\ model of the draw. The classification's label is
the same for both architectures, but its arm is the cap at twice the memory requirement, not a robust term that sees the unbounded
context: the fixed-point term keeps the ratio on Mamba but falls below 78\% clean accuracy and loses the certificates on the
minGRU-type model; the hull term keeps them on the minGRU-type model at a memory cost of 4--5 points (a dead zone) and is degenerate
in two Mamba seeds. Nothing is claimed beyond 1024 words of memory, for other tasks or larger models, and the K-uniform certificates at gap 0 do not
cover the gap that needs the memory; after 1024 trailing words they cover the substitutions in those words only (reported, no
prediction). Nothing here is a K-uniform statement about the fixed-point
term of \cref{app:remedy}, whose K-uniform reading was not pre-registered there.

\end{document}